\documentclass[a4paper,twoside]{article}

\usepackage{amssymb,amsmath}
\usepackage[dvips]{graphicx}

\numberwithin{equation}{section}
\allowdisplaybreaks[4]

\newcommand{\be}{\begin{equation}}
\newcommand{\ee}{\end{equation}}
\newcommand{\beq}{\begin{eqnarray}}
\newcommand{\eeq}{\end{eqnarray}}
\newcommand{\no}{\nonumber}
\newcommand{\bea}{\begin{array}}
\newcommand{\eea}{\end{array}}
\newcommand{\lb}{\label}
\newcommand{\mcal}{\mathcal}
\newcommand{\ve}{\varepsilon}
\newcommand{\ds}{\displaystyle}

\newcommand{\pp}{\partial}
\newcommand{\im}{\imath}
\newcommand{\ppr}{^{\prime}}

\newcommand{\wtilde}{\widetilde}
\newcommand{\ovv}{\overline}
\newcommand{\strab}{\raisebox{-4pt}{$\mbox{str}\atop {\scriptstyle \alpha,\beta}$}}
\newcommand{\STRAB}{\raisebox{-4pt}{$\mbox{STR}\atop {\scriptstyle a,\alpha;b,\beta}$}}
\newcommand{\ph}{\phantom}

\newcommand{\scz}{\scriptsize}
\newcommand{\FA}{\raisebox{0pt}{$\frac{\exp\big\{\overrightarrow{[\hat{Y},\ldots]_{-}}\big\}\;-
\;{\displaystyle\hat{1}}}
{\overrightarrow{[\hat{Y},\ldots]_{-}}}$}}

\newcommand{\fa}{\raisebox{0pt}{$\hat{F}_{1}\big(\overrightarrow{[\hat{Y},\ldots]_{-}}\big)$}}

\newcommand{\FAm}{\raisebox{0pt}{$\frac{\exp\big\{-\overrightarrow{[\hat{Y},\ldots]_{-}}\big\}\;-
\;{\displaystyle\hat{1}}}
{-\overrightarrow{[\hat{Y},\ldots]_{-}}}$}}

\newcommand{\famm}{\raisebox{0pt}{$\hat{F}_{1}\Big(-\overrightarrow{[\hat{Y},\ldots]_{-}}\Big)$}}

\begin{document}


\begin{center}
{\Large\bf Coherent state path integral and super-symmetry} \\
{\Large\bf for condensates composed of bosonic and fermionic atoms} \\
Bernhard Mieck\footnote{B. Mieck; {\sf E-mail:
"Bernhard.Mieck@t-online.de", phone: +49\,(0)203\,379\,4719, Fax: +49\,(0)203\,379\,4722}} \\
Department of Physics in Duisburg, University Duisburg-Essen, \\
Lotharstrasse 1, 47048 Duisburg, Germany \\
August 2007
\end{center}

\begin{abstract}
A super-symmetric coherent state path integral on the Keldysh time contour is considered for bosonic and
fermionic atoms which interact among each other with a common short-ranged two-body potential.
We investigate the symmetries of Bose-Einstein condensation for the equivalent
bosonic and fermionic constituents with the same interaction potential so that a super-symmetry results between
the bosonic and fermionic  components of super-fields. Apart from the super-unitary invariance \(U(L|S)\)
of the density terms, we specialize on the examination of super-symmetries for pair condensate terms.
Effective equations are derived for anomalous terms which are related
to the molecular- and BCS- condensate pairs. A Hubbard-Stratonovich transformation from 'Nambu'-doubled
super-fields leads to a generating function with super-matrices for the self-energy whose manifold is given by the
orthosympletic super-group \(Osp(S,S|2L)\). A nonlinear sigma model follows
from the spontaneous breaking of the ortho-symplectic super-group \(Osp(S,S|2L)\) to the coset decomposition
\(Osp(S,S|2L)\backslash U(L|S)\otimes U(L|S)\). The invariant subgroup \(U(L|S)\) for the vacuum or background
fields is represented by the density terms in the self-energy whereas the super-matrices on the coset space
\(Osp(S,S|2L)\backslash U(L|S)\) describe the anomalous molecular and BCS- pair condensate terms.
A change of integration measure is performed for the coset decomposition
\(Osp(S,S|2L)\backslash U(L|S)\otimes U(L|S)\) , including a separation of density and anomalous parts of the
self-energy with a gradient expansion for the Goldstone modes.
The independent anomalous fields in the actions can be transformed by the inverse square root
\(\hat{G}_{Osp\backslash U}^{-1/2}\) of the metric tensor of \(Osp(S,S|2L)\backslash U(L|S)\) so that
the non-Euclidean integration measure with super-Jacobi-determinant
\([\mbox{SDET}(\hat{G}_{Osp\backslash U})]^{1/2}\) can be removed from the coherent state path
integral and Gaussian-like integrations remain. The variations of the independent coset fields in the effective
actions result in classical field equations for a nonlinear sigma model with the anomalous terms.
The dynamics of the eigenvalues of the coset matrices is determined by Sine-Gordon equations which have a
similar meaning for the dynamics of the molecular- and BCS-pair condensates as the Gross-Pitaevskii equation
for the coherent wave function in BEC phenomena.
\end{abstract}

\vspace*{0.46cm}

\tableofcontents

\section{Introduction} \lb{s1}

\subsection{Gross-Pitaevskii equation with bosonic and fermionic fields} \lb{s11}

Phenomena of Bose-Einstein condensation (BEC) with bosonic or fermionic constituents have been realized
under various conditions \cite{mosk}-\cite{peth}. The tuning via Feshbach resonances from repulsive to attractive
interactions allows to observe effects as bright and dark solitons in the one-dimensional case and even
metastable configurations of a BEC with pure attractive bosons in three dimensions \cite{hulet3}-\cite{hulet5}.
In many cases the Gross-Pitaevskii (GP) equation is sufficient to describe these various phenomena.
In this paper we concentrate on the investigation of symmetries for BE-condensates which are composed of
bosonic and fermionic atoms with the same interaction \(V_{|\vec{x}-\vec{x}\ppr|}\) among each other. A
resulting super-symmetry is assumed between the bosonic and fermionic constituents, having the identical
trap potential $u(\vec{x})$ and the equivalent kinetic energies of similar masses. The GP-equation has
to be extended to a super-symmetric equation with complex commuting order parameters $b_{\vec{x},m}(t)$
for the bosons and anti-commuting variables $\alpha_{\vec{x},r}(t)$ for the fermions \cite{cau1}.
Apart from the super-unitary invariance of the GP-equation with super-fields, we explicitly
concentrate in this paper on the examination of super-symmetries for pair condensates whose interactions
contain effective transport coefficients determined and averaged by density terms as background fields.
In the main part spontaneous symmetry breaking (SSB) in a coherent state path integral
is analysed for the transition to bosonic molecular condensates and BCS-terms as the
corresponding Goldstone modes in a super-symmetric nonlinear sigma model
\cite{gold,nambu},\cite{hulet1}-\cite{bm1}. The common interaction potential
$V_{|\vec{x}\ppr-\vec{x}|}$ among all the bosonic and fermionic constituents is supposed to be very short-ranged in
comparison to other length scales as the range of the trap potential $u(\vec{x})$. The precise form of
$V_{|\vec{x}\ppr-\vec{x}|}$ is not specified because we only use its short-ranged character in derivations for
effective equations of anomalous terms from a coherent state path integral on the nonequilibrium Keldysh-contour. The
repulsive and attractive case of $V_{|\vec{x}\ppr-\vec{x}|}$ is studied with the \(L=2l+1\) bosonic
(\(l=0,1,2,\ldots\)) and \(S=2s+1,\;s=\frac{1}{2},\frac{3}{2},\frac{5}{2},\ldots\) fermionic angular momentum degrees
of freedom. The super-symmetric generalization of the GP-equation with super-symmetric vector field
\(\psi_{\vec{x},\alpha}(t)=(\vec{b}_{\vec{x}}(t)\;,\; \vec{\alpha}_{\vec{x}}(t))^{T}\) takes the form
\be \lb{s1_1}
\im\hbar\frac{\pp\psi_{\vec{x},\alpha}(t)}{\pp t}=-\frac{\hbar^{2}}{2m}
\vec{\nabla}^{2}\psi_{\vec{x},\alpha}(t)+u(\vec{x})\;
\psi_{\vec{x},\alpha}(t)+2\sum_{\vec{x}\ppr,\beta}\psi_{\vec{x}\ppr,\beta}^{*}(t)\;
\psi_{\vec{x}\ppr,\beta}(t)\;V_{|\vec{x}\ppr-\vec{x}|}\; \psi_{\vec{x},\alpha}(t)\;\;,
\ee
where the indices $\alpha$, $\beta$ have
values from $-l$ to $+l$ for the bosons and additionally from $-s$ to $+s$ for the fermions so that
$\psi_{\vec{x},\alpha}(t)$ is in total a \(L+S=N\) component supervector (\(L=\mbox{odd integer number}\),
\(S=\mbox{even integer number}\)) \footnote{The spatial sum $\sum_{\vec{x}}\ldots$ is dimensionless and is scaled
with the system volume so that $\sum_{\vec{x}}\ldots$ is equivalent to \(\int_{L^{d}}d^{d}x/L^{d}\).}.
The angular momentum degrees of freedom are
combined in the vectors $\vec{b}_{\vec{x}}(t)$ and anti-commuting fields $\vec{\alpha}_{\vec{x}}(t)$ of bosons
and fermions, respectively. In the following we list the precise form of the super-symmetric vector
$\psi_{\vec{x},\alpha}(t)$ and its hermitian conjugate with index $\alpha$
\beq \lb{s1_2}
\alpha&=&\underbrace{-l,\ldots,+l}_{\mbox{L bosons}}\;;\;
\underbrace{-s,\ldots,+s}_{\mbox{S fermions}} \\ \no N&=&L+S \hspace*{1.0cm}L=2l+1\hspace*{1.0cm}S=2s+1 \\ \no
\psi_{\vec{x},\alpha}(t)&=&\left( \bea{c}
\vec{b}_{\vec{x}}(t) \\
\vec{\alpha}_{\vec{x}}(t) \eea\right)\hspace*{0.5cm} \bea{rcl}
\vec{b}_{\vec{x}}(t)&=&\big\{b_{\vec{x},m}(t)\big\}=\Big(b_{\vec{x},-l}(t)\;,\;
\ldots\;,\;b_{\vec{x},+l}(t)\Big) \\
\vec{\alpha}_{\vec{x}}(t)&=&\big\{\alpha_{\vec{x},r}(t)\big\}=\Big(\alpha_{\vec{x},-s}(t)\;,\;
\ldots\;,\;\alpha_{\vec{x},+s}(t)\Big)\eea \\ \no
\psi_{\vec{x},\alpha}^{+}(t)&=&\Big(b_{\vec{x},-l}^{*}(t)\;,\;\ldots\;,\;
b_{\vec{x},+l}^{*}(t)\;;\;\alpha_{\vec{x},-s}^{*}(t)\;,\; \ldots\;,\;\alpha_{\vec{x},+s}^{*}(t)\Big)\;.
\eeq
The GP-equation (\ref{s1_1}) is invariant under global super-symmetric unitary transformations
$U(L|S)$ with matrix $\hat{U}_{\beta\alpha}$ (\ref{s1_3})
which leaves the norm of the vector $\psi_{\vec{x},\alpha}(t)$ invariant
so that one can transform from the bosonic amplitudes
$\vec{b}_{\vec{x}}(t)$ to the fermionic ones $\vec{\alpha}_{\vec{x}}(t)$ and vice versa
\footnote{The standard convention for summing over repeated indices is applied throughout the paper, unless
stated otherwise explicitly.}
\beq \lb{s1_3}
\psi_{\vec{x},\alpha}(t)\rightarrow\psi_{\vec{x},\beta}\ppr(t)=\hat{U}_{\beta\alpha}\;\;
\psi_{\vec{x},\alpha}(t)\;;\hspace*{0.5cm}\hat{U}^{+}\hat{U}=1\;;\hspace*{0.37cm} \Longrightarrow
\psi_{\vec{x},\alpha}^{+}(t)\;\psi_{\vec{x},\alpha}(t)=
\psi_{\vec{x},\beta}^{\prime+}(t)\;\psi_{\vec{x},\beta}\ppr(t)\;\;\;.
\eeq
Bright and dark soliton solutions \(\psi_{x,\alpha}^{(g<0)}(t)\), \(\psi_{x,\alpha}^{(g>0)}(t)\)
for an attractive and repulsive delta-function potential
\(V_{|\vec{x}\ppr-\vec{x}|}\)\(=g\;\delta_{\vec{x},\vec{x}\ppr}\) (with \(u(\vec{x})=\mbox{const.}\))
can be transferred from the complex commuting case
to the super-symmetric fields (\ref{s1_6},\ref{s1_7}) in one spatial dimension (\ref{s1_4})
\cite{rein1}-\cite{rein4},\cite{cau1,cau2}.
The constant density $n_{0}$ (\ref{s1_5}) follows from the constant commuting and
anti-commuting vector fields $\vec{b}$ and $\vec{\alpha}$.
The other parameters are the velocity $v$ and the derived
quantities as the length $l_{0}$, sound velocity $c_{s}$ and scale factors \(\beta_{v}=v/c_{s}\),
\(\gamma_{v}=\sqrt{1-(\beta_{v})^{2}}\)
\beq \lb{s1_4}
\im\hbar\frac{\pp\psi_{x,\alpha}(t)}{\pp t}&=&-\frac{\hbar^{2}}{2m}\frac{\pp^{2}\psi_{x,\alpha}(t)}{\pp x^{2}}
+2\;g\sum_{\beta=1}^{N=L+S}\psi_{x,\beta}^{*}(t)\;\psi_{x,\beta}(t)\;\;\psi_{x,\alpha}(t) \\ \lb{s1_5}
\psi_{\alpha}^{(0)}&=&\Big(\vec{b}\;,\;\vec{\alpha}\Big)\hspace*{1.0cm}
l_{0}=\frac{\hbar}{\sqrt{2\;|g|\;n_{0}\;m}} \\ \no
n_{0}&=&\sum_{\alpha=1}^{N=L+S}\psi_{\alpha}^{(0)*}\;\psi_{\alpha}^{(0)}=
\vec{b}^{*}\cdot\vec{b}+\vec{\alpha}^{*}\cdot\vec{\alpha}=n_{B;0}-n_{F;0} \\ \lb{s1_6}
\psi_{x,\alpha}^{(g<0)}(t)&=&
\frac{\exp\Big\{\frac{\im}{\hbar}\Big[m\;v\;x-\Big(\frac{m\;v^{2}}{2}-|g|\;n_{0}\Big)\;t\Big]\Big\}}
{\cosh\Big\{(x-v\;t)/l_{0}\Big\}}\;\;\psi_{\alpha}^{(0)} \;\;\; ; \hspace*{1.0cm}(g<0) \\ \lb{s1_7}
\psi_{x,\alpha}^{(g>0)}(t)&=&\bigg\{\im\;\beta_{v}+\gamma_{v}\;\tanh\bigg(\gamma_{v}\frac{(x-v\;t)}{l_{0}}
\bigg)\bigg\}\;\;\exp\Big\{-\frac{\im}{\hbar}\;2\;g\;n_{0}\;t\Big\}\;\psi_{\alpha}^{(0)} \;\;\; ;
\hspace*{1.0cm} (g>0) \\ \no
&& c_{s}=\sqrt{2\;g\;n_{0}/m}\;\;;\;\;\beta_{v}=\frac{v}{c_{s}}\;\;;\;\;
\gamma_{v}=\sqrt{1-\beta_{v}^{2}}\;\;.
\eeq
The invariant super-unitary \(U(L|S)\) density
\(n_{0;\vec{x}}(t)=\psi_{\vec{x},\alpha}^{*}(t)\;\psi_{\vec{x},\alpha}(t)\)
does not correspond to the sum of measured physical bosonic and fermionic densities
because the expectation values for fermionic bilinear forms averaged over a super-symmetric generating
function always involve an additional minus sign in comparison to expectation values of bosonic densities.
Therefore, the total density \(n_{0;\vec{x}}(t)=\psi_{\vec{x},\alpha}^{*}(t)\;\psi_{\vec{x},\alpha}(t)\)
of super-fields has to be regarded as the difference \(n_{0;\vec{x}}(t)=n_{B;\vec{x}}(t)-n_{F;\vec{x}}(t)\)
of the observable bosonic \(n_{B;\vec{x}}(t)=\vec{b}_{\vec{x}}^{*}(t)\cdot\vec{b}_{\vec{x}}(t)\) and
fermionic \(n_{F;\vec{x}}(t)=-\vec{\alpha}_{\vec{x}}^{*}(t)\cdot\vec{\alpha}_{\vec{x}}(t)\) density components.

However, the final derivations in this paper go beyond the super-symmetric unitary invariance
$U(L|S)$ (\ref{s1_3}) of the $L+S=N$ components
of \(\psi_{\vec{x},\alpha}(t)\) in the GP-equation (\ref{s1_1}).
In sections \ref{s2} to \ref{s5} we continue to derive and to examine
effective equations for anomalous terms
\(\langle\psi_{\vec{x},\alpha}(t)\;\psi_{\vec{x},\beta}(t)\rangle\) which are
related to the BCS- and molecular condensate  \cite{lipp}-\cite{dick}.
These equations follow from symmetry considerations
of a spontaneous breaking of an ortho-symplectic
super-group $Osp(S,S|2L)$ with the direct sum \(SO(S,S)\oplus Sp(2L)\) for the fermion-fermion and boson-boson
blocks in the even parts, respectively
\footnote{The ortho-symplectic super-group with orthogonal symmetry \(SO(S,S)\) in the fermion-fermion block
of super-matrices and symplectic symmetry \(Sp(2L)\) in the boson-boson part is usually denoted by the
symbol \(Osp(S,S|2L)\) with the first two letters '\(S,S\)' referring to the letter '\(O\)' in '\(Osp\)'
for orthogonal and '\(2L\)' related to the letters '\(sp\)' for symplectic in '\(Osp\)'. However,
the notation \(Spo(2L|S,S)\), as symplecto-ortho super-group, is also in common use, with the symplectic
symmetry \(Sp(2L)\) for the bosons and orthogonal symmetry \(SO(S,S)\) for the even fermion-fermion
section of super-matrices. In this paper we have to consider the ortho-symplectic super-groups
\(Osp(S,S|2L)\) and \(Osp(L,L|2S)\) which follow by the exchange of the orthogonal with the
symplectic symmetry in the even parts of super-matrices, e.g. \(SO(S,S)\oplus Sp(2L)\) for
\(Osp(S,S|2L)\) and \(SO(L,L)\oplus Sp(2S)\) for \(Osp(L,L|2S)\). The numbers \(L=2l+1\) and
\(S=2s+1\) of the angular momentum degrees of freedom are retained for specifying the dimension
of the prevailing group for the boson-boson and fermion-fermion parts, respectively.}
\cite{cor}-\cite{luc}. Separating the self-energy matrix into density and anomalous terms,
one can extract the Goldstone bosons and
fermions with a splitting of the ortho-symplectic super-group $Osp(S,S|2L)$.
The unitary super-group $U(L|S)$ for the density part is regarded as the invariant
subgroup for the vacuum or ground state in a SSB whereas the coset space
$Osp(S,S|2L)\backslash U(L|S)$ of the Goldstone modes is assigned to BCS- and molecular
condensate terms. The various steps for obtaining these effective equations on a coset space
$Osp(S,S|2L)\backslash U(L|S)$ are briefly summarized in the following subsection.

\subsection{Overview of the derivation of the nonlinear sigma model for the pair
condensate terms} \lb{s12}

We list the various steps to the effective equations for the anomalous terms with
reference to the corresponding sections :

\begin{itemize}
\item Supersymmetric coherent state path integral on Keldysh time contour (subsection \ref{s21})
\item Transformation of the quartic interaction to densities, anomalous terms and
the order parameter (subsection \ref{s22})
\item 'Nambu'-doubling of super-fields for anomalous terms; Hubbard-Stratonovich
transformation (HST) \cite{st,neg} from the dyadic
product of super-fields to a super-symmetric matrix for the self-energy;
integration over the remaining bilinear super-fields results in the superdeterminant composed
of the self-energy and doubled one-particle Hamiltonian (subsection \ref{s22})
\item Symmetry considerations of the actions in the coherent state path integral
and its invariance under the ortho-symplectic super-group
\(Osp(L,L|2S)\) with the direct sum of \(SO(L,L)\oplus Sp(2S)\) and its subgroup \(U(L|S)\) for the densities;
the invariant transformations \(Osp(L,L|2S)\) of the generating function differ by the number of independent
parameters from the independent field degrees of freedom in the self-energy matrix (subsection \ref{s31})
\item Comparison of the self-energy as a generator of the super-group \(Osp(S,S|2L)\) in contrast to the
invariant ortho-symplectic transformations \(Osp(L,L|2S)\) of the coherent state path integral (subsection \ref{s31});
determination of the number of independent fields in the self-energy matrix or
corresponding generator of \(Osp(S,S|2L)\) (subsection \ref{s31});
derivation of a generating function with a modified HST and anti-hermitian
anomalous parts for the self-energy so that the invariant transformations are equivalent to the group manifold
\(Osp(S,S|2L)\) of the self-energy (section \ref{s32}); assignment of the independent
parameters to the block-diagonal subgroup \(U(L|S)\) for the densities and to the
coset space \(Osp(S,S|2L)\backslash U(L|S)\) for anti-hermitian anomalous terms
including the change of integration measure (subsections \ref{s33}, and appendix \ref{sa1}-\ref{sa4})
\item Decomposition of the coherent state path integral into block-diagonal invariant
density matrices and anomalous terms including a gradient expansion
(subsections \ref{s41}-\ref{s46} and appendix \ref{sb}); elimination of the 'hinge' functions as
block diagonal subgroup \(U(L|S)\) which can be
regarded as 'massive modes' in the SSB (subsection \ref{s41}); determination of the Hilbert space
for the gradient operators acting on dyadic products of 'Nambu'-doubled super-fields (subsection \ref{s42});
rules of propagation for super-fields \(\Psi_{\vec{x},\alpha}^{a}(t_{p})\) and pair condensates
\(\delta\hat{\Sigma}_{\alpha\beta}^{12}(\vec{x},t_{p})\) within the self-energy density
\(\sigma_{D}^{(0)}(\vec{x},t_{p})\) as background field (subsections \ref{s43}, \ref{s44});
gradient expansion of the actions from the super-determinant and the BEC wave function
terms with corresponding transport coefficients averaged over the background field
\(\sigma_{D}^{(0)}(\vec{x},t_{p})\) as environment (subsections \ref{s45}, \ref{s46} and appendix \ref{sb})
\item Description and classification of dominant terms for the effective actions of
the pair condensates arising from the gradient expansion (subsection \ref{s51});
derivation of effective classical, mean field equations for pair condensates as
independent field degrees of freedom in the coset super-generators of
\(Osp(S,S|2L)\backslash U(L|S)\) (subsection \ref{s52}); relations for variations with the
coset generators of the independent anomalous terms within the exponentials
of corresponding Lie super-group elements (subsection \ref{s52});
explanation of the adjoint representation of a super-algebra from a Jacobi-identity of super-matrices
and from its structure constants with regard to simplifying
the classical nonlinear sigma equations (subsection \ref{s53})
\item Geometric meaning of the stationary nonlinear sigma actions
in d-dimensional coordinate space with transport coefficients
\(c^{ij}(\vec{x},t_{p})\), \(d^{ij}(\vec{x},t_{p})\) and the metric tensors
\(\hat{G}_{Osp\backslash U}\), \(\hat{G}_{U(L|S)}\) of the coset and subgroup manifolds
(subsection \ref{s61})
\end{itemize}

The final effective coherent state path integral for the anomalous terms is briefly outlined in
Eqs. (\ref{s1_8}-\ref{s1_16}) in anticipation of results in section \ref{s5}, leaving out details
of the derivations for the nonlinear sigma actions and the integration measure. Apart from the action
\(\mcal{A}_{\hat{J}_{\psi\psi}}\big[\hat{T}\big]\) as 'seed' for the pair condensates, the effective
generating function \(Z[\hat{\mcal{J}},J_{\psi},\im \hat{J}_{\psi\psi}]\) (\ref{s1_8})
contains 'Nambu'-doubled source fields \(J_{\psi;\alpha}^{a}(\vec{x},t_{p})\) for coherent BEC
wave functions and a source term \(\hat{\mcal{J}}_{\vec{x},\alpha;\vec{x}\ppr,\beta}^{ab}(t_{p},t_{q}\ppr)\)
within the action \(\mcal{A}\ppr\big[\hat{T};\hat{\mcal{J}}\big]\).
The derivative of (\ref{s1_8}) with respect to
\(\hat{\mcal{J}}_{\vec{x},\alpha;\vec{x}\ppr,\beta}^{ab}(t_{p},t_{q}\ppr)\) yields the
corresponding observable. We can classify the effective actions for pair condensates according to
the parameter \(\mcal{N}=\hbar\;(1/\Delta t)\cdot(L/\Delta x)^{d}\) which follows from the number
of considered discrete intervals or points of the underlying grid of coordinates. This parameter
inevitably appears as the arising super-determinant of the self-energy is transformed to the exponential of a
'Trace'-'Super-trace'-'logarithm' expression where spatial and contour time integrations have to be
introduced for defining appropriate actions in the expansion of gradient operators from the super-determinant
\beq \lb{s1_8}
Z[\hat{\mcal{J}},J_{\psi},\im \hat{J}_{\psi\psi}]&=&
\int d\big[\hat{T}^{-1}(\vec{x},t_{p})\;d\hat{T}(\vec{x},t_{p})\big]\;\;
\exp\Big\{\im\;\mcal{A}_{\hat{J}_{\psi\psi}}\big[\hat{T}\big]\Big\}  \\ \no &\times&
\exp\Big\{-\mcal{A}_{\mcal{N}^{-1}}\ppr\big[\hat{T};J_{\psi}\big]-
\mcal{A}_{\mcal{N}^{0}}\ppr\big[\hat{T};J_{\psi}\big]-
\mcal{A}_{\mcal{N}^{+1}}\ppr\big[\hat{T}\big]\Big\} \;\times\;
\exp\Big\{-\mcal{A}\ppr\big[\hat{T};\hat{\mcal{J}}\big]\Big\}_{\mbox{.}}
\eeq
The effective actions \(\mcal{A}_{\mcal{N}^{-1}}\ppr\big[\hat{T};J_{\psi}\big]\),
\(\mcal{A}_{\mcal{N}^{0}}\ppr\big[\hat{T};J_{\psi}\big]\),
\(\mcal{A}_{\mcal{N}^{+1}}\ppr\big[\hat{T}\big]\) of (\ref{s1_8})
are itemized in relations (\ref{s1_9}) to (\ref{s1_16})
with the dependence on \(\hat{T}(\vec{x},t_{p})=\exp\{-\hat{Y}(\vec{x},t_{p})\}\) and the matrix
\(\hat{Z}(\vec{x},t_{p})=\hat{T}(\vec{x},t_{p})\;\hat{S}\;\hat{T}^{-1}(\vec{x},t_{p})\) (\ref{s1_10})
for the coset decomposition \(Osp(S,S|2L)\backslash U(L|S)\). The super-matrix
\(\hat{T}(\vec{x},t_{p})=\exp\{-\hat{Y}(\vec{x},t_{p})\}\) on the coset space
\(Osp(S,S|2L)\backslash U(L|S)\) is composed of the independent molecular and BCS condensate
degrees of freedom within the generator \(\hat{Y}(\vec{x},t_{p})\). The transport coefficients
\(c^{ij}(\vec{x},t_{p})\), \(d^{ij}(\vec{x},t_{p})\) (\ref{s1_11}-\ref{s1_14})
(\(i,j=1,\ldots,d\); d-dimensional coordinate space)
follow from the average with a coherent state path integral composed of the scalar self-energy
\(\sigma_{D}^{(0)}(\vec{x},t_{p})\) as background field. The actions
\(\mcal{A}_{\mcal{N}^{-1}}\ppr\big[\hat{T};J_{\psi}\big]\),
\(\mcal{A}_{\mcal{N}^{0}}\ppr\big[\hat{T};J_{\psi}\big]\),
\(\mcal{A}_{\mcal{N}^{+1}}\ppr\big[\hat{T}\big]\) (\ref{s1_9},\ref{s1_15},\ref{s1_16})
comprise all the spatial gradients \(\wtilde{\pp}_{i}=\frac{\hbar}{\sqrt{2m}}\;\frac{\pp}{\pp x^{i}}\)
up to second order and time-derivatives \(\hat{E}_{p}=\im\hbar\frac{\pp}{\pp t_{p}}\) up to first order
which emerge from the gradient expansion of the super-determinant and BEC wave function terms
\beq \lb{s1_9}
\lefteqn{
\mcal{A}_{\mcal{N}^{-1}}\ppr\big[\hat{T};J_{\psi}\big]= \frac{1}{4}\frac{1}{\mcal{N}}\int_{C}\frac{d
t_{p}}{\hbar}\sum_{\vec{x}} c^{ij}(\vec{x},t_{p})\;\;
\mbox{STR}\Big[\Big(\wtilde{\pp}_{i}\hat{Z}(\vec{x},t_{p})\Big)\;
\Big(\wtilde{\pp}_{j}\hat{Z}(\vec{x},t_{p})\Big)\Big] - \frac{\im}{\mcal{N}} \int_{C}\frac{d
t_{p}}{\hbar}\sum_{\vec{x}}\sum_{a,b=1,2} \times }   \\ \no &\times&
\sum_{\alpha,\beta=1}^{N=L+S} d^{ij}(\vec{x},t_{p})\;
\frac{J_{\psi;\beta}^{+,b}(\vec{x},t_{p})}{\mcal{N}}\bigg(\hat{I}\;\wtilde{K}\;
\Big(\wtilde{\pp}_{i}\hat{T}(\vec{x},t_{p})\Big)\;\hat{T}^{-1}(\vec{x},t_{p})\;
\Big(\wtilde{\pp}_{j}\hat{T}(\vec{x},t_{p})\Big)\;\hat{T}^{-1}(\vec{x},t_{p})\;
\hat{I}\bigg)_{\beta\alpha}^{ba}\;\frac{J_{\psi;\alpha}^{a}(\vec{x},t_{p})}{\mcal{N}}
\eeq
\beq \lb{s1_10}
\hat{Z}(\vec{x},t_{p})&=&\hat{T}(\vec{x},t_{p})\;\hat{S}\;\hat{T}^{-1}(\vec{x},t_{p}) \;\;\;;
\hspace*{0.5cm}\breve{v}(\vec{x},t_{p})=\breve{u}(\vec{x})+\breve{\sigma}_{D}^{(0)}(\vec{x},t_{p})=
\frac{u(\vec{x})+\sigma_{D}^{(0)}(\vec{x},t_{p})}{\mcal{N}}
\\ \lb{s1_11}
c^{ij}(\vec{x},t_{p})&=&c^{(1),ij}(\vec{x},t_{p})+c^{(2),ij}(\vec{x},t_{p}) \\  \lb{s1_12}
c^{(1),ij}(\vec{x},t_{p})&=&
-2\bigg\langle\Big(\wtilde{\pp}_{i}\wtilde{\pp}_{j}\breve{v}(\vec{x},t_{p})\Big)
\bigg\rangle_{\hat{\sigma}_{D}^{(0)}} - \delta_{ij}\;\sum_{k=1}^{d}
\bigg\langle\Big(\wtilde{\pp}_{k}\wtilde{\pp}_{k}\breve{v}(\vec{x},t_{p})\Big)
\bigg\rangle_{\hat{\sigma}_{D}^{(0)}}  \\ \lb{s1_13}
c^{(2),ij}(\vec{x},t_{p})&=& 2\bigg\langle\Big(\wtilde{\pp}_{i}\breve{v}(\vec{x},t_{p})\Big)\;
\Big(\wtilde{\pp}_{j}\breve{v}(\vec{x},t_{p})\Big)\bigg\rangle_{\hat{\sigma}_{D}^{(0)}} -
\delta_{ij}\;\sum_{k=1}^{d} \bigg\langle\Big(\wtilde{\pp}_{k}\breve{v}(\vec{x},t_{p})\Big)^{2}
\bigg\rangle_{\hat{\sigma}_{D}^{(0)}}      \\    \lb{s1_14}
d^{ij}(\vec{x},t_{p}) &=& 2\;\bigg\langle 3\;\Big(\wtilde{\pp}_{i}\breve{v}(\vec{x},t_{p})\Big)\;
\Big(\wtilde{\pp}_{j}\breve{v}(\vec{x},t_{p})\Big) -
\Big(\wtilde{\pp}_{i}\wtilde{\pp}_{j}\breve{v}(\vec{x},t_{p})\Big)
\bigg\rangle_{\hat{\sigma}_{D}^{(0)}}
\eeq
\beq \no
\mcal{A}_{\mcal{N}^{0}}\ppr\big[\hat{T};J_{\psi}\big]
&=& -\frac{1}{2}\int_{C}\frac{d t_{p}}{\hbar}\sum_{\vec{x}}\Bigg\{
\mbox{STR}\bigg[\hat{T}^{-1}(\vec{x},t_{p})\;\hat{S}\;\Big(\hat{E}_{p}\hat{T}(\vec{x},t_{p})\Big)+
\hat{T}^{-1}(\vec{x},t_{p})\;\Big(\wtilde{\pp}_{i}\wtilde{\pp}_{i}\hat{T}(\vec{x},t_{p})\Big)\bigg]+ \\ \lb{s1_15} &+&
\Big(u(\vec{x})-\mu_{0}-\im\;\ve_{p}+\big\langle\sigma_{D}^{(0)}(\vec{x},t_{p})\big\rangle_{\hat{\sigma}_{D}^{(0)}}
\Big)\;\mbox{STR}\bigg[\Big(\hat{T}^{-1}(\vec{x},t_{p})\Big)^{2}-\hat{1}_{2N\times 2N}\bigg]\Bigg\} + \\ \no &-&
\frac{\im}{2}\int_{C}\frac{d t_{p}}{\hbar}\sum_{\vec{x}}\sum_{a,b=1,2}\sum_{\alpha,\beta=1}^{N=L+S}
\frac{J_{\psi;\beta}^{+,b}(\vec{x},t_{p})}{\mcal{N}}\bigg[\hat{I}\;\wtilde{K}\;
\bigg(\Big(\wtilde{\pp}_{i}\wtilde{\pp}_{i}\hat{T}(\vec{x},t_{p})\Big)\;\hat{T}^{-1}(\vec{x},t_{p})+ \\ \no &+&
\hat{T}(\vec{x},t_{p})\;\hat{S}\;\hat{T}^{-1}(\vec{x},t_{p})\;\Big(\hat{E}_{p}\hat{T}(\vec{x},t_{p})\Big)\;
\hat{T}^{-1}(\vec{x},t_{p}) + \\ \no &-&2\;
\Big(\wtilde{\pp}_{i}\hat{T}(\vec{x},t_{p})\Big)\;\hat{T}^{-1}(\vec{x},t_{p})\;
\Big(\wtilde{\pp}_{i}\hat{T}(\vec{x},t_{p})\Big)\;\hat{T}^{-1}(\vec{x},t_{p})\bigg)
\hat{I}\bigg]_{\beta\alpha}^{ba}\;\frac{J_{\psi;\alpha}^{a}(\vec{x},t_{p})}{\mcal{N}}    \\ \lb{s1_16}
\mcal{A}_{\mcal{N}^{+1}}\ppr\big[\hat{T}\big]&=& \frac{\mcal{N}}{2}\int_{C}\frac{d
t_{p}}{\hbar}\eta_{p}\sum_{\vec{x}} \mbox{STR}\bigg[\Big(\hat{T}^{-1}(\vec{x},t_{p})\Big)^{2}-
\hat{1}_{2N\times 2N}\bigg]\;\;\;.
\eeq
The coefficients \(c^{ij}(\vec{x},t_{p})\), \(d^{ij}(\vec{x},t_{p})\)  (\ref{s1_11}-\ref{s1_14})
are determined by the dimensionless, scaled trap potential \(\breve{u}(\vec{x})=u(\vec{x})/\mcal{N}\) and
scalar self-energy \(\breve{\sigma}_{D}^{(0)}(\vec{x},t_{p})=
\sigma_{D}^{(0)}(\vec{x},t_{p})/\mcal{N}\) whereas the derivatives or spatial gradients of matrices
retain a dimension as square root of the kinetic energy. Instead of an averaging procedure with
\(\sigma_{D}^{(0)}(\vec{x},t_{p})\) as background field, a saddle point equation can approximate
this scalar self-energy and the corresponding coefficients
\(c^{ij}(\vec{x},t_{p})\), \(d^{ij}(\vec{x},t_{p})\)  (\ref{s1_11}-\ref{s1_14}).
Since the eigenvalues of the effective matrix \(\hat{T}(\vec{x},t_{p})\) for the
Goldstone modes are given in terms of trigonometric functions for compact degrees of freedom
(bosons) and the hyperbolic sine-, cosine-functions  for the noncompact components (fermions),
the effective classical equations on the coset space are related to the Sine-Gordon equations.
These equations specify the dynamics of the anomalous terms in analogous manner
as the GP-equation describes the coherent condensate wave function in BEC phenomena.

There is one relation (\ref{s1_23},\ref{s3_24}) of central importance
throughout this paper which allows to perform variations of exponentials of operators or matrices, as e.g.
\(\hat{T}(\vec{x},t_{p})=\exp\{-\hat{Y}(\vec{x},t_{p})\}\) with respect to
the generator \(\hat{Y}(\vec{x},t_{p})\). This fundamental relation (\ref{s1_23}) is needed for
calculating the change of integration measure in a coset decomposition as well as for computing
the classical nonlinear sigma equations from variations of the actions for the anomalous terms
(compare reference \cite{eng} with other derivations for varying exponential as well as logarithmic or
inverse functions of matrices and operators). The variation as \(\delta\hat{A}(v\;\hat{B})\) (\ref{s1_17})
for an exponential of a matrix \(\hat{B}\) follows from differentiation with respect to a real
parameter $v$ which is finally integrated in the interval \(v\in[0,1]\) (\ref{s1_21}). We combine
the variation of \(\delta\hat{A}(v\;\hat{B})\) with respect to \(\delta\hat{B}\) with the
derivative of the real parameter $v$ and obtain relations (\ref{s1_17}-\ref{s1_20})
\beq  \lb{s1_17}
\delta\hat{A}(v\;\hat{B})&=&\exp\{v\;\hat{B}\}\;\;\delta\Big(\exp\{-v\;\hat{B}\}\Big) \\ \lb{s1_18}
\frac{d\Big(\delta\hat{A}(v\;\hat{B})\Big)}{dv} &=&
\exp\{v\;\hat{B}\}\;\hat{B}\;\delta\Big(\exp\{-v\;\hat{B}\}\Big)-
\exp\{v\;\hat{B}\}\;\;\delta\Big(\hat{B}\;\exp\{-v\;\hat{B}\}\Big)  \\ \lb{s1_19}
\delta\Big(\hat{B}\;\exp\{-v\;\hat{B}\}\Big) &=&\hat{B}\;\delta\Big(\exp\{-v\;\hat{B}\}\Big)+
\delta\hat{B}\;\;\exp\{-v\;\hat{B}\}  \\  \lb{s1_20}
\frac{d\Big(\delta\hat{A}(v\;\hat{B})\Big)}{dv} &=&
-\exp\{v\;\hat{B}\}\;\;\delta\hat{B}\;\;\exp\{-v\;\hat{B}\}\;.
\eeq
The mentioned integration of (\ref{s1_20}) within \(v\in[0,1]\) leads to the required result (\ref{s1_23})
which attributes the variation of the exponential  '\(\exp\{\hat{B}\}\;\;\delta\big(\exp\{-\hat{B}\}\big)\)'
to the variation of the matrix \(\hat{B}\) in the exponent
\beq \lb{s1_21}
\int_{0}^{1}d\Big(\delta\hat{A}(v\;\hat{B})\Big) &=&-\int_{0}^{1}dv\;\;
\exp\{v\;\hat{B}\}\;\;\delta\hat{B}\;\;\exp\{-v\;\hat{B}\}   \\  \lb{s1_22}
\delta\hat{A}(1\cdot\hat{B}) - \delta\hat{A}(0\cdot\hat{B}) &=& -\int_{0}^{1}dv\;\;
\exp\{v\;\hat{B}\}\;\;\delta\hat{B}\;\;\exp\{-v\;\hat{B}\}  \\  \lb{s1_23}
\exp\{\hat{B}\}\;\;\delta\Big(\exp\{-\hat{B}\}\Big) &=& -\int_{0}^{1}dv\;\;
\exp\{v\;\hat{B}\}\;\;\delta\hat{B}\;\;\exp\{-v\;\hat{B}\}\;\;\;.
\eeq
The integration over \(v\in[0,1]\) in Eq. (\ref{s1_23}) can be achieved in a formal manner by introducing
the exponential of the commutator-operator \(\overrightarrow{[\hat{B}\:,\:\ldots]_{-}}\).
This construction is also used in statistical mechanics in relation to the time development
of the statistical operator with the Liouville- and von-Neumann equations \cite{katz}
\be \lb{s1_24}
\exp\{\hat{B}\}\;\;\delta\Big(\exp\{-\hat{B}\}\Big) = -\int_{0}^{1}dv\;\;\Big(
\exp\Big\{v\;\overrightarrow{\big[\hat{B}\:,\:\ldots\big]_{-}}\Big\}\;\;\delta\hat{B}\Big) = -
\Bigg(\frac{\exp\Big\{\overrightarrow{\big[\hat{B}\:,\:\ldots\big]_{-}}\Big\}-\hat{1}}
{\overrightarrow{\big[\hat{B}\:,\:\ldots\big]_{-}}}\;\;\delta\hat{B}\Bigg)_{\mbox{.}}
\ee

\section{Coherent state path integral and super-matrices} \lb{s2}

\subsection{Super-symmetry and density matrix for bosonic and fermionic atoms with anomalous terms}\lb{s21}

The operators $\hat{\psi}_{\vec{x},\alpha}$, $\hat{\psi}^{+}_{\vec{x},\alpha}$
with bosonic and fermionic operator parts $\hat{\vec{b}}_{\vec{x}}$, $\hat{\vec{\alpha}}_{\vec{x}}$
are introduced in analogy to the super-fields $\psi_{\vec{x},\alpha}(t)$ (\ref{s1_2})
in order to derive the corresponding generating function for the GP-equation (\ref{s1_1})
\footnote{In the remainder operators and matrices are marked with
a hat '$\hat{\ph{\Sigma}}$' or a tilde '$\wtilde{\ph{\Sigma}}$' in this paper.
Furthermore, the dimensions of matrices are occasionally displayed,
as e.g. in \(\hat{c}_{\vec{x},L\times L}\) with dimensions \(L\times L\) or
\(\hat{\eta}_{\vec{x},S\times L}\) with dimensions \(S\times L\). Considering the dimensions and symmetry
of a matrix, one can conclude for the number of its independent variables and can remember the exact structure
of the matrix in later sections without various sets of indices.}
\be \lb{s2_1}
\hat{\psi}_{\vec{x},\alpha}=\left(\hat{\vec{b}}_{\vec{x}}\;,\; \hat{\vec{\alpha}}_{\vec{x}}\right)^{T}\hspace*{0.75cm}
\hat{\psi}_{\vec{x},\alpha}^{+}=\left(\hat{\vec{b}}_{\vec{x}}^{+}\;,\; \hat{\vec{\alpha}}_{\vec{x}}^{+}\right)\;\;.
\ee
The second quantized Hamiltonian is determined for the classical GP-equation (\ref{s1_1}) in relation (\ref{s2_2}).
Spontaneous symmetry breaking source terms $j_{\psi;N}(\vec{x},t)$
and $\wtilde{j}_{\psi\psi;N\times N}(\vec{x},t)$ are included for the macroscopic BEC wave function and
for the anomalous terms of the BCS- and molecular condensates
\beq \lb{s2_2}
\lefteqn{\hat{H}(\hat{\psi}^{+},\hat{\psi},t)= \sum_{\vec{x}}\sum_{\alpha}
\hat{\psi}^{+}_{\vec{x},\alpha}\;\; \hat{h}(\vec{x})\;\; \hat{\psi}_{\vec{x},\alpha}+
\sum_{\vec{x},\vec{x}\ppr}\sum_{\alpha,\beta}
\hat{\psi}^{+}_{\vec{x}\ppr,\beta} \hat{\psi}^{+}_{\vec{x},\alpha}\;V_{|\vec{x}\ppr-\vec{x}|}\;
\hat{\psi}_{\vec{x},\alpha} \hat{\psi}_{\vec{x}\ppr,\beta} + } \\ \no &+& \sum_{\vec{x},\alpha}
\Big(j_{\psi;\alpha}^{*}(\vec{x},t)\;\hat{\psi}_{\vec{x},\alpha} +
\hat{\psi}_{\vec{x},\alpha}^{+}(t)\;j_{\psi;\alpha}(\vec{x},t)\Big) +
\\ \no &+&\frac{1}{2}
\sum_{\vec{x}}\strab\Bigg[\;\wtilde{j}_{\psi\psi;N\times N}^{+}(\vec{x},t)\; \left( \bea{cc}
\hat{c}_{\vec{x},L\times L} & \hat{\eta}^{T}_{\vec{x},L\times S} \\
\hat{\eta}_{\vec{x},S\times L} & \hat{f}_{\vec{x},S\times S} \eea\right) + \left( \bea{cc}
\hat{c}^{+}_{\vec{x},L\times L} & \hat{\eta}^{+}_{\vec{x},L\times S} \\
\hat{\eta}^{*}_{\vec{x},S\times L} & \hat{f}^{+}_{\vec{x},S\times S}
\eea\right)\;\wtilde{j}_{\psi\psi;N\times N}(\vec{x},t)\; \Bigg]_{\mbox{.}}
\eeq
Apart from the trap potential $u(\vec{x})$, the kinetic energy
is contained in the one-particle operator $\hat{h}(\vec{x})$ (\ref{s2_2},\ref{s2_3})
where the different masses of the bosonic and fermionic atoms
are approximated to an average value. The transformation
\(\psi_{\vec{x},\alpha}(t)\rightarrow e^{-\im(\mu_{0}/\hbar)\;t}\;\psi_{\vec{x},\alpha}(t)\)
yields the chemical potential $\mu_{0}$ in the coherent state path integral for vanishing temperature.
This energy parameter $\mu_{0}$ is a reference for
the gradually varying spatial and time-like evolutions of the super-fields in the gradient expansion.
The source term $j_{\psi;\alpha}(\vec{x},t)$ (\ref{s2_2},\ref{s2_4}) is the conjugated
supervector field to $\psi_{\vec{x},\alpha}(t)$ with \(L=2l+1\) bosonic $j_{\psi;B,L}(\vec{x},t)$ and
\(S=2s+1\) fermionic $j_{\psi;F,S}(\vec{x},t)$ angular momentum degrees of freedom
\beq \lb{s2_3}
\hat{h}(\vec{x})&=&
\frac{\hat{\vec{p}}^{\;2}}{2m}+u(\hat{\vec{x}})-\mu_{0} \\ \lb{s2_4}
j_{\psi;N}(\vec{x},t)&=&\Big(j_{\psi;B,L}(\vec{x},t)\;;\;
j_{\psi;F,S}(\vec{x},t)\Big)^{T}_{\mbox{.}}
\eeq
The anomalous operator terms for spontaneous symmetry breaking are combined in the
symmetric operator matrix $\hat{c}_{\vec{x},L\times L}$ (\ref{s2_5})
for the bosonic molecular condensate, the antisymmetric
operator matrix $\hat{f}_{\vec{x},S\times S}$ (\ref{s2_6}) for the BCS-condensate
of the fermions and into the fermion-boson mixed operator $\hat{\eta}_{\vec{x},S\times L}$ (\ref{s2_7})
and its transpose $\hat{\eta}_{\vec{x},L\times S}^{T}$ (\ref{s2_8})
\beq \lb{s2_5}
\hat{c}_{\vec{x},L\times L}&=&\big\{\hat{c}_{\vec{x},mn}\big\}=\big\{\hat{b}_{\vec{x},m}\;
\hat{b}_{\vec{x},n}\big\}\hspace*{0.5cm}m,n=-l,\ldots,+l \\ \lb{s2_6}
\hat{f}_{\vec{x},S\times S}&=&\big\{\hat{f}_{\vec{x},rr\ppr}\big\}=\big\{ \hat{\alpha}_{\vec{x},r}\;
\hat{\alpha}_{\vec{x},r\ppr}\big\}\hspace*{0.5cm}r,r\ppr= -s,\ldots,+s \\ \lb{s2_7}
\hat{\eta}_{\vec{x},S\times L}&=&\big\{\hat{\eta}_{\vec{x},rm}\big\}=\big\{
\hat{\alpha}_{\vec{x},r}\;\hat{b}_{\vec{x},m}\big\}\hspace*{0.5cm}r=-s,\ldots,+s\;\;;\hspace*{0.25cm} m=-l,\ldots,+l
\\ \lb{s2_8}
\hat{\eta}_{\vec{x},L\times S}^{T}&=&\big\{\hat{\eta}_{\vec{x},mr}\big\}=\big\{
\hat{b}_{\vec{x},m}\;\hat{\alpha}_{\vec{x},r}\big\}\hspace*{0.4cm}m=-l,\ldots,+l\;\;;\hspace*{0.37cm} r=-s,\ldots,+s\;\;.
\eeq
The source term $\wtilde{j}_{\psi\psi;N\times N}(\vec{x},t)$ in (\ref{s2_2})
for the anomalous terms (\ref{s2_5}-\ref{s2_8}) is a super-symmetric matrix, consisting of
the even boson-boson block $\hat{j}_{b;L\times L}(\vec{x},t)$ (\ref{s2_9},\ref{s2_10}) in the upper
block diagonal and the even fermion-fermion block $\hat{j}_{f;S\times S}(\vec{x},t)$
(\ref{s2_9},\ref{s2_10}) in the lower block diagonal. In order to take into account the boson and
fermion statistics of the angular momentum, the boson-boson \(\hat{j}_{b;L\times L}(\vec{x},t)\)
(fermion-fermion \(\hat{j}_{f;S\times S}(\vec{x},t)\)) block is symmetric
(antisymmetric) under transposition (\ref{s2_10}). The boson-fermion
\(\hat{j}^{T}_{\eta;L\times S}(\vec{x},t)\) and fermion-boson block
\(\hat{j}_{\eta;S\times L}(\vec{x},t)\) (\ref{s2_9},\ref{s2_11},\ref{s2_12}) contain odd
Grassmann variables and are related by transposition with an additional minus sign.
The cyclic invariance for products of super-matrices in a generalized trace
relation '$\mbox{str}$' (\ref{s2_2}) requires this additional minus sign
(compare (\ref{s2_16}))
\beq \lb{s2_9}
\wtilde{j}_{\psi\psi;N\times N}(\vec{x},t)&=& \left( \bea{cc}
\hat{j}_{b;L\times L}(\vec{x},t) & -\hat{j}^{T}_{\eta;L\times S}(\vec{x},t) \\
\hat{j}_{\eta;S\times L}(\vec{x},t) & -\hat{j}_{f;S\times S}(\vec{x},t) \eea\right) \\ \lb{s2_10}
\hat{j}_{b;L\times L}(\vec{x},t)&\in&\mbox{\sf C}_{even}\hspace*{0.5cm}
\hat{j}_{b;L\times L}^{T}(\vec{x},t)=\hat{j}_{b;L\times L}(\vec{x},t) \\ \no
\hat{j}_{f;S\times S}(\vec{x},t)&\in&\mbox{\sf C}_{even}\hspace*{0.5cm}
\hat{j}_{f;S\times S}^{T}(\vec{x},t)=-\hat{j}_{f;S\times S}(\vec{x},t) \\ \lb{s2_11}
\hat{j}_{\eta;S\times L}(\vec{x},t)&=&\{j_{\eta;rm}(\vec{x},t)\}\in\mcal{C}_{odd}  \\ \lb{s2_12}
\hat{j}_{\eta;L\times S}^{+}(\vec{x},t) &=& \{j_{\eta;mr}^{*}(\vec{x},t)\}\in\mcal{C}_{odd} \\ \no &&
r=-s,\ldots,+s\hspace*{1.0cm}m=-l,\ldots,+l\;\;\;.
\eeq
In the following we define the relations for the super-algebra
applied in this paper, especially the complex conjugation of anti-commuting variables (\ref{s2_13}),
the super-transposition (\ref{s2_14},\ref{s2_15}) and
super-trace (\ref{s2_16}) of a super-matrix and its super-hermitian conjugate (\ref{s2_17}) \cite{luc}.
In this paper the complex conjugation for the product of anti-commuting variables (\ref{s2_13})
also involves an exchange of the factors to the reversed order
\footnote{The complex conjugation as in
(\ref{s2_13}) is also denoted as 'star-operation' on Grassmann variables \cite{luc}.}
\be \lb{s2_13}
\left(\xi_{1}\ldots\xi_{n}\right)^{*}=\xi_{n}^{*}\ldots\xi_{1}^{*}\;\;\;.
\ee
Under a super-transposition 'st' of the super-matrices $\hat{N}_{1}$, $\hat{N}_{2}$ (\ref{s2_14}) ,
the even block-matrices $\hat{c}_{1}$, $\hat{c}_{2}$ and $\hat{f}_{1}$, $\hat{f}_{2}$
in the boson-boson and fermion-fermion blocks are transposed
in the ordinary way. On the contrary the boson-fermion blocks
$\hat{\eta}^{T}_{1}$, $\hat{\eta}^{T}_{2}$ and fermion-boson blocks $\hat{\chi}_{1}$, $\hat{\chi}_{2}$
in (\ref{s2_14}) are exchanged by transposition with an additional minus sign
in the resulting fermion-boson block. The minus sign in the
odd parts of the transposed super-matrix $\hat{N}^{st}$ preserves the property (\ref{s2_15})
for the transpose of the product of ordinary matrices.
The super-transposition of the product of super-matrices
\(\big(\hat{N}_{1}\cdot\hat{N}_{2}\big)^{st}\) (\ref{s2_15}) gives the product of super-transposed matrices
\(\hat{N}_{2}^{st}\cdot\hat{N}_{1}^{st}\) in reversed order in analogy to the
transposition of an ordinary matrix product
\beq \lb{s2_14}
\hat{N}_{1}=\left( \bea{cc}
\hat{c}_{1} & \hat{\eta}^{T}_{1} \\
\hat{\chi}_{1} & \hat{f}_{1} \eea\right)\;;
\hspace*{0.28cm}
\hat{N}_{2}&=&\left( \bea{cc}
\hat{c}_{2} & \hat{\eta}^{T}_{2} \\
\hat{\chi}_{2} & \hat{f}_{2} \eea\right)\;;\hspace*{0.28cm}
\hat{N}^{st}_{1}=\left( \bea{cc}
\hat{c}^{T}_{1} & \hat{\chi}^{T}_{1} \\
-\hat{\eta}_{1} & \hat{f}^{T}_{1}
\eea\right)\;;\hspace*{0.28cm}
\hat{N}^{st}_{2}=\left( \bea{cc}
\hat{c}^{T}_{2} & \hat{\chi}^{T}_{2} \\
-\hat{\eta}_{2} & \hat{f}^{T}_{2}
\eea\right)
\\ \lb{s2_15}
(\hat{N}_{1}\;\cdot\;\hat{N}_{2})^{st}&=&\hat{N}_{2}^{st}\;\cdot\;\hat{N}_{1}^{st}\;\;\;.
\eeq
The corresponding super-trace '$\mbox{str}$' (\ref{s2_16}) is defined with
an additional minus sign in the trace of the fermion-fermion block matrix so that
the cyclic invariance is maintained as in a trace relation of the product of ordinary matrices.
The super-hermitian conjugation (\ref{s2_17}) of
a super-matrix as $\hat{N}$ does not involve an additional minus sign in its odd parts as for the
super-transposition (\ref{s2_14}). It is equivalent to the ordinary hermitian conjugation of matrices
except that it has to be performed on the four blocks separately with an exchange of the
boson-fermion and fermion-boson parts. It also respects the
property \((\hat{N}_{1}\;\hat{N}_{2})^{+}=\hat{N}_{2}^{+}\;\hat{N}_{1}^{+}\) (\ref{s2_17})
without a minus sign in the odd parts as
with ordinary matrices because the complex conjugation (\ref{s2_13})
has been defined with a reversed order of the
corresponding complex anti-commuting variables.
Therefore, the superhermitian conjugate of a super-matrix as $\hat{N}$ is analogous
to the ordinary hermitian conjugation, but by  consideration of the special complex conjugation
(\ref{s2_13}) for the Grassmann numbers
\beq \lb{s2_16}
\mbox{str}[\hat{N}]&=&\mbox{str}\left(
\bea{cc}
\hat{c} & \hat{\eta}^{T} \\
\hat{\chi} & \hat{f}
\eea\right)=\mbox{tr}[\hat{c}]-\mbox{tr}[\hat{f}] \hspace*{0.65cm}
\mbox{str}[\hat{N}_{1}\;\hat{N}_{2}]=\mbox{str}[\hat{N}_{2}\;\hat{N}_{1}] \\ \lb{s2_17}
\hat{N}_{1}^{+}&=&\left( \bea{cc}
\hat{c}_{1}^{+} & \hat{\chi}_{1}^{+} \\
\hat{\eta}_{1}^{*} & \hat{f}_{1}^{+} \eea\right)\;\hspace*{0.37cm}
\hat{N}_{2}^{+}=\left( \bea{cc}
\hat{c}_{2}^{+} & \hat{\chi}_{2}^{+} \\
\hat{\eta}_{2}^{*} & \hat{f}_{2}^{+} \eea\right)\;\hspace*{0.37cm}
(\hat{N}_{1}\;\hat{N}_{2})^{+}=\hat{N}_{2}^{+}\;\hat{N}_{1}^{+}\;\;.
\eeq
The relation
\(\det(\hat{M})=\exp\{\mbox{tr}\ln \hat{M}\}\) between the determinant and the exponential
trace-logarithm of ordinary matrices $\hat{M}$ can be
generalized to define a super-determinant with the super-trace (\ref{s2_16})
in an exponential super-trace-logarithm relation (\ref{s2_18}). Applying the cyclic invariance
in a trace and super-trace relation, the super-determinant (\ref{s2_18})
of the super-matrix $\hat{N}_{N\times N}$ (\ref{s2_14}) can be derived as follows
\beq \lb{s2_18}
\lefteqn{\mbox{sdet}\big(\hat{N}_{N\times N}\big)\stackrel{!}{=}\exp\left\{\mbox{str}\ln\left(
\bea{cc}
\hat{c}_{L\times L} & \hat{\eta}_{L\times S}^{T} \\
\hat{\chi}_{S\times L} & \hat{f}_{S\times S}
\eea\right)\right\} } \\ \no &=& \exp\left\{\mbox{str}\ln\left(
\bea{cc}
\hat{c}_{L\times L} & 0 \\
 0 & \hat{f}_{S\times S}
\eea\right)\left(
\bea{cc}
\hat{1}_{L\times L} & \hat{c}_{L\times L}^{-1}\;\hat{\eta}_{L\times S}^{T} \\
\hat{f}_{S\times S}^{-1}\;\hat{\chi}_{S\times L} & \hat{1}_{S\times S}
\eea\right)\right\} \\ \no &=&
\frac{\det\big(\hat{c}_{L\times L}\big)}{\det\big(\hat{f}_{S\times S}\big)}\times
\exp\left\{\mbox{str}\ln\left[ \hat{1}_{N\times N}+\left(
\bea{cc}
0 & \hat{c}_{L\times L}^{-1}\;\hat{\eta}_{L\times S}^{T} \\
\hat{f}_{S\times S}^{-1}\;\hat{\chi}_{S\times L} & 0
\eea\right)\right]\right\} \\ \no &=&
\frac{\det\big(\hat{c}_{L\times L}\big)}{\det\big(\hat{f}_{S\times S}\big)}\times
\exp\left\{-\sum_{n=1}^{\infty}\frac{1}{2n}\mbox{str}\left(
\bea{cc}
\big(\hat{c}_{L\times L}^{-1}\;\hat{\eta}_{L\times S}^{T}\hat{f}_{S\times S}^{-1}\;\hat{\chi}_{S\times L}\big)^{n} & 0 \\
0 & \big(\hat{f}_{S\times S}^{-1}\;\hat{\chi}_{S\times L}\hat{c}_{L\times L}^{-1}\;\hat{\eta}_{L\times S}^{T}\big)^{n}
\eea\right)\right\} \\ \no &=&
\frac{\det\big(\hat{c}_{L\times L}\big)}{\det\big(\hat{f}_{S\times S}\big)}\times
\exp\left\{\mbox{tr}\ln\Big(\hat{1}_{L\times L}-
\hat{c}_{L\times L}^{-1}\;\hat{\eta}_{L\times S}^{T}\hat{f}_{S\times S}^{-1}\;\hat{\chi}_{S\times L}
\Big)\right\} \\ \no &=&
\frac{\det\big(\hat{c}_{L\times L}\big)}{\det\big(\hat{f}_{S\times S}\big)}\times
\det\big(\hat{1}_{L\times L}-
\hat{c}_{L\times L}^{-1}\;\hat{\eta}_{L\times S}^{T}\hat{f}_{S\times S}^{-1}\;\hat{\chi}_{S\times L}\big) \\ \no &=&
\frac{\det\big(\hat{c}_{L\times L}-
\hat{\eta}_{L\times S}^{T}\hat{f}_{S\times S}^{-1}\;\hat{\chi}_{S\times L}\big)}{\det\big(\hat{f}_{S\times S}\big)}\;.
\eeq
The determinant of the fermion-fermion block $\hat{f}_{S\times S}$ appears in the denominator because the
super-trace includes a minus sign in the fermion-fermion block (\ref{s2_16}). Similar transformations as in the
derivation for the super-determinant (\ref{s2_18}) lead to the property
\(\mbox{sdet}(\hat{N}_{1}\cdot\hat{N}_{2})=
\mbox{sdet}(\hat{N}_{1})\cdot\mbox{sdet}(\hat{N}_{2})\) as for ordinary matrices.

Other definitions of the complex conjugation of Grassmann numbers, as the superstar-operation
'$^{\sharp}$' \cite{luc}, require a different definition of the superhermitian conjugation '$\dagger$'
of the super-matrices $\hat{N}_{1}$, $\hat{N}_{2}$ (\ref{s2_22}) which can be composed of the superstar-operation
'$^{\sharp}$' (\ref{s2_19}-\ref{s2_21}) and the super-transposition '$^{st}$' (\ref{s2_14},\ref{s2_15})
(Compare with \cite{luc},\cite{fay}-\cite{corn3},\cite{lern1})
\beq \lb{s2_19}
(\xi_{1}\ldots\xi_{N})^{\sharp}&=&\xi_{1}^{\sharp}\ldots\xi_{N}^{\sharp} \\ \lb{s2_20}
(\xi_{i}^{\sharp})^{\sharp}&=&-\xi_{i} \\ \lb{s2_21}
(\xi_{i}^{\sharp}\xi_{i})^{\sharp}&=&(\xi_{i}^{\sharp})^{\sharp}\;\xi_{i}^{\sharp}=
-\xi_{i}\;\xi_{i}^{\sharp}= \xi_{i}^{\sharp}\xi_{i}  \\ \lb{s2_22}
\hat{N}^{\dagger}&=&\hat{N}^{\sharp st}=\left( \bea{cc}
\hat{c}^{+} & \hat{\chi}^{+} \\
-\hat{\eta}^{*} & \hat{f}^{+} \eea\right)\hspace*{1.0cm}
(\hat{N}_{1}\;\hat{N}_{2})^{\dagger}=
(\hat{N}_{1}\;\hat{N}_{2})^{\sharp st}=\hat{N}_{2}^{\dagger}\;\hat{N}_{1}^{\dagger}=
\hat{N}_{2}^{\sharp st}\;\hat{N}_{1}^{\sharp st}\;\;.
\eeq
The above definition (\ref{s2_19}-\ref{s2_21})
of complex conjugation of odd variables is preferably used in disordered systems \cite{efe,lern1}
whereas the definitions (\ref{s2_13}-\ref{s2_18}), applied in this paper, retain the
hermitian properties of the coherent states in a path integral.

In the remainder we introduce the time contour integral (\ref{s2_23}) with the time variables \(t_{p}\)
on the branches \(p=\pm\) for the time development of the Hamiltonian in forward
\(\int_{-\infty}^{\infty}d t_{+}\ldots\) and backward direction
\(\int_{\infty}^{-\infty}d t_{-}\ldots\) \cite{ke}-\cite{ka}. The negative sign of the
backward propagation \(\int_{\infty}^{-\infty}d t_{-}\ldots=-\int_{-\infty}^{\infty}d t_{-}\ldots\)
will be frequently taken into account by the symbol \(\eta_{p=\pm}=p=\pm\)
\beq \lb{s2_23}
\int_{C}d t_{p}\ldots &=&\int_{-\infty}^{+\infty}d
t_{+}\ldots+ \int_{+\infty}^{-\infty}d t_{-}\ldots= \int_{-\infty}^{+\infty}d t_{+}\ldots-
\int_{-\infty}^{+\infty}d t_{-}\ldots \\ \no &=&
\sum_{p=\pm}\int_{-\infty}^{\infty}d t_{p}\;\;\eta_{p}\;\;\ldots\;\;\;\;.
\eeq
The corresponding coherent state path integral
$Z[\hat{\mcal{J}},j_{\psi},\wtilde{j}_{\psi\psi}]$ for the Hamiltonian (\ref{s2_2})
is given by relation (\ref{s2_24}) where the even matrix fields
\(\hat{c}_{L\times L}(\vec{x},t_{p})\), \(\hat{f}_{S\times S}(\vec{x},t_{p})\) and
odd matrix fields \(\hat{\eta}_{L\times S}^{T}(\vec{x},t_{p})\),
\(\hat{\eta}_{S\times L}(\vec{x},t_{p})\) correspond to the operator matrices
$\hat{c}_{\vec{x};L\times L}$, $\hat{f}_{\vec{x},S\times S}$ and
$\hat{\eta}_{\vec{x},L\times S}$, $\hat{\eta}_{\vec{x},S\times L}^{T}$
defined in relations (\ref{s2_5}-\ref{s2_8}) \cite{nag1,nag2},\cite{ke}-\cite{ka}
\beq \lb{s2_24}
\lefteqn{Z[\hat{\mcal{J}},j_{\psi},\wtilde{j}_{\psi\psi}]=
\int d[\psi_{\vec{x},\alpha}(t_{p})] \;\;
\exp\bigg\{-\frac{\im}{\hbar}\int_{C}d t_{p}\sum_{\vec{x}}\sum_{\alpha}
\psi_{\vec{x},\alpha}^{*}(t_{p})\;\;\hat{H}_{p}(\vec{x},t_{p})\;\;
\psi_{\vec{x},\alpha}(t_{p})\bigg\} } \\ \no &\times& \exp\bigg\{-\frac{\im}{\hbar}
\int_{C}dt_{p}\sum_{\vec{x},\vec{x}\ppr}\sum_{\alpha,\beta}
\psi_{\vec{x}\ppr,\beta}^{*}(t_{p})\;\psi_{\vec{x},\alpha}^{*}(t_{p})\;
V_{|\vec{x}\ppr-\vec{x}|}\;\psi_{\vec{x},\alpha}(t_{p})\;\psi_{\vec{x}\ppr,\beta}(t_{p})
\bigg\} \\ \no &\times &
\exp\bigg\{-\frac{\im}{\hbar}\int_{C}d t_{p}\sum_{\vec{x}}\sum_{\alpha}
\Big(j_{\psi;\alpha}^{*}(\vec{x},t_{p})\;\psi_{\vec{x},\alpha}(t_{p})+
\psi_{\vec{x},\alpha}^{*}(t_{p})\;j_{\psi;\alpha}(\vec{x},t_{p})\Big)\bigg\}
\\ \no &\times&
\exp\Bigg\{-\frac{\im}{2\hbar}\int_{C}d t_{p} \sum_{\vec{x}}\strab\Bigg[ \left( \bea{cc}
\hat{j}_{b}^{+}(\vec{x},t_{p}) & \hat{j}_{\eta}^{+}(\vec{x},t_{p}) \\
-\hat{j}_{\eta}^{*}(\vec{x},t_{p}) & -\hat{j}_{f}^{+}(\vec{x},t_{p}) \eea\right)\left( \bea{cc}
\hat{c}(\vec{x},t_{p}) & \hat{\eta}^{T}(\vec{x},t_{p}) \\
\hat{\eta}(\vec{x},t_{p}) & \hat{f}(\vec{x},t_{p}) \eea\right)+ \\ \no &+&\left( \bea{cc}
\hat{c}^{+}(\vec{x},t_{p}) & \hat{\eta}^{+}(\vec{x},t_{p}) \\
\hat{\eta}^{*}(\vec{x},t_{p}) & \hat{f}^{+}(\vec{x},t_{p}) \eea\right)\left( \bea{cc}
\hat{j}_{b}(\vec{x},t_{p}) & -\hat{j}^{T}_{\eta}(\vec{x},t_{p}) \\
\hat{j}_{\eta}(\vec{x},t_{p}) & -\hat{j}_{f}(\vec{x},t_{p}) \eea\right)\Bigg]\Bigg\} \\ \no &\times&
\exp\bigg\{-\frac{\im}{2\hbar}\int_{C}d t_{p_{1}}^{(1)}\; d t_{p_{2}}^{(2)}
\sum_{\vec{x},\vec{x}\ppr}\sum_{\alpha,\beta} \Psi_{\vec{x}\ppr,\beta}^{+b}(t_{p_{2}}^{(2)})\;
\hat{\mcal{J}}_{\vec{x}\ppr,\beta;\vec{x},\alpha}^{ba}(t_{p_{2}}^{(2)};
t_{p_{1}}^{(1)})\;\Psi_{\vec{x},\alpha}^{a}(t_{p_{1}}^{(1)})\bigg\}
\eeq
\be \lb{s2_25}
\hat{H}_{p}(\vec{x},t_{p}) = -\hat{E}_{p}-\im\;\ve_{p}+\hat{h}(\vec{x}) =
-\im\hbar\frac{\pp}{\pp t_{p}}-\im\;\ve_{p}+
\frac{\hat{\vec{p}}^{\;2}}{2m}+u(\hat{\vec{x}})-\mu_{0}\;\;\;.
\ee
A nonhermitian infinitesimal
part \(-\im\;\ve_{p}=-\im\;(\pm\ve)\), \((p=\pm;\;\ve>0)\) on the time contour has to be added to
\(\hat{H}_{p}(\vec{x},t_{p})\) (\ref{s2_25}) for the analytic and convergence
properties of Green functions, derived from the coherent state path integral (\ref{s2_24}).
Apart from the one-particle part \(\hat{H}_{p}(\vec{x},t_{p})\) (\ref{s2_25}) and
the common potential $V_{|\vec{x}\ppr-\vec{x}|}$ for
bosons and fermions, the source field $j_{\psi;\alpha}(\vec{x},t)$ (\ref{s2_4},\ref{s2_2})
and source matrix $\wtilde{j}_{\psi\psi;\alpha\beta}(\vec{x},t)$
(\ref{s2_9}-\ref{s2_12},\ref{s2_2}) are included
in \(Z[\hat{\mcal{J}},j_{\psi},\wtilde{j}_{\psi\psi}]\) (\ref{s2_24})
for a spontaneous symmetry breaking to a coherent condensate wave function
\(\langle\psi_{\vec{x},\alpha}(t_{p})\rangle\) and anomalous terms
\(\langle\psi_{\vec{x},\beta}(t_{p})\;\;\psi_{\vec{x},\alpha}(t_{p})\rangle\).
One can also take derivatives of
\(Z[\hat{\mcal{J}},j_{\psi},\wtilde{j}_{\psi\psi}]\) (\ref{s2_24})
with respect to the sources \(j_{\psi;\alpha}^{*}(\vec{x},t_{p})\) and
\(\wtilde{j}_{\psi\psi;\alpha\beta}^{+}(\vec{x},t_{p})\),
including a dependence on the $p=\pm$ branch of the
time contour.  The wave function \(\langle\psi_{\vec{x},\alpha}(t_{p})\rangle\) and pair condensates
\(\langle\psi_{\vec{x},\beta}(t_{p})\;\;\psi_{\vec{x},\alpha}(t_{p})\rangle\)
are considered with a possible spontaneous symmetry breaking caused by the 'condensate seeds'
\(j_{\psi;\alpha}(\vec{x},t_{\pm})=j_{\psi;\alpha}(\vec{x},t)\) (\ref{s2_26}) and
\(\wtilde{j}_{\psi\psi;\alpha\beta}(\vec{x},t_{\pm})=
\wtilde{j}_{\psi\psi;\alpha\beta}(\vec{x},t)\) (\ref{s2_27})
where the sources \(j_{\psi;\alpha}^{*}(\vec{x},t_{p})\) and
\(\wtilde{j}_{\psi\psi;\alpha\beta}^{+}(\vec{x},t_{p})\) are set to equivalent values
on the two time branches with $\hat{\mcal{J}}\equiv0$
\beq \lb{s2_26}
j_{\psi;\alpha}(\vec{x},t_{p}) &:=&j_{\psi;\alpha}(\vec{x},t)\;\;;\;\;\;\;
\mbox{'condensate seed' for } \langle\psi_{\vec{x},\alpha}(t_{p})\rangle \\ \lb{s2_27}
\wtilde{j}_{\psi\psi;\alpha\beta}(\vec{x},t_{p})&:=&
\wtilde{j}_{\psi\psi;\alpha\beta}(\vec{x},t)\;\;;\mbox{'condensate seed' for }
\langle\psi_{\vec{x},\beta}(t_{p})\;\;\psi_{\vec{x},\alpha}(t_{p})\rangle\;.
\eeq
A source term
$\hat{\mcal{J}}_{\vec{x}\ppr,\beta;\vec{x},\alpha}^{ba}(t_{p_{2}}^{(2)}; t_{p_{1}}^{(1)})$
is also inserted in (\ref{s2_24})
in order to allow for the generation of anomalous terms
\(\langle\psi_{\vec{x},\alpha}(t_{p})\;\psi_{\vec{x}\ppr,\beta}(t_{p})\rangle\) by a single differentiation with
respect to $\hat{\mcal{J}}$. However, this necessitates the doubling of the supervector
\(\psi_{\vec{x},\alpha}(t_{p})=\big(\vec{b}_{\vec{x}}(t_{p})\;;\; \vec{\alpha}_{\vec{x}}(t_{p})\big)^{T}\)
on the time contour with its complex conjugate
\(\psi_{\vec{x},\alpha}^{*}(t_{p})=\big(\vec{b}_{\vec{x}}^{*}(t_{p})\;;\;
\vec{\alpha}_{\vec{x}}^{*}(t_{p})\big)^{T}\). The vectors $\psi_{\vec{x},\alpha}(t_{p})$ and
$\psi_{\vec{x},\alpha}^{*}(t_{p})$ are combined into the supervector
$\Psi_{\vec{x},\alpha}^{a(=1/2)}(t_{p})$ (with capital '$\Psi$')
where the index $a=1$ refers to $\psi_{\vec{x},\alpha}(t_{p})$ and the index $a=2$ to the complex
conjugate field $\psi_{\vec{x},\alpha}^{*}(t_{p})$. Therefore, we obtain a supervector
$\Psi_{\vec{x},\alpha}^{a}(t_{p})$ which consists of a commuting part $\vec{b}_{\vec{x}}(t_{p})$, the anti-commuting
part $\vec{\alpha}_{\vec{x}}(t_{p})$ for the fermions, and again the complex conjugate of the commuting variables
$\vec{b}_{\vec{x}}^{*}(t_{p})$, finally followed by the complex conjugated Grassmann numbers
$\vec{\alpha}_{\vec{x}}^{*}(t_{p})$
\be \lb{s2_28}
\Psi_{\vec{x},\alpha}^{a(=1/2)}(t_{p})=\left(
\bea{c}
\psi_{\vec{x},\alpha}(t_{p}) \\
\psi_{\vec{x},\alpha}^{*}(t_{p})
\eea\right)=
\bigg(\underbrace{\vec{b}_{\vec{x}}(t_{p})\;,\;\vec{\alpha}_{\vec{x}}(t_{p})}_{a=1}\;;\;
\underbrace{\vec{b}_{\vec{x}}^{*}(t_{p})\;,\;\vec{\alpha}_{\vec{x}}^{*}(t_{p})}_{a=2}\bigg)^{T}\;\;.
\ee
According to this definition (\ref{s2_28}), the
bosonic and fermionic parts alternate in the supervector $\Psi_{\vec{x},\alpha}^{a(=1/2)}(t_{p})$.
A different order is preferred in section \ref{s31} where the doubled supervector (\ref{s2_28})
is divided into a purely bosonic part, with its first two block
components $\vec{b}_{\vec{x}}(t_{p})\;,\;\vec{b}_{\vec{x}}^{*}(t_{p})$, and a consecutively purely fermionic part
$\vec{\alpha}_{\vec{x}}(t_{p})\;,\;\vec{\alpha}_{\vec{x}}^{*}(t_{p})$. This allows to perform symmetry considerations
and to identify the invariant super-group of the coherent state path integral with the doubled supervector
$\Psi_{\vec{x},\alpha}^{a}(t_{p})$. The nonlocal order parameter
\(\hat{\Phi}_{\vec{x},\alpha;\vec{x}\ppr,\beta}^{ab}(t_{p})\) (\ref{s2_29}), also comprising the anomalous terms
\(\langle\psi_{\vec{x},\alpha}(t_{p})\;\psi_{\vec{x}\ppr,\beta}(t_{p})\rangle\) in the non-diagonal parts,
has a matrix form which follows from
the dyadic product of the doubled supervector \(\Psi_{\vec{x},\alpha}^{a(=1/2)}(t_{p})\) (\ref{s2_28})
\beq \lb{s2_29}
\hat{\Phi}_{\vec{x},\alpha;\vec{x}\ppr,\beta}^{ab}(t_{p})&=&\Psi_{\vec{x},\alpha}^{a}(t_{p})\;\otimes\;
\Psi_{\vec{x}\ppr,\beta}^{b+}(t_{p}) = \left( \bea{c}
\psi_{\vec{x},\alpha}(t_{p}) \\
\psi_{\vec{x},\alpha}^{*}(t_{p}) \eea\right)\otimes
\Big(\psi_{\vec{x}\ppr,\beta}^{*}(t_{p})\;;\;\psi_{\vec{x}\ppr,\beta}(t_{p})\Big)
\\ \no &=&\left( \bea{cc}
\langle\psi_{\vec{x},\alpha}(t_{p})\;\psi_{\vec{x}\ppr,\beta}^{*}(t_{p})\rangle &
\langle\psi_{\vec{x},\alpha}(t_{p})\;\psi_{\vec{x}\ppr,\beta}(t_{p})\rangle \\
\langle\psi_{\vec{x},\alpha}^{*}(t_{p})\;\psi_{\vec{x}\ppr,\beta}^{*}(t_{p})\rangle &
\langle\psi_{\vec{x},\alpha}^{*}(t_{p})\;\psi_{\vec{x}\ppr,\beta}(t_{p})\rangle \eea\right) = \left( \bea{cc}
\hat{\Phi}_{\vec{x},\alpha;\vec{x}\ppr,\beta}^{11}(t_{p}) &
\hat{\Phi}_{\vec{x},\alpha;\vec{x}\ppr,\beta}^{12}(t_{p}) \\
\hat{\Phi}_{\vec{x},\alpha;\vec{x}\ppr,\beta}^{21}(t_{p}) &
\hat{\Phi}_{\vec{x},\alpha;\vec{x}\ppr,\beta}^{22}(t_{p}) \eea\right)_{\mbox{.}}
\eeq
A doubling of the angular momentum space has also to be considered because of the source matrix
\(\wtilde{j}_{\psi\psi;\alpha\beta}(\vec{x},t)\) (\ref{s2_9}-\ref{s2_12},\ref{s2_2})
which causes the spontaneous symmetry breaking to the Goldstone modes for
the molecular and BCS condensates. The complete order parameter consists of four super-matrices
\(\hat{\Phi}^{11}\), \(\hat{\Phi}^{22}\), \(\hat{\Phi}^{12}\), \(\hat{\Phi}^{21}\)
so that the complete boson-boson and fermion-fermion blocks are distributed
on these four matrices by convention. The order parameter (\ref{s2_29}) is invariant
under super-unitary transformations in angular
momentum space which does not alter the block structure into densities
\(\hat{\Phi}^{11}\), \(\hat{\Phi}^{22}\) and pair
condensates \(\hat{\Phi}^{12}\), \(\hat{\Phi}^{21}\).
The order parameter \(\hat{\Phi}_{\vec{x},\alpha;\vec{x}\ppr,\beta}^{ab}(t_{p})\)
(\ref{s2_29}) also allows a global mixed form of compact and
hyperbolic super-symmetry transformations which combines densities  \(\hat{\Phi}^{11}\),
\(\hat{\Phi}^{22}\) and pair condensates \(\hat{\Phi}^{12}\), \(\hat{\Phi}^{21}\).
A complete super-symmetry group
of the path integral is spontaneously broken by the unitary subgroup
for the invariance of the densities and the source
term. This symmetry breaking leads to a nonlinear sigma model
after a gradient expansion for the anomalous terms. Note
the form of the super-matrix
\(\hat{\Phi}_{\vec{x},\alpha;\vec{x}\ppr,\beta}^{ab}(t_{p})\) which consists of the four block
super-matrices \(\hat{\Phi}_{\vec{x},\alpha;\vec{x}\ppr,\beta}^{11}(t_{p})\) ,
\(\hat{\Phi}_{\vec{x},\alpha;\vec{x}\ppr,\beta}^{22}(t_{p})\)  and
\(\hat{\Phi}_{\vec{x},\alpha;\vec{x}\ppr,\beta}^{12}(t_{p})\) ,
\(\hat{\Phi}_{\vec{x},\alpha;\vec{x}\ppr,\beta}^{21}(t_{p})\) so that
the super-transposition, super-trace and hermitian conjugation of
\(\hat{N}_{1}\) , \(\hat{N}_{2}\) in relations (\ref{s2_13}-\ref{s2_18})
have to be generalized with Eqs. (\ref{s2_30}-\ref{s2_32})
\footnote{The dependences on the vectors \(\vec{x},\;\vec{x}\ppr\) and on the time $t_{p}$
are omitted for the display of the structure of $\hat{\Phi}_{\alpha\beta}^{ab}$, but are also involved in
the super-transposition and the hermitian conjugation of $\hat{\Phi}$ in the case of a nonlocal spatial dependence.}
\beq \lb{s2_30}
\Big(\hat{\Phi}_{\alpha\beta}^{ab}\Big)^{ST}&=&\left( \bea{cc}
\hat{\Phi}_{\alpha\beta}^{11} & \hat{\Phi}_{\alpha\beta}^{12} \\
\hat{\Phi}_{\alpha\beta}^{21} &  \hat{\Phi}_{\alpha\beta}^{22} \eea\right)^{ST}= \left( \bea{cc}
\big(\hat{\Phi}_{\alpha\beta}^{11}\big)^{st} & \big(\hat{\Phi}_{\alpha\beta}^{21}\big)^{st} \\
\big(\hat{\Phi}_{\alpha\beta}^{12}\big)^{st} &  \big(\hat{\Phi}_{\alpha\beta}^{22}\big)^{st}
\eea\right) \\ \lb{s2_31}
\STRAB\Big[\hat{\Phi}_{\alpha\beta}^{ab}\Big]&=&
\strab\Big[\hat{\Phi}_{\alpha\beta}^{11}\Big]+\strab\Big[\hat{\Phi}_{\alpha\beta}^{22}\Big]
= \sum_{m=-l}^{m=+l}\hat{\Phi}_{mm}^{11}-\sum_{r=-s}^{r=+s}\hat{\Phi}_{rr}^{11}+
\sum_{m=-l}^{m=+l}\hat{\Phi}_{mm}^{22}-\sum_{r=-s}^{r=+s}\hat{\Phi}_{rr}^{22} \\ \lb{s2_32}
\Big(\hat{\Phi}_{\alpha\beta}^{ab}\Big)^{+}&=&\left( \bea{cc}
\hat{\Phi}_{\alpha\beta}^{11} & \hat{\Phi}_{\alpha\beta}^{12} \\
\hat{\Phi}_{\alpha\beta}^{21} &  \hat{\Phi}_{\alpha\beta}^{22} \eea\right)^{+}= \left( \bea{cc}
\big(\hat{\Phi}_{\alpha\beta}^{11}\big)^{+} & \big(\hat{\Phi}_{\alpha\beta}^{21}\big)^{+} \\
\big(\hat{\Phi}_{\alpha\beta}^{12}\big)^{+} &  \big(\hat{\Phi}_{\alpha\beta}^{22}\big)^{+}
\eea\right)_{\mbox{.}}
\eeq
One has to apply the super-trace '$\mbox{STR}$' (\ref{s2_31}) instead of '$\mbox{str}$' (\ref{s2_16})
in '$\mbox{sdet}$' (\ref{s2_18}) for the 'Nambu'-doubled order parameter
\(\hat{\Phi}_{\alpha\beta}^{ab}\) (\ref{s2_29})
in order to obtain the analogous doubled super-determinant '$\mbox{SDET}(\hat{\Phi}_{\alpha\beta}^{ab})$'
with the exponential super-trace-logarithm relation
\(\mbox{SDET}(\hat{\Phi}_{\alpha\beta}^{ab})=\exp\{\STRAB\ln\hat{\Phi}_{\alpha\beta}^{ab}\}\).

\subsection{Hubbard-Stratonovich transformation to a super-symmetric self-energy} \lb{s22}

The interaction with four super-fields and the short-ranged two body potential (\ref{s2_2})
has to be transformed to a relation which
is similar to the order parameter
\(\hat{\Phi}_{\vec{x},\alpha;\vec{x}\ppr,\beta}^{ab}(t_{p})\) (\ref{s2_29}). The density term
\(\psi_{\vec{x},\alpha}^{+}(t_{p})\;\psi_{\vec{x},\alpha}(t_{p})\)
can be changed to the 'Nambu'-doubled super-field
$\Psi_{\vec{x},\alpha}^{a}(t_{p})$ by doubling the product
\(\psi_{\vec{x},\alpha}^{+}(t_{p})\;\psi_{\vec{x},\alpha}(t_{p})\)
in reversed order where a minus sign has to be taken
into account for the fermionic fields
\(\vec{\alpha}^{+}_{\vec{x}}(t_{p})\cdot\vec{\alpha}_{\vec{x}}(t_{p})\rightarrow
-\vec{\alpha}^{T}_{\vec{x}}(t_{p})\cdot\vec{\alpha}_{\vec{x}}^{*}(t_{p})\)
\beq \lb{s2_33}
\lefteqn{\psi_{\vec{x},\alpha}^{+}(t_{p})\; \psi_{\vec{x},\alpha}(t_{p})= \vec{b}_{\vec{x}}^{+}(t_{p})\cdot
\vec{b}_{\vec{x}}(t_{p})+ \vec{\alpha}_{\vec{x}}^{+}(t_{p})\cdot\vec{\alpha}_{\vec{x}}(t_{p})= } \\ \no &=&
\frac{1}{2}\bigg( \vec{b}_{\vec{x}}^{+}(t_{p})\cdot\vec{b}_{\vec{x}}(t_{p})+
\vec{\alpha}_{\vec{x}}^{+}(t_{p})\cdot\vec{\alpha}_{\vec{x}}(t_{p})+
\vec{b}_{\vec{x}}^{T}(t_{p})\cdot\vec{b}_{\vec{x}}^{*}(t_{p})-
\vec{\alpha}_{\vec{x}}^{T}(t_{p})\cdot\vec{\alpha}_{\vec{x}}^{*}(t_{p})\bigg) \\
\no &=&\frac{1}{2}\bigg(\psi_{\vec{x},\alpha}^{+}(t_{p})\cdot\psi_{\vec{x},\alpha}(t_{p})+
\psi_{\vec{x},\alpha}^{T}(t_{p})\;\hat{\kappa}_{\alpha\alpha}\;\psi_{\vec{x},\alpha}^{*}(t_{p})\bigg)
= \frac{1}{2}\;\Psi_{\vec{x},\alpha}^{a+}(t_{p})\;\hat{K}\; \Psi_{\vec{x},\alpha}^{a}(t_{p}) \;\;.
\eeq
The resulting scalar product
\(\frac{1}{2}\Psi_{\vec{x},\alpha}^{+a}(t_{p})\; \hat{K}\;\Psi_{\vec{x},\alpha}^{a}(t_{p})\) contains a diagonal
matrix \(\hat{K}_{2N\times 2N}\) (\ref{s2_34}) as a metric which is defined
in the alternating form of a boson-boson 'BB' and fermion-fermion 'FF' block by the relation
\beq \lb{s2_34}
\hat{K}_{2N\times 2N}&=&\bigg\{\underbrace{\hat{1}_{L\times
L}}_{BB}\;,\; \underbrace{\hat{1}_{S\times S}}_{FF}\;;\;\underbrace{\hat{1}_{L\times L}}_{BB}\;,\;
\underbrace{-\hat{1}_{S\times S}}_{FF}\bigg\}=
\Big\{\underbrace{\hat{1}_{N\times N}}_{a=1}\;;\;\underbrace{\hat{\kappa}_{N\times N}}_{a=2}\Big\}
\\ \lb{s2_35} \hat{\kappa}_{N\times N}&=&\Big\{\hat{1}_{L\times L},-\hat{1}_{S\times S}\Big\} \;\;,
\eeq
where the diagonal submatrix \(\hat{\kappa}_{N\times N}\) (\ref{s2_35}) has been separated
as the metric for the complex valued
part \(\psi_{\vec{x},\alpha}^{*}(t_{p})\) in \(\Psi_{\vec{x},\alpha}^{a=2}(t_{p})\). Using the metric
\(\hat{K}_{2N\times 2N}\) (\ref{s2_34}), the quartic interaction of the
super-fields \(\psi_{\vec{x},\alpha}(t_{p})\) with potential
$V_{|\vec{x}\ppr-\vec{x}|}$ can be converted by dyadic products to the 'Nambu'-doubled density matrix
\(\hat{R}_{\vec{x},\alpha;\vec{x}\ppr,\beta}^{ab}(t_{p})\) in a super-trace relation
with '\(\mbox{STR}\)' (\ref{s2_31})
\beq \lb{s2_36}
\lefteqn{\sum_{\vec{x},\vec{x}\ppr}\sum_{\alpha,\beta}
\psi_{\vec{x}\ppr,\beta}^{+}(t_{p})\;\psi_{\vec{x},\alpha}^{+}(t_{p})\;
V_{|\vec{x}-\vec{x}\ppr|}\;\psi_{\vec{x},\alpha}\;\psi_{\vec{x}\ppr,\beta}=} \\
\no &=& \frac{1}{4}\sum_{\vec{x},\vec{x}\ppr}\sum_{\alpha,\beta,b}
\Psi_{\vec{x},\alpha}^{a+}(t_{p})\;\hat{K}\;\Psi_{\vec{x},\alpha}^{a}(t_{p})\;\;
\Psi_{\vec{x}\ppr,\beta}^{b+}(t_{p})\;\hat{K}\;\Psi_{\vec{x}\ppr,\beta}^{b}(t_{p})\;\; V_{|\vec{x}-\vec{x}\ppr|}= \\
\no &=&\frac{1}{4}\sum_{\vec{x},\vec{x}\ppr}
\STRAB\bigg[\hat{K}\underbrace{\left(\Psi_{\vec{x},\alpha}^{a}(t_{p})\otimes
\Psi_{\vec{x}\ppr,\beta}^{b+}(t_{p})\right)}_{ \hat{R}_{\vec{x},\alpha;\vec{x}\ppr,\beta}^{ab}(t_{p})}\hat{K}
\underbrace{\left(\Psi_{\vec{x}\ppr,\beta}^{b}(t_{p})\otimes \Psi_{\vec{x},\alpha}^{a+}(t_{p})\right)}_{
\hat{R}_{\vec{x}\ppr,\beta;\vec{x},\alpha}^{ba}(t_{p})} \bigg]\;V_{|\vec{x}-\vec{x}\ppr|}= \\ \no
&=&\frac{1}{4}\sum_{\vec{x},\vec{x}\ppr}
\STRAB\left[\hat{K}\;\hat{R}_{\vec{x},\alpha;\vec{x}\ppr,\beta}^{ab}(t_{p})\;\hat{K}\;
\hat{R}_{\vec{x}\ppr,\beta;\vec{x},\alpha}^{ba}(t_{p})\right]\;V_{|\vec{x}-\vec{x}\ppr|}\;\;\;.
\eeq
The density matrix \(\hat{R}_{\vec{x},\alpha;\vec{x}\ppr,\beta}^{ab}(t_{p})\) (\ref{s2_37}) is similar
to the order parameter \(\hat{\Phi}_{\vec{x},\alpha;\vec{x}\ppr,\beta}^{ab}(t_{p})\) (\ref{s2_29}).
Furthermore, the matrices
\(\wtilde{B},\wtilde{c}\), \(\wtilde{F},\wtilde{f}\) and Grassmann related matrix fields
\(\wtilde{\chi},\wtilde{\eta}\) in
\(\hat{R}_{\vec{x},\alpha;\vec{x}\ppr,\beta}^{ab}(t_{p})\) (\ref{s2_38}) can be
assigned to the various combinations of dyadic products of fields .
The \(2N\times 2N\) matrix and order parameter
$\hat{R}$ (\ref{s2_37}) consists of two
\((N=L+S)\times (N=L+S)\) density block super-matrices \(\hat{R}^{11}\), \(\hat{R}^{22}\)
(\ref{s2_39}), as the described super-matrix $\hat{N}$ (\ref{s2_14}-\ref{s2_18}),
and the two \((N=L+S)\times (N=L+S)\) super-matrices $\hat{R}^{12}$, $\hat{R}^{21}$ (\ref{s2_40})
for the anomalous terms as a molecular and BCS-like condensate.
Note the splitted distribution of the boson-boson and
fermion-fermion blocks according to the dyadic product of the doubled supervector fields
\(\Psi_{\vec{x},\alpha}^{a}(t_{p})\) into the matrices
$\wtilde{B}_{L\times L}$, $\wtilde{B}^{\raisebox{-3pt}{$_{T}$}}_{L\times L}$ and
$\wtilde{c}_{L\times L}$, $\wtilde{c}^{\raisebox{-3pt}{$_{+}$}}_{L\times L}$
for the bosonic part, and the even matrices
$\wtilde{F}_{S\times S}$, $-\wtilde{F}^{\raisebox{-3pt}{$_{T}$}}_{S\times S}$
with the anomalous parts $\wtilde{f}_{S\times S}$,
$\wtilde{f}_{S\times S}^{\raisebox{-3pt}{$_{+}$}}$ for the fermion-fermion part, respectively
\footnote{The spatial vectors \(\vec{x}\), \(\vec{x}\ppr\) and the time dependence $t_{p}$ of
\(\hat{R}_{\vec{x},\alpha;\vec{x}\ppr,\beta}^{ab}(t_{p})\) (\ref{s2_37}-\ref{s2_40}) have been omitted in
$\wtilde{B}$, $\wtilde{c}$, $\wtilde{F}$, $\wtilde{f}$ and
$\wtilde{\chi}$, $\wtilde{\chi}^{+}$, $\wtilde{\eta}$, $\wtilde{\eta}^{+}$ in order
to concentrate onto the 'Nambu'-doubling (\(a,b=1,2\)) for spontaneous symmetry breaking and
the angular momentum space with indices \((\alpha,\beta=-l,\ldots,+l;-s,\ldots,+s)\).}
\beq \lb{s2_37}
\hat{R}_{2N\times 2N}&=&\left( \bea{cc}
\hat{R}^{11}_{(L+S)\times(L+S)} & \hat{R}^{12}_{(L+S)\times(L+S)} \\
\hat{R}^{21}_{(L+S)\times(L+S)} & \hat{R}^{22}_{(L+S)\times(L+S)}
\eea\right) \\ \lb{s2_38}
\hat{R}_{\vec{x},\alpha;\vec{x}\ppr,\beta}^{ab}(t_{p})&=&\left( \bea{cc} \left(\bea{cc}
\wtilde{B}_{L\times L} & \wtilde{\chi}^{+}_{L\times S} \\
\wtilde{\chi}_{S\times L} & \wtilde{F}_{S\times S} \eea\right)^{11} & \left(\bea{cc}
\wtilde{c}_{L\times L} & \wtilde{\eta}^{T}_{L\times S} \\
\wtilde{\eta}_{S\times L} & \wtilde{f}_{S\times S}
\eea\right)^{12} \\
\left(\bea{cc}
\wtilde{c}^{+}_{L\times L} & \wtilde{\eta}^{+}_{L\times S} \\
\wtilde{\eta}^{*}_{S\times L} & \wtilde{f}^{+}_{S\times S} \eea\right)^{21} & \left(\bea{cc}
\wtilde{B}^{T}_{L\times L} & \wtilde{\chi}^{T}_{L\times S} \\
\wtilde{\chi}^{*}_{S\times L} & -\wtilde{F}^{T}_{S\times S} \eea\right)^{22}
\eea\right) \\ \lb{s2_39}
\hat{R}_{\vec{x},\alpha;\vec{x}\ppr,\beta}^{11}(t_{p})&=&\left( \bea{cc}
\wtilde{B}_{L\times L} & \wtilde{\chi}^{+}_{L\times S}  \\
\wtilde{\chi}_{S\times L} & \wtilde{F}_{S\times S} \eea\right) \hspace*{1.0cm}
\hat{R}_{\vec{x},\alpha;\vec{x}\ppr,\beta}^{22}(t_{p})=\left( \bea{cc}
\wtilde{B}^{T}_{L\times L} & \wtilde{\chi}^{T}_{L\times S} \\
\wtilde{\chi}^{*}_{S\times L} & -\wtilde{F}^{T}_{S\times S}
\eea\right) \\ \lb{s2_40}
\hat{R}_{\vec{x},\alpha;\vec{x}\ppr,\beta}^{12}(t_{p})&=&\left( \bea{cc}
\wtilde{c}_{L\times L} & \wtilde{\eta}^{T}_{L\times S} \\
\wtilde{\eta}_{S\times L} & \wtilde{f}_{S\times S} \\
\eea\right) \hspace*{1.3cm} \hat{R}_{\vec{x},\alpha;\vec{x}\ppr,\beta}^{21}(t_{p})=\left( \bea{cc}
\wtilde{c}^{+}_{L\times L} & \wtilde{\eta}^{+}_{L\times S} \\
\wtilde{\eta}^{*}_{S\times L} & \wtilde{f}^{+}_{S\times S} \eea\right)_{\mbox{.}}
\eeq
The commuting hermitian matrix $\wtilde{B}_{L\times L}$ (\ref{s2_41}) in
$\hat{R}^{11}_{N\times N}$, $\hat{R}^{22}_{N\times N}$ (\ref{s2_39})
is related to density terms of the bosons, and
the even hermitian matrix \(\wtilde{F}_{N\times N}\) (\ref{s2_43}) in the
fermion-fermion block corresponds to density terms of fermions.
There are also anti-commuting density terms
\(\wtilde{\chi}_{S\times L}\) (\ref{s2_42}) which result from the combination of bosons and complex
conjugated Grassmann fields. The
complex symmetric matrix \(\wtilde{c}_{L\times L}\) (\ref{s2_44})
in $\hat{R}^{12}_{N\times N}$, $\hat{R}^{21}_{N\times N}$ (\ref{s2_40}) is composed of bilinear fields
\(b_{\vec{x},m}(t_{p})\;b_{\vec{x}\ppr,n}(t_{p})\) as anomalous terms for the molecular condensate. In analogy one has
the antisymmetric matrix \(\wtilde{f}_{S\times S}\) (\ref{s2_46}) in
$\hat{R}^{12}_{N\times N}$, $\hat{R}^{21}_{N\times N}$
for the anomalous combination of two fermion fields for a BCS-condensate.
An anti-commuting anomalous term \(\wtilde{\eta}_{S\times L}\) (\ref{s2_45})
is also obtained from the coupling between a bosonic \(b_{\vec{x},m}(t_{p})\) and
fermionic field \(\alpha_{\vec{x},r}(t_{p})\)
\beq \lb{s2_41}
\wtilde{B}_{\vec{x},\vec{x}\ppr;L\times L}(t_{p})&=& \big\{b_{\vec{x},m}(t_{p})\otimes
b_{\vec{x}\ppr,n}^{*}(t_{p})\big\}\hspace*{0.75cm}
\wtilde{B}^{+}_{L\times L}=\wtilde{B}_{L\times L} \\ \lb{s2_42}
\wtilde{\chi}^{+}_{\vec{x},\vec{x}\ppr;L\times S}(t_{p})&=&
\big\{b_{\vec{x},m}(t_{p})\otimes \alpha_{\vec{x}\ppr,r\ppr}^{*}(t_{p}) \big\} \hspace*{0.75cm}
\wtilde{\chi}_{\vec{x},\vec{x}\ppr;S\times L}(t_{p})=
\big\{\alpha_{\vec{x},r}(t_{p})\otimes b_{\vec{x}\ppr,n}^{*}(t_{p})  \big\}  \\ \lb{s2_43}
\wtilde{F}_{\vec{x},\vec{x}\ppr;S\times S}(t_{p})&=&
\big\{\alpha_{\vec{x},r}(t_{p})\otimes\alpha_{\vec{x}\ppr,r\ppr}^{*}(t_{p})\big\}\hspace*{0.75cm}
\wtilde{F}^{+}_{S\times S}=\wtilde{F}_{S\times S} \\ \lb{s2_44}
\wtilde{c}_{\vec{x},\vec{x}\ppr;L\times L}(t_{p})&=&\big\{ b_{\vec{x},m}(t_{p})\otimes
b_{\vec{x}\ppr,n}(t_{p})\big\}\hspace*{0.75cm}
\wtilde{c}^{T}_{L\times L}=\wtilde{c}_{L\times L} \\ \lb{s2_45}
\wtilde{\eta}^{T}_{\vec{x},\vec{x}\ppr;L\times S}(t_{p})&=&\big\{
b_{\vec{x},m}(t_{p})\otimes \alpha_{\vec{x}\ppr,r\ppr}(t_{p})\big\}\hspace*{0.75cm}
\wtilde{\eta}_{\vec{x},\vec{x}\ppr;S\times L}(t_{p})=\big\{\alpha_{\vec{x},r}(t_{p})
\otimes b_{\vec{x}\ppr,n}(t_{p})\big\}  \\ \lb{s2_46}
\wtilde{f}_{\vec{x},\vec{x}\ppr;S\times S}(t_{p})&=&\big\{
\alpha_{\vec{x},r}(t_{p})\otimes\alpha_{\vec{x}\ppr,r\ppr}(t_{p})\big\}\hspace*{0.75cm}
\wtilde{f}^{T}_{S\times S}=-\wtilde{f}_{S\times S}\;\;.
\eeq
The symmetries of the entries in the order parameter
\(\hat{R}_{\vec{x},\alpha;\vec{x}\ppr,\beta}^{ab}(t_{p})\) (\ref{s2_37}-\ref{s2_40})
have to be maintained in the self-energy matrix
\(\hat{\Sigma}_{\vec{x},\alpha;\vec{x}\ppr,\beta}^{ab}(t_{p})\) for a Hubbard-Stratonovich transformation (HST) of the
interaction with the four super-fields and the potential \(V_{|\vec{x}\ppr-\vec{x}|}\). We introduce the nonlocal
self-energy \(\hat{\Sigma}_{\vec{x},\alpha;\vec{x}\ppr,\beta}^{ab}(t_{p})\) (\ref{s2_47},\ref{s2_48})
with analogous matrix elements as in
\(\hat{R}_{\vec{x},\alpha;\vec{x}\ppr,\beta}^{ab}(t_{p})\) (\ref{s2_37}-\ref{s2_40})
\footnote{The matrix elements of \(\hat{\Sigma}_{\vec{x},\alpha;\vec{x}\ppr,\beta}^{ab}(t_{p})\)
are marked without a tilde '$\wtilde{\ph{\Sigma}}$' in correspondence to
\(\hat{R}_{\vec{x},\alpha;\vec{x}\ppr,\beta}^{ab}(t_{p})\).}. The matrix elements in
\(\hat{\Sigma}_{\vec{x},\alpha;\vec{x}\ppr,\beta}^{ab}(t_{p})\) (\ref{s2_47},\ref{s2_48})
are not composed of dyadic products as the entries in
\(\hat{R}_{\vec{x},\alpha;\vec{x}\ppr,\beta}^{ab}(t_{p})\), but contain the same
symmetry relations (\ref{s2_51}) between the entries as in (\ref{s2_41}-\ref{s2_46})
for \(\hat{R}_{\vec{x},\alpha;\vec{x}\ppr,\beta}^{ab}(t_{p})\).
Apart from the analogous density terms \(\hat{\Sigma}_{\vec{x},\alpha;\vec{x}\ppr,\beta}^{11}(t_{p})\),
\(\hat{\Sigma}_{\vec{x},\alpha;\vec{x}\ppr,\beta}^{22}(t_{p})\) (\ref{s2_49}),
there are the anomalous terms \(\hat{\Sigma}_{\vec{x},\alpha;\vec{x}\ppr,\beta}^{12}(t_{p})\) and
\(\hat{\Sigma}_{\vec{x},\alpha;\vec{x}\ppr,\beta}^{21}(t_{p})\) (\ref{s2_50}) for the molecular and BCS condensates
\beq \lb{s2_47}
\hat{\Sigma}_{2N\times 2N}&=&\left(
\bea{cc}
\hat{\Sigma}^{11}_{(L+S)\times(L+S)} & \hat{\Sigma}^{12}_{(L+S)\times(L+S)} \\
\hat{\Sigma}^{21}_{(L+S)\times(L+S)} & \hat{\Sigma}^{22}_{(L+S)\times(L+S)}
\eea\right) \\ \lb{s2_48}
\hat{\Sigma}_{\vec{x},\alpha;\vec{x}\ppr,\beta}^{ab}(t_{p})&=&\left( \bea{cc} \left(\bea{cc}
\hat{B}_{L\times L} & \hat{\chi}^{+}_{L\times S} \\
\hat{\chi}_{S\times L} & \hat{F}_{S\times L} \eea\right)^{11} &
\left(\bea{cc}
\hat{c}_{L\times L} & \hat{\eta}^{T}_{L\times S} \\
\hat{\eta}_{S\times L} & \hat{f}_{S\times S}
\eea\right)^{12} \\
\left(\bea{cc}
\hat{c}^{+}_{L\times L} & \hat{\eta}^{+}_{L\times S} \\
\hat{\eta}^{*}_{S\times L} & \hat{f}^{+}_{S\times S}
\eea\right)^{21} & \left(\bea{cc}
\hat{B}^{T}_{L\times L} & \hat{\chi}^{T}_{L\times S} \\
\hat{\chi}^{*}_{S\times L} & -\hat{F}^{T}_{S\times S}
\eea\right)^{22} \eea\right)   \\ \lb{s2_49}
\hat{\Sigma}_{\vec{x},\alpha;\vec{x}\ppr,\beta}^{11}(t_{p})&=&\left(
\bea{cc}
\hat{B}_{L\times L} & \hat{\chi}^{+}_{L\times S}  \\
\hat{\chi}_{S\times L} & \hat{F}_{S\times S} \eea\right) \hspace*{1.3cm}
\hat{\Sigma}_{\vec{x},\alpha;\vec{x}\ppr,\beta}^{22}(t_{p})=\left( \bea{cc}
\hat{B}^{T}_{L\times L} & \hat{\chi}^{T}_{L\times S} \\
\hat{\chi}^{*}_{S\times L} & -\hat{F}^{T}_{S\times S}
\eea\right) \\ \lb{s2_50}
\hat{\Sigma}_{\vec{x},\alpha;\vec{x}\ppr,\beta}^{12}(t_{p})&=&\left( \bea{cc}
\hat{c}_{L\times L} & \hat{\eta}^{T}_{L\times S} \\
\hat{\eta}_{S\times L} & \hat{f}_{S\times S} \\
\eea\right) \hspace*{1.5cm} \hat{\Sigma}_{\vec{x},\alpha;\vec{x}\ppr,\beta}^{21}(t_{p})=\left( \bea{cc}
\hat{c}^{+}_{L\times L} & \hat{\eta}^{+}_{L\times S} \\
\hat{\eta}^{*}_{S\times L} & \hat{f}^{+}_{S\times S} \eea\right) \\ \lb{s2_51} &&
\hat{B}^{+}_{L\times L}=\hat{B}_{L\times L}\;\;;\;\;\hat{F}^{+}_{S\times S}=\hat{F}_{S\times S}
\;\;\;;\;\;\;\hat{c}^{T}_{L\times L}=\hat{c}_{L\times L}\;\;;\;\;
\hat{f}^{T}_{S\times S}=-\hat{f}_{S\times S}\;.
\eeq
The Hubbard-Stratonovich transformation exchanges the
quartic interaction of fields \(\psi_{\vec{x},\alpha}(t_{p})\) with a super-trace relation (\ref{s2_52})
of the matrices \(\hat{R}_{\vec{x},\alpha;\vec{x}\ppr,\beta}^{ab}(t_{p})\) and
\(\hat{\Sigma}_{\vec{x},\alpha;\vec{x}\ppr,\beta}^{ab}(t_{p})\)
\beq \lb{s2_52}
\lefteqn{\exp\bigg\{-\frac{\im}{\hbar}\int_{C}d t_{p}\sum_{\vec{x},\alpha;\vec{x}\ppr,\beta}
\psi_{\vec{x}\ppr,\beta}^{*}(t_{p})\;\psi_{\vec{x},\alpha}^{*}(t_{p})\;V_{|\vec{x}\ppr-\vec{x}|}\;
\psi_{\vec{x},\alpha}(t_{p})\;\psi_{\vec{x}\ppr,\beta}(t_{p})\bigg\}=} \\ \no &=&
\exp\bigg\{-\frac{\im}{4\hbar}\int_{C}d t_{p}\sum_{\vec{x},\vec{x}\ppr}
V_{|\vec{x}-\vec{x}\ppr|}\;
\STRAB\bigg[\hat{K}\;\hat{R}_{\vec{x},\alpha;\vec{x}\ppr,\beta}^{ab}(t_{p})\;\hat{K}\;
\hat{R}_{\vec{x}\ppr,\beta;\vec{x},\alpha}^{ba}(t_{p})\bigg]\bigg\}= \\ \no &=&
\int d[\hat{\Sigma}(t_{p})]\;\exp\bigg\{\frac{\im}{4\hbar}\int_{C} dt_{p}
\sum_{\vec{x},\vec{x}\ppr}\frac{1}{V_{|\vec{x}-\vec{x}\ppr|}}\;
\STRAB\bigg[\hat{\Sigma}_{\vec{x},\alpha;\vec{x}\ppr,\beta}^{ab}(t_{p})\;\hat{K}\;
\hat{\Sigma}_{\vec{x}\ppr,\beta;\vec{x},\alpha}^{ba}(t_{p})\;\hat{K}\bigg]\bigg\}
\\ \no &\times& \exp\bigg\{-\frac{\im}{2\hbar}\int_{C}d
t_{p}\sum_{\vec{x},\vec{x}\ppr} \STRAB\bigg[\hat{R}_{\vec{x},\alpha;\vec{x}\ppr,\beta}^{ab}(t_{p})\;
\hat{K}\;\hat{\Sigma}_{\vec{x}\ppr,\beta;\vec{x},\alpha}^{ba}(t_{p})\;\hat{K}\bigg]\bigg\}_{\mbox{.}}
\eeq
The quartic interaction in (\ref{s2_24}) is substituted with the HST transformation (\ref{s2_52}) and the
self-energy (\ref{s2_47}-\ref{s2_51}) so that
the generating function \(Z[\hat{\mcal{J}},j_{\psi},\wtilde{j}_{\psi\psi}]\) (\ref{s2_24}) is converted
into a path integral of \(\hat{\Sigma}_{\vec{x},\alpha;\vec{x}\ppr,\beta}^{ab}(t_{p})\)
and a bilinear relation of the doubled supervector \(\Psi_{\vec{x},\alpha}^{a}(t_{p})\).
A similar doubling of the one-particle part as for the quartic interaction (\ref{s2_33}) leads to an exchange
of the field \(\psi_{\vec{x},\alpha}(t_{p})\) with its complex valued part
\(\psi_{\vec{x}\ppr,\beta}^{*}(t_{p})\). In order to retain the same value as the
original one-particle part, one has to take
the transpose $\hat{H}_{p}^{T}(\vec{x},t_{p})$ of the one-particle operator (\ref{s2_25},\ref{s2_3}) and has to include an
additional minus for the exchange of the fermionic fields which is considered with the metric
\(\hat{\kappa}_{N\times N}\) (\ref{s2_35},\ref{s2_34})
\beq \lb{s2_53}
\lefteqn{\sum_{\vec{x}}\sum_{\alpha}
\psi_{\vec{x},\alpha}^{*}(t_{p})\;\;\hat{H}_{p}(\vec{x},t_{p})\;\;
\psi_{\vec{x},\alpha}(t_{p})=
\sum_{\vec{x}}\sum_{\alpha}
\psi_{\vec{x},\alpha}^{T}(t_{p})\;\;\hat{H}_{p}^{T}(\vec{x},t_{p})\;\;\hat{\kappa}_{\alpha\alpha}\;\;
\psi_{\vec{x},\alpha}^{*}(t_{p}) =}
 \\ \no &=& \frac{1}{2}
\sum_{\vec{x}}\sum_{\alpha}\sum_{a,b=1,2}
\Psi_{\vec{x},\alpha}^{+b}(t_{p})\;\;\left(
\bea{cc}
\hat{H}_{p}(\vec{x},t_{p}) & 0 \\
0 & \hat{H}_{p}^{T}(\vec{x},t_{p})\;\hat{\kappa}_{\alpha\alpha}
\eea\right)^{ba}\;\;\Psi_{\vec{x},\alpha}^{a}(t_{p})\;\;\;\;.
\eeq
The application of the HST
relation (\ref{s2_52}) and the doubled one-particle part (\ref{s2_53}) results in a coherent state path integral
\(Z[\hat{\mcal{J}},J_{\psi},\wtilde{J}_{\psi\psi}]\) (\ref{s2_54}) with a bilinear relation of the doubled
supervector field $\Psi_{\vec{x},\alpha}^{a}(t_{p})$. The fields $\Psi_{\vec{x},\alpha}^{a}(t_{p})$ are coupled with
the doubled one-particle part $\hat{\mcal{H}}_{\vec{x}\ppr,\beta;\vec{x},\alpha}^{ba}(t_{q}\ppr,t_{p})$
(\ref{s2_55},\ref{s2_53}), the anomalous source matrix
$\wtilde{J}_{\psi\psi;\alpha\beta}^{ab}(\vec{x},t_{p})$
and the self-energy \(\hat{\Sigma}_{\vec{x},\alpha;\vec{x}\ppr,\beta}^{ab}(t_{p})\)
(\ref{s2_47}-\ref{s2_51}). The common interaction term between bosons and fermions with potential
$V_{|\vec{x}-\vec{x}\ppr|}$ is transformed to the super-trace relation with the quadratic nonlocal self-energy
(\ref{s2_48}) and the metric \(\hat{K}_{2N\times 2N}=\big(\hat{1}_{N\times N};\hat{\kappa}_{N\times N}\big)\)
(\ref{s2_34},\ref{s2_35})
\footnote{The parameter $\mcal{N}_{x}=(L/\Delta x)^{d}$ denotes the total number of {\it spatial}
points of the underlying grid, following from discrete spatial points.}
\beq \lb{s2_54}
\lefteqn{Z[\mcal{J},J_{\psi},\wtilde{J}_{\psi\psi}]=
\int d[\hat{\Sigma}(t_{p})]\;\exp\left\{\frac{\im}{4\hbar}\int_{C}d t_{p}
\sum_{\vec{x},\vec{x}\ppr}\frac{1}{V_{|\vec{x}-\vec{x}\ppr|}}\STRAB\left[
\hat{\Sigma}_{\vec{x},\alpha;\vec{x}\ppr,\beta}^{ab}(t_{p})\;
\hat{K}\;\hat{\Sigma}_{\vec{x}\ppr,\beta;\vec{x},\alpha}^{ba}(t_{p})\;\hat{K}\right]\right\}
 } \\ \no &\times &\int d[\psi_{\vec{x},\alpha}(t_{p})]\;
\exp\Bigg\{-\frac{\im}{2\hbar}\int_{C}d t_{p}d t_{q}\ppr\sum_{\vec{x},\vec{x}\ppr}\mcal{N}_{x}\;
\Psi_{\vec{x}\ppr,\beta}^{+b}(t_{q}\ppr)
\bigg[\hat{\mcal{H}}_{\vec{x}\ppr,\beta;\vec{x},\alpha}^{ba}(t_{q}\ppr,t_{p})\hat{K}+
\frac{\hat{\mcal{J}}_{\vec{x}\ppr,\beta;\vec{x},\alpha}^{ba}(t_{q}\ppr,t_{p})}{\mcal{N}_{x}} + \\ \no &+&
\delta_{p,q}\;\eta_{p}\;\delta(t_{p}-t_{q}\ppr)\bigg[\delta_{\vec{x},\vec{x}\ppr}
\bigg( \bea{cc}
0 & \wtilde{j}_{\psi\psi;\beta\alpha}(\vec{x},t_{p}) \\
\wtilde{j}_{\psi\psi;\beta\alpha}^{+}(\vec{x},t_{p}) & 0
\eea \bigg)^{ba} +
\hat{K}\;\hat{\Sigma}_{\vec{x}\ppr,\beta;\vec{x},\alpha}^{ba}(t_{p})\;\hat{K}\bigg]\bigg]
\Psi_{\vec{x},\alpha}^{a}(t_{p})\Bigg\}
\;\times \\ \no &\times& \exp\Bigg\{-\frac{\im}{2\hbar}\int_{C}d t_{p}
\sum_{\vec{x}}\bigg(
J_{\psi;\alpha}^{+a}(\vec{x},t_{p})\;\hat{K}\;\Psi_{\vec{x},\alpha}^{a}(t_{p})+
\Psi_{\vec{x},\alpha}^{+a}(t_{p})\;\hat{K}\;J_{\psi;\alpha}^{a}(\vec{x},t_{p})\bigg)\Bigg\}
\eeq
\beq \lb{s2_55}
\hat{\mcal{H}}_{\vec{x}\ppr,\beta;\vec{x},\alpha}^{ba}(t_{q}\ppr,t_{p})&=&
\mbox{diag}\left(\hat{\mcal{H}}_{\vec{x}\ppr,\beta;\vec{x},\alpha}^{11}(t_{q}\ppr,t_{p})\;;\;
\hat{\mcal{H}}_{\vec{x}\ppr,\beta;\vec{x},\alpha}^{22}(t_{q}\ppr,t_{p})\right)= \\ \no &=&
\delta_{p,q}\;\eta_{p}\;\delta(t_{p}-t_{q}\ppr)\;\delta_{\vec{x},\vec{x}\ppr}\;\;
\mbox{diag}\left(\hat{H}_{p}(\vec{x},t_{p})\;\hat{1}_{N\times N}\;;\;
\hat{H}_{p}^{T}(\vec{x},t_{p})\;\hat{1}_{N\times N}\right) \\ \lb{s2_56}
\hat{H}_{p}(\vec{x},t_{p})&=&-\im\hbar\frac{\pp}{\pp t_{p}}-\im\;\ve_{p}+
\frac{\vec{p}^{\;2}}{2m}+u(\vec{x})-\mu_{0}  \\ \lb{s2_57}
\hat{H}_{p}^{T}(\vec{x},t_{p})&=&+\im\hbar\frac{\pp}{\pp t_{p}}-\im\;\ve_{p}+
\frac{\vec{p}^{\;2}}{2m}+u(\vec{x})-\mu_{0}\;\;\;.
\eeq
In the case of a Fourier transformation from time $t_{p}$ to frequency $\omega_{p}$, the transposed time
derivative \((-\im\hbar\pp/\pp t_{p})^{T}=+\im\hbar\pp/\pp t_{p}\) in
$\hat{H}_{p}^{T}(\vec{x},t_{p})$ (\ref{s2_57},\ref{s2_55})
acts on the complex conjugated fields
\(\Psi_{\vec{x},\alpha}^{a=2}(t_{p})=\psi_{\vec{x},\alpha}^{*}(t_{p})\) so that the same sign of the
energy is obtained as in the $\hat{\mcal{H}}^{11}$ part with
\(\Psi_{\vec{x},\alpha}^{a=1}(t_{p})=\psi_{\vec{x},\alpha}(t_{p})\). Apart from the 'Nambu'-doubling
of the one-particle part (\ref{s2_53}) and the interaction (\ref{s2_33}) in (\ref{s2_54}), the source terms
(\ref{s2_4},\ref{s2_9}-\ref{s2_12}) for a coherent condensate wave function and the anomalous terms have also to be
extended with the complex valued fields \(\psi_{\vec{x},\alpha}^{*}(t_{p})\) so that the generating
function (\ref{s2_54}) is only composed of doubled super-fields \(\Psi_{\vec{x},\alpha}^{a(=1/2)}(t_{p})\)
\beq \lb{s2_58}
J_{\psi;\alpha}^{a(=1/2)}(\vec{x},t_{p})&=&\Big(\underbrace{j_{\psi;\alpha}(\vec{x},t_{p})}_{a=1}\;;\;
\underbrace{j_{\psi;\alpha}^{*}(\vec{x},t_{p})}_{a=2}\Big) \\ \lb{s2_59}
\wtilde{J}_{\psi\psi;\alpha\beta}^{ab}(\vec{x},t_{p}) &=&
\left( \bea{cc}
0 & \wtilde{j}_{\psi\psi;\alpha\beta}(\vec{x},t_{p}) \\
\wtilde{j}_{\psi\psi;\alpha\beta}^{+}(\vec{x},t_{p}) & 0
\eea\right)^{ab} \\ \lb{s2_60}
\wtilde{j}_{\psi\psi;\alpha\beta}(\vec{x},t_{p})&=&\left( \bea{cc}
\hat{j}_{b;L\times L}(\vec{x},t_{p}) & -\hat{j}_{\eta;L\times S}^{T}(\vec{x},t_{p}) \\
\hat{j}_{\eta;S\times L}(\vec{x},t_{p}) & -\hat{j}_{f;S\times S}(\vec{x},t_{p}) \eea\right) \\ \no &&
\hat{j}_{b;L\times L}^{T}(\vec{x},t_{p})=\hat{j}_{b;L\times L}(\vec{x},t_{p})\;;\hspace*{0.75cm}
\hat{j}_{f;S\times S}^{T}(\vec{x},t_{p})=-\hat{j}_{f;S\times S}(\vec{x},t_{p})\;\;\;.
\eeq
The bilinear super-fields \(\Psi_{\vec{x},\alpha}^{a}(t_{p})\) are removed by integration in (\ref{s2_54})
so that the inverse square root of a super-determinant  is obtained. Moreover, a bilinear relation with
the source term \(J_{\psi;\alpha}^{a}(\vec{x},t_{p})\) (\ref{s2_58}) follows for a possible coherent
condensate wave function \footnote{Referring to a definition of the coherent state path integral (\ref{s2_61})
with discrete time variables, one has to introduce the frequency scale \(\Omega=1/\Delta t\) of the
applied interval \(\Delta t\) for the time steps.}
\beq \no
\lefteqn{\hspace*{-1.27cm}Z[\hat{\mcal{J}},J_{\psi},\hat{J}_{\psi\psi}]=
\int d[\hat{\Sigma}(t_{p})]\;\exp\left\{\frac{\im}{4\hbar}\int_{C}d t_{p}
\sum_{\vec{x},\vec{x}\ppr}\frac{1}{V_{|\vec{x}-\vec{x}\ppr|}}\STRAB\left[
\hat{\Sigma}_{\vec{x},\alpha;\vec{x}\ppr,\beta}^{ab}(t_{p})\;
\hat{K}\;\hat{\Sigma}_{\vec{x}\ppr,\beta;\vec{x},\alpha}^{ba}(t_{p})\;\hat{K}\right]\right\}  }
\\ \lb{s2_61} &\times& \Bigg\{\mbox{SDET}\Bigg(
\bigg[\hat{\mcal{H}}\hat{K}+\hat{\eta}
\underbrace{\left( \bea{cc}
0 & \hat{j}_{\psi\psi} \\
\hat{j}_{\psi\psi}^{+} & 0 \eea\right)}_{\hat{J}_{\psi\psi}}
+\hat{K}\;\hat{\eta}\;\frac{\hat{\mcal{J}}_{\beta\alpha}^{ba}}{\mcal{N}_{x}}\;\hat{\eta}\;\hat{K}+
\hat{\eta}\;\hat{\Sigma}(t_{p})
\bigg]_{\vec{x}\ppr,\beta;\vec{x},\alpha}^{ba}\hspace*{-0.64cm}(t_{q}\ppr,t_{p})\Bigg)
\Bigg\}^{\mathbf{-1/2}}
\\ \no &\times& \exp\Bigg\{\frac{\im}{2\hbar}\Omega^{2}\int_{C}d t_{p}\;d t_{q}\ppr
\sum_{\vec{x},\vec{x}\ppr}\mcal{N}_{x}\; J_{\psi;\beta}^{+b}(\vec{x}\ppr,t_{q}\ppr)\;\times  \\ \no &\times&
\bigg[\hat{\mcal{H}}\hat{K}+\hat{\eta}
\underbrace{\left( \bea{cc}
0 & \hat{j}_{\psi\psi} \\
\hat{j}_{\psi\psi}^{+} & 0 \eea\right)}_{\hat{J}_{\psi\psi}}
+\hat{K}\;\hat{\eta}\;\frac{\hat{\mcal{J}}_{\beta\alpha}^{ba}}{\mcal{N}_{x}}\;\hat{\eta}\;\hat{K}+
\hat{\eta}\;\hat{\Sigma}(t_{p})
\bigg]_{\vec{x}\ppr,\beta;\vec{x},\alpha}^{\mathbf{-1},\;ba}\hspace*{-0.64cm}(t_{q}\ppr,t_{p})\;\;\;
J_{\psi;\alpha}^{a}(\vec{x},t_{p})\Bigg\}\;.
\eeq
Note that the super-determinant 'SDET' in (\ref{s2_61}) is applied to the sum of the doubled
one-particle part \(\hat{\mcal{H}}\;\hat{K}\) and the self-energy \(\hat{\Sigma}_{2N\times 2N}(t_{p})\)
which has splitted boson-boson, fermion-fermion, as well as boson-fermion and fermion-boson blocks
in the super-matrices \(\hat{\Sigma}_{N\times N}^{11}(t_{p})\), \(\hat{\Sigma}_{N\times N}^{22}(t_{p})\)
for the densities and corresponding anomalous parts \(\hat{\Sigma}_{N\times N}^{12}(t_{p})\),
\(\hat{\Sigma}_{N\times N}^{21}(t_{p})\). Therefore, the 'SDET' relation in (\ref{s2_61})
generalizes the 'sdet' relation in (\ref{s2_18}) for a single super-matrix with super-trace (\ref{s2_16})
to a relation with super-trace 'STR' (\ref{s2_31}) for a block structure of four combined super-matrices
as in \(\hat{\Phi}_{\alpha\beta}^{ab}\) (\ref{s2_29},\ref{s2_31})
\be \lb{s2_62}
\mbox{sdet}\big(\hat{N}_{\alpha\beta}\big)=\exp\big\{\strab\ln \hat{N}_{\alpha\beta}\big\}\longrightarrow
\mbox{SDET}\big(\hat{\Phi}_{\alpha\beta}^{ab}\big)=\exp\big\{\STRAB\ln \hat{\Phi}_{\alpha\beta}^{ab}\big\}\;\;\;.
\ee
Furthermore, a transformation of the self-energy
\(\hat{K}\;\hat{\Sigma}(t_{p})\;\hat{K}\rightarrow\hat{\Sigma}(t_{p})\)
has been performed in \(Z[\hat{\mcal{J}},J_{\psi},\hat{J}_{\psi\psi}]\) (\ref{s2_61})
so that this transformation leads to modified source terms
\(\hat{J}_{\psi\psi;\alpha\beta}^{ab}(\vec{x},t_{p})\)
\beq \lb{s2_63}
\hat{J}_{\psi\psi;\alpha\beta}^{ab}(\vec{x},t_{p})&=&
\left( \bea{cc}
0 & \hat{j}_{\psi\psi;\alpha\beta}(\vec{x},t_{p}) \\
\hat{j}_{\psi\psi;\alpha\beta}^{+}(\vec{x},t_{p}) & 0 \eea\right)^{ab} \\ \lb{s2_64}
\hat{j}_{\psi\psi;\alpha\beta}(\vec{x},t_{p})&=&\left( \bea{cc}
\hat{j}_{b;L\times L}(\vec{x},t_{p}) & \hat{j}_{\eta;L\times S}^{T}(\vec{x},t_{p}) \\
\hat{j}_{\eta;S\times L}(\vec{x},t_{p}) & \hat{j}_{f;S\times S}(\vec{x},t_{p})
\eea\right) \\ \no && \hat{j}_{b;L\times L}^{T}(\vec{x},t_{p})=
\hat{j}_{b;L\times L}(\vec{x},t_{p})\hspace*{0.75cm}
\hat{j}_{f;S\times S}^{T}(\vec{x},t_{p})=-\hat{j}_{f;S\times S}(\vec{x},t_{p})\;.
\eeq
The original coherent state path integral
\(Z[\hat{\mcal{J}},j_{\psi},\wtilde{j}_{\psi\psi}]\) (\ref{s2_24})
with fields \(\psi_{\vec{x},\alpha}(t_{p})\), \(\psi_{\vec{x},\alpha}^{*}(t_{p})\)
has been converted to a generating function \(Z[\hat{\mcal{J}},J_{\psi},\hat{J}_{\psi\psi}]\)
(\ref{s2_61}) with spatially nonlocal self-energy matrices in 'Nambu'-doubled form (\ref{s2_47}-\ref{s2_51}).
However, the assumed short-ranged character of the interaction potential $V_{|\vec{x}-\vec{x}\ppr|}$
results in strong oscillations for the term with the super-trace of the product of two
self-energy matrices. Therefore, we suppose that the oscillating phases cancel
in \(Z[\hat{\mcal{J}},J_{\psi},\hat{J}_{\psi\psi}]\) (\ref{s2_61}) for an exceeding
interaction range \(r_{0}\) of the nonlocal spatial variables of the self-energy matrix
\(\hat{\Sigma}_{\vec{x},\alpha;\vec{x}\ppr,\beta}^{ab}(t_{p})\)
\be \lb{s2_65}
\frac{1}{V_{|\vec{x}-\vec{x}\ppr|}}\rightarrow\infty\hspace*{0.5cm}\mbox{for }|\vec{x}-\vec{x}\ppr|>r_{0}\;.
\ee
A spatially local part of the self-energy \(\hat{\Sigma}_{\alpha\beta}^{ab}(\vec{x},t_{p})\) can be
separated  by integration over self-energy matrix elements with \(|\vec{x}-\vec{x}\ppr|>r_{0}\)
in \(Z[\hat{\mcal{J}},J_{\psi},\hat{J}_{\psi\psi}]\) (\ref{s2_61}).
In the remainder we neglect the coupling between the local and nonlocal parts in the integration
over elements for \(|\vec{x}-\vec{x}\ppr|>r_{0}\) and only retain the spatially local part
of the self-energy in the generating function (\ref{s2_61})
\beq \lb{s2_66}
\hat{\Sigma}_{\vec{x},\alpha;\vec{x}\ppr,\beta}^{ab}(t_{p})
&\rightarrow&\hat{\Sigma}_{\alpha\beta}^{ab}(\vec{x},t_{p})\;\;\delta_{\vec{x},\vec{x}\ppr} \\ \lb{s2_67}
\hat{\Sigma}_{\alpha\beta}^{ab}(\vec{x},t_{p}) &=&\left( \bea{cc}
\left(\bea{cc}
\hat{B}_{L\times L}(\vec{x},t_{p}) & \hat{\chi}^{+}_{L\times S}(\vec{x},t_{p}) \\
\hat{\chi}_{S\times L}(\vec{x},t_{p}) & \hat{F}_{S\times S}(\vec{x},t_{p})
\eea\right)^{11} &
\left(\bea{cc}
\hat{c}_{L\times L}(\vec{x},t_{p}) & \hat{\eta}^{T}_{L\times S}(\vec{x},t_{p}) \\
\hat{\eta}_{S\times L}(\vec{x},t_{p}) & \hat{f}_{S\times S}(\vec{x},t_{p})
\eea\right)^{12} \\
\left(\bea{cc}
\hat{c}^{+}_{L\times L}(\vec{x},t_{p}) & \hat{\eta}^{+}_{L\times S}(\vec{x},t_{p}) \\
\hat{\eta}^{*}_{S\times L}(\vec{x},t_{p}) & \hat{f}^{+}_{S\times S}(\vec{x},t_{p})
\eea\right)^{21} &
\left(\bea{cc}
\hat{B}^{T}_{L\times L}(\vec{x},t_{p}) & \hat{\chi}^{T}_{L\times S}(\vec{x},t_{p}) \\
\hat{\chi}^{*}_{S\times L}(\vec{x},t_{p}) & -\hat{F}^{T}_{S\times S}(\vec{x},t_{p})
\eea\right)^{22}
\eea\right)  \\ \lb{s2_68} &&
\bea{rclrcl}
\hat{B}_{L\times L}^{+}(\vec{x},t_{p})&=&\hat{B}_{L\times L}(\vec{x},t_{p}) &\hspace*{0.64cm}
\hat{F}_{S\times S}^{+}(\vec{x},t_{p})&=&\hat{F}_{S\times S}(\vec{x},t_{p})  \\
\hat{c}_{L\times L}^{T}(\vec{x},t_{p})&=&\hat{c}_{L\times L}(\vec{x},t_{p}) &
\hat{f}_{S\times S}^{T}(\vec{x},t_{p})&=&-\hat{f}_{S\times S}(\vec{x},t_{p})
\eea_{\mbox{.}}
\eeq
The path integral \(Z[\hat{\mcal{J}},J_{\psi},\hat{J}_{\psi\psi}]\) (\ref{s2_69},\ref{s2_61}),
restricted to a local self-energy \(\hat{\Sigma}_{\alpha\beta}^{ab}(\vec{x},t_{p})\) (\ref{s2_66}-\ref{s2_68}),
contains the same super-unitary symmetries in the angular momentum degrees of freedom
and the additional symmetries between densities \(\hat{\Sigma}^{11}(t_{p})\), \(\hat{\Sigma}^{11}(t_{p})\)
and anomalous degrees of freedom \(\hat{\Sigma}^{12}(t_{p})\), \(\hat{\Sigma}^{21}(t_{p})\)
as the generating function (\ref{s2_61}). These symmetries are also present in
the coherent state path integral (\ref{s2_54}) with doubled super-fields
\(\Psi_{\vec{x},\alpha}^{a(=1/2)}(t_{p})\) and with the
nonlocal self-energy. Taking into account the short-ranged character of the
interaction potential \(V_{|\vec{x}-\vec{x}\ppr|}\), we finally acquire a coherent state
path integral (\ref{s2_69}) with a local self-energy \(\hat{\Sigma}_{\alpha\beta}^{ab}(\vec{x},t_{p})\)
in 'Nambu'-doubled form. The interaction potential \(V_{|\vec{x}-\vec{x}\ppr|}\) is
considered with an effective value $V_{0}$ for the spatially local super-trace relation
of quadratic self-energy matrices
\footnote{The chosen normalization with \(\mcal{N}_{x}=(L/\Delta x)^{d}\) and \(\hbar\Omega\) yields with
the integration variables \(\int_{C}(d t_{p}/\hbar)\;\eta_{p}\sum_{\vec{x}}\ldots\) and the factor \(1/2\)
the inverse square root of the super-determinant as in (\ref{s2_61}).}
(Note the additional sign \(\eta_{p}\) on the time contour $t_{p}$ (\ref{s2_70})).
\beq \lb{s2_69}
\lefteqn{Z[\hat{\mcal{J}},J_{\psi},\hat{J}_{\psi\psi}]=
\int d[\Sigma(\vec{x},t_{p})]\;\;\exp\bigg\{\frac{\im}{4\hbar} \int_{C}d t_{p}\sum_{\vec{x}}
\frac{1}{V_{0}}\STRAB\Big[\hat{\Sigma}_{\alpha\beta}^{ab}(\vec{x},t_{p})\;
\hat{K}\;\hat{\Sigma}_{\beta\alpha}^{ba}(\vec{x},t_{p})\;\hat{K}\Big]\bigg\} }
\\ \no &\times& \exp\Bigg\{-\frac{1}{2}\int_{C}
\frac{d t_{p}}{\hbar}\eta_{p}\sum_{\vec{x}}\hbar\Omega\mcal{N}_{x}\;\;
\STRAB\ln\bigg[\hat{\mcal{H}}\;\hat{K}+\hat{\eta}\;\bigg( \bea{cc}
0 & \hat{j}_{\psi\psi} \\
\hat{j}_{\psi\psi}^{+} & 0 \eea\bigg)+\hat{K}\;\hat{\eta}\;
\frac{\hat{\mcal{J}}_{\beta\alpha}^{ba}}{\mcal{N}_{x}}\;\hat{\eta}\;\hat{K}+\hat{\eta}\;
\hat{\Sigma}(\vec{x},t_{p})\bigg]\Bigg\}
\\ \no &\times& \exp\Bigg\{\frac{\im}{2\hbar}\Omega^{2}\int_{C}d t_{p}\;d t_{q}\ppr
\sum_{\vec{x},\vec{x}\ppr}\mcal{N}_{x}\; J_{\psi;\beta}^{+b}(\vec{x}\ppr,t_{q}\ppr)\;\times \\ \no &\times&
\bigg[\hat{\mcal{H}}\hat{K}+\hat{\eta}\;\left( \bea{cc}
0 & \hat{j}_{\psi\psi} \\
\hat{j}_{\psi\psi}^{+} & 0 \eea\right)+
\hat{K}\;\hat{\eta}\;\frac{\hat{\mcal{J}}_{\beta\alpha}^{ba}}{\mcal{N}_{x}}\;\hat{\eta}\;\hat{K}+
\hat{\eta}\;\hat{\Sigma}(\vec{x},t_{p})
\bigg]_{\vec{x}\ppr,\beta;\vec{x},\alpha}^{\mathbf{-1};\;ba}\hspace*{-0.64cm}(t_{q}\ppr,t_{p})\;\;\;
J_{\psi;\alpha}^{a}(\vec{x},t_{p})\Bigg\}
\eeq
\be\lb{s2_70}
\eta_{p}=\Bigg\{\bea{rcll} \eta_{+}&=&+1 \hspace*{0.5cm}&\mbox{for forward propagation with $t_{+}$} \\
\eta_{-}&=&-1 \hspace*{0.5cm}&\mbox{for backward propagation with $t_{-}$} \eea_{\mbox{.}}
\ee
Apart from the quadratic interaction term with \(\hat{\Sigma}_{\alpha\beta}^{ab}(\vec{x},t_{p})\)
and $V_{0}$, the super-matrix $\hat{M}$ (\ref{s2_71}) determines the coherent state path integral
(\ref{s2_69}) and the observables derived from it. It is composed of the doubled one-particle part
\(\hat{\mcal{H}}\;\hat{K}\), the self-energy and the source terms \(\hat{J}_{\psi\psi}\), \(\hat{\mcal{J}}\).
The super-matrix \(\hat{M}\) (\ref{s2_71}) is contained in a term with the quadratic source fields
\(J_{\psi}\), \(J_{\psi}^{+}\) for the condensate wave function and the super-determinant
 \(\big(\mbox{SDET}[\hat{M}]\big)^{-1/2}\) which is transformed to the exponential-super-trace-logarithm
relation as described in (\ref{s2_62})
\beq\no
\hat{M}_{\vec{x},\alpha;\vec{x}\ppr,\beta}^{ab}(t_{p},t_{q}\ppr)&=&
\hat{\mcal{H}}_{\vec{x},\alpha;\vec{x}\ppr,\beta}^{ab}(t_{p},t_{q}\ppr)\;\;\hat{K}+
\left( \bea{cc}
0 & \hat{j}_{\psi\psi;\alpha\beta}(\vec{x},t_{p}) \\
\hat{j}_{\psi\psi;\alpha\beta}^{+}(\vec{x},t_{p}) & 0
\eea\right)^{ab}\;\delta_{\vec{x},\vec{x}\ppr}\;\eta_{p}\;\delta_{p,q}\;\delta(t_{p}-t_{q}\ppr) + \\ \lb{s2_71}
&+& \hat{K}\;\eta_{p}\;
\frac{\hat{\mcal{J}}_{\vec{x},\alpha;\vec{x}\ppr,\beta}^{ab}(t_{p},t_{q}\ppr)}{\mcal{N}_{x}}\;\eta_{q}\;\hat{K}+
\hat{\Sigma}_{\alpha\beta}^{ab}(\vec{x},t_{p})\; \delta_{\vec{x},\vec{x}\ppr}\;
\eta_{p}\;\delta_{p,q}\;\delta(t_{p}-t_{q}\ppr)\;.
\eeq
Inserting the super-matrix (\ref{s2_71}) into (\ref{s2_69}), the coherent state path integral
takes the abbreviated form (\ref{s2_72}) with super-matrix (\ref{s2_71})
\beq \lb{s2_72}
\lefteqn{\hspace*{-1.0cm}Z[\hat{\mcal{J}},J_{\psi},\hat{J}_{\psi\psi}]=
\int d[\Sigma(\vec{x},t_{p})]\;\;\exp\bigg\{\frac{\im}{4\hbar} \int_{C}d t_{p}\sum_{\vec{x}}
\frac{1}{V_{0}}\STRAB\Big[\hat{\Sigma}_{\alpha\beta}^{ab}(\vec{x},t_{p})\;
\hat{K}\;\hat{\Sigma}_{\beta\alpha}^{ba}(\vec{x},t_{p})\;\hat{K}\Big]\bigg\} }
\\ \no &\times& \exp\Bigg\{-\frac{1}{2}\int_{C}
\frac{d t_{p}}{\hbar}\eta_{p}\sum_{\vec{x}}\hbar\Omega\mcal{N}_{x}\;\;
\STRAB\ln\bigg[\hat{M}_{\vec{x},\alpha;\vec{x}\ppr,\beta}^{ab}(t_{p},t_{q}\ppr)\bigg]\Bigg\}
\\ \no &\times& \exp\Bigg\{\frac{\im}{2\hbar}\Omega^{2}\int_{C}d t_{p}\;d t_{q}\ppr
\sum_{\vec{x},\vec{x}\ppr}\mcal{N}_{x}\; J_{\psi;\beta}^{+b}(\vec{x}\ppr,t_{q}\ppr)\;\;
\bigg[\;\hat{M}\;\bigg]_{\vec{x}\ppr,\beta;\vec{x},\alpha}^{\mathbf{-1};\;ba}
\hspace*{-0.64cm}(t_{q}\ppr,t_{p})\;\;\;
J_{\psi;\alpha}^{a}(\vec{x},t_{p})\Bigg\}_{\mbox{.}}
\eeq

\section{Symmetry considerations and parameters of the self-energy} \lb{s3}

\subsection{The ortho-symplectic Lie super-group in the coherent state path integral} \lb{s31}

Apart from the bilinear source term with \(J_{\psi;\alpha}^{a}(\vec{x},t_{p})\) (\ref{s2_58}),
the coherent state path integral (\ref{s2_69},\ref{s2_72}) consists of the trace relation with the
quadratic term of the self-energy matrix and the inverse
square root of a super-determinant, which is obtained by integration over the bilinear fields
\(\Psi_{\vec{x}\ppr,\beta}^{+b}(t_{p})\) and \(\Psi_{\vec{x},\alpha}^{a}(t_{p})\).
In order to derive effective equations for the anomalous terms, the self-energy
has to be decomposed into density terms
\(\hat{\Sigma}^{11}_{\alpha\beta}(\vec{x},t_{p})\),
\(\hat{\Sigma}^{22}_{\alpha\beta}(\vec{x},t_{p})\) (\ref{s2_66}-\ref{s2_68}),
which are related to a subgroup for spontaneous symmetry breaking, and the
appropriate coset space which should diagonalize the self-energy with the molecular and BCS-terms as
'eigenvectors' in a transferred sense.
However, it is sufficient to examine the bilinear relation
\(\Psi_{\vec{x}\ppr,\beta}^{+b}(t_{p})\;\eta_{p}\;
\hat{\mcal{H}}_{\vec{x}\ppr,\beta;\vec{x},\alpha}^{ba}(t_{p})\;\hat{K}\;
\Psi_{\vec{x},\alpha}^{a}(t_{p})\) (\ref{s3_5}) for determining the symmetries in
\(Z[\hat{\mcal{J}},J_{\psi},\hat{J}_{\psi\psi}]\) (\ref{s2_69},\ref{s2_72})
because the dyadic products (\ref{s2_36}) and the HST transformation (\ref{s2_52})
convey the symmetries from the super-fields
\(\Psi_{\vec{x},\alpha}^{a}(t_{p})\) to the matrix
\(\hat{M}_{\vec{x},\alpha;\vec{x}\ppr,\beta}^{ab}(t_{p},t_{q}\ppr)\)
(\ref{s2_71}) and to the self-energy \footnote{In the case of a disordered system one has to
include the contour times $t_{p}$, $t_{q}\ppr$ in the symmetry investigations of the self-energy.
However, the considered self-energies have only a diagonal dependence on the contour time $t_{p}$
throughout this article, due to the absence of any disorder.}.
The one-particle energy term \(\hat{\mcal{H}}\;\hat{K}\) (\ref{s2_55}) in
\(\hat{M}_{\vec{x},\alpha;\vec{x}\ppr,\beta}^{ab}(t_{p},t_{q}\ppr)\) (\ref{s2_71})
contains the operator $\hat{H}_{p}(\vec{x},t_{p})$ (\ref{s3_2},\ref{s2_56})
and its transpose $\hat{H}_{p}^{T}(\vec{x},t_{p})$ (\ref{s3_3},\ref{s2_57})
which is itself composed of the energy or time derivative \(-\im\hbar\hat{\pp}/\pp t_{p}\)
and of the operator $\hat{h}_{p}(\vec{x})$ (\ref{s3_4},\ref{s2_3})
with the kinetic energy, the trap potential \(u(\vec{x})\) and the chemical potential $\mu_{0}$.
The energy or time derivative \(\hat{\pp}/\pp t_{p}\)
changes its sign under transposition (\ref{s3_2},\ref{s3_3})
whereas $\hat{h}_{p}(\vec{x})$ (\ref{s3_4},\ref{s2_3}) is symmetric
\beq \lb{s3_1}
\hat{\mcal{H}}&=&\eta_{p}\;\left\{\hat{H}_{p}(\vec{x},t_{p})\;\hat{1}_{L\times L}\;,\;
\hat{H}_{p}(\vec{x},t_{p})\;\hat{1}_{S\times S}\;;\;
\hat{H}_{p}^{T}(\vec{x},t_{p})\;\hat{1}_{L\times L}\;,\;
\hat{H}_{p}^{T}(\vec{x},t_{p})\;\hat{1}_{S\times S}\right\} \\ \lb{s3_2}
\hat{H}_{p}(\vec{x},t_{p})&=&-\hat{E}_{p}+\hat{h}_{p}(\vec{x}) =
-\im\hbar\frac{\hat{\pp}}{\pp t_{p}}+\hat{h}_{p}(\vec{x})\hspace*{1.0cm}
\bigg(\frac{\hat{\pp}}{\pp t_{p}}\bigg)^{T}=-\frac{\hat{\pp}}{\pp t_{p}} \\ \lb{s3_3}
\hat{H}_{p}^{T}(\vec{x},t_{p})&=&+\hat{E}_{p}+\hat{h}_{p}(\vec{x}) =
+\im\hbar\frac{\hat{\pp}}{\pp t_{p}}+\hat{h}_{p}(\vec{x}) \\ \lb{s3_4}
\hat{h}_{p}(\vec{x})&=&\frac{\vec{p}^{\;2}}{2m}+u(\vec{x})-\mu_{0}-\im\;\ve_{p}
\hspace*{1.0cm}\hat{h}_{p}^{T}(\vec{x})=\hat{h}_{p}(\vec{x})\;.
\eeq
In consequence one has to separate the bilinear relation in (\ref{s3_5},\ref{s2_54}) with
\(\hat{\mcal{H}}\;\hat{K}\) into a bilinear relation
with the symmetric operator $\hat{h}_{p}(\vec{x})$ (\ref{s3_4}) and the metric \(\hat{K}\) (\ref{s3_6}),
and a bilinear relation with \(-\hat{E}_{p}=-\im\hbar\hat{\pp}/\pp t_{p}\) (\ref{s3_2},\ref{s3_3}).
The minus sign of the transpose of the time derivative transforms the metric \(\hat{K}\) (\ref{s3_6})
to a metric $\wtilde{K}$ (\ref{s3_7}).
The minus signs in the second fermion-fermion block of \(\hat{K}\) (\ref{s3_6}) are moved to
the second boson-boson block in $\wtilde{K}$ (\ref{s3_7}) in the case of the bilinear relation
with \(-\hat{E}_{p}=-\im\hbar\hat{\pp}/\pp t_{p}\)
\be \lb{s3_5}
\Psi_{\vec{x},\alpha}^{a+}(t_{p})\;\hat{\eta}\;\hat{\mcal{H}}\;\hat{K}\;\Psi_{\vec{x},\alpha}^{a}(t_{p})=
\Psi_{\vec{x},\alpha}^{a+}(t_{p})\;\hat{h}_{p}\;\hat{K}\;\Psi_{\vec{x},\alpha}^{a}(t_{p})+
\Psi_{\vec{x},\alpha}^{a+}(t_{p})\; \bigg(-\im\hbar\frac{\hat{\pp}}{\pp t_{p}}\bigg)\;\wtilde{K}\;
\Psi_{\vec{x},\alpha}^{a}(t_{p})
\ee
\beq \lb{s3_6}
\hat{K}_{2N\times 2N}&=&\bigg\{\underbrace{\hat{1}_{L\times
L}}_{\mbox{BB}}\;,\; \underbrace{\hat{1}_{S\times S}}_{\mbox{FF}}\;;\; \underbrace{\hat{1}_{L\times
L}}_{\mbox{BB}}\;,\;
\underbrace{-\hat{1}_{S\times S}}_{\mbox{FF}}\bigg\} \\ \no &=&
\bigg\{\underbrace{\hat{1}_{N\times N}}_{a=1}\;;\;
\underbrace{\hat{\kappa}_{N\times N}}_{a=2}\bigg\}\;;\hspace*{1.0cm}
\hat{\kappa}_{N\times N}=\bigg\{\underbrace{\hat{1}_{L\times L}}_{\mbox{BB}}\;,\;
\underbrace{-\hat{1}_{S\times S}}_{\mbox{FF}}\bigg\}
\\ \lb{s3_7}
\wtilde{K}_{2N\times 2N}&=&\bigg\{\underbrace{\hat{1}_{L\times L}}_{\mbox{BB}}\;,\;
\underbrace{\hat{1}_{S\times S}}_{\mbox{FF}}\;;\;
\underbrace{-\hat{1}_{L\times L}}_{\mbox{BB}}\;,\; \underbrace{\hat{1}_{S\times S}}_{\mbox{FF}}\bigg\}
\\ \no &=& \bigg\{\underbrace{\hat{1}_{N\times N}}_{a=1}\;;\;
\underbrace{\wtilde{\kappa}_{N\times N}}_{a=2}\bigg\}\;;\hspace*{1.0cm}
\wtilde{\kappa}_{N\times N}=\bigg\{\underbrace{-\hat{1}_{L\times L}}_{\mbox{BB}}\;,\;
\underbrace{\hat{1}_{S\times S}}_{\mbox{FF}}\bigg\}\;\;.
\eeq
The field \(\Psi_{\vec{x},\alpha}^{a}(t_{p})\) with its splitted boson-boson and fermion-fermion block has to
be reordered into a supervector \(\Theta_{\vec{x},i}^{M(=B/F)}(t_{p})\) (\ref{s3_8}) with a single boson-boson
\(\Theta_{\vec{x},i}^{B}(t_{p})=\big(\vec{b}_{\vec{x}}(t_{p})\;,\;\vec{b}_{\vec{x}}^{*}(t_{p})\big)\)
and a single fermion-fermion part
\(\Theta_{\vec{x},i}^{F}(t_{p})=\big(\vec{\alpha}_{\vec{x}}(t_{p})\;,\;\vec{\alpha}_{\vec{x}}^{*}(t_{p})\big)\)
so that one can identify the invariance of the two bilinear relations in (\ref{s3_5}) with common super-groups
\footnote{The upper capital indices '$M,N$' of \(\Theta_{\vec{x},i}^{M}(t_{p})\)
specify between the single bosonic part (\(M,N=B\)) and the single fermionic part (\(M,N=F\)),
each part doubled with its complex conjugated field.
Therefore, the indices '$i,j$' in \(\Theta_{\vec{x},i}^{M(=B/F)}(t_{p})\) are chosen to label the
doubled angular momentum degrees of freedom for the bosons (\(M,N=B\)), (\(i,j=1\)) to (\(i,j=2L\)) and
for the fermions (\(M,N=F\)), from (\(i,j=1\)) to (\(i,j=2S\)).}
\be \lb{s3_8}
\Psi_{\vec{x},\alpha}^{a(=1/2)}(t_{p})=\left( \bea{c}
\vec{b}_{\vec{x}}(t_{p}) \\
\vec{\alpha}_{\vec{x}}(t_{p}) \\
\vec{b}_{\vec{x}}^{*}(t_{p}) \\
\vec{\alpha}_{\vec{x}}^{*}(t_{p})
\eea\right) \Longrightarrow
\Theta_{\vec{x},i}^{M(=B/F)}(t_{p})=
\left( \bea{c}
\Theta_{\vec{x},i}^{B}(t_{p})  \\
\Theta_{\vec{x},i}^{F}(t_{p})
\eea\right) =
\left( \bea{c}
\vec{b}_{\vec{x}}(t_{p}) \\
\vec{b}_{\vec{x}}^{*}(t_{p}) \\
\vec{\alpha}_{\vec{x}}(t_{p}) \\
\vec{\alpha}_{\vec{x}}^{*}(t_{p}) \eea\right)_{\mbox{.}}
\ee
In terms of the new field \(\Theta_{\vec{x},i}^{M(=B/F)}(t_{p})\) (\ref{s3_8}),
the bilinear relation
\(\Psi_{\vec{x},\alpha}^{+a}(t_{p})\;\hat{h}_{p}(\vec{x})\;\hat{K}\;\Psi_{\vec{x},\alpha}^{a}(t_{p})\)
(\ref{s3_10}) of (\ref{s3_5}) with one-particle operator $\hat{h}_{p}(\vec{x})$
(\ref{s3_4},\ref{s2_3}) becomes a bilinear relation
with the super-transposed vector \(\Theta_{\vec{x},i}^{M(=B/F),ST}(t_{p})\) (\ref{s3_9}) and a
new metric $\hat{L}$ (\ref{s3_11})
which has only the one boson-boson block metric $\hat{G}_{2L\times 2L}$ and
fermion-fermion block metric $\hat{J}_{2S\times 2S}$
\beq \lb{s3_9}
\Theta_{\vec{x},i}^{M=(B/F),ST}(t_{p})&=&\left( \vec{b}_{\vec{x}}(t_{p}) \;,\;
\vec{b}_{\vec{x}}^{*}(t_{p}) \;;\; \vec{\alpha}_{\vec{x}}(t_{p}) \;,\; \vec{\alpha}_{\vec{x}}^{*}(t_{p})
\right) \\ \lb{s3_10}
\Psi_{\vec{x},\alpha}^{a+}(t_{p})\;\hat{h}_{p}(\vec{x})\;\hat{K}\;\Psi_{\vec{x},\alpha}^{a}(t_{p})&=&
\Theta_{\vec{x},i}^{M,ST}(t_{p})\;\hat{h}_{p}(\vec{x})\;\hat{L}\; \Theta_{\vec{x},i}^{M}(t_{p})
\eeq
\beq \lb{s3_11}
\hat{L}_{2N\times 2N}&=&
\left( \bea{cccc}
\hat{0}_{L\times L} & \hat{1}_{L\times L}&& \\
\hat{1}_{L\times L} & \hat{0}_{L\times L}&& \\
&&\hat{0}_{S\times S} & -\hat{1}_{S\times S} \\
&&\hat{1}_{S\times S} & \hat{0}_{S\times S} \\
\eea\right) = \left( \bea{cc}
\hat{G}_{2L\times 2L} & \hat{0}_{2L\times 2S} \\
\hat{0}_{2S\times 2L} & \hat{J}_{2S\times 2S} \eea\right)_{\mbox{.}}
\eeq
Applying the relation (\ref{s3_10}) of
\(\Theta_{\vec{x},i}^{M}(t_{p})\) with its super-transpose (\ref{s3_9}),
one obtains an orthogonal symmetry \(SO(L,L)\) with the indefinite
metric $\hat{G}_{2L\times 2L}$ (\ref{s3_12}) for the boson-boson block,
and a symplectic symmetry metric $\hat{J}_{2S\times 2S}$ of
\(Sp(2S)\) (\ref{s3_13}) for the fermion-fermion block
\beq \lb{s3_12}
\hat{G}_{2L\times 2L}&=&\left( \bea{cc}
\hat{0}_{L\times L} & \hat{1}_{L\times L} \\
\hat{1}_{L\times L} & \hat{0}_{L\times L} \eea\right)\propto \left( \bea{cc}
\hat{1}_{L\times L} & \hat{0}_{L\times L} \\
\hat{0}_{L\times L} & -\hat{1}_{L\times L}
\eea\right) \\ \lb{s3_13}
\hat{J}_{2S\times 2S}&=&\left( \bea{cc}
\hat{0}_{S\times S} & -\hat{1}_{S\times S} \\
\hat{1}_{S\times S} & \hat{0}_{S\times S} \eea\right)_{\mbox{.}}
\eeq
The analogous considerations for the bilinear relation (\ref{s3_14}) of (\ref{s3_5}) with
the time derivative yield a metric $\wtilde{L}$ (\ref{s3_15})
for \(\Theta_{\vec{x},i}^{M}(t_{p})\) which indicates a
symplectic symmetry \(Sp(2L)\) with
$\wtilde{J}_{2L\times 2L}$ (\ref{s3_16}) for the boson-boson part and
an indefinite orthogonal symmetry \(SO(S,S)\) with
$\wtilde{G}_{2S\times 2S}$ (\ref{s3_17}) for the fermion-fermion term
\be \lb{s3_14}
\Psi_{\vec{x},\alpha}^{a+}(t_{p})\;\bigg(-\im\hbar\frac{\hat{\pp}}{\pp t_{p}}\bigg)
\;\wtilde{K}\;\Psi_{\vec{x},\alpha}^{a}(t_{p})=
\Theta_{\vec{x},i}^{A,ST}(t_{p})\;\bigg(-\im\hbar\frac{\hat{\pp}}{\pp t_{p}}\bigg) \;\wtilde{L}\;
\Theta_{\vec{x},i}^{A}(t_{p})
\ee
\beq \lb{s3_15}
\wtilde{L}_{2N\times 2N}&=& \left( \bea{cccc}
\hat{0}_{L\times L} & -\hat{1}_{L\times L}&& \\
\hat{1}_{L\times L} & \hat{0}_{L\times L}&& \\
&&\hat{0}_{S\times S} & \hat{1}_{S\times S} \\
&&\hat{1}_{S\times S} & \hat{0}_{S\times S} \\
\eea\right) = \left( \bea{cc}
\wtilde{J}_{2L\times 2L} & \hat{0}_{2L\times 2S} \\
\hat{0}_{2S\times 2L} & \wtilde{G}_{2S\times 2S}
\eea\right) \\ \lb{s3_16}
\wtilde{J}_{2L\times 2L}&=&\left( \bea{cc}
\hat{0}_{L\times L} & -\hat{1}_{L\times L} \\
\hat{1}_{L\times L} & \hat{0}_{L\times L}
\eea\right) \\ \lb{s3_17}
\wtilde{G}_{2S\times 2S}&=&\left( \bea{cc}
\hat{0}_{S\times S} & \hat{1}_{S\times S} \\
\hat{1}_{S\times S} & \hat{0}_{S\times S} \eea\right)\propto \left( \bea{cc}
\hat{1}_{S\times S} & \hat{0}_{S\times S} \\
\hat{0}_{S\times S} & -\hat{1}_{S\times S} \eea\right)_{\mbox{.}}
\eeq
We compare the two bilinear relations  (\ref{s3_10}) and (\ref{s3_14})
containing the symmetric one-particle operator (\ref{s3_4}) and the
antisymmetric time derivative (\ref{s3_2},\ref{s3_3}), respectively.
The orthogonal symmetry $\hat{G}_{2L\times 2L}$ (\ref{s3_12}) of the boson-boson block
in (\ref{s3_10}) has been exchanged with the
symplectic metric \(\wtilde{J}_{2L\times 2L}\) (\ref{s3_16}) in the boson-boson
part of (\ref{s3_14}); similarly the symplectic metric $\hat{J}_{2S\times 2S}$ (\ref{s3_13}) of the
fermion-fermion block in (\ref{s3_10}) is replaced by
the orthogonal symmetry \(\wtilde{G}_{2S\times 2S}\)
(\ref{s3_17}) in the fermion-fermion block of the bilinear relation
(\ref{s3_14}) with \(-\hat{E}_{p}=-\im\hbar\hat{\pp}/\pp t_{p}\).
The prevailing metric $\hat{L}$ (\ref{s3_11}) or $\wtilde{L}$ (\ref{s3_15}) determines the
possible invariant transformations under corresponding super-matrices
$\hat{V}$ (\ref{s3_18}) or $\wtilde{V}$ (\ref{s3_19}) for the cases
with the one-particle energy $\hat{h}_{p}(\vec{x})$ (\ref{s3_10},\ref{s3_4}) or with the time
derivative \(-\hat{E}_{p}=-\im\hbar\hat{\pp}/\pp t_{p}\) (\ref{s3_14},\ref{s3_2},\ref{s3_3})
\beq \lb{s3_18}
\Theta_{\vec{x},i}^{M}(t_{p})&=&\hat{V}_{ij}^{MN}\;\;
\Theta_{\vec{x},j}^{N}(t_{p})\hspace*{1.0cm}\mbox{for : }\;\hat{h}_{p}(\vec{x})
\\ \lb{s3_19} \Theta_{\vec{x},i}^{M}(t_{p})&=&\wtilde{V}_{ij}^{MN}\;\;
\Theta_{\vec{x},j}^{N}(t_{p})\hspace*{1.0cm}\mbox{for :}\;
-\hat{E}_{p}=-\im\hbar\frac{\hat{\pp}}{\pp t_{p}}\;\;\;.
\eeq
One has to introduce the super-generators $\hat{W}\ppr$ (\ref{s3_20}) and
$\wtilde{W}^{\raisebox{-10pt}{$\ppr$}}$ (\ref{s3_22})
for the definitions \(\hat{V}^{ST}\;\hat{L}\;\hat{V}=\hat{L}\) (\ref{s3_21}) and
\(\wtilde{V}^{ST}\;\wtilde{L}\;\wtilde{V}=\wtilde{L}\)
(\ref{s3_23}) which yield the corresponding
constraints on the Lie generators $\hat{W}\ppr$ and $\wtilde{W}^{\raisebox{-10pt}{$\ppr$}}$
\footnote{The prime '$\ppr$' of the Lie generators $\hat{W}\ppr$ (\ref{s3_20},\ref{s3_24}) and
\(\wtilde{W}^{\raisebox{-10pt}{$\ppr$}}\) (\ref{s3_22},\ref{s3_27})
marks a separation into a single boson-boson and fermion-fermion
part, described by the bilinear relations with $\Theta_{\vec{x},i}^{M}(t_{p})$ and its transpose.
These primes are eliminated from the generators in relations after Eq. (\ref{s3_36}) because we refer
again  to splitted boson-boson, fermion-fermion, boson-fermion and fermion-boson parts or
four combined $N\times N$ super-matrices
\(\wtilde{W}^{\raisebox{-3pt}{$_{11}$}}\),
\(\wtilde{W}^{\raisebox{-3pt}{$_{22}$}}\)
and \(\wtilde{W}^{\raisebox{-3pt}{$_{12}$}}\),
\(\wtilde{W}^{\raisebox{-3pt}{$_{21}$}}\)
as in the original given order parameter $\Phi_{\alpha\beta}^{ab}$ (\ref{s2_29}-\ref{s2_32}) with \(a,b=1,2\).}.
Relations (\ref{s3_20}-\ref{s3_23}) state the invariance of the bilinear terms (\ref{s3_10},\ref{s3_14}) with
\(\Theta_{\vec{x},i}^{M}(t_{p})\), \(\Theta_{\vec{x},j}^{N,ST}(t_{p})\) and \(\hat{h}_{p}(\vec{x})\),
\(-\hat{E}_{p}=-\im\hat{\pp}/\pp t_{p}\) under super-group transformations with \(2N\times 2N\) super-matrices
\(\hat{V}\), \(\wtilde{V}\) and their super-generators \(\hat{W}\ppr\),
\(\wtilde{W}^{\raisebox{-10pt}{$\ppr$}}\)
\beq \lb{s3_20}
\hat{h}_{p}(\vec{x}) & : &
\hat{V}=\exp\{\im\;\hat{W}\ppr\}
\\ \lb{s3_21} && \hat{V}^{ST}\;\hat{L}\;\hat{V}=\hat{L}\longrightarrow
\hat{W}^{\prime ST}\;\hat{L}+\hat{L}\;\hat{W}^{\prime}=0 \\ \lb{s3_22}
-\im\hbar\frac{\hat{\pp}}{\pp t_{p}} & : & \wtilde{V}=
\exp\{\im\;\wtilde{W}^{\raisebox{-10pt}{$\ppr$}}\}
\\ \lb{s3_23} && \wtilde{V}^{ST}\;\wtilde{L}\;\wtilde{V}=\wtilde{L}
\longrightarrow \wtilde{W}^{\prime ST}\;\wtilde{L}+
\wtilde{L}\;\wtilde{W}^{\raisebox{-10pt}{$\ppr$}}=0\;\;.
\eeq
We assign
to the generator $\hat{W}\ppr_{2N\times 2N}$ (\ref{s3_24},\ref{s3_20},\ref{s3_21}) the even boson-boson part
$\hat{A}_{2L\times 2L}$ and the even fermion-fermion block $\hat{D}_{2S\times 2S}$,
including the odd parts $\hat{\beta}_{2L\times 2S}$ and $\hat{\gamma}_{2S\times 2L}$ for the
boson-fermion and fermion-boson parts, respectively. The constraint in (\ref{s3_21})
on the total generator $\hat{W}\ppr_{2N\times 2N}$ (\ref{s3_24}) with
metric $\hat{L}_{2N\times 2N}$ (\ref{s3_11}) gives the relations (\ref{s3_25})
on the even parts $\hat{A}_{2L\times 2L}$ and $\hat{D}_{2S\times 2S}$ with the
orthogonal metric $\hat{G}_{2L\times 2L}$ (\ref{s3_12}) and symplectic part
with $\hat{J}_{2S\times 2S}$ (\ref{s3_13}). The constraint on the odd parts
$\hat{\beta}_{2L\times 2S}$, $\hat{\gamma}_{2S\times 2L}$ is determined by Eq. (\ref{s3_26}) with the metrics
$\hat{G}_{2L\times 2L}$ (\ref{s3_12}) and $\hat{J}_{2S\times 2S}$ (\ref{s3_13})
\beq \lb{s3_24}
\hat{W}\ppr_{2N\times 2N}&=&\left( \bea{cc}
\underbrace{\hat{A}_{2L\times 2L}}_{\mbox{BB}} &
\underbrace{\hat{\beta}_{2L\times 2S}}_{\mbox{BF}} \\
\underbrace{\hat{\gamma}_{2S\times 2L}}_{\mbox{FB}} & \underbrace{\hat{D}_{2S\times 2S}}_{\mbox{FF}}
\eea\right) \\ \lb{s3_25}
\mbox{even parts} & : & \hat{A}^{T}_{2L\times 2L}\;\hat{G}_{2L\times 2L}+
\hat{G}_{2L\times 2L}\;\hat{A}_{2L\times 2L}=0 \\ \no &&
\hat{D}^{T}_{2S\times 2S}\;\hat{J}_{2S\times 2S}+
\hat{J}_{2S\times 2S}\;\hat{D}_{2S\times 2S}=0\hspace*{1.5cm} \\ \lb{s3_26}
\mbox{odd parts} & : & \hat{\beta}^{T}_{2S\times 2L}\;\hat{G}_{2L\times 2L}=
\hat{J}_{2S\times 2S}\;\hat{\gamma}_{2S\times 2L}\;\;\;.
\eeq
The corresponding relations for the complete
super-matrix $\wtilde{W}^{\raisebox{-10pt}{$\ppr$}}_{2N\times 2N}$
(\ref{s3_27},\ref{s3_22},\ref{s3_23}) separate  into the even boson-boson block
$\wtilde{A}_{2L\times 2L}$ and even fermion-fermion block
$\wtilde{D}_{2S\times 2S}$ (\ref{s3_28}), and the
Grassmann parts $\wtilde{\beta}_{2L\times 2S}$,
$\wtilde{\gamma}_{2S\times 2L}$ in the boson-fermion and
fermion-boson sections (\ref{s3_29}).
The even parts with $\wtilde{A}_{2L\times 2L}$ and $\wtilde{D}_{2S\times 2S}$ have to fulfill
relations (\ref{s3_28}) with metric $\wtilde{J}_{2L\times 2L}$ (\ref{s3_16}) and
$\wtilde{G}_{2S\times 2S}$ (\ref{s3_17}) in the boson-boson and fermion-fermion parts, respectively.
One has to exchange the orthogonal with the symplectic metric in (\ref{s3_26}) in order to obtain the
corresponding constraint (\ref{s3_29}) on the odd matrices $\wtilde{\beta}_{2L\times 2S}$,
$\wtilde{\gamma}_{2S\times 2L}$ with metric \(\wtilde{J}_{2L\times 2L}\) (\ref{s3_16}) and
\(\wtilde{G}_{2S\times 2S}\) (\ref{s3_17})
\beq \lb{s3_27}
\wtilde{W}^{\raisebox{-10pt}{$\ppr$}}_{2N\times 2N}&=&\left( \bea{cc}
\underbrace{\wtilde{A}_{2L\times 2L}}_{\mbox{BB}} &
\underbrace{\wtilde{\beta}_{2L\times 2S}}_{\mbox{BF}} \\
\underbrace{\wtilde{\gamma}_{2S\times 2L}}_{\mbox{FB}} &
\underbrace{\wtilde{D}_{2S\times 2S}}_{\mbox{FF}}
\eea\right) \\ \lb{s3_28}
\mbox{even parts} & : &
\wtilde{A}^{\raisebox{-10pt}{$^{T}$}}_{2L\times 2L}\;\wtilde{J}_{2L\times 2L}+
\wtilde{J}_{2L\times 2L}\; \wtilde{A}_{2L\times 2L}=0   \\ \no &&
\wtilde{D}^{\raisebox{-10pt}{$^{T}$}}_{2S\times 2S}\;
\wtilde{G}_{2S\times 2S}+\wtilde{G}_{2S\times 2S}\;
\wtilde{D}_{2S\times 2S}=0 \\ \lb{s3_29}
\mbox{odd parts} & : &
\wtilde{\beta}^{\raisebox{-10pt}{$^{T}$}}_{2S\times 2L}\;\wtilde{J}_{2L\times 2L}=
\wtilde{G}_{2S\times 2S}\; \wtilde{\gamma}_{2S\times 2L}\;\;\;.
\eeq
We construct explicit representations (\ref{s3_31},\ref{s3_34}) for the super-generators
\(\hat{W}\ppr_{2N\times 2N}\) (\ref{s3_24}) and
\(\wtilde{W}^{\raisebox{-10pt}{$\ppr$}}_{2N\times 2N}\) (\ref{s3_27})
by taking into account the constraints (\ref{s3_25},\ref{s3_26}) and (\ref{s3_28},\ref{s3_29}), respectively.
The sub-matrices \(\hat{B}_{D,L\times L}\), \(\hat{F}_{D,L\times L}\) and
\(\hat{c}_{D,L\times L}\), \(\hat{f}_{D,S\times S}\) of \(\hat{W}\ppr_{2N\times 2N}\) (\ref{s3_31})
contain only even complex elements with symmetry restrictions (\ref{s3_32}) whereas
the matrices of the boson-fermion and fermion-boson parts,
denoted by greek symbols and their dimensions, are only composed of anti-commuting
variables. Similar meaning have the corresponding even partial matrices
\(\wtilde{B}_{D,L\times L}\), \(\wtilde{F}_{D,S\times S}\) and
\(\wtilde{c}_{D,L\times L}\), \(\wtilde{f}_{D,S\times S}\)
of \(\wtilde{W}\ppr_{2N\times 2N}\) (\ref{s3_34}). These even matrices of
\(\wtilde{W}\ppr_{2N\times 2N}\) (\ref{s3_34}) are distinguished by a tilde
'\(\wtilde{\ph{W}}\)' from the even parts in \(\hat{W}\ppr_{2N\times 2N}\) (\ref{s3_31}).
Apart from partially differing signs,
the odd parts of \(\wtilde{W}\ppr_{2N\times 2N}\) (\ref{s3_34}) are ordered in
an analogous manner as the anti-commuting variables in \(\hat{W}\ppr_{2N\times 2N}\) (\ref{s3_31}).
One has also to consider for the representation of
\(\hat{W}\ppr_{2N\times 2N}\) (\ref{s3_24},\ref{s3_31}) and
\(\wtilde{W}\ppr_{2N\times 2N}\) (\ref{s3_27},\ref{s3_34})
that the fields \(\vec{b}_{\vec{x}}(t_{p})\) and \(\vec{\alpha}_{\vec{x}}(t_{p})\)
in \(\Theta_{\vec{x},i}^{M(=B/F)}(t_{p})\) of the bilinear relations (\ref{s3_30},\ref{s3_33})
are doubled with their complex conjugates
\(\vec{b}_{\vec{x}}^{*}(t_{p})\) and \(\vec{\alpha}_{\vec{x}}^{*}(t_{p})\); accordingly
the hermitian conjugates of the even and odd partial matrices (, as e.g.
\(\hat{B}_{D,L\times L}\), \(\hat{F}_{D,L\times L}\),
\(\hat{c}_{D,L\times L}\), \(\hat{f}_{D,S\times S}\) and
\(\hat{\chi}_{D,S\times L}\), \(\hat{\eta}_{D,S\times L}\) in \(\hat{W}\ppr_{2N\times 2N}\),) have also to
appear in either representation (\ref{s3_31},\ref{s3_34})
\beq \lb{s3_30}
\hat{h}_{p}&:&\Theta_{\vec{x},i}^{M,ST}(t_{p})\;
\underbrace{\Big(\exp\{\im\;\hat{W}\ppr\}\Big)^{MM^{\prime},ST}_{ii\ppr}\;
\hat{L}_{i\ppr j\ppr}^{M\ppr N\ppr}\;
\Big(\exp\{\im\;\hat{W}\ppr\}\Big)_{j\ppr j}^{N\ppr N}}_{\hat{L}_{ij}^{MN}}\;
\Theta_{\vec{x},j}^{N}(t_{p}) \\ \lb{s3_31}
\hat{W}\ppr_{2N\times 2N}&=&
\left( \bea{cccc} \hat{B}_{D,L\times L} & \hat{c}_{D,L\times L} & \hat{\chi}_{D,L\times S}^{+} &
\hat{\eta}_{D,L\times S}^{T} \\
\hat{c}_{D,L\times L}^{+} & -\hat{B}_{D,L\times L}^{T} & \hat{\eta}_{D,L\times S}^{+} &
\hat{\chi}_{D,L\times S}^{T} \\
\hat{\chi}_{D,S\times L} & \hat{\eta}_{D,S\times L} & \hat{F}_{D,S\times S} & \hat{f}_{D,S\times S} \\
-\hat{\eta}_{D,S\times L}^{*} & -\hat{\chi}_{D,S\times
L}^{*} & -\hat{f}_{D,S\times S}^{+} & -\hat{F}_{D,S\times S}^{T} \eea\right) \\ \lb{s3_32}
&& \bea{rclrcl} \hat{B}_{D,L\times L}&=&\hat{B}_{D,L\times L}^{+}\hspace*{0.75cm}&\hat{c}_{D,L\times L}&=&
-\hat{c}_{D,L\times L}^{T} \\
\hat{F}_{D,S\times S}&=&\hat{F}_{D,S\times S}^{+} & \hat{f}_{D,S\times S}&=&\hat{f}_{D,S\times S}^{T}
\eea                        \\   \lb{s3_33}
-\im\hbar\frac{\hat{\pp}}{\pp t_{p}} &:&
\Theta_{\vec{x},i}^{M,ST}(t_{p})\;
\underbrace{\Big(\exp\{\im\;\wtilde{W}\ppr\}\Big)^{MM\ppr,ST}_{ii\ppr}\;
\wtilde{L}_{i\ppr j\ppr}^{M\ppr N\ppr}\;
\Big(\exp\{\im\;\wtilde{W}\ppr\}\Big)_{j\ppr j}^{N\ppr N}}_{\wtilde{L}_{ij}^{MN}}\;
\Theta_{\vec{x},j}^{N}(t_{p}) \\ \lb{s3_34}
\wtilde{W}\ppr_{2N\times 2N}&=&
\left( \bea{cccc} \wtilde{B}_{D,L\times L} & -\wtilde{c}_{D,L\times L} &
\wtilde{\chi}_{D,L\times S}^{+} &
\wtilde{\eta}_{D,L\times S}^{T} \\
\wtilde{c}_{D,L\times L}^{+} & -\wtilde{B}_{D,L\times L}^{T} & \wtilde{\eta}_{D,L\times S}^{+} &
\wtilde{\chi}_{D,L\times S}^{T} \\
\wtilde{\chi}_{D,S\times L} & -\wtilde{\eta}_{D,S\times L} &
\wtilde{F}_{D,S\times S} & \wtilde{f}_{D,S\times S} \\
\wtilde{\eta}_{D,S\times L}^{*} & -\wtilde{\chi}_{D,S\times L}^{*} & \wtilde{f}_{D,S\times S}^{+} &
-\wtilde{F}_{D,S\times S}^{T} \eea\right)
\\ \lb{s3_35} && \bea{rclrcl}
\wtilde{B}_{D,L\times L}&=&\wtilde{B}_{D,L\times L}^{\raisebox{-10pt}{$^{+}$}}
\hspace*{0.75cm}&
\wtilde{c}_{D,L\times L}&=&
\wtilde{c}_{D,L\times L}^{\raisebox{-10pt}{$^{T}$}} \\
\wtilde{F}_{D,S\times S}&=&\wtilde{F}_{D,S\times S}^{\raisebox{-10pt}{$^{+}$}} &
\wtilde{f}_{D,S\times S}&=&-\wtilde{f}_{D,S\times S}^{\raisebox{-10pt}{$^{T}$}}
\eea_{\mbox{.}}
\eeq
The block matrix entries in $\hat{W}\ppr$ (\ref{s3_31}) and
$\wtilde{W}\ppr$ (\ref{s3_34})
have been labelled under consideration of the analogous
block matrices in the self-energy \(\hat{\Sigma}_{\alpha\beta}^{ab}(\vec{x},t_{p})\) (\ref{s2_67},\ref{s2_68}).
Therefore, $\hat{B}_{D}$, $\wtilde{B}_{D}$ and $\hat{F}_{D}$, $\wtilde{F}_{D}$
and their transposed forms in $\hat{W}\ppr$ (\ref{s3_31}) and
$\wtilde{W}\ppr$ (\ref{s3_34}) are related to the density terms
$\hat{B}_{L\times L}$, $\hat{B}_{L\times L}^{T}$ and $\hat{F}_{S\times S}$, $-\hat{F}_{S\times S}^{T}$ in the
self-energy matrix \(\hat{\Sigma}_{\alpha\beta}^{ab}(\vec{x},t_{p})\) (\ref{s2_67})
with equivalent symmetries under hermitian conjugation (\ref{s2_68},\ref{s3_32},\ref{s3_35}).
A similar relationship is given between the fermionic density terms $\hat{\chi}_{D}$, $\hat{\chi}_{D}^{+}$ in
$\hat{W}\ppr$ (\ref{s3_31}) and $\wtilde{\chi}_{D}$, \(\wtilde{\chi}_{D}^{+}\)
in $\wtilde{W}\ppr$ (\ref{s3_34}) and the density
terms $\hat{\chi}_{S\times L}$, $\hat{\chi}_{L\times S}^{+}$ of the self-energy (\ref{s2_67},\ref{s2_68}).
The anomalous terms $\hat{c}_{D}$, $\hat{f}_{D}$ and
$\wtilde{c}_{D}$, $\wtilde{f}_{D}$ in $\hat{W}\ppr$ (\ref{s3_31}) and
$\wtilde{W}\ppr$ (\ref{s3_34}) are assigned to the
molecular condensate terms $\hat{c}_{L\times L}$ and the BCS-terms $\hat{f}_{S\times S}$ in
\(\hat{\Sigma}_{\alpha\beta}^{ab}(\vec{x},t_{p})\) (\ref{s2_67},\ref{s2_68}).
However, in the case of the generator $\hat{W}\ppr$ (\ref{s3_31}) for the operator
$\hat{h}_{p}(\vec{x})$ (\ref{s3_10},\ref{s3_4}), one has an antisymmetric relation in
the molecular condensate \(\hat{c}_{D}^{T}=-\hat{c}_{D}\) (\ref{s3_32}) and a symmetric BCS-term
\(\hat{f}_{D}^{T}=\hat{f}_{D}\) (\ref{s3_32})
instead of a symmetric molecular condensate matrix \(\hat{c}^{T}=\hat{c}\)
and instead of an antisymmetric fermion condensate matrix
\(\hat{f}^{T}=-\hat{f}\) as in the self-energy
\(\Sigma_{\alpha\beta}^{ab}(\vec{x},t_{p})\) (\ref{s2_67},\ref{s2_68}).
But in the case of the bilinear expression (\ref{s3_33}) with the time derivative for the
generator $\wtilde{W}\ppr$ (\ref{s3_34}),
the same symmetries (\ref{s3_35}) for the pair condensates
$\wtilde{c}_{D}=\wtilde{c}^{T}_{D}$
and $\wtilde{f}_{D}=-\wtilde{f}^{T}_{D}$ appear as in
the corresponding parts $\hat{c}=\hat{c}^{T}$ and $\hat{f}=-\hat{f}^{T}$ of
\(\hat{\Sigma}_{\alpha\beta}^{ab}(\vec{x},t_{p})\) (\ref{s2_67},\ref{s2_68}).
Obviously the generator $\hat{W}\ppr$ (\ref{s3_31}) for the bilinear relation (\ref{s3_10},\ref{s3_30}) of
\(\Psi_{\vec{x},\alpha}^{a}(t_{p})\) with one-particle operator $\hat{h}_{p}(\vec{x})$
has a different number of parameters as
independent degrees of freedom in the selfenegy matrix (\ref{s2_67},\ref{s2_68}); but in consequence of
identical symmetries (\ref{s3_35},\ref{s2_68}) in the anomalous parts of
$\wtilde{W}\ppr$ (\ref{s3_34}) and the self-energy (\ref{s2_67},\ref{s2_68}),
the number of independent fields in $\wtilde{W}\ppr$ (\ref{s3_34})
exactly equals that in \(\hat{\Sigma}_{\alpha\beta}^{ab}(\vec{x},t_{p})\) (\ref{s2_67},\ref{s2_68}),
including the separation into the boson-boson, fermion-fermion and odd block matrices.

By exchanging second and third block rows and columns of
$\hat{W}\ppr$ (\ref{s3_31}) and \(\wtilde{W}\ppr\) (\ref{s3_34}), one returns to the
hermitian bilinear relations (\ref{s3_5},\ref{s3_40}) of \(\Psi_{\vec{x},\alpha}^{a}(t_{p})\)
with metric $\hat{K}$ (\ref{s3_37}) for the corresponding
hermitian generator $\hat{W}$ (\ref{s3_36})(marked without a prime " $\ppr$ " ).
The analogous exchanges of rows and columns in (\ref{s3_34}) result in
the related hermitian form of the super-generator $\wtilde{W}$
(\ref{s3_41})(also marked without a prime '$\ppr$' ) with metric $\wtilde{K}$ (\ref{s3_42}).
Applying the exponential to the generators $\hat{W}$ (\ref{s3_36}) and $\wtilde{W}$ (\ref{s3_41}),
one finds the elements of the ortho-symplectic super-groups in
the matrices $\hat{T}$ (\ref{s3_39}) and $\wtilde{T}$ (\ref{s3_44})
which leave the corresponding hermitian bilinear relations (\ref{s3_40},\ref{s3_45}) with metric
$\hat{K}$ and $\wtilde{K}$ invariant \cite{cor}-\cite{luc}
\beq \lb{s3_36}
\hat{W}_{\alpha\beta}^{ab}&=&
\left( \bea{cc}
\left(\bea{cc}
\hat{B}_{D,L\times L} & \hat{\chi}_{D,L\times S}^{+} \\
\hat{\chi}_{D,S\times L} & \hat{F}_{D,S\times S}
\eea\right)^{11} & \left(\bea{cc}
\hat{c}_{D,L\times L} & \hat{\eta}_{D,L\times S}^{T} \\
\hat{\eta}_{D,S\times L} & \hat{f}_{D,S\times S}
\eea\right)^{12} \\ \left(\bea{cc}
\hat{c}_{D,L\times L}^{+} & \hat{\eta}_{D,L\times S}^{+} \\
-\hat{\eta}_{D,S\times L}^{*} & -\hat{f}_{D,S\times S}^{+}
\eea\right)^{21} & \left(\bea{cc}
-\hat{B}_{D,L\times L}^{T} & \hat{\chi}_{D,L\times S}^{T} \\
-\hat{\chi}_{D,S\times L}^{*} & -\hat{F}_{D,S\times S}^{T}
\eea\right)^{22}
\eea\right) \\ \lb{s3_37}
\hat{K}_{2N\times 2N}&=&\bigg\{ \underbrace{\hat{1}_{L\times L}}_{\mbox{BB}}\;,\; \underbrace{\hat{1}_{S\times
S}}_{\mbox{FF}}\;;\; \underbrace{\hat{1}_{L\times L}}_{\mbox{BB}}\;,\; \underbrace{-\hat{1}_{S\times
S}}_{\mbox{FF}}\bigg\}  \\  \lb{s3_38} && \hat{K}\;\;\hat{W}-\hat{W}^{+}\;\;\hat{K}=0 \hspace*{0.75cm}
\hat{W}_{N\times N}^{22,st}=-\hat{W}_{N\times N}^{11}
\eeq
\beq \lb{s3_39}
\hat{T}_{\alpha\beta}^{ab}&=&
\Big(\exp\Big\{\im\;\hat{W}_{\alpha\ppr\beta\ppr}^{a\ppr b\ppr}\Big\}\Big)_{\alpha\beta}^{ab} \\ \lb{s3_40}
\Psi_{\vec{x},\alpha}^{a+}(t_{p})\;\hat{h}_{p}(\vec{x})\;\hat{K}\; \Psi_{\vec{x},\alpha}^{a}(t_{p})&=&
\Psi_{\vec{x},\alpha}^{a+}(t_{p})\; \underbrace{\hat{T}^{+}\;\hat{K}\;\hat{T}}_{=\hat{K}}\;
\hat{h}_{p}(\vec{x})\;\Psi_{\vec{x},\alpha}^{a}(t_{p})
\eeq
\beq \lb{s3_41}
\wtilde{W}_{\alpha\beta}^{ab}&=&
\left( \bea{cc}
\left(\bea{cc}
\wtilde{B}_{D,L\times L} & \wtilde{\chi}_{D,L\times S}^{+} \\
\wtilde{\chi}_{D,S\times L} & \wtilde{F}_{D,S\times S}
\eea\right)^{11} & \left(\bea{cc}
-\wtilde{c}_{D,L\times L} & \wtilde{\eta}_{D,L\times S}^{T} \\
-\wtilde{\eta}_{D,S\times L} & \wtilde{f}_{D,S\times S}
\eea\right)^{12} \\
\left(\bea{cc}
\wtilde{c}_{D,L\times L}^{+} & \wtilde{\eta}_{D,L\times S}^{+} \\
\wtilde{\eta}_{D,S\times L}^{*} & \wtilde{f}_{D,S\times S}^{+}
\eea\right)^{21} & \left(\bea{cc}
-\wtilde{B}_{D,L\times L}^{T} & \wtilde{\chi}_{D,L\times S}^{T} \\
-\wtilde{\chi}_{D,S\times L}^{*} & -\wtilde{F}_{D,S\times S}^{T}
\eea\right)^{22}
\eea\right) \\ \lb{s3_42}
\wtilde{K}_{2N\times 2N}&=&\bigg\{ \underbrace{\hat{1}_{L\times L}}_{\mbox{BB}}\;,\;
\underbrace{\hat{1}_{S\times S}}_{\mbox{FF}}\;;\; \underbrace{-\hat{1}_{L\times L}}_{\mbox{BB}}\;,\;
\underbrace{\hat{1}_{S\times S}}_{\mbox{FF}}\bigg\} \\ \lb{s3_43} &&
\wtilde{K}\;\wtilde{W}-\wtilde{W}^{+}\; \wtilde{K}=0 \hspace*{0.75cm}
\wtilde{W}_{N\times N}^{22,st}=-\wtilde{W}_{N\times N}^{11}
\eeq
\beq \lb{s3_44}
\wtilde{T}_{\alpha\beta}^{ab}&=&
\Big(\exp\Big\{\im\;\wtilde{W}_{\alpha\ppr\beta\ppr}^{a\ppr b\ppr}\Big\}\Big)_{\alpha\beta}^{ab} \\ \lb{s3_45}
\Psi_{\vec{x},\alpha}^{a+}(t_{p})\; \bigg(-\im\hbar\frac{\hat{\pp}}{\pp t_{p}}\bigg) \;\wtilde{K}\;
\Psi_{\vec{x},\alpha}^{a}(t_{p})&=& \Psi_{\vec{x},\alpha}^{a+}(t_{p})\;
\underbrace{\wtilde{T}^{+}\;\wtilde{K}\;\wtilde{T}}_{ =\wtilde{K}}\;
\bigg(-\im\hbar\frac{\hat{\pp}}{\pp t_{p}}\bigg)\; \Psi_{\vec{x},\alpha}^{a}(t_{p})\;.
\eeq
The matrix \(\hat{W}_{\alpha\beta}^{ab}\) (\ref{s3_36}), consisting of the four block super-matrices
\(\hat{W}^{11}_{N\times N}\), \(\hat{W}^{22}_{N\times N}\) and
\(\hat{W}^{12}_{N\times N}\), \(\hat{W}^{21}_{N\times N}\), is classified as
the generator of the ortho-symplectic super-group \(Osp(L,L|2S)\) with the even part
\(SO(L,L)\oplus Sp(2S)\) for the boson-boson and fermion-fermion section with
\(2L^{2}-L\) and \(2S^{2}+S\) real variables \cite{cor}-\cite{luc}.
The odd part is the \((2L,2S)\) representation
of the even part and contains \(4LS\) independent Grassmann variables so that the total
dimension of \(\hat{W}_{\alpha\beta}^{ab}\) is given by \(2(L+S)^{2}-L+S\).
However, the super-generator
\(\wtilde{W}_{\alpha\beta}^{ab}\) (\ref{s3_41}) with the four
sub-super-matrices \(\wtilde{W}^{11}_{N\times N}\), \(\wtilde{W}^{22}_{N\times N}\),
\(\wtilde{W}^{12}_{N\times N}\), \(\wtilde{W}^{21}_{N\times N}\) is assigned to the ortho-symplectic
super-group \(Osp(S,S|2L)\) with the even part \(Sp(2L)\oplus SO(S,S)\) for the boson-boson
and fermion-fermion sections, respectively. Since the orthogonal and symplectic symmetry have
been exchanged with regard to \(\hat{W}_{\alpha\beta}^{ab}\) (\ref{s3_36}) or \(Osp(L,L|2S)\),
the number of independent real even variables amounts to \(2L^{2}+L\) in the boson-boson
part with \(Sp(2L)\) symmetry and to \(2S^{2}-S\) in the fermion-fermion block
with \(SO(S,S)\) symmetry of the super-group \(Osp(S,S|2L)\). The number of odd variables of
\(Osp(S,S|2L)\) is not effected by the exchange of the orthogonal with the symplectic metric
in the even part and has also \(4SL\) independent anti-commuting variables so that the
total dimension of \(Osp(S,S|2L)\) with
\(\wtilde{W}_{\alpha\beta}^{ab}\) (\ref{s3_41})
is given by \(2(S+L)^{2}-S+L\).
The density terms, represented by the block matrices
\(\hat{W}^{11}_{N\times N}\), \(\wtilde{W}^{11}_{N\times N}\) of
\(\hat{W}_{\alpha\beta}^{ab}\) and \(\wtilde{W}_{\alpha\beta}^{ab}\)
with their transposed parts \(\hat{W}^{22}_{N\times N}\),
\(\wtilde{W}^{22}_{N\times N}\),
are equivalent to the generators of the super-unitary
group \(U(L|S)\) where the doubled density parts \(\hat{W}_{N\times N}^{22}\),
\(\wtilde{W}_{N\times N}^{22}\) are the anti-unitary
realizations of the '11' blocks \(\hat{W}^{11}_{N\times N}\),
\(\wtilde{W}^{11}_{N\times N}\) of \(U(L|S)\) \cite{wei1}.
This has to be taken into account in the gradient expansion of the
Goldstone modes in a SSB (see section \ref{s4} with subsections).
The density terms of \(\hat{W}_{\alpha\beta}^{ab}\) (\ref{s3_36}) and
\(\wtilde{W}_{\alpha\beta}^{ab}\) (\ref{s3_41}) have identical form in both cases,
\(Osp(L,L|2S)\) for \(\hat{W}_{\alpha\beta}^{ab}\) and
\(Osp(S,S|2L)\) for \(\wtilde{W}_{\alpha\beta}^{ab}\).
The super-unitary group \(U(L|S)\) for the density terms is a subgroup in both
ortho-symplectic symmetry cases and determines with its \(L+S\) diagonal elements
of \(\hat{W}^{11}\), \(\wtilde{W}^{11}\) (and their transposes
\(\hat{W}^{22}\), \(\wtilde{W}^{22}\)) the rank \(N=L+S\) of the complete
ortho-symplectic super-groups with \(\hat{W}_{\alpha\beta}^{ab}\)  (\ref{s3_36}) and
\(\wtilde{W}_{\alpha\beta}^{ab}\) (\ref{s3_41}).

Comparing the super-generator $\wtilde{W}_{\alpha\beta}^{ab}$ (\ref{s3_41}) of
\(Osp(S,S|2L)\) with the self-energy matrix
\(\hat{\Sigma}_{\alpha\beta}^{ab}(\vec{x},t_{p})\) (\ref{s2_67},\ref{s2_68}), one identifies the
super-matrix \(\hat{\Sigma}_{\alpha\beta}^{ab}(\vec{x},t_{p})\;\wtilde{K}\)
(, the self-energy multiplied with the corresponding metric  \(\wtilde{K}\) (\ref{s3_42}),) as a
super-generator of \(Osp(S,S|2L)\).
Therefore, the action of the coherent state path integral (\ref{s2_69}) consists of the supermanifold in the
group \(Osp(S,S|2L)\), represented by the self-energy \(\hat{\Sigma}_{2N\times 2N}\;\wtilde{K}\).
However, the transformations, which have different
degrees of freedom as the self-energy $\hat{\Sigma}_{2N\times 2N}$,
but leave the coherent state path integral (\ref{s2_69}) invariant,
are obtained by the super-generator $\hat{W}_{\alpha\beta}^{ab}$ (\ref{s3_36}) of \(Osp(L,L|2S)\).
The symmetry breaking term with the time derivative or energy
can be neglected in this case because slowly varying
fields with respect to the reference energy $\mu_{0}$ are only considered
for the anomalous terms. One has to distinguish between the group manifold
(\ref{s3_41},\ref{s3_44}) \(Osp(S,S|2L)\) of the self-energy matrix and the transformations
(\ref{s3_36},\ref{s3_39}) \(Osp(L,L|2S)\) which do not alter the form of
$Z[\hat{\mcal{J}},J_{\psi},\hat{J}_{\psi\psi}]$ (\ref{s2_69}).
The group manifold \(Osp(S,S|2L)\) (\ref{s3_41},\ref{s3_44})
of $\hat{\Sigma}_{2N\times 2N}\;\wtilde{K}$ differs from
the invariant transformations \(Osp(L,L|2S)\) (\ref{s3_36},\ref{s3_39}) of
$Z[\hat{\mcal{J}},J_{\psi},\hat{J}_{\psi\psi}]$ (\ref{s2_69}) by the exchange of the symplectic
symmetry with $\wtilde{J}_{2L\times 2L}$ (\ref{s3_16}) in the boson-boson block of
$\hat{\Sigma}_{\alpha\beta}^{ab}$ with the orthogonal symmetry $\hat{G}_{2L\times 2L}$
(\ref{s3_12}) in \(Osp(L,L|2S)\).
Similarly, the orthogonal symmetry with metric $\wtilde{G}_{2S\times 2S}$ (\ref{s3_17})
in the manifold of the fermion-fermion block
of $\hat{\Sigma}_{2N\times 2N}\;\wtilde{K}$ (\ref{s2_67},\ref{s2_68}) is
changed to the symplectic symmetry $\hat{J}_{2S\times 2S}$ (\ref{s3_13}) in
the invariant transformations \(Osp(L,L|2S)\) of $Z[\hat{\mcal{J}},J_{\psi},\hat{J}_{\psi\psi}]$ (\ref{s2_69}).

In order to separate the anomalous terms from the densities in a kind of 'diagonalization',
the subgroup and coset space have to be determined for the spontaneous symmetry breaking.
One can diagonalize the self-energy matrix (\ref{s3_46}) with the
coset matrix $\wtilde{T}_{\hat{Y}}$ (\ref{s3_47},\ref{s3_48}) for the anomalous terms
in the manifold of \(Osp(S,S|2L)\) and with the block diagonal density
matrix $\hat{\Sigma}_{D;2N\times 2N}$ (\ref{s3_49}-\ref{s3_51})
so that the subgroup \(U(L|S)\) remains for the manifold of the background fields or ground state in a SSB
\footnote{Compare the matrix elements and symmetries of
\(\hat{\Sigma}_{\alpha\beta}^{ab}(\vec{x},t_{p})\;\wtilde{K}\) (\ref{s2_67},\ref{s2_68}) with the generator
\(\wtilde{W}_{\alpha\beta}^{ab}\) (\ref{s3_41}) of \(Osp(S,S|2L)\).}
\beq \lb{s3_46}
\hat{\Sigma}_{\alpha\beta}^{ab}(\vec{x},t_{p})&=&\Big(\wtilde{T}_{\hat{Y}}\;
\hat{\Sigma}_{D;2N\times 2N}\;\wtilde{T}_{\hat{Y}}^{+}\Big)_{\alpha\beta}^{ab}\hspace*{1.0cm}
\wtilde{T}_{\hat{Y}}^{+}\;\wtilde{K}\;\wtilde{T}_{\hat{Y}}=\wtilde{K} \\ \lb{s3_47}
\wtilde{T}_{\hat{Y}}(\vec{x},t_{p})&=&\exp\Big\{\im\;\hat{Y}(\vec{x},t_{p})\Big\}\hspace*{1.0cm}
\hat{Y}_{\alpha\beta}^{ab}(\vec{x},t_{p})=\left( \bea{cc}
0 & \hat{X}_{\alpha\beta}(\vec{x},t_{p}) \\
\wtilde{\kappa}\;\hat{X}_{\alpha\beta}^{+}(\vec{x},t_{p}) & 0 \eea\right)^{ab} \\ \lb{s3_48}
\hat{X}_{\alpha\beta}(\vec{x},t_{p})&=&\left( \bea{cc}
-\hat{c}_{D;L\times L}(\vec{x},t_{p}) & \hat{\eta}_{D;L\times S}^{T}(\vec{x},t_{p}) \\
-\hat{\eta}_{D;S\times L}(\vec{x},t_{p}) & \hat{f}_{D;S\times S}(\vec{x},t_{p})
\eea\right) \\ \no &&
\hat{c}_{D;L\times L}^{T}(\vec{x},t_{p})=\hat{c}_{D;L\times L}(\vec{x},t_{p})\hspace*{0.55cm}
\hat{f}_{D;S\times S}^{T}(\vec{x},t_{p})=-\hat{f}_{D;S\times S}(\vec{x},t_{p}) \\ \lb{s3_49}
\hat{\Sigma}_{D;\alpha\beta}^{ab}(\vec{x},t_{p})&=&\delta_{a,b}\;\left( \bea{cc}
\hat{\Sigma}_{D;\alpha\beta}^{11}(\vec{x},t_{p}) & 0 \\
0 & \hat{\Sigma}_{D;\alpha\beta}^{22}(\vec{x},t_{p}) \eea\right) \\ \no &&
\big(\hat{\Sigma}_{D;N\times N}^{22}(\vec{x},t_{p})\;\wtilde{\kappa}\big)^{st}
=-\hat{\Sigma}_{D;N\times N}^{11}(\vec{x},t_{p}) \\ \lb{s3_50}
\hat{\Sigma}_{D;\alpha\beta}^{11}(\vec{x},t_{p})&=&\left( \bea{cc}
\hat{B}_{D;L\times L}(\vec{x},t_{p}) & \hat{\chi}_{D;L\times S}^{+}(\vec{x},t_{p}) \\
\hat{\chi}_{D;S\times L}(\vec{x},t_{p}) & \hat{F}_{D;S\times S}(\vec{x},t_{p})
\eea\right)   \\  \lb{s3_51}
\hat{\Sigma}_{D;\alpha\beta}^{22}(\vec{x},t_{p})&=&\left( \bea{cc}
\hat{B}_{D;L\times L}^{T}(\vec{x},t_{p}) & \hat{\chi}_{D;L\times S}^{T}(\vec{x},t_{p}) \\
\hat{\chi}_{D;S\times L}^{*}(\vec{x},t_{p}) & -\hat{F}_{D;S\times S}^{T}(\vec{x},t_{p})
 \eea\right)_{\mbox{.}}
\eeq
However, this form of the self-energy (\ref{s3_46}-\ref{s3_51})
with the equivalent degrees of freedom as
$\hat{\Sigma}_{2N\times 2N}\;\wtilde{K}$ in 'flat' coordinates
does not diagonalize the quadratic
interaction term \(\mbox{STR}[\hat{\Sigma}_{2N\times 2N}\;\hat{K}\;\hat{\Sigma}_{2N\times 2N}\;\hat{K}]\)
in \(Z[\hat{\mcal{J}},J_{\psi},\hat{J}_{\psi\psi}]\) (\ref{s2_69})
to the density terms \(\hat{\Sigma}_{D;N\times N}^{11}\), \(\hat{\Sigma}_{D;N\times N}^{22}\)
(\ref{s3_49}-\ref{s3_51}) because of the contained metric $\hat{K}$ (\ref{s3_37})
instead of the required metric $\wtilde{K}$ (\ref{s3_42},\ref{s3_46})
\beq \lb{s3_52} \lefteqn{\hspace*{-4.0cm}
\mbox{STR}\Big[\wtilde{T}_{\hat{Y}}(\vec{x},t_{p})\;\hat{\Sigma}_{D;\alpha\beta}^{aa}(\vec{x},t_{p})\;
\underbrace{\wtilde{T}_{\hat{Y}}^{+}(\vec{x},t_{p})\;\hat{K}\;\wtilde{T}_{\hat{Y}}(\vec{x},t_{p})}_{\neq
\hat{K}}\;\hat{\Sigma}_{D;\alpha\ppr\beta\ppr}^{a\ppr a\ppr}(\vec{x},t_{p})\;
\wtilde{T}_{\hat{Y}}^{+}(\vec{x},t_{p})\;\hat{K}\Big]  \neq }
\\ \no &\neq& \mbox{STR}\Big[\hat{\Sigma}_{D;\alpha\beta}^{ab}(\vec{x},t_{p})\;\hat{K}\;
\hat{\Sigma}_{D;\beta\alpha}^{ba}(\vec{x},t_{p})\;\hat{K}\Big]
\eeq
\be \lb{s3_53}
\hat{K}\;\hat{Y}_{\alpha\beta}^{ab}(\vec{x},t_{p})-
\hat{Y}_{\alpha\beta}^{+;ab}\;\hat{K} \neq 0\;\;\;\;.
\ee
The chosen form of \(\hat{\Sigma}_{\alpha\beta}^{ab}(\vec{x},t_{p})=
\big(\wtilde{T}_{\hat{Y}}\;\hat{\Sigma}_{D;2N\times 2N}\;\wtilde{T}_{\hat{Y}}^{+}\big)_{\alpha\beta}^{ab}\)
(\ref{s3_46}) does not decouple the anomalous
degrees of freedom from the density terms because the invariant transformations \(Osp(L,L|2S)\)
with $\hat{W}_{\alpha\beta}^{ab}$ (\ref{s3_36}) of $Z[\hat{\mcal{J}},J_{\psi},\hat{J}_{\psi\psi}]$ (\ref{s2_69})
are different from the group manifold \(Osp(S,S|2L)\) of the self-energy
(as $\wtilde{W}_{\alpha\beta}^{ab}$ (\ref{s3_41})).
Consequently one has to change the original coherent state
path integral of the fields $\psi_{\vec{x},\alpha}(t_{p})$ with a different Hubbard Stratonovich
transformation of the interaction term so that the group manifold of the self-energy also
allows invariant transformations in the coherent state path integral.

\subsection{Coset construction $Osp(S,S|2L) \backslash U(L|S)\otimes U(L|S)$ for the self-energy} \lb{s32}

In section \ref{s31} we have determined the ortho-symplectic group \(Osp(S,S|2L)\) as the manifold of the
self-energy \(\hat{\Sigma}_{2N\times 2N}\;\wtilde{K}\)
whereas the generating function \(Z[\hat{\mcal{J}},J_{\psi},\hat{J}_{\psi\psi}]\) (\ref{s2_69}), composed of
the super-trace relation (\ref{s3_52}) with two factors of $\hat{\Sigma}_{2N\times 2N}\;\hat{K}$ and of
the matrix \(\hat{M}_{\vec{x},\alpha;\vec{x}\ppr,\beta}^{ab}(t_{p},t_{q}\ppr)\) (\ref{s2_71}),
is only invariant under \(Osp(L,L|2S)\) with a different number
of independent parameters in the anomalous terms.
Various kinds of the HST can be taken for the quartic interaction term, as e.g.,
by restricting to the density terms or alternatively by only using the anomalous
parts \cite{nag1,nag2}. However, we can transform the hermitian anomalous terms
\((\hat{\Sigma}_{N\times N}^{12}(\vec{x},t_{p}))^{+}=\hat{\Sigma}_{N\times N}^{21}(\vec{x},t_{p})\) of the self-energy
in the group manifold of \(Osp(S,S|2L)\)
and the super-trace relation (\ref{s3_52}) to an anti-hermitian form
\(\hat{\Sigma}_{N\times N}^{12}(\vec{x},t_{p})\rightarrow\im\;\hat{\Sigma}_{N\times N}^{12}(\vec{x},t_{p})\),
\(\hat{\Sigma}_{N\times N}^{21}(\vec{x},t_{p})\rightarrow\im\;\hat{\Sigma}_{N\times N}^{21}(\vec{x},t_{p})\)
(Weyl-unitary trick \cite{weyl1}).
In this case the number of independent parameters does not change for the manifold \(Osp(S,S|2L)\)
of the self-energy, but the unsuitable metric $\hat{K}$ (\ref{s3_37},\ref{s3_52})
is converted to the metric $\wtilde{K}$ (\ref{s3_42},\ref{s3_53})
in the super-trace (\ref{s3_52}) with the product of two self-energy matrices
which are also composed of the manifold \(Osp(S,S|2L)\) for
\(\hat{\Sigma}_{2N\times 2N}\;\wtilde{K}\).
Therefore, the self-energy can be 'diagonalized' with the density terms as 'block' eigenvalues and
coset matrices in a similarity transformation. The invariant transformations of the generating
function are then performed according to the metric \(\wtilde{K}\) and take place
within the equivalent manifold \(Osp(S,S|2L)\) of the self-energy with the same number of
independent variables
\beq \lb{s3_54}
\lefteqn{\mbox{STR}\Bigg[\left( \bea{cc}
\hat{\Sigma}^{11} & \hat{\Sigma}^{12} \\
\hat{\Sigma}^{21} & \hat{\Sigma}^{22} \eea\right)
\underbrace{\left( \bea{cc}
\hat{1} & \\
 & \hat{\kappa}
 \eea\right)}_{\hat{K}}\left(
\bea{cc}
\hat{\Sigma}^{11} & \hat{\Sigma}^{12} \\
\hat{\Sigma}^{21} & \hat{\Sigma}^{22} \eea\right)
\underbrace{\left( \bea{cc}
\hat{1} & \\
 & \hat{\kappa}
 \eea\right)}_{\hat{K}}\Bigg]= } \\ \no &=&
\mbox{STR}\Bigg[\underbrace{\left( \bea{cc}
\hat{\Sigma}^{11} & \im\;\hat{\Sigma}^{12} \\
\im\;\hat{\Sigma}^{21} & \hat{\Sigma}^{22} \eea\right)}_{\wtilde{\Sigma}_{2N\times 2N}}
\underbrace{\left( \bea{cc}
\hat{1} & \\
 & \wtilde{\kappa}
 \eea\right)}_{\wtilde{K}}
 \underbrace{\left(
\bea{cc}
\hat{\Sigma}^{11} & \im\;\hat{\Sigma}^{12} \\
\im\;\hat{\Sigma}^{21} & \hat{\Sigma}^{22} \eea\right)}_{\wtilde{\Sigma}_{2N\times 2N}}
\underbrace{\left( \bea{cc}
\hat{1} & \\
 & \wtilde{\kappa}
 \eea\right)}_{\wtilde{K}}\Bigg]_{.}
\eeq
We decompose or 'diagonalize' the self-energy $\wtilde{\Sigma}_{2N\times 2N}$ (\ref{s3_55},\ref{s3_54})
into density terms \(\hat{\Sigma}_{D;N\times N}^{11}\), \(\hat{\Sigma}_{D;N\times N}^{22}\;\wtilde{\kappa}\)
(\ref{s3_49}-\ref{s3_51}) of the subgroup \(U(L|S)\) and cosets \(Osp(S,S|2L)\backslash U(L|S)\)
represented by the matrices $\wtilde{T}_{1}$,
$\wtilde{T}_{2}$ (\ref{s3_57},\ref{s3_58}) for the anti-hermitian anomalous terms; this
construction \(Osp(S,S|2L)\backslash U(L|S)\otimes U(L|S)\) considers spontaneous symmetry
breaking of \(Osp(S,S|2L)\) by its subgroup \(U(L|S)\)
\footnote{In the following the tilde
'$\wtilde{\ph{\Sigma}}$' of $\wtilde{\Sigma}_{2N\times 2N}$ refers to a self-energy with
anti-hermitian anomalous terms $\im\;\hat{\Sigma}_{N\times N}^{12}$,
$\im\;\hat{\Sigma}_{N\times N}^{21}$ in comparison to $\hat{\Sigma}_{2N\times 2N}$
with hermitian pair condensates $\hat{\Sigma}_{N\times N}^{12}$, $\hat{\Sigma}_{N\times N}^{21}$;
\(\hat{\Sigma}_{N\times N}^{21}=\big(\hat{\Sigma}_{N\times N}^{12}\big)^{+}\).}
\beq \lb{s3_55}
\wtilde{\Sigma}_{2N\times 2N}(\vec{x},t_{p})&=&\left( \bea{cc}
\hat{\Sigma}^{11} & \im\;\hat{\Sigma}^{12} \\
\im\;\hat{\Sigma}^{21} & \hat{\Sigma}^{22} \eea\right)=
\wtilde{T}_{1}\;\left( \bea{cc}
\hat{\Sigma}_{D}^{11} & 0 \\
0 & \hat{\Sigma}_{D}^{22} \eea\right)\;\wtilde{T}_{2} \\ \no &=&
\wtilde{T}_{1}\;\hat{\Sigma}_{D;2N\times 2N}\;\wtilde{T}_{2} \\ \lb{s3_56}
\wtilde{T}_{2}\;\wtilde{K}\;\wtilde{T}_{1}=\wtilde{K}  &&
\hat{\Sigma}_{N\times N}^{21}=\big(\hat{\Sigma}_{N\times N}^{12}\big)^{+}\hspace*{0.5cm}
\wtilde{\Sigma}_{N\times N}^{12}=\im\;\hat{\Sigma}_{N\times N}^{12}\hspace*{0.5cm}
\wtilde{\Sigma}_{N\times N}^{21}=\im\;\hat{\Sigma}_{N\times N}^{21}  \\ \lb{s3_57}
\wtilde{T}_{1}(\vec{x},t_{p})&=&\exp\Bigg\{\im\left( \bea{cc}
0 & \im\;\hat{X}_{N\times N} \\
\im\;\wtilde{\kappa}\;\hat{X}_{N\times N}^{+} & 0
\eea\right)\Bigg\}=\exp\Big\{-\hat{Y}_{2N\times 2N}\Big\} \\ \lb{s3_58}
\wtilde{T}_{2}(\vec{x},t_{p})&=&\exp\Bigg\{-\im\left( \bea{cc}
0 & \im\;\hat{X}_{N\times N}\;\wtilde{\kappa} \\
\im\;\hat{X}_{N\times N}^{+} & 0
\eea\right)\Bigg\}=\exp\Big\{\hat{Y}_{2N\times 2N}^{+}\Big\} \\ \lb{s3_59}
\hat{Y}_{2N\times 2N}(\vec{x},t_{p})&=&\left( \bea{cc}
0 & \hat{X}_{N\times N}(\vec{x},t_{p}) \\
\wtilde{\kappa}_{N\times N}\;\hat{X}_{N\times N}^{+}(\vec{x},t_{p}) & 0 \eea\right) \\ \lb{s3_60}
\hat{X}_{N\times N}(\vec{x},t_{p})&=&\left( \bea{cc}
-\hat{c}_{D;L\times L}(\vec{x},t_{p}) & \hat{\eta}_{D;L\times S}^{T}(\vec{x},t_{p}) \\
-\hat{\eta}_{D;S\times L}(\vec{x},t_{p}) & \hat{f}_{D;S\times S}(\vec{x},t_{p}) \eea\right) \\ \lb{s3_61}
\hat{c}_{D;L\times L}^{T}(\vec{x},t_{p})&\!\!=\!\!&\hat{c}_{D;L\times L}(\vec{x},t_{p})\hspace*{1.0cm}
\hat{f}_{D;S\times S}^{T}(\vec{x},t_{p})=-\hat{f}_{D;S\times S}(\vec{x},t_{p})\;\;\;.
\eeq
(Compare the definition of the super-generator $\hat{Y}_{2N\times 2N}$ (\ref{s3_59}) with
$\hat{X}_{N\times N}$ for anti-hermitian anomalous terms in $\wtilde{T}_{1}$ with
the equivalent symmetries (\ref{s3_61}) as the generators
\(\wtilde{W}_{N\times N}^{12}\),
\(\wtilde{W}_{N\times N}^{21}\) in $\wtilde{W}_{2N\times 2N}$
(\ref{s3_41}) and $\wtilde{T}$ (\ref{s3_44})).
We introduce an additional diagonal density field
\(\sigma_{D}^{(0)}(\vec{x},t_{p})\;\hat{1}_{N\times N}=
\frac{1}{2}[(\hat{\Sigma}_{\alpha\beta}^{22}(\vec{x},t_{p})\;\wtilde{\kappa})^{st}+
\hat{\Sigma}_{\alpha\beta}^{11}(\vec{x},t_{p})]\)
in the self-energy and allow for fluctuations with the densities
$\delta\hat{\Sigma}_{N\times N}^{11}$, $\delta\hat{\Sigma}_{N\times N}^{22}$
in the subgroup \(U(L|S)\) (\ref{s3_62}-\ref{s3_67}).
The subgroup \(U(L|S)\), as the invariant vacuum or ground state in the
spontaneous symmetry breaking of \(Osp(S,S|2L)\) to
\(Osp(S,S|2L)\backslash U(L|S)\), is represented
by the independent fields of $\hat{\Sigma}_{N\times N}^{11}$ (\ref{s3_62}-\ref{s3_67})
which also occur in $\hat{\Sigma}_{N\times N}^{22}$ (\ref{s3_63}-\ref{s3_67}) as a super-transposed part
\((\delta\hat{\Sigma}^{22}\;\wtilde{\kappa})^{st}=-\delta\hat{\Sigma}^{11}\),
but with an additional minus sign (\ref{s3_64}). (Compare the independent parameters of
$\delta\hat{\Sigma}_{N\times N}^{11}$, $\delta\hat{\Sigma}_{N\times N}^{22}\;\wtilde{\kappa}$
with the density terms of the '11' and '22' sections in
$\wtilde{W}_{\alpha\beta}^{aa}$ (\ref{s3_41})
as the subgroup \(U(L|S)\)  of \(Osp(S,S|2L)\).)
\beq \lb{s3_62}
\hat{\Sigma}_{N\times N}^{11}(\vec{x},t_{p})&=&
\sigma_{D}^{(0)}(\vec{x},t_{p})\;\hat{1}_{N\times N}+
\delta\hat{\Sigma}_{N\times N}^{11}(\vec{x},t_{p}) \\ \lb{s3_63}
\hat{\Sigma}_{N\times N}^{22}(\vec{x},t_{p})&=&
\sigma_{D}^{(0)}(\vec{x},t_{p})\;\wtilde{\kappa}_{N\times N}+
\delta\hat{\Sigma}_{N\times N}^{22}(\vec{x},t_{p}) \\ \lb{s3_64}
\delta\hat{\Sigma}_{N\times N}^{11}(\vec{x},t_{p}) &=&
-\Big(\delta\hat{\Sigma}_{N\times N}^{22}(\vec{x},t_{p})\;\;\wtilde{\kappa}\Big)^{st} \\ \lb{s3_65}
\sigma_{D}^{(0)}(\vec{x},t_{p})\;\hat{1}_{N\times N}&=&
\frac{1}{2}\Big[\big(\hat{\Sigma}_{N\times N}^{22}(\vec{x},t_{p})\;\;\wtilde{\kappa}\big)^{st}+
\hat{\Sigma}_{N\times N}^{11}(\vec{x},t_{p})\Big] \\ \lb{s3_66}
\delta\hat{\Sigma}^{11}_{N\times N}&=&\left( \bea{cc}
\delta \hat{B}_{L\times L} & \delta\hat{\chi}_{L\times S}^{+} \\
\delta\hat{\chi}_{S\times L} & \delta \hat{F}_{S\times S} \eea\right) \hspace*{0.75cm}
\delta\hat{\Sigma}^{22}_{N\times N} = \left( \bea{cc}
\delta \hat{B}_{L\times L}^{T} & \delta\hat{\chi}_{L\times S}^{T} \\
\delta\hat{\chi}_{S\times L}^{*} & -\delta \hat{F}_{S\times S}^{T} \eea\right) \\ \lb{s3_67}
\delta\hat{B}_{L\times L}^{+}(\vec{x},t_{p})&=&\delta\hat{B}_{L\times L}(\vec{x},t_{p})\hspace*{1.0cm}
\delta\hat{F}_{S\times S}^{+}(\vec{x},t_{p})=\delta\hat{F}_{S\times S}(\vec{x},t_{p})\;\;\;.
\eeq
Similarly, we include the mean density field $\sigma_{D}^{(0)}(\vec{x},t_{p})$ and the
corresponding fluctuations $\delta\hat{\Sigma}_{D;N\times N}^{11}$,
$\delta\hat{\Sigma}_{D;N\times N}^{22}$ in the density terms $\hat{\Sigma}_{D;N\times N}^{11}$,
$\hat{\Sigma}_{D;N\times N}^{22}$ of $\hat{\Sigma}_{D;2N\times 2N}$ (\ref{s3_55})
for 'diagonalizing' to the subgroup \(U(L|S)\) of \(Osp(S,S|2L)\)
with matrices \(\wtilde{T}_{1}\) and \(\wtilde{T}_{2}\) (\ref{s3_57}-\ref{s3_61})
\beq \lb{s3_68}
\hat{\Sigma}^{11}_{D;N\times N}(\vec{x},t_{p})&=&
\sigma_{D}^{(0)}(\vec{x},t_{p})\;\hat{1}_{N\times N}+\delta\hat{\Sigma}^{11}_{D;N\times N}(\vec{x},t_{p})
 \\ \lb{s3_69} \hat{\Sigma}^{22}_{D;N\times N}(\vec{x},t_{p})&=&
\sigma_{D}^{(0)}(\vec{x},t_{p})\;\wtilde{\kappa}_{N\times N}+
\delta\hat{\Sigma}^{22}_{D;N\times N}(\vec{x},t_{p})
\\ \lb{s3_70}
\delta\hat{\Sigma}^{11}_{D;N\times N}(\vec{x},t_{p}) &=&
-\Big(\delta\hat{\Sigma}^{22}_{D;N\times N}(\vec{x},t_{p})\;\;\wtilde{\kappa}\Big)^{st}
\\ \lb{s3_71}
\sigma_{D}^{(0)}(\vec{x},t_{p})\;\hat{1}_{N\times N}&=&
\frac{1}{2}\Big[\big(\hat{\Sigma}^{22}_{D;N\times N}(\vec{x},t_{p})\;\;\wtilde{\kappa}\big)^{st}+
\hat{\Sigma}^{11}_{D;N\times N}(\vec{x},t_{p})\Big] \\ \lb{s3_72}
\delta\hat{\Sigma}_{D;N\times N}^{11}(\vec{x},t_{p})&=&\left( \bea{cc}
\delta \hat{B}_{D;L\times L}(\vec{x},t_{p}) & \delta\hat{\chi}_{D;L\times S}^{+}(\vec{x},t_{p}) \\
\delta\hat{\chi}_{D;S\times L}(\vec{x},t_{p}) & \delta \hat{F}_{D;S\times S}(\vec{x},t_{p}) \eea\right) \\ \lb{s3_73}
\delta\hat{\Sigma}_{D;N\times N}^{22}(\vec{x},t_{p})&=&\left( \bea{cc}
\delta \hat{B}_{D;L\times L}^{T}(\vec{x},t_{p}) & \delta\hat{\chi}_{D;L\times S}^{T}(\vec{x},t_{p}) \\
\delta\hat{\chi}_{D;S\times L}^{*}(\vec{x},t_{p}) & -\delta \hat{F}_{D;S\times S}^{T}(\vec{x},t_{p}) \eea\right)
\\ \lb{s3_74} \delta\hat{B}_{D;L\times L}^{+}(\vec{x},t_{p})&=&
\delta\hat{B}_{D;L\times L}(\vec{x},t_{p})\hspace*{1.0cm}
\delta\hat{F}_{D;S\times S}^{+}(\vec{x},t_{p})=\delta\hat{F}_{D;S\times S}(\vec{x},t_{p})\;\;\;.
\eeq
In the remainder the self-energy $\wtilde{\Sigma}_{2N\times 2N}$ (\ref{s3_55})
with density fluctuations (\ref{s3_62}-\ref{s3_67}) and
coset matrices $\wtilde{T}_{1}$, $\wtilde{T}_{2}$ (\ref{s3_55}-\ref{s3_61})
is therefore set to the following form with anti-hermitian pair condensate terms
\beq \lb{s3_75}
\wtilde{\Sigma}_{2N\times 2N}(\vec{x},t_{p})&=&\left( \bea{cc}
\sigma_{D}^{(0)}\;\hat{1}_{N\times N}+\delta\hat{\Sigma}_{N\times N}^{11} &
\im\;\delta\hat{\Sigma}_{N\times N}^{12} \\
\im\;\delta\hat{\Sigma}_{N\times N}^{21} & \sigma_{D}^{(0)}\;
\wtilde{\kappa}_{N\times N}+\delta\hat{\Sigma}_{N\times N}^{22} \eea\right) \\ \no
&=&\wtilde{T}_{1}\;\left( \bea{cc}
\sigma_{D}^{(0)}\;\hat{1}_{N\times N}+\delta\hat{\Sigma}_{D;N\times N}^{11} & 0 \\
0 & \sigma_{D}^{(0)}\;\wtilde{\kappa}_{N\times N}+
\delta\hat{\Sigma}_{D;N\times N}^{22} \eea\right)\;\wtilde{T}_{2} \\ \no
&=&\sigma_{D}^{(0)}(\vec{x},t_{p})\;\wtilde{K}_{2N\times 2N}+
\underbrace{\wtilde{T}_{1}\;\;\delta\hat{\Sigma}_{D;2N\times 2N}\;\;
\wtilde{T}_{2}}_{\delta\wtilde{\Sigma}_{2N\times 2N}(\vec{x},t_{p})}
\hspace*{1.75cm}\wtilde{T}_{2}\;\wtilde{K}\;
\wtilde{T}_{1}=\wtilde{K} \\ \lb{s3_76}
\delta\wtilde{\Sigma}_{2N\times 2N}(\vec{x},t_{p})&=&\left(
\bea{cc}
\delta\hat{\Sigma}_{N\times N}^{11} & \im\;\delta\hat{\Sigma}_{N\times N}^{12} \\
\im\;\delta\hat{\Sigma}_{N\times N}^{21} & \delta\hat{\Sigma}_{N\times N}^{22}
\eea\right)=\wtilde{T}_{1}\;\;\delta\hat{\Sigma}_{D;2N\times 2N}\;\;
\wtilde{T}_{2}
\\ \lb{s3_77} \delta\hat{\Sigma}^{12}_{N\times N}(\vec{x},t_{p})&=&\left(
\bea{cc}
\delta\hat{c}_{L\times L} & \delta\hat{\eta}_{L\times S}^{T} \\
\delta\hat{\eta}_{S\times L} & \delta\hat{f}_{S\times S}
\eea\right)\hspace*{0.46cm}
\delta\hat{\Sigma}_{N\times N}^{21}=\big(\delta\hat{\Sigma}_{N\times N}^{12}\big)^{+}
\\ \no &&
\delta\hat{c}^{T}_{L\times L}=\delta\hat{c}_{L\times L}\hspace{0.35cm}
\delta\hat{f}_{S\times S}^{T}=-\delta\hat{f}_{S\times S}  \\ \lb{s3_78}
\delta\hat{\Sigma}_{D;2N\times 2N}(\vec{x},t_{p})&=&\left(
\bea{cc}
\delta\hat{\Sigma}_{D;N\times N}^{11} & 0 \\
0 & \delta\hat{\Sigma}_{D;N\times N}^{22}
\eea\right)\;\;\;.
\eeq
According to Eqs. (\ref{s3_75}), the anomalous terms or pair condensates
with coset matrices $\wtilde{T}_{1}(\vec{x},t_{p})$,
$\wtilde{T}_{2}(\vec{x},t_{p})$ (\ref{s3_57}-\ref{s3_61})
of \(Osp(S,S|2L)\backslash U(L|S)\)
are coupled to the block diagonal density fluctuations
$\delta\hat{\Sigma}_{D;2N\times 2N}(\vec{x},t_{p})$ (\ref{s3_72},\ref{s3_74})
of the subgroup \(U(L|S)\). The real density field $\sigma_{D}^{(0)}(\vec{x},t_{p})$ (\ref{s3_71},\ref{s3_65})
in the self-energy matrix $\wtilde{\Sigma}_{2N\times 2N}(\vec{x},t_{p})$
(\ref{s3_75}) is invariant with the super-unitary
transformations \(U(L|S)\) as subgroup of \(Osp(S,S|2L)\), both in the '11' block
\(\hat{\Sigma}_{N\times N}^{11}(\vec{x},t_{p})\)
and with super-transposed parameters in the '22' block
\(\hat{\Sigma}_{N\times N}^{22}(\vec{x},t_{p})\;\wtilde{\kappa}\)
(see the block diagonal super-matrices
$\wtilde{W}_{N\times N}^{11}$,
$\wtilde{W}_{N\times N}^{22}$
in $\wtilde{W}^{ab}_{\alpha\beta}$ (\ref{s3_41})). Therefore, the field
\(\sigma_{D}^{(0)}(\vec{x},t_{p})\;\wtilde{K}\) is regarded as the vacuum or ground state in
a spontaneous symmetry breaking with density fluctuations
$\delta\hat{\Sigma}_{D;2N\times 2N}(\vec{x},t_{p})$ (\ref{s3_78})
to which the anomalous fields in $\wtilde{T}_{1}(\vec{x},t_{p})$,
$\wtilde{T}_{2}(\vec{x},t_{p})$ (\ref{s3_57},\ref{s3_58}) are combined.
These density fields $\delta\hat{\Sigma}_{D;2N\times 2N}(\vec{x},t_{p})$ (\ref{s3_78}) act as
'hinge'-functions between the \(U(L|S)\) invariant vacuum field \(\sigma_{D}^{(0)}(\vec{x},t_{p})\)
and the anomalous terms in $\wtilde{T}_{1}(\vec{x},t_{p})$,
$\wtilde{T}_{2}(\vec{x},t_{p})$ (\ref{s3_57},\ref{s3_58}).

The Hubbard-Stratonovich transformation of the original interaction \(V_{|\vec{x}-\vec{x}\ppr|}\)
with fields $\psi_{\vec{x},\alpha}(t_{p})$ is divided into three parts so that the invariant
transformations of the final resulting generating function can be performed within the same
group manifold \(Osp(S,S|2L)\) as that of the self-energy
\(\wtilde{\Sigma}_{2N\times 2N}(\vec{x},t_{p})\;\;\wtilde{K}\)
(\ref{s3_55}-\ref{s3_61},\ref{s3_75}-\ref{s3_78}).
One half of the original interaction is transformed with the real density field $\sigma_{D}^{(0)}(\vec{x},t_{p})$
for the density matrices $\hat{R}_{N\times N}^{11}(\vec{x},t_{p})$, $\hat{R}_{N\times N}^{22}(\vec{x},t_{p})$,
composed of dyadic products of the fields
$\psi_{\vec{x},\alpha}(t_{p})$, $\psi_{\vec{x},\beta}^{*}(t_{p})$
as in section \ref{s22} (\ref{s2_33}-\ref{s2_46}).
We also include the approximation to a spatially local relation (\ref{s2_65}-\ref{s2_68})
with effective value $V_{0}$ of the short-ranged interaction potential $V_{|\vec{x}-\vec{x}\ppr|}$
\beq \lb{s3_79}
\lefteqn{\exp\bigg\{-\frac{\im}{2\hbar}\int_{C}d t_{p}\sum_{\vec{x},\alpha;\vec{x}\ppr,\beta}
\psi_{\vec{x},\alpha}^{*}(t_{p})\;\psi_{\vec{x}\ppr,\beta}^{*}(t_{p})\;V_{|\vec{x}\ppr-\vec{x}|}\;
\psi_{\vec{x}\ppr,\beta}(t_{p})\;\psi_{\vec{x},\alpha}(t_{p})\bigg\}
\stackrel{V_{|\vec{x}-\vec{x}\ppr|}\approx V_{0}}{\approx} } \\ \no &=&
\int\;\;d[\sigma_{D}^{(0)}(\vec{x},t_{p})]\;\; \exp\bigg\{\frac{\im}{2\hbar}
\frac{1}{V_{0}}\int_{C}d t_{p}\sum_{\vec{x}}
\sigma_{D}^{(0)}(\vec{x},t_{p})\;\;\sigma_{D}^{(0)}(\vec{x},t_{p})\bigg\} \\
\no &\times& \exp\Bigg\{-\frac{\im}{2\hbar}\int_{C}d t_{p}\sum_{\vec{x}}
\underbrace{\mbox{STR}\Bigg[\left( \bea{cc}
\hat{R}^{11} & 0 \\
0 & \hat{R}^{22} \eea\right)\left( \bea{cc}
\hat{1} & \\
 & \hat{\kappa}
 \eea\right)
\;\;\sigma_{D}^{(0)}\;\hat{K}_{2N\times 2N}\;\; \left( \bea{cc}
\hat{1} & \\
 & \hat{\kappa}
 \eea\right)\Bigg]}_{2\;\times\;\sigma_{D}^{(0)}(\vec{x},t_{p})\;
 \times\;\Big(\sum_{\alpha=1}^{N=L+S}
 \psi_{\vec{x},\alpha}^{*}(t_{p})\;\psi_{\vec{x},\alpha}(t_{p})\Big)}\Bigg\}_{.}
\eeq
The second half of the original interaction term with $V_{|\vec{x}-\vec{x}\ppr|}$
and fields $\psi_{\vec{x},\alpha}(t_{p})$, $\psi_{\vec{x},\beta}^{*}(t_{p})$ is converted
with the anti-hermitian anomalous terms \(\im\;\delta\hat{\Sigma}_{N\times N}^{12}\),
\(\im\;\delta\hat{\Sigma}_{N\times N}^{21}\) to dyadic products of anomalous terms in
$\hat{R}_{N\times N}^{12}$, $\hat{R}_{N\times N}^{21}$ (\ref{s2_33}-\ref{s2_46})
so that the super-trace of the product of two
anomalous selfenergies contains the same appropriate metric $\wtilde{K}$ (\ref{s3_42})
as the group manifold \(Osp(S,S|2L)\) of
\(\delta\wtilde{\Sigma}_{2N\times 2N}\;\wtilde{K}\) (\ref{s3_75}-\ref{s3_78})
\beq \lb{s3_80}
\lefteqn{\exp\bigg\{-\frac{\im}{2\hbar}\int_{C}d t_{p}
\sum_{\vec{x},\alpha;\vec{x}\ppr,\beta}
\psi_{\vec{x},\alpha}^{*}(t_{p})\;\psi_{\vec{x}\ppr,\beta}^{*}(t_{p})\;V_{|\vec{x}\ppr-\vec{x}|}\;
\psi_{\vec{x}\ppr,\beta}(t_{p})\;\psi_{\vec{x},\alpha}(t_{p})\bigg\}
\stackrel{V_{|\vec{x}-\vec{x}\ppr|}\approx V_{0}}{\approx} } \\ \no
&=&\int d[\delta\hat{\Sigma}^{12}(\vec{x},t_{p}),
\delta\hat{\Sigma}^{21}(\vec{x},t_{p})]\hspace*{0.5cm}
\exp\Bigg\{\frac{\im}{4\hbar}\frac{1}{V_{0}}\int_{C}d t_{p}\sum_{\vec{x}}  \\ \no &\times&
\mbox{STR}\Bigg[\left( \bea{cc}
0 & \im\;\delta\hat{\Sigma}^{12} \\
\im\;\delta\hat{\Sigma}^{21} & 0 \eea\right)\left( \bea{cc}
\hat{1} & \\
 & \wtilde{\kappa}
 \eea\right)
\left( \bea{cc}
0 & \im\;\delta\hat{\Sigma}^{12} \\
\im\;\delta\hat{\Sigma}^{21} & 0 \eea\right)\left( \bea{cc}
\hat{1} & \\
 & \wtilde{\kappa}
 \eea\right)\Bigg]\Bigg\} \\ \no &\times&
\exp\Bigg\{-\frac{\im}{2\hbar}\int_{C}d t_{p}\sum_{\vec{x}}
\mbox{STR}\Bigg[\left( \bea{cc}
0 & \hat{R}^{12} \\
\hat{R}^{21} & 0 \eea\right)\left( \bea{cc}
\hat{1} & \\
 & \hat{\kappa}
 \eea\right)\left(
\bea{cc}
0 & \delta\hat{\Sigma}^{12} \\
\delta\hat{\Sigma}^{21} & 0 \eea\right)\left( \bea{cc}
\hat{1} & \\
 & \hat{\kappa}
 \eea\right)\Bigg]\Bigg\}_{\mbox{.}}
\eeq
A third term (\ref{s3_81}) includes the density fluctuations $\delta\hat{\Sigma}_{N\times N}^{11}$,
$\delta\hat{\Sigma}_{N\times N}^{22}$ of the group \(U(L|S)\)) with Gaussian integrals.
However, the coupling term between $\hat{R}_{N\times N}^{11}$, $\hat{R}_{N\times N}^{22}$ and
$\delta\hat{\Sigma}_{N\times N}^{11}$, $\delta\hat{\Sigma}_{N\times N}^{22}$
leads by construction only to a constant factor.
The '22' block with $\delta\hat{\Sigma}_{N\times N}^{22}$
with the additional minus sign cancels the product in the '11' block
with $\hat{R}_{N\times N}^{11}$ and $\delta\hat{\Sigma}_{N\times N}^{11}$ so that the
exponent in the second factor of the integrand vanishes identically
(see third line in Eq. (\ref{s3_81})).
The density fields $\delta\hat{\Sigma}_{N\times N}^{11}$, $\delta\hat{\Sigma}_{N\times N}^{22}$
have to be introduced by the Gaussian integrals (\ref{s3_81})
because the coset matrices $\wtilde{T}_{1}$, $\wtilde{T}_{2}$ (\ref{s3_57}-\ref{s3_61})
of \(Osp(S,S|2L)\backslash U(L|S)\) couple to these fields
in a spontaneous symmetry breaking (\ref{s3_75},\ref{s3_76})
\beq \lb{s3_81}
\lefteqn{\hspace*{-0.65cm} 1 \equiv
\int d[\delta\hat{\Sigma}^{11}(t_{p}),\delta\hat{\Sigma}^{22}(t_{p})]\;\;\;
\times } \\ \no &\times&
\exp\Bigg\{\frac{\im}{4\hbar}\frac{1}{V_{0}}\int_{C}d t_{p}\sum_{\vec{x}} \mbox{STR}\Bigg[\left( \bea{cc}
\delta\hat{\Sigma}^{11} & 0 \\
0 & \delta\hat{\Sigma}^{22} \eea\right)\left( \bea{cc}
\hat{1} & \\
 & \wtilde{\kappa}
 \eea\right)\left(
\bea{cc}
\delta\hat{\Sigma}^{11} & 0 \\
0 & \delta\hat{\Sigma}^{22} \eea\right)\left( \bea{cc}
\hat{1} & \\
 & \wtilde{\kappa}
 \eea\right)\Bigg]\Bigg\} \\ \no &\times&
\exp\Bigg\{-\frac{\im}{2\hbar}\int_{C}d t_{p}\sum_{\vec{x},\vec{x}\ppr}
\underbrace{\mbox{STR}\Bigg[\left( \bea{cc}
\hat{R}^{11} & 0 \\
0 & \hat{R}^{22} \eea\right)\left( \bea{cc}
\hat{1} & \\
 & \hat{\kappa}
 \eea\right)\left(
\bea{cc}
\delta\hat{\Sigma}^{11} & 0 \\
0 & -\delta\hat{\Sigma}^{22} \eea\right)\left( \bea{cc}
\hat{1} & \\
 & \hat{\kappa}
 \eea\right)\Bigg] }_{\equiv 0}\Bigg\}_{\mbox{.}}
\eeq
Finally, the partial HST relations (\ref{s3_79},\ref{s3_80}) and (\ref{s3_81}) are combined
to an entire HST relation for the original quartic interaction of fields
\beq \lb{s3_82}
\lefteqn{\exp\bigg\{-\frac{\im}{\hbar}\int_{C}d t_{p}\sum_{\vec{x},\alpha;\vec{x}\ppr,\beta}
\psi_{\vec{x},\alpha}^{*}(t_{p})\;\psi_{\vec{x}\ppr,\beta}^{*}(t_{p})\;V_{|\vec{x}\ppr-\vec{x}|}\;
\psi_{\vec{x}\ppr,\beta}(t_{p})\;\psi_{\vec{x},\alpha}(t_{p})\bigg\}
\stackrel{V_{|\vec{x}-\vec{x}\ppr|}\approx V_{0}}{\approx}  }
\\ \no &=&
\int\;\;d[\sigma_{D}^{(0)}(\vec{x},t_{p})]\;\;
\exp\bigg\{\frac{\im}{2\hbar}\frac{1}{V_{0}}\int_{C}d t_{p}\sum_{\vec{x}}
\sigma_{D}^{(0)}(\vec{x},t_{p})\;\;\sigma_{D}^{(0)}(\vec{x},t_{p})\bigg\}
\\ \no &\times &
 \int d[\delta\wtilde{\Sigma}(\vec{x},t_{p})\;\wtilde{K}]\;\;
 \exp\Bigg\{\frac{\im}{4\hbar}\frac{1}{V_{0}}\int_{C}d t_{p}\sum_{\vec{x}}
 \mbox{STR}\Big[\delta\wtilde{\Sigma}(\vec{x},t_{p})\;\wtilde{K}\;
 \delta\wtilde{\Sigma}(\vec{x},t_{p})\;\wtilde{K}\Big]\Bigg\}
\\ \no &\times&
 \exp\Bigg\{-\frac{\im}{2\hbar}\int_{C}d t_{p}\sum_{\vec{x},\vec{x}\ppr}
 \mbox{STR}\Bigg[\left(
 \bea{cc}
 \hat{R}^{11} & \hat{R}^{12} \\
 \hat{R}^{21} & \hat{R}^{22}
 \eea\right)
 \underbrace{\left(
\bea{cc}
\hat{1} & \\
 & \hat{\kappa}
 \eea\right)}_{\hat{K}}\left(
 \bea{cc}
 \hat{\Sigma}^{11} & \delta\hat{\Sigma}^{12} \\
 \delta\hat{\Sigma}^{21} & -\hat{\Sigma}^{22}
 \eea\right)\underbrace{\left(
\bea{cc}
\hat{1} & \\
 & \hat{\kappa}
 \eea\right)}_{\hat{K}}\Bigg]\Bigg\} \\ \lb{s3_83} &&
 \hat{\Sigma}_{N\times N}^{11}=\sigma_{D}^{(0)}\;\hat{1}_{N\times N}+
 \delta\hat{\Sigma}_{N\times N}^{11}\hspace*{0.75cm}
\hat{\Sigma}_{N\times N}^{22}=\sigma_{D}^{(0)}\;\wtilde{\kappa}_{N\times N}+
\delta\hat{\Sigma}_{N\times N}^{22}\;.
\eeq
Similar conversions as in section \ref{s22} are used to obtain the
path integral \(Z[\hat{\mcal{J}},J_{\psi},\wtilde{J}_{\psi\psi}]\) (\ref{s3_84})
in terms of the fields $\delta\wtilde{\Sigma}_{2N\times 2N}$ (\ref{s3_76}) and
$\sigma_{D}^{(0)}(\vec{x},t_{p})$. The combined HST relation (\ref{s3_82}) reduces
the quartic interaction to a bilinear relation with the original coherent state fields in (\ref{s2_24})
\beq \lb{s3_84}
\lefteqn{Z[\hat{\mcal{J}},J_{\psi},\wtilde{J}_{\psi\psi}]=
\int\;\;d[\sigma_{D}^{(0)}(\vec{x},t_{p})]\;\; \exp\bigg\{\frac{\im}{2\hbar}\frac{1}{V_{0}}
\int_{C}d t_{p}\sum_{\vec{x}}
\sigma_{D}^{(0)}(\vec{x},t_{p})\;\;\sigma_{D}^{(0)}(\vec{x},t_{p})\bigg\} } \\ \no &\times &
\int d[\delta\wtilde{\Sigma}(\vec{x},t_{p})\;\wtilde{K}]\;\;
\exp\bigg\{\frac{\im}{4\hbar}\frac{1}{V_{0}}
\int_{C}d t_{p}\sum_{\vec{x}}\mbox{STR}\Big[\delta\wtilde{\Sigma}(\vec{x},t_{p})\;
\wtilde{K}\;\delta\wtilde{\Sigma}(\vec{x},t_{p})\;\wtilde{K}\Big]\bigg\}
\\ \no &\times& \int d[\psi_{\vec{x},\alpha}(t_{p})]\;
\exp\Bigg\{-\frac{\im}{2\hbar}\int_{C}d t_{p}\;d t_{q}\ppr
\sum_{\vec{x},\vec{x}\ppr}\mcal{N}_{x}\;\Psi_{\vec{x}\ppr,\beta}^{+b}(t_{q}\ppr)
\Bigg[\hat{\mcal{H}}\;\hat{K}+\hat{\eta}\;\bigg( \bea{cc}
0 & \wtilde{j}_{\psi\psi} \\
\wtilde{j}_{\psi\psi}^{+} & 0 \eea\bigg)+ \\ \no &+&
\frac{\hat{\mcal{J}}_{\beta\alpha}^{ba}}{\mcal{N}_{x}}+
\hat{K}\;\hat{\eta}\;\left(\sigma_{D}^{(0)}\;\hat{K}+\left(\bea{cc}
\delta\hat{\Sigma}^{11} & \delta\hat{\Sigma}^{12} \\
\delta\hat{\Sigma}^{21} & -\delta\hat{\Sigma}^{22}
\eea\right)\right)\;\hat{K}
\Bigg]_{\vec{x}\ppr,\beta;\vec{x},\alpha}^{ba}\hspace*{-0.64cm}(t_{q}\ppr,t_{p})\;\;\;
\Psi_{\vec{x},\alpha}^{a}(t_{p})\Bigg\} \\ \no &\times &
\exp\bigg\{-\frac{\im}{2\hbar}\int_{C}d t_{p}\sum_{\vec{x}}\bigg(
J_{\psi;\alpha}^{+a}(\vec{x},t_{p})\;\hat{K}\;\Psi_{\vec{x},\alpha}^{a}(t_{p})+
\Psi_{\vec{x},\alpha}^{+a}(t_{p})\;\hat{K}\;J_{\psi;\alpha}^{a}(\vec{x},t_{p})\bigg)\bigg\}_{\mbox{.}}
\eeq
We can integrate over the bilinear relation in (\ref{s3_84}) with the 'Nambu'-doubled fields
\(\Psi_{\vec{x}\ppr,\beta}^{+b}(t_{p})\), \(\Psi_{\vec{x},\alpha}^{a}(t_{p})\) to obtain
the generating function \(Z[\hat{\mcal{J}},J_{\psi},\wtilde{J}_{\psi\psi}]\) (\ref{s3_85})
with super-determinant and inverse of the matrix
\(\wtilde{M}_{\vec{x},\alpha;\vec{x}\ppr,\beta}^{\raisebox{-10pt}{$^{ab}$}}(t_{p},t_{q}\ppr)\)
(\ref{s3_86}) and the selfenergies as integration variables
\beq \lb{s3_85} \hspace*{-1.0cm}
\lefteqn{Z[\hat{\mcal{J}},J_{\psi},\wtilde{J}_{\psi\psi}]=
\int\;\;d[\sigma_{D}^{(0)}(\vec{x},t_{p})]\;\; \exp\bigg\{\frac{\im}{2\hbar}\frac{1}{V_{0}}
\int_{C}d t_{p}\sum_{\vec{x}}
\sigma_{D}^{(0)}(\vec{x},t_{p})\;\;\sigma_{D}^{(0)}(\vec{x},t_{p})\bigg\} } \\ \no &\times&
\int d[\delta\wtilde{\Sigma}(\vec{x},t_{p})\;\wtilde{K}]\;\;
\exp\bigg\{\frac{\im}{4\hbar}\frac{1}{V_{0}}
\int_{C}d t_{p}\sum_{\vec{x}}\mbox{STR}\Big[\delta\wtilde{\Sigma}(\vec{x},t_{p})\;
\wtilde{K}\;\delta\wtilde{\Sigma}(\vec{x},t_{p})\;\wtilde{K}\Big]\bigg\}
\\ \no &\times &
\Bigg\{\mbox{SDET}\bigg[\wtilde{M}_{\vec{x},\alpha;\vec{x}\ppr,\beta}^{ab}(t_{p},t_{q}\ppr)\bigg]
\Bigg\}^{\mathbf{-1/2}} \\ \no &\times&
\exp\bigg\{\frac{\im}{2\hbar}\Omega^{2}\int_{C}d t_{p}\;d t\ppr_{q}\sum_{\vec{x},\vec{x}\ppr}\mcal{N}_{x}\;
J_{\psi;\beta}^{+b}(\vec{x}\ppr,t_{q}\ppr)\;\hat{K}\;
\wtilde{M}_{\vec{x}\ppr,\beta;\vec{x},\alpha}^{\mathbf{-1};ba}(t_{q}\ppr,t_{p})\;\hat{K}\;
J_{\psi;\alpha}^{a}(\vec{x},t_{p})\bigg\}
\eeq
\beq \lb{s3_86}
\lefteqn{\wtilde{M}_{\vec{x},\alpha;\vec{x}\ppr,\beta}^{ab}(t_{p},t_{q}\ppr)=} \\ \no &=&
\hat{\mcal{H}}_{\vec{x},\alpha;\vec{x}\ppr,\beta}^{ab}(t_{p},t_{q}\ppr)\;\;\hat{K}+
\sigma_{D}^{(0)}(\vec{x},t_{p})\;\hat{K}\;\delta_{\vec{x},\vec{x}\ppr}\;\eta_{p}\;\delta_{p,q}\;\delta(t_{p}-t_{q}\ppr)+
\eta_{p}\;\frac{\hat{\mcal{J}}_{\vec{x},\alpha;\vec{x}\ppr,\beta}^{ab}(t_{p},t_{q}\ppr)}{\mcal{N}_{x}}\;\eta_{q}+ \\ \no &+&
\left(\wtilde{J}_{\psi\psi;\alpha\beta}^{ab}(\vec{x},t_{p}) +
\hat{K}\;\left(\bea{cc}
\delta\hat{\Sigma}_{\alpha\beta}^{11}(\vec{x},t_{p}) &
\delta\hat{\Sigma}_{\alpha\beta}^{12}(\vec{x},t_{p}) \\
\delta\hat{\Sigma}_{\alpha\beta}^{21}(\vec{x},t_{p}) &
-\delta\hat{\Sigma}_{\alpha\beta}^{22}(\vec{x},t_{p})
\eea\right)\;\hat{K}\right)\;\delta_{\vec{x},\vec{x}\ppr}\;\eta_{p}\;\delta_{p,q}\;\delta(t_{p}-t_{q}\ppr)\;.
\eeq
However, the additional transformations (\ref{s3_87})
with metric $\hat{I}_{2N\times 2N}$ (\ref{s3_88}) and $\hat{K}$ as in (\ref{s2_54},\ref{s2_61})
have to be applied to the self-energy matrix in
\(Z[\hat{\mcal{J}},J_{\psi},\wtilde{J}_{\psi\psi}]\) (\ref{s3_85},\ref{s3_86})
in order derive a coherent state path integral which is invariant under the same
transformations \(Osp(S,S|2L)\) as the group manifold of the self-energy with anti-hermitian
anomalous terms
\beq \lb{s3_87}
\hat{K}\;\left(
\bea{cc}
\hat{\Sigma}^{11} & \delta\hat{\Sigma}^{12} \\
\delta\hat{\Sigma}^{21} & -\hat{\Sigma}^{22}
\eea\right)\;\hat{K}&\rightarrow& \hat{I}\;\;
\left(
\bea{cc}
\hat{\Sigma}^{11} & \delta\hat{\Sigma}^{12} \\
\delta\hat{\Sigma}^{21} & -\hat{\Sigma}^{22}
\eea\right)\;\;\hat{I}\rightarrow
\left(
\bea{cc}
\hat{\Sigma}^{11} & \im\;\delta\hat{\Sigma}^{12} \\
\im\;\delta\hat{\Sigma}^{21} & \hat{\Sigma}^{22}
\eea\right)
\eeq
\be \lb{s3_88}
\hat{I}_{2N\times 2N}=\bigg\{\underbrace{\hat{1}_{L\times L}}_{BB}\;,\;
\underbrace{\hat{1}_{S\times S}}_{FF}\;;\;
\underbrace{\hat{\im}_{L\times L}}_{BB}\;,\;\underbrace{\hat{\im}_{S\times S}}_{FF}\bigg\}\;.
\ee
We apply various times the metric \(\hat{K}\), \(\wtilde{K}\) and
\(\hat{I}\) on both sides of the super-matrix
\(\wtilde{M}_{\vec{x},\alpha;\vec{x}\ppr,\beta}^{ab}(t_{p},t_{q}\ppr)\)
(\ref{s3_86}) so that the self-energy \(\wtilde{\Sigma}_{2N\times 2N}\;\wtilde{K}\)
(\ref{s3_55}-\ref{s3_67},\ref{s3_52}) becomes a generator of the
ortho-symplectic group \(Osp(S,S|2L)\) as
$\wtilde{W}_{\alpha\beta}^{ab}$ (\ref{s3_41}), but with
anti-hermitian anomalous terms. Finally, the matrix
\(\wtilde{M}_{\vec{x},\alpha;\vec{x}\ppr,\beta}^{ab}(t_{p},t_{q}\ppr)\)
(\ref{s3_86}) can be replaced by the super-matrix
\(\wtilde{\mcal{M}}_{\vec{x},\alpha;\vec{x}\ppr,\beta}^{ab}(t_{p},t_{q}\ppr)\)
in the generating function with only minor modifications
because the super-determinant of both forms of super-matrices is invariant under the transformations
with metrics \(\hat{K}\), \(\wtilde{K}\) and \(\hat{I}\)
\beq \lb{s3_89}
\wtilde{M}_{\vec{x},\alpha;\vec{x}\ppr,\beta}^{ab}(t_{p},t_{q}\ppr) &\rightarrow&
\hat{K}\;\Big(\hat{K}\;
\wtilde{M}_{\vec{x},\alpha;\vec{x}\ppr,\beta}^{ab}(t_{p},t_{q}\ppr)\;\hat{K}\Big)\;\hat{K}
\\ \no &\rightarrow&
\hat{K}\;\hat{I}^{-1}\;\Big(\hat{I}\;\hat{K}\;
\wtilde{M}_{\vec{x},\alpha;\vec{x}\ppr,\beta}^{ab}(t_{p},t_{q}\ppr)\;
\hat{K}\;\hat{I}\Big)\;\hat{I}^{-1}\;\hat{K} \\ \no &\rightarrow &
\hat{K}\;\hat{I}^{-1}\;\underbrace{\Big(\hat{I}\;\hat{K}\;
\wtilde{M}_{\vec{x},\alpha;\vec{x}\ppr,\beta}^{ab}(t_{p},t_{q}\ppr)\;
\hat{K}\;\hat{I}\;\wtilde{K}\Big)}_{\wtilde{\mcal{M}}_{\vec{x},\alpha;
\vec{x}\ppr,\beta}^{ab}(t_{p},t_{q}\ppr)}
\;\wtilde{K}\;\hat{I}^{-1}\;\hat{K} \\ \lb{s3_90}
\mbox{SDET}\Big\{\wtilde{M}\Big\}&=&
\mbox{SDET}\Big\{\hat{K}\;\hat{I}^{-1}\;\wtilde{\mcal{M}}\;
\wtilde{K}\;\hat{I}^{-1}\;\hat{K}\Big\}=\mbox{SDET}\Big\{\wtilde{\mcal{M}}\Big\}
\\ \lb{s3_91}
\wtilde{M}_{\vec{x},\alpha;\vec{x}\ppr,\beta}^{\mathbf{-1};ab}(t_{p},t_{q}\ppr) &=&
\hat{K}\;\hat{I}\;\wtilde{K}\;
\wtilde{\mcal{M}}_{\vec{x},\alpha;\vec{x}\ppr,\beta}^{\mathbf{-1};ab}(t_{p},t_{q}\ppr)
\;\hat{I}\;\hat{K}
\eeq
\beq \lb{s3_92}
\lefteqn{
\wtilde{\mcal{M}}_{\vec{x},\alpha;\vec{x}\ppr,\beta}^{ab}(t_{p},t_{q}\ppr)= } \\ \no &=&
\hat{\mcal{H}}_{\vec{x},\alpha;\vec{x}\ppr,\beta}^{ab}(t_{p},t_{q}\ppr)+
\sigma_{D}^{(0)}(\vec{x},t_{p})\;\hat{1}_{2N\times 2N}\;\delta_{\vec{x},\vec{x}\ppr}\;
\eta_{p}\;\delta_{p,q}\;\delta(t_{p}-t_{q}\ppr)+\eta_{p}\;\hat{I}\;\hat{K}\;
\frac{\hat{\mcal{J}}_{\vec{x},\alpha;\vec{x}\ppr,\beta}^{ab}(t_{p},t_{q}\ppr)}{\mcal{N}_{x}}\;
\hat{K}\;\hat{I}\;\wtilde{K}\;\eta_{q}+  \\ \no &+&
\left(
\im\;\hat{J}_{\psi\psi;\alpha\beta}^{ab}(\vec{x},t_{p})\;\wtilde{K} +
\left(\bea{cc}
\delta\hat{\Sigma}_{\alpha\beta}^{11}(\vec{x},t_{p}) & \im\;
\delta\hat{\Sigma}_{\alpha\beta}^{12}(\vec{x},t_{p}) \\
\im\;\delta\hat{\Sigma}_{\alpha\beta}^{21}(\vec{x},t_{p}) &
\delta\hat{\Sigma}_{\alpha\beta}^{22}(\vec{x},t_{p})
\eea\right)\;\wtilde{K}\;\right)
\delta_{\vec{x},\vec{x}\ppr}\;\eta_{p}\;\delta_{p,q}\;\delta(t_{p}-t_{q}\ppr)
\eeq
\beq  \lb{s3_93}
Z[\hat{\mcal{J}},J_{\psi},\im\hat{J}_{\psi\psi}]&=&
\int\;\;d[\sigma_{D}^{(0)}(\vec{x},t_{p})]\;\; \exp\bigg\{\frac{\im}{2\hbar}\frac{1}{V_{0}}
\int_{C}d t_{p}\sum_{\vec{x}}
\sigma_{D}^{(0)}(\vec{x},t_{p})\;\;\sigma_{D}^{(0)}(\vec{x},t_{p})\bigg\}  \\ \no &\times&
\int d[\delta\wtilde{\Sigma}(\vec{x},t_{p})\;\wtilde{K}]\;\;\exp\bigg\{\frac{\im}{4\hbar}\frac{1}{V_{0}}
\int_{C}d t_{p}\sum_{\vec{x}}\mbox{STR}\Big[\delta\wtilde{\Sigma}(\vec{x},t_{p})\;
\wtilde{K}\;\delta\wtilde{\Sigma}(\vec{x},t_{p})\;\wtilde{K}\Big]\bigg\}
\\ \no &\times &
\Bigg\{\mbox{SDET}\bigg[\wtilde{\mcal{M}}_{\vec{x},\alpha;\vec{x}\ppr,\beta}^{ab}(t_{p},t_{q}\ppr)\bigg]
\Bigg\}^{\mathbf{-1/2}} \\ \no &\times&
\exp\bigg\{\frac{\im}{2\hbar}\Omega^{2}\int_{C}d t_{p}\;d t\ppr_{q}\sum_{\vec{x},\vec{x}\ppr}\mcal{N}_{x}\;
J_{\psi;\beta}^{+b}(\vec{x}\ppr,t_{q}\ppr)\;\hat{I}\wtilde{K}\;
\wtilde{\mcal{M}}_{\vec{x}\ppr,\beta;\vec{x},\alpha}^{\mathbf{-1};ba}(t_{q}\ppr,t_{p})\;\hat{I}\;
J_{\psi;\alpha}^{a}(\vec{x},t_{p})\bigg\}_{.}
\eeq
The equivalence of the manifolds of the
invariant transformations in \(Z[\hat{\mcal{J}},J_{\psi},\im\hat{J}_{\psi\psi}]\) (\ref{s3_93})
and of the self-energy \(\wtilde{\Sigma}_{2N\times 2N}\;\wtilde{K}\)
allows the separation into density and anomalous parts in a kind of 'diagonalization'.
The analogous matrix
\(\wtilde{\mcal{M}}_{\vec{x},\alpha;\vec{x}\ppr,\beta}^{ab}(t_{p},t_{q}\ppr)\)
(\ref{s3_92}) as
\(\hat{M}_{\vec{x},\alpha;\vec{x}\ppr,\beta}^{ab}(t_{p},t_{q}\ppr)\) (\ref{s2_71})
in section \ref{s22} determines the generating function (\ref{s3_93}) with the inverse square root
of the super-determinant of the doubled one-particle part \(\hat{\mcal{H}}\;\wtilde{K}\),
the anti-hermitian source matrix $\im\hat{J}_{\psi\psi}(\vec{x},t_{p})$ (\ref{s2_63},\ref{s2_64})
for the anomalous terms and with the bilinear source field $J_{\psi;\alpha}^{a}(\vec{x},t_{p})$ (\ref{s2_58},\ref{s2_4})
for the coherent condensate wave function. The change to anti-hermitian pair condensate terms in
\(\wtilde{\mcal{M}}_{\vec{x},\alpha;\vec{x}\ppr,\beta}^{ab}(t_{p},t_{q}\ppr)\)
(\ref{s3_92}) with corresponding self-energy $\wtilde{\Sigma}_{2N\times 2N}(\vec{x},t_{p})$
(\ref{s3_75}-\ref{s3_78}) and source matrix $\im\hat{J}_{\psi\psi}(\vec{x},t_{p})$
is accompanied by the change of the metric $\hat{K}$ in
\(\hat{M}_{\vec{x},\alpha;\vec{x}\ppr,\beta}^{ab}(t_{p},t_{q}\ppr)\)
(\ref{s2_72},\ref{s2_71})
to the metric $\wtilde{K}$ in
\(\wtilde{\mcal{M}}_{\vec{x},\alpha;\vec{x}\ppr,\beta}^{ab}(t_{p},t_{q}\ppr)\)
(\ref{s3_92}). The quadratic super-trace relation with interaction parameter $V_{0}$ also contains
the metric $\wtilde{K}$ instead of $\hat{K}$ after application of the Weyl unitary trick \cite{weyl1}.
The chosen form of the HST relations (\ref{s3_79}-\ref{s3_82}) yields
a coherent state path integral \(Z[\hat{\mcal{J}},J_{\psi},\im\hat{J}_{\psi\psi}]\) (\ref{s3_93})
whose invariant transformations are equivalent to the group manifold \(Osp(S,S|2L)\) of the self-energy.
The self-energy \(\wtilde{\Sigma}_{2N\times 2N}\;\wtilde{K}\) of \(Osp(S,S|2L)\)
in the coherent state path integral (\ref{s3_93}) can therefore be decomposed (or 'diagonalized')
into density terms with $\sigma_{D}^{(0)}(\vec{x},t_{p})$ ,
\(\delta\hat{\Sigma}_{D;2N\times 2N}(\vec{x},t_{p})\) and into anomalous terms (as 'eigenvectors')
represented by the coset matrices $\wtilde{T}_{1}$, $\wtilde{T}_{2}$ (\ref{s3_55}-\ref{s3_61})
of \(Osp(S,S|2L)\backslash U(L|S)\).

\subsection{The change of integration measure to $Osp(S,S|2L)\backslash U(L|S)\otimes \;U(L|S)$} \lb{s33}

The invariant integration measure depends on the chosen parametrization of the density
and anomalous terms. We apply a coset decomposition with density terms in the super-unitary
subgroup \(U(L|S)\) and coset matrices \(Osp(S,S|2L)\backslash U(L|S)\)
with anti-hermitian terms for the pair condensates.
According to Eqs. (\ref{s3_55}-\ref{s3_61}), (\ref{s3_68}-\ref{s3_74}) and (\ref{s3_75}-\ref{s3_78}),
the generators and corresponding variables of \(Osp(S,S|2L)\), represented by the
self-energy \(\wtilde{\Sigma}_{2N\times 2N}\;\;\wtilde{K}\)
(\ref{s3_94},\ref{s3_75},\ref{s3_55}) in 'flat' coordinates,
are transformed to the density terms
\(\delta\wtilde{\Sigma}_{D;2N\times 2N}\;\;\wtilde{K}\)
(\ref{s3_68}-\ref{s3_74}),
the additional mean density field \(\sigma_{D}^{(0)}(\vec{x},t_{p})\) and anti-hermitian pair condensate
terms in the matrices \(\hat{T}(\vec{x},t_{p}):=\wtilde{T}_{1}(\vec{x},t_{p})\)
of \(Osp(S,S|2L)\backslash U(L|S)\)
with generator \(\hat{Y}_{2N\times 2N}(\vec{x},t_{p})\) (\ref{s3_95}-\ref{s3_98},\ref{s3_57}-\ref{s3_61})
\beq \lb{s3_94}
\wtilde{\Sigma}_{2N\times 2N}\;\wtilde{K}
&=&\hat{T}\;\left( \bea{cc}
\sigma_{D}^{(0)}\;\hat{1}_{N\times N}+\delta\hat{\Sigma}_{D;N\times N}^{11} & 0 \\
0 & \sigma_{D}^{(0)}\;\wtilde{\kappa}+\delta\hat{\Sigma}_{D;N\times N}^{22} \eea\right)\;\wtilde{K}\;\;
\underbrace{\hat{T}^{-1}}_{\wtilde{K}\;\wtilde{T}_{2}\;\wtilde{K}} \\ \no
&=&\sigma_{D}^{(0)}\;\hat{1}_{2N\times 2N}+
\hat{T}_{2N\times 2N}\;\;\underbrace{\left(\bea{cc}
\delta\hat{\Sigma}_{D;N\times N}^{11} & 0 \\
0 & \delta\hat{\Sigma}_{D;N\times N}^{22}\;\wtilde{\kappa}
\eea\right)}_{\delta\hat{\Sigma}_{D;2N\times 2N}\;\wtilde{K}}\;\;\hat{T}_{2N\times 2N}^{-1} \\ \lb{s3_95}
\hat{T}_{2N\times 2N}&=&\exp\Bigg\{\im\left( \bea{cc}
0 & \im\;\hat{X}_{N\times N} \\
\im\;\wtilde{\kappa}\;\hat{X}_{N\times N}^{+} & 0
\eea\right)\Bigg\}=\exp\Big\{-\hat{Y}_{2N\times 2N}\Big\} \\ \lb{s3_96}
\hat{Y}_{2N\times 2N}&=&\left( \bea{cc}
0 & \hat{X}_{N\times N} \\
\wtilde{\kappa}_{N\times N}\;\hat{X}_{N\times N}^{+} & 0 \eea\right)\;\;\;;
\hspace*{0.45cm} \hat{X}_{N\times N}=\left( \bea{cc}
-\hat{c}_{D;L\times L} & \hat{\eta}_{D;L\times S}^{T} \\
-\hat{\eta}_{D;S\times L} & \hat{f}_{D;S\times S} \eea\right) \\ \lb{s3_97} &&
\hat{c}_{D;L\times L}^{T}=\hat{c}_{D;L\times L}\hspace*{0.5cm}
\hat{f}_{D;S\times S}^{T}=-\hat{f}_{D;S\times S} \\ \lb{s3_98}
\hat{T}_{2N\times 2N}:&=&\wtilde{T}_{1;2N\times 2N}\hspace*{0.75cm}
\wtilde{T}_{2}\;\wtilde{K}\;\wtilde{T}_{1}=\wtilde{K}\longrightarrow
\hat{T}_{2N\times 2N}^{-1}:=\wtilde{K}\;
\wtilde{T}_{2;2N\times 2N}\;\wtilde{K}\;.
\eeq
The invariant integration measure for the coset decomposition is calculated by diagonalizing
the density terms \(\delta\hat{\Sigma}_{D;N\times N}^{11}\) and its super-transposed part
\((\delta\hat{\Sigma}_{D;N\times N}^{22}\;\wtilde{\kappa})^{st}
=-\delta\hat{\Sigma}_{D;N\times N}^{11}\)
(\ref{s3_70}) with the matrices \(\hat{Q}_{N\times N}^{11}\), \(\hat{Q}_{N\times N}^{22}\) to the eigenvalues
\(\delta\hat{\lambda}_{N\times N}=\big(\delta\hat{\lambda}_{B;L\times L}\;;\;
\delta\hat{\lambda}_{F;S\times S}\big)\) (\ref{s3_99}-\ref{s3_104}).
The anomalous terms with the matrices
\(\hat{X}_{N\times N}\), \(\wtilde{\kappa}\;\hat{X}_{N\times N}^{+}\) or
\(\hat{Y}_{2N\times 2N}\) (\ref{s3_96},\ref{s3_97}) are factorized with the super-unitary matrices
\(\hat{P}_{N\times N}^{11}\), \(\hat{P}_{N\times N}^{22}\) to
\(L=2l+1\) diagonal complex parameters \(\hat{\ovv{c}}_{L\times L}\)
for the bosonic molecular condensate and to complex antisymmetric quaternions
\(\hat{\ovv{f}}_{S\times S}=\big\{(\tau_{2})_{\mu\nu}\;\ovv{f}_{r}\big\}\)
(\(r=1,\ldots,S/2\)),(\(\mu,\nu=1,2\),\(\tau_{2}=\)antisymmetric standard Pauli-matrix)
along the \(2\times 2\) block diagonal of the fermion fermion part (\ref{s3_105}-\ref{s3_112})
\footnote{The range of indices for the angular momentum degrees of freedom for the
fermions and bosons is adapted from \(-s,\ldots,+s\) and \(-l,\ldots,+l\) to the range
\(1,\ldots,S=2s+1\) and \(1,\ldots,L=2l+1\). Furthermore, two notations for the index
range of the fermions are used in parallel in the remainder : (i) The first notation
labels the angular momentum degrees of freedom from \(i,j=1,\ldots,S=2s+1\).
(ii) The second one regards the quaternionic structure of the fermion-fermion part
and has a \(2\times 2\) block matrix structure with \(\mu,\nu=1,2\) and
\(r,r\ppr=1,\ldots,S/2\) so that e.g. \(\delta\lambda_{F;r\mu}\) corresponds to
\(\delta\lambda_{F;i=2(r-1)+\mu}\).}. (In the self-contained appendix \ref{sa}
we describe the diagonalization of density and anomalous terms with
super-unitary matrices and their parametrization in section \ref{sa1} in detail.
Furthermore, the various steps to the final form of the integration measure
are specified in following sections \ref{sa2} to \ref{sa4}, including lists
of partial results. However, the subsections \ref{sa3}, \ref{sa4} just catalogue the various
steps and formulas to the final integration measures, but contain every detail of the calculation.)
We list in Eqs. (\ref{s3_99}) to (\ref{s3_112}) the super-unitary matrices
\(\hat{Q}_{N\times N}^{11}\), \(\hat{Q}_{N\times N}^{22}\) and
\(\hat{P}_{N\times N}^{11}\), \(\hat{P}_{N\times N}^{22}\)
for diagonalizing density and pair condensate terms. The fields
\(\hat{\mcal{B}}_{D;mm}\), \(\hat{\mcal{F}}_{D;ii}\) and
\(\hat{\mcal{C}}_{D;mm}\), \(\hat{\mcal{G}}_{D;r\mu,r\nu}\) for
the diagonal elements of generators in \(\hat{Q}^{aa}\) and \(\hat{P}^{aa}\)
have to vanish because these degrees of freedom are already contained
in the eigenvalues \(\delta\hat{\lambda}_{N\times N}\) ,
\(\hat{\ovv{c}}_{L\times L}\) and block diagonal antisymmetric \(2\times 2\)
matrices \(\hat{\ovv{f}}_{S\times S}\) (see for details in appendix \ref{sa},
especially section \ref{sa1}.)
\beq \lb{s3_99}
\delta\hat{\Sigma}_{D;2N\times 2N}\;\wtilde{K}&=&\hat{Q}_{2N\times 2N}^{-1}\;
\delta\hat{\Lambda}_{2N\times 2N}\;\hat{Q}_{2N\times 2N} \\ \lb{s3_100}
\delta\hat{\Lambda}_{2N\times 2N}&=&\mbox{diag}\Big(\delta\hat{\lambda}_{N\times N}\;;\;
-\delta\hat{\lambda}_{N\times N}\Big) \\ \lb{s3_101}
\delta\hat{\lambda}_{N\times N}&=&\Big(\delta\hat{\lambda}_{B;1},\ldots,\delta\hat{\lambda}_{B;m},\ldots,
\delta\hat{\lambda}_{B;L}\;;\;\delta\hat{\lambda}_{F;1},\ldots,\delta\hat{\lambda}_{F;i},\ldots,
\delta\hat{\lambda}_{F;S}\Big) \\ \lb{s3_102}
\hat{Q}_{2N\times 2N}&=&\left(\bea{cc}
\hat{Q}_{N\times N}^{11} & 0 \\
0 & \hat{Q}_{N\times N}^{22}
\eea\right)\hspace*{0.35cm}\big(\hat{Q}_{N\times N}^{22}\big)^{st}=
\hat{Q}_{N\times N}^{11,+}=\hat{Q}_{N\times N}^{11,-1} \\ \lb{s3_103}
\hat{Q}^{11}_{N\times N}&=&\exp\left\{\im\left( \bea{cc}
\hat{\mcal{B}}_{D;L\times L} & \hat{\omega}_{D;L\times S}^{+} \\
\hat{\omega}_{D;S\times L} & \hat{\mcal{F}}_{D;S\times S}
\eea\right)\right\}  \hspace*{0.75cm}\hat{\mcal{B}}_{D;L\times L}^{+}=\hat{\mcal{B}}_{D;L\times L}
\\ \lb{s3_104} \hat{Q}^{22}_{N\times N}&=&\exp\left\{\im\left( \bea{cc}
-\hat{\mcal{B}}_{D;L\times L}^{T} & \hat{\omega}_{D;L\times S}^{T} \\
-\hat{\omega}_{D;S\times L}^{*} & -\hat{\mcal{F}}_{D;S\times S}^{T} \eea\right)\right\}
\hspace*{0.75cm}\hat{\mcal{F}}_{D;S\times S}^{+}=\hat{\mcal{F}}_{D;S\times S}  \\  \no &&
\hat{\mcal{B}}_{D;mm}=0\;\;(m=1,\ldots,L) \hspace*{0.5cm}
\hat{\mcal{F}}_{D;ii}=0\;\;(i=1,\ldots,S) \\ \lb{s3_105}
\hat{Y}_{2N\times 2N}&=&\hat{P}_{2N\times 2N}^{-1}\;\hat{Y}_{DD;2N\times 2N}\;
\hat{P}_{2N\times 2N} \\ \lb{s3_106}
\hat{Y}_{DD;2N\times 2N}&=&\left(\bea{cc}
0 & \hat{X}_{DD;N\times N} \\
\wtilde{\kappa}\;\hat{X}_{DD;N\times N}^{+} & 0
\eea\right) \\ \lb{s3_107}
\hat{X}_{DD;N\times N} &=&\left(\bea{cc}
-\hat{\ovv{c}}_{L\times L} & 0 \\
0 & \hat{\ovv{f}}_{S\times S}
\eea\right) \\ \lb{s3_108}
\hat{\ovv{c}}_{L\times L}&=&\mbox{diag}\Big\{\ovv{c}_{1},\ldots,
\ovv{c}_{m},\ldots,\ovv{c}_{L}\Big\}\hspace*{0.75cm}
\ovv{c}_{m}=|\ovv{c}_{m}|\;\exp\{\im\;\varphi_{m}\}\;\;\;
(\ovv{c}_{m}\in\mbox{\sf C}) \\ \lb{s3_109}
\hat{\ovv{f}}_{S\times S}&=&\mbox{diag}\Big\{\big(\tau_{2}\big)_{\mu\nu}\;\ovv{f}_{1},\ldots,
\big(\tau_{2}\big)_{\mu\nu}\;\ovv{f}_{r},\ldots,\big(\tau_{2}\big)_{\mu\nu}\;\ovv{f}_{S/2}\Big\} \\ \no &&
(r=1,\ldots,S/2),(\mu,\nu=1,2)\hspace*{0.5cm}\ovv{f}_{r}=|\ovv{f}_{r}|\;\exp\{\im\;\phi_{r}\}\;\;\;
(\ovv{f}_{r}\in\mbox{\sf C}) \\ \lb{s3_110}
\hat{P}_{2N\times 2N}&=&\left(\bea{cc}
\hat{P}_{N\times N}^{11} & 0 \\
0 & \hat{P}_{N\times N}^{22}
\eea\right)\hspace*{0.35cm}\big(\hat{P}_{N\times N}^{22}\big)^{st}=
\hat{P}_{N\times N}^{11,+}=\hat{P}_{N\times N}^{11,-1} \\ \lb{s3_111}
\hat{P}^{11}_{N\times N}&=&\exp\left\{\im\left( \bea{cc}
\hat{\mcal{C}}_{D;L\times L} & \hat{\xi}_{D;L\times S}^{+} \\
\hat{\xi}_{D;S\times L} & \hat{\mcal{G}}_{D;S\times S}
\eea\right)\right\}  \hspace*{0.75cm}\hat{\mcal{C}}_{D;L\times L}^{+}=\hat{\mcal{C}}_{D;L\times L}  \\ \lb{s3_112}
\hat{P}^{22}_{N\times N}&=&\exp\left\{\im\left( \bea{cc}
-\hat{\mcal{C}}_{D;L\times L}^{T} & \hat{\xi}_{D;L\times S}^{T} \\
-\hat{\xi}_{D;S\times L}^{*} & -\hat{\mcal{G}}_{D;S\times S}^{T} \eea\right)\right\}
\hspace*{0.75cm}\hat{\mcal{G}}_{D;S\times S}^{+}=\hat{\mcal{G}}_{D;S\times S} \\ \no &&
\hat{\mcal{C}}_{D;mm}=0\;\;\;(m=1,\ldots,L) \\ \no &&
\hat{\mcal{G}}_{D;r\mu,r\nu}=0\;\;\;(r=1,\ldots,S/2),(\mu,\nu=1,2)\;.
\eeq
In terms of the parametrization (\ref{s3_99}-\ref{s3_112}) for the density and anomalous terms,
we can finally obtain the change of integration measure from 'flat' coordinates
\(d\sigma_{D}^{(0)}(\vec{x},t_{p})\;\;
d[\delta\wtilde{\Sigma}(\vec{x},t_{p})\;\wtilde{K}]\) to the coset construction
\(Osp(S,S|2L)\backslash U(L|S)\otimes U(L|S)\) (\ref{s3_114}). The complete integration
\(d\sigma_{D}^{(0)}(\vec{x},t_{p})\;d[\delta\wtilde{\Sigma}(\vec{x},t_{p})\;\wtilde{K}]\)
in 'flat' space decomposes into the integration for the mean density field
\(d[\sigma_{D}^{(0)}(t_{p})]\), the integration over the density variables
\(d[\delta\hat{\Sigma}_{D}(\vec{x},t_{p})\;\wtilde{K}]=
d[d\hat{Q}(\vec{x},t_{p})\;\;\hat{Q}^{-1}(\vec{x},t_{p});\delta\hat{\lambda}(\vec{x},t_{p})]\)
in the subgroup \(U(L|S)\) and into the integration of the coset space
\(d[\hat{T}^{-1}(\vec{x},t_{p})\;d\hat{T}(\vec{x},t_{p});\delta\hat{\lambda}(\vec{x},t_{p})]=
\mcal{P}\big(\delta\hat{\lambda}(\vec{x},t_{p})\big)\;\;
d[\hat{T}^{-1}(\vec{x},t_{p})\;d\hat{T}(\vec{x},t_{p})]\) with additional contained products of
the eigenvalues \(\mcal{P}\big(\delta\hat{\lambda}(\vec{x},t_{p})\big)\) from the density terms
(\ref{s3_113}-\ref{s3_116}).
According to the unitary form of the considered subgroup \(U(L|S)\) of
\(Osp(S,S|2L)\) for the densities
\(d[\delta\hat{\Sigma}_{D}(\vec{x},t_{p})\;\wtilde{K}]=
d[d\hat{Q}(\vec{x},t_{p})\;\hat{Q}^{-1}(\vec{x},t_{p});\delta\hat{\lambda}(\vec{x},t_{p})]\),
the integrations with the independent variables
\((d\hat{Q}(\vec{x},t_{p})\;\hat{Q}^{-1}(\vec{x},t_{p}))_{\alpha\beta}^{aa}\),
(\(\alpha\neq\beta\)) are multiplied by the squares of
the differences of the eigenvalues (see appendix \ref{sa3}, \ref{sa4}).
The integration within the coset space is given by
the independent variables in \(d\hat{X}_{N\times N}(\vec{x},t_{p})\),
\(\wtilde{\kappa}\;d\hat{X}_{N\times N}^{+}(\vec{x},t_{p})\) (\ref{s3_96},\ref{s3_97})
which are weighted by the sine- and hyperbolic-sine functions of the modulus of the eigenvalues
\(\hat{\ovv{c}}_{L\times L}(\vec{x},t_{p})\), \(\hat{\ovv{f}}_{S\times S}(\vec{x},t_{p})\)
(\ref{s3_108},\ref{s3_109}) of \(\hat{X}_{N\times N}(\vec{x},t_{p})\),
\(\wtilde{\kappa}\;\hat{X}_{N\times N}^{+}(\vec{x},t_{p})\) and the squares of the eigenvalues
\(\delta\hat{\lambda}_{N\times N}(\vec{x},t_{p})\) of the density terms (\ref{s3_100},\ref{s3_101}).
We transfer the eigenvalues \(\mcal{P}\big(\delta\hat{\lambda}_{N\times N}(\vec{x},t_{p})\big)\)
of the density terms in the coset
measure \(d[\hat{T}^{-1}(\vec{x},t_{p})\;d\hat{T}(\vec{x},t_{p});
\delta\hat{\lambda}(\vec{x},t_{p})]\)  (\ref{s3_116},\ref{s3_119}) to the measure
\(d[\delta\hat{\Sigma}_{D}(\vec{x},t_{p})\;\wtilde{K}]=
d[d\hat{Q}(\vec{x},t_{p})\;\hat{Q}^{-1}(\vec{x},t_{p});\delta\hat{\lambda}(\vec{x},t_{p})]\) (\ref{s3_115})
of the subgroup \(U(L|S)\) so that separate integration measures
\(d[d\hat{Q}(\vec{x},t_{p})\;\hat{Q}^{-1}(\vec{x},t_{p});\delta\hat{\lambda}(\vec{x},t_{p})]\)
\(\mcal{P}\big(\delta\hat{\lambda}(\vec{x},t_{p})\big)\) and
\(d[\hat{T}^{-1}(\vec{x},t_{p})\;d\hat{T}(\vec{x},t_{p})]\)
for density and anomalous terms result instead of the integration
variables \(d[\delta\wtilde{\Sigma}_{2N\times 2N}(\vec{x},t_{p})\;\wtilde{K}]\) in 'flat' space.
The measure \(d[d\hat{Q}(\vec{x},t_{p})\;\hat{Q}^{-1}(\vec{x},t_{p});\delta\hat{\lambda}(\vec{x},t_{p})]\)
is given by relation (\ref{s3_117}) with integrations over the eigenvalues and the independent variables of
\(\big(d\hat{Q}^{11}(\vec{x},t_{p})\;\hat{Q}^{11;-1}(\vec{x},t_{p})\big)_{\alpha\beta}\) (\(\alpha\neq\beta\)).
The integration measure \(d[\hat{T}^{-1}(\vec{x},t_{p})\;d\hat{T}(\vec{x},t_{p})]\), which only contains the pair
condensates, is listed in Eq. (\ref{s3_120}). The details of the calculation for the invariant
measures of \(Osp(S,S|2L)\backslash U(L|S)\otimes U(L|S)\) are described in appendix \ref{sa}
with subsections \ref{sa1} to \ref{sa4}.

Note that the integration measures decompose into the even parts with the boson-boson 'BB' and
fermion-fermion 'FF' parts of the super-unitary matrices. The square root of the determinant
for the considered sub-metric-matrix enters as weight function into the integration
measure of the independent variables in the even boson-boson or fermion-fermion parts (\ref{A141}-\ref{A167})
(Compare appendix \ref{sa3}, esp. Eqs. (\ref{A141}) to (\ref{A149}) for the metric
of the boson-boson part and (\ref{A150}) to (\ref{A167}) for the even fermion-fermion part
in \(d[\hat{T}^{-1}\;d\hat{T};\delta\hat{\lambda}]\)). The prevailing metric tensor for the odd variables
in the boson-fermion 'BF' and fermion-boson 'FB' sections has to be taken into account by
the {\it inverse} square root of the determinant of the sub-metric for the assigned
subspace of anti-commuting variables (last terms in Eqs. (\ref{s3_117},\ref{s3_120}))
(see also appendix \ref{sa3} Eqs. (\ref{A168}) to (\ref{A174})).
In general the square root of the super-determinant
of the super-symmetric metric tensor determines the change of
integration measure for commuting and Grassmann variables (see appendices \ref{sa1} to \ref{sa4}
for the definition of a distance with metric tensors of \(Osp(S,S|2L)\backslash U(L|S)\otimes U(L|S)\);
compare also with the invariant volume element in the theory of general relativity \cite{hobson,hua}.)
\beq \lb{s3_113}
\delta\hat{\Sigma}_{D;2N\times 2N}\;\wtilde{K}&=&\hat{Q}_{2N\times 2N}^{-1}\;\;
\delta\hat{\Lambda}_{2N\times 2N}\;\;\hat{Q}_{2N\times 2N} \\ \lb{s3_114}
d\big[\sigma_{D}^{(0)}(t_{p})\big]\;\;
d\big[\delta\wtilde{\Sigma}_{2N\times 2N}(t_{p})\;\wtilde{K}\big]&=&
d\big[\sigma_{D}^{(0)}(t_{p})\big]\;\;
d\big[d\hat{Q}\;\hat{Q}^{-1};\delta\hat{\lambda}\big]\;\;
d\big[\hat{T}^{-1}\;d\hat{T};\delta\hat{\lambda}\big] \\ \lb{s3_115}
d\big[d\hat{Q}\;\hat{Q}^{-1};\delta\hat{\lambda}\big] &=&
d\big[\delta\hat{\Sigma}_{D}\;\wtilde{K}\big]  \\ \lb{s3_116}
d\big[\hat{T}^{-1}\;d\hat{T};\delta\hat{\lambda}\big] &=&\mcal{P}\big(\delta\hat{\lambda}\big)\;\;
d\big[\hat{T}^{-1}\;d\hat{T}\big]
\eeq
\beq \lb{s3_117}
\lefteqn{d\big[d\hat{Q}\;\hat{Q}^{-1},\delta\hat{\Lambda}\big]=
d\big[\delta\hat{\Sigma}_{D}\;\wtilde{K}\big]= } \\ \no &=&
\prod_{\{\vec{x},t_{p}\}}\bigg\{2^{(L+S)/2}\;
\bigg(\prod_{m=1}^{L}d\big(\delta\lambda_{B;m}\big)\bigg)
\bigg(\prod_{i=1}^{S}d\big(\delta\lambda_{F;i}\big)\bigg)\bigg\}\times \\ \no &\times&
\prod_{\{\vec{x},t_{p}\}}\bigg\{
\prod_{m=1}^{L}\prod_{n=m+1}^{L}\bigg(4\;
\frac{\big(d\hat{Q}^{11}\;\hat{Q}^{11,-1}\big)_{BB;mn}\wedge
\big(d\hat{Q}^{11}\;\hat{Q}^{11,-1}\big)_{BB;nm}}{2\;\im}\;
\big(\delta\hat{\lambda}_{B;n}-\delta\hat{\lambda}_{B;m}\big)^{2}\bigg)\bigg\}
\\ \no &\times&
\prod_{\{\vec{x},t_{p}\}}\bigg\{
\prod_{i=1}^{S}\prod_{i\ppr=i+1}^{S}\bigg(4\;
\frac{\big(d\hat{Q}^{11}\;\hat{Q}^{11,-1}\big)_{FF;ii\ppr}\wedge
\big(d\hat{Q}^{11}\;\hat{Q}^{11,-1}\big)_{FF;i\ppr i}}{2\;\im}\;
\big(\delta\hat{\lambda}_{F;i\ppr}-\delta\hat{\lambda}_{F;i}\big)^{2}\bigg)\bigg\}
\\ \no &\times&
\prod_{\{\vec{x},t_{p}\}}\bigg\{
\prod_{m=1}^{L}\prod_{i\ppr=1}^{S}\bigg(\frac{1}{4}\;
\big(d\hat{Q}^{11}\;\hat{Q}^{11,-1}\big)_{BF;mi\ppr}\;
\big(d\hat{Q}^{11}\;\hat{Q}^{11,-1}\big)_{FB;i\ppr m}\;
\big(\delta\hat{\lambda}_{F;i\ppr}-\delta\hat{\lambda}_{B;m}\big)^{\mbox{\boldmath{$^{-2}$}}}\bigg)\bigg\}
\eeq
\beq \lb{s3_118}
\lefteqn{d\big[\delta\hat{\Sigma}_{D}\;\wtilde{K}\big]=
d\big[d\hat{Q}\;\hat{Q}^{-1},\delta\hat{\Lambda}\big]=
} \\ \no &=&
\prod_{\{\vec{x},t_{p}\}}\bigg\{2^{(L+S)/2}\;
\bigg(\prod_{m=1}^{L}d\big(\delta\hat{B}_{D;mm}\big)\bigg)
\bigg(\prod_{i=1}^{S}d\big(\delta\hat{F}_{D;ii}\big)\bigg)\bigg\}\times \\ \no &\times&
\prod_{\{\vec{x},t_{p}\}}\bigg\{
\prod_{m=1}^{L}\prod_{n=m+1}^{L}\bigg(4\;\;
\frac{d\big(\delta\hat{B}_{D;mn}^{*}\big)\;\wedge\;
d\big(\delta\hat{B}_{D;mn}\big)}{2\;\im}\bigg)\bigg\}
\\ \no &\times&
\prod_{\{\vec{x},t_{p}\}}\bigg\{
\prod_{i=1}^{S}\prod_{i\ppr=i+1}^{S}\bigg(4\;\;
\frac{d\big(\delta\hat{F}_{D;ii\ppr}^{*}\big)\;\wedge\;
d\big(\delta\hat{F}_{D;ii\ppr}\big)}{2\;\im}\bigg)\bigg\}
\\ \no &\times&
\prod_{\{\vec{x},t_{p}\}}\bigg\{
\prod_{m=1}^{L}\prod_{i\ppr=1}^{S}\bigg(\frac{1}{4}\;\;
d\big(\delta\hat{\chi}_{D;mi\ppr}^{*}\big)\;\;
d\big(\delta\hat{\chi}_{D;i\ppr m}\big)\bigg)\bigg\}
\eeq
\beq \lb{s3_119}
\lefteqn{\mcal{P}\big(\delta\hat{\lambda}\big)=\prod_{\{\vec{x},t_{p}\}}\bigg\{
\bigg(\prod_{m=1}^{L}\big(\delta\hat{\lambda}_{B;m}\big)^{2}\bigg)
\bigg(\prod_{r=1}^{S/2}\big(\delta\hat{\lambda}_{F;r1}+\delta\hat{\lambda}_{F;r2}\big)^{2}\bigg)\bigg\} }
\\ \no &\times&
\prod_{\{\vec{x},t_{p}\}}\bigg\{
\prod_{m=1}^{L}\prod_{n=m+1}^{L}\bigg(
\big(\delta\hat{\lambda}_{B;n}+\delta\hat{\lambda}_{B;m}\big)^{2}\bigg)\bigg\}
\\ \no &\times&
\prod_{\{\vec{x},t_{p}\}}\bigg\{
\prod_{r=1}^{S/2}\prod_{r\ppr=r+1}^{S/2}\bigg(
\big(\delta\hat{\lambda}_{F;r1}+\delta\hat{\lambda}_{F;r\ppr 1}\big)^{2}\;\;
\big(\delta\hat{\lambda}_{F;r2}+\delta\hat{\lambda}_{F;r\ppr 2}\big)^{2} \times \\ \no &\times&
\big(\delta\hat{\lambda}_{F;r2}+\delta\hat{\lambda}_{F;r\ppr 1}\big)^{2}\;\;
\big(\delta\hat{\lambda}_{F;r1}+\delta\hat{\lambda}_{F;r\ppr 2}\big)^{2} \bigg)\bigg\} \\ \no &\times&
\prod_{\{\vec{x},t_{p}\}}\bigg\{
\prod_{m=1}^{L}\prod_{r\ppr=1}^{S/2}\bigg(
\big(\delta\hat{\lambda}_{F;r\ppr 1}+\delta\hat{\lambda}_{B;m}\big)^{2}\;\;
\big(\delta\hat{\lambda}_{F;r\ppr 2}+\delta\hat{\lambda}_{B;m}\big)^{2} \bigg)^{\mbox{\boldmath{$-1$}}}\bigg\}_{.}
\eeq
One has to distinguish between the diagonal integration variables
\(d\hat{c}_{D;mm}^{*}\; d\hat{c}_{D;mm}\),
\(d\hat{f}_{D;rr}^{(2)*}\; d\hat{f}_{D;rr}^{(2)}\) and off-diagonal integration parameters
\(d\hat{c}_{D;mn}^{*}\; d\hat{c}_{D;mn}\), (\(m\neq n\)),
\(d\hat{f}_{D;rr\ppr}^{(k)*}\; d\hat{f}_{D;rr\ppr}^{(0)}\), (\(r\neq r\ppr\), \(k=0,\ldots,3\))
for the measure of the coset space. However, a limit process of the eigenvalues
\((|\ovv{c}_{m}|-|\ovv{c}_{n}|)\rightarrow 0\), \((|\ovv{f}_{r}|-|\ovv{f}_{r\ppr}|)\rightarrow 0\)
can be performed in the partial integration measures of the off-diagonal variables
so that the integration measure follows for the diagonal integration parameters
\(d\hat{c}_{D;mm}^{*}\; d\hat{c}_{D;mm}\),
\(d\hat{f}_{D;rr}^{(2)*}\; d\hat{f}_{D;rr}^{(2)}\)
\beq \lb{s3_120}
\lefteqn{d\big[\hat{T}^{-1}\;d\hat{T}\big]=} \\ \no &=&
\prod_{\{\vec{x},t_{p}\}}\Bigg\{\prod_{m=1}^{L}\bigg(
\frac{d\hat{c}_{D;mm}^{*}\wedge d\hat{c}_{D;mm}}{2\;\im}\;\;2\;
\left|\frac{\sin\big(2\;|\ovv{c}_{m}|\big)}{|\ovv{c}_{m}|}\right|\bigg)\Bigg\} \\ \no &\times&
\prod_{\{\vec{x},t_{p}\}}\Bigg\{\prod_{m=1}^{L}\prod_{n=m+1}^{L}\bigg(
\frac{d\hat{c}_{D;mn}^{*}\wedge d\hat{c}_{D;mn}}{2\;\im}\;\;2\;
\left|\frac{\sin\big(|\ovv{c}_{m}|+|\ovv{c}_{n}|\big)}{|\ovv{c}_{m}|+|\ovv{c}_{n}|}\right|\;\;
\left|\frac{\sin\big(|\ovv{c}_{m}|-|\ovv{c}_{n}|\big)}{|\ovv{c}_{m}|-|\ovv{c}_{n}|}\right| \bigg)\Bigg\}
\\ \no &\times&
\prod_{\{\vec{x},t_{p}\}}\Bigg\{\prod_{r=1}^{S/2}\bigg(
\frac{d\hat{f}_{D;rr}^{(2)*}\wedge d\hat{f}_{D;rr}^{(2)}}{2\;\im}\;\;
\frac{\sinh\big(2\;|\ovv{f}_{r}|\big)}{|\ovv{f}_{r}|}\bigg)\Bigg\} \\ \no &\times&
\prod_{\{\vec{x},t_{p}\}}\Bigg\{\prod_{r=1}^{S/2}\prod_{r\ppr=r+1}^{S/2}\prod_{k=0}^{3}\bigg(
\frac{d\hat{f}_{D;rr\ppr}^{(k)*}\wedge d\hat{f}_{D;rr\ppr}^{(k)}}{2\;\im}\;4\;
\left|\frac{\sinh\big(|\ovv{f}_{r}|+|\ovv{f}_{r\ppr}|\big)}{
|\ovv{f}_{r}|+|\ovv{f}_{r\ppr}|}\right|\;
\left|\frac{\sinh\big(|\ovv{f}_{r}|-|\ovv{f}_{r\ppr}|\big)}{
|\ovv{f}_{r}|-|\ovv{f}_{r\ppr}|}\right|\bigg)\Bigg\}   \\ \no &\times&
\prod_{\{\vec{x},t_{p}\}}\Bigg\{\prod_{m=1}^{L}\prod_{r\ppr=1}^{S/2}
\frac{d\hat{\eta}_{D;r\ppr 1,m}^{*}\; d\hat{\eta}_{D;r\ppr 1,m}\;\;
d\hat{\eta}_{D;r\ppr 2,m}^{*}\; d\hat{\eta}_{D;r\ppr 2,m}}{ {\ds
\left(2\;\left|\frac{\sinh\big(|\ovv{f}_{r\ppr}|+\im\;|\ovv{c}_{m}|\big)}{|\ovv{f}_{r\ppr}|+
\im\;|\ovv{c}_{m}|}\right|\;\;
\left|\frac{\sinh\big(|\ovv{f}_{r\ppr}|-\im\;|\ovv{c}_{m}|\big)}{|\ovv{f}_{r\ppr}|-\im\;|\ovv{c}_{m}|}\right|
\right)^{\mbox{\boldmath{$2$}}}  }}   \Bigg\}_{.}
\eeq
The complete 'flat' integration measure
\(d[\sigma_{D}^{(0)}]\;d[\delta\wtilde{\Sigma}_{2N\times 2N}\;\wtilde{K}]\) of the
coherent state path integral  (\ref{s3_93},\ref{s3_92}) \(Z[\hat{\mcal{J}},J_{\psi},\im\hat{J}_{\psi\psi}]\)
factorizes into the subgroup \(U(L|S)\) with the densities
\(d[d\hat{Q}\;\hat{Q}^{-1};\delta\hat{\lambda}]\) (\ref{s3_117},\ref{s3_118})
and into the coset part \(d[\hat{T}^{-1}\;d\hat{T}]\) (\ref{s3_120}) \(Osp(S,S|2L)\backslash U(L|S)\) for
the condensate pair terms where the eigenvalues \(\mcal{P}\big(\delta\hat{\lambda}\big)\) (\ref{s3_119}) in
\(d[\hat{T}^{-1}\;d\hat{T};\delta\hat{\lambda}]\) (\ref{s3_116}) have been shifted to the part
\(d[\delta\hat{\Sigma}_{D}\;\wtilde{K}]\) (\ref{s3_115}) within the \(U(L|S)\) group as additional factors.
In consequence we can decompose
\(Z[\hat{\mcal{J}},J_{\psi},\im\hat{J}_{\psi\psi}]\) (\ref{s3_93},\ref{s3_92})
with a gradient expansion of the super-matrix
\(\wtilde{\mcal{M}}_{\vec{x},\alpha;\vec{x}\ppr,\beta}^{ab}(t_{p},t_{q}\ppr)\) (\ref{s3_92})
in the super-determinant and the remaining actions in the exponentials of (\ref{s3_93}).
The density and anomalous terms separate in the integrands according to spontaneous symmetry breaking of
\(Osp(S,S|2L)\backslash U(L|S)\otimes U(L|S)\). The Goldstone modes are then specified by the
lowest orders of gradients of the coset matrices \(\hat{T}^{-1}\;(\pp_{i}\hat{T})\) ,
\(\hat{T}^{-1}\;(\pp_{t_{p}}\hat{T})\) , (\(i=1,\ldots,d\)).

\section{Gradient expansion into densities and anomalous terms} \lb{s4}

\subsection{Removal of densities $\delta\hat{\Sigma}_{D;\alpha\beta}^{aa}(\vec{x},t_{p})$ or 'hinge'-functions
from the coherent state path integral} \lb{s41}

It is the aim of this subsection \ref{s41} to eliminate the block diagonal densities
\(\delta\hat{\Sigma}_{D;\alpha\beta}^{aa}(\vec{x},t_{p})\) (\ref{s3_68}-\ref{s3_74})
from the coherent state path integral. They have been originally
included in the HST transformations (\ref{s3_79}-\ref{s3_83})
with 'superfluous' Gaussian integrals, but are necessary for the coset decomposition with the anomalous
terms. They act as 'hinges' when the matrix \(\hat{T}_{\alpha\beta}^{ab}(\vec{x},t_{p})\)
for anomalous terms (\ref{s3_94}-\ref{s3_98}) is shifted from the total self-energy,
decomposed into densities and anti-hermitian pair condensates,
to the 'Nambu'-doubled one-particle Hamilton operator \(\hat{\mcal{H}}\) (\ref{s2_55}-\ref{s2_57})
in the super-determinant and BEC coherent wave function part with matrix
\(\wtilde{\mcal{M}}_{\vec{x},\alpha;\vec{x}\ppr,\beta}^{ab}(t_{p},t_{q}\ppr)\)
(\ref{s3_93},\ref{s3_92},\ref{s4_15}-\ref{s4_19}).

In section \ref{s3} we have determined the symmetries of the generating function and have identified
the ortho-symplectic group \(Osp(S,S|2L)\) as the manifold of the self-energy. By using the Weyl unitary
trick \cite{weyl1}, we have performed HST transformations with the 'hinge' functions
\(\delta\hat{\Sigma}_{D;\alpha\beta}^{aa}(\vec{x},t_{p})\) so that the resulting path integral
\(Z[\hat{\mcal{J}},J_{\psi},\im\hat{J}_{\psi\psi}]\) (\ref{s3_93},\ref{s3_92}) has symmetries which
coincide with the manifold of the self-energy (or the corresponding \(Osp(S,S|2L)\) group)
\beq \lb{s4_1}
\lefteqn{Z[\hat{\mcal{J}},J_{\psi},\im\hat{J}_{\psi\psi}]=
\int\;\;d[\sigma_{D}^{(0)}(\vec{x},t_{p})]\;\; \exp\bigg\{\frac{\im}{2\hbar}\frac{1}{V_{0}}
\int_{C}d t_{p}\sum_{\vec{x}}
\sigma_{D}^{(0)}(\vec{x},t_{p})\;\;\sigma_{D}^{(0)}(\vec{x},t_{p})\bigg\} } \\ \no &\times&
\int d[\delta\wtilde{\Sigma}(\vec{x},t_{p})\;\wtilde{K}]\;\;
\exp\bigg\{\frac{\im}{4\hbar}\frac{1}{V_{0}}
\int_{C}d t_{p}\sum_{\vec{x}}\mbox{STR}\Big[\delta\wtilde{\Sigma}(\vec{x},t_{p})\;
\wtilde{K}\;\delta\wtilde{\Sigma}(\vec{x},t_{p})\;\wtilde{K}\Big]\bigg\}
\\ \no &\times &
\Bigg\{\mbox{SDET}\bigg[\wtilde{\mcal{M}}_{\vec{x},\alpha;\vec{x}\ppr,\beta}^{ab}(t_{p},t_{q}\ppr)\bigg]
\Bigg\}^{\mathbf{-1/2}} \; \times  \\ \no &\times&
\exp\bigg\{\frac{\im}{2\hbar}\Omega^{2}\int_{C}d t_{p}\;d t\ppr_{q}\sum_{\vec{x},\vec{x}\ppr}\mcal{N}_{x}\;
J_{\psi;\beta}^{+b}(\vec{x}\ppr,t_{q}\ppr)\;\hat{I}\wtilde{K}\;
\wtilde{\mcal{M}}_{\vec{x}\ppr,\beta;\vec{x},\alpha}^{\mathbf{-1};ba}(t_{q}\ppr,t_{p})\;\hat{I}\;
J_{\psi;\alpha}^{a}(\vec{x},t_{p})\bigg\}
\eeq
\beq \lb{s4_2}
\lefteqn{
\wtilde{\mcal{M}}_{\vec{x},\alpha;\vec{x}\ppr,\beta}^{ab}(t_{p},t_{q}\ppr)=
\hat{\mcal{H}}_{\vec{x},\alpha;\vec{x}\ppr,\beta}^{ab}(t_{p},t_{q}\ppr)+
\sigma_{D}^{(0)}(\vec{x},t_{p})\;\hat{1}_{2N\times 2N}\;\delta_{\vec{x},\vec{x}\ppr}\;
\eta_{p}\;\delta_{p,q}\;\delta(t_{p}-t_{q}\ppr) + } \\ \no &+&
\left( \im\;\hat{J}_{\psi\psi;\alpha\beta}^{ab}(\vec{x},t_{p})\;\wtilde{K} +
\left(\bea{cc}
\delta\hat{\Sigma}_{\alpha\beta}^{11}(\vec{x},t_{p}) & \im\;
\delta\hat{\Sigma}_{\alpha\beta}^{12}(\vec{x},t_{p}) \\
\im\;\delta\hat{\Sigma}_{\alpha\beta}^{21}(\vec{x},t_{p}) &
\delta\hat{\Sigma}_{\alpha\beta}^{22}(\vec{x},t_{p})
\eea\right)\;\wtilde{K}\;\right)
\delta_{\vec{x},\vec{x}\ppr}\;\eta_{p}\;\delta_{p,q}\;\delta(t_{p}-t_{q}\ppr) + \\ \no &+&
\underbrace{\hat{I}\;\hat{K}\;\eta_{p}\;
\frac{\hat{\mcal{J}}_{\vec{x},\alpha;\vec{x}\ppr,\beta}^{ab}(t_{p},t_{q}\ppr)}{\mcal{N}_{x}}\;
\eta_{q}\;\hat{K}\;\hat{I}\;\wtilde{K}
}_{\wtilde{\mcal{J}}_{\vec{x},\alpha;\vec{x}\ppr,\beta}^{ab}(t_{p},t_{q}\ppr)} \;\;\;.
\eeq
The self-energy \(\wtilde{\Sigma}_{\alpha\beta}^{ab}(\vec{x},t_{p})\;\wtilde{K}\),
as a generator of \(Osp(S,S|2L)\) and with anti-hermitian anomalous terms in (\ref{s4_1},\ref{s4_2}),
can therefore be decomposed into the block diagonal densities
\(\delta\hat{\Sigma}_{D;\alpha\beta}^{aa}(\vec{x},t_{p})\;\wtilde{K}\)
as a subgroup \(U(L|S)\) of \(Osp(S,S|2L)\) and into the coset part \(\hat{T}_{\alpha\beta}^{ab}(\vec{x},t_{p})\),
\(Osp(S,S|2L)\backslash U(L|S)\).
The densities \(\delta\hat{\Sigma}_{D;\alpha\beta}^{aa}(\vec{x},t_{p})\;\wtilde{K}\)
are further factorized into doubled 'eigenvalues'
\(\delta\hat{\Lambda}_{\alpha}^{a}(\vec{x},t_{p})=\big(
\delta\hat{\lambda}_{\alpha}^{a}(\vec{x},t_{p})\;;\;-
\delta\hat{\lambda}_{\alpha}^{a}(\vec{x},t_{p})\big)\) as the maximal Abelian Cartan subalgebra
of \(Osp(S,S|2L)\) and 'eigenvectors' \(\hat{Q}_{\alpha\beta}^{aa}(\vec{x},t_{p})\) composed
of the shift operators \(\hat{\mcal{B}}_{D;L\times L}(\vec{x},t_{p})\),
\(\hat{\mcal{F}}_{D;S\times S}(\vec{x},t_{p})\), \(\hat{\mcal{\omega}}_{D;S\times L}(\vec{x},t_{p})\)
\(\hat{\mcal{\omega}}_{D;L\times S}^{+}(\vec{x},t_{p})\) of \(U(L|S)\) (\ref{s3_99}-\ref{s3_104}).
The self-energy \(\wtilde{\Sigma}_{\alpha\beta}^{ab}(\vec{x},t_{p})\;\wtilde{K}\)
consists therefore of parameters according to following Eq. (\ref{s4_3})
\beq \lb{s4_3}
\lefteqn{\wtilde{\Sigma}_{2N\times 2N}(\vec{x},t_{p})\;\wtilde{K}=\sigma_{D}^{(0)}(\vec{x},t_{p})\;
\hat{1}_{2N\times 2N}+\delta\wtilde{\Sigma}_{2N\times 2N}(\vec{x},t_{p})\;\wtilde{K} } \\ \no &=&
\hat{T}(\vec{x},t_{p})\;\Big(\sigma_{D}^{(0)}(\vec{x},t_{p})\;
\hat{1}_{2N\times 2N}+\delta\hat{\Sigma}_{D;2N\times 2N}(\vec{x},t_{p})\;\wtilde{K}\Big)\;
\hat{T}^{-1}(\vec{x},t_{p}) \\ \no &=&
\hat{T}(\vec{x},t_{p})\;\Big(\sigma_{D}^{(0)}(\vec{x},t_{p})\;
\hat{1}_{2N\times 2N}+
\hat{Q}_{2N\times 2N}^{-1}(\vec{x},t_{p})\;
\delta\hat{\Lambda}_{2N\times 2N}(\vec{x},t_{p})\;\hat{Q}_{2N\times 2N}(\vec{x},t_{p})\Big)\;
\hat{T}^{-1}(\vec{x},t_{p}) \\ \no &=&
\hat{T}_{0}(\vec{x},t_{p})\;\Big(\sigma_{D}^{(0)}(\vec{x},t_{p})\;
\hat{1}_{2N\times 2N}+\delta\hat{\Lambda}_{2N\times 2N}(\vec{x},t_{p})\Big)\;
\hat{T}_{0}^{-1}(\vec{x},t_{p})\;,
\eeq
where we have also introduced the modified coset part
\(\hat{T}_{0;\alpha\beta}^{ab}(\vec{x},t_{p})=\hat{T}_{\alpha\gamma}^{ab}(\vec{x},t_{p})\;
\hat{Q}_{\gamma\beta}^{-1;bb}(\vec{x},t_{p})\) (\ref{s4_4}) as in the calculation of the
invariant measure in order to diagonalize the density parts
\(\delta\hat{\Sigma}_{D;\alpha\beta}^{aa}(\vec{x},t_{p})\;\wtilde{K}\)
\beq \lb{s4_4}
\hat{T}_{0}(\vec{x},t_{p})&=&\hat{T}_{2N\times 2N}(\vec{x},t_{p})\;
\hat{Q}_{2N\times 2N}^{-1}(\vec{x},t_{p})   \\ \lb{s4_5}
\delta\hat{\Sigma}_{D;2N\times 2N}(\vec{x},t_{p})\;\wtilde{K}&=&
\hat{Q}_{2N\times 2N}^{-1}(\vec{x},t_{p})\;
\delta\hat{\Lambda}_{2N\times 2N}(\vec{x},t_{p})\;\hat{Q}_{2N\times 2N}(\vec{x},t_{p})
\\ \lb{s4_6} \hat{Q}_{2N\times 2N}(\vec{x},t_{p})&=&\left(
\bea{cc}
\hat{Q}_{N\times N}^{11}(\vec{x},t_{p}) & \\
 & \hat{Q}_{N\times N}^{22}(\vec{x},t_{p})
\eea\right)_{\mbox{.}}
\eeq
The exponential
\(\big(\exp\{-\hat{Y}_{\alpha\ppr\beta\ppr}^{a\ppr b\ppr}(\vec{x},t_{p})\}\big)_{\alpha\beta}^{ab}\)
(\ref{s4_7},\ref{s3_95}) of the coset part \(Osp(S,S|2L)\backslash U(L|S)\) can be performed
by using the representation with \(\hat{X}_{\alpha\beta}(\vec{x},t_{p})\) and
\(\wtilde{\kappa}\;\hat{X}_{\alpha\beta}^{+}(\vec{x},t_{p})\) (\ref{s3_96},\ref{s3_97})
so that the hyperbolic sine- and cosine-functions are obtained from these matrices
\(\hat{X}_{\alpha\beta}(\vec{x},t_{p})\) and
\(\wtilde{\kappa}\;\hat{X}_{\alpha\beta}^{+}(\vec{x},t_{p})\). However,
the submetric \(\wtilde{\kappa}=\big\{-\hat{1}_{L\times L}\;;\;+\hat{1}_{S\times S}\big\}\)
in the boson-boson part changes the hyperbolic sine- and cosine- functions to its analogous
trigonometric parts. Moreover, we apply  the diagonalized forms of \(\hat{Y}_{\alpha\beta}^{ab}(\vec{x},t_{p})\)
and \(\hat{X}_{\alpha\beta}(\vec{x},t_{p})\) (\ref{s3_105}-\ref{s3_112}) with the complex valued eigenvalues in
\(\hat{X}_{DD;\alpha\beta}(\vec{x},t_{p})\) and with the shift generators of \(U(L|S)\) in the matrix
\(\hat{P}_{\alpha\beta}^{aa}(\vec{x},t_{p})\) so that the matrix \(\hat{T}_{D;\alpha\beta}^{ab}(\vec{x},t_{p})\)
(\ref{s4_8}) is acquired with corresponding diagonal blocks
of \(\hat{T}_{\alpha\beta}^{ab}(\vec{x},t_{p})\) (\ref{s4_7})
\beq \lb{s4_7}
\hat{T}_{2N\times 2N}(\vec{x},t_{p}) &=&
\exp\Big\{-\hat{Y}_{2N\times 2N}(\vec{x},t_{p})\Big\}  \\  \no  &=&
\left( \bea{cc} \cosh(\sqrt{\hat{X}\;\wtilde{\kappa}\;\hat{X}^{+}})  &
-\frac{{\ds\sinh(\sqrt{\hat{X}\;\wtilde{\kappa}\;\hat{X}^{+}})}}{{\ds\sqrt{\hat{X}\;
\wtilde{\kappa}\;\hat{X}^{+}}}}\;\hat{X} \\
-\wtilde{\kappa}\;\hat{X}^{+}\;\frac{{\ds\sinh(\sqrt{\hat{X}\;
\wtilde{\kappa}\;\hat{X}^{+}})}}{{\ds\sqrt{\hat{X}\;\wtilde{\kappa}\;\hat{X}^{+}}}}
& \cosh(\sqrt{\wtilde{\kappa}\;\hat{X}^{+}\;\hat{X}}) \eea \right)_{2N\times 2N}   \\  \no &=&
\hat{P}_{2N\times 2N}^{-1}\;\exp\Big\{-\hat{Y}_{DD;2N\times 2N}\Big\}\;
\hat{P}_{2N\times 2N} \\ \no &=&\hat{P}_{2N\times 2N}^{-1}\;\hat{T}_{D;2N\times 2N}\;\hat{P}_{2N\times 2N}
\\ \lb{s4_8} \hat{T}_{D;2N\times 2N}&=&
\left( \bea{cc} \cosh(\sqrt{\hat{X}_{DD}\;\wtilde{\kappa}\;\hat{X}_{DD}^{+}})  &
-\frac{{\ds\sinh(\sqrt{\hat{X}_{DD}\;\wtilde{\kappa}\;\hat{X}_{DD}^{+}})}}{{\ds\sqrt{\hat{X}_{DD}\;
\wtilde{\kappa}\;\hat{X}_{DD}^{+}}}}\;\hat{X}_{DD} \\
-\wtilde{\kappa}\;\hat{X}_{DD}^{+}\;\frac{{\ds\sinh(\sqrt{\hat{X}_{DD}\;
\wtilde{\kappa}\;\hat{X}_{DD}^{+}})}}{{\ds\sqrt{\hat{X}_{DD}\;\wtilde{\kappa}\;\hat{X}_{DD}^{+}}}}
& \cosh(\sqrt{\wtilde{\kappa}\;\hat{X}_{DD}^{+}\;\hat{X}_{DD}}) \eea \right)_{2N\times 2N\;\;\; .}
\eeq
After a shift of the self-energy \(\delta\wtilde{\Sigma}_{\alpha\beta}^{ab}(\vec{x},t_{p})\) by
\(\im\;\hat{J}_{\psi\psi;\alpha\beta}^{a\neq b}(\vec{x},t_{p})\) (\ref{s2_63},\ref{s2_64}),
we have moved the source matrix for the anomalous terms from the matrix
\(\wtilde{\mcal{M}}_{\vec{x},\alpha;\vec{x}\ppr,\beta}^{ab}(t_{p},t_{q}\ppr)\) (\ref{s4_2},\ref{s3_92})
to the quadratic term \(\mcal{A}_{2}\big[\hat{T},\delta\hat{\Sigma}_{D};\im\hat{J}_{\psi\psi}\big]\)
(\ref{s4_11}) with the self-energy and the interaction parameter \(1/V_{0}\). The actions
\(\mcal{A}_{SDET}\big[\hat{T},\delta\hat{\Sigma}_{D},\hat{\sigma}_{D}^{(0)};\hat{\mcal{J}}\big]\) (\ref{s4_12}),
\(\mcal{A}_{J_{\psi}}\big[\hat{T},\delta\hat{\Sigma}_{D},\hat{\sigma}_{D}^{(0)};
\hat{\mcal{J}}\big]\) (\ref{s4_13}) of the super-determinant and the bilinear source term of the coherent
BEC wave function do therefore not contain the source matrix \(\im\;\hat{J}_{\psi\psi;\alpha\beta}^{a\neq b}(\vec{x},t_{p})\)
as a 'seed' for the pair condensates \footnote{The trace over spatial and time-like variables in
\(\mcal{A}_{SDET}\big[\hat{T},\delta\hat{\Sigma}_{D},\hat{\sigma}_{D}^{(0)};\hat{\mcal{J}}\big]\) (\ref{s4_12})
has to be normalized by the parameter \(\mcal{N}=\hbar\Omega\;\mcal{N}_{x}\)
with \(\mcal{N}_{x}=\big(L/\Delta x\big)^{d}\) and \(\Omega=1/\Delta t\) which are determined by the discrete
finite intervals of the coherent state path integral. Similarly, the action
\(\mcal{A}_{J_{\psi}}\big[\hat{T},\delta\hat{\Sigma}_{D},\hat{\sigma}_{D}^{(0)};
\hat{\mcal{J}}\big]\) (\ref{s4_13}) has to be scaled by \(\Omega^{2}\) and \(\mcal{N}_{x}\)
for normalization.} (see also footnote
\footnote{Note that in the remainder of this paper (sections \ref{s4}-\ref{s6} and appendix \ref{sb}) any
density-like operator terms as $\hat{\sigma}_{D}^{(0)}$ or $\delta\hat{\Lambda}$ already include the contour time
metric $\eta_{p}$ as one considers the spatial and contour time representation of such density terms. We even
define in subsection \ref{s42} these density terms with the additional contour metric $\eta_{p}$
(\ref{s4_53},\ref{s4_61},\ref{s4_62}).})
\be \lb{s4_9}
\delta\wtilde{\Sigma}_{2N\times 2N}(\vec{x},t_{p})+\im\;\hat{J}_{\psi\psi}(\vec{x},t_{p})\rightarrow
\delta\wtilde{\Sigma}_{2N\times 2N}(\vec{x},t_{p})
\ee
\beq \lb{s4_10}
\wtilde{\mcal{M}}_{\vec{x},\alpha;\vec{x}\ppr,\beta}^{ab}(t_{p},t_{q}\ppr)
\mbox{ (\ref{s4_2},\ref{s3_92}) }  &\rightarrow&  \\ \no
\wtilde{\mcal{M}}_{\vec{x},\alpha;\vec{x}\ppr,\beta}^{ab}(t_{p},t_{q}\ppr)   &=&
\Big[\hat{\mcal{H}}+\hat{\sigma}_{D}^{(0)}\;\hat{1}_{2N\times 2N}+
\wtilde{\mcal{J}}_{\alpha\beta}^{ab}+\hat{T}_{0}\;\delta\hat{\Lambda}\;
\hat{T}_{0}^{-1}\Big]_{\vec{x},\alpha;\vec{x}\ppr,\beta}^{ab}\!\!\!\!(t_{p},t_{q}\ppr)
\eeq
\beq \lb{s4_11}
\lefteqn{\mcal{A}_{2}\big[\hat{T},\delta\hat{\Sigma}_{D};\im\hat{J}_{\psi\psi}\big]  = }  \\ \no &=&
\frac{1}{4\hbar\;V_{0}}\int_{C}d t_{p}\sum_{\vec{x}}
\mbox{STR}\Big[\big(\delta\wtilde{\Sigma}(\vec{x},t_{p})-\im\;\hat{J}_{\psi\psi}(\vec{x},t_{p})\big)\;
\wtilde{K}\;\big(\delta\wtilde{\Sigma}(\vec{x},t_{p})-\im\;\hat{J}_{\psi\psi}(\vec{x},t_{p})\big)\;
\wtilde{K}\Big]
\\ \no  &=&\frac{1}{4\hbar\;V_{0}}\int_{C}d t_{p}\sum_{\vec{x}}\bigg\{
2\;\mbox{str}\Big[\big(\delta\hat{\lambda}_{N\times N}(\vec{x},t_{p})\big)^{2}\Big]+ \\ \no &-&
2\;\mbox{STR}\Big[\im\;\hat{J}_{\psi\psi}(\vec{x},t_{p})\;\wtilde{K}\;
\underbrace{\hat{T}(\vec{x},t_{p})\;\hat{Q}^{-1}(\vec{x},t_{p})}_{\hat{T}_{0}(\vec{x},t_{p})}\;
\delta\hat{\Lambda}(\vec{x},t_{p})\;
\underbrace{\hat{Q}(\vec{x},t_{p})\;\hat{T}^{-1}(\vec{x},t_{p})}_{\hat{T}_{0}^{-1}(\vec{x},t_{p})}
\Big] + \\ \no &+&\mbox{STR}\Big[\im\;\hat{J}_{\psi\psi}(\vec{x},t_{p})\;\wtilde{K}\;\im\;
\hat{J}_{\psi\psi}(\vec{x},t_{p})\;\wtilde{K}\Big]\bigg\}
\eeq
\beq \lb{s4_12}
\mcal{A}_{SDET}\big[\hat{T},\delta\hat{\Sigma}_{D},\hat{\sigma}_{D}^{(0)};\hat{\mcal{J}}\big]&=&
\frac{1}{2}\int_{C}\frac{d t_{p}}{\hbar}\eta_{p}\sum_{\vec{x}}\mcal{N}\;
\mbox{STR}\Big[\ln\Big(\wtilde{\mcal{M}}_{\vec{x},\alpha;\vec{x}\ppr,\beta}^{ab}(t_{p},t_{q}\ppr)
\Big)\Big]   \\   \lb{s4_13}
\mcal{A}_{J_{\psi}}\big[\hat{T},\delta\hat{\Sigma}_{D},\hat{\sigma}_{D}^{(0)};
\hat{\mcal{J}}\big]
&=&\frac{\Omega^{2}}{2\hbar}\!\!\int_{C}\!\!d t_{p}\;d t_{q}\ppr
\sum_{\vec{x},\vec{x}\ppr}\mcal{N}_{x}\;
J_{\psi;\beta}^{+b}(\vec{x}\ppr,t_{q}\ppr)\;\hat{I}\;\wtilde{K}\;
\wtilde{\mcal{M}}_{\vec{x}\ppr,\beta;\vec{x},\alpha}^{-1;ba}(t_{q}\ppr,t_{p})\;\hat{I}\;
J_{\psi;\alpha}^{a}(\vec{x},t_{p})_{\mbox{.}}
\eeq
We insert the invariant measure (\ref{s3_113}-\ref{s3_120}) for the coset decomposition
\(Osp(S,S|2L)\backslash U(L|S)\otimes U(L|S)\) and the actions (\ref{s4_11}-\ref{s4_13}) with matrix
\(\wtilde{\mcal{M}}_{\vec{x},\alpha;\vec{x}\ppr,\beta}^{ab}(t_{p},t_{q}\ppr)\) (\ref{s4_10})
into (\ref{s4_1},\ref{s3_93}). This leads to the generating function
\(Z[\hat{\mcal{J}},J_{\psi},\im\;\hat{J}_{\psi\psi}]\) (\ref{s4_14}) which still comprises
the densities \(\delta\hat{\Sigma}_{D;\alpha\beta}^{aa}(\vec{x},t_{p})\;\wtilde{K}_{\beta}^{a}\)
or 'hinge' functions
\beq \lb{s4_14}
\lefteqn{\hspace*{-0.46cm}Z[\hat{\mcal{J}},J_{\psi},\im\;\hat{J}_{\psi\psi}]= } \\ \no &=&
\int d[\hat{\sigma}_{D}^{(0)}(\vec{x},t_{p})]\;\;
\exp\bigg\{\frac{\im}{2\hbar}\frac{1}{V_{0}}\int_{C}d t_{p}\sum_{\vec{x}}
\sigma_{D}^{(0)}(\vec{x},t_{p})\;\sigma_{D}^{(0)}(\vec{x},t_{p})\bigg\} \;\;
\int d\big[\hat{T}^{-1}(\vec{x},t_{p})\;d\hat{T}(\vec{x},t_{p})\big]  \\ \no &\times&
\bigg(\int d\big[d\hat{Q}(\vec{x},t_{p})\;\hat{Q}^{-1}(\vec{x},t_{p});
\delta\hat{\lambda}(\vec{x},t_{p})\big]\;\;
\mcal{P}\big(\delta\hat{\lambda}(\vec{x},t_{p})\big)\;\;
\exp\Big\{\im\;\mcal{A}_{2}\big[\hat{T},\delta\hat{\Sigma}_{D};\im\hat{J}_{\psi\psi}\big]\Big\} \bigg)
\\ \no &\times &
\exp\Big\{-\mcal{A}_{SDET}\big[\hat{T},\delta\hat{\Sigma}_{D},\hat{\sigma}_{D}^{(0)};\hat{\mcal{J}}\big]\Big\}\;\;
\exp\Big\{\im\;\mcal{A}_{J_{\psi}}\big[\hat{T},\delta\hat{\Sigma}_{D},
\hat{\sigma}_{D}^{(0)};\hat{\mcal{J}}\big]\Big\}
\\ \no &=&
\int d\big[\hat{\sigma}_{D}^{(0)}(\vec{x},t_{p})\big]\;\;
\exp\bigg\{\frac{\im}{2\hbar}\frac{1}{V_{0}}\int_{C}d t_{p}\sum_{\vec{x}}
\sigma_{D}^{(0)}(\vec{x},t_{p})\;\sigma_{D}^{(0)}(\vec{x},t_{p})\bigg\}
\\ \no &\times &
\int d\big[\hat{T}^{-1}(\vec{x},t_{p})\;d\hat{T}(\vec{x},t_{p})\big]\bigg(
\int d\big[\delta\hat{\Sigma}_{D}(\vec{x},t_{p})\;\wtilde{K}\big]\;\;
\mcal{P}\big(\delta\hat{\lambda}(\vec{x},t_{p})\big)\;\;
\exp\Big\{\im\;\mcal{A}_{2}\big[\hat{T},\delta\hat{\Sigma}_{D};\im\hat{J}_{\psi\psi}\big]\Big\} \bigg)
 \\ \no &\times &
\exp\Big\{-\mcal{A}_{SDET}\big[\hat{T},\delta\hat{\Sigma}_{D},\hat{\sigma}_{D}^{(0)};\hat{\mcal{J}}\big]\Big\}\;\;
\exp\Big\{\im\;\mcal{A}_{J_{\psi}}\big[\hat{T},\delta\hat{\Sigma}_{D},
\hat{\sigma}_{D}^{(0)};\hat{\mcal{J}}\big]\Big\}_{\mbox{.}}
\eeq
Finally, we can perform the step where the block diagonal self-energy densities
\(\delta\hat{\Sigma}_{D;\alpha\beta}^{aa}(\vec{x},t_{p})\;\wtilde{K}_{\beta}^{a}\) act as
'hinge' functions for the pair condensates with coset matrix \(\hat{T}(\vec{x},t_{p})\). The matrices
\(\hat{T}(\vec{x},t_{p})\), \(\hat{T}^{-1}(\vec{x},t_{p})\) in the coset decomposition of the total
self-energy (\ref{s4_3}) are moved out of the square brackets of the matrix
\(\wtilde{\mcal{M}}_{\vec{x},\alpha;\vec{x}\ppr,\beta}^{ab}(t_{p},t_{q}\ppr)\) (\ref{s4_10}),
consisting of the doubled one-particle operator
\(\hat{\mcal{H}}_{\vec{x},\alpha;\vec{x}\ppr,\beta}^{ab}(t_{p},t_{q}\ppr)\),
the self-energy density field \(\sigma_{D}^{(0)}(\vec{x},t_{p})\) and source term
\(\wtilde{\mcal{J}}_{\vec{x},\alpha;\vec{x}\ppr,\beta}^{ab}(t_{p},t_{q}\ppr)\). In consequence
the coset matrices couple to the spatial and time-like gradient operators of
\(\hat{\mcal{H}}\), (\(\hat{T}^{-1}\;\hat{\mcal{H}}\;\hat{T}\)), and to the source matrix
\(\wtilde{\mcal{J}}(\hat{T}^{-1},\hat{T})=\hat{T}^{-1}\;\wtilde{\mcal{J}}\;\hat{T}\).
After adding and subtracting the one-particle part \(\hat{\mcal{H}}\), one obtains the diagonal
term \(\hat{\mcal{H}}+\hat{\sigma}_{D}^{(0)}\;\hat{1}_{2N\times 2N}\),
the source matrix \(\wtilde{\mcal{J}}(\hat{T}^{-1},\hat{T})\)
and gradient terms of
\(\delta\hat{\mcal{H}}(\hat{T}^{-1},\hat{T})=\hat{T}^{-1}\;\hat{\mcal{H}}\;\hat{T}-\hat{\mcal{H}}\)
which are combined into the super-matrix
\(\wtilde{\mcal{N}}_{\vec{x},\alpha;\vec{x}\ppr,\beta}^{ab}(t_{p},t_{q}\ppr;\delta\hat{\Sigma}_{D})\)
as replacement of \(\wtilde{\mcal{M}}_{\vec{x},\alpha;\vec{x}\ppr,\beta}^{ab}(t_{p},t_{q}\ppr)\)
\beq \lb{s4_15}
\wtilde{\mcal{M}}_{\vec{x},\alpha;\vec{x}\ppr,\beta}^{ab}(t_{p},t_{q}\ppr) &=&
\Big[\hat{\mcal{H}}+\wtilde{\mcal{J}}_{\alpha\beta}^{ab}+\hat{\sigma}_{D}^{(0)}\;
\hat{1}_{2N\times 2N}+\hat{T}_{0}\;\delta\hat{\Lambda}\;
\hat{T}_{0}^{-1}\Big]_{\vec{x},\alpha;\vec{x}\ppr,\beta}^{ab}\!\!\!\!(t_{p},t_{q}\ppr)
\\ \no &=&
\hat{T}(\vec{x},t_{p})\Bigg[\Big(\hat{\mcal{H}}+\hat{\sigma}_{D}^{(0)}\;\hat{1}_{2N\times 2N}\Big)+
\Big(\hat{T}^{-1}\hat{\mcal{H}}\hat{T}-\hat{\mcal{H}}\Big)+
\wtilde{\mcal{J}}(\hat{T}^{-1},\hat{T}) + \\ \no &+&
\left(\bea{cc} \delta\hat{\Sigma}_{D;N\times N}^{11} & 0 \\
0 & \delta\hat{\Sigma}_{D;N\times N}^{22}\;\wtilde{\kappa}
\eea\right)\Bigg]_{\vec{x},\alpha;\vec{x}\ppr,\beta}^{ab}\!\!\hspace*{-0.46cm}(t_{p},t_{q}\ppr)\;\;\;\;
\hat{T}^{-1}(\vec{x}\ppr,t_{q}\ppr) \\ \no &=&
\hat{T}(\vec{x},t_{p})\;\;
\wtilde{\mcal{N}}_{\vec{x},\alpha;\vec{x}\ppr,\beta}^{ab}(t_{p},t_{q}\ppr;\delta\hat{\Sigma}_{D})\;\;
\hat{T}^{-1}(\vec{x}\ppr,t_{q}\ppr)
\eeq
\beq \lb{s4_16}
\hat{\mcal{H}}_{\vec{x},\alpha;\vec{x}\ppr,\beta}^{ab}(t_{p},t_{q}\ppr)&=&\delta_{p,q}\;\eta_{p}\;
\delta(t_{p}-t_{q}\ppr)\;\delta_{\vec{x},\vec{x}\ppr}\left(\bea{cc}
\hat{H}_{p}(\vec{x},t_{p}) & \\ & \hat{H}_{p}^{T}(\vec{x},t_{p})
\eea\right)     \\ \lb{s4_17}
\hat{H}_{p}(\vec{x},t_{p})&=&-\hat{E}_{p}+\hat{h}_{p}(\vec{x}) \hspace*{1.0cm}
\hat{h}_{p}(\vec{x})=\frac{\hat{\vec{p}}^{\;2}}{2m}+u(\vec{x})-\mu_{0}-\im\;\ve_{p}  \\ \lb{s4_18}
\wtilde{\mcal{J}}_{\vec{x},\alpha;\vec{x}\ppr,\beta}^{ab}(t_{p},t_{q}\ppr)&=&
\hat{I}\;\hat{K}\;\eta_{p}\;
\frac{\hat{\mcal{J}}_{\vec{x},\alpha;\vec{x}\ppr,\beta}^{ab}(t_{p},t_{q}\ppr)}{\mcal{N}_{x}}\;\eta_{q}\;
\hat{K}\;\hat{I}\;\wtilde{K}  \\ \lb{s4_19}
\wtilde{\mcal{J}}_{\vec{x},\alpha;\vec{x}\ppr,\beta}^{ab}(\hat{T}^{-1}(t_{p}),\hat{T}(t_{q}\ppr)) &=&
\hat{T}_{\alpha\alpha\ppr}^{-1;aa\ppr}(\vec{x},t_{p})\;\;
\hat{I}\;\hat{K}\;\eta_{p}\;\frac{\hat{\mcal{J}}_{\vec{x},\alpha\ppr;
\vec{x}\ppr,\beta\ppr}^{a\ppr b\ppr}(t_{p},t_{q}\ppr)}{\mcal{N}_{x}}\;\eta_{q}\;
\hat{K}\;\hat{I}\;\wtilde{K}\;\;\hat{T}_{\beta\ppr\beta}^{b\ppr b}(\vec{x}\ppr,t_{q}\ppr)\;.
\eeq
In order to remove the self-energy densities
\(\delta\hat{\Sigma}_{D;\alpha\beta}^{aa}(\vec{x},t_{p})\;\wtilde{K}_{\beta}^{a}\) or 'hinge' functions,
we execute the reverse operations which have originally transformed the matrix
\(\wtilde{M}_{\vec{x},\alpha;\vec{x}\ppr,\beta}^{ab}(t_{p},t_{q}\ppr)\) (\ref{s3_86}) to
\(\wtilde{\mcal{M}}_{\vec{x},\alpha;\vec{x}\ppr,\beta}^{ab}(t_{p},t_{q}\ppr)\) (\ref{s3_92})
using the metrics $\wtilde{K}$, $\hat{I}$, $\hat{I}^{-1}$, $\hat{K}$ in relations (\ref{s3_89}-\ref{s3_93}).
These reverse operations lead from the matrix
\(\wtilde{\mcal{M}}_{\vec{x},\alpha;\vec{x}\ppr,\beta}^{ab}(t_{p},t_{q}\ppr)\) (\ref{s4_15}) to
\(\wtilde{\mcal{N}}_{\vec{x},\alpha;\vec{x}\ppr,\beta}^{ab}(t_{p},t_{q}\ppr;\delta\hat{\Sigma}_{D})\)
(\ref{s4_20}) and finally to \(\wtilde{N}_{\vec{x},\alpha;\vec{x}\ppr,\beta}^{ab}(t_{p},t_{q}\ppr;
\hat{I}^{-1}\delta\hat{\Sigma}_{D}\;\hat{I}^{-1})\) (\ref{s4_21},\ref{s4_20}), still including the 'hinge'
functions. Furthermore, we define the super-matrix
\(\hat{\mcal{O}}_{\vec{x},\alpha;\vec{x}\ppr,\beta}^{ab}(t_{p},t_{q}\ppr)\) (\ref{s4_22}) which is
derived from \(\wtilde{N}_{\vec{x},\alpha;\vec{x}\ppr,\beta}^{ab}(t_{p},t_{q}\ppr)\) in (\ref{s4_21})
without the self-energy densities \(\delta\hat{\Sigma}_{D;\alpha\beta}^{aa}(\vec{x},t_{p})\)
\beq \lb{s4_20}
\wtilde{\mcal{M}}_{\vec{x},\alpha;\vec{x}\ppr,\beta}^{ab}(t_{p},t_{q}\ppr)  &=&
\hat{T}(\vec{x},t_{p})\;\hat{I}\;\hat{K}
\underbrace{\bigg(\hat{K}\;\hat{I}^{-1}\;
\wtilde{\mcal{N}}_{\vec{x},\alpha;\vec{x}\ppr,\beta}^{ab}(t_{p},t_{q}\ppr;\delta\hat{\Sigma}_{D})\;
\wtilde{K}\;\hat{I}^{-1}\;\hat{K}\bigg)}_{
\wtilde{N}_{\vec{x},\alpha;\vec{x}\ppr,\beta}^{ab}(t_{p},t_{q}\ppr;
\hat{I}^{-1}\delta\hat{\Sigma}_{D}\;\hat{I}^{-1})}\;
\hat{K}\;\hat{I}\;\wtilde{K}\;\hat{T}^{-1}(\vec{x}\ppr,t_{q}\ppr)
\eeq
\beq  \lb{s4_21}
\lefteqn{\wtilde{N}_{\vec{x},\alpha;\vec{x}\ppr,\beta}^{ab}(t_{p},t_{q}\ppr;
\hat{I}^{-1}\;\delta\hat{\Sigma}_{D}\;\hat{I}^{-1}) = } \\ \no  &=&
\wtilde{N}_{\vec{x},\alpha;\vec{x}\ppr,\beta}^{ab}(t_{p},t_{q}\ppr) +
\left[\hat{K}\;\;
\left(\bea{cc} \delta\hat{\Sigma}_{D;N\times N}^{11} & 0 \\
0 & -\delta\hat{\Sigma}_{D;N\times N}^{22}
\eea\right)\;\;\hat{K}\right]_{\vec{x},\alpha;\vec{x}\ppr,\beta}^{ab}\!\!\!\!(t_{p},t_{q}\ppr)
\eeq
\beq \lb{s4_22}
\lefteqn{\wtilde{N}_{\vec{x},\alpha;\vec{x}\ppr,\beta}^{ab}(t_{p},t_{q}\ppr)=
\hat{K}\;\hat{I}^{-1}\;\;
\hat{\mcal{O}}_{\vec{x},\alpha;\vec{x}\ppr,\beta}^{ab}(t_{p},t_{q}\ppr)\;\;\hat{I} = } \\ \no &=&
\hat{K}\;\Bigg\{
\Big(\hat{\mcal{H}}+\hat{\sigma}_{D}^{(0)}\;\hat{1}_{2N\times 2N}\Big) +
\Big[\Big(\hat{T}(\vec{x},t_{p})\;\hat{I}\Big)^{-1}\;\hat{\mcal{H}}\;
\Big(\hat{T}(\vec{x}\ppr,t_{q}\ppr)\;\hat{I}\Big)-\hat{\mcal{H}}\Big] + \\ \no &+&
\Big(\hat{T}(\vec{x},t_{p})\;\hat{I}\Big)^{-1}\;
\hat{I}\;\hat{K}\;
\hat{\eta}\;\frac{\hat{\mcal{J}}_{\alpha\ppr\beta\ppr}^{a\ppr b\ppr}}{\mcal{N}_{x}}\;\hat{\eta}\;
\;\hat{K}\;\hat{I}\;\wtilde{K}\;
\Big(\hat{T}(\vec{x}\ppr,t_{q}\ppr)\;\hat{I}\Big)\Bigg\}_{\vec{x},\alpha;\vec{x}\ppr,\beta}^{ab}
\hspace*{-0.64cm}(t_{p},t_{q}\ppr)\;\;\;.
\eeq
The value of the super-determinant \(\mbox{SDET}\big\{\wtilde{\mcal{M}}\big\}\) (\ref{s4_23})
with matrix \(\wtilde{\mcal{M}}_{\alpha\beta}^{ab}\) (\ref{s4_20}) is not affected by the
transformation to \(\wtilde{N}_{\vec{x},\alpha;\vec{x}\ppr,\beta}^{ab}(t_{p},t_{q}\ppr;
\hat{I}^{-1}\;\delta\hat{\Sigma}_{D}\;\hat{I}^{-1})\) (\ref{s4_21}). The Green function of
\(\wtilde{\mcal{M}}_{\alpha\beta}^{ab}\) (\ref{s4_25}) can also be given in terms of
\(\wtilde{N}_{\vec{x},\alpha;\vec{x}\ppr,\beta}^{ab}(t_{p},t_{q}\ppr;
\hat{I}^{-1}\;\delta\hat{\Sigma}_{D}\;\hat{I}^{-1})\) (\ref{s4_21}) so that the action
\(\mcal{A}_{J_{\psi}}\big[\hat{T},\delta\hat{\Sigma}_{D},
\hat{\sigma}_{D}^{(0)};\hat{\mcal{J}}\big]\) (\ref{s4_26}) follows
with transformed source fields \(\wtilde{J}_{\psi;\alpha}^{a}(\vec{x},t_{p})\),
\(\wtilde{J}_{\psi;\beta}^{+b}(\vec{x}\ppr,t_{q}\ppr)\) for the coherent BEC wave function
\beq \lb{s4_23}
\mbox{SDET}\Big\{\wtilde{\mcal{M}}\Big\}&=&
\mbox{SDET}\Big\{\wtilde{N}_{\vec{x},\alpha;\vec{x}\ppr,\beta}^{ab}(t_{p},t_{q}\ppr;
\hat{I}^{-1}\;\delta\hat{\Sigma}_{D}\;\hat{I}^{-1})
\Big\} \\  \lb{s4_24} \lefteqn{ \hspace*{-1.0cm}
\mcal{A}_{SDET}\big[\hat{T},\delta\hat{\Sigma}_{D},\hat{\sigma}_{D}^{(0)};\hat{\mcal{J}}\big]
  = } \\ \no &=& \frac{1}{2}\int_{C}\frac{d t_{p}}{\hbar}\eta_{p}\sum_{\vec{x}}\mcal{N}\;
\mbox{STR}\Bigg[\ln\bigg(\wtilde{N}_{\vec{x},\alpha;\vec{x}\ppr,\beta}^{ab}(t_{p},t_{q}\ppr)+
\hat{K}\;\;
\left(\bea{cc} \delta\hat{\Sigma}_{D;N\times N}^{11} & 0 \\
0 & -\delta\hat{\Sigma}_{D;N\times N}^{22}
\eea\right)\;\;\hat{K} \bigg)\Bigg]
\eeq
\beq \lb{s4_25}
\wtilde{\mcal{M}}_{\vec{x},\alpha;\vec{x}\ppr,\beta}^{
\mbox{\small\boldmath{$-1$}};ab}(t_{p},t_{q}\ppr) &=&
\hat{T}(\vec{x},t_{p})\;\wtilde{K}\;\hat{I}^{-1}\;\hat{K}\;\;
\wtilde{N}_{\vec{x},\alpha;\vec{x}\ppr,\beta}^{\mbox{\small\boldmath{$-1$}};ab}(t_{p},t_{q}\ppr;
\hat{I}^{-1}\delta\hat{\Sigma}_{D}\;\hat{I}^{-1})\;\;
\hat{K}\;\hat{I}^{-1}\;\hat{T}^{-1}(\vec{x}\ppr,t_{q}\ppr)
\eeq
\beq \lb{s4_26}
\lefteqn{\mcal{A}_{J_{\psi}}\big[\hat{T},\delta\hat{\Sigma}_{D},
\hat{\sigma}_{D}^{(0)};\hat{\mcal{J}}\big]
=\frac{\Omega^{2}}{2\hbar}\int_{C}d t_{p}\;d t_{q}\ppr
\sum_{\vec{x},\vec{x}\ppr}\mcal{N}_{x}\; \underbrace{J_{\psi;\beta}^{+b}(\vec{x}\ppr,t_{q}\ppr)\;
\Big(\hat{I}\;\wtilde{K}\;\hat{T}(\vec{x}\ppr,t_{q}\ppr)\;\wtilde{K}\;\hat{I}^{-1}\Big)}_{
\wtilde{J}_{\psi;\beta}^{+b}(\vec{x}\ppr,t_{q}\ppr)}
\times }  \\ \no &\times&
\hat{K}\;\;
\wtilde{N}_{\vec{x},\alpha\ppr;\vec{x}\ppr,\beta\ppr}^{
\mbox{\small\boldmath{$-1$}};a\ppr b\ppr}(t_{p},t_{q}\ppr;
\hat{I}^{-1}\delta\hat{\Sigma}_{D}\;\hat{I}^{-1})\;\;
\hat{K}\;\;\underbrace{\Big(\hat{I}^{-1}\;\hat{T}^{-1}(\vec{x}\ppr,t_{q}\ppr)\;\hat{I}\Big)
J_{\psi;\alpha}^{a}(\vec{x},t_{p})}_{\wtilde{J}_{\psi;\alpha}^{a}(\vec{x},t_{p})}\;\;\;.
\eeq
The actions \(\mcal{A}_{SDET}\), \(\mcal{A}_{J_{\psi}}\) (\ref{s4_24},\ref{s4_26})
originally result from the Gaussian integrations with doubled super-fields
\(\Psi_{\vec{x},\alpha}^{a}(t_{p})\), \(\Psi_{\vec{x}\ppr,\beta}^{+b}(t_{q}\ppr)\) so that we can
transform \(Z[\hat{\mcal{J}},J_{\psi},\im\hat{J}_{\psi\psi}]\) (\ref{s4_14}) with relations
(\ref{s4_15}-\ref{s4_26}) to (\ref{s4_27}) with BEC source term
\(\wtilde{J}_{\psi;\alpha}^{+a}(\vec{x},t_{p})\;\hat{K}\;\Psi_{\vec{x},\alpha}^{a}(t_{p})+
\Psi_{\vec{x},\alpha}^{+a}(t_{p})\;\hat{K}\;\wtilde{J}_{\psi;\alpha}^{a}(\vec{x},t_{p})\)
\beq  \lb{s4_27}
\lefteqn{\hspace*{-1.0cm}Z[\hat{\mcal{J}},J_{\psi},\im\;\hat{J}_{\psi\psi}]=
\int d[\hat{\sigma}_{D}^{(0)}(\vec{x},t_{p})]\;\;
\exp\bigg\{\frac{\im}{2\hbar}\frac{1}{V_{0}}\int_{C}d t_{p}\sum_{\vec{x}}
\sigma_{D}^{(0)}(\vec{x},t_{p})\;\sigma_{D}^{(0)}(\vec{x},t_{p})\bigg\}  } \\ \no &\times &
\int d\big[\hat{T}^{-1}(\vec{x},t_{p})\;d\hat{T}(\vec{x},t_{p})\big]\;\;\Bigg(\int
d\big[\delta\hat{\Sigma}_{D}(\vec{x},t_{p})\;\wtilde{K}\big]\;\;
\mcal{P}\big(\delta\hat{\lambda}(\vec{x},t_{p})\big)\;\;\times \\ \no &\times&
\exp\Bigg\{\frac{\im}{4\hbar\;V_{0}}\int_{C}d t_{p}\sum_{\vec{x}}\bigg(
\mbox{STR}\Big[\delta\hat{\Sigma}_{D;2N\times 2N}(\vec{x},t_{p})\;\wtilde{K}\;
\delta\hat{\Sigma}_{D;2N\times 2N}(\vec{x},t_{p})\;\wtilde{K}\Big] + \\ \no &-&
2\;\mbox{STR}\Big[\im\;\hat{J}_{\psi\psi}(\vec{x},t_{p})\;\wtilde{K}\;\hat{T}(\vec{x},t_{p})\;
\delta\hat{\Sigma}_{D;2N\times 2N}(\vec{x},t_{p})\;\wtilde{K}\;
\hat{T}^{-1}(\vec{x},t_{p})\Big] + \\ \no &+&
\mbox{STR}\Big[\im\;\hat{J}_{\psi\psi}(\vec{x},t_{p})\;\wtilde{K}\;
\im\;\hat{J}_{\psi\psi}(\vec{x},t_{p})\;\wtilde{K}\Big]\bigg)\Bigg\}\Bigg)\times
\\ \no &\times& \int d\big[\psi_{\vec{x},\alpha}(t_{p})\big]\;\;
\exp\Bigg\{-\frac{\im}{2\hbar}\int_{C}d t_{p}\;d t_{q}\ppr\sum_{\vec{x},\vec{x}\ppr}\mcal{N}_{x}\;
\Psi_{\vec{x}\ppr,\beta}^{+b}(t_{q}\ppr)\;\bigg[
\wtilde{N}_{\vec{x}\ppr,\beta;\vec{x},\alpha}^{ba}(t_{q}\ppr,t_{p})+ \\ \no &+&
\hat{K}\;\;
\left(\bea{cc} \delta\hat{\Sigma}_{D;N\times N}^{11} & 0 \\
0 & -\delta\hat{\Sigma}_{D;N\times N}^{22}
\eea\right)\;\;\hat{K}\bigg]_{\vec{x}\ppr,\beta;\vec{x},\alpha}^{ba}\!\!\!\!(t_{q}\ppr,t_{p})\;\;
\Psi_{\vec{x},\alpha}^{a}(t_{p})\Bigg\} \times \\ \no &\times&
\exp\Bigg\{-\frac{\im}{2\hbar}\int_{C}d t_{p}\sum_{\vec{x}} \bigg(
\underbrace{J_{\psi;\beta}^{+b}(\vec{x},t_{p})\;
\Big(\hat{I}\;\wtilde{K}\;\hat{T}(\vec{x},t_{p})\;\wtilde{K}\;\hat{I}^{-1}\Big)}_{
\wtilde{J}_{\psi;\alpha}^{+a}(\vec{x},t_{p})}\;\;\hat{K}\;\Psi_{\vec{x},\alpha}^{a}(t_{p})\; +
 \\ \no &+&
\Psi_{\vec{x},\alpha}^{+a}(t_{p})\;\hat{K}\;\;
\underbrace{\Big(\hat{I}^{-1}\;\hat{T}^{-1}(\vec{x},t_{p})\;\hat{I}\Big)\;
J_{\psi;\beta}^{b}(\vec{x},t_{p})}_{\wtilde{J}_{\psi;\alpha}^{a}(\vec{x},t_{p})} \bigg)\Bigg\}_{\mbox{.}}
\eeq
However, we have to verify that the fields \(\wtilde{J}_{\psi;\alpha}^{a}(\vec{x},t_{p})\),
\(\wtilde{J}_{\psi;\alpha}^{+a}(\vec{x},t_{p})\), defined in the couplings to
\(\Psi_{\vec{x},\alpha}^{+a}(t_{p})\), \(\Psi_{\vec{x},\alpha}^{a}(t_{p})\) (\ref{s4_27}), are
related by superhermitian conjugation as stated. This can be proved by application of
Eqs. (\ref{s4_28},\ref{s4_29}) for the coset matrices
\beq  \lb{s4_28}
\hat{I}^{-1}\;\hat{T}^{-1}\;\hat{I}&=&
\exp\left\{\im\Bigg(\bea{cc} 0 & \hat{X} \\ -\wtilde{\kappa}\;\hat{X}^{+} & 0 \eea\Bigg)\right\}
\\ \lb{s4_29}
\Big(\hat{I}^{-1}\;\hat{T}^{-1}\;\hat{I}\Big)^{+}&=&
\exp\left\{\im\Bigg(\bea{cc} 0 & \hat{X}\;\wtilde{\kappa} \\ -\hat{X}^{+} & 0 \eea\Bigg)\right\}
= \hat{I}\;\wtilde{K}\;
\exp\left\{-\Bigg(\bea{cc} 0 & \hat{X} \\ \wtilde{\kappa}\;\hat{X}^{+} & 0 \eea\Bigg)\right\}
\;\wtilde{K}\;\hat{I}^{-1} \\ \no &=&
\hat{I}\;\wtilde{K}\;\hat{T}\;\wtilde{K}\;\hat{I}^{-1}\;,
\eeq
so that the super-hermitian conjugation of \(\wtilde{J}_{\psi;\alpha}^{a}(\vec{x},t_{p})\) (\ref{s4_30})
yields \(\wtilde{J}_{\psi;\alpha}^{+a}(\vec{x},t_{p})\) (\ref{s4_31})
\beq \lb{s4_30}
\lefteqn{\bigg(\Big(\hat{I}^{-1}\;\hat{T}^{-1}(\vec{x},t_{p})\;\hat{I}\Big)_{\alpha\beta}^{ab}\;
J_{\psi;\beta}^{b}(\vec{x},t_{p})\bigg)_{\vec{x},\alpha}^{+a} =
\Big(\wtilde{J}_{\psi;\alpha}^{a}(\vec{x},t_{p})\Big)^{+} } \\ \lb{s4_31} &\Longrightarrow&
\bigg(J_{\psi;\beta}^{+b}(\vec{x},t_{p})\;
\Big(\hat{I}\;\wtilde{K}\;\hat{T}(\vec{x},t_{p})\;\wtilde{K}\;\hat{I}^{-1}\Big)_{\beta\alpha}^{ba}
\bigg)_{\vec{x},\alpha}^{a}=
\wtilde{J}_{\psi;\alpha}^{+a}(\vec{x},t_{p})\;\;\;.
\eeq
Eventually, we observe that the coherent state path integral (\ref{s4_27}) contains in the coupling to
the bilinear fields \(\Psi_{\vec{x}\ppr,\beta}^{+b}(t_{q}\ppr)\), \(\Psi_{\vec{x},\alpha}^{a}(t_{p})\)
the term (\ref{s4_32}) with densities \(\delta\hat{\Sigma}_{D;\alpha\beta}^{aa}(\vec{x},t_{p})\) which
is definitely constant because of the negative sign before
\(\delta\hat{\Sigma}_{D;\alpha\beta}^{22}(\vec{x},t_{p})\), thereby canceling the '11' part with
\(\delta\hat{\Sigma}_{D;\alpha\beta}^{11}(\vec{x},t_{p})\). This has been accomplished by the
reverse operations (\ref{s4_20}) from matrix \(\wtilde{\mcal{M}}_{\alpha\beta}^{ab}\) (\ref{s4_15})
to \(\wtilde{N}_{\vec{x},\alpha;\vec{x}\ppr,\beta}^{ab}(t_{p},t_{q}\ppr)\) (\ref{s4_21})
with the metrics $\wtilde{K}$, $\hat{I}^{-1}$, $\hat{K}$ (compare (\ref{s3_89}))
\be \lb{s4_32}
\exp\Bigg\{-\frac{\im}{2\hbar}\int_{C}d t_{p}\sum_{\vec{x}}
\underbrace{\Psi_{\vec{x},\beta}^{+b}(t_{p})\;\hat{K}\;
\left(\bea{cc} \delta\hat{\Sigma}_{D;\beta\alpha}^{11}(\vec{x},t_{p}) & 0 \\
0 & -\delta\hat{\Sigma}_{D;\beta\alpha}^{22}(\vec{x},t_{p})
\eea\right)^{ba}\;\hat{K}\;\Psi_{\vec{x},\alpha}^{a}(t_{p})}_{\equiv 0}\Bigg\} \equiv 1\;.
\ee
Therefore, we acquire the generating function (\ref{s4_33}) where the block diagonal
densities \(\delta\hat{\Sigma}_{D;\alpha\beta}^{aa}(\vec{x},t_{p})\) enter only as
integration variables for the resulting action \(\mcal{A}_{\hat{J}_{\psi\psi}}\big[\hat{T}\big]\)
(\ref{s4_34}). We assume formal performance of these integrations with
\(\delta\hat{\Sigma}_{D;\alpha\beta}^{aa}(\vec{x},t_{p})\) (\ref{s4_34}-\ref{s4_36})
to some unknown action \(\mcal{A}_{\hat{J}_{\psi\psi}}\big[\hat{T}\big]\) (\ref{s4_34})
because these integrations also involve a suitable detailed parametrization of the super-unitary
group \(U(L|S)\), concerning the shift generators in \(\hat{Q}_{\alpha\beta}^{aa}(\vec{x},t_{p})\)
(\ref{s3_102}-\ref{s3_104}). The integrations with the eigenvalues
\(\delta\hat{\lambda}_{\alpha}(\vec{x},t_{p})\), including the polynomial
\(\mcal{P}\big(\delta\hat{\lambda}(\vec{x},t_{p})\big)\) (\ref{s3_119}), can be achieved
using properties of Vandermonde determinants and related Hermite polynomials with
Gaussian weight functions \cite{mehta}
\beq \lb{s4_33}
\lefteqn{\hspace*{-1.0cm}Z[\hat{\mcal{J}},J_{\psi},\im\;\hat{J}_{\psi\psi}]=
\int d[\hat{\sigma}_{D}^{(0)}(\vec{x},t_{p})]\;\;
\exp\bigg\{\frac{\im}{2\hbar}\frac{1}{V_{0}}\int_{C}d t_{p}\sum_{\vec{x}}
\sigma_{D}^{(0)}(\vec{x},t_{p})\;\sigma_{D}^{(0)}(\vec{x},t_{p})\bigg\}  } \\ \no &\times &
\int d\big[\hat{T}^{-1}(\vec{x},t_{p})\;d\hat{T}(\vec{x},t_{p})\big]\;\;
\exp\Big\{\im\;\mcal{A}_{\hat{J}_{\psi\psi}}\big[\hat{T}\big]\Big\}\;\;\times
\\ \no &\times&
\exp\Big\{-\mcal{A}_{SDET}\big[\hat{T},\hat{\sigma}_{D}^{(0)};\hat{\mcal{J}}\big]\Big\}\;\;
\exp\Big\{\im\;\mcal{A}_{J_{\psi}}\big[\hat{T},\hat{\sigma}_{D}^{(0)};\hat{\mcal{J}}\big]
\Big\}
\eeq
\beq \lb{s4_34}
\lefteqn{
\exp\Big\{\im\;\mcal{A}_{\hat{J}_{\psi\psi}}\big[\hat{T}\big]\Big\}  =
\int d\big[\delta\hat{\Sigma}_{D}(\vec{x},t_{p})\;\wtilde{K}\big]\;\;
\mcal{P}\big(\delta\hat{\lambda}(\vec{x},t_{p})\big)\;\;
\exp\Big\{\im\;\mcal{A}_{2}\big[\hat{T},\delta\hat{\Sigma}_{D};\im\hat{J}_{\psi\psi}\big]\Big\} = }
\\ \no &=&
\int d\big[d\hat{Q}(\vec{x},t_{p})\;\hat{Q}^{-1}(\vec{x},t_{p});
\delta\hat{\lambda}(\vec{x},t_{p})\big]\;\;
\mcal{P}\big(\delta\hat{\lambda}(\vec{x},t_{p})\big)\;\;
\exp\Big\{\im\;\mcal{A}_{2}\big[\hat{T},\hat{Q}^{-1}\;\delta\hat{\Lambda}\;\hat{Q};
\im\hat{J}_{\psi\psi}\big]\Big\}
\eeq
\beq \lb{s4_35}
\mbox{sdet}\Big\{\delta\hat{\Sigma}_{D;\alpha\beta}^{11}-\delta\lambda\;\;\delta_{\alpha\beta}\Big\}
&=&0\hspace*{1.5cm}
\mbox{sdet}\Big\{\delta\hat{\Sigma}_{D;\alpha\beta}^{22}\;\wtilde{\kappa}-
\big(-\delta\lambda\big)\;\;\delta_{\alpha\beta}\Big\}=0  \\  \lb{s4_36}
\delta\hat{\Sigma}_{D;N\times N}^{11}(\vec{x},t_{p})&=&-
\Big(\delta\hat{\Sigma}_{D;N\times N}^{22}(\vec{x},t_{p})\;\;\wtilde{\kappa}\Big)^{st}
\eeq
\beq \lb{s4_37}
\lefteqn{\hspace*{-1.8cm}\mcal{A}_{2}\big[\hat{T},\delta\hat{\Sigma}_{D};\im\hat{J}_{\psi\psi}\big] =
\frac{1}{4\hbar\;V_{0}}\int_{C}d t_{p}\sum_{\vec{x}} \bigg\{
\mbox{STR}\Big[\delta\hat{\Sigma}_{D;2N\times 2N}(\vec{x},t_{p})\;
\wtilde{K}\;\delta\hat{\Sigma}_{D;2N\times 2N}(\vec{x},t_{p})\;
\wtilde{K}\Big]+  } \\ \no &-&
2\;\mbox{STR}\Big[\im\;\hat{J}_{\psi\psi}(\vec{x},t_{p})\;\wtilde{K}\;
\hat{T}(\vec{x},t_{p})\;\delta\hat{\Sigma}_{D;2N\times 2N}(\vec{x},t_{p})\;\wtilde{K}\;
\hat{T}^{-1}(\vec{x},t_{p}) \Big] +
\\ \no &+&\mbox{STR}\Big[\im\;\hat{J}_{\psi\psi}(\vec{x},t_{p})\;\wtilde{K}\;\im\;
\hat{J}_{\psi\psi}(\vec{x},t_{p})\;\wtilde{K}\Big]\bigg\}
\eeq
\beq \lb{s4_38}
\lefteqn{\mcal{A}_{2}\big[\hat{T},\hat{Q}^{-1}\;\delta\hat{\Lambda}\;\hat{Q};
\im\hat{J}_{\psi\psi}\big]  = }  \\ \no &=&
\frac{1}{4\hbar\;V_{0}}\int_{C}d t_{p}\sum_{\vec{x}}
\mbox{STR}\Big[\big(\delta\wtilde{\Sigma}(\vec{x},t_{p})-\im\;\hat{J}_{\psi\psi}(\vec{x},t_{p})\big)\;
\wtilde{K}\;\big(\delta\wtilde{\Sigma}(\vec{x},t_{p})-\im\;\hat{J}_{\psi\psi}(\vec{x},t_{p})\big)\;
\wtilde{K}\Big]
\\ \no  &=&\frac{1}{4\hbar\;V_{0}}\int_{C}d t_{p}\sum_{\vec{x}}\bigg\{
2\;\mbox{str}\Big[\big(\delta\hat{\lambda}_{N\times N}(\vec{x},t_{p})\big)^{2}\Big]+ \\ \no &-&
2\;\mbox{STR}\Big[\im\;\hat{J}_{\psi\psi}(\vec{x},t_{p})\;\wtilde{K}\;
\hat{T}(\vec{x},t_{p})\;\hat{Q}^{-1}(\vec{x},t_{p})\;
\delta\hat{\Lambda}(\vec{x},t_{p})\;
\hat{Q}(\vec{x},t_{p})\;\hat{T}^{-1}(\vec{x},t_{p})
\Big] + \\ \no &+&\mbox{STR}\Big[\im\;\hat{J}_{\psi\psi}(\vec{x},t_{p})\;\wtilde{K}\;\im\;
\hat{J}_{\psi\psi}(\vec{x},t_{p})\;\wtilde{K}\Big]\bigg\}\;.
\eeq
The actions \(\mcal{A}_{SDET}\big[\hat{T},\hat{\sigma}_{D}^{(0)};\hat{\mcal{J}}\big]\),
\(\mcal{A}_{J_{\psi}}\big[\hat{T},\hat{\sigma}_{D}^{(0)};\hat{\mcal{J}}\big]\)
(\ref{s4_39},\ref{s4_40}) of the super-determinant and bilinear source fields in (\ref{s4_33})
are determined by the super-matrix
\(\hat{\mcal{O}}_{\vec{x},\alpha;\vec{x}\ppr,\beta}^{ab}(t_{p},t_{q}\ppr)\) (\ref{s4_41},\ref{s4_22})
with diagonal part \(\hat{\mcal{H}}+\hat{\sigma}_{D}^{(0)}\;\hat{1}_{2N\times 2N}\), the source matrix
\(\wtilde{\mcal{J}}(\hat{T}^{-1},\hat{T})\)
and the gradient operator \(\delta\hat{\mcal{H}}(\hat{T}^{-1},\hat{T})\)
\beq \lb{s4_39}
\mcal{A}_{SDET}\big[\hat{T},\hat{\sigma}_{D}^{(0)};\hat{\mcal{J}}\big]
&=&\frac{1}{2}\int_{C}\frac{d t_{p}}{\hbar}\eta_{p}\sum_{\vec{x}}\mcal{N}\;
\mbox{STR}\Big[\ln\Big(\hat{\mcal{O}}_{\vec{x},\alpha;\vec{x}\ppr,\beta}^{ab}(t_{p},t_{q}\ppr)
\Big)\Big]
\eeq
\beq \lb{s4_40}
\lefteqn{\mcal{A}_{J_{\psi}}\big[\hat{T},\hat{\sigma}_{D}^{(0)};\hat{\mcal{J}}\big]=
\frac{\Omega^{2}}{2\hbar}\int_{C}d t_{p}\;d t_{q}\ppr\sum_{\vec{x},\vec{x}\ppr}\mcal{N}_{x} \times}
\\ \no &\times &
J_{\psi;\beta}^{+b}(\vec{x}\ppr,t_{q}\ppr)\;\hat{I}\;\wtilde{K}\;
\bigg(\hat{T}(\vec{x}\ppr,t_{q}\ppr)\;\;
\hat{\mcal{O}}_{\vec{x}\ppr,\beta\ppr;\vec{x},\alpha\ppr}^{-1;b\ppr a\ppr}(t_{q}\ppr,t_{p})\;\;
\hat{T}^{-1}(\vec{x},t_{p})\bigg)_{\vec{x}\ppr,\beta;\vec{x},\alpha}^{ba}\;\hat{I}\;
J_{\psi;\alpha}^{a}(\vec{x},t_{p})
\eeq
\beq\lb{s4_41}
\hat{\mcal{O}}_{\vec{x},\alpha;\vec{x}\ppr,\beta}^{ab}(t_{p},t_{q}\ppr)&=&
\bigg[\Big(\hat{\mcal{H}}+\hat{\sigma}_{D}^{(0)}\;\hat{1}_{2N\times 2N}\Big)+
\Big(\hat{T}^{-1}(\vec{x},t_{p})\;\hat{\mcal{H}}\;\hat{T}(\vec{x}\ppr,t_{q}\ppr)-\hat{\mcal{H}}\Big) +
\\  \no     &+& \underbrace{
\hat{T}^{-1}(\vec{x},t_{p})\;\hat{I}\;\hat{K}\;\eta_{p}\;
\frac{\hat{\mcal{J}}_{\alpha\ppr\beta\ppr}^{a\ppr b\ppr}}{\mcal{N}_{x}}\;\eta_{q}\;\hat{K}\;\hat{I}\;
\wtilde{K}\;\hat{T}(\vec{x}\ppr,t_{q}\ppr)}_{\wtilde{\mcal{J}}(\hat{T}^{-1},\hat{T})}
\bigg]_{\vec{x},\alpha;\vec{x}\ppr,\beta}^{ab}\hspace*{-0.64cm}(t_{p},t_{q}\ppr)\;\;\;\;\;\;.
\eeq
In order to perform the gradient expansion with \(\delta\hat{\mcal{H}}(\hat{T}^{-1},\hat{T})\),
we have to consider the appropriate Hilbert space of the gradient operators
\(\pp_{\mu}\), \(\im\hbar\;\pp/\pp t_{p}\)
in combination with the Hilbert space of the self-energy following from the dyadic product of the
doubled super-fields \(\Psi_{\vec{x},\alpha}^{a}(t_{p})\).

\subsection{Hilbert space of gradient operators and their representations} \lb{s42}

We have to determine how the gradient operators  {\boldmath\(\hat{\pp}_{i}\)} and
{\boldmath\(\im\hbar\hat{\pp}_{t_{p}}\)} act on the coset matrices $\hat{T}$ and on the
self-energy $\hat{\sigma}_{D}^{(0)}$ in the operator $\hat{\mcal{O}}$ (\ref{s4_41}) with
\(\delta\hat{\mcal{H}}(\hat{T}^{-1},\hat{T})\). One can consider some representation as the
time-coordinate space where the operators $\hat{T}$, $\hat{Q}$, $\delta\hat{\lambda}$,
$\hat{\sigma}_{D}^{(0)}$ are defined to be diagonal in the time-like and spatial variables
and are given by the super-matrix fields \(\hat{T}_{\alpha\beta}^{ab}(\vec{x},t_{p})\),
\(\hat{Q}_{\alpha\beta}^{aa}(\vec{x},t_{p})\) and the scalar fields
\(\delta\hat{\lambda}_{\alpha}(\vec{x},t_{p})\), \(\sigma_{D}^{(0)}(\vec{x},t_{p})\).
However, it has to be taken into account
that the inverse square root of the super-determinant follows
from integration over the bilinear fields which are doubled by their complex conjugates
\(\psi_{\vec{x},\alpha}^{*}(t_{p})\). Consequently a Hilbert space for
\(\psi_{\vec{x},\alpha}(t_{p})\) with 'ket' \(|\psi_{\alpha}\rangle\) has also to be doubled
by its 'dual' space \(\ovv{|\psi_{\alpha}\rangle}=\langle\psi_{\alpha}|\) the 'bra'.
The unsaturated operators {\boldmath\(\hat{\pp}_{i}\)}, {\boldmath\(\im\hbar\hat{\pp}_{t_{p}}\)}
are printed in boldmath in order to distinguish from the matrix functions
\(\big(\pp_{\mu}\hat{T}\big)\), \(\big(\im\hbar\;\pp_{t_{p}}\hat{T}\big)\) embraced in brackets,
denoting the limited action of the derivatives on the prevailing coset matrix \(\hat{T}(\vec{x},t_{p})\).
These 'saturated' gradient operators are not involved in further
derivative actions on matrices or fields outside the braces and are therefore not printed in bold letters.

The detailed structure of the Hilbert space with its doubled dual part
\(\langle\psi_{\alpha}|=\ovv{|\psi_{\alpha}\rangle}\) is important because the doubled operator
\(\hat{\mcal{H}}\) (\ref{s2_55}-\ref{s2_57}) applies the transpose
\(\hat{H}_{p}^{T}(\vec{x},t_{p})\) in the '22' block
instead of \(\hat{H}_{p}(\vec{x},t_{p})\) as in the '11' part. An operator in quantum mechanics
is defined by the mapping and the space on which it acts. Completely different results can follow
if one considers for one and the same mapping of an operator different spaces where the operator
transforms the prevailing states.
The coherent state path integral (\ref{s4_33}) follows by integration over
the doubled super-fields \(\Psi_{\vec{x},\alpha}^{a}(t_{p})\) from (\ref{s4_27}).
The corresponding doubled abstract states \(\widehat{|\psi_{\alpha}\rangle}^{a(=1/2)}\)
with angular momentum label \(\alpha,\;\beta,\;\gamma,\ldots\) are defined in (\ref{s4_42})
\be\lb{s4_42}
\Psi_{\vec{x},\alpha}^{a(=1/2)}(t_{p})=\left(
\bea{c}
\psi_{\vec{x},\alpha}(t_{p}) \\
\psi_{\vec{x},\alpha}^{*}(t_{p})
\eea\right)\propto \widehat{|\Psi_{\alpha}\rangle}^{a(=1/2)}=
\left(\bea{c}
|\psi_{\alpha}\rangle^{a=1} \vspace*{0.15cm} \\
\ovv{|\psi_{\alpha}\rangle}^{a=2}
\eea\right)_{\mbox{.}}
\ee
The appropriate abstract Hilbert space has to be introduced for the definition of the operators
{\boldmath\(\hat{\pp}_{i}\)}, {\boldmath\(\im\hbar\hat{\pp}_{t_{p}}\)} in the
super-determinant \(\mcal{A}_{SDET}\) and \(\mcal{A}_{J_{\psi}}\) (\ref{s4_39}-\ref{s4_41}).
According to the doubling with the dual part \(\ovv{|\psi_{\alpha}\rangle}=\langle\psi_{\alpha}|\),
we have an antilinear property in the second part \(\widehat{|\psi_{\alpha}\rangle}^{a=2}\)
\be \lb{s4_43}
\widehat{|c\;\Psi_{\alpha}\rangle}=\left(
\bea{c}
c\;|\psi_{\alpha}\rangle \vspace*{0.15cm} \\
c^{*}\;\ovv{|\psi_{\alpha}\rangle}
\eea\right)\hspace*{1.0cm}c\in\mbox{\sf C}  \;\;\;.
\ee
Furthermore, we simultaneously have the super-unitary and 'anti'-super-unitary representation
of \(U(L|S)\) in the '11' and '22' block, respectively. This is in accordance with a theorem of Wigner
that a symmetry in quantum mechanics can have unitary or anti-unitary realizations \cite{wei1}.
The corresponding Hilbert space for spatial and time variables has therefore also to be
doubled with the antilinear part
\be \lb{s4_44}
\widehat{|\vec{x},t_{p}\rangle}^{a(=1/2)}=
\left(\bea{c}
|\vec{x},t_{p}\rangle^{a=1} \vspace*{0.15cm} \\
\ovv{|\vec{x},t_{p}\rangle}^{a=2}
\eea\right)
\ee
\be \lb{s4_45}
\bea{rclrcl}
\langle t_{p}|t_{q}\ppr\rangle&=&\delta_{p,q}\;\delta_{t_{p},t_{q}\ppr} &
\hspace*{1.25cm} \langle\vec{x}|\vec{x}\ppr\rangle&=&\delta_{\vec{x},\vec{x}\ppr}  \\
 \langle\omega_{p}|\omega_{q}\ppr\rangle&=&
\delta_{p,q}\;\delta_{\omega_{p},\omega_{q}\ppr} &
\langle\vec{k}|\vec{k}\ppr\rangle&=&\delta_{\vec{k},\vec{k}\ppr} \\
\langle t_{p}|\omega_{q}\rangle&=&\delta_{p,q}\;\exp\{-\im\;\omega_{p}\;t_{p}\} &
\langle\vec{x}|\vec{k}\rangle&=&\exp\{\im\;\vec{k}\cdot\vec{x}\}
\eea
\ee
\beq \lb{s4_46} \sum_{a=1,2}
\widehat{\langle\vec{x},t_{p}}^{a}|\widehat{\vec{k},\omega_{q}\rangle}^{a}&=&
\langle\vec{x},t_{p}|\vec{k},\omega_{q}\rangle +
\ovv{\langle\vec{x},t_{p}}|\ovv{\vec{k},\omega_{q}\rangle} \\ \lb{s4_47}
\langle\vec{x},t_{p}|\vec{k},\omega_{q}\rangle &=&\delta_{p,q}\;
\exp\{\im(\vec{k}\cdot\vec{x}-\omega_{p}\cdot t_{p})\} \\ \lb{s4_48}
\ovv{\langle\vec{x},t_{p}}|\ovv{\vec{k},\omega_{q}\rangle}&=&
\langle\vec{k},\omega_{q}|\vec{x},t_{p}\rangle=
\big(\langle\vec{x},t_{p}|\vec{k},\omega_{q}\rangle\big)^{*} = \delta_{p,q}\;
\exp\{-\im(\vec{k}\cdot\vec{x}-\omega_{p}\cdot t_{p})\}\;\;\;.
\eeq
The total unit operators with the unitary and anti-unitary parts are listed in Eqs. (\ref{s4_49},\ref{s4_50})
for time and energy states with spatial coordinates. We have to combine the contour integrals of
forward and backward propagation with the contour metric \(\eta_{p=\pm}=\pm\) (\ref{s2_23}) so that
the defining relations (\ref{s4_44}-\ref{s4_48}) exactly match with the properties of the unit operators
(\ref{s4_49},\ref{s4_50})
\beq \lb{s4_49}
\hat{1}&=&\left(
\bea{cc}
\hat{1}^{11} & \\
 & \hat{1}^{22}
 \eea\right)=\int_{C}\frac{d t_{p}}{\hbar}\eta_{p}\sum_{\vec{x}}\mcal{N}
 \left(\bea{cc}
|\vec{x},t_{p}\rangle\langle\vec{x},t_{p}| & \\
 & \ovv{|\vec{x},t_{p}\rangle}\;\ovv{\langle\vec{x},t_{p}|}
 \eea\right) = \\ \no &=&\int_{C}\frac{d t_{p}}{\hbar}\eta_{p}\sum_{\vec{x}}
 \sum_{a=1,2}\mcal{N}\;\;
 \widehat{|\vec{x},t_{p}\rangle}^{a(=1/2)}\;\widehat{\langle\vec{x},t_{p}|}^{a(=1/2)}\;; \hspace*{1.0cm}
\bigg( \mcal{N}=\frac{\hbar}{\Delta t}\;\bigg(\frac{L}{\Delta x}\bigg)^{d} \bigg) \\ \lb{s4_50}
\hat{1}&=&\left(
\bea{cc}
\hat{1}^{11} & \\
& \hat{1}^{22}
\eea\right)=\int_{C}\hbar d\omega_{p}\eta_{p}\sum_{\vec{x}}\mcal{N}\ppr
\left(\bea{cc}
|\vec{x},\omega_{p}\rangle\langle\vec{x},\omega_{p}| & \\
& \ovv{|\vec{x},\omega_{p}\rangle}\;\ovv{\langle\vec{x},\omega_{p}|}
\eea\right) = \\ \no &=&\int_{C}\hbar d\omega_{p}\eta_{p}\sum_{\vec{x}}\sum_{a=1,2}\mcal{N}\ppr\;\;
\widehat{|\vec{x},\omega_{p}\rangle}^{a(=1/2)}\;\widehat{\langle\vec{x},\omega_{p}|}^{a(=1/2)}
\;;\hspace*{1.0cm}\bigg(\mcal{N}\ppr=\frac{T_{\infty}}{2\pi\;\hbar}\;
\bigg(\frac{L}{\Delta x}\bigg)^{d}\bigg)_{\mbox{.}}
\eeq
The space-time and space-energy representation of the abstract doubled 'super'-Hilbert states (\ref{s4_43})
are given in (\ref{s4_51},\ref{s4_52}) where the relations (\ref{s4_44}-\ref{s4_48}) are applied,
including the anti-unitary second part
\beq \lb{s4_51}
\widehat{\langle\vec{x},t_{p}}^{a}|\widehat{\Psi_{\alpha}\rangle}&=&
\left(\bea{c}
\langle\vec{x},t_{p}|\psi_{\alpha}\rangle \vspace*{0.15cm}\\
\ovv{\langle\vec{x},t_{p}}|\ovv{\psi_{\alpha}\rangle}
\eea\right)^{a(=1/2)}=
\left(\bea{c}
\psi_{\vec{x},\alpha}(t_{p}) \\
\langle\psi_{\alpha}|\vec{x},t_{p}\rangle
\eea\right)^{a(=1/2)}=
\left(\bea{c}
\psi_{\vec{x},\alpha}(t_{p}) \\
\psi_{\vec{x},\alpha}^{*}(t_{p})
\eea\right)^{a(=1/2)}  \\ \lb{s4_52}
\widehat{\langle\vec{x},\omega_{p}}^{a}|\widehat{\Psi_{\alpha}\rangle}&=&
\left(\bea{c}
\langle\vec{x},\omega_{p}|\psi_{\alpha}\rangle \vspace*{0.15cm}\\
\ovv{\langle\vec{x},\omega_{p}}|\ovv{\psi_{\alpha}\rangle}
\eea\right)^{a(=1/2)}=
\left(\bea{c}
\psi_{\vec{x},\alpha}(\omega_{p}) \\
\langle\psi_{\alpha}|\vec{x},\omega_{p}\rangle
\eea\right)^{a(=1/2)}=
\left(\bea{c}
\psi_{\vec{x},\alpha}(\omega_{p}) \\
\psi_{\vec{x},\alpha}^{*}(\omega_{p})
\eea\right)^{a(=1/2)}\;.
\eeq
The scalar density $\hat{\sigma}_{D}^{(0)}$ operates on the doubled space-time state
\(|\widehat{\vec{x}\ppr,t_{q}\ppr\rangle}^{a}\) as in the well-known case of an
annihilation operator on coherent states, but one has to incorporate the contour metric
$\eta_{q}$ and has to consider that the resulting coherent state field
\(\sigma_{D}^{(0)}(\vec{x}\ppr,t_{q}\ppr)\) only takes real values (\ref{s4_53}).
This additional contour metric \(\eta_{p}\) has to be taken into account
for the one-particle operator $\hat{\mcal{H}}$, the self-energy density \(\hat{\sigma}_{D}^{(0)}\)
and \(\delta\hat{\Sigma}_{\alpha\beta}^{ab}\) because these operators appear in the
original coherent state path integrals and only lead to diagonal matrix elements
in the time contour due to a missing disorder. An ensemble average with a random potential
would include non-diagonal terms in the total self-energy concerning the time contour \cite{bmdis}.
However, the abstract operator action with the anomalous parts \(\hat{Y}_{\alpha\beta}^{ab}\)
in \(\hat{T}_{\alpha\beta}^{ab}\) does not involve an additional contour metric in the
considered case without disorder
\beq \lb{s4_53}
\hat{\sigma}_{D}^{(0)}\;\widehat{|\vec{x}\ppr,t_{q}\ppr\rangle}^{a}&=& \eta_{q}\;
\sigma_{D}^{(0)}(\vec{x}\ppr,t_{q}\ppr)\;\widehat{|\vec{x}\ppr,t_{q}\ppr\rangle}^{a}
\hspace*{1.0cm}\sigma_{D}^{(0)}(\vec{x}\ppr,t_{q}\ppr)\in\mbox{\sf R} \\ \lb{s4_54}
\hat{T}_{\alpha\beta}^{ab}\;\widehat{|\vec{x}\ppr,t_{q}\ppr\rangle}^{b}&=&
\hat{T}_{\alpha\beta}^{ab}(\vec{x}\ppr,t_{q}\ppr)\;\widehat{|\vec{x}\ppr,t_{q}\ppr\rangle}^{b}\;\;\;.
\eeq
Using the definitions (\ref{s4_44}-\ref{s4_48}) of the abstract doubled Hilbert space, we can
pursue the various steps for calculating matrix elements
\(\widehat{\langle\vec{x},t_{p}}^{a}|\hat{T}_{\alpha\beta}^{ab}|\widehat{\vec{x}\ppr,t_{q}\ppr\rangle}^{b}\)
from the generating operators \(\hat{Y}_{\alpha\beta}^{ab}\), \(\hat{X}_{\alpha\beta}\),
\(\hat{X}_{\alpha\beta}^{+}\) in the exponential of \(\hat{T}\). However, the operator
\(\hat{X}_{\alpha\beta}\) of the anomalous parts is constructed from two field operators
\(\hat{\psi}_{\alpha}\;\hat{\psi}_{\beta}\) so that matrix elements
\(\langle\vec{x},t_{p}|\hat{X}_{\alpha\beta}|\ovv{\vec{x}\ppr,t_{q}\ppr\rangle}\)
(\ref{s4_56},\ref{s4_57}) result in the expansion of \(\hat{T}_{\alpha\beta}^{ab}\)
(\ref{s4_55}), combining also Hilbert states with linear and antilinear parts
\beq \lb{s4_55}
\lefteqn{\widehat{\langle\vec{x},t_{p}}^{a}|\hat{T}_{\alpha\beta}^{ab}|
\widehat{\vec{x}\ppr,t_{q}\ppr\rangle}^{b}=
\widehat{\langle\vec{x},t_{p}}^{a}|\big(\hat{1}-\hat{Y}_{\alpha\beta}^{ab}\pm\ldots\big)
|\widehat{\vec{x}\ppr,t_{q}\ppr\rangle}^{b}=} \\ \no &=&
\delta_{a,b}\;\delta_{\alpha,\beta}\;\delta_{\vec{x},\vec{x}\ppr}\;\delta_{p,q}\;
\delta_{t_{p},t_{q}\ppr}-
\left(\bea{c}
\langle\vec{x},t_{p}| \vspace*{0.15cm} \\ \ovv{\langle\vec{x},t_{p}|} \eea\right)^{T}
\left(\bea{cc}
0 & \hat{X}_{\alpha\beta} \\
\wtilde{\kappa}\;\hat{X}^{+}_{\alpha\beta} & 0
\eea\right)\left(
\bea{c}
|\vec{x}\ppr,t_{q}\ppr\rangle \vspace*{0.15cm} \\
\ovv{|\vec{x}\ppr,t_{q}\ppr\rangle} \eea\right)\pm\ldots \\ \no &=&
\delta_{\vec{x},\vec{x}\ppr}\;\delta_{p,q}\;\delta_{t_{p},t_{q}\ppr}\;
\left[\hat{1}\;\delta_{a,b}\;\delta_{\alpha,\beta}-\left(
\bea{cc}
0 & \hat{X}_{\alpha\beta}(\vec{x},t_{p}) \\
\wtilde{\kappa}\;\hat{X}_{\alpha\beta}^{+}(\vec{x},t_{p}) & 0
\eea\right)^{ab}\pm\ldots\right]
\eeq
\beq \lb{s4_56}
\langle\vec{x},t_{p}|\hat{X}_{\alpha\beta}|\ovv{\vec{x}\ppr,t_{q}\ppr\rangle}&=&
\hat{X}_{\alpha\beta}(\vec{x},t_{p})\;\delta_{\vec{x},\vec{x}\ppr}\;\delta_{p,q}\;
\delta_{t_{p},t_{q}\ppr} \\ \lb{s4_57}
\ovv{\langle\vec{x},t_{p}}|\hat{X}_{\alpha\beta}^{+}|\vec{x}\ppr,t_{q}\ppr\rangle&=&
\hat{X}_{\alpha\beta}^{+}(\vec{x},t_{p})\;\delta_{\vec{x},\vec{x}\ppr}\;\delta_{p,q}\;
\delta_{t_{p},t_{q}\ppr} \\ \lb{s4_58}
\Big(\langle\vec{x},t_{p}|\hat{X}_{\alpha\beta}|\ovv{\vec{x}\ppr,t_{q}\ppr\rangle}\Big)^{*}&=&
\ovv{\langle\vec{x}\ppr,t_{q}\ppr}|\hat{X}_{\alpha\beta}^{+}|\vec{x},t_{p}\rangle=
\ovv{\langle\vec{x},t_{p}}|\hat{X}_{\alpha\beta}^{+}|\vec{x}\ppr,t_{q}\ppr\rangle \\ \lb{s4_59}
\hat{X}_{\alpha\beta}(\vec{x},t_{p})&\propto & \big(\psi_{\vec{x},\alpha}(t_{p})\;
\psi_{\vec{x},\beta}(t_{p})\big)\;\wtilde{\kappa} \\ \lb{s4_60}
\hat{X}_{\alpha\beta}^{+}(\vec{x},t_{p})&\propto &\wtilde{\kappa}\;
\big(\psi_{\vec{x},\alpha}^{*}(t_{p})\;\psi_{\vec{x},\beta}^{*}(t_{p})\big) \;\;\;.
\eeq
Summarizing the effect of the doubling of Hilbert states with the anti-unitary extension,
we list the matrix elements of the density parts (\ref{s4_61},\ref{s4_62}), always
containing the contour metric \(\eta_{p}\), and the pair condensates
(\ref{s4_63}), as derived with the properties (\ref{s4_56}-\ref{s4_60}) in the
expansion (\ref{s4_55})
\beq \lb{s4_61}
\widehat{\langle\vec{x},t_{p}}^{a}|\hat{\mcal{H}}|\widehat{\vec{x}\ppr,t_{q}\ppr\rangle}^{b}&=&\!\!
\delta_{a,b}\;\delta_{\alpha,\beta}\;\delta_{\vec{x},\vec{x}\ppr}\;
\eta_{p}\;\delta_{p,q}\;\delta_{t_{p},t_{q}\ppr} \!\left(
\bea{cc}
\hat{H}_{p}(\vec{x},t_{p}) & 0 \\ 0 & \hat{H}_{p}^{T}(\vec{x},t_{p})
\eea\right)^{ab} \\ \lb{s4_62}
\widehat{\langle\vec{x},t_{p}}^{a}|\big(\hat{u}+\hat{\sigma}_{D}^{(0)}\big)\;\hat{1}_{2N\times 2N}+
\delta\hat{\Sigma}_{\alpha\beta}^{ab}|\widehat{\vec{x}\ppr,t_{q}\ppr\rangle}^{b}&=&
\delta_{\vec{x},\vec{x}\ppr}\;
\eta_{p}\;\delta_{p,q}\;\delta_{t_{p},t_{q}\ppr}\;\times \\ \no &\times&
\Big(\big(u(\vec{x})+\sigma_{D}^{(0)}(\vec{x},t_{p})\big)\;\delta_{a,b}\;\delta_{\alpha,\beta} +
\delta\hat{\Sigma}_{\alpha\beta}^{ab}(\vec{x},t_{p})\Big) \\ \lb{s4_63}
\widehat{\langle\vec{x},t_{p}}^{a}|\hat{T}_{\alpha\beta}^{ab}
|\widehat{\vec{x}\ppr,t_{q}\ppr\rangle}^{b}&=&
\hat{T}_{\alpha\beta}^{ab}(\vec{x},t_{p})\;\;\delta_{\vec{x},\vec{x}\ppr}\;
\delta_{p,q}\;\delta_{t_{p},t_{q}\ppr}\;.
\eeq
The source field $J_{\psi;\alpha}^{a}(\vec{x},t_{p})$ for the BEC wave function and
the source matrix \(\wtilde{\mcal{J}}_{\alpha\beta}^{ab}\) for generating observables
are defined in (\ref{s4_64},\ref{s4_65}) for corresponding doubled super-states
and super-matrices
\beq \lb{s4_64}
J_{\psi;\alpha}^{a}(\vec{x},t_{p})&=&\widehat{\langle\vec{x},t_{p}}^{a}|
\widehat{J_{\psi;\alpha}^{a}\rangle}=
\left(\bea{c}
j_{\psi;\alpha}(\vec{x},t_{p}) \\
j_{\psi;\alpha}^{*}(\vec{x},t_{p})
\eea\right)^{a}   \\  \lb{s4_65}
\wtilde{\mcal{J}}_{\vec{x},\alpha;\vec{x}\ppr,\beta}^{ab}(t_{p},t_{q}\ppr)&=&
\widehat{\langle\vec{x},t_{p}}^{a}|\wtilde{\mcal{J}}_{\alpha\beta}^{ab}|
\widehat{\vec{x}\ppr,t_{q}\ppr\rangle}^{b} =
\hat{I}\;\hat{K}\;\eta_{p}\;
\frac{\hat{\mcal{J}}_{\vec{x},\alpha;\vec{x}\ppr,\beta}^{ab}(t_{p},t_{q}\ppr)}{\mcal{N}_{x}}\;
\eta_{q}\;\hat{K}\;\hat{I}\;\wtilde{K}\;\;\;\;\;\;.
\eeq
The definition of the unit operators (\ref{s4_49},\ref{s4_50}) with contour integration {\it and}
additional contour metric can be transformed to a trace relation as one can change the
unit operator \(\hat{1}=\sum_{n}|n\rangle\langle n|\) of a complete set of states \(|n\rangle\)
in ordinary quantum mechanics to a trace relation \(\mbox{tr}[\ldots]=\sum_{n}\langle n|\ldots|n\rangle\).
However, we have to distinguish between the 'Nambu'-doubled trace \(\mbox{Tr}[\ldots]\) (\ref{s4_66})
with the anti-unitary second part \(\ovv{\langle\vec{x},t_{p}}|\ldots|\ovv{\vec{x},t_{p}\rangle}\)
and the ordinary trace \(\mbox{tr}[\ldots]\) also with contour integration, but without
the 'Nambu'-doubled anti-unitary part
\beq \lb{s4_66}
\mbox{Tr}\Big[\ldots\Big]&:=&\int_{C}\frac{d t_{p}}{\hbar}\eta_{p}\sum_{\vec{x}}\mcal{N}\;
\sum_{a=1,2}\widehat{\langle\vec{x},t_{p}}^{a}|\ldots|\widehat{\vec{x},t_{p}\rangle}^{a} \\ \no &=&
\int_{-\infty}^{\infty}\frac{d t_{+}}{\hbar}\sum_{\vec{x}}\mcal{N}\;
\sum_{a=1,2}\widehat{\langle\vec{x},t_{+}}^{a}|\ldots|\widehat{\vec{x},t_{+}\rangle}^{a} +
\int_{-\infty}^{\infty}\frac{d t_{-}}{\hbar}\sum_{\vec{x}}\mcal{N}\;
\sum_{a=1,2}\widehat{\langle\vec{x},t_{-}}^{a}|\ldots|\widehat{\vec{x},t_{-}\rangle}^{a} \\ \lb{s4_67}
\mbox{tr}\Big[\ldots\Big]&:=&\int_{C}\frac{d t_{p}}{\hbar}\eta_{p}\sum_{\vec{x}}\mcal{N}\;
\langle\vec{x},t_{p}|\ldots|\vec{x},t_{p}\rangle \\ \no &=&
\int_{-\infty}^{\infty}\frac{d t_{+}}{\hbar}\sum_{\vec{x}}\mcal{N}\;
\langle\vec{x},t_{+}|\ldots|\vec{x},t_{+}\rangle +
\int_{-\infty}^{\infty}\frac{d t_{-}}{\hbar}\sum_{\vec{x}}\mcal{N}\;
\langle\vec{x},t_{-}|\ldots|\vec{x},t_{-}\rangle\;\;\;.
\eeq
The actions \(\mcal{A}_{SDET}\), \(\mcal{A}_{J_{\psi}}\) (\ref{s4_68}-\ref{s4_70})
of the super-determinant and the bilinear source fields consist
of the operator \(\hat{\mcal{O}}\) (\ref{s4_41}) with
one-particle part \(\hat{\mcal{H}}\), the scalar, real self-energy density \(\hat{\sigma}_{D}^{(0)}\)
and the part \(\Delta\hat{\mcal{O}}\) for the gradient expansion
\(\delta\hat{\mcal{H}}(\hat{T}^{-1},\hat{T})\) and the source term
\(\wtilde{\mcal{J}}(\hat{T}^{-1},\hat{T})\) (\ref{s4_71}-\ref{s4_74}).
Note that the actions \(\mcal{A}_{SDET}\),
\(\mcal{A}_{J_{\psi}}\) in (\ref{s4_69},\ref{s4_70}) are given in terms of traces of the
doubled abstract Hilbert space and the corresponding doubled scalar Hilbert product
\beq \lb{s4_68}
Z[\hat{\mcal{J}},J_{\psi},\im\;\hat{J}_{\psi\psi}]&=&
\int d[\hat{\sigma}_{D}^{(0)}(\vec{x},t_{p})]\;\;
\exp\bigg\{\frac{\im}{2\hbar}\frac{1}{V_{0}}\int_{C}d t_{p}\sum_{\vec{x}}
\sigma_{D}^{(0)}(\vec{x},t_{p})\;\sigma_{D}^{(0)}(\vec{x},t_{p})\bigg\}    \\ \no &\times &
\int d[\hat{T}^{-1}(\vec{x},t_{p})\;d\hat{T}(\vec{x},t_{p})]\;\;
\exp\Big\{\im\;\mcal{A}_{\hat{J}_{\psi\psi}}\big[\hat{T}\big]\Big\}
\\ \no &\times &
\exp\Big\{-\mcal{A}_{SDET}\big[\hat{T},\hat{\sigma}_{D}^{(0)};\hat{\mcal{J}}\big]\Big\}  \;\;
\exp\Big\{\im\;\mcal{A}_{J_{\psi}}\big[\hat{T},\hat{\sigma}_{D}^{(0)};\hat{\mcal{J}}\big]\Big\}
\eeq
\beq \lb{s4_69}
\mcal{A}_{SDET}\big[\hat{T},\hat{\sigma}_{D}^{(0)};\hat{\mcal{J}}\big]&=&\frac{1}{2}
\mbox{Tr}\bigg\{\mbox{STR}\Big[\ln\Big(\hat{\mcal{O}}/\mcal{N}\Big)\Big]\bigg\} \\ \lb{s4_70}
\mcal{A}_{J_{\psi}}\big[\hat{T},\hat{\sigma}_{D}^{(0)};\hat{\mcal{J}}\big]
&=&\frac{1}{2}\sum_{a,b=1,2}\sum_{\alpha,\beta=1}^{N=L+S}
\frac{1}{\mcal{N}^{2}}\;
\widehat{\langle J_{\psi;\beta}^{b}}|\hat{\eta}\;\hat{I}\;\wtilde{K}\;\hat{T}\;
\Big(\hat{\mcal{O}}/\mcal{N}\Big)_{\beta\alpha}^{-1;ba}\;\hat{T}^{-1}\;\hat{I}\;\hat{\eta}|
\widehat{J_{\psi;\alpha}^{a}\rangle} \\ \lb{s4_71}
\hat{\mcal{O}}/\mcal{N}&=&
\Big(\hat{\mcal{H}}+\hat{\sigma}_{D}^{(0)}\;
\hat{1}_{2N\times 2N}+\Delta\hat{\mcal{O}}\Big)\Big/\mcal{N} \\ \lb{s4_72}
\Delta\hat{\mcal{O}} &=& \delta\hat{\mcal{H}}(\hat{T}^{-1},\hat{T})+
\wtilde{\mcal{J}}(\hat{T}^{-1},\hat{T})  \\ \lb{s4_73}
\delta\hat{\mcal{H}}(\hat{T}^{-1},\hat{T}) &=& \hat{T}^{-1}\;\;\hat{\mcal{H}}\;\;\hat{T}-\hat{\mcal{H}}
\\  \lb{s4_74} \wtilde{\mcal{J}}(\hat{T}^{-1},\hat{T}) &=&
\hat{T}^{-1}\;\hat{I}\;\hat{K}\;\hat{\eta}\;
\frac{\hat{\mcal{J}}}{\mcal{N}_{x}}\;\hat{\eta}\;
\hat{K}\;\hat{I}\;\wtilde{K}\;\hat{T} \;\;\;.
\eeq
One has to remember the distinction between \(\mbox{{\boldmath$\hat{E}_{p}$}}\hat{T}\) with
unsaturated derivative operator \(\mbox{{\boldmath$\hat{E}_{p}$}}\) which acts further to the right
or left and the saturated derivative $E_{p}$ in \((E_{p}\hat{T})\) where \((E_{p}\hat{T})\) is
just the time derivative of \(\hat{T}\) without further action to the right or left. Similar
results hold for \(\mbox{{\boldmath$\hat{\pp}_{i}$}}\hat{T}\) and the spatial derivative
\((\pp_{i}\hat{T})\) of $\hat{T}$. The appropriate steps for the gradient operator
\(\delta\hat{\mcal{H}}(\hat{T}^{-1},\hat{T})\) from Eq. (\ref{s4_75}) to (\ref{s4_81}) are
briefly tabulated with the needed commutators (\ref{s4_80}), the additional metric \(\hat{S}\)
(\ref{s4_76}) and the operators (\ref{s4_77}-\ref{s4_79})
\beq \lb{s4_75}
\delta\hat{\mcal{H}}(\hat{T}^{-1},\hat{T})&=&\hat{T}^{-1}\;\hat{\mcal{H}}\;\hat{T}-
\hat{\mcal{H}} =
\hat{T}^{-1}\;\hat{\eta}\left(\bea{cc} \hat{H}_{p} & \\ & \hat{H}_{p}^{T} \eea\right) \hat{T}-\hat{\eta}
\left(\bea{cc} \hat{H}_{p} & \\ & \hat{H}_{p}^{T} \eea\right) \\ \no &=&
\hat{T}^{-1}\;\hat{\eta}\Big(-\hat{S}\;\mbox{{\boldmath$\hat{E}_{p}$}}-
\mbox{{\boldmath$\wtilde{\pp}_{i}\wtilde{\pp}_{i}$}}\Big)\;\hat{T}-\hat{\eta}
\Big(-\hat{S}\;\mbox{{\boldmath$\hat{E}_{p}$}}-
\mbox{{\boldmath$\wtilde{\pp}_{i}\wtilde{\pp}_{i}$}}\Big)  \\  \lb{s4_76}
\hat{S} &=&\Big\{\hat{S}_{\alpha\beta}^{ab}\Big\}=\Big\{\delta_{a,b}\;
\Big(\underbrace{\hat{1}_{N\times N}}_{a=1}\;;\;\underbrace{-\hat{1}_{N\times N}}_{a=2}\Big)\Big\}
 \\ \lb{s4_77}
\hat{H}_{p}&=&-\hat{E}_{p}+\hat{h}_{p}\hspace*{0.75cm}\hat{E}_{p}^{T}=-\hat{E}_{p}\hspace*{0.75cm}
\hat{h}_{p}^{T}=\hat{h}_{p} \\ \lb{s4_78}
\mbox{{\boldmath$\hat{E}_{p}$}} &=& \mbox{{\boldmath${\ds \im\hbar\frac{\hat{\pp}}{\pp t_{p}}}$}}=
\mbox{{\boldmath$\im\hbar\;\hat{\pp}_{t_{p}}$}}\hspace{0.82cm}
\mbox{{\boldmath$\wtilde{\pp}_{i}$}}=
\mbox{{\boldmath${\ds\frac{\hbar}{\sqrt{2m}}\frac{\hat{\pp}}{\pp x^{i}}}$}}  \\ \lb{s4_79}
\hat{h}_{p}(\vec{x})&=&\frac{\hat{\vec{p}}^{\;2}}{2m}+u(\hat{\vec{x}})-\mu_{0}-\im\;\ve_{p}
\eeq
\be \lb{s4_80}
\bea{rclrcl}
\Big[\mbox{{\boldmath$\wtilde{\pp}_{i}$}}\;,\;\hat{T}\Big]&=&\big(\wtilde{\pp}_{i}\hat{T}\big) &
\Big[\hat{T}^{-1}\;,\;\mbox{{\boldmath$\wtilde{\pp}_{i}$}}\Big]&=&
-\big(\wtilde{\pp}_{i}\hat{T}^{-1}\big)=\hat{T}^{-1}\;\big(\wtilde{\pp}_{i}\hat{T}\big)\;\hat{T}^{-1} \\
\mbox{{\boldmath$\wtilde{\pp}_{i}$}}\;\hat{T}&=&\big(\wtilde{\pp}_{i}\hat{T}\big)+
\hat{T}\;\mbox{{\boldmath$\wtilde{\pp}_{i}$}}\hspace*{1.0cm} &
\hat{T}^{-1}\;\mbox{{\boldmath$\wtilde{\pp}_{i}$}}&=&-\big(\wtilde{\pp}_{i}\hat{T}^{-1}\big) +
\mbox{{\boldmath$\wtilde{\pp}_{i}$}}\;\hat{T}^{-1}
\eea
\ee
\be \lb{s4_81}
\delta\hat{\mcal{H}}(\hat{T}^{-1},\hat{T}) = -\hat{\eta}\Big(\hat{T}^{-1}\;\hat{S}\;\big(E_{p}\hat{T}\big)
+\hat{T}^{-1}\;\big(\wtilde{\pp}_{i}\wtilde{\pp}_{i}\hat{T}\big)  +
\big(\hat{T}^{-1}\;\hat{S}\;\hat{T}-\hat{S}\big)\;\mbox{{\boldmath$\hat{E}_{p}$}} +
2\;\hat{T}^{-1}\;\big(\wtilde{\pp}_{i}\hat{T}\big)\;\mbox{{\boldmath$\wtilde{\pp}_{i}$}}\Big)\;.
\ee
We have to perform an expansion with operator \(\Delta\hat{\mcal{O}}\) (\ref{s4_72}-\ref{s4_74})
of the actions \(\mcal{A}_{SDET}\), \(\mcal{A}_{J_{\psi}}\) (\ref{s4_69},\ref{s4_70})
and take into account terms up to \((\Delta\hat{\mcal{O}})^{2}\). We exclude the doubled operator
\(\big(\hat{\mcal{H}}+\hat{\sigma}_{D}^{(0)}\;\hat{1}_{2N\times 2N}\big)\) from the
operator \(\hat{\mcal{O}}\) (\ref{s4_82}) and thereby introduce the Green function (\ref{s4_83}), composed of
two block diagonal Green function parts, including the transposed form
\(\big(\hat{g}^{(0)}[\hat{\sigma}_{D}^{(0)}]_{N\times N}\big)^{T}\) for the
anti-unitary section. The transposed Green function in the '22' section differs from the
Green function in the '11' part by the opposite sign of the 'time' operator \(\hat{E}_{p}\),
due to the anti-symmetry under transposition
\beq \lb{s4_82}
\hat{\mcal{O}}/\mcal{N}&=&\frac{1}{\mcal{N}}\Big(\hat{\mcal{H}}+\hat{\sigma}_{D}^{(0)}\;
\hat{1}_{2N\times 2N}+\Delta\hat{\mcal{O}}\Big) \\ \no
&=&\frac{1}{\mcal{N}}\Big(\hat{\mcal{H}}+\hat{\sigma}_{D}^{(0)}\;
\hat{1}_{2N\times 2N}\Big)\times
\bigg[\hat{1}+\hat{G}^{(0)}[\hat{\sigma}_{D}^{(0)}]\;\;
\underbrace{\Big(\wtilde{\mcal{J}}(\hat{T}^{-1},\hat{T})+
\delta\hat{\mcal{H}}(\hat{T}^{-1},\hat{T})\Big)}_{\Delta\hat{\mcal{O}}}\bigg] \\ \lb{s4_83}
\hat{G}^{(0)}[\hat{\sigma}_{D}^{(0)}]&=&
\Big(\hat{\mcal{H}}+\hat{\sigma}_{D}^{(0)}\;\hat{1}_{2N\times 2N}\Big)^{-1} =
\Big(-\hat{\eta}\hat{S}\;\hat{E}_{p}+
\big(\hat{\eta}\hat{h}_{p}+\hat{\sigma}_{D}^{(0)}\big)\;\hat{1}_{2N\times 2N}\Big)^{-1}
\\ \no &=& \left(\bea{cc}
\hat{g}^{(0)}[\hat{\sigma}_{D}^{(0)}]_{N\times N} & \\
 & \big(\hat{g}^{(0)}[\hat{\sigma}_{D}^{(0)}]_{N\times N}\big)^{T}
\eea\right)^{ab} \\ \lb{s4_84}
\hat{g}^{(0)}[\hat{\sigma}_{D}^{(0)}]_{N\times N}&=&\Big(\hat{\eta}\big(-\hat{E}_{p}+\hat{h}_{p}\big)+
\hat{\sigma}_{D}^{(0)}\Big)^{-1}
\;\hat{1}_{N\times N} \\ \lb{s4_85}
\big(\hat{g}^{(0)}[\hat{\sigma}_{D}^{(0)}]_{N\times N}\big)^{T}&=&
\Big(\hat{\eta}\big(\hat{E}_{p}+\hat{h}_{p}\big)+
\hat{\sigma}_{D}^{(0)}\Big)^{-1}
\;\hat{1}_{N\times N}\;\;\;.
\eeq
We consider the first three terms in the expansion of the logarithm of the super-determinant
with \(\Delta\hat{\mcal{O}}\) and with the Green function \(\hat{G}^{(0)}[\hat{\sigma}_{D}^{(0)}]\).
The expansion of the inverse operator \(\big(\hat{\mcal{O}}/\mcal{N}\big)^{-1}\) (\ref{s4_87})
up to second order in \(\Delta\hat{\mcal{O}}\) is listed in Eq. (\ref{s4_88}) for the
action \(\mcal{A}_{J_{\psi}}\) of a coherent BEC wave function
\beq \lb{s4_86}
\lefteqn{\mcal{A}_{SDET}\big[\hat{T},\hat{\sigma}_{D}^{(0)};\hat{\mcal{J}}\big]
=\frac{1}{2}\mbox{Tr}\bigg\{\mbox{STR}\bigg[\ln\Big(
\big(\hat{\mcal{H}}+\hat{\sigma}_{D}^{(0)}\big)/\mcal{N}\times
\Big[\hat{1}+\hat{G}^{(0)}[\hat{\sigma}_{D}^{(0)}]\;\;\;\Delta\hat{\mcal{O}}\Big]\Big)\bigg]
\bigg\} } \\ \no &=& \frac{1}{2}
\mbox{Tr}\bigg\{\mbox{STR}\bigg[\ln\Big(\big(\hat{\mcal{H}}+\hat{\sigma}_{D}^{(0)}\big)/\mcal{N}\Big)
\bigg]\bigg\}  +  \frac{1}{2}\mbox{Tr}\Big[\mbox{STR}\Big(\Delta\hat{\mcal{O}}\;\;\;
\hat{G}^{(0)}[\hat{\sigma}_{D}^{(0)}]\Big)\Big] + \\ \no &-& \frac{1}{4}
\mbox{Tr}\Big[\mbox{STR}\Big(\Delta\hat{\mcal{O}}\;\;\;
\hat{G}^{(0)}[\hat{\sigma}_{D}^{(0)}]\;\;\Delta\hat{\mcal{O}}\;\;\;
\hat{G}^{(0)}[\hat{\sigma}_{D}^{(0)}]\Big)\Big]\pm\ldots
\eeq
\be \lb{s4_87}
\Big(\hat{\mcal{O}}\Big/\mcal{N}\Big)^{-1}=\mcal{N}\;\;
\Big[\hat{1}+\hat{G}^{(0)}[\hat{\sigma}_{D}^{(0)}]\;\;\Delta\hat{\mcal{O}}\Big]^{-1}\;\;
\hat{G}^{(0)}[\hat{\sigma}_{D}^{(0)}]
\ee
\beq \no
\lefteqn{\hspace*{-1.9cm}
\mcal{A}_{J_{\psi}}\big[\hat{T},\hat{\sigma}_{D}^{(0)};\hat{\mcal{J}}\big]
=\frac{1}{2}\sum_{a,b=1,2}\sum_{\alpha,\beta=1}^{N=L+S}
\frac{1}{\mcal{N}}
\widehat{\langle J_{\psi;\beta}^{b}}|\hat{\eta}\Big(\hat{I}\;\wtilde{K}\;\hat{T}
\Big(\hat{1}+\hat{G}^{(0)}[\hat{\sigma}_{D}^{(0)}]\;\;\Delta\hat{\mcal{O}}\Big)^{-1}
\hat{G}^{(0)}[\hat{\sigma}_{D}^{(0)}]\;\;\hat{T}^{-1}\;\hat{I}\Big)_{\beta\alpha}^{ba}\hat{\eta}|
\widehat{J_{\psi;\alpha}^{a}\rangle}  }
\\ \lb{s4_88} &=& \frac{1}{2}\sum_{a,b=1,2}\sum_{\alpha,\beta=1}^{N=L+S}
\frac{1}{\mcal{N}} \Bigg\{\widehat{\langle J_{\psi;\beta}^{b}}|\hat{\eta}\Big(\hat{I}\;\wtilde{K}\;\hat{T}
\;\;\hat{G}^{(0)}[\hat{\sigma}_{D}^{(0)}]\;\;\hat{T}^{-1}\;\hat{I}\Big)_{\beta\alpha}^{ba}\hat{\eta}|
\widehat{J_{\psi;\alpha}^{a}\rangle}+ \\ \no &-&
\widehat{\langle J_{\psi;\beta}^{b}}|\hat{\eta}\Big(\hat{I}\;\wtilde{K}\;\hat{T}\;\;
\hat{G}^{(0)}[\hat{\sigma}_{D}^{(0)}]\;\;\Delta\hat{\mcal{O}}\;\;
\hat{G}^{(0)}[\hat{\sigma}_{D}^{(0)}]\;\;\hat{T}^{-1}\;\hat{I}\Big)_{\beta\alpha}^{ba}\hat{\eta}|
\widehat{J_{\psi;\alpha}^{a}\rangle}+ \\ \no &+&
\widehat{\langle J_{\psi;\beta}^{b}}|\hat{\eta}\Big(\hat{I}\;\wtilde{K}\;\hat{T}\;\;
\hat{G}^{(0)}[\hat{\sigma}_{D}^{(0)}]\;\;\Big(\Delta\hat{\mcal{O}}\;\;
\hat{G}^{(0)}[\hat{\sigma}_{D}^{(0)}]\Big)^{2}\;\;
\hat{T}^{-1}\;\hat{I}\Big)_{\beta\alpha}^{ba}\hat{\eta}|
\widehat{J_{\psi;\alpha}^{a}\rangle}\mp\ldots\Bigg\} \;\;\;.
\eeq
In the following sections \ref{s43}-\ref{s46} the gradient operators
\(\delta\hat{\mcal{H}}(\hat{T}^{-1},\hat{T})\) in \(\Delta\hat{\mcal{O}}\) of
\(\mcal{A}_{SDET}\), \(\mcal{A}_{J_{\psi}}\) have to act on the
super-matrices \(\hat{T}\), \(\hat{T}^{-1}\) and the real self-energy density
field so that the gradient operators saturate and derivatives of the matrices
and scalar density appear. The given definitions of the doubled Hilbert space
have to be applied with anti-unitary second part; however, we also introduce
an averaging of the gradient actions where the coefficients, multiplying the
derivatives \(\big(\pp_{i}\hat{T}\big)\), \(\big(E_{p}\hat{T}\big)\) of coset matrices,
result in terms of the scalar density field \(\sigma_{D}^{(0)}(\vec{x},t_{p})\)
and trap potential. These coefficients, determined by \(\sigma_{D}^{(0)}(\vec{x},t_{p})\)
and \(u(\vec{x})\), are averaged by a coherent state path integral with
the real self-energy acting as a background field.

\subsection{Effective actions for the anomalous terms with
averaged Green functions of $\sigma_{D}^{(0)}(\vec{x},t_{p})$} \lb{s43}

We shift the first term in the expansion of the logarithm in \(\mcal{A}_{SDET}\) (\ref{s4_86}) to a
coherent state path integral \(Z[j_{\psi};\hat{\sigma}_{D}^{(0)}]\) with the
scalar density field \(\sigma_{D}^{(0)}(\vec{x},t_{p})\) and also include a term
with \(j_{\psi;\alpha}(\vec{x},t_{p})\) for the generation of a coherent
BEC wave function (last line and last factor in (\ref{s4_90}))
\be \lb{s4_89}
\frac{1}{2}\mbox{Tr}\bigg\{\mbox{STR}\Big[\ln\big(\hat{\mcal{H}}+\hat{\sigma}_{D}^{(0)}\big)
\big/\mcal{N}\Big]\bigg\}=  (L-S)\;\mbox{tr}\bigg[\ln\bigg(\hat{\eta}\big(
-\hat{E}_{p}\underbrace{-\im\;\ve_{p}+\frac{\hat{\vec{p}}^{\;2}}{2m}+
u(\hat{\vec{x}})-\mu_{0}}_{\hat{h}_{p}}\big)
+\hat{\sigma}_{D}^{(0)} \bigg)\bigg]
\ee
\beq \lb{s4_90}
Z[j_{\psi};\hat{\sigma}_{D}^{(0)}] &=& \int d[\hat{\sigma}_{D}^{(0)}(\vec{x},t_{p})]\;\;
\exp\bigg\{\frac{\im}{2\hbar}\frac{1}{V_{0}}\int_{C}d t_{p}\sum_{\vec{x}}
\sigma_{D}^{(0)}(\vec{x},t_{p})\;\sigma_{D}^{(0)}(\vec{x},t_{p})\bigg\}    \\ \no &\times &
\exp\bigg\{-(L-S)\;\mbox{tr}\bigg[\ln\bigg(\hat{\eta}\big(-\hat{E}_{p}
\underbrace{-\im\;\ve_{p}+\frac{\hat{\vec{p}}^{\;2}}{2m}+u(\hat{\vec{x}})-\mu_{0}}_{\hat{h}_{p}}\big)
+\hat{\sigma}_{D}^{(0)} \bigg)\bigg]\bigg\}  \\ \no &\times&
\exp\bigg\{\im\sum_{\alpha=1}^{N=L+S}\frac{1}{\mcal{N}}\;
\langle j_{\psi;\alpha}|\hat{\eta}\;\hat{g}^{(0)}[\hat{\sigma}_{D}^{(0)}]\;\hat{\eta}|
j_{\psi;\alpha}\rangle\bigg\}_{\mbox{.}}
\eeq
The scalar density field \(\sigma_{D}^{(0)}(\vec{x},t_{p})\) in
\(Z[\hat{\mcal{J}},J_{\psi},\im\hat{J}_{\psi\psi}]\)
(\ref{s4_90}) functions as a background field for the effective coherent state path integral
\(Z[\hat{\mcal{J}},J_{\psi},\im\hat{J}_{\psi\psi};\hat{T},\hat{\sigma}_{D}^{(0)}]\) with gradients
of the anomalous terms. This averaging process is formally denoted by relation (\ref{s4_91})
for the coefficients of gradient terms with field
\(\hat{\sigma}_{D}^{(0)}\)
\be \lb{s4_91}
Z[\hat{\mcal{J}},J_{\psi},\im\hat{J}_{\psi\psi}] =
\bigg(Z[j_{\psi};\hat{\sigma}_{D}^{(0)}]\times
Z[\hat{\mcal{J}},J_{\psi},\im\hat{J}_{\psi\psi};\hat{T},\hat{\sigma}_{D}^{(0)}]\bigg) =
\bigg\langle Z[\hat{\mcal{J}},J_{\psi},\im\hat{J}_{\psi\psi};\hat{T}]
\bigg\rangle_{\hat{\sigma}_{D}^{(0)}}\;\;\;.
\ee
Since we have moved the lowest order terms of \(\mcal{A}_{SDET}\),
\(\mcal{A}_{J_{\psi}}\) (\ref{s4_86}-\ref{s4_88})
to a coherent state path integral \(Z[j_{\psi};\hat{\sigma}_{D}^{(0)}]\) (\ref{s4_90}) for
a background field \(\sigma_{D}^{(0)}(\vec{x},t_{p})\), the actions \(\mcal{A}_{SDET}\ppr\),
\(\mcal{A}_{J_{\psi}}\ppr\) (\ref{s4_92},\ref{s4_93}) and also
\(\mcal{A}_{\hat{J}_{\psi\psi}}[\hat{T}]\) remain to be averaged by
\(Z[j_{\psi};\hat{\sigma}_{D}^{(0)}]\) (\ref{s4_91})
\beq \lb{s4_92}
\mcal{A}_{SDET}\ppr \big[\hat{T};\hat{\mcal{J}}\big]&=&
\bigg\langle\mcal{A}_{SDET}\ppr
\big[\hat{T},\hat{\sigma}_{D}^{(0)};\hat{\mcal{J}}\big]\bigg\rangle_{\hat{\sigma}_{D}^{(0)}} =
\frac{1}{2}\bigg\langle\mbox{Tr}\Big[\mbox{STR}\Big(\Delta\hat{\mcal{O}}\;\;\;
\hat{G}^{(0)}[\hat{\sigma}_{D}^{(0)}]\Big)\Big]
\bigg\rangle_{\hat{\sigma}_{D}^{(0)}} + \\ \no & - & \frac{1}{4}
\bigg\langle\mbox{Tr}\Big[\mbox{STR}\Big(\Delta\hat{\mcal{O}}\;\;\;
\hat{G}^{(0)}[\hat{\sigma}_{D}^{(0)}]\;\;\Delta\hat{\mcal{O}}\;\;\;
\hat{G}^{(0)}[\hat{\sigma}_{D}^{(0)}]\Big)\Big]
\bigg\rangle_{\hat{\sigma}_{D}^{(0)}} \pm\ldots
\eeq
\beq \lb{s4_93}
\lefteqn{\mcal{A}_{J_{\psi}}\ppr\big[\hat{T};\hat{\mcal{J}}\big] =
\bigg\langle\mcal{A}_{J_{\psi}}\ppr\big[\hat{T},\hat{\sigma}_{D}^{(0)};
\hat{\mcal{J}}\big]\bigg\rangle_{\hat{\sigma}_{D}^{(0)}} = }
\\ \no &=& \underbrace{\bigg\langle\frac{1}{\mcal{N}}\bigg(\frac{1}{2}
\widehat{\langle J_{\psi;\beta}^{b}}|\hat{\eta}\Big(\hat{I}\;\wtilde{K}\;\hat{T}
\;\;\hat{G}^{(0)}[\hat{\sigma}_{D}^{(0)}]\;\;\hat{T}^{-1}\;\hat{I}\Big)_{\beta\alpha}^{ba}\hat{\eta}|
\widehat{J_{\psi;\alpha}^{a}\rangle} -
\langle j_{\psi;\alpha}|\hat{\eta}\;\hat{g}^{(0)}[\hat{\sigma}_{D}^{(0)}]\;\hat{\eta}
|j_{\psi;\alpha}\rangle
\bigg)\bigg\rangle_{\hat{\sigma}_{D}^{(0)}}  }_{\approx 0} + \\ \no &-&\frac{1}{2}
\bigg\langle\frac{1}{\mcal{N}}
\widehat{\langle J_{\psi;\beta}^{b}}|\hat{\eta}\Big(\hat{I}\;\wtilde{K}\;\hat{T}\;\;
\hat{G}^{(0)}[\hat{\sigma}_{D}^{(0)}]\;\;\Delta\hat{\mcal{O}}\;\;
\hat{G}^{(0)}[\hat{\sigma}_{D}^{(0)}]\;\;\hat{T}^{-1}\;\hat{I}\Big)_{\beta\alpha}^{ba}\hat{\eta}|
\widehat{J_{\psi;\alpha}^{a}\rangle}\bigg\rangle_{\hat{\sigma}_{D}^{(0)}}
+ \\ \no &+&\frac{1}{2}\bigg\langle\frac{1}{\mcal{N}}
\widehat{\langle J_{\psi;\beta}^{b}}|\hat{\eta}\Big(\hat{I}\;\wtilde{K}\;\hat{T}\;\;
\hat{G}^{(0)}[\hat{\sigma}_{D}^{(0)}]\;\;\Big(\Delta\hat{\mcal{O}}\;\;
\hat{G}^{(0)}[\hat{\sigma}_{D}^{(0)}]\Big)^{2}\;\;
\hat{T}^{-1}\;\hat{I}\Big)_{\beta\alpha}^{ba}\hat{\eta}|
\widehat{J_{\psi;\alpha}^{a}\rangle}\bigg\rangle_{\hat{\sigma}_{D}^{(0)}}\mp\ldots
\eeq
\beq \lb{s4_94}
Z[\hat{\mcal{J}},J_{\psi},\im\hat{J}_{\psi\psi}] &=&
\Bigg(Z[j_{\psi};\hat{\sigma}_{D}^{(0)}]\;\;
\int d\big[\hat{T}^{-1}(\vec{x},t_{p})\;d\hat{T}(\vec{x},t_{p})\big]\;
\exp\Big\{\im\;\mcal{A}_{\hat{J}_{\psi\psi}}\big[\hat{T}\big]\Big\} \times \\ \no &\times&
\exp\Big\{-\mcal{A}_{SDET}\ppr\big[\hat{T},\hat{\sigma}_{D}^{(0)};\hat{\mcal{J}}\big]\Big\}\;\;
\exp\Big\{\im\;\mcal{A}_{J_{\psi}}\ppr\big[\hat{T},
\hat{\sigma}_{D}^{(0)};\hat{\mcal{J}}\big]\Big\} \Bigg)_{\mbox{.}}
\eeq
As already introduced in relation (\ref{s4_91}), the braces \(\langle\ldots\rangle_{\hat{\sigma}_{D}^{(0)}}\)
are used to mark the averaging process with the coherent state path integral (\ref{s4_90}) of the
density field \(\sigma_{D}^{(0)}(\vec{x},t_{p})\)
\beq \lb{s4_95}
\lefteqn{
\bigg\langle\bigg(\mbox{functional of }\sigma_{D}^{(0)}(\vec{x},t_{p})\mbox{ with gradient terms}\bigg)
\bigg\rangle_{\hat{\sigma}_{D}^{(0)}}= } \\ \no &=&
\int d[\hat{\sigma}_{D}^{(0)}(\vec{x},t_{p})] \;\;
\exp\bigg\{\frac{\im}{2\hbar}\frac{1}{V_{0}}\int_{C}d t_{p}\sum_{\vec{x}}
\sigma_{D}^{(0)}(\vec{x},t_{p})\;\sigma_{D}^{(0)}(\vec{x},t_{p})\bigg\}  \\ \no &\times&
\exp\bigg\{-(L-S)\;\mbox{tr}\ln\bigg(\hat{\eta}\big(-\hat{E}_{p}
\underbrace{-\im\;\ve_{p}+\frac{\hat{\vec{p}}^{2}}{2m}+u(\hat{\vec{x}})-\mu_{0}}_{\hat{h}_{p}}\big)+
\sigma_{D}^{(0)}\bigg)\bigg\}  \\ \no &\times&
\exp\bigg\{\im\sum_{\alpha=1}^{N=L+S}\frac{1}{\mcal{N}}
\langle j_{\psi;\alpha}|\hat{\eta}\;\hat{g}^{(0)}[\hat{\sigma}_{D}^{(0)}]\;\hat{\eta}
|j_{\psi;\alpha}\rangle\bigg\}
\;\times\;\bigg(\mbox{functional of }\sigma_{D}^{(0)}(\vec{x},t_{p})\mbox{ with gradient terms}\bigg)_{\mbox{.}}
\eeq
Finally, the average of the exponentials of \(\mcal{A}_{SDET}\ppr\),
\(\mcal{A}_{J_{\psi}}\ppr\) is approximated by the exponentials
of the averages over the actions (\ref{s4_92},\ref{s4_93})
\beq \lb{s4_96}
\lefteqn{Z[\hat{\mcal{J}},J_{\psi},\im\hat{J}_{\psi\psi}]=
\int d[\hat{T}^{-1}(\vec{x},t_{p})\;d\hat{T}(\vec{x},t_{p})]\;\;
\exp\Big\{\im\;\mcal{A}_{\hat{J}_{\psi\psi}}\big[\hat{T}\big]\Big\}  }  \\ \no &\times &
\bigg\langle
\exp\Big\{-\mcal{A}_{SDET}\ppr\big[\hat{T},\hat{\sigma}_{D}^{(0)};
\hat{\mcal{J}}\big]\Big\}  \;\;
\exp\Big\{\im\;\mcal{A}_{J_{\psi}}\ppr\big[\hat{T},\hat{\sigma}_{D}^{(0)};
\hat{\mcal{J}}\big]\Big\} \bigg\rangle_{\hat{\sigma}_{D}^{(0)}}
\\ \no &\approx &
\int d[\hat{T}^{-1}(\vec{x},t_{p})\;d\hat{T}(\vec{x},t_{p})]\;\;
\exp\Big\{\im\;\mcal{A}_{\hat{J}_{\psi\psi}}\big[\hat{T}\big]\Big\}   \\ \no &\times &
\exp\Big\{-\Big\langle\mcal{A}_{SDET}\ppr\big[\hat{T},\hat{\sigma}_{D}^{(0)};
\hat{\mcal{J}}\big]\Big\rangle_{\hat{\sigma}_{D}^{(0)}}\Big\}  \;\;
\exp\Big\{\im\;\Big\langle\mcal{A}_{J_{\psi}}\ppr\big[\hat{T},\hat{\sigma}_{D}^{(0)};
\hat{\mcal{J}}\big]\Big\rangle_{\hat{\sigma}_{D}^{(0)}}\Big\}_{\mbox{.}}
\eeq
One has to distinguish between unaveraged quantum fields \(\sigma_{D}^{(0)}(\vec{x},t_{p})\)
as integration variables on the time contour where it holds in general that
\(\sigma_{D}^{(0)}(\vec{x},t_{+})\neq\sigma_{D}^{(0)}(\vec{x},t_{-})\) and averaged
'classical' fields
\(\ovv{\sigma}_{D}^{(0)}(\vec{x},t)=\langle\sigma_{D}^{(0)}(\vec{x},t_{+})\rangle_{\hat{\sigma}_{D}^{(0)}}=
\langle\sigma_{D}^{(0)}(\vec{x},t_{-})\rangle_{\hat{\sigma}_{D}^{(0)}}\)
with equivalent fields on the time contour for hermitian actions. Furthermore, there is the
analogous difference between the unaveraged Green function term
\(\widehat{\langle\vec{x},t_{p}}^{a}|\hat{G}^{(0)}[\hat{\sigma}_{D}^{(0)}]|
\widehat{\vec{x}\ppr,t_{q}\ppr\rangle}^{b}\) as a term in a path integral of
\(\sigma_{D}^{(0)}(\vec{x},t_{p})\) and the averaged correspondence
\(\langle\widehat{\langle\vec{x},t_{p}}^{a}|\hat{G}^{(0)}[\hat{\sigma}_{D}^{(0)}]|
\widehat{\vec{x}\ppr,t_{q}\ppr\rangle}^{b}\rangle_{\hat{\sigma}_{D}^{(0)}}\) as a classical term;
instead of averaging the pair condensate path integral
\(Z[\hat{\mcal{J}},J_{\psi},\im\hat{J}_{\psi\psi};\hat{T},\hat{\sigma}_{D}^{(0)}]\) (\ref{s4_91})
over the mean density, one can also use a saddle point approximation for
\(Z[j_{\psi};\hat{\sigma}_{D}^{(0)}]\) with \(\hat{\sigma}_{D}^{(0)}\)
leading to a classical solution \(\ovv{\sigma}_{D}^{(0)}(\vec{x},t)\) as the result
of a mean field equation. This saddle point approximation is particular applicable
in \(d=2\) spatial dimensions for the determination of averaged coefficients with
the trap potential \(u(\vec{x})\) and the density field \(\sigma_{D}^{(0)}(\vec{x},t_{p})\).
The classical field solution \(\ovv{\sigma}_{D}^{(0)}(\vec{x},t)\) and the
averaged 'classical' Green function
\(\langle\widehat{\langle\vec{x},t_{p}}^{a}|\hat{G}^{(0)}[\hat{\sigma}_{D}^{(0)}]|
\widehat{\vec{x}\ppr,t_{q}\ppr\rangle}^{b}\rangle_{\hat{\sigma}_{D}^{(0)}}\) can be
improved by considering quadratic fluctuations \(\delta\sigma_{D}^{(0)}(\vec{x},t_{p})\)
around \(\ovv{\sigma}_{D}^{(0)}(\vec{x},t)\), e.g. as
second order variation of the actions in \(Z[j_{\psi};\hat{\sigma}_{D}^{(0)}]\)
with respect to \(\delta\sigma_{D}^{(0)}(\vec{x},t_{p})\). The first order variation has to vanish
because $\ovv{\sigma}_{D}^{(0)}(\vec{x},t)$ is the result of the saddle point equation
\be\lb{s4_97}
\sigma_{D}^{(0)}(\vec{x},t_{p})=\langle\sigma_{D}^{(0)}(\vec{x},t_{p})\rangle_{\hat{\sigma}_{D}^{(0)}}+
\delta\sigma_{D}^{(0)}(\vec{x},t_{p})=
\ovv{\sigma}_{D}^{(0)}(\vec{x},t)+ \delta\sigma_{D}^{(0)}(\vec{x},t_{p})\;\;\;.
\ee

\subsection{Propagation of fields $\Psi_{\vec{x},\alpha}^{a}(t_{p})$ and pair
condensates $\delta\hat{\Sigma}_{\alpha\beta}^{12}(\vec{x},t_{p})$ with
$\sigma_{D}^{(0)}(\vec{x},t_{p})$ as environment}  \lb{s44}

In subsection \ref{s41} we have removed the density parts
\(\delta\hat{\Sigma}_{D;\alpha\beta}^{aa}(\vec{x},t_{p})\) or hinge functions as subgroup
\(U(L|S)\) of \(Osp(S,S|2L)\) from the super-determinant \(\mcal{A}_{SDET}\) (\ref{s4_39})
and from the term \(\mcal{A}_{J_{\psi}}\) (\ref{s4_40}) for the BEC
wave function. Subsection \ref{s42} contains the definitions for a Hilbert space of
gradient operators and the doubled abstract states \(|\widehat{\psi_{\alpha}\rangle}^{a}\)
with antilinear second parts (\ref{s4_42},\ref{s4_43}). In subsection \ref{s43} we have
introduced the coherent state path integral \(Z[j_{\psi};\hat{\sigma}_{D}^{(0)}]\)
(\ref{s4_90}) with the real, scalar density \(\sigma_{D}^{(0)}(\vec{x},t_{p})\) as
background field for the gradient terms in \(\mcal{A}_{SDET}\ppr\), \(\mcal{A}_{J_{\psi}}\ppr\)
(\ref{s4_92},\ref{s4_93}) where the averaging with \(Z[j_{\psi};\hat{\sigma}_{D}^{(0)}]\)
(\ref{s4_90}) is formally denoted by the braces \(\langle\ldots\rangle_{\hat{\sigma}_{D}^{(0)}}\).

In continuation of principles for a gradient expansion, we state that a 'Nambu'-doubled field
\(\Psi_{\vec{x},\alpha}^{a}(t_{p})\) propagates with the block diagonal, doubled Green function
\(G^{(0)}[\hat{\sigma}_{D}^{(0)}]\), averaged by \(Z[j_{\psi};\hat{\sigma}_{D}^{(0)}]\) (\ref{s4_90}).
The reasons for this property of propagation can be verified by differentiating
\(Z[j_{\psi};\hat{\sigma}_{D}^{(0)}]\) (\ref{s4_90}) with respect to the super-field
\(\psi_{\vec{x}\ppr,\beta}^{*,b}(t_{q}\ppr)\) and by using a representation of (\ref{s4_90})
where the bilinear terms \(\psi_{\vec{x}\ppr,\beta}^{+,b}(t_{q}\ppr)\ldots\psi_{\vec{x},\alpha}^{a}(t_{p})\)
are still present with the quadratic term of the self-energy from the HST
\be \lb{s4_98}
\Psi_{\vec{x},\alpha}^{a(=1/2)}(t_{p})=
\left(\bea{c} \psi_{\vec{x},\alpha}(t_{p}) \\
\psi_{\vec{x},\alpha}^{*}(t_{p})
\eea\right)^{a} =
\int_{C}\frac{d t_{q}\ppr}{\hbar}\mcal{N}^{2}\sum_{\vec{x}\ppr}\sum_{b=1,2}
\bigg\langle\widehat{\langle\vec{x},t_{p}}^{a}|\hat{G}^{(0)}[\hat{\sigma}_{D}^{(0)}]|
\widehat{\vec{x}\ppr,t_{q}\ppr\rangle}^{b}\bigg\rangle_{\hat{\sigma}_{D}^{(0)}}\;
\left(\bea{c} \psi_{\vec{x}\ppr,\alpha}(t_{q}\ppr) \\
\psi_{\vec{x}\ppr,\alpha}^{*}(t_{q}\ppr) \eea\right)^{b}_{\mbox{.}}
\ee
This principle has to be used in the expansion of \(\mcal{A}_{J_{\psi}}\ppr\) (\ref{s4_93})
where one starts to propagate with the source field \(J_{\psi;\alpha}^{a}(\vec{x},t_{p})\)
on the right hand-side of the action for the BEC wave function. It replaces the wave function
\(\Psi_{\vec{x},\alpha}^{a}(t_{p})\) in (\ref{s4_98}). However, the propagation of fields
with relation (\ref{s4_98}) is not applicable for the action \(\mcal{A}_{SDET}\ppr\) (\ref{s4_92})
because of the cyclic invariance of traces, both of the super-trace '\(\mbox{STR}\)' and the
Hilbert state trace '\(\mbox{Tr}\)' of 'doubled' quantum mechanics. The trace '\(\mbox{Tr}\)'
means a propagation back to the same space point. This property is not included in rule
(\ref{s4_98}) where we have a propagation of the source field \(J_{\psi;\alpha}^{a}(\vec{x},t_{p})\)
(as a 'condensate seed') with repeated application of \(G^{(0)}[\hat{\sigma}_{D}^{(0)}]\) to the
left hand-side \(J_{\psi;\beta}^{+b}(\vec{x}\ppr,t_{q}\ppr)\) in the action
\(\mcal{A}_{J_{\psi}}\ppr\) (\ref{s4_93}).
We generalize rule (\ref{s4_98}) for the propagation
of arbitrary fields \(\big(f_{\alpha}(\vec{x},t_{p})\;;\;f_{\alpha}^{*}(\vec{x},t_{p})\big)^{T,a}\)
(\(\alpha=1,\ldots,N=L+S\)) and list in Eqs. (\ref{s4_99}-\ref{s4_101}) also the propagation
for the splitted parts \(f_{\alpha}(\vec{x},t_{p})\) and its complex conjugated field
\(f_{\alpha}^{*}(\vec{x},t_{p})\). The doubled fields \(\big(f_{\alpha}(\vec{x},t_{p})\;;\;
f_{\alpha}^{*}(\vec{x},t_{p})\big)^{T,a}\) can be identified with the fields in the various steps
of propagation in \(\mcal{A}_{J_{\psi}}\ppr\) (\ref{s4_93})
with \(G^{(0)}[\hat{\sigma}_{D}^{(0)}]\), starting from
\(J_{\psi;\alpha}^{a}(\vec{x},t_{p})\) on the right hand-side
\beq \lb{s4_99}
f_{\alpha}(\vec{x},t_{p})&=&\int_{C}\frac{d t_{q}\ppr}{\hbar}\mcal{N}^{2}\sum_{\vec{x}\ppr}
\bigg\langle\langle\vec{x},t_{p}|\hat{g}^{(0)}[\hat{\sigma}_{D}^{(0)}]|\vec{x}\ppr,t_{q}\ppr\rangle
\bigg\rangle_{\hat{\sigma}_{D}^{(0)}}\;f_{\alpha}(\vec{x}\ppr,t_{q}\ppr) \\ \lb{s4_100}
f_{\alpha}^{*}(\vec{x},t_{p})&=&\int_{C}
\frac{d t_{q}\ppr}{\hbar}\mcal{N}^{2}\sum_{\vec{x}\ppr}
\bigg\langle\ovv{\langle\vec{x},t_{p}}|\big(\hat{g}^{(0)}[\hat{\sigma}_{D}^{(0)}]\big)^{T}|
\ovv{\vec{x}\ppr,t_{q}\ppr\rangle}
\bigg\rangle_{\hat{\sigma}_{D}^{(0)}}\;f_{\alpha}^{*}(\vec{x}\ppr,t_{q}\ppr) \\ \no &=&
\int_{C}\frac{d t_{q}\ppr}{\hbar}\mcal{N}^{2}\sum_{\vec{x}\ppr}
f_{\alpha}^{*}(\vec{x}\ppr,t_{q}\ppr)\;
\bigg\langle\langle\vec{x}\ppr,t_{q}\ppr|\hat{g}^{(0)}[\hat{\sigma}_{D}^{(0)}]|
\vec{x},t_{p}\rangle \bigg\rangle_{\hat{\sigma}_{D}^{(0)}}  \\    \lb{s4_101}
\left(\bea{c}
f_{\alpha}(\vec{x},t_{p}) \\ f_{\alpha}^{*}(\vec{x},t_{p})
\eea\right)^{a} &=& \int_{C}\frac{d t_{q}\ppr}{\hbar}
\mcal{N}^{2}\sum_{\vec{x}\ppr}\sum_{b=1,2}
\bigg\langle\widehat{\langle\vec{x},t_{p}}^{a}|\hat{G}^{(0)}[\hat{\sigma}_{D}^{(0)}]|
\widehat{\vec{x}\ppr,t_{q}\ppr\rangle}^{b}
\bigg\rangle_{\hat{\sigma}_{D}^{(0)}}\;
\left(\bea{c} f_{\alpha}(\vec{x}\ppr,t_{q}\ppr) \\ f_{\alpha}^{*}(\vec{x}\ppr,t_{q}\ppr)
\eea\right)^{b}_{\mbox{.}}
\eeq
Since the determining operator \(\hat{\mcal{O}}=\hat{\mcal{H}}+\hat{\sigma}_{D}^{(0)}\cdot\hat{1}_{2N\times 2N}+
\Delta\hat{\mcal{O}}\) is normalized by the energy parameter
\(\mcal{N}=\hbar\Omega\;\mcal{N}_{x}\) throughout the gradient expansion (\ref{s4_71},\ref{s4_82}),
the appearance of every Green function involves a normalizing factor \(\mcal{N}\).

Apart from the average of a single Green function in (\ref{s4_98}-\ref{s4_101}), we
have to find principles for averages of products of Green functions with the
self-energy field \(\sigma_{D}^{(0)}(\vec{x},t_{p})\) and possible factorizations.
Since the field \(\sigma_{D}^{(0)}(\vec{x},t_{p})\propto \sum_{\alpha=1}^{N}
\psi_{\vec{x},\alpha}^{*}(t_{p})\;\psi_{\vec{x},\alpha}(t_{p})\) has the property
of a \(U(1)\) invariant density, it can only describe the noncondensed, incoherent
parts of the atomic constituents
\be \lb{s4_102}
\sigma_{D}^{(0)}(\vec{x},t_{p}) \propto \mbox{number of }\;
\Big(\mbox{noncondensed bosons }\;n_{B}(\vec{x},t_{p}) -
\mbox{noncondensed fermions }\;n_{F}(\vec{x},t_{p}) \Big)\;.
\ee
Therefore, we only assume spatially short-ranged correlations for \(\sigma_{D}^{(0)}(\vec{x},t_{p})\),
due to its property as being related to the noncondensed parts. Furthermore, we extend the central
limit theorem for the variance of the assumed independent fields \(\sigma_{D}^{(0)}(\vec{x},t_{p})\)
to the averages over products of Green functions with \(Z[j_{\psi};\hat{\sigma}_{D}^{(0)}]\).
The expansion of \(\mcal{A}_{SDET}\ppr\) (\ref{s4_92}) with trace '\(\mbox{Tr}\)'
only contains terms of the type which start at a particular time \(t_{p}+\delta t_{p}\) and
propagate back to the same time \(t_{p}\) over meantime \(t_{q}\ppr\), \(t_{q}\ppr+\delta t_{q}\ppr\),
including averages over products of Green functions (\ref{s4_103}).
Due to the approximate independence of averages with \(\sigma_{D}^{(0)}(\vec{x},t_{p})\)
in \(Z[j_{\psi};\hat{\sigma}_{D}^{(0)}]\), we factorize the average of the product of
the two time contour Green functions in (\ref{s4_103}). Since each averaged, single Green function
consists of a contour Heaviside function \(\Theta_{C}(t_{p}-t_{q}\ppr-\delta t_{q}\ppr)\),
\(\Theta_{C}(t_{q}\ppr-t_{p}-\delta t_{p})\), but with counter propagating time contour arguments,
the total propagation from \(t_{p}+\delta t_{p}\) back to \(t_{p}\) in Eq. (\ref{s4_103}) can
be neglected in the action \(\mcal{A}_{SDET}\ppr\) (\ref{s4_92})
with Hilbert trace terms '\(\mbox{Tr}\)'
\beq \lb{s4_103}
\lefteqn{\bigg\langle\langle\vec{x},t_{p}|\hat{g}^{(0)}[\hat{\sigma}_{D}^{(0)}]|
\vec{x}\ppr,t_{q}\ppr+\delta t_{q}\ppr\rangle\;\;
\langle\vec{x}\ppr,t_{q}\ppr|\hat{g}^{(0)}[\hat{\sigma}_{D}^{(0)}]|
\vec{x},t_{p}+\delta t_{p}\rangle
\bigg\rangle_{\hat{\sigma}_{D}^{(0)}} \approx } \\ \no &=&
\underbrace{\bigg\langle\langle\vec{x},t_{p}|\hat{g}^{(0)}[\hat{\sigma}_{D}^{(0)}]|
\vec{x}\ppr,t_{q}\ppr+\delta t_{q}\ppr\rangle\bigg\rangle_{\hat{\sigma}_{D}^{(0)}}}_{
\Theta_{C}(t_{p}-t_{q}\ppr-\delta t_{q}\ppr)}   \;\;\;
\underbrace{\bigg\langle\langle\vec{x}\ppr,t_{q}\ppr|\hat{g}^{(0)}[\hat{\sigma}_{D}^{(0)}]|
\vec{x},t_{p}+\delta t_{p}\rangle\bigg\rangle_{\hat{\sigma}_{D}^{(0)}}}_{
\Theta_{C}(t_{q}\ppr-t_{p}-\delta t_{p})}
\equiv 0\;\;\;.
\eeq
Although we have removed the density or hinge functions
\(\delta\hat{\Sigma}_{D;\alpha\beta}^{aa}(\vec{x},t_{p})\) from \(\mcal{A}_{SDET}\) and
\(\mcal{A}_{J_{\psi}}\) in subsection \ref{s41}, there can appear density-like
terms in the gradient expansion, in analogy denoted by \(\delta\hat{\Sigma}_{\alpha\beta}^{aa}(\vec{x},t_{p})\),
but which are composed of the matrices \(\hat{T}\), \(\hat{T}^{-1}\), as e.g. in
\(\delta\hat{\mcal{H}}(\hat{T}^{-1},\hat{T})\). We relate a {\it self-energy-like} density
matrix \(\delta\hat{\Sigma}_{\alpha\beta}^{11}(\vec{x}_{1},t_{q}\ppr)\) to the dyadic product
of generalized fields \(f_{\alpha}(\vec{x}_{1},t_{q}\ppr)\otimes f_{\beta}^{*}(\vec{x}_{2},t_{q}\ppr)\)
for short-ranged correlations \(\vec{x}_{1}\approx \vec{x}_{2}\) and propagate this density with
the average of the product of two Green functions according to the rules (\ref{s4_99}-\ref{s4_101})
and the approximation (\ref{s4_103}). The opposite time contour arguments in the Heaviside functions
lead to a vanishing propagation of density terms in the action
\(\mcal{A}_{SDET}\ppr\) (\ref{s4_92}) with the Hilbert trace terms '\(\mbox{Tr}\)'
\be \lb{s4_104}
\langle\vec{x}_{1},t_{q}\ppr|\delta\hat{\Sigma}_{\alpha\beta}^{11}|\vec{x}_{2},t_{q}\ppr\rangle=\eta_{q}\;
\delta_{\vec{x}_{1},\vec{x}_{2}}\;
\delta\hat{\Sigma}_{\alpha\beta}^{11}(\vec{x}_{1},t_{q}\ppr)=\eta_{q}\;
\delta\hat{\Sigma}_{\vec{x}_{1},\alpha;\vec{x}_{2},\beta}^{11}(t_{q}\ppr)
\Big|_{\vec{x}_{1}=\vec{x}_{2}}\propto\eta_{q}\;
f_{\alpha}(\vec{x}_{1},t_{q}\ppr)\;
f_{\beta}^{*}(\vec{x}_{2},t_{q}\ppr)\Big|_{\vec{x}_{1}=\vec{x}_{2}}
\ee
\beq \lb{s4_105}
\lefteqn{\delta\hat{\Sigma}_{\alpha\beta}^{11}(\vec{x},t_{p})=
\delta\hat{\Sigma}_{\vec{x},\alpha;\vec{x}\ppr,\beta}^{11}(t_{p})
\Big|_{\vec{x}=\vec{x}\ppr}\propto
f_{\alpha}(\vec{x},t_{p})\;
f_{\beta}^{*}(\vec{x}\ppr,t_{p})\Big|_{\vec{x}=\vec{x}\ppr}=
\int_{C}\frac{d t_{q}\ppr}{\hbar}\eta_{q}
\mcal{N}^{3}\sum_{\vec{x}_{1},\vec{x}_{2}} \times } \\ \no &\times&
\bigg\langle\langle\vec{x},t_{p}|\hat{g}^{(0)}[\hat{\sigma}_{D}^{(0)}]|
\vec{x}_{1},t_{q}\ppr\rangle\;\;
\langle\vec{x}_{2},t_{q}\ppr|\hat{g}^{(0)}[\hat{\sigma}_{D}^{(0)}]|
\vec{x}\ppr,t_{p}\rangle\bigg\rangle_{\hat{\sigma}_{D}^{(0)}} \;
\underbrace{\eta_{q}\;f_{\alpha}(\vec{x}_{1},t_{q}\ppr)\;f_{\beta}^{*}(\vec{x}_{2},t_{q}\ppr)}_{\propto
\;\eta_{q}\;\delta\hat{\Sigma}_{\vec{x}_{1},\alpha;\vec{x}_{2},\beta}^{11}(t_{q}\ppr)}
\Big|_{\vec{x}_{1}=\vec{x}_{2}}\Big|_{\vec{x}=\vec{x}\ppr}\equiv 0
\eeq
\beq \lb{s4_106}
\lefteqn{\bigg\langle\langle\vec{x},t_{p}|\hat{g}^{(0)}[\hat{\sigma}_{D}^{(0)}]|
\vec{x}_{1},t_{q}\ppr+\delta t_{q}\ppr\rangle\;\;
\langle\vec{x}_{2},t_{q}\ppr|\hat{g}^{(0)}[\hat{\sigma}_{D}^{(0)}]|
\vec{x}\ppr,t_{p}+\delta t_{p}\rangle
\bigg\rangle_{\hat{\sigma}_{D}^{(0)}} \approx } \\ \no &=&
\underbrace{\bigg\langle\langle\vec{x},t_{p}|\hat{g}^{(0)}[\hat{\sigma}_{D}^{(0)}]|
\vec{x}_{1},t_{q}\ppr+\delta t_{q}\ppr\rangle\bigg\rangle_{\hat{\sigma}_{D}^{(0)}}}_{
\Theta_{C}(t_{p}-t_{q}\ppr-\delta t_{q}\ppr)}
\;\;\;\underbrace{\bigg\langle\langle\vec{x}_{2},t_{q}\ppr|\hat{g}^{(0)}[\hat{\sigma}_{D}^{(0)}]|
\vec{x}\ppr,t_{p}+\delta t_{p}\rangle\bigg\rangle_{\hat{\sigma}_{D}^{(0)}}}_{
\Theta_{C}(t_{q}\ppr-t_{p}-\delta t_{p})}
\equiv 0\;\;\;.
\eeq
Conequently we can ignore pure density-like terms
{\it combined with the average of two Green functions} in the gradient expansion of
\(\mcal{A}_{SDET}\ppr\) (\ref{s4_92}).
However, it is supposed that anomalous-like terms, related to
self-energy-like terms \(\delta\hat{\Sigma}_{\alpha\beta}^{12}(\vec{x},t_{p})\), remain
in the gradient expansion. We regard the pair-condensate-like terms
\(\delta\hat{\Sigma}_{\alpha\beta}^{12}(\vec{x},t_{p})\) as being composed of
the dyadic product of fields \(f_{\alpha}(\vec{x}_{1},t_{q}\ppr)\otimes f_{\beta}(\vec{x}_{2},t_{q}\ppr)\)
with short-ranged correlations (\(\vec{x}_{1}\approx\vec{x}_{2}\)) (\ref{s4_107}).
Since the fields \(f_{\alpha}(\vec{x}_{1},t_{q}\ppr)\otimes f_{\beta}(\vec{x}_{2},t_{q}\ppr)\)
propagate to the same direction \(t_{p}\), we have the equivalent contour time arguments
in the Heaviside functions of Green functions (\ref{s4_109}). The propagation of
\(f_{\beta}(\vec{x}_{2},t_{q}\ppr)\) refers to the transposed Green function with antilinear
Hilbert space states so that anomalous-like terms, following from
\(\delta\hat{\mcal{H}}(\hat{T}^{-1},\hat{T})\), propagate in the trace terms '\(\mbox{Tr}\)'
of \(\mcal{A}_{SDET}\ppr\) without vanishing, due to the anti-unitary property in the '22'
blocks
\be \lb{s4_107}
\langle\vec{x}_{1},t_{q}\ppr|\delta\hat{\Sigma}_{\alpha\beta}^{12}|\vec{x}_{2},t_{q}\ppr\rangle=
\eta_{q}\;\delta_{\vec{x}_{1},\vec{x}_{2}}\;
\delta\hat{\Sigma}_{\alpha\beta}^{12}(\vec{x}_{1},t_{q}\ppr)=
\eta_{q}\;\delta\hat{\Sigma}_{\vec{x}_{1},\alpha;\vec{x}_{2},\beta}^{12}(t_{q}\ppr)
\Big|_{\vec{x}_{1}=\vec{x}_{2}}\propto
\eta_{q}\;f_{\alpha}(\vec{x}_{1},t_{q}\ppr)\;
f_{\beta}(\vec{x}_{2},t_{q}\ppr)\Big|_{\vec{x}_{1}=\vec{x}_{2}}
\ee
\beq \lb{s4_108}
\lefteqn{\delta\hat{\Sigma}_{\alpha\beta}^{12}(\vec{x},t_{p})=
\delta\hat{\Sigma}_{\vec{x},\alpha;\vec{x}\ppr,\beta}^{12}(t_{p})
\Big|_{\vec{x}=\vec{x}\ppr}\propto
f_{\alpha}(\vec{x},t_{p})\;
f_{\beta}(\vec{x}\ppr,t_{p})\Big|_{\vec{x}=\vec{x}\ppr} =
\int_{C}\frac{d t_{q}\ppr}{\hbar}\eta_{q}
 \mcal{N}^{3}\sum_{\vec{x}_{1},\vec{x}_{2}} \times} \\ \no &\times&
\bigg\langle\langle\vec{x},t_{p}|\hat{g}^{(0)}[\hat{\sigma}_{D}^{(0)}]|
\vec{x}_{1},t_{q}\ppr\rangle\;\;
\underbrace{\langle\vec{x}\ppr,t_{p}|\hat{g}^{(0)}[\hat{\sigma}_{D}^{(0)}]|
\vec{x}_{2},t_{q}\ppr\rangle}_{\ovv{\langle\vec{x}_{2},t_{q}\ppr}|
\big(\hat{g}^{(0)}[\hat{\sigma}_{D}^{(0)}]\big)^{T}|\ovv{\vec{x}\ppr,t_{p}\rangle}}
\bigg\rangle_{\hat{\sigma}_{D}^{(0)}} \;
\underbrace{\eta_{q}\;f_{\alpha}(\vec{x}_{1},t_{q}\ppr)\;f_{\beta}(\vec{x}_{2},t_{q}\ppr)}_{\propto
\;\eta_{q}\;\delta\hat{\Sigma}_{\vec{x}_{1},\alpha;\vec{x}_{2},\beta}^{12}(t_{q}\ppr)}
\Big|_{\vec{x}_{1}=\vec{x}_{2}}\Big|_{\vec{x}=\vec{x}\ppr}
\eeq
\beq \lb{s4_109}
\lefteqn{\bigg\langle\langle\vec{x},t_{p}|\hat{g}^{(0)}[\hat{\sigma}_{D}^{(0)}]|
\vec{x}_{1},t_{q}\ppr+\delta t_{q}\ppr\rangle \;\;
\ovv{\langle\vec{x}_{2},t_{q}\ppr+\delta t_{q}\ppr}|
\big(\hat{g}^{(0)}[\hat{\sigma}_{D}^{(0)}]\big)^{T}|
\ovv{\vec{x}\ppr,t_{p}\rangle} \bigg\rangle_{\hat{\sigma}_{D}^{(0)}} = }  \\ \no &=&
\bigg\langle\langle\vec{x},t_{p}|\hat{g}^{(0)}[\hat{\sigma}_{D}^{(0)}]|
\vec{x}_{1},t_{q}\ppr+\delta t_{q}\ppr\rangle\;\;
\langle\vec{x}\ppr,t_{p}|\hat{g}^{(0)}[\hat{\sigma}_{D}^{(0)}]|
\vec{x}_{2},t_{q}\ppr+\delta t_{q}\ppr\rangle
\bigg\rangle_{\hat{\sigma}_{D}^{(0)}}\;\;\approx \\ \no &=&
\underbrace{\bigg\langle\langle\vec{x},t_{p}|\hat{g}^{(0)}[\hat{\sigma}_{D}^{(0)}]|
\vec{x}_{1},t_{q}\ppr+\delta t_{q}\ppr\rangle\bigg\rangle_{\hat{\sigma}_{D}^{(0)}} }_{
\propto \Theta_{C}(t_{p}-t_{q}\ppr-\delta t_{q}\ppr);\;
\mbox{\small of order }N_{\sigma}\;\langle\sigma\rangle }
\;\;\underbrace{
\bigg\langle\langle\vec{x}\ppr,t_{p}|\hat{g}^{(0)}[\hat{\sigma}_{D}^{(0)}]|
\vec{x}_{2},t_{q}\ppr+\delta t_{q}\ppr\rangle\bigg\rangle_{\hat{\sigma}_{D}^{(0)}} }_{
\propto \Theta_{C}(t_{p}-t_{q}\ppr-\delta t_{q}\ppr);\;
\mbox{\small of order }N_{\sigma}\;\langle\sigma\rangle} \neq  0\;\;\;.
\eeq
The propagation of the vanishing density-like terms and nonvanishing pair-condensates can be
summarized by the dyadic products of 'Nambu'-doubled fields (\ref{s4_103}-\ref{s4_109}).
We introduce the doubled Green function \(\hat{G}^{(0)}[\hat{\sigma}_{D}^{(0)}]\) for
\(\hat{g}^{(0)}[\hat{\sigma}_{D}^{(0)}]\) and its transpose
\(\big(\hat{g}^{(0)}[\hat{\sigma}_{D}^{(0)}]\big)^{T}\) and obtain the combined
relation (\ref{s4_110}) where the anomalous-like parts (\(a\neq b\), \(a_{1}\neq b_{1}\))
are the only relevant terms for the propagation
\beq\lb{s4_110}
\lefteqn{\delta\hat{\Sigma}_{\alpha\beta}^{a\neq b}(\vec{x},t_{p})\Big|^{a\neq b}=
\delta\hat{\Sigma}_{\vec{x},\alpha;\vec{x}\ppr,\beta}^{a\neq b}(t_{p})
\Big|_{\vec{x}=\vec{x}\ppr}\propto
\left(
\bea{c} f_{\alpha}(\vec{x},t_{p}) \\ f_{\alpha}^{*}(\vec{x},t_{p}) \eea\right)^{a}\otimes
\Big( f_{\beta}^{*}(\vec{x}\ppr,t_{p})\;;\;
f_{\beta}(\vec{x}\ppr,t_{p}) \Big)^{b}\Big|_{\vec{x}=\vec{x}\ppr}^{a\neq b}= } \\  \no &=&
\int_{C}\frac{d t_{q}\ppr}{\hbar}\eta_{q}\mcal{N}^{3}\sum_{\vec{x}_{1},\vec{x}_{2}}
\sum_{a_{1},b_{1}=1,2}^{a_{1}\neq b_{1}}
\bigg\langle\widehat{\langle\vec{x},t_{p}}^{a}|\hat{G}^{(0)}[\hat{\sigma}_{D}^{(0)}]|
\widehat{\vec{x}_{1},t_{q}\ppr\rangle}^{a_{1}}\;\;
\widehat{\langle\vec{x}_{2},t_{q}\ppr}^{b_{1}}|\hat{G}^{(0)}[\hat{\sigma}_{D}^{(0)}]|
\widehat{\vec{x}\ppr,t_{p}\rangle}^{b} \bigg\rangle_{\hat{\sigma}_{D}^{(0)}}  \\ \no &\times&
\underbrace{\eta_{q}\;\left(\bea{c} f_{\alpha}(\vec{x}_{1},t_{q}\ppr) \\
f_{\alpha}^{*}(\vec{x}_{1},t_{q}\ppr) \eea\right)^{a_{1}}\otimes
\Big( f_{\beta}^{*}(\vec{x}_{2},t_{q}\ppr)\;;\;
f_{\beta}(\vec{x}_{2},t_{q}\ppr) \Big)^{b_{1}}}_{\propto\;\eta_{q}\;
\delta\hat{\Sigma}_{\vec{x}_{1},\alpha;\vec{x}_{2},\beta}^{a_{1}b_{1}}
(t_{q}\ppr);\;\;(a_{1}\neq b_{1})}\bigg|_{\vec{x}_{1}=\vec{x}_{2}}^{a_{1}\neq b_{1}}
\bigg|_{\vec{x}=\vec{x}\ppr}^{a\neq b} \;\;\;.
\eeq
Apart from self-energy-like matrices in \(\delta\hat{\mcal{H}}(\hat{T}^{-1},\hat{T})\)
as \(\delta\hat{\Sigma}_{\alpha\beta}^{ab}(\vec{x},t_{p})\) (\ref{s4_110}), the
unsaturated gradient operators \(\mbox{{\boldmath$\wtilde{\pp}_{i}$}}\),
\(\mbox{{\boldmath$\hat{E}_{p}$}}\) also act onto the Green functions
\(\hat{G}^{(0)}[\hat{\sigma}_{D}^{(0)}]\). In the following we make repeated use of
the commutation relation (\ref{s4_111}) which can also be transferred in analogous
manner to the commutator with \(\mbox{{\boldmath$\hat{E}_{p}$}}\) (\ref{s4_112})
\beq \lb{s4_111}
\Big[\mbox{{\boldmath$\wtilde{\pp}_{i}$}}\;,\;\hat{G}^{(0)}[\hat{\sigma}_{D}^{(0)}]\Big]&=&
-\hat{G}^{(0)}[\hat{\sigma}_{D}^{(0)}]\;\Big(\wtilde{\pp}_{i}\hat{u}+\wtilde{\pp}_{i}
\hat{\sigma}_{D}^{(0)}\Big)\;\hat{G}^{(0)}[\hat{\sigma}_{D}^{(0)}]  \\ \lb{s4_112}
\Big[\mbox{{\boldmath$\hat{E}_{p}$}}\;,\;\hat{G}^{(0)}[\hat{\sigma}_{D}^{(0)}]\Big]&=&
-\hat{G}^{(0)}[\hat{\sigma}_{D}^{(0)}]\;\Big(\hat{E}_{p}
\hat{\sigma}_{D}^{(0)}\Big)\;\hat{G}^{(0)}[\hat{\sigma}_{D}^{(0)}]\;\;\;.
\eeq
The derivative of a local self-energy-like matrix
\(\big(\wtilde{\pp}_{i}\delta\hat{\Sigma}_{\alpha\beta}^{ab}(\vec{x},t_{p})\big)\) is
considered as the derivative of a dyadic product (\ref{s4_110}) where we only keep the diagonal
term \(\vec{x}\approx\vec{x}\ppr\) after action of the spatial gradient. We start out from the
propagation in Eq. (\ref{s4_110}) for the nonvanishing, anomalous-like terms and include
the gradient operator in the propagation of the averages of the product with two Green
functions so that relation (\ref{s4_113}) follows for the time-like extension from time $t_{q}\ppr$ to
\(\big(\wtilde{\pp}_{i}\delta\hat{\Sigma}_{\alpha\beta}^{ab}(\vec{x},t_{p})\big)\)
\beq \lb{s4_113}
\lefteqn{\Big(\wtilde{\pp}_{i}
\delta\hat{\Sigma}_{\alpha\beta}^{a\neq b}(\vec{x},t_{p})\Big)\Big|^{a\neq b}=
\Big(\wtilde{\pp}_{i}
\delta\hat{\Sigma}_{\vec{x},\alpha;\vec{x}\ppr,\beta}^{a\neq b}(t_{p})+
\wtilde{\pp}_{i}\ppr
\delta\hat{\Sigma}_{\vec{x},\alpha;\vec{x}\ppr,\beta}^{a\neq b}(t_{p})\Big)
\Big|_{\vec{x}=\vec{x}\ppr} =} \\ \no &=&
\int_{C}\frac{d t_{q}\ppr}{\hbar}\eta_{q}\;\mcal{N}^{3}\sum_{\vec{x}_{1},\vec{x}_{2}}
\Bigg(\bigg\langle\widehat{\langle\vec{x},t_{p}}^{a}|\mbox{{\boldmath$\wtilde{\pp}_{i}$}}\;
\hat{G}^{(0)}[\hat{\sigma}_{D}^{(0)}]|\widehat{\vec{x}_{1},t_{q}\ppr\rangle}^{a_{1}}\;\;
\widehat{\langle\vec{x}_{2},t_{q}\ppr}^{b_{1}}|\hat{G}^{(0)}[\hat{\sigma}_{D}^{(0)}]|
\widehat{\vec{x}\ppr,t_{p}\rangle}^{b}\bigg\rangle_{\hat{\sigma}_{D}^{(0)}}+ \\ \no &-&
\bigg\langle\widehat{\langle\vec{x},t_{p}}^{a}|
\hat{G}^{(0)}[\hat{\sigma}_{D}^{(0)}]|\widehat{\vec{x}_{1},t_{q}\ppr\rangle}^{a_{1}}\;\;
\widehat{\langle\vec{x}_{2},t_{q}\ppr}^{b_{1}}|\hat{G}^{(0)}[\hat{\sigma}_{D}^{(0)}]\;
\mbox{{\boldmath$\wtilde{\pp}_{i}$}}|
\widehat{\vec{x}\ppr,t_{p}\rangle}^{b}\bigg\rangle_{\hat{\sigma}_{D}^{(0)}} \Bigg)\times
\\ \no &\times&
\underbrace{\eta_{q}\;\left(\bea{c} f_{\alpha}(\vec{x}_{1},t_{q}\ppr) \\
f_{\alpha}^{*}(\vec{x}_{1},t_{q}\ppr) \eea\right)^{a_{1}}
\otimes \left(\bea{c}
f_{\beta}^{*}(\vec{x}_{2},t_{q}\ppr)\;;\;  f_{\beta}(\vec{x}_{2},t_{q}\ppr) \eea\right)^{b_{1}}}_{
\propto\;\eta_{q}\;
\delta\hat{\Sigma}_{\vec{x}_{1},\alpha;\vec{x}_{2},\beta}^{a_{1}b_{1}}(t_{q}\ppr);\;(a_{1}\neq b_{1})}
\bigg|_{\vec{x}_{1}=\vec{x}_{2}}^{a_{1}\neq b_{1}}
\bigg|_{\vec{x}=\vec{x}\ppr}^{a\neq b}\;\;\;.
\eeq
One has to apply the commutator (\ref{s4_111}) two times so that the gradient operator
\(\mbox{{\boldmath$\wtilde{\pp}_{i}$}}\) functions onto the dyadic product of
'Nambu'-doubled fields with arguments (\(\vec{x}_{1},t_{q}\ppr\)) and (\(\vec{x}_{2},t_{q}\ppr\))
which propagate to (\(\vec{x},t_{p}\)) and (\(\vec{x}\ppr,t_{p}\)). We also only take the
spatially local terms with (\(\vec{x}\approx\vec{x}\ppr\)) and (\(\vec{x}_{1}\approx\vec{x}_{2}\))
and find relations (\ref{s4_114},\ref{s4_115}) for the propagation of
\(\big(\wtilde{\pp}_{i}\delta\hat{\Sigma}_{\alpha\beta}^{a\neq b}(\vec{x},t_{p})\big)\big|^{a\neq b}\)
in Hilbert state traces '\(\mbox{Tr}\)' of \(\mcal{A}_{SDET}\ppr\) (\ref{s4_92})
where the density terms completely vanish in the case of averages
of the product with two Green functions
\beq \lb{s4_114}
\lefteqn{\Big(\wtilde{\pp}_{i}
\delta\hat{\Sigma}_{\alpha\beta}^{a\neq b}(\vec{x},t_{p})\Big)\Big|^{a\neq b}=
\Big(\wtilde{\pp}_{i}
\delta\hat{\Sigma}_{\vec{x},\alpha;\vec{x}\ppr,\beta}^{a\neq b}(t_{p})+
\wtilde{\pp}_{i}\ppr
\delta\hat{\Sigma}_{\vec{x},\alpha;\vec{x}\ppr,\beta}^{a\neq b}(t_{p})\Big)
\Big|_{\vec{x}=\vec{x}\ppr} =} \\ \no &=&
\int_{C}\frac{d t_{q}\ppr}{\hbar}\eta_{q}\;\mcal{N}^{3}\sum_{\vec{x}_{1},\vec{x}_{2}}
\Bigg(\bigg\langle\widehat{\langle\vec{x},t_{p}}^{a}|\hat{G}^{(0)}[\hat{\sigma}_{D}^{(0)}]\;
\mbox{{\boldmath$\wtilde{\pp}_{i}$}}-\hat{G}^{(0)}[\hat{\sigma}_{D}^{(0)}]\;
\big(\wtilde{\pp}_{i}\hat{u}+\wtilde{\pp}_{i}\hat{\sigma}_{D}^{(0)}\big)\;
\hat{G}^{(0)}[\hat{\sigma}_{D}^{(0)}]|\widehat{\vec{x}_{1},t_{q}\ppr\rangle}^{a_{1}} \times \\ \no &\times&
\widehat{\langle\vec{x}_{2},t_{q}\ppr}^{b_{1}}|\hat{G}^{(0)}[\hat{\sigma}_{D}^{(0)}]|
\widehat{\vec{x}\ppr,t_{p}\rangle}^{b}\bigg\rangle_{\hat{\sigma}_{D}^{(0)}} -
\bigg\langle\widehat{\langle\vec{x},t_{p}}^{a}|
\hat{G}^{(0)}[\hat{\sigma}_{D}^{(0)}]|\widehat{\vec{x}_{1},t_{q}\ppr\rangle}^{a_{1}}\times \\ \no &\times&
\widehat{\langle\vec{x}_{2},t_{q}\ppr}^{b_{1}}|\mbox{{\boldmath$\wtilde{\pp}_{i}$}}\;
\hat{G}^{(0)}[\hat{\sigma}_{D}^{(0)}] +\hat{G}^{(0)}[\hat{\sigma}_{D}^{(0)}]\;
\big(\wtilde{\pp}_{i}\hat{u}+\wtilde{\pp}_{i}\hat{\sigma}_{D}^{(0)}\big)\;
\hat{G}^{(0)}[\hat{\sigma}_{D}^{(0)}]|
\widehat{\vec{x}\ppr,t_{p}\rangle}^{b}\bigg\rangle_{\hat{\sigma}_{D}^{(0)}} \Bigg)\times
\\ \no &\times&
\underbrace{\eta_{q}\;\left(\bea{c} f_{\alpha}(\vec{x}_{1},t_{q}\ppr) \\
f_{\alpha}^{*}(\vec{x}_{1},t_{q}\ppr) \eea\right)^{a_{1}}
\otimes \left(\bea{c}
f_{\beta}^{*}(\vec{x}_{2},t_{q}\ppr)\;;\;  f_{\beta}(\vec{x}_{2},t_{q}\ppr) \eea\right)^{b_{1}}}_{
\propto\;\eta_{q}\;
\delta\hat{\Sigma}_{\vec{x}_{1},\alpha;\vec{x}_{2},\beta}^{a_{1}b_{1}}(t_{q}\ppr);\;(a_{1}\neq b_{1})}
\bigg|_{\vec{x}_{1}=\vec{x}_{2}}^{a_{1}\neq b_{1}}\bigg|_{\vec{x}=\vec{x}\ppr}^{a\neq b}
\eeq
\beq \lb{s4_115}
\lefteqn{\Big(\wtilde{\pp}_{i}
\delta\hat{\Sigma}_{\alpha\beta}^{a\neq b}(\vec{x},t_{p})\Big)\Big|^{a\neq b}=
\Big(\wtilde{\pp}_{i}
\delta\hat{\Sigma}_{\vec{x},\alpha;\vec{x}\ppr,\beta}^{a\neq b}(t_{p})+
\wtilde{\pp}_{i}\ppr
\delta\hat{\Sigma}_{\vec{x},\alpha;\vec{x}\ppr,\beta}^{a\neq b}(t_{p})\Big)
\Big|_{\vec{x}=\vec{x}\ppr} =} \\ \no &=&
\int_{C}\frac{d t_{q}\ppr}{\hbar}\eta_{q}\mcal{N}^{3}\sum_{\vec{x}_{1},\vec{x}_{2}}
\bigg\langle\widehat{\langle\vec{x},t_{p}}^{a}|\hat{G}^{(0)}[\hat{\sigma}_{D}^{(0)}]|
\widehat{\vec{x}_{1},t_{q}\ppr\rangle}^{a_{1}}\;\;
\widehat{\langle\vec{x}_{2},t_{q}\ppr}^{b_{1}}|\hat{G}^{(0)}[\hat{\sigma}_{D}^{(0)}]|
\widehat{\vec{x}\ppr,t_{p}\rangle}^{b}\bigg\rangle_{\hat{\sigma}_{D}^{(0)}} \times
\\ \no &\times&\eta_{q}\;
\bigg(\wtilde{\pp}_{i}^{(1)}
\delta\hat{\Sigma}_{\vec{x}_{1},\alpha;\vec{x}_{2},\beta}^{a_{1}\neq b_{1}}(t_{q}\ppr)+
\wtilde{\pp}_{i}^{(2)}
\delta\hat{\Sigma}_{\vec{x}_{1},\alpha;\vec{x}_{2},\beta}^{a_{1}\neq b_{1}}(t_{q}\ppr)\bigg)
\bigg|_{\vec{x}_{1}=\vec{x}_{2}}^{a_{1}\neq b_{1}}
\bigg|_{\vec{x}=\vec{x}\ppr}^{a\neq b} + \\ \no &-&
\int_{C}\frac{d t_{q}\ppr}{\hbar}\eta_{q}\mcal{N}^{3}\sum_{\vec{x}_{1},\vec{x}_{2}}
\Bigg(\bigg\langle\widehat{\langle\vec{x},t_{p}}^{a}|\hat{G}^{(0)}[\hat{\sigma}_{D}^{(0)}]\;
\big(\wtilde{\pp}_{i}\hat{u}+\wtilde{\pp}_{i}\hat{\sigma}_{D}^{(0)}\big)\;
\hat{G}^{(0)}[\hat{\sigma}_{D}^{(0)}]|\widehat{\vec{x}_{1},t_{q}\ppr\rangle}^{a_{1}} \;\;\times
\\ \no &\times&
\widehat{\langle\vec{x}_{2},t_{q}\ppr}^{b_{1}}|\hat{G}^{(0)}[\hat{\sigma}_{D}^{(0)}]|
\widehat{\vec{x}\ppr,t_{p}\rangle}^{b}\bigg\rangle_{\hat{\sigma}_{D}^{(0)}} +
\bigg\langle\widehat{\langle\vec{x},t_{p}}^{a}|
\hat{G}^{(0)}[\hat{\sigma}_{D}^{(0)}]|\widehat{\vec{x}_{1},t_{q}\ppr\rangle}^{a_{1}} \;\; \times
\\ \no &\times&
\widehat{\langle\vec{x}_{2},t_{q}\ppr}^{b_{1}}|\hat{G}^{(0)}[\hat{\sigma}_{D}^{(0)}]\;
\big(\wtilde{\pp}_{i}\hat{u}+\wtilde{\pp}_{i}\hat{\sigma}_{D}^{(0)}\big)\;
\hat{G}^{(0)}[\hat{\sigma}_{D}^{(0)}]|
\widehat{\vec{x}\ppr,t_{p}\rangle}^{b}\bigg\rangle_{\hat{\sigma}_{D}^{(0)}} \Bigg) \times\;\eta_{q}\;
\delta\hat{\Sigma}_{\vec{x}_{1},\alpha;\vec{x}_{2},\beta}^{a_{1}\neq b_{1}}(t_{q}\ppr)
\bigg|_{\vec{x}_{1}=\vec{x}_{2}}^{a_{1}\neq b_{1}}\bigg|_{\vec{x}=\vec{x}\ppr}^{a\neq b}\;\;\;.
\eeq
We can summarize the propagation of anomalous-like matrices with the average over the product
of two Green functions by using the definitions in subsection \ref{s42} for the doubled Hilbert
space states with anti-unitary part. Relation (\ref{s4_110}) is abbreviated by (\ref{s4_116}) with
the average of \(Z[j_{\psi};\hat{\sigma}_{D}^{(0)}]\) for the two Green functions
\beq \lb{s4_116}
\frac{1}{\mcal{N}^{2}}\;\delta\hat{\Sigma}_{\alpha\beta}^{a\neq b}(\vec{x},t_{p})\Big|^{a\neq b} &=&
\bigg\langle\widehat{\langle\vec{x},t_{p}}^{a}|\hat{G}^{(0)}[\hat{\sigma}_{D}^{(0)}]
\;\delta\hat{\Sigma}_{\alpha\beta}^{a\neq b}\;\hat{G}^{(0)}[\hat{\sigma}_{D}^{(0)}]|
\widehat{\vec{x},t_{p}\rangle}^{b} \bigg\rangle_{\hat{\sigma}_{D}^{(0)}} \;\;\;.
\eeq
Similarly, we reduce the rule (\ref{s4_115}) for the propagation of derivatives of anomalous-like
terms to its abbreviated form (\ref{s4_117}) with the definitions of the doubled Hilbert space
in subsection \ref{s42}
\beq \lb{s4_117}
\lefteqn{\frac{1}{\mcal{N}^{2}}\;
\Big(\wtilde{\pp}_{i}\delta\hat{\Sigma}_{\alpha\beta}^{a\neq b}(\vec{x},t_{p})\Big)\Big|^{a\neq b} =
\bigg\langle\widehat{\langle\vec{x},t_{p}}^{a}|\hat{G}^{(0)}[\hat{\sigma}_{D}^{(0)}]
\;\Big(\wtilde{\pp}_{i}\delta\hat{\Sigma}_{\alpha\beta}^{a\neq b}\Big)\;
\hat{G}^{(0)}[\hat{\sigma}_{D}^{(0)}]|
\widehat{\vec{x},t_{p}\rangle}^{b} \bigg\rangle_{\hat{\sigma}_{D}^{(0)}} +   }  \\ \no &-&
\bigg\langle\widehat{\langle\vec{x},t_{p}}^{a}|\hat{G}^{(0)}[\hat{\sigma}_{D}^{(0)}]\;
\big(\wtilde{\pp}_{i}\hat{u}+\wtilde{\pp}_{i}\hat{\sigma}_{D}^{(0)}\big)\;
\hat{G}^{(0)}[\hat{\sigma}_{D}^{(0)}]\;\delta\hat{\Sigma}_{\alpha\beta}^{a\neq b}\;
\hat{G}^{(0)}[\hat{\sigma}_{D}^{(0)}]|
\widehat{\vec{x},t_{p}\rangle}^{b}\bigg\rangle_{\hat{\sigma}_{D}^{(0)}} +
\\ \no &-&
\bigg\langle\widehat{\langle\vec{x},t_{p}}^{a}|
\hat{G}^{(0)}[\hat{\sigma}_{D}^{(0)}]\;\delta\hat{\Sigma}_{\alpha\beta}^{a\neq b}\;
\hat{G}^{(0)}[\hat{\sigma}_{D}^{(0)}]\;
\big(\wtilde{\pp}_{i}\hat{u}+\wtilde{\pp}_{i}\hat{\sigma}_{D}^{(0)}\big)\;
\hat{G}^{(0)}[\hat{\sigma}_{D}^{(0)}]|
\widehat{\vec{x},t_{p}\rangle}^{b}\bigg\rangle_{\hat{\sigma}_{D}^{(0)}}  \;\;\;.
\eeq
It is of crucial importance to distinguish between matrix elements of the self-energy
\(\widehat{\langle\vec{x},t_{p}}^{a}|\delta\hat{\Sigma}_{\alpha\beta}^{ab}|
\widehat{\vec{x}\ppr,t_{q}\ppr\rangle}^{b}=\mbox{{\boldmath$\eta_{p}$}}\;
\delta_{p,q}\;\delta_{t_{p},t_{q}\ppr}\;\delta_{\vec{x},\vec{x}\ppr}\;\;
\delta\hat{\Sigma}_{\alpha\beta}^{ab}(\vec{x},t_{p})\)
and matrix elements of the coset matrix
\(\widehat{\langle\vec{x},t_{p}}^{a}|\big[\hat{T}^{-1}\big(\wtilde{\pp}_{i}\hat{T}\big)\big]_{\alpha\beta}^{ab}|
\widehat{\vec{x}\ppr,t_{q}\ppr\rangle}^{b}=
\delta_{p,q}\;\delta_{t_{p},t_{q}\ppr}\;\delta_{\vec{x},\vec{x}\ppr}\;\;
\big[\hat{T}^{-1}(\vec{x},t_{p})\;\big(\wtilde{\pp}_{i}\hat{T}(\vec{x},t_{p})\big)\big]_{\alpha\beta}^{ab}\).
The latter do not involve the contour metric $\hat{\eta}$ as the matrix elements of the self-energy.
However, the gradient term
\(\delta\hat{\mcal{H}}(\hat{T}^{-1},\hat{T})=\hat{T}^{-1}\;\hat{\mcal{H}}\;\hat{T}-\hat{\mcal{H}}\)
(\ref{s4_75},\ref{s4_81}) contains the 'Nambu'-doubled one-particle operator \(\hat{\mcal{H}}\)
whose matrix elements include the time contour metric $\hat{\eta}$ (\ref{s4_61},\ref{s2_55}-\ref{s2_57}).
Therefore, we have to substitute in relations (\ref{s4_104}-\ref{s4_117})
the (derivatives of the) operator \(\delta\hat{\Sigma}_{\alpha\beta}^{ab}\) of the self-energy
by the operator \(\hat{\eta}\;\big[\hat{T}^{-1}\big(\wtilde{\pp}_{i}\hat{T}\big)\big]_{\alpha\beta}^{ab}\)
of the coset matrices having an additional time contour metric \(\hat{\eta}\).

\subsection{Expansion of the super-determinant term
$\big\langle\mcal{A}_{SDET}\ppr\big[\hat{T},\hat{\sigma}_{D}^{(0)};
\hat{\mcal{J}}\big]\big\rangle_{\hat{\sigma}_{D}^{(0)}}$} \lb{s45}

The rules of the previous subsection are applied to simplify the action
\(\big\langle\mcal{A}_{SDET}\ppr\big[\hat{T},\hat{\sigma}_{D}^{(0)};
\hat{\mcal{J}}\big]\big\rangle_{\hat{\sigma}_{D}^{(0)}}\) with the
averaging of \(Z[j_{\psi};\hat{\sigma}_{D}^{(0)}]\) and the
density \(\sigma_{D}^{(0)}(\vec{x},t_{p})\) as background field. The
gradient expansion results in an action \(\mcal{A}_{SDET}\ppr\big[\hat{T};
\hat{\mcal{J}}\big]\) for the Goldstone modes with gradients up to second
order in \(\Delta\hat{\mcal{O}}=\delta\hat{\mcal{H}}(\hat{T}^{-1},\hat{T}) +
\wtilde{\mcal{J}}(\hat{T}^{-1},\hat{T})\), including coefficient functions
which follow from the average with \(Z[j_{\psi};\hat{\sigma}_{D}^{(0)}]\) (\ref{s4_90})
over the background field \(\sigma_{D}^{(0)}(\vec{x},t_{p})\) and its derivative
\beq \lb{s4_118}
\lefteqn{\hspace*{-1.0cm}
\mcal{A}_{SDET}\ppr\big[\hat{T};
\hat{\mcal{J}}\big]=
\Big\langle\mcal{A}_{SDET}\ppr\big[\hat{T},\hat{\sigma}_{D}^{(0)};
\hat{\mcal{J}}\big]\Big\rangle_{\hat{\sigma}_{D}^{(0)}} \approx
\frac{1}{2}\;\mbox{Tr}\bigg[\mbox{STR}\bigg(\Delta\hat{\mcal{O}}\;\;
\Big\langle\hat{G}^{(0)}[\hat{\sigma}_{D}^{(0)}]\Big\rangle_{\hat{\sigma}_{D}^{(0)}}
\bigg)\bigg] +   }   \\ \no &-& \frac{1}{4}\;
\bigg\langle\mbox{Tr}\bigg[\mbox{STR}\bigg(
\Delta\hat{\mcal{O}}\;\;\hat{G}^{(0)}[\hat{\sigma}_{D}^{(0)}]\;\;
\Delta\hat{\mcal{O}}\;\;\hat{G}^{(0)}[\hat{\sigma}_{D}^{(0)}]
\bigg)\bigg]\bigg\rangle_{\hat{\sigma}_{D}^{(0)}} \pm \ldots
\eeq
\beq \lb{s4_119}
\Delta\hat{\mcal{O}}&=&\delta\hat{\mcal{H}}(\hat{T}^{-1},\hat{T}) +
\wtilde{\mcal{J}}(\hat{T}^{-1},\hat{T})     \\ \lb{s4_120}
\delta\hat{\mcal{H}}(\hat{T}^{-1},\hat{T}) &=&
-\hat{\eta}\Big(\hat{T}^{-1}\;\hat{S}\;\big(E_{p}\hat{T}\big)+\hat{T}^{-1}\;
\big(\wtilde{\pp}_{i}\wtilde{\pp}_{i}\hat{T}\big)  +
\big(\hat{T}^{-1}\;\hat{S}\;\hat{T}-\hat{S}\big)\;\mbox{{\boldmath$\hat{E}_{p}$}}+
2\;\hat{T}^{-1}\;\big(\wtilde{\pp}_{i}\hat{T}\big)\;
\mbox{{\boldmath$\wtilde{\pp}_{i}$}}\Big).
\eeq
We describe in this subsection details of the gradient expansion which may be left out and
refer the uninterested reader to the resulting equations
(\ref{s4_143}-\ref{s4_146},\ref{s4_138}),(\ref{s4_150}-\ref{s4_153}) for the second order term
\(\big(\Delta\hat{\mcal{O}}\big)^{2}\) and to the equations (\ref{s4_162},\ref{s4_163}-\ref{s4_169}) for the
first order term. The mentioned results for the expansion (\ref{s4_118}) are obtained
by consequent application of the rules for propagation, listed in subsection \ref{s44}.
At first we consider the expansion of the second order term
\(\big(\Delta\hat{\mcal{O}}\big)^{2}\) in \(\mcal{A}_{SDET}\ppr[\hat{T};\hat{\mcal{J}}]\) (\ref{s4_118})
\beq\lb{s4_121}
\lefteqn{\bigg\langle\mbox{Tr}\bigg[\mbox{STR}\bigg(
\Delta\hat{\mcal{O}}\;\;\hat{G}^{(0)}[\hat{\sigma}_{D}^{(0)}]\;\;
\Delta\hat{\mcal{O}}\;\;\hat{G}^{(0)}[\hat{\sigma}_{D}^{(0)}]
\bigg)\bigg]\bigg\rangle_{\hat{\sigma}_{D}^{(0)}}  = } \\ \no &=&
\bigg\langle\mbox{Tr}\bigg[\mbox{STR}\bigg(
\delta\hat{\mcal{H}}(\hat{T}^{-1},\hat{T}) \;\;\hat{G}^{(0)}[\hat{\sigma}_{D}^{(0)}]\;\;
\delta\hat{\mcal{H}}(\hat{T}^{-1},\hat{T})\;\;\hat{G}^{(0)}[\hat{\sigma}_{D}^{(0)}]
\bigg)\bigg]\bigg\rangle_{\hat{\sigma}_{D}^{(0)}} + \\ \no &+& 2\;
\bigg\langle\mbox{Tr}\bigg[\mbox{STR}\bigg(
\delta\hat{\mcal{H}}(\hat{T}^{-1},\hat{T}) \;\;\hat{G}^{(0)}[\hat{\sigma}_{D}^{(0)}]\;\;
\wtilde{\mcal{J}}(\hat{T}^{-1},\hat{T})\;\;\hat{G}^{(0)}[\hat{\sigma}_{D}^{(0)}]
\bigg)\bigg]\bigg\rangle_{\hat{\sigma}_{D}^{(0)}} +  \\ \no &+&
\bigg\langle\mbox{Tr}\bigg[\mbox{STR}\bigg(
\wtilde{\mcal{J}}(\hat{T}^{-1},\hat{T})\;\;\hat{G}^{(0)}[\hat{\sigma}_{D}^{(0)}]\;\;
\wtilde{\mcal{J}}(\hat{T}^{-1},\hat{T})\;\;\hat{G}^{(0)}[\hat{\sigma}_{D}^{(0)}]
\bigg)\bigg]\bigg\rangle_{\hat{\sigma}_{D}^{(0)}} \;\;\;.
\eeq
In the following we concentrate on the second order term
\(\big(\delta\hat{\mcal{H}}(\hat{T}^{-1},\hat{T})\big)^{2}\) and keep only gradients up
to second order in spatial derivatives and up to first order in the time derivative
\beq \lb{s4_122}
\lefteqn{\bigg\langle\mbox{Tr}\bigg[\mbox{STR}\bigg(
\delta\hat{\mcal{H}}(\hat{T}^{-1},\hat{T}) \;\;\hat{G}^{(0)}[\hat{\sigma}_{D}^{(0)}]\;\;
\delta\hat{\mcal{H}}(\hat{T}^{-1},\hat{T})\;\;\hat{G}^{(0)}[\hat{\sigma}_{D}^{(0)}]
\bigg)\bigg]\bigg\rangle_{\hat{\sigma}_{D}^{(0)}}  = } \\ \no &=&
4\;\bigg\langle\mbox{Tr}\bigg[\mbox{STR}\bigg(\hat{\eta}\;
\hat{T}^{-1}\;\big(\wtilde{\pp}_{i}\hat{T}\big)\;\mbox{{\boldmath$\wtilde{\pp}_{i}$}}\;
\hat{G}^{(0)}[\hat{\sigma}_{D}^{(0)}]\;\hat{\eta}\;
\hat{T}^{-1}\;\big(\wtilde{\pp}_{j}\hat{T}\big)\;\mbox{{\boldmath$\wtilde{\pp}_{j}$}}\;
\hat{G}^{(0)}[\hat{\sigma}_{D}^{(0)}] \bigg)\bigg]\bigg\rangle_{\hat{\sigma}_{D}^{(0)}} +
\\ \no &+& 4\;
\bigg\langle\mbox{Tr}\bigg[\mbox{STR}\bigg(\hat{\eta}\;
\hat{T}^{-1}\;\big(\wtilde{\pp}_{i}\hat{T}\big)\;\mbox{{\boldmath$\wtilde{\pp}_{i}$}}\;
\hat{G}^{(0)}[\hat{\sigma}_{D}^{(0)}]\;\hat{\eta}\;
\big(\hat{T}^{-1}\;\hat{S}\;\hat{T}-\hat{S}\big)\;\mbox{{\boldmath$\hat{E}_{p}$}}\;
\hat{G}^{(0)}[\hat{\sigma}_{D}^{(0)}] \bigg)\bigg]\bigg\rangle_{\hat{\sigma}_{D}^{(0)}} +
\\ \no &+& 2\;
\bigg\langle\mbox{Tr}\bigg[\mbox{STR}\bigg(\hat{\eta}\;
\hat{T}^{-1}\;\big(\wtilde{\pp}_{i}\wtilde{\pp}_{i}\hat{T}\big)\;
\hat{G}^{(0)}[\hat{\sigma}_{D}^{(0)}]\;\hat{\eta}\;
\big(\hat{T}^{-1}\;\hat{S}\;\hat{T}-\hat{S}\big)\;\mbox{{\boldmath$\hat{E}_{p}$}}\;
\hat{G}^{(0)}[\hat{\sigma}_{D}^{(0)}] \bigg)\bigg]\bigg\rangle_{\hat{\sigma}_{D}^{(0)}} +
\\ \no &+& 2\;
\bigg\langle\mbox{Tr}\bigg[\mbox{STR}\bigg(\hat{\eta}\;
\hat{T}^{-1}\;\hat{S}\;\big(E_{p}\hat{T}\big)\;
\hat{G}^{(0)}[\hat{\sigma}_{D}^{(0)}]\;\hat{\eta}\;
\big(\hat{T}^{-1}\;\hat{S}\;\hat{T}-\hat{S}\big)\;\mbox{{\boldmath$\hat{E}_{p}$}}\;
\hat{G}^{(0)}[\hat{\sigma}_{D}^{(0)}] \bigg)\bigg]\bigg\rangle_{\hat{\sigma}_{D}^{(0)}} +
\\  \no &+&
\bigg\langle\mbox{Tr}\bigg[\mbox{STR}\bigg(\hat{\eta}\;
\big(\hat{T}^{-1}\;\hat{S}\;\hat{T}-\hat{S}\big)\;\mbox{{\boldmath$\hat{E}_{p}$}}\;
\hat{G}^{(0)}[\hat{\sigma}_{D}^{(0)}]\;\hat{\eta}\;
\big(\hat{T}^{-1}\;\hat{S}\;\hat{T}-\hat{S}\big)\;\mbox{{\boldmath$\hat{E}_{p}$}}\;
\hat{G}^{(0)}[\hat{\sigma}_{D}^{(0)}] \bigg)\bigg]\bigg\rangle_{\hat{\sigma}_{D}^{(0)}} + \ldots\;.
\eeq
We list again the commutators of \(\mbox{{\boldmath$\wtilde{\pp}_{i}$}}\),
\(\mbox{{\boldmath$\hat{E}_{p}$}}\) with \(\hat{G}^{(0)}[\hat{\sigma}_{D}^{(0)}]\)
(\ref{s4_123},\ref{s4_124}) and have to specify the transpose of gradient operators which we
abbreviate by a common symbol \(\mbox{{\boldmath$\hat{\mcal{D}}$}}\) (\ref{s4_125},\ref{s4_126}).
It has already been described in subsection \ref{s44} that the propagation of matrices with the
product of two Green functions has only contributions for the anomalous terms in the Hilbert
state traces '\(\mbox{Tr}\)' as of \(\mcal{A}_{SDET}\ppr\) (\ref{s4_118}).
Therefore, matrix elements with a gradient operator and Green function can be changed
from index $a$ to $b$ with \(a\neq b\) as in Eq. (\ref{s4_127})
where the commutator in (\ref{s4_127}) can be simplified by relations
(\ref{s4_123}-\ref{s4_126})
\beq \lb{s4_123}
\Big[\mbox{{\boldmath$\wtilde{\pp}_{i}$}}\;,\;\hat{G}^{(0)}[\hat{\sigma}_{D}^{(0)}]\Big]&=&
-\hat{G}^{(0)}[\hat{\sigma}_{D}^{(0)}]\;\Big(\wtilde{\pp}_{i}\hat{u}+\wtilde{\pp}_{i}
\hat{\sigma}_{D}^{(0)}\Big)\;\hat{G}^{(0)}[\hat{\sigma}_{D}^{(0)}]  \\ \lb{s4_124}
\Big[\mbox{{\boldmath$\hat{E}_{p}$}}\;,\;\hat{G}^{(0)}[\hat{\sigma}_{D}^{(0)}]\Big]&=&
-\hat{G}^{(0)}[\hat{\sigma}_{D}^{(0)}]\;
\Big(E_{p}\hat{\sigma}_{D}^{(0)}\Big)\;\hat{G}^{(0)}[\hat{\sigma}_{D}^{(0)}]
\eeq
\beq \lb{s4_125}
\mbox{{\boldmath$\hat{\mcal{D}}$}} &:=&\mbox{{\boldmath$\wtilde{\pp}_{i}$}},\;
\mbox{{\boldmath$\hat{E}_{p}$}},\;\mbox{{\boldmath$\wtilde{\pp}_{i}$}}
\mbox{{\boldmath$\wtilde{\pp}_{j}$}}  \\ \lb{s4_126}
\Big(\mbox{{\boldmath$\wtilde{\pp}_{i}$}}\Big)^{T} &=& -
\mbox{{\boldmath$\wtilde{\pp}_{i}$}}  \hspace*{0.75cm}
\Big(\mbox{{\boldmath$\hat{E}_{p}$}}\Big)^{T}=-\mbox{{\boldmath$\hat{E}_{p}$}}\hspace*{0.75cm}
\Big(\mbox{{\boldmath$\wtilde{\pp}_{i}$}}\mbox{{\boldmath$\wtilde{\pp}_{j}$}}\Big)^{T} =
\mbox{{\boldmath$\wtilde{\pp}_{i}$}}\mbox{{\boldmath$\wtilde{\pp}_{j}$}}
\eeq
\beq \lb{s4_127}
\lefteqn{\widehat{\langle\vec{x}\ppr,t_{q}\ppr}^{a}|\Big(\mbox{{\boldmath$\hat{\mcal{D}}$}}\;
\hat{G}^{(0)}[\hat{\sigma}_{D}^{(0)}]\Big)^{aa}|\widehat{\vec{x},t_{p}\rangle}^{a}=
\widehat{\langle\vec{x},t_{p}}^{b}|\Big(\mbox{{\boldmath$\hat{\mcal{D}}$}}\;
\hat{G}^{(0)}[\hat{\sigma}_{D}^{(0)}]\Big)^{aa,T}|
\widehat{\vec{x}\ppr,t_{q}\ppr\rangle}^{b}\Big|^{b\neq a}= } \\ \no &=&
\widehat{\langle\vec{x},t_{p}}^{b}|\Big(\hat{G}^{(0)}[\hat{\sigma}_{D}^{(0)}]\Big)^{bb}\;
\mbox{{\boldmath$\hat{\mcal{D}}$}}^{T}|\widehat{\vec{x}\ppr,t_{q}\ppr\rangle}^{b}\Big|^{b\neq a}=
\widehat{\langle\vec{x},t_{p}}^{b}|\mbox{{\boldmath$\hat{\mcal{D}}$}}^{T}\;
\hat{G}^{(0)}[\hat{\sigma}_{D}^{(0)}]+
\Big[\hat{G}^{(0)}[\hat{\sigma}_{D}^{(0)}]\;,\;
\mbox{{\boldmath$\hat{\mcal{D}}$}}^{T}\Big]|
\widehat{\vec{x}\ppr,t_{q}\ppr\rangle}^{b}\Big|^{b\neq a}\;\;\;.
\eeq
The most important part for the expansion of the Goldstone modes is determined by the first term
with two spatial derivatives in (\ref{s4_122})
\beq \lb{s4_128}
\lefteqn{\bigg\langle\mbox{Tr}\bigg[\mbox{STR}\bigg(\hat{\eta}\;\hat{T}^{-1}\;
\big(\wtilde{\pp}_{i}\hat{T}\big)\;
\mbox{{\boldmath$\wtilde{\pp}_{i}$}}\;\hat{G}^{(0)}[\hat{\sigma}_{D}^{(0)}]\;\hat{\eta}\;
\hat{T}^{-1}\;\big(\wtilde{\pp}_{j}\hat{T}\big)\;
\mbox{{\boldmath$\wtilde{\pp}_{j}$}}\;\hat{G}^{(0)}[\hat{\sigma}_{D}^{(0)}]
\bigg)\bigg]\bigg\rangle_{\hat{\sigma}_{D}^{(0)}} = } \\ \no &=&
\int_{C}\frac{d t_{p}}{\hbar}\;\frac{d t_{q}\ppr}{\hbar}\;\mcal{N}^{2}\;
\sum_{\vec{x},\vec{x}\ppr}\sum_{a,b=1,2}^{(a\neq b)}
\strab\bigg(\Big[\hat{T}^{-1}(\vec{x},t_{p})\;
\big(\wtilde{\pp}_{i}\hat{T}(\vec{x},t_{p})\big)\Big]_{\alpha\beta}^{a\neq b} \;\;
\Big[\hat{T}^{-1}(\vec{x}\ppr,t_{q}\ppr)\;
\big(\wtilde{\pp}_{j}\hat{T}(\vec{x}\ppr,t_{q}\ppr)\big)\Big]_{\beta\alpha}^{b\neq a} \bigg)\times
\\ \no &\times&
\bigg\langle\widehat{\langle\vec{x},t_{p}}^{b}|\;\mbox{{\boldmath$\wtilde{\pp}_{i}$}}\;
\hat{G}^{(0)}[\hat{\sigma}_{D}^{(0)}]|\widehat{\vec{x}\ppr,t_{q}\ppr\rangle}^{b}\;\;
\widehat{\langle\vec{x}\ppr,t_{q}\ppr}^{a}|\;\mbox{{\boldmath$\wtilde{\pp}_{j}$}}\;
\hat{G}^{(0)}[\hat{\sigma}_{D}^{(0)}]|
\widehat{\vec{x},t_{p}\rangle}^{a}\bigg\rangle_{\hat{\sigma}_{D}^{(0)}}\bigg|^{a\neq b}\;\;\;.
\eeq
The second matrix element with spatial gradient operator
\(\mbox{{\boldmath$\wtilde{\pp}_{j}$}}\) and Green function in (\ref{s4_128}) is transformed
with relations (\ref{s4_123}-\ref{s4_127})
\beq \lb{s4_129}
\lefteqn{\widehat{\langle\vec{x}\ppr,t_{q}\ppr}^{a}|\;\mbox{{\boldmath$\wtilde{\pp}_{j}$}}\;
\hat{G}^{(0)}[\hat{\sigma}_{D}^{(0)}]|\widehat{\vec{x},t_{p}\rangle}^{a} = } \\ \no &=& -\;
\widehat{\langle\vec{x},t_{p}}^{b}|\;\mbox{{\boldmath$\wtilde{\pp}_{j}$}}\;
\hat{G}^{(0)}[\hat{\sigma}_{D}^{(0)}] +
\hat{G}^{(0)}[\hat{\sigma}_{D}^{(0)}]\;
\Big(\wtilde{\pp}_{j}\hat{u}+\wtilde{\pp}_{j}\hat{\sigma}_{D}^{(0)}\Big)\;
\hat{G}^{(0)}[\hat{\sigma}_{D}^{(0)}]\;
|\widehat{\vec{x}\ppr,t_{q}\ppr\rangle}^{b}\bigg|^{b\neq a}  \;\;\;.
\eeq
Substitution of Eq. (\ref{s4_129}) into (\ref{s4_128}) yields relation (\ref{s4_130}) where we
exchange the doubled, diagonal matrix element, given by the first line in (\ref{s4_131}) with index
$b$, to the transposed form having diagonal index $a$ (with \(a\neq b\), \(a,b=1,2\))
\beq \lb{s4_130}
\lefteqn{\bigg\langle\mbox{Tr}\bigg[\mbox{STR}\bigg(\hat{\eta}\;\hat{T}^{-1}\;\big(\wtilde{\pp}_{i}\hat{T}\big)\;
\mbox{{\boldmath$\wtilde{\pp}_{i}$}}\;\hat{G}^{(0)}[\hat{\sigma}_{D}^{(0)}]\;\hat{\eta}\;
\hat{T}^{-1}\;\big(\wtilde{\pp}_{j}\hat{T}\big)\;
\mbox{{\boldmath$\wtilde{\pp}_{j}$}}\;\hat{G}^{(0)}[\hat{\sigma}_{D}^{(0)}]
\bigg)\bigg] \bigg\rangle_{\hat{\sigma}_{D}^{(0)}} = } \\ \no &=&-
\int_{C}\frac{d t_{p}}{\hbar}\;\frac{d t_{q}\ppr}{\hbar}\;\mcal{N}^{2}\;
\sum_{\vec{x},\vec{x}\ppr}\sum_{a,b=1,2}^{(a\neq b)}
\strab\bigg(\Big[\hat{T}^{-1}(\vec{x},t_{p})\;
\big(\wtilde{\pp}_{i}\hat{T}(\vec{x},t_{p})\big)\Big]_{\alpha\beta}^{a\neq b} \;\;
\Big[\hat{T}^{-1}(\vec{x}\ppr,t_{q}\ppr)\;
\big(\wtilde{\pp}_{j}\hat{T}(\vec{x}\ppr,t_{q}\ppr)\big)\Big]_{\beta\alpha}^{b\neq a} \bigg)\times
\\ \no &\times&\hspace*{-0.37cm}
\bigg\langle\widehat{\langle\vec{x},t_{p}}^{b}|\;\mbox{{\boldmath$\wtilde{\pp}_{i}$}}\;
\hat{G}^{(0)}[\hat{\sigma}_{D}^{(0)}]|\widehat{\vec{x}\ppr,t_{q}\ppr\rangle}^{b}\;\;
\widehat{\langle\vec{x},t_{p}}^{b}|\;\mbox{{\boldmath$\wtilde{\pp}_{j}$}}\;
\hat{G}^{(0)}[\hat{\sigma}_{D}^{(0)}]+\hat{G}^{(0)}[\hat{\sigma}_{D}^{(0)}]\;
\Big(\wtilde{\pp}_{j}\hat{u}+\wtilde{\pp}_{j}\hat{\sigma}_{D}^{(0)}\Big)\;
\hat{G}^{(0)}[\hat{\sigma}_{D}^{(0)}]\;
|\widehat{\vec{x}\ppr,t_{q}\ppr\rangle}^{b}\bigg\rangle_{\hat{\sigma}_{D}^{(0)}}
\eeq
\beq \lb{s4_131}
\lefteqn{\bigg\langle\widehat{\langle\vec{x},t_{p}}^{b}|\;\hat{G}^{(0)}[\hat{\sigma}_{D}^{(0)}]\;
\Big(\wtilde{\pp}_{j}\hat{u}+\wtilde{\pp}_{j}\hat{\sigma}_{D}^{(0)}\Big)\;
\hat{G}^{(0)}[\hat{\sigma}_{D}^{(0)}]\;|\widehat{\vec{x}\ppr,t_{q}\ppr\rangle}^{b}
\bigg\rangle_{\hat{\sigma}_{D}^{(0)}} =} \\ \no &=&
\bigg\langle\widehat{\langle\vec{x}\ppr,t_{q}\ppr}^{a}|\;\hat{G}^{(0)}[\hat{\sigma}_{D}^{(0)}]\;
\Big(\wtilde{\pp}_{j}\hat{u}+\wtilde{\pp}_{j}\hat{\sigma}_{D}^{(0)}\Big)\;
\hat{G}^{(0)}[\hat{\sigma}_{D}^{(0)}]\;|
\widehat{\vec{x},t_{p}\rangle}^{a}\bigg\rangle_{\hat{\sigma}_{D}^{(0)}}
\bigg|^{a\neq b}\;\;\;.
\eeq
The matrix element of Eq.(\ref{s4_131}) can be reduced to
\(\big\langle\big(\wtilde{\pp}_{j}\ppr u(\vec{x}\ppr)+
\wtilde{\pp}_{j}\ppr\sigma_{D}^{(0)}(\vec{x}\ppr,t_{q}\ppr)\big)\;
\widehat{\langle\vec{x}\ppr,t_{q}\ppr}^{a}|\;\hat{G}^{(0)}[\hat{\sigma}_{D}^{(0)}]\;|
\widehat{\vec{x},t_{p}\rangle}^{a}\big\rangle_{\hat{\sigma}_{D}^{(0)}}\) or to
\(\big\langle\widehat{\langle\vec{x}\ppr,t_{q}\ppr}^{a}|\;\hat{G}^{(0)}[\hat{\sigma}_{D}^{(0)}]\;|
\widehat{\vec{x},t_{p}\rangle}^{a}\;\big(\wtilde{\pp}_{j} u(\vec{x})+
\wtilde{\pp}_{j}\sigma_{D}^{(0)}(\vec{x},t_{p})\big)\big\rangle_{\hat{\sigma}_{D}^{(0)}}\)
according to the rules in subsection \ref{s44} with remaining Green functions for
propagation of derivatives of the coset matrices \(\hat{T}^{-1}\), \(\hat{T}\). Apart from
matrix element (\ref{s4_131}), the gradient expansion for
\(\big(\delta\hat{\mcal{H}}(\hat{T}^{-1},\hat{T})\big)^{2}\) contains the average of two
Green functions each multiplied by a spatial gradient operator (\ref{s4_132}).
We also transform this matrix element with (\ref{s4_127}) and have to take
into account that the range of saturated operators \(\wtilde{\pp}_{i}\),
\(\wtilde{\pp}_{j}\) (without bold letters) is restricted to the prevailing braces, as e.g. in
\(\big(\wtilde{\pp}_{i}\;\wtilde{\pp}_{j}\ldots\big)\)
\beq \lb{s4_132}
\lefteqn{\bigg\langle\widehat{\langle\vec{x},t_{p}}^{b}|\;\mbox{{\boldmath$\wtilde{\pp}_{i}$}}\;
\hat{G}^{(0)}[\hat{\sigma}_{D}^{(0)}]|\widehat{\vec{x}\ppr,t_{q}\ppr\rangle}^{b}\;\;
\widehat{\langle\vec{x},t_{p}}^{b}|\;\mbox{{\boldmath$\wtilde{\pp}_{j}$}}\;
\hat{G}^{(0)}[\hat{\sigma}_{D}^{(0)}]|
\widehat{\vec{x}\ppr,t_{q}\ppr\rangle}^{b}\bigg\rangle_{\hat{\sigma}_{D}^{(0)}} = } \\ \no &=&
\bigg\langle\frac{1}{2}\;\bigg(\wtilde{\pp}_{i}\wtilde{\pp}_{j}\;\Big(
\widehat{\langle\vec{x},t_{p}}^{b}|\;\hat{G}^{(0)}[\hat{\sigma}_{D}^{(0)}]\;
|\widehat{\vec{x}\ppr,t_{q}\ppr\rangle}^{b}\Big)^{2}\bigg) + \\ \no &-&
\widehat{\langle\vec{x},t_{p}}^{b}|\;\mbox{{\boldmath$\wtilde{\pp}_{i}$}}\;
\mbox{{\boldmath$\wtilde{\pp}_{j}$}}\;\hat{G}^{(0)}[\hat{\sigma}_{D}^{(0)}]\;
|\widehat{\vec{x}\ppr,t_{q}\ppr\rangle}^{b}\;\;
\widehat{\langle\vec{x},t_{p}}^{b}|\;\hat{G}^{(0)}[\hat{\sigma}_{D}^{(0)}]\;
|\widehat{\vec{x}\ppr,t_{q}\ppr\rangle}^{b} \bigg\rangle_{\hat{\sigma}_{D}^{(0)}}= \\ \no &\stackrel{(a\neq b)}{=}&
\bigg\langle\frac{1}{2}\;\bigg(\wtilde{\pp}_{i}\wtilde{\pp}_{j}\;\Big(
\widehat{\langle\vec{x},t_{p}}^{b}|\;\hat{G}^{(0)}[\hat{\sigma}_{D}^{(0)}]\;
|\widehat{\vec{x}\ppr,t_{q}\ppr\rangle}^{b}\Big)\;
\Big(\widehat{\langle\vec{x}\ppr,t_{q}\ppr}^{a}|\;\hat{G}^{(0)}[\hat{\sigma}_{D}^{(0)}]\;
|\widehat{\vec{x},t_{p}\rangle}^{a}\Big)
\bigg) + \\ \no &-& \frac{1}{2}
\widehat{\langle\vec{x},t_{p}}^{b}|\;\mbox{{\boldmath$\wtilde{\pp}_{i}$}}\;
\mbox{{\boldmath$\wtilde{\pp}_{j}$}}\;\hat{G}^{(0)}[\hat{\sigma}_{D}^{(0)}]\;
|\widehat{\vec{x}\ppr,t_{q}\ppr\rangle}^{b}\;\;
\widehat{\langle\vec{x}\ppr,t_{q}\ppr}^{a}|\;\hat{G}^{(0)}[\hat{\sigma}_{D}^{(0)}]\;
|\widehat{\vec{x},t_{p}\rangle}^{a} + \\ \no &-& \frac{1}{2}
\widehat{\langle\vec{x},t_{p}}^{b}|\;\hat{G}^{(0)}[\hat{\sigma}_{D}^{(0)}]\;
|\widehat{\vec{x}\ppr,t_{q}\ppr\rangle}^{b}\;\;
\widehat{\langle\vec{x}\ppr,t_{q}\ppr}^{a}|\;\hat{G}^{(0)}[\hat{\sigma}_{D}^{(0)}]\;
\mbox{{\boldmath$\wtilde{\pp}_{i}$}}\;
\mbox{{\boldmath$\wtilde{\pp}_{j}$}}\;|\widehat{\vec{x},t_{p}\rangle}^{a}
\bigg\rangle_{\hat{\sigma}_{D}^{(0)}}\bigg|^{a\neq b}\;\;\;.
\eeq
Insertion of (\ref{s4_131},\ref{s4_132}) into (\ref{s4_130}) leads to Eq. (\ref{s4_133}) which
consists of three super-trace parts '\(\mbox{str}\)' or three 'Nambu'-doubled super-traces
'\(\mbox{STR}\)' with \(a\neq b\).
The doubled Hilbert space states of matrix elements in (\ref{s4_133}) are classified by
the number of {\it unsaturated} spatial operators (see (\ref{s4_134}-\ref{s4_136})).
The first super-trace term in (\ref{s4_133}) with matrix element \(C_{\beta\alpha;j}^{b\neq a}\)
(\ref{s4_134}) does not involve an unsaturated spatial gradient
whereas the second super-trace has one
(\(C_{\beta\alpha}^{b\neq a}(\mbox{{\boldmath$\wtilde{\pp}_{i}$}})\), (\ref{s4_135})) and the
third super-trace part incorporates two spatial unsaturated operators
(\(C_{\beta\alpha;j}^{b\neq a}(\mbox{{\boldmath$\wtilde{\pp}_{i}$}},
\mbox{{\boldmath$\wtilde{\pp}_{j}$}})\), (\ref{s4_136})).
We list these matrix elements in (\ref{s4_133})
with increasing order of unsaturated spatial gradients in following Eqs. (\ref{s4_134}-\ref{s4_136})
\beq \lb{s4_133}
\lefteqn{\bigg\langle\mbox{Tr}\bigg[\mbox{STR}\bigg(\hat{\eta}\;\hat{T}^{-1}\;
\big(\wtilde{\pp}_{i}\hat{T}\big)\;
\mbox{{\boldmath$\wtilde{\pp}_{i}$}}\;\hat{G}^{(0)}[\hat{\sigma}_{D}^{(0)}]\;\hat{\eta}\;
\hat{T}^{-1}\;\big(\wtilde{\pp}_{j}\hat{T}\big)\;
\mbox{{\boldmath$\wtilde{\pp}_{j}$}}\;\hat{G}^{(0)}[\hat{\sigma}_{D}^{(0)}]
\bigg)\bigg]\bigg\rangle_{\hat{\sigma}_{D}^{(0)}} = -
\int_{C}\frac{d t_{p}}{\hbar}\;\mcal{N}\;
\sum_{\vec{x}}\sum_{a,b=1,2}^{(a\neq b)}  }  \\ \no &\times& \mbox{{\boldmath${\ds\Bigg[}$}}
\strab\bigg\{\Big[\hat{T}^{-1}(\vec{x},t_{p})\;
\big(\wtilde{\pp}_{i}\hat{T}(\vec{x},t_{p})\big)\Big]_{\alpha\beta}^{a\neq b} \;
\Big[\frac{1}{2}\;\Big(\wtilde{\pp}_{i}\wtilde{\pp}_{j}\;C_{\beta\alpha;j}^{b\neq a} \Big)
+C_{\beta\alpha}^{b\neq a}(\mbox{{\boldmath$\wtilde{\pp}_{i}$}})
-\frac{1}{2}\;C_{\beta\alpha;j}^{b\neq a}(\mbox{{\boldmath$\wtilde{\pp}_{i}$}},
\mbox{{\boldmath$\wtilde{\pp}_{j}$}}) \Big]_{\beta\alpha}^{b\neq a}
\bigg\} \mbox{{\boldmath${\ds\Bigg]}$}}\Bigg|^{a\neq b}
\eeq
\be \lb{s4_134}
C_{\beta\alpha;j}^{b\neq a}=
\bigg\langle\widehat{\langle\vec{x},t_{p}}^{b}|\;\hat{G}^{(0)}[\hat{\sigma}_{D}^{(0)}]\;\hat{\eta}\;
\Big[\hat{T}^{-1}\;\big(\wtilde{\pp}_{j}\hat{T}\big)\Big]_{\beta\alpha}^{b\neq a}\;
\hat{G}^{(0)}[\hat{\sigma}_{D}^{(0)}]\;
|\widehat{\vec{x},t_{p}\rangle}^{a}\bigg\rangle_{\hat{\sigma}_{D}^{(0)}}=\mbox{ see Eq. (\ref{B14})}
\ee
\beq \no
\lefteqn{\hspace*{-5.5cm}C_{\beta\alpha}^{b\neq a}(\mbox{{\boldmath$\wtilde{\pp}_{i}$}}) =
\bigg\langle\widehat{\langle\vec{x},t_{p}}^{b}|\;\mbox{{\boldmath$\wtilde{\pp}_{i}$}}\;
\hat{G}^{(0)}[\hat{\sigma}_{D}^{(0)}]\;\hat{\eta}\;
\Big[\hat{T}^{-1}\;\big(\wtilde{\pp}_{j}\hat{T}\big)\Big]_{\beta\alpha}^{b\neq a}\;
\hat{G}^{(0)}[\hat{\sigma}_{D}^{(0)}]\;
\Big(\wtilde{\pp}_{j}\hat{u}+\wtilde{\pp}_{j}\hat{\sigma}_{D}^{(0)}\Big)\;
\hat{G}^{(0)}[\hat{\sigma}_{D}^{(0)}]\;
|\widehat{\vec{x},t_{p}\rangle}^{a}\bigg\rangle_{\hat{\sigma}_{D}^{(0)}}  } \\ \lb{s4_135} &=&
\mbox{ see Eq. (\ref{B16})}
\eeq
\beq \lb{s4_136}
\lefteqn{C_{\beta\alpha;j}^{b\neq a}(\mbox{{\boldmath$\wtilde{\pp}_{i}$}},
\mbox{{\boldmath$\wtilde{\pp}_{j}$}})=
\bigg\langle\widehat{\langle\vec{x},t_{p}}^{b}|\;\mbox{{\boldmath$\wtilde{\pp}_{i}$}}\;
\mbox{{\boldmath$\wtilde{\pp}_{j}$}}\;\hat{G}^{(0)}[\hat{\sigma}_{D}^{(0)}]\;\hat{\eta}\;
\Big[\hat{T}^{-1}\;\big(\wtilde{\pp}_{j}\hat{T}\big)\Big]_{\beta\alpha}^{b\neq a}\;
\hat{G}^{(0)}[\hat{\sigma}_{D}^{(0)}]\;
|\widehat{\vec{x},t_{p}\rangle}^{a} + } \\ \no &+&
\widehat{\langle\vec{x},t_{p}}^{b}|\;\hat{G}^{(0)}[\hat{\sigma}_{D}^{(0)}]\;\hat{\eta}\;
\Big[\hat{T}^{-1}\;\big(\wtilde{\pp}_{j}\hat{T}\big)\Big]_{\beta\alpha}^{b\neq a}\;
\hat{G}^{(0)}[\hat{\sigma}_{D}^{(0)}]\;\mbox{{\boldmath$\wtilde{\pp}_{i}$}}\;
\mbox{{\boldmath$\wtilde{\pp}_{j}$}}\;|\widehat{\vec{x},t_{p}\rangle}^{a}
\bigg\rangle_{\hat{\sigma}_{D}^{(0)}}=\mbox{ see Eqs. (\ref{B20},\ref{B21})} \;\;\;.
\eeq
In appendix \ref{sb} we describe how to simplify these matrix elements (\ref{s4_134}-\ref{s4_136})
with the rules for propagation defined in subsection \ref{s44} and
with the corresponding commutators between gradient operators and Green functions.
We include the results of transformations (\ref{B14},\ref{B16},\ref{B20}-\ref{B21}) for the
matrix elements (\ref{s4_134}-\ref{s4_136}) from appendix
\ref{sb} and obtain the approximation (\ref{s4_137},\ref{B28}) for the expansion of Goldstone modes
which is restricted to second order derivatives of
\(\big(\wtilde{\pp}_{i}\hat{T}(\vec{x},t_{p})\big)\), \(\big(\wtilde{\pp}_{j}\hat{T}(\vec{x},t_{p})\big)\).
The trap potential $u(\vec{x})$ and self-energy field $\sigma_{D}^{(0)}(\vec{x},t_{p})$ are summed up to the
field \(v(\vec{x},t_{p})=u(\vec{x})+\sigma_{D}^{(0)}(\vec{x},t_{p})\) which appears as dimensionless
quantity \(\breve{v}(\vec{x},t_{p})=v(\vec{x},t_{p})\:/\mcal{N}\)
in the coefficient functions, scaled by the parameter \(\mcal{N}=(\hbar/\Delta t)\cdot(L/\Delta x)^{d}\)
\beq \lb{s4_137}
\lefteqn{\bigg\langle\mbox{Tr}\bigg[\mbox{STR}\bigg(\hat{\eta}\;\hat{T}^{-1}\;
\big(\wtilde{\pp}_{i}\hat{T}\big)\;
\mbox{{\boldmath$\wtilde{\pp}_{i}$}}\;\hat{G}^{(0)}[\hat{\sigma}_{D}^{(0)}]\;\hat{\eta}\;
\hat{T}^{-1}\;\big(\wtilde{\pp}_{j}\hat{T}\big)\;
\mbox{{\boldmath$\wtilde{\pp}_{j}$}}\;\hat{G}^{(0)}[\hat{\sigma}_{D}^{(0)}]
\bigg)\bigg]\bigg\rangle_{\hat{\sigma}_{D}^{(0)}} \approx } \\ \no &=&
\int_{C}\frac{d t_{p}}{\hbar}\;\frac{1}{\mcal{N}}\;
\sum_{\vec{x}}\sum_{a,b=1,2}^{(a\neq b)}
\Bigg[\strab\bigg\{\Big[\hat{T}^{-1}(\vec{x},t_{p})\;
\big(\wtilde{\pp}_{i}\hat{T}(\vec{x},t_{p})\big)\Big]_{\alpha\beta}^{a\neq b} \;\;
\Big[\hat{T}^{-1}(\vec{x},t_{p})\;
\big(\wtilde{\pp}_{j}\hat{T}(\vec{x},t_{p})\big)\Big]_{\beta\alpha}^{b\neq a}\bigg\}\;\times
\\ \no &\times &2\;
\bigg\langle\Big(\wtilde{\pp}_{i}\breve{v}(\vec{x},t_{p})\Big)
\Big(\wtilde{\pp}_{j}\breve{v}(\vec{x},t_{p})\Big) -
 \Big(\wtilde{\pp}_{i}\wtilde{\pp}_{j}\breve{v}(\vec{x},t_{p})\Big)
\bigg\rangle_{\hat{\sigma}_{D}^{(0)}} +  \\ \no &-&
\strab\bigg\{\Big[\hat{T}^{-1}(\vec{x},t_{p})\;
\big(\wtilde{\pp}_{i}\hat{T}(\vec{x},t_{p})\big)\Big]_{\alpha\beta}^{a\neq b} \;\;
\Big[\hat{T}^{-1}(\vec{x},t_{p})\;
\big(\wtilde{\pp}_{i}\hat{T}(\vec{x},t_{p})\big)\Big]_{\beta\alpha}^{b\neq a}\bigg\}\;\times
\\ \no &\times &
\bigg\langle\sum_{k=1}^{d}\bigg(\Big(\wtilde{\pp}_{k}\breve{v}(\vec{x},t_{p})\Big)^{2} +
\Big(\wtilde{\pp}_{k}\wtilde{\pp}_{k}\breve{v}(\vec{x},t_{p})\Big)\bigg)
\bigg\rangle_{\hat{\sigma}_{D}^{(0)}}  \Bigg]_{\mbox{.}}
\eeq
In Eq. (\ref{s4_137}) the total doubled super-trace '\(\mbox{STR}\)' is limited to the super-trace '\(\mbox{str}\)'
over angular momentum states of bosons and fermions and to the summation over the
anomalous terms with \(a\neq b\). This is in accordance to the introduced rules of
propagation for products of Green functions in subsection \ref{s44}. In order to
replace the super-trace '\(\mbox{str}\)' by '\(\mbox{STR}\)', the super-matrix
\(\hat{Z}(\vec{x},t_{p})\) (\ref{s4_138}) is defined with metric
\(\hat{S}^{a}=\big\{+1\mbox{ for }a=1\;;\;-1\mbox{ for }a=2\big\}\) (\ref{s4_76})
\be \lb{s4_138}
\hat{Z}(\vec{x},t_{p}) = \hat{T}(\vec{x},t_{p})\;\hat{S}\;\hat{T}^{-1}(\vec{x},t_{p})\;\;\;.
\ee
Using the derivative (\ref{s4_139}) of \(\hat{Z}(\vec{x},t_{p})\), the partial super-trace
'\(\mbox{str}\)' in (\ref{s4_137}) is related to a complete super-trace '\(\mbox{STR}\)'
with \(\big(\wtilde{\pp}_{i}\hat{Z}(\vec{x},t_{p})\big)\) whose summations are also lastly confined
to the anomalous terms or pair condensates
\beq \lb{s4_139}
\Big(\wtilde{\pp}_{i}\hat{Z}(\vec{x},t_{p})\Big) &=&
\Big(\wtilde{\pp}_{i}\hat{T}(\vec{x},t_{p})\Big)\;\hat{S}\;\hat{T}^{-1}(\vec{x},t_{p})-
\hat{T}(\vec{x},t_{p})\;\hat{S}\;\hat{T}^{-1}(\vec{x},t_{p})\;
\Big(\wtilde{\pp}_{i}\hat{T}(\vec{x},t_{p})\Big)\;\hat{T}^{-1}(\vec{x},t_{p})
\eeq
\beq \lb{s4_140}
\lefteqn{\STRAB\bigg[\Big(\wtilde{\pp}_{i}\hat{Z}(\vec{x},t_{p})\Big)\;
\Big(\wtilde{\pp}_{j}\hat{Z}(\vec{x},t_{p})\Big)\bigg] = } \\ \no &=&
2\;\STRAB\bigg[\hat{T}^{-1}(\vec{x},t_{p})\;
\Big(\wtilde{\pp}_{i}\hat{T}(\vec{x},t_{p})\Big)\;\hat{S}\;
\hat{T}^{-1}(\vec{x},t_{p})\;
\Big(\wtilde{\pp}_{j}\hat{T}(\vec{x},t_{p})\Big)\;\hat{S}\bigg] + \\ \no &-&
2\;\STRAB\bigg[\hat{T}^{-1}(\vec{x},t_{p})\;
\Big(\wtilde{\pp}_{i}\hat{T}(\vec{x},t_{p})\Big)\;
\hat{T}^{-1}(\vec{x},t_{p})\;
\Big(\wtilde{\pp}_{j}\hat{T}(\vec{x},t_{p})\Big)\bigg] = \\ \no &=&
-4\sum_{a,b=1,2}^{(a\neq b)}\strab\bigg\{
\Big[\hat{T}^{-1}(\vec{x},t_{p})\;
\big(\wtilde{\pp}_{i}\hat{T}(\vec{x},t_{p})\big)\Big]_{\alpha\beta}^{a\neq b}\;\;
\Big[\hat{T}^{-1}(\vec{x},t_{p})\;
\big(\wtilde{\pp}_{j}
\hat{T}(\vec{x},t_{p})\big)\Big]_{\beta\alpha}^{b\neq a}\bigg\}\;\;\;.
\eeq
The matrix \(\hat{Z}(\vec{x},t_{p})\) of coset matrices \(\hat{T}(\vec{x},t_{p})\)
is invariant under subgroup transformations \(U(L|S)\) with \(\hat{Q}(\vec{x},t_{p})\)
because \(\hat{Q}(\vec{x},t_{p})\) commutes with the coset metric \(\hat{S}\) (\ref{s4_141}). This
verifies the assumed spontaneous symmetry breaking of the group
\(Osp(S,S|2L)\) to \(Osp(S,S|2L)\backslash U(L|S)\otimes U(L|S)\) and the extraction
of the corresponding Goldstone modes in a gradient expansion
\beq \lb{s4_141}
\hat{Z}_{0}(\vec{x},t_{p}) &=& \hat{T}_{0}(\vec{x},t_{p})\;\hat{S}\;\hat{T}_{0}^{-1}(\vec{x},t_{p}) \\
\no &=& \hat{T}(\vec{x},t_{p})\;\hat{Q}^{-1}(\vec{x},t_{p})\;\hat{S}\;
\hat{Q}(\vec{x},t_{p})\;\hat{T}^{-1}(\vec{x},t_{p}) \\ \no &=&
\hat{Z}(\vec{x},t_{p})\;\;\;; \hspace*{1.75cm}
\Big[\hat{Q}(\vec{x},t_{p})\;,\;\hat{S}\Big]=0  \\  \lb{s4_142}
\Big(\wtilde{\pp}_{i}\hat{Z}_{0}(\vec{x},t_{p})\Big) &=&
\Big(\wtilde{\pp}_{i}\hat{Z}(\vec{x},t_{p})\Big)  \\ \no &=&
\Big(\wtilde{\pp}_{i}\hat{T}_{0}(\vec{x},t_{p})\Big)\;\hat{S}\;\hat{T}_{0}^{-1}(\vec{x},t_{p})-
\hat{T}_{0}(\vec{x},t_{p})\;\hat{S}\;\hat{T}_{0}^{-1}(\vec{x},t_{p})\;
\Big(\wtilde{\pp}_{i}\hat{T}_{0}(\vec{x},t_{p})\Big)\;\hat{T}_{0}^{-1}(\vec{x},t_{p})\;\;\;.
\eeq
We summarize the result of the purely spatial gradient expansion of order
\(\big(\delta\hat{\mcal{H}}(\hat{T}^{-1},\hat{T})\big)^{2}\) in \(\mcal{A}_{SDET}\ppr\)
with the averaging over a background field \(\sigma_{D}^{(0)}(\vec{x},t_{p})\) in
the coherent state path integral \(Z[j_{\psi};\hat{\sigma}_{D}^{(0)}]\). Finally, we
obtain relation (\ref{s4_143}) with coset matrices \(\hat{Z}(\vec{x},t_{p})\) for the pair
condensates and coefficients \(c^{ij}(\vec{x},t_{p})\) following from averages over the background
field \(\sigma_{D}^{(0)}(\vec{x},t_{p})\) in the trap potential \(u(\vec{x})\).
These are scaled to dimensionless quantities \(\breve{u}(\vec{x})\),
\(\breve{\sigma}_{D}^{(0)}(\vec{x},t_{p})\) with energy parameter \(\mcal{N}=\hbar\Omega\;
\mcal{N}_{x}\) , (\(\mcal{N}_{x}=\big(L/\Delta x\big)^{d}\))
\beq \lb{s4_143}
\lefteqn{-\frac{1}{4}\Bigg\langle\mbox{Tr}\bigg[\mbox{STR}\bigg(
\delta\hat{\mcal{H}}\big(\hat{T}^{-1},\hat{T}\big)\;
\hat{G}^{(0)}\big[\hat{\sigma}_{D}^{(0)}\big]\;
\delta\hat{\mcal{H}}\big(\hat{T}^{-1},\hat{T}\big)\;
\hat{G}^{(0)}\big[\hat{\sigma}_{D}^{(0)}\big]\bigg)\bigg]\Bigg\rangle_{\hat{\sigma}_{D}^{(0)}}\approx }
\\ \no &=& \frac{1}{4}\int_{C}\frac{d t_{p}}{\hbar}\;\frac{1}{\mcal{N}}\sum_{\vec{x}}
\mbox{STR}\bigg[\Big(\wtilde{\pp}_{i}\hat{Z}(\vec{x},t_{p})\Big)\;\;
\Big(\wtilde{\pp}_{j}\hat{Z}(\vec{x},t_{p})\Big)\bigg]\;\;
c^{ij}(\vec{x},t_{p})
\eeq
\beq \lb{s4_144}
c^{ij}(\vec{x},t_{p}) &=& c^{(1),ij}(\vec{x},t_{p}) +
c^{(2),ij}(\vec{x},t_{p})   \hspace*{1.0cm}(i,j,k=1,\ldots,d) \\ \lb{s4_145}
c^{(1),ij}(\vec{x},t_{p}) &=&
-2\bigg\langle\Big(\wtilde{\pp}_{i}\wtilde{\pp}_{j}\breve{v}(\vec{x},t_{p})\Big)
\bigg\rangle_{\hat{\sigma}_{D}^{(0)}} - \delta_{ij}\sum_{k=1}^{d}
\bigg\langle\Big(\wtilde{\pp}_{k}\wtilde{\pp}_{k}\breve{v}_{D}^{(0)}(\vec{x},t_{p})\Big)
\bigg\rangle_{\hat{\sigma}_{D}^{(0)}}   \\  \lb{s4_146}
c^{(2),ij}(\vec{x},t_{p}) &=&
2\bigg\langle\Big(\wtilde{\pp}_{i}\breve{v}(\vec{x},t_{p})\Big)\;
\Big(\wtilde{\pp}_{j}\breve{v}(\vec{x},t_{p})
\Big)\bigg\rangle_{\hat{\sigma}_{D}^{(0)}} - \delta_{ij}\sum_{k=1}^{d}
\bigg\langle\Big(\wtilde{\pp}_{k}\breve{v}(\vec{x},t_{p})\Big)^{2}
\bigg\rangle_{\hat{\sigma}_{D}^{(0)}}  \\ \no
\breve{v}(\vec{x},t_{p}) &=& \Big(u(\vec{x})+\sigma_{D}^{(0)}(\vec{x},t_{p})\Big)/\mcal{N}\;\;\;.
\eeq
We have separated the effective coefficients \(c^{ij}(\vec{x},t_{p})\) into one-point
parts \(c^{(1),ij}(\vec{x},t_{p})\) with the average over a single field
\(\sigma_{D}^{(0)}(\vec{x},t_{p})\) and transport- or two-point- functions with the
average over two background fields.

In the two-dimensional isotropic case \(d=2\) of spatial coordinates, one can neglect the
transport coefficients \(c^{(2),ij}(\vec{x},t_{p})\approx 0\) and has only to
consider the one-point coefficients \(c^{(1),ij}(\vec{x},t_{p})\). These can be
acquired by a saddle point or mean field equation so that a complete average with
\(Z[j_{\psi};\hat{\sigma}_{D}^{(0)}]\) (\ref{s4_90}) is not necessary. Therefore, the
two-dimensional case allows further approximations for determining the coefficients
\(c^{ij}(\vec{x},t_{p})\).

Apart from simplifying the coefficients \(c^{ij}(\vec{x},t_{p})\), the
two-dimensional case of the action (\ref{s4_143}) is conformal invariant under
spatial transformations, even including the measure
\(d\big[\hat{T}^{-1}(\vec{x},t_{p})\;d\hat{T}(\vec{x},t_{p})\big]\).
The measure of the coset space is invariant under subgroup transformations with
\(\hat{Q}(\vec{x},t_{p})\) in analogy to the invariance of \(\hat{Z}(\vec{x},t_{p})\)
for \(U(L|S)\). However, the spatially two-dimensional case has even more structures
between conformal invariant action with \(\big(\wtilde{\pp}_{i}\hat{Z}(\vec{x},t_{p})\big)\)
(\ref{s4_143}) and the invariant coset measure because both are determined by the metric tensor
\(\hat{G}_{Osp\backslash U}\) of the coset space \(Osp(S,S|2L)\backslash U(L|S)\).
The invariant measure for the coset space is given by the square root of the super-determinant
of the metric tensor \(\big(\mbox{SDET}(\hat{G}_{Osp\backslash U})\big)^{1/2}\),
and the action (\ref{s4_143}) is specified by a bilinear relation with the metric
tensor \(\hat{G}_{Osp\backslash U}\) and the independent parameters of the coset matrix
in the two-dimensional case of conformal invariance
\beq \lb{s4_147}
\hat{T}_{0}(\vec{x},t_{p}) &\rightarrow & \hat{T}(\vec{x},t_{p}) \\ \lb{s4_148}
d\big[\hat{T}_{0}^{-1}(\vec{x},t_{p})\;d\hat{T}_{0}(\vec{x},t_{p})\big] &=&
d\big[\hat{T}^{-1}(\vec{x},t_{p})\;d\hat{T}(\vec{x},t_{p})\big]  \\ \lb{s4_149}
\Big[\hat{T}^{-1}(\vec{x},t_{p})\;d\hat{T}(\vec{x},t_{p})\Big]_{\beta\alpha}^{ba}\bigg|^{b\neq a}
&=&\Big[-\Big(ds_{\kappa}(\vec{x},t_{p})\Big)\;\;
\hat{Y}^{(\kappa)}\Big]_{\beta\alpha}^{ba}\bigg|^{b\neq a}\;\;\;.
\eeq
Therefore, concerning the second order gradients in the expansion, the approximated generating function
with \(\mcal{A}_{SDET}\ppr\) can be reduced to Gaussian integrals by combining the metric tensor
in the action (\ref{s4_143}) with the metric tensor as a super-determinant
for the invariant coset integration measure. This dependence has to be expected for extracting
Goldstone modes in a SSB.

It remains to disentangle the terms with an unsaturated operator \(\mbox{{\boldmath$\hat{E}_{p}$}}\)
for the time development in the second order expansion of
\(\big(\delta\hat{\mcal{H}}(\hat{T}^{-1},\hat{T})\big)^{2}\) in \(\mcal{A}_{SDET}\ppr\)
(compare (\ref{s4_122},\ref{s4_121}) and (\ref{s4_118})). Using the commutators (\ref{s4_123},\ref{s4_124}),
we can approximate the terms with gradient operator \(\mbox{{\boldmath$\hat{E}_{p}$}}\)
in (\ref{s4_122}) to relations (\ref{s4_150}-\ref{s4_153}) where one has to apply similar
transformations as in the derivation of the term with two spatial operators
\(\mbox{{\boldmath$\wtilde{\pp}_{i}$}}\)\(\mbox{{\boldmath$\wtilde{\pp}_{j}$}}\)
\beq \lb{s4_150}
\lefteqn{\hspace*{-0.64cm}\bigg\langle\mbox{Tr}\bigg[\mbox{STR}\bigg(\hat{\eta}\;
\hat{T}^{-1}\;\big(\wtilde{\pp}_{i}\hat{T}\big)\;\mbox{{\boldmath$\wtilde{\pp}_{i}$}}\;
\hat{G}^{(0)}[\hat{\sigma}_{D}^{(0)}]\;\hat{\eta}\;
\big(\hat{T}^{-1}\;\hat{S}\;\hat{T}-\hat{S}\big)\;\mbox{{\boldmath$\hat{E}_{p}$}}\;
\hat{G}^{(0)}[\hat{\sigma}_{D}^{(0)}] \bigg)\bigg]\bigg\rangle_{\hat{\sigma}_{D}^{(0)}} \approx
-\frac{1}{2}\int_{C}\frac{d t_{p}}{\hbar}\;\frac{1}{\mcal{N}}
\sum_{\vec{x}}    } \\ \no &\times&\sum_{a,b=1,2}^{(a\neq b)}
\bigg\langle E_{p}\breve{\sigma}_{D}^{(0)}(\vec{x},t_{p})\bigg\rangle_{\hat{\sigma}_{D}^{(0)}} \;
\strab\bigg\{\Big[\hat{T}^{-1}(\vec{x},t_{p})\;
\big(\wtilde{\pp}_{i}\hat{T}(\vec{x},t_{p})\big)\Big]_{\alpha\beta}^{a\neq b}\;
\Big[\wtilde{\pp}_{i}\hat{T}^{-1}(\vec{x},t_{p})\;\hat{S}\;\hat{T}(\vec{x},t_{p})\Big]_{\beta\alpha}^{b\neq a}
\bigg\}
\eeq
\beq \lb{s4_151}
\lefteqn{\hspace*{-0.64cm}\bigg\langle\mbox{Tr}\bigg[\mbox{STR}\bigg(\hat{\eta}\;
\hat{T}^{-1}\;\big(\wtilde{\pp}_{i}\wtilde{\pp}_{i}\hat{T}\big)\;
\hat{G}^{(0)}[\hat{\sigma}_{D}^{(0)}]\;\hat{\eta}\;
\big(\hat{T}^{-1}\;\hat{S}\;\hat{T}-\hat{S}\big)\;\mbox{{\boldmath$\hat{E}_{p}$}}\;
\hat{G}^{(0)}[\hat{\sigma}_{D}^{(0)}] \bigg)\bigg]\bigg\rangle_{\hat{\sigma}_{D}^{(0)}} \approx
-\int_{C}\frac{d t_{p}}{\hbar}\;\frac{1}{\mcal{N}}
\sum_{\vec{x}}    }  \\ \no &\times&\sum_{a,b=1,2}^{(a\neq b)}
\bigg\langle E_{p}\breve{\sigma}_{D}^{(0)}(\vec{x},t_{p})\bigg\rangle_{\hat{\sigma}_{D}^{(0)}} \;
\strab\bigg\{\Big[\hat{T}^{-1}(\vec{x},t_{p})\;
\big(\wtilde{\pp}_{i}\wtilde{\pp}_{i}\hat{T}(\vec{x},t_{p})\big)\Big]_{\alpha\beta}^{a\neq b}\;
\Big[\hat{T}^{-1}(\vec{x},t_{p})\;\hat{S}\;\hat{T}(\vec{x},t_{p})\Big]_{\beta\alpha}^{b\neq a}
\bigg\}
\eeq
\beq \lb{s4_152}
\lefteqn{\hspace*{-0.64cm}\bigg\langle\mbox{Tr}\bigg[\mbox{STR}\bigg(\hat{\eta}\;
\big(\hat{T}^{-1}\;\hat{S}\;(E_{p}\hat{T})\big)\;
\hat{G}^{(0)}[\hat{\sigma}_{D}^{(0)}]\;\hat{\eta}\;
\big(\hat{T}^{-1}\;\hat{S}\;\hat{T}-\hat{S}\big)\;\mbox{{\boldmath$\hat{E}_{p}$}}\;
\hat{G}^{(0)}[\hat{\sigma}_{D}^{(0)}] \bigg)\bigg]\bigg\rangle_{\hat{\sigma}_{D}^{(0)}} \approx
-\int_{C}\frac{d t_{p}}{\hbar}\;\frac{1}{\mcal{N}}
\sum_{\vec{x}}  }  \\ \no &\times&   \sum_{a,b=1,2}^{(a\neq b)}
\bigg\langle E_{p}\breve{\sigma}_{D}^{(0)}(\vec{x},t_{p})\bigg\rangle_{\hat{\sigma}_{D}^{(0)}} \;
\strab\bigg\{
\Big[\hat{T}^{-1}(\vec{x},t_{p})\;\hat{S}\;\big(E_{p}\hat{T}(\vec{x},t_{p})\big)\Big]_{\alpha\beta}^{a\neq b}
\Big[\hat{T}^{-1}(\vec{x},t_{p})\;\hat{S}\;\hat{T}(\vec{x},t_{p})\Big]_{\beta\alpha}^{b\neq a}
\bigg\}
\eeq
\beq \no
\lefteqn{\hspace*{-0.64cm}\bigg\langle\mbox{Tr}\bigg[\mbox{STR}\bigg(\hat{\eta}\;
\big(\hat{T}^{-1}\;\hat{S}\;\hat{T}-\hat{S}\big)\;\mbox{{\boldmath$\hat{E}_{p}$}}\;
\hat{G}^{(0)}[\hat{\sigma}_{D}^{(0)}]\;\hat{\eta}\;
\big(\hat{T}^{-1}\;\hat{S}\;\hat{T}-\hat{S}\big)\;\mbox{{\boldmath$\hat{E}_{q}$}}\;
\hat{G}^{(0)}[\hat{\sigma}_{D}^{(0)}] \bigg)\bigg]\bigg\rangle_{\hat{\sigma}_{D}^{(0)}} \approx
-\frac{1}{2}\int_{C}\frac{d t_{p}}{\hbar}\;\frac{1}{\mcal{N}}
\sum_{\vec{x}}   }    \\    \lb{s4_153}      &\times& \sum_{a,b=1,2}^{(a\neq b)}
\bigg\langle E_{p}\breve{\sigma}_{D}^{(0)}(\vec{x},t_{p})\bigg\rangle_{\hat{\sigma}_{D}^{(0)}} \;
\strab\bigg\{
\Big[\hat{T}^{-1}(\vec{x},t_{p})\;\hat{S}\;\hat{T}(\vec{x},t_{p})\Big]_{\alpha\beta}^{a\neq b}
\Big[E_{p}\;\hat{T}^{-1}(\vec{x},t_{p})\;\hat{S}\;\hat{T}(\vec{x},t_{p})\Big]_{\beta\alpha}^{b\neq a}
\bigg\}_{\mbox{.}}
\eeq
However, these terms with \(\mbox{{\boldmath$\hat{E}_{p}$}}\) can be neglected in the case of
a stationary background field where the averaged time-like derivative of \(\sigma_{D}^{(0)}(\vec{x},t_{p})\)
vanishes
(\(\big\langle E_{p}\breve{\sigma}_{D}^{(0)}(\vec{x},t_{p})\big\rangle_{\hat{\sigma}_{D}^{(0)}}\approx 0\)).
This can be required for a time independent trap potential and for source fields of the BEC wave function
which allow the real, scalar self-energy density \(\sigma_{D}^{(0)}(\vec{x},t_{p})\) to reach
an equilibrium state within \(Z[j_{\psi};\hat{\sigma}_{D}^{(0)}]\) (\ref{s4_90}).

The expansion of the first order term \(\delta\hat{\mcal{H}}(\hat{T}^{-1},\hat{T})\) with one
Green function \(\big\langle \hat{G}^{(0)}[\hat{\sigma}_{D}^{(0)}]\big\rangle_{\hat{\sigma}_{D}^{(0)}}\)
in \(\mcal{A}_{SDET}\ppr\) does not involve the complication of propagation of two Green functions.
One returns from time \(t_{p}+\delta t_{p}\) back to the same contour time \(t_{p}\).
We directly act with \(\delta\hat{\mcal{H}}(\hat{T}^{-1},\hat{T})\)
onto the averaged Green function in (\ref{s4_154}) and obtain averaged matrix elements composed of the
unsaturated gradients \(\mbox{{\boldmath${\ds\hat{E}_{p}}$}}\),
\(\mbox{{\boldmath${\ds\wtilde{\pp}_{i}}$}}\) with a single Green function
\beq \lb{s4_154}
\lefteqn{\frac{1}{2}\mbox{Tr}\bigg[\mbox{STR}\bigg(\delta\hat{\mcal{H}}(\hat{T}^{-1},\hat{T})\;\;
\Big\langle\hat{G}^{(0)}[\hat{\sigma}_{D}^{(0)}]\Big\rangle_{\hat{\sigma}_{D}^{(0)}}\bigg)\bigg] =
-\frac{1}{2}\int_{C}\frac{d t_{p}}{\hbar}\;\mcal{N}\sum_{\vec{x}}
\sum_{a,b=1,2}\sum_{\alpha,\beta=1}^{N=L+S}
\delta_{a,b}\;\;\delta_{\alpha,\beta}\;\; \times } \\ \no &\times & \hspace*{-0.28cm}\Bigg[
\STRAB\bigg\{\Big[\hat{T}^{-1}(\vec{x},t_{p})\;\hat{S}\;\big(E_{p}\hat{T}(\vec{x},t_{p})\big)+
\hat{T}^{-1}(\vec{x},t_{p})\;
\big(\wtilde{\pp}_{i}\wtilde{\pp}_{i}\hat{T}(\vec{x},t_{p})\big)\Big]_{\alpha\beta}^{ab}\bigg\}\;
\Big\langle\widehat{\langle\vec{x},t_{p}}^{b}|\hat{G}^{(0)}[\hat{\sigma}_{D}^{(0)}]|
\widehat{\vec{x},t_{p}\rangle}^{a}\Big\rangle_{\hat{\sigma}_{D}^{(0)}} + \\ \no &+&
\STRAB\bigg\{\Big[\hat{T}^{-1}(\vec{x},t_{p})\;\hat{S}\;
\hat{T}(\vec{x},t_{p})-\hat{S}\Big]_{\alpha\beta}^{ab}\bigg\}\;
\Big\langle\widehat{\langle\vec{x},t_{p}}^{b}|\mbox{{\boldmath${\ds\hat{E}_{p}}$}}\;
\hat{G}^{(0)}[\hat{\sigma}_{D}^{(0)}]|
\widehat{\vec{x},t_{p}\rangle}^{a}\Big\rangle_{\hat{\sigma}_{D}^{(0)}} + \\ \no &+& 2\;
\STRAB\bigg\{\Big[\hat{T}^{-1}(\vec{x},t_{p})\;\big(\wtilde{\pp}_{i}\hat{T}(\vec{x},t_{p})\big)
\Big]_{\alpha\beta}^{ab}\bigg\}\;
\Big\langle\widehat{\langle\vec{x},t_{p}}^{b}|\mbox{{\boldmath${\ds\wtilde{\pp}_{i}}$}}\;
\hat{G}^{(0)}[\hat{\sigma}_{D}^{(0)}]|
\widehat{\vec{x},t_{p}\rangle}^{a}\Big\rangle_{\hat{\sigma}_{D}^{(0)}} \Bigg]_{\mbox{.}}
\eeq
However, the spacetime matrix elements of the averages of a single Green function (also with gradients)
in (\ref{s4_154}) only enter with their diagonal parts, leading to reduced and simplified equations.
The average of the Green function without gradients results in diagonal terms of Kronecker-deltas
for the 'Nambu'-doubled space (\(a=b\)), the angular monentum degrees of freedom (\(\alpha=\beta\))
and also for the space and time contour variables (\ref{s4_155})
\be \lb{s4_155}
\Big\langle\widehat{\langle\vec{x}\ppr,t_{p}}^{a}|\hat{G}^{(0)}[\hat{\sigma}_{D}^{(0)}]|
\widehat{\vec{x},t_{p}\rangle}^{b}\Big\rangle_{\hat{\sigma}_{D}^{(0)}}=\delta_{a,b}\;
\delta_{\alpha,\beta}\;\frac{1}{\mcal{N}}\;\delta_{\vec{x},\vec{x}\ppr}\;
\delta_{t_{p},t_{p}}\;\;\;.
\ee
In order to disentangle the matrix element with \(\mbox{{\boldmath${\ds\hat{E}_{p}}$}}\),
we consider the defining Eq. (\ref{s4_156}) for the doubled Green function
\(\hat{G}^{(0)}[\hat{\sigma}_{D}^{(0)}]\) and solve for
\(\mbox{{\boldmath${\ds\hat{E}_{p}}$}}\;\hat{G}^{(0)}[\hat{\sigma}_{D}^{(0)}]\)  so that
the corresponding matrix element (\ref{s4_158}) is transformed to the diagonal spacetime element
of \(\big(\hat{\eta}\;\hat{h}_{p}(\hat{\vec{x}})+\hat{\sigma}_{D}^{(0)}\big)\;
\hat{S}\;\hat{G}^{(0)}[\hat{\sigma}_{D}^{(0)}]-\hat{S}\). We can completely remove the matrix elements
with spatial gradient operators \(\mbox{{\boldmath${\ds\wtilde{\pp}_{i}}$}}\;
\mbox{{\boldmath${\ds\wtilde{\pp}_{i}}$}}\) and also
\(\mbox{{\boldmath${\ds\wtilde{\pp}_{i}}$}}\) because the diagonal
matrix elements of these only appear in Eqs. (\ref{s4_158}-\ref{s4_160}) and (\ref{s4_154})
\be \lb{s4_156}
\bigg[-\hat{\eta}\;\hat{S}\;\mbox{{\boldmath${\ds\hat{E}_{p}}$}}+
\underbrace{\hat{\eta}\;\Big(-\mbox{{\boldmath${\ds\wtilde{\pp}_{i}}$}}\;
\mbox{{\boldmath${\ds\wtilde{\pp}_{i}}$}}+
u(\hat{\vec{x}})-\mu_{0}-\im\;\ve_{p}\Big)}_{\hat{\eta}\;\hat{h}_{p}}+\hat{\sigma}_{D}^{(0)}\bigg]\;
\hat{G}^{(0)}[\hat{\sigma}_{D}^{(0)}]=\hat{1}
\ee
\be  \lb{s4_157}
\mbox{{\boldmath${\ds\hat{E}_{p}}$}}\;\hat{G}^{(0)}[\hat{\sigma}_{D}^{(0)}]=
\Big(-\mbox{{\boldmath${\ds\wtilde{\pp}_{i}}$}}\;
\mbox{{\boldmath${\ds\wtilde{\pp}_{i}}$}}+
u(\hat{\vec{x}})-\mu_{0}-\im\;\ve_{p}+\hat{\eta}\;\hat{\sigma}_{D}^{(0)}\Big)\;\hat{S}\;
\hat{G}^{(0)}[\hat{\sigma}_{D}^{(0)}] -\hat{\eta}\;\hat{S}
\ee
\beq \lb{s4_158}
\lefteqn{\Big\langle\widehat{\langle\vec{x},t_{p}}^{a}|\mbox{{\boldmath${\ds\hat{E}_{p}}$}}
\;\hat{G}^{(0)}[\hat{\sigma}_{D}^{(0)}]|
\widehat{\vec{x},t_{p}\rangle}^{b}\Big\rangle_{\hat{\sigma}_{D}^{(0)}} = } \\ \no &=&
\Big\langle\widehat{\vec{x},t_{p}}^{a}|
\Big[\Big(-\mbox{{\boldmath${\ds\wtilde{\pp}_{i}}$}}\;
\mbox{{\boldmath${\ds\wtilde{\pp}_{i}}$}}+
u(\hat{\vec{x}})-\mu_{0}-\im\;\ve_{p}+\hat{\eta}\;\hat{\sigma}_{D}^{(0)}\Big)\;\hat{S}\;
\hat{G}^{(0)}[\hat{\sigma}_{D}^{(0)}] -\hat{\eta}\;\hat{S}\Big]|
\widehat{\vec{x},t_{p}\rangle}^{b}\Big\rangle_{\hat{\sigma}_{D}^{(0)}} \approx \\ \no &=&
\delta_{a,b}\;\hat{S}^{a}\;\delta_{\alpha,\beta}\;\delta_{\vec{x},\vec{x}}\;\delta_{t_{p},t_{p}}\;\;
\Big[\Big(\breve{u}(\vec{x})-\breve{\mu}_{0}+
\big\langle\breve{\sigma}_{D}^{(0)}(\vec{x},t_{p})\big\rangle_{\hat{\sigma}_{D}^{(0)}}\Big)-
\eta_{p}\Big]
\eeq
\beq \lb{s4_159}
\lefteqn{\hspace*{-1.54cm}
\Big\langle\widehat{\langle\vec{x},t_{p}}^{b}|\mbox{{\boldmath${\ds\wtilde{\pp}_{i}}$}}\;
\hat{G}^{(0)}[\hat{\sigma}_{D}^{(0)}]|
\widehat{\vec{x},t_{p}\rangle}^{a}\Big\rangle_{\hat{\sigma}_{D}^{(0)}}=\bigg(\wtilde{\pp}_{i}\;
\Big\langle\widehat{\langle\vec{x},t_{p}}^{b}|
\hat{G}^{(0)}[\hat{\sigma}_{D}^{(0)}]|
\widehat{\vec{x},t_{p}\rangle}^{a}\Big\rangle_{\hat{\sigma}_{D}^{(0)}}\bigg)\approx } \\ \no &=&
\bigg(\wtilde{\pp}_{i}\;\delta_{a,b}\;\delta_{\alpha,\beta}\;\frac{1}{\mcal{N}}\;
\delta_{\vec{x},\vec{x}}\;\delta_{t_{p},t_{p}}\bigg)\equiv 0
\eeq
\be\lb{s4_160}
\Big\langle\widehat{\langle\vec{x},t_{p}}^{b}|\mbox{{\boldmath${\ds\wtilde{\pp}_{i}}$}}\;
\mbox{{\boldmath${\ds\wtilde{\pp}_{i}}$}}\;\hat{G}^{(0)}[\hat{\sigma}_{D}^{(0)}]|
\widehat{\vec{x},t_{p}\rangle}^{a}\Big\rangle_{\hat{\sigma}_{D}^{(0)}} \equiv 0
\ee
\be \lb{s4_161}
\Big\langle\widehat{\langle\vec{x},t_{p}}^{a}|\;\hat{\eta}\;\hat{\sigma}_{D}^{(0)}\;
\hat{G}^{(0)}[\hat{\sigma}_{D}^{(0)}]|
\widehat{\vec{x},t_{p}\rangle}^{b}\Big\rangle_{\hat{\sigma}_{D}^{(0)}} \approx
\big\langle\breve{\sigma}_{D}^{(0)}(\vec{x},t_{p})\big\rangle_{\hat{\sigma}_{D}^{(0)}}\;\;
\Big\langle\widehat{\langle\vec{x},t_{p}}^{a}|\hat{G}^{(0)}[\hat{\sigma}_{D}^{(0)}]|
\widehat{\vec{x},t_{p}\rangle}^{b}\Big\rangle_{\hat{\sigma}_{D}^{(0)}} \;\;\;.
\ee
Insertion of (\ref{s4_155}-\ref{s4_161}) into (\ref{s4_154}) yields the relation (\ref{s4_162})
for the first order term \(\delta\hat{\mcal{H}}(\hat{T}^{-1},\hat{T})\) in the gradient expansion
of \(\mcal{A}_{SDET}\ppr\)
\beq \lb{s4_162}
\lefteqn{\frac{1}{2}\mbox{Tr}\bigg[\mbox{STR}\bigg(\delta\hat{\mcal{H}}(\hat{T}^{-1},\hat{T})\;\;
\Big\langle\hat{G}^{(0)}[\hat{\sigma}_{D}^{(0)}]\Big\rangle_{\hat{\sigma}_{D}^{(0)}}\bigg)\bigg] = }
\\ \no &=& -\frac{1}{2}\int_{C}\frac{d t_{p}}{\hbar}\sum_{\vec{x}} \Bigg[
\STRAB\bigg[\hat{T}^{-1}(\vec{x},t_{p})\;\hat{S}\;\big(E_{p}\hat{T}(\vec{x},t_{p})\big)+
\hat{T}^{-1}(\vec{x},t_{p})\;
\big(\wtilde{\pp}_{i}\wtilde{\pp}_{i}\hat{T}(\vec{x},t_{p})\big)\bigg] + \\ \no &+&
\STRAB\bigg[\hat{T}^{-1}(\vec{x},t_{p})\;\hat{S}\;
\hat{T}(\vec{x},t_{p})\;\hat{S}-\hat{1}_{2N\times 2N}\bigg]\;\times \;
\bigg(\Big(u(\vec{x})-\mu_{0}-\im\;\ve_{p}+\big\langle\sigma_{D}^{(0)}(\vec{x},t_{p})
\big\rangle_{\hat{\sigma}_{D}^{(0)}}\Big)-\mcal{N}\;\eta_{p}\bigg) \Bigg]_{\mbox{.}}
\eeq
We further investigate the first order term of \(\delta\hat{\mcal{H}}(\hat{T}^{-1},\hat{T})\)
(\ref{s4_162}) by using the transformations (\ref{s4_163},\ref{s4_164}) and acquire Eq. (\ref{s4_165})
\beq \lb{s4_163}
\hat{T}^{-1}(\vec{x},t_{p})\;\hat{S}\;\hat{T}(\vec{x},t_{p})\;\hat{S}-\hat{1}_{2N\times 2N}&=&
\Big(\hat{T}^{-1}(\vec{x},t_{p})\Big)^{2}-\hat{1}_{2N\times 2N}  \\ \no &=&
\exp\Big\{2\;\hat{Y}_{2N\times 2N}(\vec{x},t_{p})\Big\}-\hat{1}_{2N\times 2N}  \\  \lb{s4_164}
\STRAB\Big[\exp\Big\{2\;\hat{Y}_{2N\times 2N}(\vec{x},t_{p})\Big\}-\hat{1}_{2N\times 2N}\Big]&=&
\STRAB\Big[\cosh\Big\{2\;\hat{Y}_{2N\times 2N}(\vec{x},t_{p})\Big\}-\hat{1}_{2N\times 2N}\Big]
\eeq
\beq \lb{s4_165}
\lefteqn{\frac{1}{2}\mbox{Tr}\bigg[\mbox{STR}\bigg(\delta\hat{\mcal{H}}(\hat{T}^{-1},\hat{T})\;\;
\Big\langle\hat{G}^{(0)}[\hat{\sigma}_{D}^{(0)}]\Big\rangle_{\hat{\sigma}_{D}^{(0)}}\bigg)\bigg] = }
\\ \no &=& -\frac{1}{2}\int_{C}\frac{d t_{p}}{\hbar}\sum_{\vec{x}} \Bigg[
\STRAB\bigg\{\hat{T}^{-1}(\vec{x},t_{p})\;\hat{S}\;\big(E_{p}\hat{T}(\vec{x},t_{p})\big)+
\hat{T}^{-1}(\vec{x},t_{p})\;
\big(\wtilde{\pp}_{i}\wtilde{\pp}_{i}\hat{T}(\vec{x},t_{p})\big)\bigg\} + \\ \no &+&
\STRAB\bigg\{\Big(\hat{T}^{-1}(\vec{x},t_{p})\Big)^{2}-\hat{1}_{2N\times 2N}\bigg\}\;
\Big(u(\vec{x})-\mu_{0}-\im\;\ve_{p}+\big\langle\sigma_{D}^{(0)}(\vec{x},t_{p})
\big\rangle_{\hat{\sigma}_{D}^{(0)}}\Big) \Bigg]+ \\ \no &+&\frac{1}{2}
\int_{C}\frac{d t_{p}}{\hbar}\eta_{p}\mcal{N}\sum_{\vec{x}}
\STRAB\Big[\Big(\hat{T}^{-1}(\vec{x},t_{p})\Big)^{2}-\hat{1}_{2N\times 2N}\Big]\;\;\;.
\eeq
We expand \(\hat{T}^{-1}(\vec{x},t_{p})\) and \(\hat{T}^{-2}(\vec{x},t_{p})\) in the generator
\(\hat{Y}(\vec{x},t_{p})\) for small amplitudes of the pair condensates. We achieve the
expectation value of amplitudes \(\hat{Y}(\vec{x},t_{p})\) with one-particle operator
consisting of the trap and chemical potential, the averaged self-energy density and the kinetic
energy with the {\it doubled mass value \(2m\)} of the atoms with short-ranged interaction.
\beq \lb{s4_166}
\hat{T}^{-1}(\vec{x},t_{p})&=&\hat{1}_{2N\times 2N}+\hat{Y}_{2N\times 2N}(\vec{x},t_{p})+
\frac{1}{2!}\Big[\hat{Y}_{2N\times 2N}(\vec{x},t_{p})\Big]^{2}+\ldots \\ \lb{s4_167} \hspace*{-0.73cm}
\Big(\hat{T}^{-1}(\vec{x},t_{p})\Big)^{2}&=&\exp\Big\{2\;\hat{Y}_{2N\times 2N}(\vec{x},t_{p})\Big\} =
\hat{1}_{2N\times 2N}+2\;\hat{Y}_{2N\times 2N}(\vec{x},t_{p})+2\;
\Big[\hat{Y}_{2N\times 2N}(\vec{x},t_{p})\Big]^{2}+\ldots
\eeq
\beq  \lb{s4_168}
\lefteqn{\frac{1}{2}\mbox{Tr}\bigg[\mbox{STR}\bigg(\delta\hat{\mcal{H}}(\hat{T}^{-1},\hat{T})\;\;
\Big\langle\hat{G}^{(0)}[\hat{\sigma}_{D}^{(0)}]\Big\rangle_{\hat{\sigma}_{D}^{(0)}}\bigg)\bigg] \approx }
\\ \no &\hspace*{-0.46cm}=& \hspace*{-0.46cm}-\frac{1}{2}\int_{C}\frac{d t_{p}}{\hbar}\sum_{\vec{x}}
\STRAB\Bigg\{\hat{Y}(\vec{x},t_{p})\left(\bigg[-\hat{S}\;\hat{E}_{p}+2\bigg(
\frac{\hat{\vec{p}}^{\;2}}{2\;(2m)}+u(\vec{x})-\mu_{0}+\big\langle\sigma_{D}^{(0)}(\vec{x},t_{p})
\big\rangle_{\hat{\sigma}_{D}^{(0)}}\bigg)\bigg]\hat{Y}(\vec{x},t_{p})\right)\Bigg\}+
\\ \no &+&\frac{1}{2}\int_{C}\frac{d t_{p}}{\hbar}\eta_{p}\mcal{N}\sum_{\vec{x}}
\STRAB\Big[\cosh\Big(2\;\hat{Y}(\vec{x},t_{p})\Big)-\hat{1}_{2N\times 2N}\Big]\;\;\; .
\eeq
Since the coset generator \(\hat{Y}(\vec{x},t_{p})\) is only composed of the non-diagonal block
super-matrices \(\hat{X}(\vec{x},t_{p})\), \(\wtilde{\kappa}\;\hat{X}^{+}(\vec{x},t_{p})\),
one can even further reduce the expectation value with \(\hat{Y}(\vec{x},t_{p})\) in (\ref{s4_168})
to relation (\ref{s4_169}) where one also has the {\it doubled atomic mass value \(2m\)} in the kinetic
energy
\beq \lb{s4_169}
\lefteqn{\frac{1}{2}\mbox{Tr}\bigg[\mbox{STR}\bigg(\delta\hat{\mcal{H}}(\hat{T}^{-1},\hat{T})\;\;
\Big\langle\hat{G}^{(0)}[\hat{\sigma}_{D}^{(0)}]\Big\rangle_{\hat{\sigma}_{D}^{(0)}}\bigg)\bigg] \approx }
\\ \no &=& -\int_{C}\frac{d t_{p}}{\hbar}\sum_{\vec{x}}
\strab\Bigg\{\wtilde{\kappa}\hat{X}^{+}(\vec{x},t_{p})\left(\bigg[-\hat{E}_{p}+2\bigg(
\frac{\hat{\vec{p}}^{\;2}}{2\;(2m)}+u(\vec{x})-\mu_{0}+\big\langle\sigma_{D}^{(0)}(\vec{x},t_{p})
\big\rangle_{\hat{\sigma}_{D}^{(0)}}\bigg)\bigg]\hat{X}(\vec{x},t_{p})\right)\Bigg\}+
\\ \no &+& 2 \int_{C}\frac{d t_{p}}{\hbar}\eta_{p}\mcal{N}\sum_{\vec{x}}
\strab\Big[\wtilde{\kappa}\hat{X}^{+}(\vec{x},t_{p})\;\;
\hat{X}(\vec{x},t_{p})\Big]\;\;\; .
\eeq
Comparing the second order and first order term of \(\delta\hat{\mcal{H}}(\hat{T}^{-1},\hat{T})\)
in \(\mcal{A}_{SDET}\ppr\), we notice various orders of the energy scale variable
\(\mcal{N}=\hbar\Omega\;\mcal{N}_{x}\). The derived action (\ref{s4_143}) for the second order of
\(\delta\hat{\mcal{H}}(\hat{T}^{-1},\hat{T})\) is scaled by \(\mcal{N}^{-1}\) and is therefore
the dominant part in the calculation of correlation functions.
The derived action term (\ref{s4_165}) with first order of \(\delta\hat{\mcal{H}}(\hat{T}^{-1},\hat{T})\)
from \(\mcal{A}_{SDET}\ppr\) contains the order \(\mcal{N}^{0}\) for the energy expectation values of the pair
condensates and even the order \(\mcal{N}^{+1}\) for a fluctuation term.

\subsection{Expansion of the condensate wave function term
$\big\langle\mcal{A}_{J_{\psi}}\ppr\big[\hat{T},\hat{\sigma}_{D}^{(0)};
\hat{\mcal{J}}\big]\big\rangle_{\hat{\sigma}_{D}^{(0)}}$} \lb{s46}

In comparison to the expansion of \(\mcal{A}_{SDET}\ppr[\hat{T};\hat{\mcal{J}}]\), the gradient
expansion of the condensate wave function can be attained by less complicated steps
because the propagation and average of two or more Green functions \(\hat{G}^{(0)}[\hat{\sigma}_{D}^{(0)}]\)
in \(\mcal{A}_{J_{\psi}}\ppr\) does not involve a forward and backward direction with
a return to the same spacetime point in a doubled Hilbert trace relation '\(\mbox{Tr}\)'. The
propagation in \(\mcal{A}_{J_{\psi}}\ppr[\hat{T};\hat{\mcal{J}}]\) (\ref{s4_93}) starts
on the right hand-side with source field state \(|\widehat{J_{\psi;\alpha}^{a}\rangle}\) which is
multiplied by the coset matrix \(\hat{T}^{-1}\) and metric \(\hat{I}\) yielding some other
'Nambu'-doubled field. This field can be identified with
\(\big(f_{\alpha}(\vec{x}\ppr,t_{q}\ppr)\;;\;f_{\alpha}^{*}(\vec{x}\ppr,t_{q}\ppr)\big)^{T,b}\)
in (\ref{s4_99}-\ref{s4_101}). Repeated application of rules (\ref{s4_99}-\ref{s4_101})
transfers the fields to the left hand-side to the source field state
\(\widehat{\langle J_{\psi;\beta}^{b}}|\). We list in (\ref{s4_170}-\ref{s4_173}) the gradient
terms \(\delta\hat{\mcal{H}}(\hat{T}^{-1},\hat{T})\) and source operator
\(\wtilde{\mcal{J}}(\hat{T}^{-1},\hat{T})\) in the expansion of \(\Delta\hat{\mcal{O}}\)
and assume a quasi-stationary case of the background field \(\sigma_{D}^{(0)}(\vec{x},t_{p})\)
in \(Z[j_{\psi};\hat{\sigma}_{D}^{(0)}]\) (\ref{s4_90}). Therefore, the time-like derivative
\(\big(E_{p}\hat{\sigma}_{D}^{(0)}\big)\approx 0\) is disregarded in the first order term of
\(\delta\hat{\mcal{H}}(\hat{T}^{-1},\hat{T})\) acting onto \(\hat{G}^{(0)}[\hat{\sigma}_{D}^{(0)}]\)
and further to the right
\beq \lb{s4_170}
\Delta\hat{\mcal{O}}&=&\delta\hat{\mcal{H}}(\hat{T}^{-1},\hat{T})+
\wtilde{\mcal{J}}(\hat{T}^{-1},\hat{T})    \\  \lb{s4_171}
\delta\hat{\mcal{H}}(\hat{T}^{-1},\hat{T}) &=&
\hat{T}^{-1}\;\hat{\mcal{H}}\;\hat{T}-\hat{\mcal{H}} \\ \no &=&
-\hat{\eta}\;
\Big(\big(\hat{T}^{-1}\;\hat{S}\;(E_{p}\hat{T})\big)+
\hat{T}^{-1}\;\big(\wtilde{\pp}_{i}\wtilde{\pp}_{i}
\hat{T}\big) +  \big(\hat{T}^{-1}\;\hat{S}\;\hat{T}-\hat{S}\big)\;
\mbox{{\boldmath${\ds\hat{E}_{p}}$}}  +  2\;\hat{T}^{-1}\;\big(\wtilde{\pp}_{i}\hat{T}\big)\;
\mbox{{\boldmath${\ds\wtilde{\pp}_{i}}$}} \Big) \\ \lb{s4_172}
\wtilde{\mcal{J}}(\hat{T}^{-1},\hat{T}) &=&
\hat{T}^{-1}\;\hat{I}\;\hat{K}\;\hat{\eta}\;\hat{\mcal{J}}\;\hat{\eta}\;
\hat{K}\;\hat{I}\;\wtilde{K}\;\hat{T}
\eeq
\beq \lb{s4_173}
\lefteqn{\delta\hat{\mcal{H}}(\hat{T}^{-1},\hat{T})\;\hat{G}^{(0)}[\hat{\sigma}_{D}^{(0)}]=
-\hat{\eta}\;\Big[\hat{T }^{-1}\;\hat{S}\;\big(E_{p}\hat{T}\big)+\hat{T}^{-1}\;
\big(\wtilde{\pp}_{i}\wtilde{\pp}_{i}\hat{T}\big)\Big]\;\hat{G}^{(0)}[\hat{\sigma}_{D}^{(0)}] + } \\ \no &+&
\hat{\eta}\;\big(\hat{T}^{-1}\;\hat{S}\;\hat{T}-\hat{S}\big)\;\hat{G}^{(0)}[\hat{\sigma}_{D}^{(0)}]\;
\underbrace{\Big(E_{p}\hat{\sigma}_{D}^{(0)}\Big)}_{\approx 0}\;
\hat{G}^{(0)}[\hat{\sigma}_{D}^{(0)}] +
\\ \no &+&\hat{\eta}\;2\;\hat{T}^{-1}\;\big(\wtilde{\pp}_{i}\hat{T}\big)\;
\hat{G}^{(0)}[\hat{\sigma}_{D}^{(0)}]\;\Big(\wtilde{\pp}_{i}\hat{u}+\wtilde{\pp}_{i}\hat{\sigma}_{D}^{(0)}
\Big)\;\hat{G}^{(0)}[\hat{\sigma}_{D}^{(0)}] +  \\ \no &-&\hat{\eta}\;
\big(\hat{T}^{-1}\;\hat{S}\;\hat{T}-\hat{S}\big)\;\hat{G}^{(0)}[\hat{\sigma}_{D}^{(0)}]\;
\mbox{{\boldmath${\ds\hat{E}_{p}}$}}-\hat{\eta}\;2\;\hat{T}^{-1}\;\big(\wtilde{\pp}_{i}\hat{T}\big)\;
\hat{G}^{(0)}[\hat{\sigma}_{D}^{(0)}]\;\mbox{{\boldmath${\ds\wtilde{\pp}_{i}}$}} \;\;\;.
\eeq
The expansion with first and second order of \(\Delta\hat{\mcal{O}}\) is again given for the
condensate wave function term \(\mcal{A}_{J_{\psi}}\ppr[\hat{T};\hat{\mcal{J}}]\)
in (\ref{s4_174}). One has to use the commutators between unsaturated gradient operators and
Green functions so that the unsaturated gradients are shifted to the right and finally act onto
the source field state \(|\widehat{J_{\psi;\alpha}^{a}\rangle}\). We suppose slowly varying states
\(|\widehat{J_{\psi;\alpha}^{a}\rangle}\) in space and time and introduce the approximations (\ref{s4_175})
for the expansion with \(\Delta\hat{\mcal{O}}\), \(\big(\Delta\hat{\mcal{O}}\big)^{2}\) in (\ref{s4_174}).
The terms (\ref{s4_176}) are only kept in the expansion with up to second order in
spatial derivatives and first order time derivative
\beq \lb{s4_174}
\lefteqn{\mcal{A}_{J_{\psi}}\ppr[\hat{T};\hat{\mcal{J}}]=
\Big\langle\mcal{A}_{J_{\psi}}\ppr[\hat{T},\hat{\sigma}_{D}^{(0)};
\hat{\mcal{J}}]\Big\rangle_{\hat{\sigma}_{D}^{(0)}} \approx } \\ \no &=&-\frac{1}{2}\frac{1}{\mcal{N}}
\Big\langle\widehat{\langle J_{\psi;\beta}^{b}}|\hat{\eta}\;\Big(\hat{I}\;\wtilde{K}\;\hat{T}\;
\hat{G}^{(0)}[\hat{\sigma}_{D}^{(0)}]\;\Delta\hat{\mcal{O}}\;\hat{G}^{(0)}[\hat{\sigma}_{D}^{(0)}]\;
\hat{T}^{-1}\;\hat{I}\Big)_{\beta\alpha}^{ba}\;\hat{\eta}|\widehat{J_{\psi;\alpha}^{a}\rangle}
\Big\rangle_{\hat{\sigma}_{D}^{(0)}} + \\ \no &+&\frac{1}{2}\frac{1}{\mcal{N}}
\Big\langle\widehat{\langle J_{\psi;\beta}^{b}}|\hat{\eta}\;\Big(\hat{I}\;\wtilde{K}\;\hat{T}\;
\hat{G}^{(0)}[\hat{\sigma}_{D}^{(0)}]\;
\Big(\Delta\hat{\mcal{O}}\;\hat{G}^{(0)}[\hat{\sigma}_{D}^{(0)}]\Big)^{2}\;
\hat{T}^{-1}\;\hat{I}\Big)_{\beta\alpha}^{ba}\;\hat{\eta}|\widehat{J_{\psi;\alpha}^{a}\rangle}
\Big\rangle_{\hat{\sigma}_{D}^{(0)}} \mp \ldots
\eeq
\be \lb{s4_175}
\mbox{{\boldmath${\ds\wtilde{\pp}_{i}}$}}\;|\widehat{J_{\psi;\alpha}^{a}\rangle}\approx 0
\hspace*{1.0cm}\mbox{{\boldmath${\ds\hat{E}_{p}}$}}\;|\widehat{J_{\psi;\alpha}^{a}\rangle}\approx 0
\hspace*{1.0cm}\Big\langle E_{p}\hat{\sigma}_{D}^{(0)}\Big\rangle_{\hat{\sigma}_{D}^{(0)}}\approx 0
\ee
\be \lb{s4_176}
\hat{T},\;\hat{T}^{-1},\;\big(\wtilde{\pp}_{i}\hat{T}\big),\;
\big(\wtilde{\pp}_{i}\wtilde{\pp}_{i}\hat{T}\big),\;
\big(\wtilde{\pp}_{i}\hat{T}\big)\cdot\big(\wtilde{\pp}_{j}\hat{T}\big),\;
\big(E_{p}\hat{T}\big)\;\;\;.
\ee
The result of the second and first order expansion of \(\delta\hat{\mcal{H}}(\hat{T}^{-1},\hat{T})\)
in \(\mcal{A}_{J_{\psi}}\ppr[\hat{T};\hat{\mcal{J}}]\) (\ref{s4_174}) leads to
(\ref{s4_177},\ref{s4_179}) with averaged coefficients \(d^{ij}(\vec{x},t_{p})\) (\ref{s4_178}).
In order to obtain these results, one has to apply consequently the commutators (\ref{s4_123},\ref{s4_124})
and rules of propagation (\ref{s4_99}-\ref{s4_101}) so that the unsaturated gradient operators are
conveyed to the right with final action onto \(|\widehat{J_{\psi;\alpha}^{a}\rangle}\).
We ignore these final gradients according to (\ref{s4_175}) with the assumption of slowly varying fields
\(|\widehat{J_{\psi;\alpha}^{a}\rangle}\) and of slowly varying background field and further assume
an approximate translation invariance in the trap potential (\ref{s4_180}) excluding boundary effects
\beq \lb{s4_177}
\lefteqn{
\frac{1}{2}\frac{1}{\mcal{N}}
\Big\langle\widehat{\langle J_{\psi;\beta}^{b}}|\hat{\eta}\;\Big(\hat{I}\;\wtilde{K}\;\hat{T}\;
\hat{G}^{(0)}[\hat{\sigma}_{D}^{(0)}]\;
\Big(\Delta\hat{\mcal{O}}\;\hat{G}^{(0)}[\hat{\sigma}_{D}^{(0)}]\Big)^{2}\;
\hat{T}^{-1}\;\hat{I}\Big)_{\beta\alpha}^{ba}\;\hat{\eta}|\widehat{J_{\psi;\alpha}^{a}\rangle}
\Big\rangle_{\hat{\sigma}_{D}^{(0)}}  \approx
\int_{C}\frac{d t_{p}}{\hbar}\;\frac{1}{\mcal{N}}\sum_{\vec{x}}  }
\\ \no &\hspace*{-0.64cm}\times&\hspace*{-0.64cm}
\sum_{a,b=1,2}\sum_{\alpha,\beta=1}^{N=L+S} d^{ij}(\vec{x},t_{p}) \;
\frac{J_{\psi;\beta}^{+,b}(\vec{x},t_{p})}{\mcal{N}}\bigg(\hat{I}\;\wtilde{K}\;
\big(\wtilde{\pp}_{i}\hat{T}(\vec{x},t_{p})\big)\;\hat{T}^{-1}(\vec{x},t_{p})\;
\big(\wtilde{\pp}_{j}\hat{T}(\vec{x},t_{p})\big)\;\hat{T}^{-1}(\vec{x},t_{p})\;\hat{I}\bigg)_{\beta\alpha}^{ba}
\frac{J_{\psi;\alpha}^{a}(\vec{x},t_{p})}{\mcal{N}}
\eeq
\be \lb{s4_178}
d^{ij}(\vec{x},t_{p})=2\;\bigg\langle
3\;\Big(\wtilde{\pp}_{i}\breve{v}(\vec{x},t_{p})\Big)
\Big(\wtilde{\pp}_{j}\breve{v}(\vec{x},t_{p})\Big)
- \Big(\wtilde{\pp}_{i}\wtilde{\pp}_{j}\breve{v}(\vec{x},t_{p})\Big)
\bigg\rangle_{\hat{\sigma}_{D}^{(0)}} \;\;\;;\hspace*{0.5cm}
\breve{v}(\vec{x},t_{p})=\breve{u}(\vec{x})+\breve{\sigma}_{D}^{(0)}(\vec{x},t_{p})
\ee
\beq \no
\lefteqn{-\frac{1}{2}\frac{1}{\mcal{N}}
\Big\langle\widehat{\langle J_{\psi;\beta}^{b}}|\hat{\eta}\;\Big(\hat{I}\;\wtilde{K}\;\hat{T}\;
\hat{G}^{(0)}[\hat{\sigma}_{D}^{(0)}]\;\Delta\hat{\mcal{O}}\;\hat{G}^{(0)}[\hat{\sigma}_{D}^{(0)}]\;
\hat{T}^{-1}\;\hat{I}\Big)_{\beta\alpha}^{ba}\;\hat{\eta}|\widehat{J_{\psi;\alpha}^{a}\rangle}
\Big\rangle_{\hat{\sigma}_{D}^{(0)}} \approx  -\frac{1}{2}
\;\int_{C}\frac{d t_{p}}{\hbar}\;\sum_{\vec{x}}
\sum_{a,b=1,2}\sum_{\alpha,\beta=1}^{N=L+S}  }   \\ \no &\times&
\frac{J_{\psi;\beta}^{+,b}(\vec{x},t_{p})}{\mcal{N}}\bigg[\hat{I}\;\wtilde{K}\bigg(
-\big(\wtilde{\pp}_{i}\wtilde{\pp}_{i}\hat{T}(\vec{x},t_{p})\big)\;\hat{T}^{-1}(\vec{x},t_{p}) -
\hat{T}(\vec{x},t_{p})\;\hat{S}\;\hat{T}^{-1}(\vec{x},t_{p})\;
\big(E_{p}\hat{T}(\vec{x},t_{p})\big)\;\hat{T}^{-1}(\vec{x},t_{p})+  \\ \lb{s4_179}  &+& 2\;
\big(\wtilde{\pp}_{i}\hat{T}(\vec{x},t_{p})\big)\;\hat{T}^{-1}(\vec{x},t_{p})\;
\big(\wtilde{\pp}_{i}\hat{T}(\vec{x},t_{p})\big)\;\hat{T}^{-1}(\vec{x},t_{p})\;
\bigg)\hat{I}\bigg]_{\beta\alpha}^{ba}
\frac{J_{\psi;\alpha}^{a}(\vec{x},t_{p})}{\mcal{N}}
\eeq
\be \lb{s4_180}
\Big\langle E_{p}\breve{\sigma}_{D}^{(0)}(\vec{x},t_{p})\Big\rangle_{\hat{\sigma}_{D}^{(0)}}
\approx 0 \hspace*{1.5cm}
\Big\langle \wtilde{\pp}_{i}\breve{u}(\vec{x})+
\wtilde{\pp}_{i}\breve{\sigma}_{D}^{(0)}(\vec{x},t_{p})\Big\rangle_{\hat{\sigma}_{D}^{(0)}}
\approx 0 \;\;\; .
\ee
Compared to the result of the gradient expansion for \(\mcal{A}_{SDET}\ppr[\hat{T};\hat{\mcal{J}}]\)
(\ref{s4_143}-\ref{s4_146}), (\ref{s4_162}-\ref{s4_169}), there are no possible simplifying
relations for the spatial \(2D\) case in \(\mcal{A}_{J_{\psi}}\ppr[\hat{T};\hat{\mcal{J}}]\)
concerning the coefficients \(d^{ij}(\vec{x},t_{p})\).
A conformal invariance can be derived for \(\mcal{A}_{J_{\psi}}\ppr[\hat{T};\hat{\mcal{J}}]\)
under the assumption of spatially constant or at least slowly varying source fields for the
BEC wave function (\(\big(\wtilde{\pp}_{i}J_{\psi;\alpha}^{a}(\vec{x},t_{p})\big)\approx 0\)).
The universal properties of conformal invariance in \(\mcal{A}_{SDET}\ppr[\hat{T};\hat{\mcal{J}}]\)
follow from the simultaneous appearance of the metric tensor \(\hat{G}_{Osp\backslash U}\) in the
invariant coset measure and the action (\ref{s4_143}) and lead to Gaussian integrals
for the pair condensates in \(2D\) coordinate space. In the case of
\(\mcal{A}\ppr_{J_{\psi}}[\hat{T};\hat{\mcal{J}}]\) one has also to include
the functions \(\big(\wtilde{\pp}_{i}g_{\kappa}(\vec{x},t_{p})\big)\) for the subgroup
\(U(L|S)\) with corresponding metric tensor \(\hat{G}_{U(L|S)}\).
But according to the different metric tensor \(\hat{G}_{Osp\backslash U}\) in the invariant
coset integration measure, simple Gaussian-like integrals do not result as for the expansion with the action
\(\mcal{A}_{SDET}\ppr[\hat{T};\hat{\mcal{J}}]\) (compare appendix \ref{sa2} with Eqs. (\ref{A65}-\ref{A67})
and subsection \ref{s61}).

\section{Effective actions for the super-symmetric pair condensates} \lb{s5}

\subsection{Classification of the actions for pair condensates
according to the energy scale $\mcal{N}=\hbar\Omega\;\mcal{N}_{x}$} \lb{s51}

In section \ref{s4} the gradient expansion has been performed in the actions
\(\mcal{A}_{SDET}\ppr[\hat{T};\hat{\mcal{J}}]\), \(\mcal{A}_{J_{\psi}}\ppr[\hat{T};\hat{\mcal{J}}]\)
of the super-determinant and the coherent BEC wave function term. The gradient terms
\(\delta\hat{\mcal{H}}(\hat{T}^{-1},\hat{T})\) are retained up to second order in
\(\Delta\hat{\mcal{O}}=\delta\hat{\mcal{H}}(\hat{T}^{-1},\hat{T})+
\wtilde{\mcal{J}}(\hat{T}^{-1},\hat{T})\) with a maximum of two spatial derivatives and
a single time derivative of the coset matrices \(\hat{T}(\vec{x},t_{p})\),
\(\hat{T}^{-1}(\vec{x},t_{p})\)
\beq \lb{s5_1}
\mcal{A}_{SDET}\ppr\big[\hat{T};\hat{\mcal{J}}\big]&\approx&-\frac{1}{4}
\bigg\langle\mbox{Tr}\Big[\mbox{STR}\Big(\Delta\hat{\mcal{O}}\;\hat{G}^{(0)}[\hat{\sigma}_{D}^{(0)}]\;
\Delta\hat{\mcal{O}}\;\hat{G}^{(0)}[\hat{\sigma}_{D}^{(0)}]\Big)\Big]\bigg\rangle_{\hat{\sigma}_{D}^{(0)}}+
\\ \no &+& \frac{1}{2}
\mbox{Tr}\Big[\mbox{STR}\Big(\Delta\hat{\mcal{O}}\;\Big\langle\hat{G}^{(0)}[\hat{\sigma}_{D}^{(0)}]
\Big\rangle_{\hat{\sigma}_{D}^{(0)}}\Big)\Big]    \\   \lb{s5_2}
\mcal{A}_{J_{\psi}}\ppr\big[\hat{T};\hat{\mcal{J}}\big]&\approx&
\frac{1}{2}\frac{1}{\mcal{N}}\bigg\langle\widehat{\langle J_{\psi;\beta}^{b}}|\hat{\eta}\;
\Big(\hat{I}\;\wtilde{K}\;\hat{T}\;\hat{G}^{(0)}[\hat{\sigma}_{D}^{(0)}]\;
\Big(\Delta\hat{\mcal{O}}\;\hat{G}^{(0)}[\hat{\sigma}_{D}^{(0)}]\Big)^{2}\;
\hat{T}^{-1}\;\hat{I}\Big)_{\beta\alpha}^{ba}\hat{\eta}|
\widehat{J_{\psi;\alpha}^{a}\rangle}\bigg\rangle_{\hat{\sigma}_{D}^{(0)}} + \\ \no &-&
\frac{1}{2}\frac{1}{\mcal{N}}\bigg\langle\widehat{\langle J_{\psi;\beta}^{b}}|\hat{\eta}\;
\Big(\hat{I}\;\wtilde{K}\;\hat{T}\;\hat{G}^{(0)}[\hat{\sigma}_{D}^{(0)}]\;
\Delta\hat{\mcal{O}}\;\hat{G}^{(0)}[\hat{\sigma}_{D}^{(0)}]\; \hat{T}^{-1}\;\hat{I}\Big)_{\beta\alpha}^{ba}\hat{\eta}|
\widehat{J_{\psi;\alpha}^{a}\rangle}\bigg\rangle_{\hat{\sigma}_{D}^{(0)}}\;\;\;.
\eeq
We include the average of the Green functions \(\hat{G}^{(0)}[\hat{\sigma}_{D}^{(0)}]\) in (\ref{s5_1},\ref{s5_2})
with the background field \(\sigma_{D}^{(0)}(\vec{x},t_{p})\) of the coherent state path integral
\(Z[j_{\psi};\hat{\sigma}_{D}^{(0)}]\) (\ref{s4_89}-\ref{s4_91}). This averaging procedure
with \(Z[j_{\psi};\hat{\sigma}_{D}^{(0)}]\) (\ref{s4_89}-\ref{s4_91}) takes into account
the noncondensed atoms, but also the condensed, coherent BEC wave function. The resulting
effective coherent state path integral (\ref{s5_3}) only contains the field degrees of freedom for the
pair condensates in the coset matrix \(\hat{T}(\vec{x},t_{p})=\exp\{-\hat{Y}(\vec{x},t_{p})\}\)
\beq \lb{s5_3}
Z[\hat{\mcal{J}},J_{\psi},\im\hat{J}_{\psi\psi}]&=&
\int d[\hat{T}^{-1}(\vec{x},t_{p})\;d\hat{T}(\vec{x},t_{p})]\;\;
\exp\Big\{\im\;\mcal{A}_{\hat{J}_{\psi\psi}}\big[\hat{T}\big]\Big\}  \\ \no &\times&
\exp\Big\{-\mcal{A}_{SDET}\ppr\big[\hat{T};\hat{\mcal{J}}\big]\Big\}\;\;\;
\exp\Big\{\im\;\mcal{A}_{J_{\psi}}\ppr\big[\hat{T};\hat{\mcal{J}}\big]\Big\} \;.
\eeq
The effective path integral (\ref{s5_3}) consists of the coset measure
\(d[\hat{T}^{-1}(\vec{x},t_{p})\;d\hat{T}(\vec{x},t_{p})]\)(\(Osp(S,S|2L)\backslash U(L|S)\)),
being invariant under \(U(L|S)\) subgroup transformations,
and the source term \(\mcal{A}_{\hat{J}_{\psi\psi}}\big[\hat{T}\big]\)
following from integrations over the 'hinge' degrees of freedom
\(\delta\hat{\Sigma}_{D;N\times N}^{aa}(\vec{x},t_{p})\;\wtilde{K}=
\big(\hat{Q}^{-1}(\vec{x},t_{p})\;\delta\hat{\Lambda}(\vec{x},t_{p})\;
\hat{Q}(\vec{x},t_{p})\big)_{N\times N}^{aa}\).
The field \(\hat{J}_{\psi\psi;\alpha\beta}^{a\neq b}(\vec{x},t_{p})\) has the effect of a
'seed' for the pair condensates in \(\hat{T}(\vec{x},t_{p})\), starting the time development
of \(\hat{Y}(\vec{x},t_{p})\) which is determined by the actions
\(\mcal{A}_{SDET}\ppr[\hat{T};\hat{\mcal{J}}]\) and
\(\mcal{A}_{J_{\psi}}[\hat{T};\hat{\mcal{J}}]\).

In fact, the various parts of the actions \(\mcal{A}_{SDET}\ppr[\hat{T};\hat{\mcal{J}}]\),
\(\mcal{A}_{J_{\psi}}\ppr[\hat{T};\hat{\mcal{J}}]\) can be classified according to the
parameter \(\mcal{N}=\hbar\Omega\;\mcal{N}_{x}=\hbar\big(1/\Delta t\big)\cdot\big(L/\Delta x\big)^{d}\).
It describes the total number of spatial points of the underlying grid and the maximum possible
energy due to the finite time intervals. This parameter \(\mcal{N}\) has inevitably to appear for
normalization as one transforms the super-determinant, just being a factor of eigenvalues
e. g. \(\big[\mbox{SDET}(\wtilde{\mcal{M}})\big]^{-1/2}\) with exponent \(-1/2\) (\ref{s4_10}),
to an action \(\mcal{A}_{SDET}[\hat{T},\delta\hat{\Sigma}_{D},\hat{\sigma}_{D}^{(0)};\hat{\mcal{J}}]\)
(\ref{s4_12}) with spatial integral \(\sum_{\vec{x}}\ldots=
\int_{L^{d}}(d^{d}x/L^{d})\ldots\) and positive time contour integration with additional metric
\(\eta_{p=\pm}\) for the trace '\(\mbox{Tr}\)'
\beq\lb{s5_4}
\Big[\mbox{SDET}\big(\wtilde{\mcal{M}}\big)\Big]^{-1/2}&=&
\exp\Big\{-\mcal{A}_{SDET}[\hat{T},\delta\hat{\Sigma}_{D},\hat{\sigma}_{D}^{(0)};
\hat{\mcal{J}}]\Big\}  \\  \lb{s5_5}
\mcal{A}_{SDET}[\hat{T},\delta\hat{\Sigma}_{D},\hat{\sigma}_{D}^{(0)};\hat{\mcal{J}}] &=&\frac{1}{2}\;
\mbox{Tr}\;\mbox{STR}\;\ln\Big(\wtilde{\mcal{M}}\Big)  \\ \no &=&\frac{1}{2}
\int_{C}\frac{dt_{p}}{\hbar}\eta_{p}\underbrace{\int_{L^{d}}\frac{d^{d}x}{L^{d}}}_{\sum_{\vec{x}}}\;
\underbrace{\Big(\hbar\Omega\cdot \mcal{N}_{x}\Big)}_{\mcal{N}}\STRAB\;\ln\;\Big(
\wtilde{\mcal{M}}_{\vec{x},\alpha;\vec{x}\ppr,\beta}^{ab}(t_{p},t_{q}\ppr)\Big)\;.
\eeq
This normalization also arises for the terms in the gradient expansion with the Green functions
\(\hat{G}^{(0)}[\hat{\sigma}_{D}^{(0)}]\) in \(Z[j_{\psi};\hat{\sigma}_{D}^{(0)}]\)
(\ref{s4_89}-\ref{s4_91}) and is consequently considered within all transformations of this paper
(also according to the doubled Hilbert space definitions in subsection \ref{s42}). Consequently,
we obtain from the second order gradient expansion
three effective actions from orders \(\mcal{N}^{-1}\) to \(\mcal{N}^{+1}\).
The action \(\mcal{A}\ppr\big[\hat{T};\hat{\mcal{J}}\big]\) is listed in appendix \ref{sb}
(Eq. (\ref{B31})) for the expansion of the generating source field
\(\hat{\mcal{J}}_{\vec{x},\alpha;\vec{x}\ppr,\beta}(t_{p},t_{q}\ppr)\) up to quadratic order
\beq \lb{s5_6}
Z[\hat{\mcal{J}},J_{\psi},\im \hat{J}_{\psi\psi}]&=&
\int d\big[\hat{T}^{-1}(\vec{x},t_{p})\;d\hat{T}(\vec{x},t_{p})\big]\;\;
\exp\Big\{\im\;\mcal{A}_{\hat{J}_{\psi\psi}}\big[\hat{T}\big]\Big\}  \\ \no &\times&
\exp\Big\{-\mcal{A}_{\mcal{N}^{-1}}\ppr\big[\hat{T};J_{\psi}\big]-
\mcal{A}_{\mcal{N}^{0}}\ppr\big[\hat{T};J_{\psi}\big]-
\mcal{A}_{\mcal{N}^{+1}}\ppr\big[\hat{T}\big]\Big\} \;\times\;
\exp\Big\{-\mcal{A}\ppr\big[\hat{T};\hat{\mcal{J}}\big]\Big\}_{\mbox{.}}
\eeq
We have to point out that
all spatial and time-like variables within the gradients and within the arguments of the coset matrix
\(\hat{T}(\vec{x},t_{p})=\exp\big\{-\hat{Y}(\vec{x},t_{p})\big\}\)
are not scaled by \(\mcal{N}\) or \(\mcal{N}^{\pm 1/2}\) to some dimensionless quantities
whereas energy values as \(\breve{u}(\vec{x})\) or \(\breve{\sigma}_{D}^{(0)}(\vec{x},t_{p})\)
have as reference the overall energy and number of spatial points within the parameter \(\mcal{N}\).
A re-scaling of the arguments in \(\hat{T}(\vec{x},t_{p})=\exp\big\{-\hat{Y}(\vec{x},t_{p})\big\}\)
according to \(\wtilde{\pp}_{i}=\frac{\hbar}{\sqrt{2m}}\frac{\pp}{\pp x^{\mu}}
\;\to\;\frac{1}{\sqrt{\mcal{N}}}\wtilde{\pp}_{i}=
\frac{1}{\sqrt{\mcal{N}}}\frac{\hbar}{\sqrt{2m}}\frac{\pp}{\pp x^{\mu}}\)
would contract the spatial and time-like dependence of
\(\hat{T}(\frac{\hbar}{\sqrt{2m\:\mcal{N}}}\;x^{\prime\mu},\frac{\hbar}{\mcal{N}}\;t_{p}\ppr)=
\exp\big\{-\hat{Y}(\frac{\hbar}{\sqrt{2m\:\mcal{N}}}\;x^{\prime\mu},\frac{\hbar}{\mcal{N}}\;t_{p}\ppr)\big\}\)
to vanishing values in the continuum limit \(\mcal{N}\to\infty\) after the gradient expansion.
Therefore, the scales of the spatial and time-like variables in the arguments of
\(\hat{T}(\vec{x},t_{p})=\exp\big\{-\hat{Y}(\vec{x},t_{p})\big\}\) are retained unchanged as in the
original coherent state path integral with super-fields \(\psi_{\vec{x},\alpha}(t_{p})\)
before the HST transformation. The gradient expansion of slowly varying Goldstone modes has to
keep the original physical scale of spatial and time-like variables in the path integrals
whereas the remaining physical quantities are averaged by the background field and are finally maintained
as dimensionless parameters for the gradients of
\(\hat{T}(\vec{x},t_{p})=\exp\big\{-\hat{Y}(\vec{x},t_{p})\big\}\). These gradients of
\(\hat{T}(\vec{x},t_{p})\) are the only relevant physical
variables for an effective theory of the pair condensates \(\hat{Y}(\vec{x},t_{p})\).

The action \(\mcal{A}_{\mcal{N}^{-1}}\ppr[\hat{T};J_{\psi}]\) describes the dominant time development
as instant response to the 'seed' \(\hat{J}_{\psi\psi;\alpha\beta}^{a\neq b}(\vec{x},t_{p})\)
in \(\mcal{A}_{\hat{J}_{\psi\psi}}[\hat{T}]\) for the pair condensates. The action
\(\mcal{A}_{\mcal{N}^{0}}\ppr[\hat{T};J_{\psi}]\) of order \(\mcal{N}^{0}\) determines the effective
time development of the condensate pairs without the transport coefficients
\(c^{ij}(\vec{x},t_{p})\) and \(d^{ij}(\vec{x},t_{p})\) and can be understood from
mean field equations similar to the Gross-Pitaevskii equation in many-particle physics \cite{lipp}-\cite{dick}.
We list in relations (\ref{s5_7}) to (\ref{s5_12}) the relevant
definitions for the action \(\mcal{A}_{\mcal{N}^{-1}}\ppr[\hat{T};J_{\psi}]\) which represents
the Goldstone modes of the SSB \(Osp(S,S|2L)\backslash U(L|S)\otimes U(L|S)\)
\beq \lb{s5_7}
\lefteqn{
\mcal{A}_{\mcal{N}^{-1}}\ppr\big[\hat{T};J_{\psi}\big]= \frac{1}{4}\frac{1}{\mcal{N}}\int_{C}\frac{d
t_{p}}{\hbar}\sum_{\vec{x}} c^{ij}(\vec{x},t_{p})\;\;
\mbox{STR}\Big[\Big(\wtilde{\pp}_{i}\hat{Z}(\vec{x},t_{p})\Big)\;
\Big(\wtilde{\pp}_{j}\hat{Z}(\vec{x},t_{p})\Big)\Big] - \frac{\im}{\mcal{N}} \int_{C}\frac{d
t_{p}}{\hbar}\sum_{\vec{x}}\sum_{a,b=1,2} \times }   \\ \no &\times&
\sum_{\alpha,\beta=1}^{N=L+S} d^{ij}(\vec{x},t_{p})\;
\frac{J_{\psi;\beta}^{+,b}(\vec{x},t_{p})}{\mcal{N}}\bigg(\hat{I}\;\wtilde{K}\;
\Big(\wtilde{\pp}_{i}\hat{T}(\vec{x},t_{p})\Big)\;\hat{T}^{-1}(\vec{x},t_{p})\;
\Big(\wtilde{\pp}_{j}\hat{T}(\vec{x},t_{p})\Big)\;\hat{T}^{-1}(\vec{x},t_{p})\;
\hat{I}\bigg)_{\beta\alpha}^{ba}\;\frac{J_{\psi;\alpha}^{a}(\vec{x},t_{p})}{\mcal{N}}
\eeq
\beq \lb{s5_8}
\hat{Z}(\vec{x},t_{p})&=&\hat{T}(\vec{x},t_{p})\;\hat{S}\;\hat{T}^{-1}(\vec{x},t_{p}) \;\;\;;
\hspace*{0.5cm}\breve{v}(\vec{x},t_{p})=\breve{u}(\vec{x})+\breve{\sigma}_{D}^{(0)}(\vec{x},t_{p})=
\frac{u(\vec{x})+\sigma_{D}^{(0)}(\vec{x},t_{p})}{\mcal{N}} \\ \lb{s5_9}
c^{ij}(\vec{x},t_{p})&=&c^{(1),ij}(\vec{x},t_{p})+c^{(2),ij}(\vec{x},t_{p}) \\ \lb{s5_10}
c^{(1),ij}(\vec{x},t_{p})&=&
-2\bigg\langle\Big(\wtilde{\pp}_{i}\wtilde{\pp}_{j}\breve{v}(\vec{x},t_{p})\Big)
\bigg\rangle_{\hat{\sigma}_{D}^{(0)}} - \delta_{ij}\;\sum_{k=1}^{d}
\bigg\langle\Big(\wtilde{\pp}_{k}\wtilde{\pp}_{k}\breve{v}(\vec{x},t_{p})\Big)
\bigg\rangle_{\hat{\sigma}_{D}^{(0)}}  \\    \lb{s5_11}
c^{(2),ij}(\vec{x},t_{p})&=& 2\bigg\langle\Big(\wtilde{\pp}_{i}\breve{v}(\vec{x},t_{p})\Big)\;
\Big(\wtilde{\pp}_{j}\breve{v}(\vec{x},t_{p})\Big)\bigg\rangle_{\hat{\sigma}_{D}^{(0)}} -
\delta_{ij}\;\sum_{k=1}^{d} \bigg\langle\Big(\wtilde{\pp}_{k}\breve{v}(\vec{x},t_{p})\Big)^{2}
\bigg\rangle_{\hat{\sigma}_{D}^{(0)}}        \\ \lb{s5_12}
d^{ij}(\vec{x},t_{p}) &=& 2\;\bigg\langle 3\;\Big(\wtilde{\pp}_{i}\breve{v}(\vec{x},t_{p})\Big)\;
\Big(\wtilde{\pp}_{j}\breve{v}(\vec{x},t_{p})\Big) -
\Big(\wtilde{\pp}_{i}\wtilde{\pp}_{j}\breve{v}(\vec{x},t_{p})\Big)
\bigg\rangle_{\hat{\sigma}_{D}^{(0)}}\;.
\eeq
The action \(\mcal{A}_{\mcal{N}^{0}}\ppr[\hat{T};J_{\psi}]\) is given by Eq. (\ref{s5_13}) whose corresponding
mean field equation, following from first order variation with \(\delta\hat{Y}(\vec{x},t_{p})\),
is analogous to the Gross-Pitaevskii equation for the coherent BEC wave function, being invariant
under \(U(L|S)\) subgroup transformations of \(Osp(S,S|2L)\)
\beq \no
\mcal{A}_{\mcal{N}^{0}}\ppr\big[\hat{T};J_{\psi}\big]
&=& -\frac{1}{2}\int_{C}\frac{d t_{p}}{\hbar}\sum_{\vec{x}}\Bigg\{
\mbox{STR}\bigg[\hat{T}^{-1}(\vec{x},t_{p})\;\hat{S}\;\Big(\hat{E}_{p}\hat{T}(\vec{x},t_{p})\Big)+
\hat{T}^{-1}(\vec{x},t_{p})\;\Big(\wtilde{\pp}_{i}\wtilde{\pp}_{i}\hat{T}(\vec{x},t_{p})\Big)\bigg]+ \\ \lb{s5_13} &+&
\Big(u(\vec{x})-\mu_{0}-\im\;\ve_{p}+\big\langle\sigma_{D}^{(0)}(\vec{x},t_{p})\big\rangle_{\hat{\sigma}_{D}^{(0)}}
\Big)\;\mbox{STR}\bigg[\Big(\hat{T}^{-1}(\vec{x},t_{p})\Big)^{2}-\hat{1}_{2N\times 2N}\bigg]\Bigg\} + \\ \no &-&
\frac{\im}{2}\int_{C}\frac{d t_{p}}{\hbar}\sum_{\vec{x}}\sum_{a,b=1,2}\sum_{\alpha,\beta=1}^{N=L+S}
\frac{J_{\psi;\beta}^{+,b}(\vec{x},t_{p})}{\mcal{N}}\bigg[\hat{I}\;\wtilde{K}\;
\bigg(\Big(\wtilde{\pp}_{i}\wtilde{\pp}_{i}\hat{T}(\vec{x},t_{p})\Big)\;\hat{T}^{-1}(\vec{x},t_{p})+ \\ \no &+&
\hat{T}(\vec{x},t_{p})\;\hat{S}\;\hat{T}^{-1}(\vec{x},t_{p})\;\Big(\hat{E}_{p}\hat{T}(\vec{x},t_{p})\Big)\;
\hat{T}^{-1}(\vec{x},t_{p}) + \\ \no &-&2\;
\Big(\wtilde{\pp}_{i}\hat{T}(\vec{x},t_{p})\Big)\;\hat{T}^{-1}(\vec{x},t_{p})\;
\Big(\wtilde{\pp}_{i}\hat{T}(\vec{x},t_{p})\Big)\;\hat{T}^{-1}(\vec{x},t_{p})\bigg)
\hat{I}\bigg]_{\beta\alpha}^{ba}\;\frac{J_{\psi;\alpha}^{a}(\vec{x},t_{p})}{\mcal{N}}_{\mbox{.}}
\eeq
The action \(\mcal{A}_{\mcal{N}^{+1}}[\hat{T}]\) of order \(\mcal{N}^{+1}\) is the fluctuation
part whose first order and all higher odd-number variations of \(\delta\hat{Y}(\vec{x},t_{p})\)
vanish completely so that the first relevant variation is of quadratic order in the fields of the
pair condensates. This is caused by the additional presence of the metric \(\eta_{p}\) in the time
contour integral \(\int_{C}\big(dt_{p}/\hbar\big)\;\eta_{p}\ldots=
\int_{-\infty}^{+\infty}\big(dt_{+}/\hbar\big)\ldots+\int_{-\infty}^{+\infty}\big(dt_{-}/\hbar\big)\ldots\)
leading to the {\it sum} of the two branches of the contour integrals
\beq \lb{s5_14}
\mcal{A}_{\mcal{N}^{+1}}\ppr\big[\hat{T}\big]&=& \frac{\mcal{N}}{2}\int_{C}\frac{d
t_{p}}{\hbar}\eta_{p}\sum_{\vec{x}} \mbox{STR}\bigg[\Big(\hat{T}^{-1}(\vec{x},t_{p})\Big)^{2}-
\hat{1}_{2N\times 2N}\bigg]\;\;\;.
\eeq
Note that up to second order of the gradient expansion there are no coefficients as
\(c^{ij}(\vec{x},t_{p})\), \(d^{ij}(\vec{x},t_{p})\) or other parametric dependencies
in \(\mcal{A}_{\mcal{N}^{+1}}\ppr[\hat{T}]\) so that this fluctuation term
\(\mcal{A}_{\mcal{N}^{+1}}\ppr[\hat{T}]\) (\ref{s5_14}) only depends on the coset symmetry
\(Osp(S,S|2L)\backslash U(L|S)\) and therefore contributes universal fluctuations to the
action \(\mcal{A}_{\mcal{N}^{0}}\ppr[\hat{T};J_{\psi}]\) regarded as a mean field part similar to the
GP-equation.

\subsection{Classical nonlinear sigma model from variations
of the action $\mcal{A}_{\mcal{N}^{-1}}\ppr[\hat{T};J_{\psi}]$} \lb{s52}

In this subsection \ref{s52} we give a detailed description for classical effective equations
of the pair condensates as Goldstone modes from variation of
\(\mcal{A}_{\mcal{N}^{-1}}\ppr[\hat{T};J_{\psi}]\). The variation of
\(\mcal{A}_{\mcal{N}^{-1}}\ppr[\hat{T};J_{\psi}]\) has to be performed with respect to the
independent parameters of the coset generator \(\hat{Y}(\vec{x},t_{p})\) in the exponential
\(\hat{T}(\vec{x},t_{p})\). In Eqs. (\ref{s5_15}-\ref{s5_20}) we list again the matrix
elements \(\hat{c}_{D;mn}(\vec{x},t_{p})\), \(\hat{f}_{D;r\mu,r\ppr\nu}(\vec{x},t_{p})\) and
\(\hat{\eta}_{D;r\mu,n}(\vec{x},t_{p})\) for the super-symmetric pair condensates of the
matrices \(\hat{X}_{\alpha\beta}(\vec{x},t_{p})\) and \(\big(\hat{Y}(\vec{x},t_{p})\big)_{\alpha\beta}^{ab}\)
because these determine with their additional symmetries (\ref{s5_20}) the variations and the final
mean field equations
\beq \lb{s5_15}
\Big(\hat{T}(\vec{x},t_{p})\Big)_{\alpha\beta}^{ab}&=&
\bigg(\exp\Big\{-\big(\hat{Y}(\vec{x},t_{p})
\big)_{\alpha\ppr\beta\ppr}^{a\ppr b\ppr}\Big\}\bigg)_{\alpha\beta}^{ab}  \\  \lb{s5_16}
\Big(\hat{Y}(\vec{x},t_{p})\Big)_{\alpha\beta}^{ab}&=&
\left(\bea{cc} 0 & \hat{X}_{\alpha\beta}(\vec{x},t_{p})  \\
\wtilde{\kappa}_{\alpha}\;\hat{X}_{\alpha\beta}^{+}(\vec{x},t_{p}) & 0
\eea\right)^{ab}
\hspace*{0.64cm}\hat{Y}_{\alpha\beta}^{aa}\equiv 0\hspace*{0.46cm}\hat{Y}_{\alpha\beta}^{a\neq b}\neq 0
\\  \lb{s5_17}  \hat{X}_{\alpha\beta}(\vec{x},t_{p}) &=&
\left(\bea{cc}
-\hat{c}_{D;mn}(\vec{x},t_{p})  &  \hat{\eta}_{D;m,r\ppr\nu}^{T}(\vec{x},t_{p})   \\
-\hat{\eta}_{D;r\mu,n}(\vec{x},t_{p})  &  \hat{f}_{D;r\mu,r\ppr\nu}(\vec{x},t_{p})
\eea\right)_{\alpha\beta}    \\   \lb{s5_18}
\alpha &=& \Big\{\underbrace{m=1,\ldots,L}_{\mbox{\scz index for bosons}}\;;\;
\underbrace{r=1\:(\mu=1,2)\:,\ldots,\:S/2\:(\mu=1,2)}_{\mbox{\scz index for fermions}}\Big\} \\  \lb{s5_19}
\beta &=& \Big\{\underbrace{n=1,\ldots,L}_{\mbox{\scz index for bosons}}\;;\;
\underbrace{r\ppr=1\:(\nu=1,2)\:,\ldots,\:S/2\:(\nu=1,2)}_{\mbox{\scz index for fermions}}\Big\} \\ \lb{s5_20}
\hat{c}_{D;L\times L}^{T}(\vec{x},t_{p})&=&\hat{c}_{D;L\times L}(\vec{x},t_{p})\hspace*{1.0cm}
\hat{f}_{D;S\times S}^{T}(\vec{x},t_{p})\;=\;-\hat{f}_{D;S\times S}(\vec{x},t_{p})  \;\;\; .
\eeq
Furthermore, it is appropriate to consider the eigenvalues of \(\hat{X}_{\alpha\beta}(\vec{x},t_{p})\)
in the coset matrix \(\hat{Y}(\vec{x},t_{p})\) which appears in the combination of matrices
\(\hat{T}^{-1}(\vec{x},t_{p})\;\:d\big(\hat{T}(\vec{x},t_{p})\big)\) (\ref{s5_21}).
The transformation of \(\hat{Y}(\vec{x},t_{p})\) (or respectively of \(\hat{X}_{\alpha\beta}(\vec{x},t_{p})\))
to diagonal form is achieved by the \(U(L|S)\) matrix \(\hat{P}(\vec{x},t_{p})\) as subgroup of the
total \(Osp(S,S|2L)\) symmetry
\beq \lb{s5_21}
\hat{T}^{-1}(\vec{x},t_{p})\;\;d\big(\hat{T}(\vec{x},t_{p})\big)&=&
\exp\Big\{\hat{Y}(\vec{x},t_{p})\Big\}\;\;d\Big(\exp\Big\{-\hat{Y}(\vec{x},t_{p})\Big\}\Big)  \\ \lb{s5_22}
\hat{Y}(\vec{x},t_{p}) &=& \hat{P}^{-1}(\vec{x},t_{p})\;\;\hat{Y}_{DD}(\vec{x},t_{p})\;\;
\hat{P}(\vec{x},t_{p})   \;\;\;.
\eeq
The eigenvalues \(\hat{Y}_{DD;\alpha\beta}^{ab}(\vec{x},t_{p})\) (respectively
\(\hat{X}_{DD;\alpha\beta}(\vec{x},t_{p})\)) of \(\hat{Y}(\vec{x},t_{p})\) (and
\(\hat{X}(\vec{x},t_{p})\)) are grouped in relations (\ref{s5_23}-\ref{s5_29}).
It has to be noted that the eigenvalues \(\ovv{f}_{r\mu,r\nu}(\vec{x},t_{p})\)
(\(\mu,\nu=1,2\)) of the fermion-fermion part have quaternion structures with
the antisymmetric Pauli-matrix \(\big(\tau\big)_{\mu\nu}\) because the
complete matrix \(\hat{f}_{D;r\mu,r\ppr\nu}(\vec{x},t_{p})\) of the fermion-fermion section
is antisymmetric due to the property of pair condensates for the fermions.
The eigenvalues of the boson-boson and fermion-fermion parts of \(\hat{X}(\vec{x},t_{p})\)
are also separated into the modulus and phase of the diagonal elements
\(\ovv{c}_{m}(\vec{x},t_{p})\) and \(\ovv{f}_{r}(\vec{x},t_{p})\)
\beq  \lb{s5_23}
\hat{Y}_{DD;\alpha\beta}^{ab}(\vec{x},t_{p}) &=&
\left(\bea{cc} 0 & \hat{X}_{DD;\alpha\beta}(\vec{x},t_{p})  \\
\wtilde{\kappa}_{\alpha}\;\hat{X}_{DD;\alpha\beta}^{+}(\vec{x},t_{p})  &   0
\eea\right)^{ab}    \\   \lb{s5_24}
\hat{X}_{DD;\alpha\beta}(\vec{x},t_{p}) &=&
\left(\bea{cc}  -\ovv{c}_{m}(\vec{x},t_{p})\;\;\delta_{mn}    &    0  \\
0   & \ovv{f}_{r\mu,r\ppr\nu}(\vec{x},t_{p})\;\;\delta_{rr\ppr}
\eea\right)_{\alpha\beta}   \\   \lb{s5_25}
\hat{\ovv{c}}_{L\times L}(\vec{x},t_{p}) &=&\mbox{diag}\Big\{\ovv{c}_{1}(\vec{x},t_{p}),\ldots,
\ovv{c}_{m}(\vec{x},t_{p}),\ldots,
\ovv{c}_{L}(\vec{x},t_{p})\Big\}     \\  \lb{s5_26}
\ovv{c}_{m}(\vec{x},t_{p})  &=& \big|\ovv{c}_{m}(\vec{x},t_{p})\big|\;\;
\exp\big\{\im\;\varphi_{m}(\vec{x},t_{p})\Big\}\hspace*{1.0cm}\Big(\ovv{c}_{m}\in\mbox{\sf C}\Big)
  \\  \lb{s5_27}   \wtilde{f}_{S\times S}(\vec{x},t_{p}) &=&
\mbox{diag}\Big\{\big(\tau_{2}\big)_{\mu\nu}\;\ovv{f}_{1}(\vec{x},t_{p}),\ldots,
\big(\tau_{2}\big)_{\mu\nu}\;\ovv{f}_{r}(\vec{x},t_{p}),\ldots,
\big(\tau_{2}\big)_{\mu\nu}\;\ovv{f}_{S/2}(\vec{x},t_{p})\Big\}   \\    \lb{s5_28}
\ovv{f}_{r\mu,r\nu}(\vec{x},t_{p})  &=& \big(\tau_{2}\big)_{\mu\nu}\;\ovv{f}_{r}(\vec{x},t_{p}) = \im\;
\left(\bea{cc}  0 & -\ovv{f}_{r}(\vec{x},t_{p})   \\  \ovv{f}_{r}(\vec{x},t_{p})   &  0 \eea\right)_{\mu\nu}
    \\    \lb{s5_29}  \ovv{f}_{r}(\vec{x},t_{p})   &=& \big|\ovv{f}_{r}(\vec{x},t_{p})\big|\;\;
\exp\big\{\im\;\phi_{r}(\vec{x},t_{p})\big\}  \hspace*{1.0cm}\Big(\ovv{f}_{r}\in\mbox{\sf C}\Big) \;\;\;.
\eeq
The parameters of the diagonalizing matrix \(\hat{P}(\vec{x},t_{p})\) of \(\hat{Y}(\vec{x},t_{p})\)
(\ref{s5_22}) are tabulated in Eqs. (\ref{s5_30}-\ref{s5_35}). Since the eigenvalues
\(\hat{Y}_{DD;\alpha\beta}^{ab}(\vec{x},t_{p})\) (respectively \(\hat{X}_{DD;\alpha\beta}(\vec{x},t_{p})\))
already contain independent degrees of freedom of the pair condensate elements of \(\hat{Y}(\vec{x},t_{p})\),
\(\hat{X}(\vec{x},t_{p})\), the diagonal elements of the super-generators \(U(L|S)\) for
\(\hat{P}_{\alpha\beta}^{11}(\vec{x},t_{p})\), \(\hat{P}_{\alpha\beta}^{22}(\vec{x},t_{p})\) have to
vanish completely (compare symmetry restrictions and definitions of zero elements in the
boson-boson and fermion-fermion parts of the generators for \(\hat{P}(\vec{x},t_{p})\)
(\ref{s5_34},\ref{s5_35}))
\beq  \lb{s5_30}
\hat{P}_{\alpha\beta}^{ab}(\vec{x},t_{p})  &=&
\left(\bea{cc}  \hat{P}_{\alpha\beta}^{11}(\vec{x},t_{p})  &   0  \\
0  &  \hat{P}_{\alpha\beta}^{22}(\vec{x},t_{p})  \eea \right)^{ab}   \\  \lb{s5_31}
\Big(\hat{P}_{N\times N}^{22}(\vec{x},t_{p})\Big)^{st}&=&\hat{P}_{N\times N}^{11,+}(\vec{x},t_{p})=
\hat{P}_{N\times N}^{11,-1}(\vec{x},t_{p})  \\  \lb{s5_32}
\hat{P}_{\alpha\beta}^{11}(\vec{x},t_{p})  &=&
\exp\left\{\im\left(\bea{cc}
\hat{\mcal{C}}_{D;mn}(\vec{x},t_{p})  &  \hat{\xi}_{D;m,r\ppr\nu}^{+}(\vec{x},t_{p})   \\
\hat{\xi}_{D;r\mu,n}(\vec{x},t_{p})  &  \hat{\mcal{G}}_{D;r\mu,r\ppr\nu}(\vec{x},t_{p})
\eea\right)\right\}   \\     \lb{s5_33}
\hat{P}_{\alpha\beta}^{22}(\vec{x},t_{p})  &=&
\exp\left\{\im\left(\bea{cc}
-\hat{\mcal{C}}_{D;mn}^{T}(\vec{x},t_{p})  &  \hat{\xi}_{D;m,r\ppr\nu}^{T}(\vec{x},t_{p})   \\
-\hat{\xi}_{D;r\mu,n}^{*}(\vec{x},t_{p})  &  -\hat{\mcal{G}}_{D;r\mu,r\ppr\nu}^{T}(\vec{x},t_{p})
\eea\right)\right\}   \\    \lb{s5_34}
\hat{\mcal{C}}_{D;L\times L}^{+}(\vec{x},t_{p})  &=&  \hat{\mcal{C}}_{D;L\times L}(\vec{x},t_{p})\hspace*{1.0cm}
\mcal{C}_{D;mm}(\vec{x},t_{p})=0\;\;\;(m=1,\ldots,L)    \\  \lb{s5_35}
\hat{\mcal{G}}_{D;S\times S}^{+}(\vec{x},t_{p}) &=&  \hat{\mcal{G}}_{D;S\times S}(\vec{x},t_{p})\hspace*{1.0cm}
\mcal{G}_{D;r\mu,r\nu}(\vec{x},t_{p})=0\;\;\;(r=1,\ldots,S/2),(\mu,\nu=1,2)\;.
\eeq
The action \(\mcal{A}_{\mcal{N}^{-1}}\ppr[\hat{T};J_{\psi}]\) consists in essence of the combination
of matrices \(\exp\{\hat{Y}\}\;d\big(\exp\{-\hat{Y}\}\big)\) where the differential
\(d\big(\exp\{-\hat{Y}\}\big)\) of the exponential of \(\hat{Y}\) can be extended to a spatial
derivative \(\big(\wtilde{\pp}_{i}\exp\{-\hat{Y}\}\big)\) or to the necessary variations
\(\delta\big(\exp\{-\hat{Y}\}\big)\) with \(\delta\hat{Y}(\vec{x},t_{p})\) in order to derive the
mean field equations. Therefore, the relation (\ref{s5_36}), which has already been applied in the
calculation of the invariant integration measure, is of central importance for all further
variational steps to the classical nonlinear sigma model equations
\beq  \lb{s5_36}
\exp\Big\{\hat{Y}\Big\}\;d\Big(\exp\Big\{-\hat{Y}\Big\}\Big)
&=&-\int_{0}^{1}dv\;\;\exp\Big\{+v\;\hat{Y}\Big\}\;d\hat{Y}\;\exp\Big\{-v\;\hat{Y}\Big\}  \\ \no
&=&-\int_{0}^{1}dv\;\;\bigg(\exp\Big\{v\;\overrightarrow{[\hat{Y},\ldots]_{-}}\Big\}\;\;d\hat{Y}\bigg)
= -\bigg(\frac{\exp\Big\{
\overrightarrow{[\hat{Y},\ldots]_{-}}\Big\}\;-\;\hat{1}}{\overrightarrow{[\hat{Y},\ldots]_{-}}}\;\;
d\hat{Y}\bigg)   \\  \lb{s5_37}   \fa &=& \FA\;\;\;.
\eeq
The integration over the parameter \(v\in[0,1]\) is acquired from the operator
{\scz \(\exp\big\{v\;\overrightarrow{[\hat{Y},\ldots]_{-}}\big\}\)} which contains the commutator
{\scz \(\overrightarrow{[\hat{Y},\ldots]_{-}}\)} acting onto the differential \(d\hat{Y}\) within
the outer braces (see second line in (\ref{s5_36})). The transformation of the first to the
second line in (\ref{s5_36}) is known from statistical mechanics where one introduces
the Liouville operator acting on other operators in the fundamental von Neumann equation
with the statistical operator. After formal integration over \(v\in[0,1]\), a closed
relation follows for \(\exp\{\hat{Y}\}\;d\big(\exp\{-\hat{Y}\}\big)\) which contains
the commutator function {\scz\(\fa\)} (\ref{s5_37}) acting onto \(d\hat{Y}\) within the outer
braces in (\ref{s5_36}). We record in Eqs. (\ref{s5_38}-\ref{s5_40}) modifications of
relation (\ref{s5_36}) with \(\exp\{\hat{Y}\}\;d\big(\exp\{-\hat{Y}\}\big)\) and operator
{\scz\(\fa\)} (\ref{s5_37}) because they have to be used in the various steps of the
variation of \(\mcal{A}_{\mcal{N}^{-1}}\ppr[\hat{T};J_{\psi}]\) with respect to
\(\delta\hat{Y}(\vec{x},t_{p})\)
\beq   \lb{s5_38}
\exp\Big\{-\hat{Y}\Big\}\;d\Big(\exp\Big\{\hat{Y}\Big\}\Big)
&=&\int_{0}^{1}dv\;\;\exp\Big\{-v\;\hat{Y}\Big\}\;d\hat{Y}\;\exp\Big\{+v\;\hat{Y}\Big\}
 \\ \no &=& \bigg(\FAm\;\;d\hat{Y}\bigg)=\bigg(\famm\;\;d\hat{Y}\bigg)   \\   \lb{s5_39}
d\Big(\exp\Big\{-\hat{Y}\Big\}\Big)\;\exp\Big\{\hat{Y}\Big\} &=&-
\exp\Big\{-\hat{Y}\Big\}\;d\Big(\exp\Big\{\hat{Y}\Big\}\Big) \\ \no &=&-
\int_{0}^{1}dv\;\;\exp\Big\{-v\;\hat{Y}\Big\}\;d\hat{Y}\;\exp\Big\{+v\;\hat{Y}\Big\}  \\ \no &=&
-\bigg(\FAm\;\;d\hat{Y}\bigg)=-\bigg(\famm\;\;d\hat{Y}\bigg)    \\     \lb{s5_40}
d\Big(\exp\Big\{\hat{Y}\Big\}\Big)\;\exp\Big\{-\hat{Y}\Big\} &=&-
\exp\Big\{\hat{Y}\Big\}\;d\Big(\exp\Big\{-\hat{Y}\Big\}\Big) \\ \no &=&
\int_{0}^{1}dv\;\;\exp\Big\{+v\;\hat{Y}\Big\}\;d\hat{Y}\;\exp\Big\{-v\;\hat{Y}\Big\}  \\ \no &=&
\bigg(\FA\;\;d\hat{Y}\bigg)=\bigg(\fa\;\;d\hat{Y}\bigg)_{\mbox{.}}
\eeq
We combine relations of the type (\ref{s5_36}-\ref{s5_40}) with diagonal parameters or eigenvalues
of \(\hat{Y}(\vec{x},t_{p})\), consisting of \(\hat{Y}_{DD;\alpha\beta}^{ab}(\vec{x},t_{p})\)
and 'eigenvectors' \(\hat{P}_{\alpha\beta}^{aa}(\vec{x},t_{p})\), so that the commutator argument
of the operator {\scz\(\fa\)} only involves the diagonal part {\scz\(\overrightarrow{[\hat{Y}_{DD},\ldots]_{-}}\)}
of the coset operator.
These disentangling relations are given for both parts of \(\mcal{A}_{\mcal{N}^{-1}}\ppr[\hat{T};J_{\psi}]\)
with coefficients \(c^{ij}(\vec{x},t_{p})\) and \(d^{ij}(\vec{x},t_{p})\). However, the spatial
derivative \(\big(\wtilde{\pp}_{i}\hat{Y}(\vec{x},t_{p})\big)\) has to be rotated by
\(\hat{P}(\vec{x},t_{p})\), \(\hat{P}^{-1}(\vec{x},t_{p})\) to the matrix
\(\big(\wtilde{\pp}_{i}\hat{Y}(\vec{x},t_{p})\big)_{P}\) (\ref{s5_48}) so that the
matrices \(\hat{P}\), \(\hat{P}^{-1}\) are removed from the operator {\scz\(\fa\)} and the diagonal
coset commutator argument {\scz\(\overrightarrow{[\hat{Y}_{DD},\ldots]_{-}}\)} results
\beq \no
\lefteqn{\hat{T}^{-1}(\vec{x},t_{p})\;\Big(\wtilde{\pp}_{i}\hat{T}(\vec{x},t_{p})\Big)=-
\int_{0}^{1}dv\;\;\exp\Big\{+v\;\hat{Y}(\vec{x},t_{p})\Big\}\;
\Big(\wtilde{\pp}_{i}\hat{Y}(\vec{x},t_{p})\Big)\;
\exp\Big\{-v\;\hat{Y}(\vec{x},t_{p})\Big\}   =  }  \\  \lb{s5_46}  &=&
-\hat{P}^{-1}\;\int_{0}^{1}dv\;\;\exp\Big\{+v\;\hat{Y}_{DD}\Big\}\;
\bigg(\hat{P}\;\Big(\wtilde{\pp}_{i}\hat{P}^{-1}\;\hat{Y}_{DD}\;\hat{P}\Big)\;
\hat{P}^{-1}\bigg)\;\exp\Big\{-v\;\hat{Y}_{DD}\Big\}\;\hat{P}
\\ \no &=&-\bigg(\FA\;\Big(\wtilde{\pp}_{i}\hat{Y}\Big)\bigg) \\ \no &=&
-\hat{P}^{-1}(\vec{x},t_{p})\;\;
\bigg({\raisebox{0pt}{$\frac{\exp\big\{\overrightarrow{[\hat{Y}_{DD},\ldots]_{-}}\big\}\;-
\;{\displaystyle\hat{1}}}{\overrightarrow{[\hat{Y}_{DD},\ldots]_{-}}}$}}
\;\underbrace{\Big(\hat{P}(\vec{x},t_{p})\;\Big(\wtilde{\pp}_{i}\hat{P}^{-1}\;\hat{Y}_{DD}\;\hat{P}\Big)\;
\hat{P}^{-1}(\vec{x},t_{p})\Big)}_{\big(\wtilde{\pp}_{i}\hat{Y}(\vec{x},t_{p})\big)_{\hat{P}}}\bigg)
\;\;\hat{P}(\vec{x},t_{p})
\eeq
\beq \no
\lefteqn{\Big(\wtilde{\pp}_{i}\hat{T}(\vec{x},t_{p})\Big)\;\hat{T}^{-1}(\vec{x},t_{p})=-
\int_{0}^{1}dv\;\;\exp\Big\{-v\;\hat{Y}(\vec{x},t_{p})\Big\}\;
\Big(\wtilde{\pp}_{i}\hat{Y}(\vec{x},t_{p})\Big)\;
\exp\Big\{+v\;\hat{Y}(\vec{x},t_{p})\Big\}   =  }
\\   \lb{s5_47}    &=&-\bigg(\FAm\;\Big(\wtilde{\pp}_{i}\hat{Y}\Big)\bigg) \\ \no &=&
-\hat{P}^{-1}(\vec{x},t_{p})\;\;
\bigg({\raisebox{0pt}{$\frac{\exp\big\{\overrightarrow{[\hat{Y}_{DD},\ldots]_{-}}\big\}\;-
\;{\displaystyle\hat{1}}}{-\overrightarrow{[\hat{Y}_{DD},\ldots]_{-}}}$}}
\;\underbrace{\Big(\hat{P}(\vec{x},t_{p})\;\Big(\wtilde{\pp}_{i}\hat{P}^{-1}\;\hat{Y}_{DD}\;\hat{P}\Big)\;
\hat{P}^{-1}(\vec{x},t_{p})\Big)}_{\big(\wtilde{\pp}_{i}\hat{Y}(\vec{x},t_{p})\big)_{\hat{P}}}\bigg)
\;\;\hat{P}(\vec{x},t_{p})
\eeq
\be \lb{s5_48}
\Big(\hat{P}(\vec{x},t_{p})\;\Big(\wtilde{\pp}_{i}\hat{P}^{-1}(\vec{x},t_{p})\;
\hat{Y}_{DD}(\vec{x},t_{p})\;\hat{P}(\vec{x},t_{p})\Big)\;
\hat{P}^{-1}(\vec{x},t_{p})\Big) = \Big(\wtilde{\pp}_{i}\hat{Y}(\vec{x},t_{p})\Big)_{\hat{P}}\;.
\ee
It still seems to be a problem to perform powers of the commutator
{\scz\(\overrightarrow{[\hat{Y}_{DD},\ldots]_{-}}\)} in {\scz\(\fa\)} onto the rotated derivatives
of pair condensate fields
\(\big(\wtilde{\pp}_{i}\hat{Y}(\vec{x},t_{p})\big)_{P}\) by repeated application; but it has to be
noted  that the adjoint representation of a group or super-group as \(Osp(S,S|2L)\), determined
by its structure constants, reduces the commutators (or respectively anticommutators) to the
multiplication of matrices composed of the structure elements (see subsection \ref{s53} for a
detailed description).

Using relations (\ref{s5_15}-\ref{s5_48}) as preparation, we start with the first order variation
of the part \(\mcal{A}_{\mcal{N}^{-1}}^{\prime(1)}[\hat{Z}=\hat{T}\;\hat{S}\;\hat{T}^{-1}]\) of
\(\mcal{A}_{\mcal{N}^{-1}}\ppr[\hat{T};J_{\psi}]\) from the coherent state path integral
\(Z_{\mcal{N}^{-1}}[\hat{\mcal{J}},J_{\psi},\im\hat{J}_{\psi\psi}]\) restricted to actions
of order \(\mcal{N}^{-1}\). The first part \(\mcal{A}_{\mcal{N}^{-1}}^{\prime(1)}[\hat{Z}]=
\mcal{A}_{\mcal{N}^{-1}}^{\prime(1)}[\hat{T}]\) comprises the matrix
\(\hat{Z}(\vec{x},t_{p})=\hat{T}(\vec{x},t_{p})\;\hat{S}\;\hat{T}^{-1}(\vec{x},t_{p})\) and
the coefficients \(c^{ij}(\vec{x},t_{p})\), resulting from averages over the background field
\(\sigma_{D}^{(0)}(\vec{x},t_{p})\). The introduction of the matrix \(\hat{Z}(\vec{x},t_{p})\)
limits the super-trace '\(\mbox{STR}\)' in \(\mcal{A}_{\mcal{N}^{-1}}^{\prime(1)}[\hat{T}]\) to the
non-diagonal 'Nambu'-block parts or the coset elements of \(Osp(S,S|2L)\backslash U(L|S)\)
for the pair condensates without any elements in the block diagonal subgroup parts \(U(L|S)\)
\beq \lb{s5_49}
Z_{\mcal{N}^{-1}}[\hat{\mcal{J}},J_{\psi},\im\hat{J}_{\psi\psi}]&=&\int
d[\hat{T}^{-1}(\vec{x},t_{p})\;d\hat{T}(\vec{x},t_{p})]\;\;
\exp\Big\{\im\;\mcal{A}_{\hat{J}_{\psi\psi}}[\hat{T}]\Big\} \\ \no &\times &
\exp\Big\{-\mcal{A}_{\mcal{N}^{-1}}\ppr[\hat{T};J_{\psi}]\Big\}\;\;\;
\exp\Big\{-\mcal{A}\ppr[\hat{T};\hat{\mcal{J}}]\Big\}   \\     \lb{s5_50}
\mcal{A}_{\mcal{N}^{-1}}\ppr[\hat{T};J_{\psi}]&=&
\mcal{A}_{\mcal{N}^{-1}}^{\prime(1)}[\hat{Z}=\hat{T}\;\hat{S}\;\hat{T}^{-1}]+
\mcal{A}_{\mcal{N}^{-1}}^{\prime(2)}[\hat{T};J_{\psi}]   \\  \lb{s5_51}
\mcal{A}_{\mcal{N}^{-1}}^{\prime(1)}[\hat{Z}=\hat{T}\;\hat{S}\;\hat{T}^{-1}] &=&
\frac{1}{4}\frac{1}{\mcal{N}}\int_{C}\frac{dt_{p}}{\hbar}\sum_{\vec{x}}c^{ij}(\vec{x},t_{p})\;\;
\mbox{STR}\Big[\Big(\wtilde{\pp}_{i}\hat{Z}(\vec{x},t_{p})\Big)\;
\Big(\wtilde{\pp}_{j}\hat{Z}(\vec{x},t_{p})\Big)\Big]_{\mbox{.}}
\eeq
The second part \(\mcal{A}_{\mcal{N}^{-1}}^{\prime(2)}[\hat{T};J_{\psi}]\) of
\(\mcal{A}_{\mcal{N}^{-1}}\ppr[\hat{T};J_{\psi}]\) with source field \(J_{\psi;\alpha}^{a}(\vec{x},t_{p})\)
for a macroscopic BEC wave function is transformed to a relation with a super-trace (\ref{s5_52})
by including the dyadic product of the source fields (\ref{s5_53}). This leads to a super-matrix
\(\wtilde{J}_{J_{\psi},J_{\psi}^{+};\alpha\beta}^{ab}(\vec{x},t_{p})\;\wtilde{K}\)
as generator of the ortho-symplectic super-algebra (\ref{s5_54})
\beq \lb{s5_52}
\lefteqn{\mcal{A}_{\mcal{N}^{-1}}^{\prime(2)}[\hat{T};J_{\psi}] =
\mcal{A}_{\mcal{N}^{-1}}^{\prime(2)}[\hat{T};\wtilde{J}_{J_{\psi},J_{\psi}^{+}}\;\wtilde{K}]=  }
\\ \no &=& -\frac{\im}{\mcal{N}}\int_{C}\frac{dt_{p}}{\hbar}\sum_{\vec{x}}d^{ij}(\vec{x},t_{p})\;\;
\mbox{STR}\bigg[\Big(\wtilde{\pp}_{i}\hat{T}\Big)\;\hat{T}^{-1}\;\;
\underbrace{\hat{I}\;\Big(\breve{J}_{\psi;\alpha}^{a}(\vec{x},t_{p})\otimes
\breve{J}_{\psi;\beta}^{+,b}(\vec{x},t_{p})\Big)\;\hat{I}}_{
\wtilde{J}_{J_{\psi},J_{\psi}^{+};\alpha\beta}^{ab}(\vec{x},t_{p})}\;\wtilde{K}\;\;
\Big(\wtilde{\pp}_{j}\hat{T}\Big)\;\hat{T}^{-1}\bigg]
\eeq
\beq  \lb{s5_53}
\wtilde{J}_{J_{\psi},J_{\psi}^{+};\alpha\beta}^{ab}(\vec{x},t_{p}) &=&
\hat{I}_{\alpha}^{a}\;\Big(\breve{J}_{\psi;\alpha}^{a}(\vec{x},t_{p})\otimes
\breve{J}_{\psi;\beta}^{+,b}(\vec{x},t_{p})\Big)\;\hat{I}_{\beta}^{b}\;\hspace*{1.0cm}
\breve{J}_{\psi;\alpha}^{a}(\vec{x},t_{p})=\frac{J_{\psi;\alpha}^{a}(\vec{x},t_{p})}{\mcal{N}}
 \\   \lb{s5_54}
\wtilde{J}_{J_{\psi},J_{\psi}^{+}}(\vec{x},t_{p})\;\wtilde{K}&\in& osp(S,S|2L)\;\;\;.
\eeq
Finally, we acquire the first order variation of
\(\mcal{A}_{\mcal{N}^{-1}}^{\prime(1)}[\hat{T}]\) with respect to \(\delta\hat{Z}\) and
\(\big(\delta\hat{T}\big)\;\hat{T}^{-1}\). We use relation (\ref{s5_39}) to reduce the
variation with \(\big(\delta\hat{T}\big)\;\hat{T}^{-1}\)
to the variation with \(\delta\hat{Y}\) (\ref{s5_57})
\beq   \lb{s5_55}
\delta\mcal{A}_{\mcal{N}^{-1}}^{\prime(1)}[\hat{T}]&=&-\frac{1}{2}\frac{1}{\mcal{N}}
\int_{C}\frac{dt_{p}}{\hbar}\sum_{\vec{x}}
\mbox{STR}\bigg[\delta\hat{Z}(\vec{x},t_{p})\;\Big(\wtilde{\pp}_{i}\;
c^{ij}(\vec{x},t_{p})\;\Big(\wtilde{\pp}_{j}\hat{Z}(\vec{x},t_{p})\Big)\Big)\bigg]  \\ \no
\hat{Z}(\vec{x},t_{p})&=&\hat{T}(\vec{x},t_{p})\;\hat{S}\;\hat{T}^{-1}(\vec{x},t_{p})=
\Big(\exp\Big\{-\overrightarrow{[\hat{Y}(\vec{x},t_{p}),\ldots]_{-}}\Big\}\;\;\hat{S}\Big) \\ \no &=&
\hat{S}\;\;\hat{T}^{-2}(\vec{x},t_{p})=\hat{S}\;\;\exp\big\{2\;\hat{Y}(\vec{x},t_{p})\big\} \\ \lb{s5_56}
\delta \hat{Z} &=& \big(\delta\hat{T}\big)\;\hat{S}\;\hat{T}^{-1}-\hat{T}\;\hat{S}\;\hat{T}^{-1}\;
\big(\delta\hat{T}\big)\;\hat{T}^{-1}  =
\Big(\big(\delta\hat{T}\big)\;\hat{T}^{-1} \Big)\;\;\hat{Z}-\hat{Z}\;\;
\Big(\big(\delta\hat{T}\big)\;\hat{T}^{-1}\Big) \\ \no &=&
\Big[\big(\delta\hat{T}\big)\;\hat{T}^{-1} \;,\;\hat{Z}\Big]_{-}  \\  \lb{s5_57}
\big(\delta\hat{T}\big)\;\hat{T}^{-1} &=& \delta\Big(\exp\{-\hat{Y}\}\Big)\;\;\exp\{\hat{Y}\}
=-\int_{0}^{1}dv\;\;\exp\{-v\;\hat{Y}\}\;\;\delta\hat{Y}\;\;\exp\{+v\;\hat{Y}\} \;\;\;.
\eeq
This allows to confine the variation of \(\delta\mcal{A}_{\mcal{N}^{-1}}^{\prime(1)}[\hat{T}]\)
to the variation of \(\delta\hat{Y}(\vec{x},t_{p})\) as the original independent degrees of freedom
of the pair condensates
\beq  \lb{s5_58}
\lefteqn{\delta\mcal{A}_{\mcal{N}^{-1}}^{\prime(1)}[\hat{T}]=\frac{1}{2}\frac{1}{\mcal{N}}
\int_{C}\frac{dt_{p}}{\hbar}\sum_{\vec{x}}  } \\ \no &\times&
\mbox{STR}\bigg\{\delta\hat{Y}(\vec{x},t_{p})\int_{0}^{1}dv\;\;e^{+v\;\hat{Y}(\vec{x},t_{p})}\;
\bigg[\hat{Z}(\vec{x},t_{p})\;,\;\Big(\wtilde{\pp}_{i}\;
c^{ij}(\vec{x},t_{p})\;\Big(\wtilde{\pp}_{j}\hat{Z}(\vec{x},t_{p})\Big)\Big)\bigg]_{-}
\;e^{-v\;\hat{Y}(\vec{x},t_{p})}\bigg\}_{\mbox{.}}
\eeq
The variation of the second part \(\mcal{A}_{\mcal{N}^{-1}}^{\prime(2)}[\hat{T};J_{\psi}]\)
is given in Eq. (\ref{s5_59}) with the
coefficients \(d^{ij}(\vec{x},t_{p})\) and the dyadic source matrix
\(\wtilde{J}_{J_{\psi},J_{\psi}^{+};\alpha\beta}^{ab}(\vec{x},t_{p})\;\wtilde{K}\).
It consists of an anticommutator (\ref{s5_60}) between the term
\(\big(d^{ij}(\vec{x},t_{p})\;\big(\wtilde{\pp}_{j}\hat{T}\big)\;\hat{T}^{-1}\big)\)
of anomalous fields and the dyadic source matrix
\(\wtilde{J}_{J_{\psi},J_{\psi}^{+};\alpha\beta}^{ab}(\vec{x},t_{p})\;\wtilde{K}\)
as an element of the \(osp(S,S|2L)\) super-algebra
\beq  \lb{s5_59}
\lefteqn{\delta\mcal{A}_{\mcal{N}^{-1}}^{\prime(2)}[\hat{T};J_{\psi}]=-\frac{\im}{\mcal{N}}
\int_{C}\frac{dt_{p}}{\hbar}\sum_{\vec{x}}   }    \\ \no &\times &
\mbox{STR}\bigg[\delta\Big(\big(\wtilde{\pp}_{i}\hat{T}\big)\;\hat{T}^{-1}\Big)\;\;
\Big\{\Big(d^{ij}(\vec{x},t_{p})\;\big(\wtilde{\pp}_{j}\hat{T}\big)\;\hat{T}^{-1}\Big)\;,\;
\wtilde{J}_{J_{\psi},J_{\psi}^{+}}(\vec{x},t_{p})\;\wtilde{K}\Big\}_{+}\bigg]\;.
\eeq
\beq \lb{s5_60}
\lefteqn{\Big\{\Big(d^{ij}(\vec{x},t_{p})\;\big(\wtilde{\pp}_{j}\hat{T}\big)\;\hat{T}^{-1}\Big)\;,\;
\wtilde{J}_{J_{\psi},J_{\psi}^{+}}(\vec{x},t_{p})\;\wtilde{K}\Big\}_{+}  =  } \\ \no &=&
\Big(d^{ij}(\vec{x},t_{p})\;\big(\wtilde{\pp}_{j}\hat{T}\big)\;\hat{T}^{-1}\Big)\;\;
\wtilde{J}_{J_{\psi},J_{\psi}^{+}}(\vec{x},t_{p})\;\wtilde{K} \;+\;
\wtilde{J}_{J_{\psi},J_{\psi}^{+}}(\vec{x},t_{p})\;\wtilde{K}\;\;
\Big(d^{ij}(\vec{x},t_{p})\;\big(\wtilde{\pp}_{j}\hat{T}\big)\;\hat{T}^{-1}\Big)_{\mbox{.}}
\eeq
The variation in (\ref{s5_59}) with respect to
\(\delta\big(\big(\wtilde{\pp}_{i}\hat{T}\big)\;\hat{T}^{-1}\big)\) has to be attributed to
the variation with \(\delta\hat{Y}(\vec{x},t_{p})\) of the original independent degrees of freedom
for the pair condensates. Repeated application of (\ref{s5_36}-\ref{s5_40}) yields relation (\ref{s5_64})
for \(\delta\big(\big(\wtilde{\pp}_{i}\hat{T}\big)\;\hat{T}^{-1}\big)\) with variation of
\(\delta\hat{Y}(\vec{x},t_{p})\) and the spatial derivative \(\big(\wtilde{\pp}_{i}\;\delta\hat{Y}\big)\)
\beq  \lb{s5_61}
\delta\Big(\big(\wtilde{\pp}_{i}\hat{T}\big)\;\hat{T}^{-1}\Big)&=&-\int_{0}^{1}dv\;\;
\bigg[\delta\Big(e^{-v\;\hat{Y}}\Big) \;\Big(\wtilde{\pp}_{i}\hat{Y}\Big)\;e^{+v\;\hat{Y}}+
e^{-v\;\hat{Y}}\;\Big(\wtilde{\pp}_{i}\hat{Y}\Big)\;\delta\Big(e^{+v\;\hat{Y}}\Big) +  \\ \no &+&
e^{-v\;\hat{Y}}\;\Big(\wtilde{\pp}_{i}\;\delta\hat{Y}\Big)\;e^{+v\;\hat{Y}}\bigg]    \\   \lb{s5_62}
\delta\Big(e^{-v\;\hat{Y}}\Big) &=& e^{-v\;\hat{Y}}\;\Big(e^{+v\;\hat{Y}}\;\;
\delta\Big(e^{-v\;\hat{Y}}\Big)\Big) = - e^{-v\;\hat{Y}}\;
\int_{0}^{1}du\;\;e^{+u\;v\;\hat{Y}}\;v\;\delta\hat{Y}\;e^{-u\;v\;\hat{Y}}  \\  \lb{s5_63}
\delta\Big(e^{+v\;\hat{Y}}\Big) &=&  e^{+v\;\hat{Y}}\;
\int_{0}^{1}du\;\;e^{-u\;v\;\hat{Y}}\;v\;\delta\hat{Y}\;e^{+u\;v\;\hat{Y}}   \\  \lb{s5_64}
\delta\Big(\big(\wtilde{\pp}_{i}\hat{T}\big)\;\hat{T}^{-1}\Big)&=&-\int_{0}^{1}dv\;\;
e^{-v\;\hat{Y}}\;\Big(\wtilde{\pp}_{i}\;\delta\hat{Y}\Big)\;e^{+v\;\hat{Y}}  +  \\ \no &+&
\int_{0}^{1}dv\int_{0}^{1}du\;v\bigg[e^{-v\;\hat{Y}}\;e^{+u\;v\;\hat{Y}}\;\delta\hat{Y}\;
e^{-u\;v\;\hat{Y}}\;\Big(\wtilde{\pp}_{i}\hat{Y}\Big)\;e^{+v\;\hat{Y}} + \\ \no &-&
e^{-v\;\hat{Y}}\;\Big(\wtilde{\pp}_{i}\hat{Y}\Big)\;e^{+v\;\hat{Y}}\;
e^{-u\;v\;\hat{Y}}\;\delta\hat{Y}\;e^{+u\;v\;\hat{Y}}\bigg]_{\mbox{.}}
\eeq
The combination of Eqs. (\ref{s5_59}-\ref{s5_64}) finally leads to the variation of
\(\delta\mcal{A}_{\mcal{N}^{-1}}^{\prime(2)}[\hat{T};J_{\psi}]\) (\ref{s5_65}) where we have
substituted the real integration variable $u$  in (\ref{s5_64}) by \(w=u\cdot v\)
so that the second super-trace term in (\ref{s5_65}) has the double integrations
\(\int_{0}^{1}dv\int_{0}^{v}dw\ldots\)
\beq  \lb{s5_65}
\lefteqn{\delta\mcal{A}_{\mcal{N}^{-1}}^{\prime(2)}[\hat{T};J_{\psi}]=-\frac{\im}{\mcal{N}}
\int_{C}\frac{dt_{p}}{\hbar}\sum_{\vec{x}}   }     \\    \no &\times& \Bigg(
\mbox{STR}\bigg\{\delta\hat{Y}(\vec{x},t_{p})
\bigg(\wtilde{\pp}_{i}\int_{0}^{1}dv\;\;e^{+v\;\hat{Y}}\;
\Big\{d^{ij}\;\big(\wtilde{\pp}_{j}\hat{T}\big)\;\hat{T}^{-1}\;,\;
\wtilde{J}_{J_{\psi},J_{\psi}^{+}}\;\wtilde{K}\Big\}_{+}\;e^{-v\;\hat{Y}}\bigg)\bigg\} + \\ \no &+&
\mbox{STR}\bigg\{\delta\hat{Y}(\vec{x},t_{p})\bigg[\int_{0}^{1}dv\int_{0}^{v}dw\bigg(
e^{-w\;\hat{Y}}\;\Big(\wtilde{\pp}_{i}\hat{Y}\Big)\;e^{+v\;\hat{Y}}\;
\Big\{d^{ij}\;\big(\wtilde{\pp}_{j}\hat{T}\big)\;\hat{T}^{-1}\;,\;
\wtilde{J}_{J_{\psi},J_{\psi}^{+}}\;\wtilde{K}\Big\}_{+}\;e^{-v\;\hat{Y}}\;e^{+w\;\hat{Y}}\bigg) + \\ \no &-&
\bigg(e^{+w\;\hat{Y}}\;\Big\{d^{ij}\;\big(\wtilde{\pp}_{j}\hat{T}\big)\;\hat{T}^{-1}\;,\;
\wtilde{J}_{J_{\psi},J_{\psi}^{+}}\;\wtilde{K}\Big\}_{+}\;e^{-v\;\hat{Y}}\;
\Big(\wtilde{\pp}_{i}\hat{Y}\Big)\;e^{+v\;\hat{Y}}\;e^{-w\;\hat{Y}}\bigg)\bigg]\bigg\}
\Bigg)_{\mbox{.}}
\eeq
The variation of the independent fields in \(\delta\hat{Y}(\vec{x},t_{p})\) on the two branches of
the time contour is separated into the field \(\hat{Y}(\vec{x},t)\) with a mean time $t$ without
contour index and into \(\pm\frac{1}{2}\;\delta\hat{Y}(\vec{x},t)\) where the $\pm$ sign refers to
the corresponding time branch in \(\hat{Y}(\vec{x},t_{p=\pm})\)
\beq \lb{s5_66}
\hat{Y}(\vec{x},t_{p=\pm})  &=& \hat{Y}(\vec{x},t)+\delta\hat{Y}(\vec{x},t_{p=\pm})=
\hat{Y}(\vec{x},t)\pm\frac{1}{2}\;\delta\hat{Y}(\vec{x},t)  \\ \no
\hat{Y}(\vec{x},t_{p=\pm})  &=& y_{\kappa}(\vec{x},t_{p=\pm})\;\;\hat{Y}^{(\kappa)}  \\ \no
y_{\kappa}(\vec{x},t_{p=\pm}) &=& y_{\kappa}(\vec{x},t)+\delta y_{\kappa}(\vec{x},t_{p=\pm})=
y_{\kappa}(\vec{x},t)\pm\frac{1}{2}\;\delta y_{\kappa}(\vec{x},t)\;\;\;.
\eeq
The total variation (\ref{s5_67}) of the sum \(\delta\mcal{A}_{\mcal{N}^{-1}}^{\prime(1)}[\hat{T}]\)
(\ref{s5_58}) and \(\delta\mcal{A}_{\mcal{N}^{-1}}^{\prime(2)}[\hat{T};J_{\psi}]\) (\ref{s5_65})
with respect to \(\delta\hat{Y}(\vec{x},t_{p})=\eta_{p}\;\frac{1}{2}\;\delta y_{\kappa}(\vec{x},t)\;\hat{Y}^{(\kappa)}\)
results in equation (\ref{s5_68}) for the matrices
\(\hat{Z}(\vec{x},t_{p})\), \(\hat{T}(\vec{x},t_{p})\) and also in the dependence on
\(\hat{Y}(\vec{x},t_{p})\) and \(\big(\wtilde{\pp}_{i}\hat{Y}(\vec{x},t_{p})\big)\)
\beq \lb{s5_67}
\delta\mcal{A}_{\mcal{N}^{-1}}\ppr[\hat{T}\;J_{\psi}] &=&
\delta\mcal{A}_{\mcal{N}^{-1}}^{\prime(1)}[\hat{T}]+
\delta\mcal{A}_{\mcal{N}^{-1}}^{\prime(2)}[\hat{T};J_{\psi}]\stackrel{!}{=}0   \\  \lb{s5_68}
0 &=& \mbox{STR}\bigg\{\hat{Y}^{(\kappa)}\bigg[\frac{1}{2}\int_{0}^{1}dv
\bigg(\mbox{$e^{v\;\overrightarrow{[\hat{Y},\ldots]_{-}}}$}\;
\Big[\hat{Z}\;,\;\Big(\wtilde{\pp}_{i}c^{ij}\;
\Big(\wtilde{\pp}_{j}\hat{Z}\Big)\Big)\Big]_{-}\bigg)^{a\neq b} +  \\ \no &-&\im\;
\int_{0}^{1}dv\bigg(\wtilde{\pp}_{i}\;\mbox{$e^{v\;\overrightarrow{[\hat{Y}(\vec{x},t_{p}),\ldots]_{-}}}$}\;
\Big\{d^{ij}\;\big(\wtilde{\pp}_{j}\hat{T}\big)\;\hat{T}^{-1}\;,\;
\wtilde{J}_{J_{\psi},J_{\psi}^{+}}\;\wtilde{K}\Big\}_{+}\bigg)^{a\neq b}  + \\ \no &-&\im\;
\int_{0}^{1}dv\int_{0}^{v}dw\;
\bigg(\mbox{$e^{-w\;\overrightarrow{[\hat{Y},\ldots]_{-}}}$}\;
\Big(\wtilde{\pp}_{i}\hat{Y}\Big)\;e^{+v\;\hat{Y}}\;
\Big\{d^{ij}\;\big(\wtilde{\pp}_{j}\hat{T}\big)\;\hat{T}^{-1}\;,\;
\wtilde{J}_{J_{\psi},J_{\psi}^{+}}\;\wtilde{K}\Big\}_{+}\;e^{-v\;\hat{Y}}\bigg)^{a\neq b} +  \\ \no &+&\im\;
\int_{0}^{1}dv\int_{0}^{v}dw\;
\bigg(\mbox{$e^{+w\;\overrightarrow{[\hat{Y},\ldots]_{-}}}$}\;
\Big\{d^{ij}\;\big(\wtilde{\pp}_{j}\hat{T}\big)\;\hat{T}^{-1}\;,\;
\wtilde{J}_{J_{\psi},J_{\psi}^{+}}\;\wtilde{K}\Big\}_{+}\;e^{-v\;\hat{Y}}
\Big(\wtilde{\pp}_{i}\hat{Y}\Big)\;e^{+v\;\hat{Y}}\bigg)^{a\neq b}\bigg]\bigg\}_{\mbox{.}}
\eeq
We have still to perform the integrations over the real parameters \(dw\), \(dv\) in (\ref{s5_68}).
The formal integration of the exponential of {\scz\(\pm w\;\overrightarrow{[\hat{Y},\ldots]_{-}}\)}
is given in the third and fourth line of (\ref{s5_68}). These integrations over $w$ have to be
calculated within the limits of \(w\in[0,v]\). The other integrations over \(dv\) in the first
and second line of (\ref{s5_68}) can also be accomplished as in relations (\ref{s5_36}-\ref{s5_40}),
except that the matrix \(\delta\hat{Y}\) has to be replaced by the corresponding terms in the first
and second term of (\ref{s5_68}). We find the Eq. (\ref{s5_69}) as intermediate step for the
mean field equations of the pair condensates where the integrations over $dv$ in the third and
fourth term remain for further simplification
\beq \lb{s5_69}
0&=&\mbox{STR}\bigg\{\hat{Y}^{(\kappa)}\bigg[\frac{1}{2}\bigg(\FA\;\Big[\hat{Z}\;,\;\Big(\wtilde{\pp}_{i}c^{ij}\;
\Big(\wtilde{\pp}_{j}\hat{Z}\Big)\Big)\Big]_{-}\bigg)^{a\neq b}  + \\ \no &-&\im\;
\bigg(\wtilde{\pp}_{i}\;\FA\;
\Big\{d^{ij}\;\big(\wtilde{\pp}_{j}\hat{T}\big)\;\hat{T}^{-1}\;,\;
\wtilde{J}_{J_{\psi},J_{\psi}^{+}}\;\wtilde{K}\Big\}_{+}\bigg)^{a\neq b}  +  \\ \no &+&\im\;
\int_{0}^{1}dv\bigg(
\mbox{$\frac{\exp\big\{-v\;\overrightarrow{[\hat{Y},\ldots]_{-}}\big\}\;-
\;{\displaystyle\hat{1}}}{\overrightarrow{[\hat{Y},\ldots]_{-}}}$}\;
\Big(\wtilde{\pp}_{i}\hat{Y}\Big)\;e^{+v\;\hat{Y}}\;
\Big\{d^{ij}\;\big(\wtilde{\pp}_{j}\hat{T}\big)\;\hat{T}^{-1}\;,\;
\wtilde{J}_{J_{\psi},J_{\psi}^{+}}\;\wtilde{K}\Big\}_{+}\;e^{-v\;\hat{Y}}\bigg)^{a\neq b} +  \\ \no &+&\im\;
\int_{0}^{1}dv\bigg(
\mbox{$\frac{\exp\big\{+v\;\overrightarrow{[\hat{Y},\ldots]_{-}}\big\}\;-
\;{\displaystyle\hat{1}}}{\overrightarrow{[\hat{Y},\ldots]_{-}}}$}\;
\Big\{d^{ij}\;\big(\wtilde{\pp}_{j}\hat{T}\big)\;\hat{T}^{-1}\;,\;
\wtilde{J}_{J_{\psi},J_{\psi}^{+}}\;\wtilde{K}\Big\}_{+}\;e^{-v\;\hat{Y}}\;
\Big(\wtilde{\pp}_{i}\hat{Y}\Big)\;e^{+v\;\hat{Y}}\bigg)^{a\neq b}\bigg]\bigg\}_{\mbox{.}}
\eeq
We separately consider the remaining integrations over $dv$ in (\ref{s5_69}) and obtain equations
(\ref{s5_70}) for the third term and (\ref{s5_71}) for the fourth term
\beq\no
\lefteqn{\hspace*{-1.36cm}\Bigg[\Big(\overrightarrow{[\hat{Y},\ldots]_{-}}\Big)^{-1}\;\im\;\int_{0}^{1}dv\;
\bigg(\Big(e^{-v\;\overrightarrow{[\hat{Y},\ldots]_{-}}}-\hat{1}\Big)\;
\Big(\wtilde{\pp}_{i}\hat{Y}\Big)\;e^{+v\;\hat{Y}}\;
\Big\{d^{ij}\;\big(\wtilde{\pp}_{j}\hat{T}\big)\;\hat{T}^{-1}\;,\;
\wtilde{J}_{J_{\psi},J_{\psi}^{+}}\;\wtilde{K}\Big\}_{+}\;e^{-v\;\hat{Y}}\bigg)\Bigg]^{a\neq b}= } \\ \lb{s5_70} &=&
\Bigg[\Big(\overrightarrow{[\hat{Y},\ldots]_{-}}\Big)^{-1}\;\im\;\int_{0}^{1}dv\;\bigg(e^{-v\;\hat{Y}}\;
\Big(\wtilde{\pp}_{i}\hat{Y}\Big)\;e^{+v\;\hat{Y}}\;
\Big\{d^{ij}\;\big(\wtilde{\pp}_{j}\hat{T}\big)\;\hat{T}^{-1}\;,\;
\wtilde{J}_{J_{\psi},J_{\psi}^{+}}\;\wtilde{K}\Big\}_{+} + \\ \no &-&
\Big(\wtilde{\pp}_{i}\hat{Y}\Big)\;e^{+v\;\hat{Y}}\;
\Big\{d^{ij}\;\big(\wtilde{\pp}_{j}\hat{T}\big)\;\hat{T}^{-1}\;,\;
\wtilde{J}_{J_{\psi},J_{\psi}^{+}}\;\wtilde{K}\Big\}_{+}\;e^{-v\;\hat{Y}}\bigg)\Bigg]^{a\neq b}  =  \\ \no &=&
\im\;\Bigg[\Big(\overrightarrow{[\hat{Y},\ldots]_{-}}\Big)^{-1}\;
\bigg[\bigg(\FAm\;\Big(\wtilde{\pp}_{i}\hat{Y}\Big)\bigg)\;
\Big\{d^{ij}\;\big(\wtilde{\pp}_{j}\hat{T}\big)\;\hat{T}^{-1}\;,\;
\wtilde{J}_{J_{\psi},J_{\psi}^{+}}\;\wtilde{K}\Big\}_{+} +  \\ \no &-&
\Big(\wtilde{\pp}_{i}\hat{Y}\Big)\;
\bigg(\FA\;\Big\{d^{ij}\;\big(\wtilde{\pp}_{j}\hat{T}\big)\;\hat{T}^{-1}\;,\;
\wtilde{J}_{J_{\psi},J_{\psi}^{+}}\;\wtilde{K}\Big\}_{+}\bigg)\bigg]\Bigg]^{a\neq b}
\eeq
\beq\no
\lefteqn{\hspace*{-1.36cm}\Bigg[\Big(\overrightarrow{[\hat{Y},\ldots]_{-}}\Big)^{-1}\;\im\;\int_{0}^{1}dv\;
\bigg(\Big(e^{+v\;\overrightarrow{[\hat{Y},\ldots]_{-}}}-\hat{1}\Big)\;
\Big\{d^{ij}\;\big(\wtilde{\pp}_{j}\hat{T}\big)\;\hat{T}^{-1}\;,\;
\wtilde{J}_{J_{\psi},J_{\psi}^{+}}\;\wtilde{K}\Big\}_{+}\;e^{-v\;\hat{Y}}
\Big(\wtilde{\pp}_{i}\hat{Y}\Big)\;e^{+v\;\hat{Y}}\; \bigg)\Bigg]^{a\neq b}= } \\    \lb{s5_71}   &=&
\Bigg[\Big(\overrightarrow{[\hat{Y},\ldots]_{-}}\Big)^{-1}\;\im\;\int_{0}^{1}dv\;\bigg(e^{+v\;\hat{Y}}\;
\Big\{d^{ij}\;\big(\wtilde{\pp}_{j}\hat{T}\big)\;\hat{T}^{-1}\;,\;
\wtilde{J}_{J_{\psi},J_{\psi}^{+}}\;\wtilde{K}\Big\}_{+}\;e^{-v\;\hat{Y}}\;
\Big(\wtilde{\pp}_{i}\hat{Y}\Big) + \\ \no &-&
\Big\{d^{ij}\;\big(\wtilde{\pp}_{j}\hat{T}\big)\;\hat{T}^{-1}\;,\;
\wtilde{J}_{J_{\psi},J_{\psi}^{+}}\;\wtilde{K}\Big\}_{+}\;e^{-v\;\hat{Y}}
\Big(\wtilde{\pp}_{i}\hat{Y}\Big)\;e^{+v\;\hat{Y}} \bigg) \Bigg]^{a\neq b} =  \\ \no &=&
\im\;\Bigg[\Big(\overrightarrow{[\hat{Y},\ldots]_{-}}\Big)^{-1}\;
\bigg[\bigg(\FA\;\Big\{d^{ij}\;\big(\wtilde{\pp}_{j}\hat{T}\big)\;\hat{T}^{-1}\;,\;
\wtilde{J}_{J_{\psi},J_{\psi}^{+}}\;\wtilde{K}\Big\}_{+}\bigg)\;
\Big(\wtilde{\pp}_{i}\hat{Y}\Big) +  \\ \no &-&
\Big\{d^{ij}\;\big(\wtilde{\pp}_{j}\hat{T}\big)\;\hat{T}^{-1}\;,\;
\wtilde{J}_{J_{\psi},J_{\psi}^{+}}\;\wtilde{K}\Big\}_{+}\;
\bigg(\FAm\;\Big(\wtilde{\pp}_{i}\hat{Y}\Big)\bigg)\bigg]\Bigg]^{a\neq b}_{\mbox{.}}
\eeq
Inserting (\ref{s5_70},\ref{s5_71}) into (\ref{s5_69}), the final mean field equation
(\ref{s5_72}) is accomplished for the super-symmetric pair condensates \(\hat{Y}(\vec{x},t_{p})\).
We introduce the operator {\scz\(\fa\)} (\ref{s5_73})
in order to simplify the classical equations (\ref{s5_72})
for the pair condensates
\beq \lb{s5_72}
0&=&\mbox{STR}\bigg\{\hat{Y}^{(\kappa)}\bigg[\frac{1}{2}\Bigg(\fa\;\Big[\hat{Z}\;,\;\Big(\wtilde{\pp}_{i}c^{ij}\;
\Big(\wtilde{\pp}_{j}\hat{Z}\Big)\Big)\Big]_{-}\Bigg)^{a\neq b}  + \\ \no &-&\im\;
\Bigg(\wtilde{\pp}_{i}\;\fa\;
\Big\{d^{ij}\;\big(\wtilde{\pp}_{j}\hat{T}\big)\;\hat{T}^{-1}\;,\;
\wtilde{J}_{J_{\psi},J_{\psi}^{+}}\;\wtilde{K}\Big\}_{+}\Bigg)^{a\neq b}  +  \\ \no &+&\im\;
\Bigg(\Big(\overrightarrow{[\hat{Y},\ldots]_{-}}\Big)^{-1}\;
\bigg[\bigg(\fa\;\Big\{d^{ij}\;\big(\wtilde{\pp}_{j}\hat{T}\big)\;\hat{T}^{-1}\;,\;
\wtilde{J}_{J_{\psi},J_{\psi}^{+}}\;\wtilde{K}\Big\}_{+}\bigg)\;,\;
\Big(\wtilde{\pp}_{i}\hat{Y}\Big)\bigg]_{-}\Bigg)^{a\neq b} +  \\ \no &-& \im\;
\Bigg(\Big(\overrightarrow{[\hat{Y},\ldots]_{-}}\Big)^{-1}\;
\bigg[\Big\{d^{ij}\;\big(\wtilde{\pp}_{j}\hat{T}\big)\;\hat{T}^{-1}\;,\;
\wtilde{J}_{J_{\psi},J_{\psi}^{+}}\;\wtilde{K}\Big\}_{+}\;,\;
\bigg(\famm\;\Big(\wtilde{\pp}_{i}\hat{Y}\Big)\bigg)\bigg]_{-}\Bigg)^{a\neq b}\bigg]\bigg\}
\eeq
\beq  \lb{s5_73}
\fa &=& \FA \;\;\;.
\eeq
The first term of the classical nonlinear sigma equations (\ref{s5_72}) with matrix
\(\hat{Z}(\vec{x},t_{p})\) determines the spatial distribution of the pair condensates as elements
within the coset manifold \(Osp(S,S|2L)\backslash U(L|S)\).
However, since we consider both the pair condensates and the coherent BEC wave function by
source terms \(\hat{J}_{\psi\psi;\alpha\beta}^{a\neq b}(\vec{x},t_{p})\) and
\(J_{\psi;\alpha}^{a}(\vec{x},t_{p})\) with inclusion of 'Nambu'-doubled fields
\(\Psi_{\vec{x},\alpha}^{a}(t_{p})\) and doubled HST transformation, the second to fourth term
in (\ref{s5_72}) describes the interaction of the pair condensates \(\hat{Y}(\vec{x},t_{p})\)
with the coherent BEC wave function.
This 'macroscopic' coherent BEC wave function follows from the source matrix
\(\wtilde{J}_{J_{\psi},J_{\psi}^{+}}(\vec{x},t_{p})\;\wtilde{K}=
\hat{I}_{\alpha}^{a}\;\big(\breve{J}_{\psi;\alpha}^{a}(\vec{x},t_{p})\otimes
\breve{J}_{\psi;\beta}^{+,b}(\vec{x},t_{p})\big)\;\hat{I}_{\beta}^{b}\;\wtilde{K}\)
(\ref{s5_53},\ref{s5_54}) as dyadic product of the source fields \(J_{\psi;\alpha}^{a}(\vec{x},t_{p})\).
The noncondensed parts of the densities are contained in the coefficients
\(c^{ij}(\vec{x},t_{p})\), \(d^{ij}(\vec{x},t_{p})\) whose averages over the background field
\(\sigma_{D}^{(0)}(\vec{x},t_{p})\) also regard coherent BEC wave function effects due to the
presence of the source field \(j_{\psi;\alpha}(\vec{x},t_{p})\) in the generating
function \(Z[j_{\psi};\hat{\sigma}_{D}^{(0)}]\) (\ref{s4_89}-\ref{s4_91}) for the self-energy
field \(\sigma_{D}^{(0)}(\vec{x},t_{p})\).

\subsection{The adjoint representation of a super-algebra} \lb{s53}

The adjoint representation of a group or super-group (respectively their Lie algebras) is entirely determined
by its structure constants. We have considered the fundamental representation of the ortho-symplectic
super-group \(Osp(S,S|2L)\) in the derivation of the nonlinear sigma model actions.
The fundamental representation has also been used for the invariant integration
measures and equations, starting from symmetry investigations (section \ref{s3}) and the gradient expansion
(section \ref{s4}). However, since the coherent state path integrals are given by actions of invariant
super-traces of \(Osp(S,S|2L)\), one can choose different representations for the \(Osp(S,S|2L)\) matrices
within the super-traces of actions or within the classical nonlinear sigma equations of
\(Osp(S,S|2L)\backslash U(L|S)\) and \(U(L|S)\) super-matrices.

In this subsection \ref{s53} the adjoint representation of a super-algebra with super-generators
\(\hat{\Sigma}^{(\kappa)}\) is examined, having structure constants \(f^{\kappa\lambda}_{\ph{\kappa\lambda}\mu}\)
(The super- and sub-scripts \(\kappa\), \(\lambda\), \(\mu\) label the super-generators, e.g. the coset generators
\(\hat{Y}^{(\kappa)}\) of \(Osp(S,S|2L)\backslash U(L|S)\) or the subgroup generators \(\hat{H}^{(\kappa)}\) of
\(U(L|S)\).). The considered generators \(\hat{\Sigma}^{(\kappa)}\),
\(\hat{\Sigma}^{(\lambda)}\), \(\hat{\Sigma}^{(\mu)}\) of a Lie super-algebra
are multiplied by three sets of corresponding parameters \(d\sigma_{1,\kappa}\), \(d\sigma_{2,\lambda}\),
\(d\sigma_{3,\mu}\) to form the super-matrices \(\hat{\Sigma}_{1}\), \(\hat{\Sigma}_{2}\), \(\hat{\Sigma}_{3}\).
The three sets of parameters \(d\sigma_{1,\kappa}\), \(d\sigma_{2,\lambda}\), \(d\sigma_{3,\mu}\) contain
commuting as well as anti-commuting elements which can be arranged with the super-generators to result in
super-matrices as the self-energy \(\delta\wtilde{\Sigma}\;\wtilde{K}\) in the coherent
state path integrals with 'Nambu'-doubled, separated boson-boson, fermion-fermion, fermion-boson and
boson-fermion parts. (Note that one has to distinguish between  lower and upper indices or
contravariant and covariant components.) However, as the parameters \(d\sigma_{1,\kappa}\),
\(d\sigma_{2,\lambda}\) are shifted out from the commutator of two super-matrices
\(\hat{\Sigma}_{1}\), \(\hat{\Sigma}_{2}\) (e.g. two self-energies) {\it in a specific order},
one has to include an additional minus sign in the case of two anti-commuting parameters so that
a supercommutator \([\ldots\;,\;\ldots\}\) has to be defined. The supercommutator takes into account
the even or odd degree of the commuting or the Grassmann variables, respectively
\beq \lb{s5_77}
\hat{\Sigma}_{1}&=&d\sigma_{1,\kappa}\;\;\hat{\Sigma}^{(\kappa)}\hspace*{1.0cm}
\hat{\Sigma}_{2}\;=\;d\sigma_{2,\lambda}\;\;\hat{\Sigma}^{(\lambda)}\hspace*{1.0cm}
\hat{\Sigma}_{3}\;=\;d\sigma_{3,\mu}\;\;\hat{\Sigma}^{(\mu)}   \\  \lb{s5_78}
\Big[\hat{\Sigma}_{1}\:,\:\hat{\Sigma}_{2}\Big]_{-} &=&
d\sigma_{1,\kappa}\;d\sigma_{2,\lambda}\;\Big(\hat{\Sigma}^{(\kappa)}\;
\hat{\Sigma}^{(\lambda)}-(-1)^{\chi(\kappa)\cdot\chi(\lambda)}\;\;
\hat{\Sigma}^{(\lambda)}\;\hat{\Sigma}^{(\kappa)}\Big) = \\ \no &=&
d\sigma_{1,\kappa}\;\;d\sigma_{2,\lambda}\;\;
\Big[\hat{\Sigma}^{(\kappa)}\:,\:\hat{\Sigma}^{(\lambda)}\Big\} =
d\sigma_{1,\kappa}\;\;d\sigma_{2,\lambda}\;\;
\im\;f^{\kappa\lambda}_{\ph{\kappa\lambda}\mu}\;\;\hat{\Sigma}^{(\mu)}   \\\lb{s5_79}
\Big[\hat{\Sigma}^{(\kappa)}\:,\:\hat{\Sigma}^{(\lambda)}\Big\} &:=&
\Big(\hat{\Sigma}^{(\kappa)}\;
\hat{\Sigma}^{(\lambda)}-(-1)^{\chi(\kappa)\cdot\chi(\lambda)}\;\;
\hat{\Sigma}^{(\lambda)}\;\hat{\Sigma}^{(\kappa)}\Big)  =
\im\;f^{\kappa\lambda}_{\ph{\kappa\lambda}\mu}\;\;\hat{\Sigma}^{(\mu)} \;\;\;.
\eeq
The definition of the supercommutator (\ref{s5_79}) regards the degrees of the generators
\(\hat{\Sigma}^{(\kappa)}\), \(\hat{\Sigma}^{(\lambda)}\) whether being combined with a
commuting or anti-commuting parameter \(d\sigma_{1,\kappa}\), \(d\sigma_{2,\lambda}\). The degree
\(\chi(\kappa)\) of a parameter \(d\sigma_{\kappa}\)
for super-generator \(\hat{\Sigma}^{(\kappa)}\) is plus one for
Grassmann variables and zero for commuting complex or real parameters
\beq  \lb{s5_80}
\mbox{degree of }\;\big(d\sigma_{\kappa}\big) &:=&\chi(\kappa)=
\Bigg\{\bea{l} +1\;\mbox{ for odd or Grassmann variables} \\
\;\;0\;\mbox{ for even or commuting variables} \eea \\  \lb{s5_81}
\chi(\mu) &=&\big(\chi(\kappa) +\chi(\lambda)\big)\;\mbox{ modulo }\;2 \\  \lb{s5_82}
f^{\lambda\kappa}_{\ph{\lambda\kappa}\mu} &=&-\;\;
f^{\kappa\lambda}_{\ph{\kappa\lambda}\mu}\;\;(-1)^{\chi(\kappa)\cdot\chi(\lambda)}\;\;\; .
\eeq
Therefore, the resulting degree \(\chi(\mu)\) of \(\hat{\Sigma}^{(\mu)}\) in the supercommutator
\(\big[\hat{\Sigma}^{(\kappa)}\:,\:\hat{\Sigma}^{(\lambda)}\big\}\) (\ref{s5_79}) is given by
\(\big(\chi(\kappa) +\chi(\lambda)\big)\;\mbox{ modulo }\;2\); the structure constants
\(f^{\kappa\lambda}_{\ph{\kappa\lambda}\mu}\) are antisymmetric, except for the encounter of two
odd degree generators in the supercommutator leading to a symmetric relation in the two first upper
indices of the structure constants (\ref{s5_82}).

Starting from the Jacobi identity of super-matrices \(\hat{\Sigma}_{1}\), \(\hat{\Sigma}_{2}\),
\(\hat{\Sigma}_{3}\) with cyclic permutations in the commutators, one can derive the super
Jacobi identity of the three corresponding super-generators
\(\hat{\Sigma}^{(\kappa)}\), \(\hat{\Sigma}^{(\lambda)}\), \(\hat{\Sigma}^{(\mu)}\).
As one moves out the parameters \(d\sigma_{1,\kappa}\), \(d\sigma_{2,\lambda}\),
\(d\sigma_{3,\mu}\) from the commutators with the super-matrices
\(\hat{\Sigma}_{1}\), \(\hat{\Sigma}_{2}\), \(\hat{\Sigma}_{3}\) in the Jacobi identity,
one obtains the supercommutators of the generators
\(\hat{\Sigma}^{(\kappa)}\), \(\hat{\Sigma}^{(\lambda)}\), \(\hat{\Sigma}^{(\mu)}\);
however, one has to correct the cyclic permutations with the degree of the first and
third super-generator in the prevailing summand for a consistent super Jacobi identity
\be \lb{s5_83}
\Big[\hat{\Sigma}_{1}\:,\:\Big[\hat{\Sigma}_{2}\:,\:\hat{\Sigma}_{3}\Big]_{-}\Big]_{-}+
\Big[\hat{\Sigma}_{2}\:,\:\Big[\hat{\Sigma}_{3}\:,\:\hat{\Sigma}_{1}\Big]_{-}\Big]_{-}+
\Big[\hat{\Sigma}_{3}\:,\:\Big[\hat{\Sigma}_{1}\:,\:\hat{\Sigma}_{2}\Big]_{-}\Big]_{-} =0
\ee
\beq \lb{s5_84}
\lefteqn{\hspace*{-2.8cm}\Big[d\sigma_{1,\kappa}\;\hat{\Sigma}^{(\kappa)}\:,\:
\Big[d\sigma_{2,\lambda}\;\hat{\Sigma}^{(\lambda)}\:,
\:d\sigma_{3,\mu}\;\hat{\Sigma}^{(\mu)}\Big]_{-}\Big]_{-}+
\Big[d\sigma_{2,\lambda}\;\hat{\Sigma}^{(\lambda)}\:,\:
\Big[d\sigma_{3,\mu}\;\hat{\Sigma}^{(\mu)}\:,\:
d\sigma_{1,\kappa}\;\hat{\Sigma}^{(\kappa)}\Big]_{-}\Big]_{-} + } \\ \no &+&
\Big[d\sigma_{3,\mu}\;\hat{\Sigma}^{(\mu)}\:,\:
\Big[d\sigma_{1,\kappa}\;\hat{\Sigma}^{(\kappa)}\:,\:
d\sigma_{2,\lambda}\;\hat{\Sigma}^{(\lambda)}\Big]_{-}\Big]_{-} =0
\eeq
\beq  \lb{s5_85}
\lefteqn{\hspace*{-2.8cm}d\sigma_{1,\kappa}\;d\sigma_{2,\lambda}\;d\sigma_{3,\mu}\;
\Big[\hat{\Sigma}^{(\kappa)}\:,\:
\Big[\hat{\Sigma}^{(\lambda)}\:,
\:\hat{\Sigma}^{(\mu)}\Big\}\Big\}+
d\sigma_{2,\lambda}\;d\sigma_{3,\mu}\;d\sigma_{1,\kappa}\;
\Big[\hat{\Sigma}^{(\lambda)}\:,\:\Big[\hat{\Sigma}^{(\mu)}\:,\:
\hat{\Sigma}^{(\kappa)}\Big\}\Big\} + } \\ \no &+&
d\sigma_{3,\mu}\;d\sigma_{1,\kappa}\;d\sigma_{2,\lambda}\;
\Big[\hat{\Sigma}^{(\mu)}\:,\:\Big[\hat{\Sigma}^{(\kappa)}\:,\:
\hat{\Sigma}^{(\lambda)}\Big\}\Big\} =0
\eeq
\beq  \lb{s5_86}
\lefteqn{\hspace*{-2.8cm}(-1)^{\chi(\kappa)\cdot\chi(\mu)}\;\;
\Big[\hat{\Sigma}^{(\kappa)}\:,\:
\Big[\hat{\Sigma}^{(\lambda)}\:,
\:\hat{\Sigma}^{(\mu)}\Big\}\Big\}+
(-1)^{\chi(\lambda)\cdot\chi(\kappa)}\;\;
\Big[\hat{\Sigma}^{(\lambda)}\:,\:\Big[\hat{\Sigma}^{(\mu)}\:,\:
\hat{\Sigma}^{(\kappa)}\Big\}\Big\} + } \\ \no &+&
(-1)^{\chi(\mu)\cdot\chi(\lambda)}\;\;
\Big[\hat{\Sigma}^{(\mu)}\:,\:\Big[\hat{\Sigma}^{(\kappa)}\:,\:
\hat{\Sigma}^{(\lambda)}\Big\}\Big\} =0  \;\;\;.
\eeq
The super Jacobi identity (\ref{s5_86}) allows to define the adjoint operator
\(\mbox{ad}_{\hat{\Sigma}^{(\kappa)}} \propto \big[\hat{\Sigma}^{(\kappa)}\:,\ldots\big\}\)
of the super-algebra with structure constants \(f^{\kappa\lambda}_{\ph{\kappa\lambda}\mu}\).
One can transform the identity (\ref{s5_86}) to a supercommutator relation of the
operators \(\big[\hat{\Sigma}^{(\kappa)}\:,\ldots\big\}\),
\(\big[\hat{\Sigma}^{(\lambda)}\:,\ldots\big\}\). These supercommutator operators
\(\big[\hat{\Sigma}^{(\kappa)}\:,\ldots\big\}\), \(\big[\hat{\Sigma}^{(\lambda)}\:,\ldots\big\}\)
form by themselves (or by the identity (\ref{s5_86})) the defining supercommutator with the
resulting operator \(\big[[\hat{\Sigma}^{(\kappa)}\:,\:\hat{\Sigma}^{(\lambda)}\}\:,\ldots\big\}\)
in place of the structure constants. The abstract supercommutator for the adjoint representation
is usually abbreviated by the operator symbol
\(\mbox{ad}_{\hat{\Sigma}^{(\kappa)}} \propto \big[\hat{\Sigma}^{(\kappa)}\:,\ldots\big\}\)
\be \lb{s5_87}
\Big[\hat{\Sigma}^{(\kappa)}\:,\:\Big[\hat{\Sigma}^{(\lambda)}\:,\ldots\Big\}\Big\}-
(-1)^{\chi(\kappa)\cdot\chi(\lambda)}\;\;
\Big[\hat{\Sigma}^{(\lambda)}\:,\:\Big[\hat{\Sigma}^{(\kappa)}\:,\ldots\Big\}\Big\}=
\Big[\Big[\hat{\Sigma}^{(\kappa)}\:,\:\hat{\Sigma}^{(\lambda)}\Big\}\:,\ldots\Big\}
\ee
\beq \lb{s5_88}
\Big[\mbox{ad}_{\hat{\Sigma}^{(\kappa)}}\;,\;\mbox{ad}_{\hat{\Sigma}^{(\lambda)}}\Big\} &=&
\mbox{ad}_{[\hat{\Sigma}^{(\kappa)}\:,\:\hat{\Sigma}^{(\lambda)}\}}  \\  \lb{s5_89}
\mbox{ad}_{\hat{\Sigma}^{(\kappa)}} &\propto &
\Big[\hat{\Sigma}^{(\kappa)}\:,\ldots\Big\}  \;\;\; .
\eeq
We apply the definition (\ref{s5_79}) with structure constants
\(f^{\kappa\lambda}_{\ph{\kappa\lambda}\mu}\) and the super Jacobi identity (\ref{s5_86},\ref{s5_87})
to construct a representation with the structure constants
\(\big(\im\;f^{\mu(\kappa)}_{\ph{\mu(\kappa)}\nu}\big)\)
for the generator \(^{\mu}\big(\hat{\Sigma}^{(\kappa)}\big)_{\nu}\)
with label '\((\kappa)\)' and with row and column indices \(\mu\), \(\nu\)
\be  \lb{s5_90}
\Big(\im\;f^{\mu(\kappa)}_{\ph{\mu(\kappa)}\nu}\Big)
\Big(\im\;f^{\nu(\lambda)}_{\ph{\nu(\lambda)}\rho}\Big) -
(-1)^{\chi(\kappa)\cdot\chi(\lambda)}\;
\Big(\im\;f^{\mu(\lambda)}_{\ph{\mu(\lambda)}\nu}\Big)
\Big(\im\;f^{\nu(\kappa)}_{\ph{\nu(\kappa)}\rho}\Big)=
\im\;f^{\kappa\lambda}_{\ph{\kappa\lambda}\nu}\;\;
\Big(\im\;f^{\mu(\nu)}_{\ph{\mu(\nu)}\rho}\Big)_{\mbox{.}}
\ee
The equations (\ref{s5_87}-\ref{s5_90}) define the adjoint representation for \(\hat{\Sigma}^{(\kappa)}\)
with the supercommutator \(\big[\hat{\Sigma}^{(\kappa)}\:,\ldots\big\}\)
and its corresponding matrix elements
\(\big(\im\;f^{\mu(\kappa)}_{\ph{\mu(\kappa)}\nu}\big)\)
with row and column indices \(\mu\), \(\nu\). This matrix representation is obtained
by the action of the operator
{\scz\(\overrightarrow{\big[(\im\;\hat{f}^{(\kappa)})\:,\ldots\big\}}\)}
onto the state \((\im\;\hat{f}^{(\lambda)})\) whose coefficients
\(\im\;f^{\kappa\lambda}_{\ph{\kappa\lambda}\nu}\) of the resulting states
\((\im\;\hat{f}^{(\nu)})\) yield the matrices for the adjoint
representation. Using the super-trace '\(\mbox{STR}\)' to construct a scalar product
between 'basis states' of generators \(\big(\im\;\hat{f}^{(\mu)}\big)\) ,
\(\big(\im\;\hat{f}^{(\nu)}\big)\) , one finds the matrix representation of the
operator {\scz\(\overrightarrow{\big[(\im\;\hat{f}^{(\kappa)})\:,\ldots\big\}}\)}
with the metric tensor \(\hat{g}^{\mu\nu}=\hat{g}^{\nu\mu}=
\mbox{STR}\big[\big(\im\;\hat{f}^{(\mu)}\big)\;\big(\im\;\hat{f}^{(\nu)}\big)\big]\).
The structure constants \(\im\;f^{\kappa\lambda}_{\ph{\kappa\lambda}\nu}\) can therefore
be regarded as the operators and also as the corresponding states in the adjoint representation
\beq  \lb{s5_92}
\Big[\Big(\im\;\hat{f}^{(\kappa)}\Big)\:,\ldots\Big\}
&\propto & \Big[\hat{\Sigma}^{(\kappa)}\:,\ldots\Big\} \propto
\mbox{ad}_{\hat{\Sigma}^{(\kappa)}} \\  \lb{s5_93}
\underbrace{\overrightarrow{\Big[\Big(\im\;\hat{f}^{(\kappa)}\Big)\:,\ldots\Big\}}}_{
\mbox{\scz Operator}}\;\;\underbrace{\Big(\im\;\hat{f}^{(\lambda)}\Big)}_{\mbox{\scz state}} &=&
\underbrace{\im\;f^{\kappa\lambda}_{\ph{\kappa\lambda}\nu}}_{\mbox{\scz coefficients}}\;\;
\underbrace{\Big(\im\;\hat{f}^{(\nu)}\Big)}_{\mbox{\scz state}} \\  \lb{s5_93_a}
\mbox{STR}\bigg[\underbrace{\Big(\im\;\hat{f}^{(\mu)}\Big)}_{\mbox{\scz state}}\;\;
\underbrace{\overrightarrow{\Big[\Big(\im\;\hat{f}^{(\kappa)}\Big)\:,\ldots\Big\}}}_{
\mbox{\scz Operator}}\;\;\underbrace{\Big(\im\;\hat{f}^{(\lambda)}\Big)}_{\mbox{\scz state}}\bigg] &=&
\underbrace{\im\;f^{\kappa\lambda}_{\ph{\kappa\lambda}\nu}}_{\mbox{\scz coefficients}}\;\;
\underbrace{\mbox{STR}\bigg[\Big(\im\;\hat{f}^{(\mu)}\Big)\;\;
\Big(\im\;\hat{f}^{(\nu)}\Big)\bigg]}_{\mbox{\scz scalar product of states }\mu,\nu}  \\ \no &=&
\im\;f^{\kappa\lambda}_{\ph{\kappa\lambda}\nu}\;\;\hat{g}^{\mu\nu}=
\im\;f^{\kappa\lambda\mu}\;.
\eeq
The ambiguity of the structure constants, as being operators and states in the adjoint representation, can be applied
to simplify arbitrary powers of {\scz\(\overrightarrow{[\hat{Y},\ldots]_{-}}\)} in {\scz\(\fa\)}
within the classical nonlinear sigma equations.
A subsequent scalar product with state \(\big(\im\;\hat{f}^{(\sigma)}\big)\) and super-trace '\(\mbox{STR}\)'
allows to simplify the effective actions of the coherent state path integrals which only consist of
super-traces '\(\mbox{STR}\)' with the coset decomposition.
We list in Eq. (\ref{s5_94}) the action of a quadratic power
of operators in the adjoint representation onto another generator of structure constants, being considered as a
state. According to rule (\ref{s5_92},\ref{s5_93}), we have just to multiply and sum the states
\(\big(\im\;\hat{f}^{(\rho)}\big)\) with coefficients
\(\im\;f^{\lambda\mu}_{\ph{\lambda\mu}\nu}\;\;\im\;f^{\kappa\nu}_{\ph{\kappa\nu}\rho}\)
in reversed order instead of calculating powers of {\scz\(\overrightarrow{[\hat{Y},\ldots]_{-}}\)}
with \(\hat{Y}=y_{\kappa}\;\;\hat{Y}^{(\kappa)}\)
\beq  \lb{s5_94}
\lefteqn{
\underbrace{\overrightarrow{\Big[\Big(\im\;\hat{f}^{(\kappa)}\Big)\:,\ldots\Big\}}}_{
\mbox{\scz Operator}}\;\;
\underbrace{\overrightarrow{\Big[\Big(\im\;\hat{f}^{(\lambda)}\Big)\:,\ldots\Big\}}}_{
\mbox{\scz Operator}}\;\;\underbrace{\Big(\im\;\hat{f}^{(\mu)}\Big)}_{\mbox{\scz state}} = }
\\ \no &=&\underbrace{\overrightarrow{\Big[\Big(\im\;\hat{f}^{(\kappa)}\Big)\:,\ldots\Big\}}}_{
\mbox{\scz Operator}}\;\;\underbrace{\im\;f^{\lambda\mu}_{\ph{\lambda\mu}\nu}}_{\mbox{\scz coefficients}}\;\;
\underbrace{\Big(\im\;\hat{f}^{(\nu)}\Big)}_{\mbox{\scz state}} =
\underbrace{\im\;f^{\lambda\mu}_{\ph{\lambda\mu}\nu}}_{\mbox{\scz coefficients}}\;\;
\underbrace{\im\;f^{\kappa\nu}_{\ph{\kappa\nu}\rho}}_{\mbox{\scz coefficients}}\;\;
\underbrace{\Big(\im\;\hat{f}^{(\rho)}\Big)}_{\mbox{\scz state}}
\eeq
\beq  \lb{s5_94_b}
\lefteqn{\mbox{STR}\bigg[\underbrace{\Big(\im\;\hat{f}^{(\sigma)}\Big)}_{\mbox{\scz state}}\;\;
\underbrace{\overrightarrow{\Big[\Big(\im\;\hat{f}^{(\kappa)}\Big)\:,\ldots\Big\}}}_{
\mbox{\scz Operator}}\;\;
\underbrace{\overrightarrow{\Big[\Big(\im\;\hat{f}^{(\lambda)}\Big)\:,\ldots\Big\}}}_{
\mbox{\scz Operator}}\;\;\underbrace{\Big(\im\;\hat{f}^{(\mu)}\Big)}_{\mbox{\scz state}}\bigg] = }
\\ \no &=&\underbrace{\im\;f^{\lambda\mu}_{\ph{\lambda\mu}\nu}}_{\mbox{\scz coefficients}}\;\;
\underbrace{\im\;f^{\kappa\nu}_{\ph{\kappa\nu}\rho}}_{\mbox{\scz coefficients}}\;\;
\underbrace{\mbox{STR}\bigg[\Big(\im\;\hat{f}^{(\sigma)}\Big)\;\;
\Big(\im\;\hat{f}^{(\rho)}\Big)\bigg]}_{\mbox{\scz scalar product of states }\sigma,\rho}  =
\underbrace{\im\;f^{\lambda\mu}_{\ph{\lambda\mu}\nu}}_{\mbox{\scz coefficients}}\;\;
\underbrace{\im\;f^{\kappa\nu}_{\ph{\kappa\nu}\rho}}_{\mbox{\scz coefficients}}\;\;\hat{g}^{\sigma\rho}=
\im\;f^{\lambda\mu}_{\ph{\lambda\mu}\nu}\;\;
\im\;f^{\kappa\nu\sigma}  \;.
\eeq
The relations (\ref{s5_94},\ref{s5_94_b}) for a quadratic power of {\scz\(\overrightarrow{[\hat{Y},\ldots]_{-}}\)}
can be generalized to an arbitrary power $n$ where one has to perform the multiplication
of the structure constants for the operators in reversed order with an additional summation
of states \(\big(-\im\;\hat{f}^{(\nu_{1})}\big)\)
\beq  \lb{s5_95}
\lefteqn{
\underbrace{\overrightarrow{\Big[\Big(\im\;\hat{f}^{(\kappa_{1})}\Big)\:,\ldots\Big\}}
\ldots\ldots\overrightarrow{\Big[\Big(\im\;\hat{f}^{(\kappa_{n})}\Big)\:,\ldots\Big\}}
}_{\mbox{\scz Operators (n - times)}}\;\;
\underbrace{\Big(\im\;\hat{f}^{(\mu)}\Big)}_{\mbox{\scz state}} = }  \\ \no &=&
\underbrace{\im\;f^{\kappa_{n},\mu}_{\ph{\kappa_{n}\mu}\nu_{n}}\;\;
\im\;f^{\kappa_{n-1},\nu_{n}}_{\ph{\kappa_{n-1}\nu_{n}}\nu_{n-1}}\;\;\ldots\;\;
\im\;f^{\kappa_{2},\nu_{3}}_{\ph{\kappa_{2}\nu_{3}}\nu_{2}}\;\;
\im\;f^{\kappa_{1},\nu_{2}}_{\ph{\kappa_{1}\nu_{2}}\nu_{1}}}_{\mbox{\scz coefficients (n-times)}}\;\;
\underbrace{\Big(\im\;\hat{f}^{(\nu_{1})}\Big)}_{\mbox{\scz state}}
\eeq
\beq  \lb{s5_95_c}
\lefteqn{\mbox{STR}\bigg[\underbrace{\Big(\im\;\hat{f}^{(\nu)}\Big)}_{\mbox{\scz state}}\;\;
\underbrace{\overrightarrow{\Big[\Big(\im\;\hat{f}^{(\kappa_{1})}\Big)\:,\ldots\Big\}}
\ldots\ldots\overrightarrow{\Big[\Big(\im\;\hat{f}^{(\kappa_{n})}\Big)\:,\ldots\Big\}}
}_{\mbox{\scz Operators (n - times)}}\;\;
\underbrace{\Big(\im\;\hat{f}^{(\mu)}\Big)}_{\mbox{\scz state}}\bigg] = }  \\ \no &=&
\underbrace{\im\;f^{\kappa_{n},\mu}_{\ph{\kappa_{n}\mu}\nu_{n}}\;\;
\im\;f^{\kappa_{n-1},\nu_{n}}_{\ph{\kappa_{n-1}\nu_{n}}\nu_{n-1}}\;\;\ldots\;\;
\im\;f^{\kappa_{2},\nu_{3}}_{\ph{\kappa_{2}\nu_{3}}\nu_{2}}\;\;
\im\;f^{\kappa_{1},\nu_{2}}_{\ph{\kappa_{1}\nu_{2}}\nu_{1}}}_{\mbox{\scz coefficients (n-times)}}\;\;
\underbrace{\mbox{STR}\bigg[\underbrace{\Big(\im\;\hat{f}^{(\nu)}\Big)}_{\mbox{\scz state}}\;\;
\underbrace{\Big(\im\;\hat{f}^{(\nu_{1})}\Big)}_{\mbox{\scz state}}\bigg]}_{\hat{g}^{\nu,\nu_{1}}=
\hat{g}^{\nu_{1},\nu}} =  \\ \no &=&
\im\;f^{\kappa_{n},\mu}_{\ph{\kappa_{n}\mu}\nu_{n}}\;\;
\im\;f^{\kappa_{n-1},\nu_{n}}_{\ph{\kappa_{n-1}\nu_{n}}\nu_{n-1}}\;\;\ldots\;\;
\im\;f^{\kappa_{2},\nu_{3}}_{\ph{\kappa_{2}\nu_{3}}\nu_{2}}\;\;
\im\;f^{\kappa_{1},\nu_{2},\nu}   \;\;\;.
\eeq
The invariant integration measure of \(Osp(S,S|2L)\backslash U(L|S)\otimes U(L|S)\) has been determined
in the fundamental representation (see appendix \ref{sa}), but it can also be calculated in the adjoint
representation by using the Killing form. However, the Killing form may become degenerate for certain
super-algebras so that one has to rely to a different representation (as e.g. the fundamental one)
instead of the adjoint representation for computing integration measures. It has already been
mentioned in subsection \ref{s52} that the matrix \(\hat{Y}(\vec{x},t_{p})\) of the pair condensates
can be transformed to a diagonal form \(\hat{Y}_{DD}(\vec{x},t_{p})\) in the fundamental representation.
This transformation definitely simplifies the classical nonlinear sigma equations (\ref{s5_72})
which also have a geometric meaning with the metric tensors
\(\hat{G}_{Osp\backslash U}^{\lambda\lambda\ppr}\) and \(\hat{G}_{U(L|S)}^{\lambda\lambda\ppr}\)
(see following subsection \ref{s61}). The specific use of representations for
\(Osp(S,S|2L)\backslash U(L|S)\otimes U(L|S)\) can reduce the expenditure for determining the
integration measure or the classical nonlinear sigma equations (compare e.g. the Cartan-Weyl basis
or the Chevalley basis for simplifying the structure constants)\cite{luc}.

\section{Summary and discussion} \lb{s6}

\subsection{Geometric meaning of the $Osp(S,S|2L)\backslash U(L|S)$ nonlinear sigma model} \lb{s61}

At first we investigate the action \(\mcal{A}_{\mcal{N}^{-1}}^{\prime(1)}[\hat{Z}=\hat{T}\;\hat{S}\;\hat{T}^{-1}]\)
(\ref{s6_1}) following from the gradient expansion of the super-determinant. This action (\ref{s6_1})
only contains parts in the anomalous sectors \(a\neq b\) of the matrix fields
\(\big[\hat{T}^{-1}(\vec{x},t_{p})\;\big(\wtilde{\pp}_{i}\hat{T}(\vec{x},t_{p})\big)\big]_{\alpha\beta}^{ab}\)
which decompose into the \(U(L|S)\) subgroup terms
\(\big[\im\;\big(\wtilde{\pp}_{i}g_{\kappa}\big)\;\hat{H}^{(\kappa)}\big]_{\alpha\beta}^{aa}\)
and the \(Osp(S,S|2L)\backslash U(L|S)\) coset part
\(\big[-\big(\wtilde{\pp}_{i}s_{\kappa}\big)\;\hat{Y}^{(\kappa)}\big]_{\alpha\beta}^{a\neq b}\)
((\ref{s6_2}), see also appendix \ref{sa}, especially \ref{sa2} with Eq. (\ref{A66})). The functions
\(\big(\wtilde{\pp}_{i}g_{\kappa}\big)\) and \(\big(\wtilde{\pp}_{i}s_{\kappa}\big)\)
depend on the pair condensate fields of the coset generator
\(\hat{Y}(\vec{x},t_{p})\) of \(\hat{T}(\vec{x},t_{p})=\exp\big\{-\hat{Y}(\vec{x},t_{p})\big\}\).
The explicit form of \(\big[\hat{T}^{-1}\;(\wtilde{\pp}_{i}\hat{T})\big]_{\alpha\beta}^{ab}\)
can be read from the relations in appendix \ref{sc1} for \(a\neq b\) (Eqs. (\ref{C26}-\ref{C36}));
the subgroup parts with \(a=b\) follow straightforwardly with application of Eqs. (\ref{s1_17}-\ref{s1_24})
and similar calculations and are also specified by Eqs. (\ref{C12}-\ref{C25}) in appendix \ref{sc1}.
The comparison of
\(\big[\hat{T}^{-1}\;(\wtilde{\pp}_{i}\hat{T})\big]_{\alpha\beta}^{ab}\) with the independent
entries of \(\wtilde{W}_{\alpha\beta}^{ab}\) ((\ref{s3_41}) in subsection \ref{s31}) in
\(Osp(S,S|2L)\) allows to extract the generators \(\hat{H}^{(\kappa)}\), \(\hat{Y}^{(\kappa)}\)
and corresponding functions \(\big(\wtilde{\pp}_{i}g_{\kappa}\big)\), \(\big(\wtilde{\pp}_{i}s_{\kappa}\big)\)
\beq \lb{s6_1}
\lefteqn{\mcal{A}_{\mcal{N}^{-1}}^{\prime(1)}[\hat{Z}=\hat{T}\;\hat{S}\;\hat{T}^{-1}]=
\mcal{A}_{\mcal{N}^{-1}}^{\prime(1)}[\hat{T}]=-\frac{1}{\mcal{N}}\int_{C}\frac{dt_{p}}{\hbar}
\sum_{\vec{x}}\sum_{a,b=1,2}^{(a\neq b)}\;c^{ij}(\vec{x},t_{p})\;\times}  \\ \no &\times&
\strab\bigg\{\Big[\hat{T}^{-1}(\vec{x},t_{p})\;
\Big(\wtilde{\pp}_{i}\hat{T}(\vec{x},t_{p})\Big)\Big]_{\alpha\beta}^{a\neq b}\;
\Big[\hat{T}^{-1}(\vec{x},t_{p})\;
\Big(\wtilde{\pp}_{j}\hat{T}(\vec{x},t_{p})\Big)\Big]_{\beta\alpha}^{b\neq a}\bigg\}_{\mbox{.}}
\eeq
\be \lb{s6_2}
\Big[\hat{T}^{-1}(\vec{x},t_{p})\;
\Big(\wtilde{\pp}_{i}\hat{T}(\vec{x},t_{p})\Big)\Big]_{\alpha\beta}^{ab} =
\Big[\im\;\big(\wtilde{\pp}_{i}g_{\kappa}\big)\;\hat{H}^{(\kappa)}\Big]_{\alpha\beta}^{ab}\;
\delta_{ab}+\Big[-\big(\wtilde{\pp}_{i}s_{\kappa}\big)\;\hat{Y}^{(\kappa)}\Big]_{\alpha\beta}^{a\neq b}\;.
\ee
Using the chain rule of differentiation, we introduce the left and right derivatives of even and odd
parameters of the pair condensate fields \(y_{\mu}\) onto the functions
\(s_{\kappa}(y_{\mu})\) and \(g_{\kappa}(y_{\mu})\) in the coset and subgroup parts (\ref{s6_3},\ref{s6_4}).
In consequence the spatial derivatives \(\wtilde{\pp}_{i}\) act on the pair condensate fields
\(y_{\lambda}\) of the super-generator \(\hat{Y}(\vec{x},t_{p})=y_{\lambda}(\vec{x},t_{p})\;\hat{Y}^{(\lambda)}\)
of \(Osp(S,S|2L)\backslash U(L|S)\) and are therefore separated from the functions
\(s_{\kappa}(y_{\mu})\), \(g_{\kappa}(y_{\mu})\)
\beq \lb{s6_3}
\big(\wtilde{\pp}_{i}s_{\kappa}\big)&=&\big(\wtilde{\pp}_{i}y_{\lambda}\big)\;\;
\bigg(\frac{\pp}{\pp\overrightarrow{y}_{\lambda}}\;s_{\kappa}(y_{\mu})\bigg)=
\bigg(s_{\kappa}(y_{\mu})\;\frac{\pp}{\pp\overleftarrow{y}_{\lambda}}\bigg)\;\;
\big(\wtilde{\pp}_{i}y_{\lambda}\big)   \\   \lb{s6_4}
\big(\wtilde{\pp}_{i}g_{\kappa}\big)&=&\big(\wtilde{\pp}_{i}y_{\lambda}\big)\;\;
\bigg(\frac{\pp}{\pp\overrightarrow{y}_{\lambda}}\;g_{\kappa}(y_{\mu})\bigg)=
\bigg(g_{\kappa}(y_{\mu})\;\frac{\pp}{\pp\overleftarrow{y}_{\lambda}}\bigg)\;\;
\big(\wtilde{\pp}_{i}y_{\lambda}\big)\;\;\;.
\eeq
After insertion of (\ref{s6_3},\ref{s6_2}) into \(\mcal{A}_{\mcal{N}^{-1}}^{\prime(1)}[\hat{T}]\)
(\ref{s6_1}) with \(a\neq b\), we obtain the corresponding action (\ref{s6_5}) with the metric
tensor \(\hat{G}_{Osp\backslash U}^{\lambda\lambda\ppr}(y_{\mu})\) (\ref{s6_6}) of
\(Osp(S,S|2L)\backslash U(L|S)\) depending only on the pair condensate fields \(y_{\mu}(\vec{x},t_{p})\).
The spatial dependence is restricted to the derivatives \(\big(\wtilde{\pp}_{i}y_{\lambda}(\vec{x},t_{p})\big)\),
\(\big(\wtilde{\pp}_{j}y_{\lambda\ppr}(\vec{x},t_{p})\big)\) which are combined with the transport coefficients
\(c^{ij}(\vec{x},t_{p})\). These transport coefficients can itself be regarded as a metric tensor
of spacetime coordinates
\beq \lb{s6_5}
\mcal{A}_{\mcal{N}^{-1}}^{\prime(1)}[\hat{T}]&=&-\frac{2}{\mcal{N}}\int_{C}\frac{dt_{p}}{\hbar}
\sum_{\vec{x}}c^{ij}(\vec{x},t_{p})\;\times  \\ \no &\times&
\strab\bigg\{\bigg[\big(\wtilde{\pp}_{i}y_{\lambda}\big)\;\;
\bigg(\frac{\pp}{\pp\overrightarrow{y}_{\lambda}}\;s_{\kappa}(y_{\mu})\bigg)\;
\hat{Y}^{(\kappa)}\bigg]_{\alpha\beta}^{12}\;\;
\bigg[\hat{Y}^{(\kappa\ppr)}\;\;\bigg(s_{\kappa\ppr}(y_{\mu\ppr})\;
\frac{\pp}{\pp\overleftarrow{y}_{\lambda\ppr}}\bigg)\;\;
\big(\wtilde{\pp}_{j}y_{\lambda\ppr}\big)\bigg]_{\beta\alpha}^{21}\bigg\}   \\ \no &=&
-\frac{2}{\mcal{N}}\int_{C}\frac{dt_{p}}{\hbar}\sum_{\vec{x}}\;c^{ij}(\vec{x},t_{p})\;\;
\big(\wtilde{\pp}_{i}y_{\lambda}\big)\;\;\hat{G}_{Osp\backslash U}^{\lambda\lambda\ppr}\;\;
\big(\wtilde{\pp}_{j}y_{\lambda\ppr}\big)   \\ \lb{s6_6}
\hat{G}_{Osp\backslash U}^{\lambda\lambda\ppr}(y_{\mu}) &=&
\strab\bigg\{\bigg[\bigg(\frac{\pp}{\pp\overrightarrow{y}_{\lambda}}\;s_{\kappa}(y_{\mu})\bigg)\;
\hat{Y}^{(\kappa)}\bigg]_{\alpha\beta}^{12}\;\;
\bigg[\hat{Y}^{(\kappa\ppr)}\;\;\bigg(s_{\kappa\ppr}(y_{\mu\ppr})\;
\frac{\pp}{\pp\overleftarrow{y}_{\lambda\ppr}}\bigg)\bigg]_{\beta\alpha}^{21}\bigg\}_{\mbox{.}}
\eeq
As one varies the action (\ref{s6_5},\ref{s6_6}) with respect to
\(y_{\lambda}(\vec{x},t_{p})=y_{\lambda}(x^{1})\) in the spatial \(d=1\) stationary case
for classical equations, one acquires a geodesic curve \(y_{\lambda}^{(0)}(x^{1})\) as result
from a extremal principle for the squared distance \(\big(ds_{Osp\backslash U}\big)^{2}\)
\be \lb{s6_7}
\big(ds_{Osp\backslash U}\big)^{2}\propto dx^{1}\;\;c^{11}(x^{1})\;\;
\big(\wtilde{\pp}_{1}y_{\lambda}(x^{1})\big)\;\;\hat{G}_{Osp\backslash U}^{\lambda\lambda\ppr}(y_{\mu})\;\;
\big(\wtilde{\pp}_{1}y_{\lambda\ppr}(x^{1})\big)\;.
\ee
This is similar to the extremal principle
\(\int ds\propto\int d\tau\;\big(\frac{dx_{\lambda}(\tau)}{d\tau}\;\hat{g}^{\lambda\lambda\ppr}(x_{\mu})\;
\frac{dx_{\lambda\ppr}(\tau)}{d\tau}\big)^{1/2}\) in the theory of relativity \cite{hobson},
apart from the square of the distance in (\ref{s6_7}).
The pair condensate fields \(y_{\lambda}\) are analogous to the spacetime coordinates \(x_{\lambda}\)
in the theory of relativity and the metric tensor \(\hat{G}_{Osp\backslash U}^{\lambda\lambda\ppr}(y_{\mu})\)
of the coset part is replaced by the metric tensor \(\hat{g}^{\lambda\lambda\ppr}(x_{\mu})\)
of the spacetime variables.

As one proceeds to the \(d=2\) stationary case in \(\mcal{A}_{\mcal{N}^{-1}}^{\prime(1)}[\hat{T}]\) (\ref{s6_5})
(with \(y_{\lambda}(\vec{x},t_{p})=y_{\lambda}(x^{1},x^{2})\)), a conformal invariance can be achieved with
the transport coefficients or 'world-sheet' metric \(c^{ij}(x^{1},x^{2})\) (\(i,j=1,2\)) following from the
average with the background field \(\sigma_{D}^{(0)}(\vec{x},t_{p})\).
Applying the extremal principle with respect to \(y_{\lambda}(x^{1},x^{2})\), one considers an area element
$dS$ (\ref{s6_8}) in the space of \(y_{\lambda}\), \(\hat{G}_{Osp\backslash U}^{\lambda\lambda\ppr}(y_{\mu})\)
which depends on conformal 'world-sheet' parameters \(dz\), \(dz^{*}\)
\beq \lb{s6_8}
\int dS &\propto & \int dx^{1}\;\:dx^{2}\;\;c^{ij}(x^{1},x^{2})\;\;
\big(\wtilde{\pp}_{i}y_{\lambda}\big)\;\;\hat{G}_{Osp\backslash U}^{\lambda\lambda\ppr}(y_{\mu})\;\;
\big(\wtilde{\pp}_{j}y_{\lambda\ppr}\big)\;\;;\hspace*{0.37cm}(i,j=1,2)  \\ \no &\propto&
\int dz\;\:dz^{*}\;\;c(z,z^{*})\;\;
\big(\wtilde{\pp}_{z}y_{\lambda}\big)\;\;\hat{G}_{Osp\backslash U}^{\lambda\lambda\ppr}(y_{\mu})\;\;
\big(\wtilde{\pp}_{z^{*}}y_{\lambda\ppr}\big)  \\ \no &\propto&
\int \;\:\;\;c(z,z^{*})\;\;
dy_{\lambda}\;\;\;\hat{G}_{Osp\backslash U}^{\lambda\lambda\ppr}(y_{\mu})\;\;\;dy_{\lambda\ppr}\;.
\eeq
This conformal invariant case (\ref{s6_8}) allows to absorb the integration measure
\(\big[\mbox{SDET}(\hat{G}_{Osp\backslash U})\big]^{1/2}\) of the coherent state path integral
for pair condensates into the action (\ref{s6_5}) so that Gaussian integrals are contained among the
various other terms of the second order gradient expansion.
Comparing to string or super-string theory, one can notice a close analogy with {\it the
'world-sheet' parameters} '\(x^{1}\:,\:x^{2}\)' and {\it the complex valued spacetime variables}
'\(y_{\lambda}(x^{1},x^{2})\)' in a transferred sense \cite{szabo}. In fact the coset integration
measure can be eliminated in the coherent state path integral and be transformed to the anomalous
fields for arbitrary spatial dimensions. In general one has to apply the transformation
\be \lb{s6_9}
\int dy_{\lambda} = \int
\big[\hat{G}_{Osp\backslash U}^{-1/2}\big]_{\lambda}^{\ph{\lambda}\lambda\ppr}\;\;
d\wtilde{y}_{\lambda\ppr}\;\;\;\Longrightarrow \;\;\;
y_{\lambda}(\vec{x},t_{p})=y_{\lambda}\big(\wtilde{y}_{\mu}(\vec{x},t_{p})\big)\;\;\;,
\ee
so that the super-Jacobi-determinant \(\big[\mbox{SDET}(\hat{G}_{Osp\backslash U})\big]^{1/2}\)
is canceled in the path integral. One obtains an additional functional relation
\(y_{\lambda}=y_{\lambda}\big(\wtilde{y}_{\mu}\big)\)  (\ref{s6_9}) in the actions with new variables
\(\wtilde{y}_{\lambda}(\vec{x},t_{p})\), but 'flat' Euclidean integration measures and integration variables
\(d\wtilde{y}_{\lambda}(\vec{x},t_{p})\) in the coherent state path integral. Assuming constant
coefficients \(c^{ij}(\vec{x},t_{p})\) of the background field \(\sigma_{D}^{(0)}(\vec{x},t_{p})\) ,
Gaussian-like integrations remain due to the restriction to second order spatial
derivatives in the gradient expansion. Note that the derivation of the classical nonlinear
sigma model equations (\ref{s5_72}-\ref{s5_73})
in subsection \ref{s52} does not alter with the inclusion of the
coset integration measure and the transformation (\ref{s6_9}) to the new pair condensate
variables \(y_{\lambda}=y_{\lambda}\big(\wtilde{y}_{\mu}\big)\). Instead of varying
with respect to \(\delta y_{\lambda}(\vec{x},t_{p})\) in
\(\delta\hat{Y}(\vec{x},t_{p})=\delta y_{\kappa}(\vec{x},t_{p})\;\hat{Y}^{(\kappa)}\),
variations can be performed with the relations
\(\delta y_{\kappa}=\frac{dy_{\kappa}}{d\wtilde{y}_{\kappa\ppr}}\;\;
\delta\wtilde{y}_{\kappa\ppr}\) which do not change the derivation of the nonlinear
sigma model equations in subsection \ref{s52}. The inclusion of the coset integration measure
can be regarded as introducing quantum properties into the nonlinear sigma model. This means
in the case of the resulting nonlinear sigma model equations in subsection \ref{s52} that
the functional dependence \(y_{\lambda}(\vec{x},t_{p})\) of the coset generator
\(\hat{Y}(\vec{x},t_{p})=y_{\lambda}(\vec{x},t_{p})\;\hat{Y}^{(\lambda)}\)
in the effective equations (\ref{s5_72}-\ref{s5_73}) is corrected according
to \(y_{\lambda}(\vec{x},t_{p})=y_{\lambda}\big(\wtilde{y}_{\mu}(\vec{x},t_{p})\big)\)
for the new 'quantum-like' pair condensate variables \(\wtilde{y}_{\mu}(\vec{x},t_{p})\).

Similar considerations apply for the action
\(\mcal{A}_{\mcal{N}^{-1}}^{\prime(2)}[\hat{T};\wtilde{J}_{J_{\psi},J_{\psi}^{+}}\;
\wtilde{K}]\) (\ref{s6_10}) in the spatial \(d=1\) or \(d=2\) stationary case.
The variation in the \(d=1\) spatial case can be compared to the derivation of a geodesic
curve in the theory of relativity apart from the 'square' of the distance
\(\big(ds_{U(L|S)}\big)^{2}\). The action
\(\mcal{A}_{\mcal{N}^{-1}}^{\prime(2)}[\hat{T};\wtilde{J}_{J_{\psi},J_{\psi}^{+}}\;
\wtilde{K}]\) (\ref{s6_10}) is composed of the dyadic source matrix
\(\wtilde{J}_{J_{\psi},J_{\psi}^{+}}(\vec{x},t_{p})\;\wtilde{K}\) (\ref{s5_53},\ref{s5_54}) and the matrix fields
\(\big[\hat{T}^{-1}(\vec{x},t_{p})\;\big(\wtilde{\pp}_{i}\hat{T}(\vec{x},t_{p})\big)\big]_{\alpha\beta}^{ab}\)
of the coset generator \(\hat{Y}(\vec{x},t_{p})\). The latter matrix fields can be attributed to relation
(\ref{s6_2}) with subgroup generators \(\hat{H}^{(\kappa)}\) and functions \(\big(\wtilde{\pp}_{i}g_{\kappa}\big)\)
and coset parts \(\hat{Y}^{(\kappa)}\) with functions \(\big(\wtilde{\pp}_{i}s_{\kappa}\big)\)
(see Eq. (\ref{s6_11}))
\beq \lb{s6_10}
\lefteqn{\mcal{A}_{\mcal{N}^{-1}}^{\prime(2)}[\hat{T};\wtilde{J}_{J_{\psi},J_{\psi}^{+}}\;
\wtilde{K}]=-\frac{\im}{\mcal{N}}\int_{C}\frac{dt_{p}}{\hbar}\sum_{\vec{x}}\;
d^{ij}(\vec{x},t_{p})\;\times } \\ \no &\times&
\STRAB\bigg[\Big(\wtilde{\pp}_{i}\hat{T}(\vec{x},t_{p})\Big)\;\hat{T}^{-1}(\vec{x},t_{p})\;\;
\wtilde{J}_{J_{\psi},J_{\psi}^{+}}(\vec{x},t_{p})\;\wtilde{K}\;\;
\Big(\wtilde{\pp}_{j}\hat{T}(\vec{x},t_{p})\Big)\;\hat{T}^{-1}(\vec{x},t_{p})\bigg]
\eeq
\beq  \lb{s6_11}
\lefteqn{\Big[\Big(\wtilde{\pp}_{i}\hat{T}(\vec{x},t_{p})\Big)\;
\hat{T}^{-1}(\vec{x},t_{p})\Big]_{\alpha\beta}^{ab} =
\underbrace{\Big[\Big(\wtilde{\pp}_{i}\hat{T}^{\prime,-1}(\vec{x},t_{p})\Big)\;
\hat{T}^{\prime}(\vec{x},t_{p})\Big]_{\alpha\beta}^{ab}}_{\hat{T}\ppr=\exp\{\hat{Y}\}=\hat{T}^{-1}} = } \\ \no &=&
-\underbrace{\Big[\hat{T}^{\prime,-1}(\vec{x},t_{p})\;\Big(\wtilde{\pp}_{i}\hat{T}\ppr(\vec{x},t_{p})\Big)
\Big]_{\alpha\beta}^{ab}}_{\hat{Y}\leftrightarrow -\hat{Y}} =
\Big[-\im\;\big(\wtilde{\pp}_{i}g_{\kappa}\big)\;\hat{H}^{(\kappa)}\Big]_{\alpha\beta}^{ab}\;
\delta_{ab}+\Big[-\big(\wtilde{\pp}_{i}s_{\kappa}\big)\;\hat{Y}^{(\kappa)}\Big]_{\alpha\beta}^{a\neq b}\;.
\eeq
Using the properties of the dyadic source matrix and (\ref{s6_11}), we can reduce relation (\ref{s6_10})
to an action (\ref{s6_12}) which is composed of the metric tensors
\(\hat{G}_{Osp\backslash U}^{\lambda\lambda\ppr}\big(y_{\mu};\wtilde{J}_{J_{\psi},J_{\psi}^{+}}\;\wtilde{K}\big)\),
\(\hat{G}_{U(L|S)}^{\lambda\lambda\ppr}\big(y_{\mu};\wtilde{J}_{J_{\psi},J_{\psi}^{+}}\;\wtilde{K}\big)\)
(\ref{s6_14},\ref{s6_13}) of the coset \(Osp(S,S|2L)\backslash U(L|S)\) and \(U(L|S)\) parts
with the dependence on the dyadic source matrix
\(\big(\breve{j}_{\psi;\beta}(\vec{x},t_{p})\otimes
\breve{j}_{\psi;\gamma}^{*}(\vec{x},t_{p})\big)_{\beta\gamma}^{11}\)
\beq \lb{s6_12}
\mcal{A}_{\mcal{N}^{-1}}^{\prime(2)}[\hat{T};\wtilde{J}_{J_{\psi},J_{\psi}^{+}}\wtilde{K}] &=&
-\frac{2\;\im}{\mcal{N}}\int_{C}\frac{dt_{p}}{\hbar}\sum_{\vec{x}}\;d^{ij}(\vec{x},t_{p})\;\times \\ \no &\times&
\Big(\wtilde{\pp}_{i}y_{\lambda}\Big)\;\;\bigg[
\hat{G}_{Osp\backslash U}^{\lambda\lambda\ppr}\big(y_{\mu};
\wtilde{J}_{J_{\psi},J_{\psi}^{+}}(\vec{x},t_{p})\;\wtilde{K}\big) \;-\;
\hat{G}_{U(L|S)}^{\lambda\lambda\ppr}\big(y_{\mu};\wtilde{J}_{J_{\psi},J_{\psi}^{+}}(\vec{x},t_{p})\;\wtilde{K}\big)\bigg]\;\;
\Big(\wtilde{\pp}_{j}y_{\lambda\ppr}\Big)
\eeq
\beq \lb{s6_13}
\lefteqn{\hat{G}_{U(L|S)}^{\lambda\lambda\ppr}\big(y_{\mu};\wtilde{J}_{J_{\psi},J_{\psi}^{+}}(\vec{x},t_{p})\;
\wtilde{K}\big)=  }  \\ \no &=&
\strab\bigg\{\bigg[\bigg(\frac{\pp}{\pp\overrightarrow{y}_{\lambda}}g_{\kappa}(y_{\mu})\bigg)\;
\hat{H}^{(\kappa)}\bigg]_{\alpha\beta}^{11}\;\;\Big(
\breve{j}_{\psi;\beta}(\vec{x},t_{p})\otimes \breve{j}_{\psi;\gamma}^{*}(\vec{x},t_{p})\Big)_{\beta\gamma}^{11}\;\;
\bigg[\hat{H}^{(\kappa\ppr)}\;\bigg(g_{\kappa\ppr}(y_{\mu\ppr})
\frac{\pp}{\pp\overleftarrow{y}_{\lambda\ppr}}\bigg)\bigg]_{\gamma\alpha}^{11}\bigg\}_{\mbox{.}}
\eeq
\beq \lb{s6_14}
\lefteqn{\hat{G}_{Osp\backslash U}^{\lambda\lambda\ppr}\big(y_{\mu};\wtilde{J}_{J_{\psi},J_{\psi}^{+}}(\vec{x},t_{p})\;
\wtilde{K}\big)=  }  \\ \no &=&
\strab\bigg\{\bigg[\bigg(\frac{\pp}{\pp\overrightarrow{y}_{\lambda}}s_{\kappa}(y_{\mu})\bigg)\;
\hat{Y}^{(\kappa)}\bigg]_{\alpha\beta}^{21}\;\;\Big(
\breve{j}_{\psi;\beta}(\vec{x},t_{p})\otimes \breve{j}_{\psi;\gamma}^{*}(\vec{x},t_{p})\Big)_{\beta\gamma}^{11}\;\;
\bigg[\hat{Y}^{(\kappa\ppr)}\;\bigg(s_{\kappa\ppr}(y_{\mu\ppr})
\frac{\pp}{\pp\overleftarrow{y}_{\lambda\ppr}}\bigg)\bigg]_{\gamma\alpha}^{12}\bigg\}_{\mbox{.}}
\eeq
A conformal invariant action in the \(d=2\) stationary case can also be derived for (\ref{s6_13})
with the dyadic source matrix. The integration measure
\(\big[\mbox{SDET}(\hat{G}_{Osp\backslash U})\big]^{1/2}\) of the coherent state path integral
of pair condensate fields can be transformed to the action
\(\mcal{A}_{\mcal{N}^{-1}}^{\prime(2)}[\hat{T};\wtilde{J}_{J_{\psi},J_{\psi}^{+}}\wtilde{K}]\);
however, one does not achieve simple Gaussian integrals with constant variances in the exponent
as in (\ref{s6_8},\ref{s6_5},\ref{s6_6}) for \(d=2\)
because the transformation with the inverse square root of the metric tensor
\(\big(\hat{G}_{Osp\backslash U}\big)^{-1/2}\) from the integration measure does not cancel
the metric tensor \(\hat{G}_{U(L|S)}\) in the action (\ref{s6_12},\ref{s6_13}) of the subgroup part.
Therefore, trivial Gaussian integrals as in (\ref{s6_8},\ref{s6_5},\ref{s6_6}) with the coset part
do not result for the spatial \(d=2\) stationary case of (\ref{s6_12},\ref{s6_13})
with subgroup \(U(L|S)\) and metric
\(\hat{G}_{U(L|S)}^{\lambda\lambda\ppr}\big(y_{\mu};\wtilde{J}_{J_{\psi},J_{\psi}^{+}}(\vec{x},t_{p})\;
\wtilde{K}\big)\). In the general case of arbitrary spatial dimension $d$, the transformation (\ref{s6_9})
can be applied for the actions (\ref{s6_1},\ref{s6_5}-\ref{s6_6},
\ref{s6_10},\ref{s6_12}-\ref{s6_14}) of the coset and subgroup parts. The integrations
\(\int dy_{\lambda} = \int
\big[\hat{G}_{Osp\backslash U}^{-1/2}\big]_{\lambda}^{\ph{\lambda}\lambda\ppr}\;\;
d\wtilde{y}_{\lambda\ppr}\) (\ref{s6_9}) have to be performed for determining new functions
\(y_{\lambda}(\vec{x},t_{p})=y_{\lambda}\big(\wtilde{y}_{\mu}(\vec{x},t_{p})\big)\)
which replace the coset integration measure in the coherent state path integral with new
Euclidean integration variables \(d\big(\wtilde{y}_{\mu}(\vec{x},t_{p})\big)\)
for the pair condensates. According to the restriction to second order spatial gradients,
Gaussian-like integrations follow for arbitrary spatial dimensions, but with nontrivial
variances which depend on the new
'quantum-corrected' pair condensates \(\wtilde{y}_{\mu}(\vec{x},t_{p})\).
The transformation to the new pair condensate fields \(\wtilde{y}_{\mu}(\vec{x},t_{p})\) (\ref{s6_9})
includes the 'quantum-corrections' of the coset integration measure into the
derived classical nonlinear sigma model equations (\ref{s5_72}-\ref{s5_73})

\subsection{Conclusion and outlook} \lb{s62}

Although the assumed super-symmetry of bosonic and fermionic atoms may appear fantastic, there are
experimental and theoretical attempts to incorporate fermionic degrees into BE-condensates of
bosonic atoms in trap potentials \cite{hulet1}-\cite{schreck},\cite{tim1}. Even when the assumed
super-symmetry cannot be realized, this article describes the various steps for
extracting Goldstone modes of pair condensates with either symplectic \(Sp(2L)\) or orthogonal
\(SO(S,S)\) symmetry for bosonic or fermionic constituents. One can also transfer the symmetry
considerations of section \ref{s3} to the case with disorder and with an ensemble average over
a random potential \cite{bmdis}. The HST transformations have to be modified by 'hinge' functions
so that the group manifold of the self-energy coincides with the group of the invariant
transformations of the coherent state path integral. The separation into subgroup and coset
parts is possible due to the factorization of the integration measure and the parametrization
of the total self-energy into factors of density and pair condensate matrices. The gradient
expansion extracts the Goldstone modes from the super-determinant and we obtain an
additional term in the nonlinear sigma model caused by the source fields or 'BEC wave function
seeds' \(j_{\psi;\alpha}(\vec{x},t_{p})\). There is a close analogy of the actions with
pair condensates to the theory of relativity and string theory in the spatial \(d=1\) and \(d=2\)
stationary case. It definitely seems fantastic, but the production of ordinary paired
bosonic atoms for molecular condensates or the creation of paired fermions to BCS terms
can be regarded as a realization of either a boson-boson or fermion-fermion
string theory in spatial \(d=2\) experiments with anomalous terms as Goldstone modes.

Further investigations are persecuted for the problem of integrability concerning
the nonlinear sigma models \cite{abl1,abl2},\cite{gra1,gra2} and also the Skyrme model
\cite{sky1}-\cite{sky4}. The Lax pair for various nonlinear
sigma models is definitely known, as in the case of e.g. the nonlinear
Schr\"{o}dinger equation (NLS), so that an infinite series of conserved quantities
can be generated (see for example \cite{abl1,abl2}). However, one has to prove
the independence or involution of these conserved quantities. In the case of
the NLS equation or other nonlinear equations, this is accomplished by the
'fundamental Poisson bracket relation' and the r-matrix which has to fulfill
a compatibility relation (a kind of Jacobi-identity), the so-called
'Classical Yang-Baxter Equation' (CYBE). This CYBE allows a group theoretical
investigation and classification of possible r-matrices and corresponding
integrable nonlinear equations \cite{ma}. Eventually, extensions of the CYBE may also allow
statements concerning non-integrable systems as one includes deviations into the integrable
equations for chaotic behavior. However, in the case of the nonlinear sigma models, the
'fundamental Poisson bracket relations' have to include Schwinger-terms
(derivatives of spatial delta functions as result of commutator relations) with an
additional so-called s-matrix apart from the mentioned r-matrix \cite{mai1,bo}.
Therefore, we examine extensions of the CYBE with the r-s matrices and
Schwinger terms. A similar classification of these r-s matrices may be possible
by modified CYBE of 'non-ultra local form'. The extensions of ordinary algebras
and groups to the case with Schwinger-terms have been investigated \cite{kerf1}-\cite{schaper2}
and may be applied to our case of the \(Osp(S,S|2L)\backslash U(L|S)\)
nonlinear sigma model with r-s matrices and modified CYBE's.

\vspace*{1.0cm}

\noindent {\bf Acknowledgement}\newline \noindent I would like to acknowledge comfortable working
conditions of the department of physics at the University Duisburg-Essen and would like to thank
especially R. Graham, K. Krutitsky, M. Motzko, A. Pelster and the other members of the theory group.
\vspace*{0.5cm}

\begin{appendix}

\section{The Integration measure for $\boldsymbol{Osp(S,S|2L)\backslash U(L|S)$ $\otimes\;U(L|S)}$} \lb{sa}

\subsection{Diagonal density and anomalous self-energies with
super-unitary matrices of $U(L|S)$} \lb{sa1}

The self-energy $\wtilde{\Sigma}_{2N\times 2N}(\vec{x},t_{p})\;\wtilde{K}$
(\ref{s3_94}-\ref{s3_98}) with metric
$\wtilde{K}$ (\ref{s3_7}) and anti-hermitian anomalous terms is a generator of \(Osp(S,S|2L)\)
as \(\wtilde{W}_{\alpha\beta}^{ab}\) (\ref{s3_41}) in subsection \ref{s31}.
In subsection \ref{s32} a coherent state
path integral (\ref{s3_93}) with super-matrix
\(\wtilde{\mcal{M}}_{\vec{x},\alpha;\vec{x}\ppr,\beta}^{ab}(t_{p},t_{q}\ppr)\)
(\ref{s3_92}) has been derived so that a decomposition or 'diagonalization'
into density terms $\delta\hat{\Sigma}_{D;2N\times 2N}(\vec{x},t_{p})$
and mean field $\sigma_{D}^{(0)}(\vec{x},t_{p})$ (\ref{s3_68}-\ref{s3_74}) can be achieved
with coset matrices $\wtilde{T}_{1}(\vec{x},t_{p})$,
$\wtilde{T}_{2}(\vec{x},t_{p})$ (\ref{s3_55}-\ref{s3_61}) \(Osp(S,S|2L)\backslash U(L|S)\) in
\(Z[\hat{\mcal{J}},J_{\psi},\im\hat{J}_{\psi\psi}]\) (\ref{s3_93})
\footnote{In the following the tilde
'$\wtilde{\ph{\Sigma}}$' of $\wtilde{\Sigma}_{2N\times 2N}$
(and also of \(\wtilde{\mcal{M}}_{\vec{x},\alpha;\vec{x}\ppr,\beta}^{ab}(t_{p},t_{q}\ppr)\)
(\ref{s3_92})) refers to a self-energy with
anti-hermitian anomalous terms
\(\wtilde{\Sigma}_{N\times N}^{12}=\im\;\hat{\Sigma}_{N\times N}^{12}\),
\(\wtilde{\Sigma}_{N\times N}^{21}=\im\;\hat{\Sigma}_{N\times N}^{21}\),
in comparison to $\hat{\Sigma}_{2N\times 2N}$ with hermitian pair condensates $\hat{\Sigma}_{N\times N}^{12}$,
$\hat{\Sigma}_{N\times N}^{21}$; \(\hat{\Sigma}_{N\times N}^{21}=\big(\hat{\Sigma}_{N\times N}^{12}\big)^{+}\).}.
We choose the following transformation (\ref{A1}) from the 'flat' form of the super-generator
\(\wtilde{\Sigma}_{2N\times 2N}(\vec{x},t_{p})\;\wtilde{K}\) of
\(Osp(S,S|2L)\) to a coset decomposition
\(Osp(S,S|2L)\backslash U(L|S)\otimes U(L|S)\) where we define the matrix
\(\hat{T}(\vec{x},t_{p})=\wtilde{T}_{1}(\vec{x},t_{p})\) and its inverse form
\(\hat{T}^{-1}(\vec{x},t_{p})=\wtilde{K}\;\wtilde{T}_{2}(\vec{x},t_{p})\;\wtilde{K}\) from
\(\wtilde{T}_{2}(\vec{x},t_{p})\) (\ref{A8}). We also introduce the transposed form
\((\delta\hat{\Sigma}_{D;N\times N}^{22}(\vec{x},t_{p})\;\wtilde{\kappa})^{st}
=-\delta\hat{\Sigma}_{D;N\times N}^{11}(\vec{x},t_{p})\)
(\ref{A7}) from the '11' block of \(\delta\hat{\Sigma}_{D;2N\times 2N}(\vec{x},t_{p})\;\wtilde{K}\)
 with the metric $\wtilde{\kappa}$ (\ref{s3_7}) in the '22' block
\beq \lb{A1}
\wtilde{\Sigma}_{2N\times 2N}(\vec{x},t_{p})\;\wtilde{K}&=&\left( \bea{cc}
\sigma_{D}^{(0)}\;\hat{1}_{N\times N}+
\delta\hat{\Sigma}^{11} & \im\;\delta\hat{\Sigma}^{12} \\
\im\;\delta\hat{\Sigma}^{21} &
\sigma_{D}^{(0)}\;\wtilde{\kappa}+\delta\hat{\Sigma}^{22}
\eea\right)\;\left(
\bea{cc}
\hat{1}_{N\times N} & 0 \\
0 & \wtilde{\kappa}_{N\times N}
\eea\right) \\ \no &=&
\wtilde{T}_{1}\;\left( \bea{cc}
\sigma_{D}^{(0)}\;\hat{1}_{N\times N}+\delta\hat{\Sigma}^{11}_{D} & 0 \\
0 & \sigma_{D}^{(0)}\;\wtilde{\kappa}_{N\times N}+
\delta\hat{\Sigma}^{22}_{D} \eea\right)\;\wtilde{K}\;
\underbrace{\wtilde{K}\;\wtilde{T}_{2}\;\wtilde{K}}_{\wtilde{T}_{1}^{-1}}
\\  \no &=& \sigma_{D}^{(0)}\;\hat{1}_{2N\times 2N}+\hat{T}(\vec{x},t_{p})\;\underbrace{\left(
\bea{cc}
\delta\hat{\Sigma}_{D;N\times N}^{11} & 0 \\
0 & \delta\hat{\Sigma}_{D;N\times N}^{22}\;\wtilde{\kappa} \eea\right)}_{
\delta\hat{\Sigma}_{D;2N\times 2N}(\vec{x},t_{p})\;\wtilde{K}}\;\hat{T}^{-1}(\vec{x},t_{p})
\\ \lb{A2}
\delta\hat{\Sigma}_{N\times N}^{21}&=&
\big(\delta\hat{\Sigma}_{N\times N}^{12}\big)^{+} \hspace*{0.5cm}
\delta\wtilde{\Sigma}_{N\times N}^{\raisebox{-10pt}{$^{12}$}}=
\im\;\delta\hat{\Sigma}_{N\times N}^{12} \hspace*{0.5cm}
\delta\wtilde{\Sigma}_{N\times N}^{\raisebox{-10pt}{$^{21}$}}=
\im\;\delta\hat{\Sigma}_{N\times N}^{21}    \\ \lb{A3}
\delta\hat{\Sigma}_{N\times N}^{12}(\vec{x},t_{p})&=&\left(
\bea{cc}
\delta\hat{c}_{L\times L} & \delta\hat{\eta}_{L\times S}^{T} \\
\delta\hat{\eta}_{S\times L} & \delta\hat{f}_{S\times S}
\eea\right)   \\ \no &&\big(\delta\hat{c}_{L\times L}\big)^{T}=\delta\hat{c}_{L\times L}
\hspace*{0.5cm}\big(\delta\hat{f}_{S\times S}\big)^{T}=-\delta\hat{f}_{S\times S} \\ \lb{A4}
\delta\hat{\Sigma}_{D;2N\times 2N}(\vec{x},t_{p})\;\wtilde{K}&=&\left(
\bea{cc}
\delta\hat{\Sigma}_{D;N\times N}^{11}(\vec{x},t_{p}) & 0 \\
0 & \delta\hat{\Sigma}_{D;N\times N}^{22}(\vec{x},t_{p})\;\;\wtilde{\kappa}_{N\times N}
\eea\right) \\ \lb{A5}
\delta\hat{\Sigma}_{D;N\times N}^{11}(\vec{x},t_{p})&=&\left(
\bea{cc}
\delta\hat{B}_{D,L\times L} & \delta\hat{\chi}_{D,L\times S}^{+} \\
\delta\hat{\chi}_{D,S\times L} & \delta\hat{F}_{D,S\times S}
\eea\right)    \\ \lb{A6} &&
\big(\delta\hat{B}_{D,L\times L}\big)^{+}=\delta\hat{B}_{D,L\times L}\hspace*{0.75cm}
\big(\delta\hat{F}_{D,L\times L}\big)^{+}=\delta\hat{F}_{D,L\times L} \\ \lb{A7}
\big(\delta\hat{\Sigma}^{22}_{D;N\times N}(\vec{x},t_{p})\;\;\wtilde{\kappa}\big)^{st}&=&-
\delta\hat{\Sigma}_{D;N\times N}^{11}(\vec{x},t_{p})       \\ \lb{A8}
\hat{T}(\vec{x},t_{p})&:=&\wtilde{T}_{1}(\vec{x},t_{p})\hspace*{1.27cm}
\wtilde{T}_{2}\;\wtilde{K}\;\wtilde{T}_{1}=
\wtilde{K}\longrightarrow
\hat{T}^{-1}(\vec{x},t_{p})=\wtilde{K}\;\wtilde{T}_{2}\;\wtilde{K}\;.
\eeq
One has to calculate the super-symmetric metric tensor $\hat{G}_{Osp}^{\mu\nu}$ for the transformation
from the 'flat' local coordinates \(d(\wtilde{\Sigma}_{2N\times 2N}(\vec{x},t_{p})\;\wtilde{K})\)
(see the first line in (\ref{A1})) to curvilinear coordinates
\(d\sigma_{D}^{0}\;\hat{1}_{2N\times 2N}+
d\big(\hat{T}\;\delta\hat{\Sigma}_{D;2N\times 2N}\;
\wtilde{K}\;\hat{T}^{-1}\big)\)
with mean density field $\sigma_{D}^{(0)}(\vec{x},t_{p})$
and coset matrices \(\hat{T}\), \(\hat{T}^{-1}\) coupled to the densities
in a coset decomposition \(Osp(S,S|2L)\backslash U(L|S)\otimes U(L|S)\)
so that the local distance \((ds\ppr)^{2}\) (\ref{A9}) is invariant.
(Greek indices label commuting, as well as anti-commuting variables; the summation convention
over repeated multiple indices is always implied.) The super-Jacobian
\(\mu(\sigma_{\mu})\) for the transformation to general curvilinear coordinates $d\sigma_{\mu}$
(\(\mu,\nu=\) indices for the independent even and odd parameters of \(Osp(S,S|2L)\)) is given by the square root
of the super-determinant of the metric tensor $\hat{G}_{Osp}^{\mu\nu}$ (Compare with the
invariant volume element in the theory of general relativity \cite{hobson,hua})
\beq \lb{A9}
(d s\ppr)^{2}&=&\mbox{STR}\Big[d\big(\wtilde{\Sigma}_{2N\times 2N}\;\wtilde{K}\big)\;\;
d\big(\wtilde{\Sigma}_{2N\times 2N}\;\wtilde{K}\big)\Big]
\\ \no &=& 2(L-S)\;\Big(d\sigma_{D}^{(0)}\Big)^{2}+
\mbox{STR}\Big[d\big(\hat{T}\;\delta\hat{\Sigma}_{D;2N\times 2N}\;\wtilde{K}\;\hat{T}^{-1}\big)\;
\;d\big(\hat{T}\;\delta\hat{\Sigma}_{D;2N\times 2N}\;\wtilde{K}\;\hat{T}^{-1}\big)\Big]
\\ \no &+& 2\;d\sigma_{D}^{(0)}\;\;
\underbrace{\mbox{STR}\Big[\hat{T}\;
\delta\hat{\Sigma}_{D;2N\times 2N}\;\wtilde{K}\;\hat{T}^{-1}\Big]}_{\equiv 0} \hspace*{0.75cm}
\delta\hat{\Sigma}_{D;N\times N}^{11}
=-\big(\delta\hat{\Sigma}_{D;N\times N}^{22}\;\wtilde{\kappa}\big)^{ST} \\ \lb{A10}
(d s_{Osp})^{2}&=&d\sigma_{\mu}\;\;\hat{G}_{Osp}^{\mu\nu}\;\;d\sigma_{\nu}\Longrightarrow
\mu(\sigma_{\mu})=\sqrt{\mbox{SDET}\big(\hat{G}_{Osp}\big)}\;\;\;\;.
\eeq
The mean density field $d\sigma_{D}^{(0)}$ decouples from
\(d(\hat{T}\;\delta\hat{\Sigma}_{D;2N\times 2N}\;\wtilde{K}\;\hat{T}^{-1})\)
in (\ref{A9}) so that we
obtain as the first independent integration variable the field $d\sigma_{D}^{(0)}$ and a remaining
invariant length \((ds)^{2}\) (\ref{A11}) with the coset decomposition in (\ref{A1})
\be \lb{A11}
(d s_{Osp})^{2}= \mbox{STR}\Big[d\big(\hat{T}\;\delta\hat{\Sigma}_{D;2N\times 2N}\;
\wtilde{K}\;\hat{T}^{-1}\big)\;
\;d\big(\hat{T}\;\delta\hat{\Sigma}_{D;2N\times 2N}\;\wtilde{K}\;\hat{T}^{-1}\big)\Big]\;.
\ee
In order to obtain the invariant measure,
we have to diagonalize the density term \(\delta\hat{\Sigma}_{D;2N\times 2N}\;\wtilde{K}\)
(\ref{A12}-\ref{A14}) with the '11' block \(\delta\hat{\Sigma}_{D;N\times N}^{11}\) and its super-transposed part
\(\delta\hat{\Sigma}_{D;N\times N}^{22}\;\wtilde{\kappa}\) in the '22' block with the additional minus
sign (\ref{A7}). (The index range for the angular momentum degrees of freedom is changed from
\(-l,\ldots,+l\) and \(-s,\ldots,+s\) to \(1,\ldots,L=2l+1\) and \(1,\ldots,S=2s+1\) for the
bosons and fermions, respectively.) This diagonalization is performed with the
super-unitary matrices \(\hat{Q}_{N\times N}^{11}\), \(\hat{Q}_{N\times N}^{22}\) (\ref{A15}-\ref{A19})
of \(U(L|S)\) with the same number of independent parameters for \(\delta\hat{\Sigma}_{D;N\times N}^{11}\),
\(\delta\hat{\Sigma}_{D;N\times N}^{22}\;\wtilde{\kappa}\), respectively.
The eigenvalues \(\delta\hat{\lambda}_{N\times N}\) (\ref{A14}) for
the '11' and '22' block \(\delta\hat{\Sigma}_{D;N\times N}^{11}\),
\(\delta\hat{\Sigma}_{D;N\times N}^{22}\;\wtilde{\kappa}\) differ only by a minus sign and determine
the maximal abelian Cartan subalgebra of \(U(L|S)\) with rank $L$ and rank $S$ in the boson-boson and
fermion-fermion parts. Therefore, the eigenvectors \(\hat{Q}_{N\times N}^{22}\) (\ref{A16}) are related to
\(\hat{Q}_{N\times N}^{11,-1}\) by super-transposition (\ref{A19}).
The eigenvectors \(\hat{Q}_{N\times N}^{11}\) (\ref{A15}) and \(\hat{Q}_{N\times N}^{22}\) (\ref{A16})
are specified by the angles \(\hat{\mcal{B}}_{D,L\times L}\), \(\hat{\mcal{F}}_{D,S\times S}\)
and \(\hat{\omega}_{D,S\times L}\), \(\hat{\omega}_{D,L\times S}^{+}\) which form a super-hermitian
matrix. Since \(N=L+S\) degrees of freedom are already contained in the eigenvalues
\(\delta\hat{\lambda}_{N\times N}\) (\ref{A14}), we have to require that $L$ ($S$) parameters have to vanish
in the hermitian matrix $\hat{\mcal{B}}_{D,L\times L}$ ($\hat{\mcal{F}}_{D,S\times S}$)
of the boson-boson (fermion-fermion) part (\ref{A17},\ref{A18}). It is convenient to choose vanishing diagonal
matrix elements \(\hat{\mcal{B}}_{D,mm}=0\); \(m=1,\ldots,L\)
(\(\hat{\mcal{F}}_{D,ii}=0\); \(i=1,\ldots,S\))
\beq \lb{A12}
\delta\hat{\Sigma}_{D;N\times N}^{11} &=&
\underbrace{Q_{N\times N}^{11,+}}_{Q_{N\times N}^{11,-1}}\;\underbrace{\left( \bea{cc}
\delta\hat{\lambda}_{B,L\times L} & 0 \\
0 & \delta\hat{\lambda}_{F,S\times S}
\eea\right)_{N\times N}}_{\delta\hat{\lambda}_{N\times N}}\;Q_{N\times N}^{11} \\ \no &=&
Q_{N\times N}^{11,-1}\;\;\delta\hat{\lambda}_{N\times N}\;\;Q_{N\times N}^{11} \\ \lb{A13}
\delta\hat{\Sigma}_{D;N\times N}^{22}\;\wtilde{\kappa} &=&
\underbrace{\big(\wtilde{\kappa}\;Q_{N\times N}^{22,+}\;\wtilde{\kappa}\big)}_{Q_{N\times N}^{22,-1}}\;
\underbrace{\left( \bea{cc} \delta\hat{\lambda}_{B} & 0 \\
0 & \delta\hat{\lambda}_{F} \eea\right)_{N\times N}\; \underbrace{\hat{\kappa}\;\wtilde{\kappa}}_{
-\hat{1}_{N\times N}}}_{-\delta\hat{\lambda}_{N\times N}}\;Q_{N\times N}^{22}
\\ \no &=&Q_{N\times N}^{22,-1}\;\;\big(-\delta\hat{\lambda}_{N\times N})\;\;Q_{N\times N}^{22}
\eeq
\beq  \lb{A14}
\delta\hat{\lambda}_{N\times N}&=&\Big(\delta\hat{\lambda}_{B;1},\ldots,\delta\hat{\lambda}_{B;m},\ldots,
\delta\hat{\lambda}_{B;L}\;;\;\delta\hat{\lambda}_{F;1},\ldots,\delta\hat{\lambda}_{F;i},\ldots,
\delta\hat{\lambda}_{F;S}\Big)
\\ \lb{A15}
\hat{Q}^{11}_{N\times N}&=&\exp\left\{\im\left( \bea{cc}
\hat{\mcal{B}}_{D,L\times L} & \hat{\omega}_{D,L\times S}^{+} \\
\hat{\omega}_{D,S\times L} & \hat{\mcal{F}}_{D,S\times S}
\eea\right)\right\}  \hspace*{0.75cm}\hat{\mcal{B}}_{D,L\times L}^{+}=\hat{\mcal{B}}_{D,L\times L}  \\ \lb{A16}
\hat{Q}^{22}_{N\times N}&=&\exp\left\{\im\left( \bea{cc}
-\hat{\mcal{B}}_{D,L\times L}^{T} & \hat{\omega}_{D,L\times S}^{T} \\
-\hat{\omega}_{D,S\times L}^{*} & -\hat{\mcal{F}}_{D,S\times S}^{T} \eea\right)\right\}
\hspace*{0.75cm}\hat{\mcal{F}}_{D,S\times S}^{+}=\hat{\mcal{F}}_{D,S\times S}  \\ \lb{A17} &&
\hat{\mcal{B}}_{D;mm}=0\;\;,(m=1,\ldots,L) \\ \lb{A18} &&
\hat{\mcal{F}}_{D;ii}=0\;\;(i=1,\ldots,S)  \\   \lb{A19}
\big(\hat{Q}_{N\times N}^{22}\big)^{st}&=&\hat{Q}_{N\times N}^{11,+}=\hat{Q}_{N\times N}^{11,-1}\;.
\eeq
Considering relations (\ref{A12}-\ref{A19}), the block diagonal density term
\(\delta\hat{\Sigma}_{D;2N\times 2N}\;\wtilde{K}\) is diagonalized with
the block diagonal super-unitary matrix $\hat{Q}_{2N\times 2N}$ and doubled
eigenvectors \(\delta\hat{\Lambda}_{\alpha}^{a}=(\delta\hat{\lambda}_{\alpha}\;;\;
-\delta\hat{\lambda}_{\alpha})\) (\(\alpha=1,\ldots,L+S=N\); \(a=1,2\))
as Cartan subalgebra of \(Osp(S,S|2L)\)
\beq \no
\delta\hat{\Sigma}_{D;2N\times 2N}\;\wtilde{K}&=&\underbrace{\left( \bea{cc}
\hat{Q}_{N\times N}^{11,-1}  & 0 \\
0 & \hat{Q}_{N\times N}^{22,-1} \eea\right)}_{\hat{Q}_{2N\times 2N}^{-1}}\; \underbrace{\left( \bea{cc}
\delta\hat{\lambda}_{N\times N} & 0  \\
0 & -\delta\hat{\lambda}_{N\times N} \eea\right)}_{\delta\hat{\Lambda}_{2N\times 2N}}\;\underbrace{\left( \bea{cc}
\hat{Q}_{N\times N}^{11} & 0 \\
0 & \hat{Q}_{N\times N}^{22} \eea\right)}_{\hat{Q}_{2N\times 2N}} \\ \lb{A20} &=&
\hat{Q}_{2N\times 2N}^{-1}\;\;\delta\hat{\Lambda}_{2N\times 2N}\;\;\hat{Q}_{2N\times 2N} \\ \lb{A21}
\delta\hat{\Lambda}_{2N\times 2N}&=&\mbox{diag}\Big(\delta\hat{\lambda}_{N\times N}\;;\;
-\delta\hat{\lambda}_{N\times N}\Big)\;\;\;.
\eeq
The coset matrix $\hat{T}$ (\ref{A1},\ref{A8})
\(Osp(S,S|2L)\backslash U(L|S)\) has to include an additional imaginary factor
in the super-generator $\hat{Y}_{2N\times 2N}$ (\ref{A22}-\ref{A24})
so that anti-hermitian pair condensate terms are attained in
\(\wtilde{\Sigma}_{2N\times 2N}\;\wtilde{K}\) (\ref{A1}). We define the super-matrix
\(\hat{Y}_{2N\times 2N}\) as generator of $\hat{T}_{2N\times 2N}$ with the matrices \(\hat{X}_{N\times N}\),
\(\wtilde{\kappa}_{N\times N}\;\hat{X}^{+}_{N\times N}\) and their symmetry restrictions
(\ref{A24}) which are determined by the '12' and '21'
blocks of the derived generator \(\wtilde{W}_{\alpha\beta}^{ab}\) (\ref{s3_41}) of \(Osp(S,S|2L)\)
in section \ref{s31}
\beq \lb{A22}
\hat{T}_{2N\times 2N}&=&\exp\left\{\im\left( \bea{cc}
0 & \im\;\hat{X}_{N\times N} \\
\im\;\wtilde{\kappa}\;\hat{X}_{N\times N}^{+} & 0
\eea\right)\right\}=\exp\Big\{-\hat{Y}_{2N\times 2N}\Big\} \\ \lb{A23}
\hat{Y}_{2N\times 2N}&=&\left( \bea{cc}
0 & \hat{X}_{N\times N} \\
\wtilde{\kappa}\;\hat{X}_{N\times N}^{+} & 0 \eea\right)
\hspace*{0.75cm} \hat{X}_{N\times N}=\left( \bea{cc}
-\hat{c}_{D,L\times L} & \hat{\eta}_{D,L\times S}^{T} \\
-\hat{\eta}_{D,S\times L} & \hat{f}_{D,S\times S} \eea\right)  \\ \lb{A24} &&
\hat{c}_{D,L\times L}^{T}=\hat{c}_{D,L\times L}\hspace*{1.0cm}
\hat{f}_{D,S\times S}^{T}=-\hat{f}_{D,S\times S}\;\;\;.
\eeq
Similarly to the diagonalization of \(\delta\hat{\Sigma}_{D;2N\times 2N}\;\wtilde{K}\)
(\ref{A20},\ref{A21}) with matrix \(\hat{Q}_{2N\times 2N}\) of
\(U(L|S)\) in relations (\ref{A12}) to (\ref{A21}), the coset generator
\(\hat{X}_{N\times N}\) (\ref{A23}) can be factorized into diagonal complex terms
\((-\hat{\ovv{c}}_{L\times L}\;;\;\hat{\ovv{f}}_{S\times S})\) of the matrix
\(\hat{X}_{DD;N\times N}\) (\ref{A25})
(with \(\hat{\ovv{f}}_{S\times S}\) composed of antisymmetric quaternions
or standard Pauli-matrix $\tau_{2}$) and related super-unitary matrices
\(\hat{P}_{N\times N}^{11}\), \(\hat{P}_{N\times N}^{22}\) with equivalent
independent parameters. A similar relation (\ref{A28}) exists for super-transposition between
\(\hat{P}_{N\times N}^{11}\) and \(\hat{P}_{N\times N}^{22}\)
as for \(\hat{Q}_{N\times N}^{11}\) and \(\hat{Q}_{N\times N}^{22}\) (\ref{A19})
\beq \lb{A25}
\hat{X}_{N\times N}&=&\hat{P}_{N\times N}^{11,+}\;\underbrace{\left( \bea{cc}
-\hat{\ovv{c}}_{L\times L} & 0 \\
0 & \hat{\ovv{f}}_{S\times S} \eea\right)}_{\hat{X}_{DD;N\times N}}\;\hat{P}_{N\times N}^{22}=
\hat{P}_{N\times N}^{11,+}\;\;\hat{X}_{DD;N\times N}\;\;\hat{P}_{N\times N}^{22}
\\   \lb{A26}
\hat{P}^{11}_{N\times N}&=&\exp\left\{\im\left( \bea{cc}
\hat{\mcal{C}}_{D;L\times L} & \hat{\xi}_{D;L\times S}^{+} \\
\hat{\xi}_{D;S\times L} & \hat{\mcal{G}}_{D;S\times S}
\eea\right)\right\}  \hspace*{0.75cm}\hat{\mcal{C}}_{D,L\times L}^{+}=\hat{\mcal{C}}_{D,L\times L}  \\ \lb{A27}
\hat{P}^{22}_{N\times N}&=&\exp\left\{\im\left( \bea{cc}
-\hat{\mcal{C}}_{D;L\times L}^{T} & \hat{\xi}_{D;L\times S}^{T} \\
-\hat{\xi}_{D;S\times L}^{*} & -\hat{\mcal{G}}_{D;S\times S}^{T} \eea\right)\right\}
\hspace*{0.75cm}\hat{\mcal{G}}_{D,S\times S}^{+}=\hat{\mcal{G}}_{D,S\times S} \\ \lb{A28}
\big(\hat{P}_{N\times N}^{22}\big)^{st}&=&\hat{P}_{N\times N}^{11,+}=\hat{P}_{N\times N}^{11,-1}\;\;\;\;.
\eeq
The symmetric molecular condensate matrix \(\hat{c}_{D,L\times L}\) (\ref{A23},\ref{A24})
consists of \(L(L+1)\) independent real parameters
which have also to be contained with the equivalent degrees of
freedom in the diagonalization (\ref{A25}-\ref{A28}). This is accomplished by vanishing
diagonal elements \(\hat{\mcal{C}}_{D,mm}=0\) in the hermitian boson-boson part
\(\hat{\mcal{C}}_{D,L\times L}\) of the super-unitary \(U(L|S)\) generator of  \(\hat{P}_{N\times N}^{11}\),
\(\hat{P}_{N\times N}^{22}\) and by choosing complex eigenvalues
\(\ovv{c}_{1},\ldots,\ovv{c}_{L}\) in \(\hat{X}_{DD;N\times N}\) (\ref{A25})
\beq \lb{A29}
\hat{\mcal{C}}_{D,mm}&=&0\hspace*{0.5cm}(m=1,\ldots,L) \\ \lb{A30}
\hat{\ovv{c}}_{L\times L}&=&\mbox{diag}\Big\{\ovv{c}_{m}\Big\}=
\mbox{diag}\Big\{\ovv{c}_{1},\ldots,\ovv{c}_{m},\ldots,\ovv{c}_{L}
\Big\}\hspace*{0.5cm}\ovv{c}_{m}\in\mbox{\sf C}
\eeq
\be \lb{A31}
\bea{rcl}
\mbox{parameters of :}\Big\{\hat{c}_{D,L\times L}\Big\} &=& L(L+1)\mbox{ independent real entries} \\
\mbox{parameters of :}\Big\{\hat{\mcal{C}}_{D,L\times L}\Big\} &=& L^{2}-L\mbox{ independent real entries} \\
\mbox{parameters of :}\Big\{\hat{\ovv{c}}_{L\times L}\Big\} &=& 2L\mbox{ independent real entries}
\eea_{.}
\ee
The choice of parameters for the fermion-fermion part in \(\hat{P}_{N\times N}^{11}\),
\(\hat{P}_{N\times N}^{22}\) (\ref{A26}-\ref{A28}) is more complicated
as in the boson-boson block (\ref{A29}-\ref{A31}) because one has to
introduce quaternions or the \(2\times 2\) Pauli-matrices $\tau_{j}$ (\(j=1,2,3\)) including
the unit matrix \(\tau_{0}=\hat{1}_{2\times 2}\)
for the parameters in \(\hat{\mcal{G}}_{D,S\times S}\) (\ref{A27})
and for the complex quaternionic eigenvalues \(\hat{\ovv{f}}_{S\times S}\) (\ref{A25}).
The eigenvalues \(\hat{\ovv{f}}_{S\times S}\)
are block diagonal quaternions \(\hat{\ovv{f}}_{r\mu;r\nu}=(\tau_{2})_{\mu\nu}\;\ovv{f}_{r}\)
(\(r,r\ppr=1,\ldots,S/2\); \(\mu,\nu=1,2\)) and have to be antisymmetric
(or have to be assigned to the standard
Pauli-matrix $\tau_{2}$) so that the BCS-terms \(\hat{f}_{D,S\times S}\) in $\hat{X}_{N\times N}$,
$\hat{Y}_{2N\times 2N}$ (\ref{A23},\ref{A24}) are also antisymmetric as \(\hat{f}_{S\times S}\) in
\(\wtilde{\Sigma}_{2N\times 2N}\;\wtilde{K}\)
(Compare with the anomalous parts in the generator
\(\wtilde{W}_{\alpha\beta}^{ab}\) (\ref{s3_41}) of \(Osp(S,S|2L)\) in section \ref{s31}).
The matrix \(\hat{\mcal{G}}_{D;r\mu,r\ppr\nu}\) (\ref{A32}) is
therefore composed of \(2\times 2\) quaternions with complex fields
\(\mcal{G}_{D;rr\ppr}^{(j)}\) whose indices $r$, $r\ppr$ range only from $1$ to $S/2$ because of the
division into $2\times 2$ submatrix elements $(\tau_{j})_{\mu\nu}$, (\(j=0,\ldots,3\); \(\mu,\nu=1,2\)).
Since the quaternions are represented by hermitian matrices, the elements \(\mcal{G}_{D;rr\ppr}^{(j)}\)
are also hermitian (\ref{A33}). The antisymmetric matrix \(\hat{f}_{D,S\times S}\) in
\(\hat{X}_{N\times N}\), \(\hat{Y}_{2N\times 2N}\) (\ref{A22}-\ref{A24}) for the BCS-condensate
has \(S^{2}-S\) independent real entries (\ref{A37}). Since the complex quaternionic eigenvalues
\(\hat{\ovv{f}}_{r\mu;r\nu}=(\tau_{2})_{\mu\nu}\;\ovv{f}_{r}\)
(\(r,r\ppr=1,\ldots,S/2\)) contain $S$ independent real parameters, the block diagonal quaternions
or \(2\times 2\) matrices \(\hat{\mcal{G}}_{D;r\mu,r\nu}\) (\ref{A34}) have to vanish in the hermitian
fermion-fermion part of \(\hat{P}_{N\times N}^{11}\), \(\hat{P}_{N\times N}^{22}\) so that the
same number of \(S^{2}-S\) real degrees of freedom is retained in the parametrization (\ref{A25}-\ref{A28})
for the fermion-fermion section (\ref{A37})
\beq \lb{A32}
\hat{\mcal{G}}_{D;r\mu,r\ppr\nu}&=&\sum_{j=0}^{3}(\tau_{j})_{\mu\nu}\;
\mcal{G}_{D;rr\ppr}^{(j)}\hspace*{0.5cm} (r,r\ppr=1,\ldots,S/2) \\
\no && (\mu,\nu=1,2),(\tau_{0}=\hat{1}_{2\times 2}\;\;,\tau_{1},\tau_{2},\tau_{3}=\mbox{Pauli-matrices})
\\ \lb{A33}  \hat{\mcal{G}}_{D;r\mu,r\ppr\nu}^{+}&=&\mcal{G}_{D;r\mu,r\ppr\nu}\hspace*{0.75cm}
\mcal{G}_{D;rr\ppr}^{(j)*}=\mcal{G}_{D;r\ppr r}^{(j)}\;\;\;(j=0,\ldots,3)
\\ \lb{A34}
\hat{\mcal{G}}_{D;r\mu,r\nu}&\equiv& 0\;\;\;\;\mbox{for }(r=1,\ldots,S/2),(\mu,\nu=1,2)
\\ \lb{A35} \hat{\ovv{f}}_{S\times S}&=&\mbox{diag}\Big\{\ovv{f}_{r\mu;r\nu}\Big\}=
\mbox{diag}\Big\{(\tau_{2})_{\mu\nu}\;\ovv{f}_{1},\ldots,(\tau_{2})_{\mu\nu}\;\ovv{f}_{r},\ldots,
(\tau_{2})_{\mu\nu}\;\ovv{f}_{S/2}\Big\}
\\ \no && (r=1,\ldots,S/2),(\mu,\nu=1,2)  \hspace*{0.5cm}\ovv{f}_{r}\in\mbox{\sf C}   \\ \lb{A36}
\hat{\ovv{f}}_{S\times S}^{T}&=&-\hat{\ovv{f}}_{S\times S}
\eeq
\be \lb{A37}
\bea{rcl}
\mbox{parameters of :}\Big\{\hat{f}_{D,S\times S}\Big\} &=& S^{2}-S\mbox{ independent real entries} \\
\mbox{parameters of :}\Big\{\hat{\mcal{G}}_{D;r\mu,r\ppr\nu}\Big\} &=& S^{2}-2S\mbox{ independent real entries} \\
\mbox{parameters of :}\Big\{\hat{\ovv{f}}_{S\times S}\Big\} &=& S\mbox{ independent real entries}
\eea_{.}
\ee
Since the generators \(\hat{X}_{N\times N}\), \(\hat{Y}_{2N\times 2N}\) (\ref{A22}-\ref{A24})
can be transformed to the
diagonal elements \(\hat{\ovv{c}}_{L\times L}\) for the boson-boson part (\ref{A25}), and to the
block diagonal antisymmetric quaternions \(\hat{\ovv{f}}_{S\times S}\) (\ref{A25}),
the coset matrix \(\hat{T}_{2N\times 2N}\)
(\ref{A22}) of \(Osp(S,S|2L)\backslash U(L|S)\) can finally be decomposed into
the block diagonal eigenvectors
\(\hat{P}_{N\times N}^{11}\), \(\hat{P}_{N\times N}^{22}\) of \(\hat{P}_{2N\times 2N}\) and
the coset matrix \(\hat{T}_{D;2N\times 2N}\) with generators
\(\hat{Y}_{DD;2N\times 2N}\), \(\hat{X}_{DD;N\times N}\)
\beq \lb{A38}
\hat{X}_{N\times N}&=&\hat{P}_{N\times N}^{11,-1}\;\hat{X}_{DD;N\times N}\;\hat{P}_{N\times N}^{22}=
\hat{P}_{N\times N}^{11,-1}\;\left(
\bea{cc}
-\hat{\ovv{c}}_{L\times L} & 0 \\
0 & \hat{\ovv{f}}_{S\times S}
\eea\right)\;\hat{P}_{N\times N}^{22} \\ \lb{A39}
\wtilde{\kappa}\;\;\hat{X}_{N\times N}^{+}&=&
\underbrace{\wtilde{\kappa}\;\hat{P}_{N\times N}^{22,+}\;\wtilde{\kappa}}_{
\hat{P}_{N\times N}^{22,-1}}\;
\underbrace{\wtilde{\kappa}\;\left( \bea{cc}
-\hat{\ovv{c}}_{L\times L}^{+} &  0 \\
0 & \hat{\ovv{f}}_{S\times S}^{+}
\eea\right)}_{\wtilde{\kappa}\;\hat{X}_{DD;N\times N}^{+}}\;\hat{P}_{N\times N}^{11} \\ \no
\hat{Y}_{2N\times 2N}&=&\underbrace{\left( \bea{cc}
\hat{P}_{N\times N}^{11,-1} & 0 \\
0 & \hat{P}_{N\times N}^{22,-1} \eea\right)}_{\hat{P}_{2N\times 2N}^{-1}}\; \underbrace{\left( \bea{cc}
0 & \hat{X}_{DD;N\times N} \\
\wtilde{\kappa}\;\hat{X}_{DD;N\times N}^{+} & 0 \eea\right)}_{\hat{Y}_{DD;2N\times 2N}}\;
\underbrace{\left( \bea{cc}
\hat{P}_{N\times N}^{11} & 0 \\
0 & \hat{P}_{N\times N}^{22}
\eea\right)}_{\hat{P}_{2N\times 2N}} \\ \lb{A40} &=&\hat{P}^{-1}_{2N\times 2N}\;\hat{Y}_{DD;2N\times 2N}\;
\hat{P}_{2N\times 2N} \\ \lb{A41}
\hat{T}_{D;2N\times 2N}&=&\exp\Big\{-\hat{Y}_{DD;2N\times 2N}\Big\} \\ \no & \longrightarrow &
\hat{T}_{2N\times 2N}=\exp\{-\hat{Y}_{2N\times 2N}\}=
\hat{P}_{2N\times 2N}^{-1}\;\;\hat{T}_{D;2N\times 2N}\;\;\hat{P}_{2N\times 2N}\;.
\eeq
In subsection \ref{sa3} the exponential \(\exp\{-\hat{Y}_{DD;2N\times 2N}\}\) (\ref{A41}) is expanded
with the diagonal matrices \(\hat{X}_{DD;N\times N}\),
\(\wtilde{\kappa}\;\hat{X}_{DD;N\times N}^{+}\)
(\ref{A38},\ref{A39}) in the generator \(\hat{Y}_{DD;2N\times 2N}\) (\ref{A40})
so that the even (boson-boson, fermion-fermion) parts and
the odd elements in the boson-fermion and fermion-boson sections can be computed
in terms of the eigenvalues \(\ovv{c}_{m}\) (\(m=1,\ldots,L\)) and
\(\ovv{f}_{r}\;(\tau_{2})_{\mu\nu}\) (\(r=1,\ldots,S/2\), \(\mu,\nu=1,2\)).
The relation (\ref{A41}) then leads from the expanded
exponential $\hat{T}_{D;2N\times 2N}$ with the eigenvalues in
\(\hat{X}_{DD;N\times N}\) to the coset matrix
$\hat{T}_{2N\times 2N}$ with block diagonal matrix $\hat{P}_{2N\times 2N}$ and its inverse.

The density term \(\delta\hat{\Sigma}_{D;2N\times 2N}\;\wtilde{K}\) (\ref{A20},\ref{A21}) is diagonalized
with matrix \(\hat{Q}_{2N\times 2N}\) and the doubled eigenvalues $\delta\hat{\Lambda}_{2N\times 2N}$
so that the invariant length \((d s_{Osp})^{2}\) (\ref{A11}) takes the form (\ref{A42})
with coset matrix \(\hat{T}_{2N\times 2N}\)
\beq \lb{A42}
(d s_{Osp})^{2}&=&\mbox{STR}\Big[d\big(\hat{T}\;\delta\hat{\Sigma}_{D;2N\times 2N}\;
\wtilde{K}\;\hat{T}^{-1}\big)\;\;
d\big(\hat{T}\;\delta\hat{\Sigma}_{D;2N\times 2N}\;
\wtilde{K}\;\hat{T}^{-1}\big)\Big]
\\ \no &=&\mbox{STR}\Big[d\big(\underbrace{\hat{T}\;\hat{Q}^{-1}}_{\hat{T}_{0}}\;
\delta\hat{\Lambda}_{2N\times 2N}\;
\underbrace{\hat{Q}\;\hat{T}^{-1}}_{\hat{T}_{0}^{-1}}\big)\;\;
d\big(\underbrace{\hat{T}\;\hat{Q}^{-1}}_{\hat{T}_{0}}\;\delta\hat{\Lambda}_{2N\times 2N}\;
\underbrace{\hat{Q}\;\hat{T}^{-1}}_{\hat{T}_{0}^{-1}}\big)\Big]_{,}
\eeq
where \(\hat{T}_{0;2N\times 2N}\) is defined by the product of
\(\hat{T}_{2N\times 2N}\) (\ref{A22}) and the super-unitary matrix
\(\hat{Q}_{2N\times 2N}\) (\ref{A15}-\ref{A20}) of \(U(L|S)\)
\be \lb{A43}
\hat{T}_{0;2N\times 2N}=\hat{T}_{2N\times 2N}\;\hat{Q}_{2N\times 2N}^{-1}\;\;\;.
\ee
The expansion of curvilinear elements \(d(\hat{T}_{0}\;\delta\hat{\Lambda}\;\hat{T}_{0}^{-1})\) (\ref{A44})
results in a commutator between \(\hat{T}_{0}^{-1}\;d\hat{T}_{0}\) and
the diagonal doubled eigenvalue density term
$\delta\hat{\Lambda}_{2N\times 2N}$, the Cartan subalgebra elements of \(Osp(S,S|2L)\),
and in addition in the local eigenvalue increments \(d(\delta\hat{\Lambda}_{2N\times 2N})\).
The sum of these two terms is rotated by the matrices $\hat{T}_{0}$, $\hat{T}_{0}^{-1}$ which act as a
similarity transformation (\ref{A44}) so that the invariant length $(d s_{Osp})^{2}$ (\ref{A42})
is transformed to the square of the sum of the commutator and
the local eigenvalue elements \(d(\delta\hat{\Lambda})\) (\ref{A45})
\beq \lb{A44}
d\big(\hat{T}_{0}\;\delta\hat{\Lambda}\;\hat{T}_{0}^{-1}\big)&=&\hat{T}_{0}\;\;
\bigg(\Big[\hat{T}_{0}^{-1}\;d\hat{T}_{0}\;,\;\delta\hat{\Lambda}\Big]+
d\big(\delta\hat{\Lambda}\big)\bigg)\;\;\hat{T}_{0}^{-1}
\\ \lb{A45} (d s_{Osp})^{2}&=&\mbox{STR}\bigg[\;\bigg(
\Big[\hat{T}_{0}^{-1}\;d\hat{T}_{0}\;,\;\delta\hat{\Lambda}\Big]+
d\big(\delta\hat{\Lambda}\big)\bigg)^{2}\;\bigg]\;\;\;.
\eeq
The diagonal matrix elements
\(\big[(\hat{T}_{0}^{-1}\;d\hat{T}_{0})\:,\:\delta\hat{\Lambda}\big]_{\alpha\alpha}^{aa}\)
(\(\alpha=1,\ldots,N=L+S\); \(a=1,2\)) of the commutator in $(ds)^{2}$ (\ref{A45},\ref{A42},\ref{A11})
vanish so that the local eigenvalue increments \(d(\delta\hat{\Lambda}_{\alpha}^{a})\)
decouple from the commutator term in (\ref{A45}) and yield another decomposition of the metric (\ref{A46}).
Since the '11' block \(\delta\hat{\Sigma}_{D;N\times N}^{11}\) follows by super-transposition from
the '22' part \(\delta\hat{\Sigma}_{D;N\times N}^{22}\;\wtilde{\kappa}\) (\ref{A7})
(\(\delta\hat{\Sigma}_{D;N\times N}^{11}
=-\big(\delta\hat{\Sigma}_{D;N\times N}^{22}\;\wtilde{\kappa}\big)^{st}\))
and since the super-traces are
cyclic invariant for the matrices of the pair condensates, the corresponding terms in the density and
anomalous parts can be summarized by a factor of two.
The invariant length $(d s_{Osp})^{2}$ (\ref{A45}) therefore reduces to the following relation
(The metric \(\hat{\kappa}_{N\times N}=\big(\hat{1}_{L\times L}\:;\:-\hat{1}_{S\times S}\big)\)
considers the additional minus sign in the fermion-fermion part of a super-trace relation
(Compare with (\ref{s2_16},\ref{s2_31})).)
\beq \lb{A46}
\lefteqn{(d s_{Osp})^{2}=\mbox{STR}\bigg[\bigg(\big[\hat{T}_{0}^{-1}\;
d\hat{T}_{0}\;,\;\delta\hat{\Lambda}\big]+
d\big(\delta\hat{\Lambda}\big)\bigg)^{2}\bigg]  =} \\ \no &=&
2\;\mbox{str}\Big[d\big(\delta\hat{\lambda}_{\alpha}\big)\;\;
d\big(\delta\hat{\lambda}_{\alpha}\big)\Big]-
 \underbrace{\sum_{\alpha,\beta=1}^{N=L+S}\hat{\kappa}_{\alpha\alpha}}_{\mbox{str}}\;\;
\Big[\big(\hat{T}_{0}^{-1}\;d\hat{T}_{0}\big)_{\alpha\beta}^{11}\;\;
\big(\hat{T}_{0}^{-1}\;d\hat{T}_{0}\big)_{\beta\alpha}^{11}\,\;
\big(\delta\hat{\lambda}_{\beta}-\delta\hat{\lambda}_{\alpha}\big)^{2}\Big]
\\ \no &-& \underbrace{\sum_{\alpha,\beta=1}^{N=L+S}\hat{\kappa}_{\alpha\alpha}}_{\mbox{str}}\;\;
\Big[\big(\hat{T}_{0}^{-1}\;d\hat{T}_{0}\big)_{\alpha\beta}^{22}\;\;
\big(\hat{T}_{0}^{-1}\;d\hat{T}_{0}\big)_{\beta\alpha}^{22}\,\;
\big(\delta\hat{\lambda}_{\beta}-\delta\hat{\lambda}_{\alpha}\big)^{2}\Big]
\\ \no &-& \underbrace{\sum_{\alpha,\beta=1}^{N=L+S}\hat{\kappa}_{\alpha\alpha}}_{\mbox{str}}\;\;
\Big[\big(\hat{T}_{0}^{-1}\;d\hat{T}_{0}\big)_{\alpha\beta}^{12}\;\;
\big(\hat{T}_{0}^{-1}\;d\hat{T}_{0}\big)_{\beta\alpha}^{21}\,\;
\big(\delta\hat{\lambda}_{\beta}+\delta\hat{\lambda}_{\alpha}\big)^{2}\Big]
\\ \no &-& \underbrace{\sum_{\alpha,\beta=1}^{N=L+S}\hat{\kappa}_{\alpha\alpha}}_{\mbox{str}}\;\;
\Big[\big(\hat{T}_{0}^{-1}\;d\hat{T}_{0}\big)_{\alpha\beta}^{21}\;\;
\big(\hat{T}_{0}^{-1}\;d\hat{T}_{0}\big)_{\beta\alpha}^{12}\,\;
\big(\delta\hat{\lambda}_{\alpha}+\delta\hat{\lambda}_{\beta}\big)^{2}\Big]=
\\ \no &=&
2\;\mbox{str}\Big[d\big(\delta\hat{\lambda}_{\alpha}\big)\;\;
d\big(\delta\hat{\lambda}_{\alpha}\big)\Big]
-2 \underbrace{\sum_{\alpha,\beta=1}^{N=L+S}\hat{\kappa}_{\alpha\alpha}}_{\mbox{str}}\;\;
\Big[\big(\hat{T}_{0}^{-1}\;d\hat{T}_{0}\big)_{\alpha\beta}^{11}\;\;
\big(\hat{T}_{0}^{-1}\;d\hat{T}_{0}\big)_{\beta\alpha}^{11}\,\;
\big(\delta\hat{\lambda}_{\beta}-\delta\hat{\lambda}_{\alpha}\big)^{2}\Big]
\\ \no &-&2 \underbrace{\sum_{\alpha,\beta=1}^{N=L+S}\hat{\kappa}_{\alpha\alpha}}_{\mbox{str}}\;\;
\Big[\big(\hat{T}_{0}^{-1}\;d\hat{T}_{0}\big)_{\alpha\beta}^{12}\;\;
\big(\hat{T}_{0}^{-1}\;d\hat{T}_{0}\big)_{\beta\alpha}^{21}\,\;
\big(\delta\hat{\lambda}_{\beta}+\delta\hat{\lambda}_{\alpha}\big)^{2}\Big] \;\;.
\eeq
It remains to determine \(\big(\hat{T}_{0}^{-1}\;d\hat{T}_{0}\big)^{ab}\) in \((d s_{Osp})^{2}\) (\ref{A46})
with block diagonal matrix $\hat{Q}_{2N\times 2N}$ (\ref{A15}-\ref{A20}) of \(U(L|S)\)
and generator \(\hat{Y}_{2N\times 2N}\) (\ref{A22}-\ref{A24}) which has been diagonalized
with $\hat{P}_{2N\times 2N}$ (\ref{A25}-\ref{A28}) to $\hat{Y}_{DD;2N\times 2N}$ (\ref{A38}-\ref{A41}).
In the next section \ref{sa2} we derive general relations about the variation of matrices
as \(\hat{T}_{0}^{-1}\;d\hat{T}_{0}\) and \(\hat{Q}^{-1}\;d\hat{Q}\)
in terms of the generators where one has to apply the group structure of
these matrices.

\subsection{Determination of $\hat{T}_{0}^{-1}\;d\hat{T}_{0}$ and
$\hat{Q}^{-1}\;d\hat{Q}$ with generators of
$Osp(S,S|2L)$ and $U(L|S)$} \lb{sa2}

In order to compute \(\hat{T}_{0}^{-1}\;d\hat{T}_{0}\) and \(\hat{Q}^{-1}\;d\hat{Q}\),
we have to apply the Lie group properties of $\hat{T}_{0}$, $\hat{Q}$ which are not any
set of arbitrary matrices, but consist of a closed Lie super-algebra  \(osp(S,S|2L)\)
and \(u(L|S)\). By way of example we consider a general Lie super-group with elements $\hat{G}_{\hat{A}}$
(\ref{A47}) composed of a basis of generators $\hat{A}^{(\kappa)}$ with a closed Lie super-algebra
(\ref{A48}), following from supercommutators \(\big[\ldots\;,\;\ldots\big\}\)
with corresponding structure constants \(f^{\kappa\lambda}_{\ph{\kappa\lambda}\mu}\)
(compare section \ref{s53}). The super-generators
$\hat{A}^{(\kappa)}$ are combined with even or odd parameters $a_{\kappa}$, depending on the
even or odd degree of $\hat{A}^{(\kappa)}$, so that
the exponential of \(\hat{A}=\hat{A}^{(\kappa)}\;a_{\kappa}\)
results in a group element $\hat{G}_{\hat{A}}$ (\ref{A47}) which can be multiplied with an
element \(\hat{G}_{\hat{A}\ppr}=\exp\{\hat{A}^{(\lambda)}\;a_{\lambda}\ppr\}\) with a different set
of parameters $a_{\lambda}\ppr$ in order to lead to another Lie group element
$\hat{G}_{\hat{A}^{\prime\prime}}$ with parameters $a_{\mu}^{\prime\prime}$ (\ref{A49})
\beq \lb{A47}
\hat{G}_{\hat{A}}&=&\exp\big\{\hat{A}\big\}
\hspace*{1.5cm}\hat{A}=\hat{A}^{(\kappa)}\;a_{\kappa} \\  \lb{A48}
\Big[\hat{A}^{(\kappa)}\;,\;\hat{A}^{(\lambda)}\Big\}&=&
\im\;f^{\kappa\lambda}_{\ph{\kappa\lambda}\mu}\;\;\hat{A}^{(\mu)} \\ \lb{A49}
\hat{G}_{\hat{A}}\;\hat{G}_{\hat{A}\ppr}=\hat{G}_{\hat{A}^{\prime\prime}} &\Longrightarrow &
\exp\big\{\hat{A}^{(\kappa)}\;a_{\kappa}\big\}\;\;
\exp\big\{\hat{A}^{(\lambda)}\;a_{\lambda}\ppr\big\}=
\exp\big\{\hat{A}^{(\mu)}\;a_{\mu}^{\prime\prime}\big\} \;\;;\hspace*{0.55cm}
a_{\mu}^{\prime\prime}=a_{\mu}^{\prime\prime}(a_{\kappa},a_{\lambda}\ppr)\;\;\;\;.
\eeq
The general composition \(\hat{G}_{\hat{A}}^{-1}\;d\hat{G}_{\hat{A}}\) as
\(\hat{T}_{0}^{-1}\;d\hat{T}_{0}\) in the invariant length $(ds)^{2}$ (\ref{A46})
implies a variation with the generators
\(d\hat{A}=\hat{A}^{(\kappa)}\;da_{\kappa}\) of the infinitesimal
parameters $da_{\kappa}$
\beq \lb{A50}
\hat{G}_{\hat{A}}^{-1}\;\;d\hat{G}_{\hat{A}}&=&\exp\big\{-\hat{A}\big\}\;\;
d\big(\exp\big\{\hat{A}\big\}\big)=
\exp\big\{-\hat{A}\big\}\;
\Big(\exp\big\{\hat{A}+d\hat{A}\big\}-\exp\big\{\hat{A}\big\}\Big) \\ \lb{A51}
d\hat{A}&=&\hat{A}^{(\kappa)}\;\;da_{\kappa}\;\;\;\;.
\eeq
However, we have to consider that the exponential \(\exp\big\{\hat{A}+d\hat{A}\big\}\) in (\ref{A50})
is again an element of the underlying Lie super-group which can therefore be obtained from multiplication
of \(\hat{G}_{\hat{A}}=\exp\big\{\hat{A}\big\}\) with an element $\hat{G}_{d\hat{B}}$ of the Lie group,
consisting of the variations \(db_{\kappa}(a_{\lambda})\) of yet undetermined functions
$b_{\kappa}(a_{\lambda})$ of the original parameters $a_{\kappa}$ of $\hat{A}$
\beq \lb{A52}
\exp\big\{\hat{A}+d\hat{A}\big\}&=&\hat{G}_{\hat{A}+d\hat{A}}=\hat{G}_{\hat{A}}\;\hat{G}_{d\hat{B}}
\\ \no &=&\exp\big\{\hat{A}\big\}\;\;\exp\big\{d\hat{B}\big\}\hspace*{1.5cm}
d\hat{B}=\hat{A}^{(\kappa)}\;\Big(b_{\kappa}(a_{\lambda}+da_{\lambda})-b_{\kappa}(a_{\lambda})\Big)\;\;.
\eeq
By substituting (\ref{A52}) into (\ref{A50}), we derive the general relation (\ref{A53}) for the
variation of a Lie (super)group.
The variation \(\exp\big\{-\hat{A}\big\}\;\;d\big(\exp\big\{\hat{A}\big\}\big)\)
is equivalent to the sum of the (super)generators $\hat{A}^{(\kappa)}$, each multiplied with the
derivative \((\pp b_{\kappa})/(\pp a_{\lambda})\) of functions $b_{\kappa}(a_{\lambda})$ and the infinitesimal
variations $da_{\lambda}$ of the parameters in $d\hat{A}$ (\ref{A51})
\beq \lb{A53}
\exp\big\{-\hat{A}\big\}\;\;d\big(\exp\big\{\hat{A}\big\}\big)&=&
\exp\big\{-\hat{A}\big\}\;\;\Big(\exp\big\{\hat{A}\big\}\;\;
\exp\big\{d\hat{B}\big\}-\exp\big\{\hat{A}\big\}\Big)
\\ \no &=& \exp\big\{d\hat{B}\big\}-\hat{1}     \approx
d\hat{B}=\hat{A}^{(\kappa)}\;\;db_{\kappa}(a_{\lambda})=\hat{A}^{(\kappa)}\;\;
\frac{db_{\kappa}(a_{\lambda})}{da_{\lambda}}\;\;da_{\lambda}\;\;\;.
\eeq
The functions $b_{\kappa}(a_{\lambda})$ (\ref{A52},\ref{A53})
depend on the considered group and remain to be computed by a different relation
(\ref{A80},\ref{A83}) which has already been introduced and proved in subsection \ref{s12}
(compare Eqs. (\ref{s1_17}-\ref{s1_24})).

The Eqs. (\ref{A47}) to (\ref{A53}) have to be assigned to the relations
\(\hat{T}_{0}^{-1}\;d\hat{T}_{0}\) and \(\hat{Q}^{-1}\;d\hat{Q}\) in the invariant length
\((ds_{Osp})^{2}\) (\ref{A46}). The maximal commuting Cartan subalgebra with generators
$\hat{\Lambda}^{(\alpha)}$  of \(Osp(S,S|2L)\) is given by the doubled eigenvalues
\(\delta\hat{\Lambda}_{2N\times 2N}=
\big(\delta\hat{\lambda}_{N\times N}\;;\;-\delta\hat{\lambda}_{N\times N}\big)\) (\ref{A54},\ref{A55}).
The super-unitary subgroup \(U(L|S)\) of \(Osp(S,S|2L)\) consists of the diagonal commuting elements
\(\delta\hat{\lambda}_{\alpha}=(\delta\hat{\lambda}_{B;m}\:;\:\delta\hat{\lambda}_{F;i})\)
and the ladder operators which are determined by
the direct sum of the block diagonal generators $\hat{h}^{(\kappa)}$ in the '11' and '22' parts
of the matrices \(\hat{Q}_{2N\times 2N}\), \(\hat{P}_{2N\times 2N}\) (\ref{A56},\ref{A57}), respectively.
The independent parameters $q_{\kappa}$, $p_{\kappa}$ (\ref{A56},\ref{A57}) for the generators
\(\hat{h}^{(\kappa)}\) in \(\hat{Q}_{2N\times 2N}\), \(\hat{P}_{2N\times 2N}\) follow from
the defining relations (\ref{A15}-\ref{A19}) and (\ref{A25}-\ref{A37}) in subsection \ref{sa1}.
(Note that the diagonal elements \(\hat{\mcal{B}}_{D,mm}=0\), \(\hat{\mcal{F}}_{D,ii}=0\) and
\(\hat{\mcal{C}}_{D,mm}=0\), \(\hat{\mcal{G}}_{D;r\mu,r\nu}=0\) vanish; compare with Eqs.
(\ref{A17},\ref{A18}) and (\ref{A29},\ref{A34})).
The coset elements \(\hat{T}_{2N\times 2N}\) (\ref{A22}-\ref{A24})
\(Osp(S,S|2L)\backslash U(L|S)\) for the antihermtian
anomalous terms are given by the generators $\hat{Y}^{(\kappa)}$ (\ref{A58})
of the matrix $\hat{Y}_{2N\times 2N}$ (\ref{A23})
with the set of the independent variables \(y_{\kappa}\) of \(\hat{c}_{D,L\times L}\), \(\hat{c}_{D,L\times L}^{+}\),
\(\hat{f}_{D,S\times S}\), \(\hat{f}_{D,S\times S}^{+}\) and
\(\hat{\eta}_{D,S\times L}\), \(\hat{\eta}_{D,L\times S}^{+}\). The ladder operators \(\hat{h}^{\prime(\kappa)}\)
of \(\hat{P}_{2N\times 2N}^{aa}\) (\ref{A57}) ('note the prime') have the structure of a Clifford algebra
in the fermion-fermion block; therefore, the ladder operators \(\hat{h}^{\prime(\kappa)}\) are not completely
identical to those \(\hat{h}^{(\kappa)}\) (\ref{A56}) within \(\hat{Q}_{2N\times 2N}^{aa}\) for the \(U(L|S)\)
subgroup (e.g. compare \cite{corn3} for details)
\beq \lb{A54}
\delta\hat{\Lambda}_{2N\times 2N}&=&\left( \bea{cc}
\delta\hat{\lambda}_{N\times N} & 0 \\
0 & -\delta\hat{\lambda}_{N\times N}
\eea\right)=\hat{\Lambda}^{(\alpha)}\;\;\delta\lambda_{\alpha} \\ \lb{A55}
\big\{\delta\lambda_{\alpha}\big\}&=&\Big\{\delta\lambda_{B;1},\ldots,
\delta\lambda_{B;L};\delta\lambda_{F;1},\ldots,\delta\lambda_{F;S}\Big\}
\\ \lb{A56}
\hat{Q}_{2N\times 2N}&=&\exp\big\{\im\;q_{\kappa}\;\hat{h}^{(\kappa)}\big\}   \\ \no
q_{\kappa}&=&\mbox{independent variables of :}\Big\{\mcal{B}_{D;L\times L},\mcal{F}_{D;S\times
S};\omega_{D;S\times L},\omega_{D;L\times S}^{+}\Big\} \\ \no &&
\hat{\mcal{B}}_{D,mm}=0\;\;\;(m=1,\ldots,L)\hspace*{1.0cm}\hat{\mcal{F}}_{D,ii}=0\;\;\;
(i=1,\ldots,S)
\\ \lb{A57}
\hat{P}_{2N\times 2N}&=&\exp\big\{\im\;p_{\kappa}\;\hat{h}^{\prime(\kappa)}\big\}  \\ \no
p_{\kappa}&=&\mbox{independent variables of :}\Big\{\mcal{C}_{D;L\times L},\mcal{G}_{D;S\times S};
\xi_{D;S\times L},\xi_{D;L\times S}^{+}\Big\} \\ \no &&
\hat{\mcal{C}}_{D,mm}=0\;\;\;(m=1,\ldots,L)\hspace*{1.0cm}\hat{\mcal{G}}_{D;r\mu,r\nu}=0\;\;\;
(r=1,\ldots,S/2;\;\;\mu,\nu=1,2)
\\ \lb{A58}
\hat{T}_{2N\times 2N}&=&\exp\big\{-y_{\kappa}\;\hat{Y}^{(\kappa)}\big\}   \\ \no
y_{\kappa}&=&\mbox{independent variables of :}
\Big\{\hat{c}_{D},\hat{f}_{D},\hat{c}_{D}^{+},\hat{f}_{D}^{+},\eta_{D},\eta_{D}^{+}\Big\} \\ \no &&
\hat{c}_{D,L\times L}^{T}=\hat{c}_{D,L\times L}\hspace*{1.0cm}
\hat{f}_{D,S\times S}^{T}=-\hat{f}_{D,S\times S}\;\;.
\eeq
The ortho-symplectic Lie super-algebra \(osp(S,S|2L)\) therefore separates into the
super-unitary subalgebra \(u(L|S)\) with doubled generators $\hat{H}^{(\kappa)}$ of the '11' and the '22' parts
of the density terms which can be divided into the maximal abelian Cartan subalgebra with doubled
'eigenvalue' generators $\hat{\Lambda}^{(\alpha)}$ (\ref{A54},\ref{A55},\ref{A21},
\ref{A14}) and ladder operators $\hat{h}^{(\kappa)}$
of \(u(L|S)\) (\ref{A59},\ref{A61},\ref{A62}).
The remaining generators $\hat{Y}^{(\kappa)}$ (\ref{A60}) of $\hat{Y}_{2N\times 2N}$ in \(\hat{T}_{2N\times 2N}\)
define the coset space \(Osp(S,S|2L)\backslash U(L|S)\) and are derived from the '12' and '21' parts with
matrices \(\hat{X}_{N\times N}\) and \(\wtilde{\kappa}\;\hat{X}_{N\times N}^{+}\) (\ref{A22}-\ref{A24}).
The Lie super-algebra \(osp(S,S|2L)\) contains closed supercommutator relations for the subalgebra
\(u(L|S)\) (\ref{A62}) with maximal abelian Cartan subalgebra (\ref{A61})
of the doubled 'eigenvalue' generators $\hat{\Lambda}^{(\alpha)}$ (\ref{A54}).
The subalgebra \(u(L|S)\) is represented by the direct sum of block diagonal density terms
\(\delta\hat{\Sigma}^{11}\), \(\delta\hat{\Sigma}^{22}\;\wtilde{\kappa}\)
whose commutators therefore result again into block diagonal structures of density terms (\ref{A62}).
The commutators between $\hat{Y}^{(\kappa)}$ (having entries in the '12' and '21' sections with
matrices \(\hat{X}_{N\times N}\), \(\wtilde{\kappa}\;\hat{X}_{N\times N}^{+}\)) and
the \(u(L|S)\) subalgebra generators $\hat{H}^{(\kappa)}$ lead again to coset elements $\hat{Y}^{(\kappa)}$
(\ref{A63}) whereas the supercommutator of two coset elements yields  elements $\{\hat{H}\}$ of
the subalgebra \(u(L|S)\) of \(osp(S,S|2L)\) (The \(\hat{h}^{(\kappa)}\) generators
denote the ladder operators of \(u(L|S)\).)
\beq \lb{A59}
\Big\{\hat{H}^{(\kappa)}\Big\}&=&\Big\{\hat{\Lambda}^{(\alpha)},\hat{h}^{(\kappa)}\Big\}\hspace*{0.75cm}
\mbox{subgroup $U(L|S)$ of $Osp(S,S|2L)$}  \\ \lb{A60}
\Big\{\hat{Y}^{(\kappa)}\Big\}&=&\mbox{coset space $Osp(S,S|2L)\backslash U(L|S)$}
\\  \lb{A61}
\Big[\hat{\Lambda}^{(\alpha)}\;,\;\hat{\Lambda}^{(\beta)}\Big\}&=&0 \\ \lb{A62}
\Big[\big\{\hat{H}\big\}\;,\;\big\{\hat{H}\big\}\Big\}&=&\big\{\hat{H}\big\} \\ \lb{A63}
\Big[\big\{\hat{H}\big\}\;,\;\big\{\hat{Y}\big\}\Big\}&=&\big\{\hat{Y}\big\} \\ \lb{A64}
\Big[\big\{\hat{Y}\big\}\;,\;\big\{\hat{Y}\big\}\Big\}&=&\big\{\hat{H}\big\}\;\;\;.
\eeq
Applying the general considerations (\ref{A47}-\ref{A53}) and (\ref{A54}-\ref{A64}),
we conclude for the variations \(\hat{T}_{0}^{-1}\;d\hat{T}_{0}\) and
\(\hat{Q}^{-1}\;d\hat{Q}\) (\ref{A65}) the relations (\ref{A66},\ref{A67}).
In consequence of Eq. (\ref{A53}) for a general Lie group, we acquire for the variation
\(\hat{Q}^{-1}\;d\hat{Q}\) (\ref{A65},\ref{A66}) in the subgroup \(U(L|S)\) the sum of
super-unitary generators \(\hat{H}^{(\kappa)}\) \(u(L|S)\), each multiplied with
corresponding functions \(df_{\kappa}(q_{\lambda})\) (in place of \(db_{\kappa}(a_{\lambda})\)
in Eqs. (\ref{A52},\ref{A53})). According to the commutation relations
(\ref{A63},\ref{A64}), the variation \(\hat{T}^{-1}\;d\hat{T}\) consists of coset algebra elements
\(\{\hat{Y}^{(\kappa)}\}\) with functions $ds_{\kappa}(y_{\lambda})$,
which have to be calculated, and also subalgebra generators
\(\{\hat{H}^{(\kappa)}\}\) of \(u(L|S)\) with corresponding unspecified increments $dg_{\kappa}(y_{\lambda})$
(\(\hat{T}_{0;2N\times 2N}=\hat{T}_{2N\times 2N}\;\hat{Q}_{2N\times 2N}^{-1}\))
\beq \lb{A65}
\hat{T}_{0}^{-1}\;\;d\hat{T}_{0}&=&
\hat{Q}\;\Big(\hat{T}^{-1}\;\;d\hat{T}-\hat{Q}^{-1}\;\;d\hat{Q}\Big)\;\hat{Q}^{-1} \\ \lb{A66}
\hat{Q}^{-1}\;d\hat{Q}&=&\im\;df_{\kappa}(\hat{\mcal{B}}_{D},\hat{\mcal{F}}_{D},
\hat{\omega}_{D},\hat{\omega}_{D}^{+})\;\;\;\hat{H}^{(\kappa)} \\ \no
\hat{T}^{-1}\;\;d\hat{T}&=&\im\;dg_{\kappa}(\hat{c}_{D},\hat{f}_{D},
\hat{c}_{D}^{+},\hat{f}_{D}^{+},\eta_{D},\eta_{D}^{+})\;\;\;\hat{H}^{(\kappa)} -
ds_{\kappa}(\hat{c}_{D},\hat{f}_{D},\hat{c}_{D}^{+},\hat{f}_{D}^{+},\eta_{D},\eta_{D}^{+})\;\;\hat{Y}^{(\kappa)}
\\ \lb{A67}
\hat{T}_{0}^{-1}\;d\hat{T}_{0}&=&-ds_{\kappa}\;\;\hat{Q}\;\hat{Y}^{(\kappa)}\;\hat{Q}^{-1}
-\im\;\big(df_{\kappa}-dg_{\kappa}\big)\;\;\hat{Q}\;\hat{H}^{(\kappa)}\;\hat{Q}^{-1}\;.
\eeq
The structure of the supercommutators (\ref{A61}-\ref{A64}) with the super-unitary subalgebra elements
\(\{\hat{H}^{(\kappa)}\}\) of \(u(L|S)\) and the coset generators \(\{\hat{Y}^{(\kappa)}\}\) of
\(Osp(S,S|2L)\backslash U(L|S)\) is invariant under {\it local} subgroup transformations with the
block diagonal matrices \((\hat{P}\;\hat{Q}^{-1})_{2N\times 2N}\) and
\((\hat{P}\;\hat{Q}^{-1})_{2N\times 2N}^{-1}\) with their
doubled form of \(U(L|S)\), given by \(\hat{P}^{11}_{N\times N}\), \(\hat{P}^{22}_{N\times N}\)
(\ref{A26}-\ref{A40}) and \(\hat{Q}_{N\times N}^{11}\), \(\hat{Q}_{N\times N}^{22}\)
(\ref{A15}-\ref{A20}). We have to choose a {\it locally} invariant
version of a similarity transformation with a dependence on the parameters
\(p_{\kappa}=\big\{\mcal{C}_{D;L\times L},\mcal{G}_{D;S\times S};
\xi_{D;S\times L},\xi_{D;L\times S}^{+}\big\}\) (\ref{A57}) and
\(q_{\kappa}=\big\{\mcal{B}_{D;L\times L},\mcal{F}_{D;S\times
S};\omega_{D;S\times L},\omega_{D;L\times S}^{+}\big\}\) (\ref{A56}) so that the variations
\(\hat{T}_{0}^{-1}\;d\hat{T}_{0}\), \(\hat{Q}^{-1}\;d\hat{Q}\) and \(\hat{T}^{-1}\;d\hat{T}\)
(\ref{A65}-\ref{A67}) can be partially diagonalized. This implies a {\it local} similarity transformation
for the generators \(\hat{\Lambda}^{(\alpha)}\), \(\hat{H}^{(\kappa)}\) and \(\hat{Y}^{(\kappa)}\)
in (\ref{A65}-\ref{A67})
to the generators \(\wtilde{\Lambda}^{(\alpha)}(p_{\kappa})\),
\(\wtilde{H}^{(\kappa)}(p_{\kappa})\),
\(\wtilde{Y}^{(\kappa)}(p_{\kappa})\)
which leave {\it locally} (with a dependence on
$p_{\kappa}$) the structure of the supercommutators (\ref{A61}-\ref{A64}) invariant
\footnote{The tilde '$\wtilde{\ph{\Lambda}}$' above the generators
\(\wtilde{\Lambda}^{(\alpha)}(p_{\kappa})\),
\(\wtilde{H}^{(\kappa)}(p_{\kappa})\),
\(\wtilde{Y}^{(\kappa)}(p_{\kappa})\) denotes their local
dependence on the parameters \(p_{\kappa}\) (\ref{A57}).}
\beq\lb{A68}
\hat{Q}\;\hat{\Lambda}^{(\alpha)}\;\hat{Q}^{-1} &\rightarrow&
(\hat{P}\hat{Q}^{-1})\;
\hat{Q}\;\hat{\Lambda}^{(\alpha)}\;\hat{Q}^{-1}\;(\hat{P}\hat{Q}^{-1})^{-1}=
\wtilde{\Lambda}^{(\alpha)}(p_{\kappa})
=\hat{P}(p_{\kappa})\;\;\hat{\Lambda}^{(\alpha)}\;\;\hat{P}^{-1}(p_{\kappa}) \\ \lb{A69}
\hat{Q}\;\hat{H}^{(\kappa)}\;\hat{Q}^{-1} &\rightarrow&
(\hat{P}\hat{Q}^{-1})\;\hat{Q}\;\hat{H}^{(\kappa)}\;\hat{Q}^{-1}
\;(\hat{P}\hat{Q}^{-1})^{-1}=
\wtilde{H}^{(\kappa)}(p_{\kappa})
=\hat{P}(p_{\kappa})\;\;\hat{H}^{(\kappa)}\;\;\hat{P}^{-1}(p_{\kappa}) \\ \lb{A70}
\hat{Q}\;\hat{Y}^{(\kappa)}\;\hat{Q}^{-1}&\rightarrow&
(\hat{P}\hat{Q}^{-1})\;\hat{Q}\;\hat{Y}^{(\kappa)}\;\hat{Q}^{-1}
\;(\hat{P}\hat{Q}^{-1})^{-1}=\wtilde{Y}^{(\kappa)}(p_{\kappa})
=\hat{P}(p_{\kappa})\;\;\hat{Y}^{(\kappa)}\;\;\hat{P}^{-1}(p_{\kappa})  \\ \no
\hat{T}_{0}^{-1}\;d\hat{T}_{0}&\rightarrow&
(\hat{P}\hat{Q}^{-1})\;\hat{T}_{0}^{-1}\;d\hat{T}_{0}
\;(\hat{P}\hat{Q}^{-1})^{-1}=
\wtilde{T}_{0}^{\raisebox{-10pt}{$^{-1}$}}\;d\wtilde{T}_{0}=
\big(\hat{P}\hat{Q}^{-1}\big)\;\hat{T}_{0}^{-1}\;d\hat{T}_{0}\;\big(\hat{P}\hat{Q}^{-1}\big)^{-1}
= \\ \lb{A71} && \hspace*{2.8cm}=
\hat{P}\;\hat{T}^{-1}\;d\hat{T}\;\hat{P}^{-1}-\hat{P}\;\hat{Q}^{-1}\;d\hat{Q}\;\hat{P}^{-1}
\eeq
This invariance under the {\it local} transformations with \((\hat{P}\;\hat{Q}^{-1})\)
of the generators of \(Osp(S,S|2L)\) (\ref{A68}-\ref{A70}) also holds for the metric length
\((ds_{Osp})^{2}\) with the super-trace in (\ref{A46}) where the variations
\(\hat{T}_{0}^{-1}\;d\hat{T}_{0}\), \(d(\delta\hat{\Lambda})\) in \((ds_{Osp})^{2}\) are changed
to a sum of terms for the infinitesimal increments \(df_{\kappa}(q_{\lambda})\), \(dg_{\kappa}(y_{\lambda})\),
\(ds_{\kappa}(y_{\lambda})\)
as in (\ref{A65}-\ref{A67}) , but with a {\it local} dependence of the generators
\(\wtilde{\Lambda}^{(\alpha)}(p_{\kappa})\),
\(\wtilde{H}^{(\kappa)}(p_{\kappa})\),
\(\wtilde{Y}^{(\kappa)}(p_{\kappa})\) (\ref{A68}-\ref{A70})
\beq \lb{A72}
(d s_{Osp})^{2}&=&
\mbox{STR}\bigg[\bigg(\big[\hat{T}_{0}^{-1}\;d\hat{T}_{0}\;,\;
\delta\hat{\Lambda}\big]+
d\big(\delta\hat{\Lambda}\big)\bigg)^{2}\bigg] \\ \no &=&
\mbox{STR}\bigg[\bigg(\big[\wtilde{T}_{0}^{-1}\;d\wtilde{T}_{0}\;,\;
\delta\wtilde{\Lambda}\big]+
d\big(\delta\wtilde{\Lambda}\big)\bigg)^{2}\bigg] \\ \no &=&
2\mbox{str}\Big[d\big(\delta\wtilde{\lambda}_{\alpha}\big)\;
d\big(\delta\wtilde{\lambda}_{\alpha}\big)\Big]-2\;\;\strab
\Big[\Big(\wtilde{T}_{0}^{-1}\;d\wtilde{T}_{0}\Big)_{\alpha\beta}^{11}\;\;
\Big(\wtilde{T}_{0}^{-1}\;d\wtilde{T}_{0}\Big)_{\beta\alpha}^{11}\;\;
\Big(\delta\wtilde{\lambda}_{\beta}-\delta\wtilde{\lambda}_{\alpha}\Big)^{2}\Big] \\ \no &-&
2\;\;\strab\Big[\Big(\wtilde{T}_{0}^{-1}\;d\wtilde{T}_{0}\Big)_{\alpha\beta}^{12}\;\;
\Big(\wtilde{T}_{0}^{-1}\;d\wtilde{T}_{0}\Big)_{\beta\alpha}^{21}\;\;
\Big(\delta\wtilde{\lambda}_{\beta}+\delta\wtilde{\lambda}_{\alpha}\Big)^{2}\Big] \\ \lb{A73}
d\big(\delta\wtilde{\lambda}_{\alpha}\big) &=&\underbrace{
\hat{P}^{11}(p_{\kappa})\;\hat{\lambda}^{(\alpha)}\;\hat{P}^{11,-1}(p_{\kappa})}_{
\wtilde{\lambda}^{(\alpha)}(p_{\kappa})}\;\;\;
d\big(\delta\lambda_{\alpha}\big)
\eeq
The invaraince of \((ds_{Osp})^{2}\) (\ref{A72}) under {\it local} subgroup transformations \(U(L|S)\)
with \((\hat{P}\;\hat{Q}^{-1})\) involves a transformation of the variation
\(\hat{T}_{0}^{-1}\;d\hat{T}_{0}\) to
\(\wtilde{T}_{0}^{-1}\;d\wtilde{T}_{0}\) (\ref{A72}). The generators in (\ref{A66},\ref{A67})
are converted to their local transforms (\ref{A68}-\ref{A70}). Furthermore, one can notice a separation
of the metric \((ds_{Osp})^{2}\) (\ref{A72},\ref{A71}) into coset elements with functions \(ds_{\kappa}(y_{\lambda})\)
and corresponding generators \(\wtilde{Y}^{(\kappa)}(p_{\kappa})\)
and super-unitary subgroup elements with shifted functions \(d\wtilde{f}_{\kappa}(q_{\lambda})\) (\ref{A74}).
This follows because the part of the variation \(\hat{P}\;\hat{T}^{-1}\;d\hat{T}\;\hat{P}^{-1}\)
with the {\it local} \(u(L|S)\) subalgebra generators
\(\wtilde{H}^{(\kappa)}(p_{\kappa})\) and functions \(dg_{\kappa}(y_{\lambda})\)
is absorbed by the variation of the density terms \(\hat{P}\;\hat{Q}^{-1}\;d\hat{Q}\;\hat{P}^{-1}\) in $U(L|S)$
with local increments \(d\wtilde{f}_{\kappa}(q_{\lambda})=df_{\kappa}(q_{\lambda})-dg_{\kappa}(y_{\lambda})\)
\beq \lb{A74}
\hat{T}_{0}^{-1}\;\;d\hat{T}_{0}&\rightarrow&\big(\hat{P}\;\hat{Q}^{-1}\big)\;(\hat{T}_{0}^{-1}\;d\hat{T}_{0})\;
\big(\hat{P}\;\hat{Q}^{-1}\big)^{-1}=\wtilde{T}_{0}^{-1}\;d\wtilde{T}_{0}=
\\ \no
&=&\hat{P}\;\;\hat{T}^{-1}\;d\hat{T}\;\;\hat{P}^{-1}-\hat{P}\;\hat{Q}^{-1}\;d\hat{Q}\;\hat{P}^{-1}=
\\ \no &=&-ds_{\kappa}(\hat{c}_{D},\hat{f}_{D},\hat{c}_{D}^{+},
\hat{f}_{D}^{+},\eta_{D},\eta_{D}^{+})\;\;\wtilde{Y}^{(\kappa)}(p_{\lambda})
\\ \no &-&\im\;\;\underbrace{\Big(df_{\kappa}(\hat{\mcal{B}}_{D},
\hat{\mcal{F}}_{D},\hat{\omega}_{D},\hat{\omega}_{D}^{+})-
dg_{\kappa}(\hat{c}_{D},\hat{f}_{D},\hat{c}_{D}^{+},\hat{f}_{D}^{+},\eta_{D},\eta_{D}^{+})\Big)}_{
d\wtilde{f}_{\kappa}(\hat{\mcal{B}}_{D},\hat{\mcal{F}}_{D},\hat{\omega}_{D},\hat{\omega}_{D}^{+})}\;\;
\wtilde{H}^{(\kappa)}(p_{\lambda}).
\eeq
Since the super-trace of the product of a coset generator
\(\wtilde{Y}^{(\mu)}(p_{\kappa})\) and a subalgebra element
\(\wtilde{H}^{(\lambda)}(p_{\kappa})\) vanishes
for all possible pairs \(\mu,\;\lambda\) of indices, the
integration measure splits into a part for the anomalous terms, representing the coset space
\(Osp(S,S|2L)\backslash U(L|S)\), and a part for the density terms with the super-unitary subgroup
\(U(L|S)\). In subsection \ref{sa3} we finally calculate the coset measure for the anomalous terms
with the relation
\((\wtilde{T}_{0}^{-1}\;d\wtilde{T}_{0})^{ab}=
(\hat{P}\;\hat{T}^{-1}\;d\hat{T}\;\hat{P}^{-1})^{ab}\), (\(a\neq b\), \(a,b=1,2\))
by applying the similarity transformation with the eigenvectors
\(\hat{P}_{2N\times 2N}\), \(\hat{P}_{2N\times 2N}^{-1}\)
of the generators \(\hat{Y}_{2N\times 2N}\), \(\hat{X}_{N\times N}\),
\(\wtilde{\kappa}\;\hat{X}_{N\times N}^{+}\) in
\(\hat{T}_{2N\times 2N}=\exp\{-\hat{Y}_{2N\times 2N}\}\). The integration measure for the
density terms with the \(U(L|S)\) group can be obtained separately because
the variations \(\hat{P}\;\hat{Q}^{-1}\;d\hat{Q}\;\hat{P}^{-1}\) of the density terms
absorb the part of the variations \(\hat{P}\;\hat{T}^{-1}\;d\hat{T}\;\hat{P}^{-1}\)
within the super-unitary subgroup (\(d\wtilde{f}_{\kappa}=df_{\kappa}-dg_{\kappa}\) ,
(\ref{A67}-\ref{A74})).

\subsection{Calculation of $\hat{P}\;\hat{T}^{-1}\;d\hat{T}\;\hat{P}^{-1}$ with diagonal generator
$\hat{Y}=\hat{P}^{-1}\;\hat{Y}_{DD}\;\hat{P}$} \lb{sa3}

Subsections \ref{sa3}, \ref{sa4} are composed of various intermediate steps for calculating the
integration measures of the coset decomposition \(Osp(S,S|2L)\backslash U(L|S) \otimes U(L|S)\). However,
we do not describe all these in detail and just give a short overview.

The first aim is to compute Eq. (\ref{A80})
where the underlying integral relation with parameter \(v\in[0,1]\) has already
been proved in (\ref{s1_17}-\ref{s1_24}). We diagonalize the coset matrix \(\hat{Y}(\vec{x},t_{p})\)
with \(\hat{P}(\vec{x},t_{p})\) to \(\hat{Y}_{DD}(\vec{x},t_{p})\) and introduce rotated
infinitesimal increments \(d\hat{Y}\ppr(\vec{x},t_{p})=\hat{P}(\vec{x},t_{p})\;
d\hat{Y}(\vec{x},t_{p})\;\hat{P}^{-1}(\vec{x},t_{p})\) which have the same properties as
integration variables as the unrotated infinitesimal matrix \(d\hat{Y}(\vec{x},t_{p})\)
(\ref{A84}-\ref{A86}). The symmetries between the matrix elements of
\(d\hat{Y}\ppr(\vec{x},t_{p})\) are the same as for \(d\hat{Y}(\vec{x},t_{p})\)
and are not affected by the rotations with \(\hat{P}(\vec{x},t_{p})\)
(\ref{A87}-\ref{A90})\footnote{A prime "'" of the parameters
\(d\hat{c}_{D,L\times L}\ppr\) , \(d\hat{f}_{D,S\times S}\ppr\) , \(d\hat{\eta}_{D,S\times L}\ppr\)
denotes the corresponding independent entries of
\(d\hat{Y}\ppr(\vec{x},t_{p})=\hat{P}(\vec{x},t_{p})\;d\hat{Y}(\vec{x},t_{p})\;
\hat{P}^{-1}(\vec{x},t_{p})\) with rotations \(\hat{P}\:,\:\hat{P}^{-1}\).}.

We determine the invariant length \(\big(ds_{Osp}\big)^{2}\) and find for the chosen
parametrization a separation into coset and subgroup integration measures according to
the corresponding separation of the invariant length (\ref{A91}-\ref{A95}). The invariant length
is computed for the coset integration measure with \(\big(ds_{Osp\backslash U}(\delta\hat{\lambda})\big)^{2}\)
and diagonal coset generator \(\hat{Y}_{DD}(\vec{x},t_{p})\) (\ref{A97}-\ref{A105}).
One has to calculate the exponentials \(\exp\{\pm v\;\hat{Y}_{DD}\}\) of \(\hat{Y}_{DD}(\vec{x},t_{p})\)
for the application in formula (\ref{A98}),(\ref{s1_17}-\ref{s1_24}). According to the block diagonal
form of \(\hat{Y}_{DD}(\vec{x},t_{p})\), this computation is possible and we list the various steps
in Eqs. (\ref{A106}-\ref{A118}).

Relation (\ref{A119}) describes the separation of the invariant length
\(\big(ds_{Osp\backslash U}(\delta\hat{\lambda})\big)^{2}\) into boson-boson, fermion-fermion,
boson-fermion and fermion-boson parts. Using the exponentials of \(\hat{Y}_{DD}(\vec{x},t_{p})\)
(\ref{A106}-\ref{A118}), the fundamental integral relations
\(\big(d\wtilde{Y}\big)_{\alpha\beta}^{\prime\: a\neq b}=\int_{0}^{1}dv\;\big(
\exp\{v\;\hat{Y}_{DD}\}\;d\hat{Y}\ppr\;\exp\{-v\;\hat{Y}_{DD}\}\big)_{\alpha\beta}^{a\neq b}\)
are finally determined and inserted into (\ref{A119}). Relations (\ref{A120}-\ref{A136}) contain the various steps
for reducing the boson-boson 'BB', the fermion-fermion 'FF', boson-fermion 'BF' and fermion-boson 'FB' sections
with the integral over \(v\in[0,1]\). According to (\ref{A119},\ref{A137},\ref{A141},\ref{A150},
\ref{A168}), we can finally compute the square root and inverse square root (for the odd variables)
of the diagonalized metric tensor \(\hat{G}_{Osp\backslash U}\) with the eigenvalues of
\(\hat{X}_{DD}(\vec{x},t_{p})\), \(\wtilde{\kappa}\;\hat{X}_{DD}^{+}(\vec{x},t_{p})\).
\vspace*{0.46cm}

In relations (\ref{A75}-\ref{A79}) we describe again the structure of
\(\big(\wtilde{T}_{0}^{-1}\;d\wtilde{T}_{0}\big)^{ab}\) (\ref{A75}) where we have already
shifted the dependence in the \(U(L|S)\) subgroup parts of the anomalous fields to the
density terms \(d\wtilde{f}_{\kappa}^{(aa)}(\hat{\mcal{B}}_{D},\hat{\mcal{F}}_{D},
\hat{\omega}_{D},\hat{\omega}_{D}^{+})\) (Compare with section \ref{sa2} and Eqs. (\ref{A65})-(\ref{A74})).
Therefore, we can restrict in this sub-appendix \ref{sa3} to the computation of the coset integration
measure for the anomalous fields
\beq \lb{A75}
\big(\wtilde{T}_{0}^{-1}\;d\wtilde{T}_{0}\big)^{ab}&=&\left( \bea{cc}
\big(\wtilde{T}_{0}^{-1}\;d\wtilde{T}_{0}\big)^{11} &
\big(\wtilde{T}_{0}^{-1}\;d\wtilde{T}_{0}\big)^{12}
\\ \big(\wtilde{T}_{0}^{-1}\;d\wtilde{T}_{0}\big)^{21} &
\big(\wtilde{T}_{0}^{-1}\;d\wtilde{T}_{0}\big)^{22}
\eea\right) \\   \lb{A76}
\big(\wtilde{T}_{0}^{-1}\;d\wtilde{T}_{0}\big)^{aa} &=&-\im\;\;
d\wtilde{f}_{\kappa}^{(aa)}(\hat{\mcal{B}}_{D},\hat{\mcal{F}}_{D},
\hat{\omega}_{D},\hat{\omega}_{D}^{+})\;\;
\big(\wtilde{H}^{(\kappa)}\big)^{aa}\hspace*{0.75cm} a=1,2 \\  \lb{A77}
\big(\wtilde{T}_{0}^{-1}\;d\wtilde{T}_{0}\big)^{12} &=&
-ds_{\kappa}^{(12)}(\hat{c}_{D},\hat{f}_{D},\hat{c}_{D}^{+},\hat{f}_{D}^{+},\eta_{D},\eta_{D}^{+})\;\;
\big(\wtilde{Y}^{(\kappa)}\big)^{12} \\  \lb{A78}
\big(\wtilde{T}_{0}^{-1}\;d\wtilde{T}_{0}\big)^{21} &=&
-ds_{\kappa}^{(21)}(\hat{c}_{D},\hat{f}_{D},\hat{c}_{D}^{+},\hat{f}_{D}^{+},\eta_{D},\eta_{D}^{+})\;\;
\big(\wtilde{Y}^{(\kappa)}\big)^{21}   \\   \lb{A79}
\wtilde{T}_{0}^{-1}\;\;d\wtilde{T}_{0}&=&\hat{P}\;\;\hat{T}^{-1}\;d\hat{T}\;\;\hat{P}^{-1}-
\hat{P}\;\;\hat{Q}^{-1}\;d\hat{Q}\;\;\hat{P}^{-1} \;\;\;.
\eeq
Relation (\ref{A80}) is of fundamental importance because it relates the infinitesimal increment
of an exponential of a super-matrix to its infinitesimal increment of the generator
\beq \lb{A80}
\hat{T}^{-1}\;\;d\hat{T}&=&\exp\{\hat{Y}\}\;\;d\big(\exp\{-\hat{Y}\}\big) =
-\int_{0}^{1}d v\;\;\exp\{v\;\hat{Y}\}\;\;d\hat{Y}\;\;\exp\{-v\;\hat{Y}\}
\\  \lb{A81} d\hat{Y}_{2N\times 2N}&=&\left(
\bea{cc}
0 & d\hat{X} \\
\wtilde{\kappa}\;d\hat{X}^{+} & 0 \eea\right) \hspace*{0.75cm} d\hat{X}_{N\times N}=\left( \bea{cc}
-d\hat{c}_{D,L\times L} & d\hat{\eta}_{D,L\times S}^{T} \\
-d\hat{\eta}_{D,S\times L} & d\hat{f}_{D,S\times S} \eea\right)  \\ \lb{A82} &&
d\hat{c}_{D,L\times L}^{T}=d\hat{c}_{D,L\times L}\hspace*{1.0cm}
d\hat{f}_{D,S\times S}^{T}=-d\hat{f}_{D,S\times S} \;\;\;.
\eeq
Although relation (\ref{A80}) already simplifies the infinitesimal increment of an exponential of
a matrix, one cannot even calculate the exponential of a matrix, completely occupied in all its
entries. Therefore, we transform the coset generator \(\hat{Y}(\vec{x},t_{p})\) to the
diagonal form \(\hat{Y}_{DD}(\vec{x},t_{p})\)
((\ref{A38}-\ref{A41}), \(\hat{Y}=\hat{P}^{-1}\;\hat{Y}_{DD}\;\hat{P}\))
and apply the matrices \(\hat{P}\), \(\hat{P}^{-1}\) to diagonalize
\(\hat{T}^{-1}\;d\hat{T}\)
\beq\no
\hat{P}\;\hat{T}^{-1}\;\;d\hat{T}\;\hat{P}^{-1}&=&\hat{P}\;\exp\{\hat{Y}\}\;\;
d\big(\exp\{-\hat{Y}\}\big)\;\hat{P}^{-1}
=-\hat{P}\;\Big(\int_{0}^{1}d v\;\;\exp\{v\;\hat{Y}\}\;\;
d\hat{Y}\;\;\exp\{-v\;\hat{Y}\}\Big)\;\hat{P}^{-1}
\\ \lb{A83} &=&-\hat{P}\;\Big(\int_{0}^{1}d v\;\;\hat{P}^{-1}\;\exp\{v\;\hat{Y}_{DD}\}\;\;
\underbrace{\hat{P}\;d\hat{Y}\;\hat{P}^{-1}}_{d\hat{Y}\ppr}\;\;\exp\{-v\;\hat{Y}_{DD}\}\;
\hat{P}\Big)\;\hat{P}^{-1}
\\ \no &=&  -\int_{0}^{1}d v\;\;\exp\{v\;\hat{Y}_{DD}\}\;\;d\hat{Y}\ppr\;\;\exp\{-v\;\hat{Y}_{DD}\}
=  -\int_{0}^{1}d v\;\;\exp\Big\{v\;[\hat{Y}_{DD}\;,\;\;\ldots]\Big\}\;\;d\hat{Y}\ppr
\\ \no
&=&-\left(\frac{\exp\Big\{[\hat{Y}_{DD}\;,\;\;\ldots]\Big\}-\hat{1}}{\Big[\hat{Y}_{DD}\;,\;\;\ldots\Big]}
\right)_{2N\times 2N}\;\;\;d\hat{Y}\ppr_{2N\times 2N}\;.
\eeq
However, the transformation with \(\hat{P}\), \(\hat{P}^{-1}\) also acts onto the independent elements
of \(d\hat{Y}\). Since the transformation of \(d\hat{Y}\) to \(d\hat{Y}\ppr=\hat{P}\;d\hat{Y}\;\hat{P}^{-1}\)
does not change the symmetry relations (\ref{A87}-\ref{A90})between the transformed entries \(d\hat{c}_{D;mn}\ppr\),
\(d\hat{f}_{D;r\mu,r\ppr\nu}\ppr\), \(d\hat{\eta}_{D;r\mu,n}\ppr\), due to the group properties, and since
this transformation retains the invariance of the super-trace (\ref{A86}), one can determine the coset
integration measure in terms of the rotated parameters of \(d\hat{Y}\ppr\) (\ref{A84},\ref{A85})
\beq \lb{A84}
d\hat{Y}\ppr&=&\hat{P}\;d\hat{Y}\;\hat{P}^{-1}= \left( \bea{cc}
0 & d\hat{X}\ppr \\
\wtilde{\kappa}\;d\hat{X}^{\prime +} & 0 \eea\right)= \\ \no &=& \left( \bea{cc}
0 & \hat{P}^{11}\;d\hat{X}\;\hat{P}^{22,-1} \\
\wtilde{\kappa}\;\big(\wtilde{\kappa}\;\hat{P}^{22}\;\wtilde{\kappa}\big)\;d\hat{X}^{+}\;\hat{P}^{11,-1}
& 0
\eea\right) \\   \lb{A85}
d\hat{X}\ppr&=&\left( \bea{cc}
-d\hat{c}\ppr_{D;mn} & d\hat{\eta}\ppr_{D;m,r\ppr\nu} \\
-d\hat{\eta}\ppr_{D;r\mu,n} & d\hat{f}\ppr_{D;r\mu,r\ppr\nu}
\eea\right) \\   \lb{A86}
\mbox{STR}\Big[d\hat{Y}\ppr\;\;d\hat{Y}\ppr\Big]&=&
\mbox{STR}\Big[\hat{P}\;d\hat{Y}\;\hat{P}^{-1}\;\;\hat{P}\;d\hat{Y}\;\hat{P}^{-1}\Big]=
\mbox{STR}\Big[d\hat{Y}\;\;d\hat{Y}\Big]    \\  \lb{A87}
d\hat{c}_{D,L\times L}^{\prime T}=d\hat{c}_{D,L\times L}\ppr
&\Longrightarrow&d\hat{c}\ppr_{D;mn}=d\hat{c}\ppr_{D;nm}\hspace*{0.75cm}m,n=1,\ldots,L \\  \lb{A88}
d\hat{f}\ppr_{D;r\mu,r\ppr\nu}&=&\sum_{k=0}^{3}\big(\tau_{k}\big)_{\mu\nu}\;d\hat{f}_{D;rr\ppr}^{\prime(k)}
\\ \no &&
\tau_{0}=\hat{1}_{2\times 2}\;;\;\tau_{1}\;,\;\tau_{2}\;,\;\tau_{3}\;=\mbox{Pauli-matrices} \\ \lb{A89}
d\hat{f}_{D,S\times S}^{\prime T}=-d\hat{f}\ppr_{D,S\times S} &\Longrightarrow&
d\hat{f}_{D;rr\ppr}^{\prime(k)}=-d\hat{f}_{D;r\ppr r}^{\prime(k)}\hspace*{0.5cm}\mbox{for }k=0,1,3 \\ \lb{A90}
d\hat{f}_{D;rr\ppr}^{\prime(2)}&=&d\hat{f}_{D;r\ppr r}^{\prime (2)}
\hspace*{0.75cm}r,r\ppr=1,\ldots,S/2\hspace*{0.5cm} \mu,\nu=1,2
\\ \no &&\mbox{since }\;\big(\tau_{2}\big)^{T}=-\tau_{2}\;\mbox{and }\;
\big(\tau_{k}\big)^{T}=\tau_{k}\;\mbox{for }k=0,1,3 \;.
\eeq
The invariant length \((ds_{Osp})^{2}\) (\ref{A91}) separates into density or the subgroup part
\((ds_{U})^{2}\) (\ref{A93}) and the coset part
\(\big(d s_{Osp\backslash U}\big(\delta\hat{\lambda}\big)\big)^{2}\) (\ref{A94}).
However, we have to point out that the coset part
\(\big(d s_{Osp\backslash U}\big(\delta\hat{\lambda}\big)\big)^{2}\) (\ref{A94})
for the anomalous fields also contains the eigenvalues
\(\big(\delta\wtilde{\lambda}_{\beta}+\delta\wtilde{\lambda}_{\alpha}\big)^{2}\) of the
density or subgroup part. This is caused by the transformation (\ref{A42}-\ref{A46})
so that the final integration measure for \(\big(d s_{Osp\backslash U}\big(\delta\hat{\lambda}\big)\big)^{2}\)
contains apart from the anomalous fields a polynomial of the eigenvalues \(\delta\hat{\lambda}_{\alpha}\)
of the density terms. However, this polynomial factorizes from the coset integration measure and can be moved
to the integration measure of density terms, following from \(\big(d s_{U}\big)^{2}\) (\ref{A95}), thereby
modifying the eigenvalue integration measure in the \(U(L|S)\) subgroup part
\beq \lb{A91}
\Big(d s_{Osp}\Big)^{2}&=&
\mbox{STR}\bigg[\bigg(\big[\wtilde{T}_{0}^{-1}\;d\wtilde{T}_{0}\;,\;
\delta\wtilde{\Lambda}\big]+
d\big(\delta\wtilde{\Lambda}\big)\bigg)^{2}\bigg] =
\Big(d s_{U}\Big)^{2}+\Big(d s_{Osp\backslash U}\big(\delta\hat{\lambda}\big)\Big)^{2} \\ \lb{A92}
d\big(\delta\wtilde{\Lambda}\big)&=&\wtilde{\Lambda}^{(\alpha)}
(p_{\kappa},q_{\kappa})\;\;
d\big(\delta\lambda_{\alpha}\big) \\  \lb{A93}
\Big(d s_{U}\Big)^{2}&=&
2\;\mbox{str}\Big[d\big(\delta\wtilde{\lambda}_{\alpha}\big)\;\;
d\big(\delta\wtilde{\lambda}_{\alpha}\big)\Big] -
2\;\strab\Big[\big(\wtilde{T}_{0}^{-1}\;d\wtilde{T}_{0}\big)_{\alpha\beta}^{11}\;\;
\big(\wtilde{T}_{0}^{-1}\;d\wtilde{T}_{0}\big)_{\beta\alpha}^{11}\;\;
\big(\delta\wtilde{\lambda}_{\beta}-\delta\wtilde{\lambda}_{\alpha}\big)^{2}\Big] \\  \lb{A94}
\Big(d s_{Osp\backslash U}\big(\delta\hat{\lambda}\big)\Big)^{2}&=&-2\;
\strab\Big[\big(\wtilde{T}_{0}^{-1}\;d\wtilde{T}_{0}\big)_{\alpha\beta}^{12}\;\;
\big(\wtilde{T}_{0}^{-1}\;d\wtilde{T}_{0}\big)_{\beta\alpha}^{21}\,\;
\big(\delta\wtilde{\lambda}_{\beta}+\delta\wtilde{\lambda}_{\alpha}\big)^{2}\Big]  \\  \lb{A95}
\Big(d s_{U}\Big)^{2}&=&2\;
\strab\bigg[d\Big(\hat{Q}_{N\times N}^{11,-1}\;\delta\hat{\lambda}_{N\times N}\;
\hat{Q}_{N\times N}^{11}\Big)_{\alpha\beta}\;\;
d\Big(\hat{Q}_{N\times N}^{11,-1}\;\delta\hat{\lambda}_{N\times N}\;\;
\hat{Q}_{N\times N}^{11}\Big)_{\beta\alpha}\bigg]
\\ \no &=&2\;\strab\bigg[d\Big(\delta\hat{\Sigma}_{D;N\times N}^{11}\Big)_{\alpha\beta}\;\;
d\Big(\delta\hat{\Sigma}_{D;N\times N}^{11}\Big)_{\beta\alpha}\bigg] \\ \no &=&
\STRAB\bigg[d\Big(\delta\hat{\Sigma}_{D;2N\times 2N}\;\wtilde{K}\Big)_{\alpha\beta}^{ab}\;\;
d\Big(\delta\hat{\Sigma}_{D;2N\times 2N}\;\wtilde{K}\Big)_{\beta\alpha}^{ba}\bigg]\;.
\eeq
The super-group \(Osp(S,S|2L)\) incorporates the subgroup \(U(L|S)\) with the same parameters
in a double form so that one can relate the increments \(d\hat{Q}^{11}\;\hat{Q}^{11,-1}\)
in the '11' block to the increments in the '22' block. The complete invariant length
\(\big(d s_{U}\big)^{2}\) of the \(U(L|S)\) subgroup can therefore be obtained from the
super-trace with \(N\times N\)-matrices of the '11' block with additional factor of 'two'
for consideration of the '22' block
\beq \lb{A96}
d\hat{Q}_{N\times N}^{11}\;\;\hat{Q}_{N\times N}^{11,-1}&=&
\Big(d\hat{Q}_{N\times N}^{22,-1}\Big)^{st}\;\;\Big(\hat{Q}_{N\times N}^{22}\Big)^{st} =
-\Big(\hat{Q}_{N\times N}^{22,-1}\;\;d\hat{Q}_{N\times N}^{22}\;\;\hat{Q}_{N\times N}^{22,-1}\Big)^{st}\;\;
\Big(\hat{Q}_{N\times N}^{22}\Big)^{st} \\ \no &=&
-\Big(\hat{Q}_{N\times N}^{22,-1}\Big)^{st}\;\;\Big(d\hat{Q}_{N\times N}^{22}\Big)^{st}\;\;\;.
\eeq
It remains to determine the exponential \(\exp\{v\;\hat{Y}_{DD}\}\) with diagonal coset generator
\(\hat{Y}_{DD}\) in (\ref{A83}) and to compute the integrations over \(dv\) (\(v\in[0,1]\))
in the fundamental relation (\ref{A83}) for the increments of the exponential of matrices.
The final result depends on the rotated increments \(d\hat{Y}\ppr\) (\ref{A84}-\ref{A90})
and the eigenvalues \(\ovv{c}_{m}\), \(\ovv{f}_{r}\) and is abbreviated by
\((d\wtilde{Y})^{\prime ab}_{\alpha\beta}\) (\ref{A98}). This quantity (\ref{A98})
directly enters into the invariant coset length
\(\big(d s_{Osp\backslash U}\big(\delta\hat{\lambda}\big)\big)^{2}\) (\ref{A99})
with its off-diagonal block parts \(a\neq b\) and separates the invariant length
\(\big(d s_{Osp\backslash U}\big(\delta\hat{\lambda}\big)\big)^{2}\) with
sub-metric tensors of the eigenvalues into quadratic parts with the rotated parameters
of \(d\hat{Y}\ppr\)
\beq \lb{A97}
\Big(\wtilde{T}_{0}^{-1}\;\;d\wtilde{T}_{0}\Big)_{\alpha\beta}^{ab}&=&
\Big(\hat{P}\;\;\hat{T}^{-1}\;d\hat{T}\;\;\hat{P}^{-1}\Big)_{\alpha\beta}^{ab}-
\Big(\hat{P}\;\;\hat{Q}^{-1}\;d\hat{Q}\;\;\hat{P}^{-1}\Big)_{\alpha\beta}^{ab} \\  \lb{A98}
\Big(\hat{P}\;\hat{T}^{-1}\;d\hat{T}\;\hat{P}^{-1}\Big)_{\alpha\beta}^{ab}&=&
\Big(\hat{P}\;\exp\{\hat{Y}\}\;d\big(\exp\{-\hat{Y}\}\big)\;\hat{P}^{-1}\Big)_{\alpha\beta}^{ab}
\\ \no &=& -
\int_{0}^{1}d v\;\;\Big(\exp\{v\;\hat{Y}_{DD}\}\;d\hat{Y}\ppr\;
\exp\{-v\;\hat{Y}_{DD}\}\Big)_{\alpha\beta}^{ab}  =
-\Big(d\wtilde{Y}\Big)^{\prime ab}_{\alpha\beta}  \hspace*{1.9cm}(a\neq b)   \\  \lb{A99}
\Big(d s_{Osp\backslash U}\big(\delta\hat{\lambda}\big)\Big)^{2}&=&-2\;
\strab\bigg[\Big(d\wtilde{Y}\Big)^{\prime 12}_{\alpha\beta}\;\;
\Big(d\wtilde{Y}\Big)^{\prime 21}_{\beta\alpha}\;\;
\Big(\delta\wtilde{\lambda}_{\beta}+\delta\wtilde{\lambda}_{\alpha}\Big)^{2}
\bigg]     \\    \lb{A100}
\hat{Y}_{DD}&=&\left( \bea{cc}
0 & \hat{X}_{DD} \\
\wtilde{\kappa}\;\hat{X}_{DD}^{+} & 0
\eea\right) \\ \lb{A101}
\hat{X}_{DD}&=&\left( \bea{cc}
-\ovv{c}_{m}\;\delta_{mn} & 0 \\
0 & \big(\tau_{2})_{\mu\nu}\;\ovv{f}_{r}\;\delta_{rr\ppr} \eea\right) \\ \lb{A102} &&
\ovv{c}_{m}=|\ovv{c}_{m}|\;\exp\{\im\;\varphi_{m}\}\hspace*{1.0cm}
\ovv{f}_{r}=|\ovv{f}_{r}|\;\exp\{\im\;\phi_{r}\} \\  \lb{A103}
&& m,n=1,\ldots,L\hspace*{1.0cm}r,r\ppr=1,\ldots,S/2\hspace*{0.5cm}\mu,\nu=1,2
\eeq
\beq  \lb{A104}
\Big\{\alpha=1,\ldots,N=L+S\Big\}\hspace*{-0.25cm}&=&\hspace*{-0.25cm}\Big\{
\underbrace{m=1,\ldots,L}_{bosonic};\underbrace{r=1(\mu=1,2),\ldots,r=S/2(\mu=1,2)}_{fermionic} \Big\} \\ \lb{A105}
\Big\{\beta=1,\ldots,N=L+S\Big\}\hspace*{-0.25cm}&=&\hspace*{-0.25cm}\Big\{
\underbrace{n=1,\ldots,L}_{bosonic};\underbrace{r\ppr=1(\nu=1,2),\ldots,r\ppr=S/2(\nu=1,2)}_{fermionic} \Big\}_{\mbox{.}}
\eeq
In relations (\ref{A106}-\ref{A109}) we list the expansion of the exponential
\(\exp\{v\;\hat{Y}_{DD}\}\) (\ref{A106}) with diagonal coset generator \(\hat{Y}_{DD}\) (\ref{A100}-\ref{A105})
for the block diagonal parts '11', '22' and especially for the off-diagonal coset parts (\ref{A108},\ref{A109}).
Note how these off-diagonal parts separate into the diagonal with sine-functions of the modulus of eigenvalues
for the bosonic molecular condensate matrix and into the quaternion diagonal \((\tau_{2})_{\mu\nu}\) with the
hyperbolic sine-functions of the absolute values of eigenvalues for the BCS condensate matrix of fermions
\beq \lb{A106}
\lefteqn{\left(\exp\Big\{v\;\hat{Y}_{DD}\Big\}\right)^{ab}=\left(\exp\left\{v\left( \bea{cc}
0 & \hat{X}_{DD} \\
\wtilde{\kappa}\;\hat{X}_{DD}^{+} & 0 \eea\right)\right\}\right)^{ab}=  } \\ \no &=& \left(\bea{cc}
\left( \cosh\Big(v\sqrt{\hat{X}_{DD}\;\wtilde{\kappa}\;\hat{X}_{DD}^{+}}\Big) \right)_{\alpha\beta}^{11} &
\left( \frac{{\ds\sinh\Big(v\sqrt{\hat{X}_{DD}\;\wtilde{\kappa}\;\hat{X}_{DD}^{+}}\Big)}}{{\ds\sqrt{
\hat{X}_{DD}\;\wtilde{\kappa}\;\hat{X}_{DD}^{+}}}}\; \hat{X}_{DD} \right)_{\alpha\beta}^{12} \\
\left( \wtilde{\kappa}\;\hat{X}_{DD}^{+}\;
\frac{{\ds\sinh\Big(v\sqrt{\hat{X}_{DD}\;\wtilde{\kappa}\;\hat{X}_{DD}^{+}}\Big)}}{{\ds
\sqrt{\hat{X}_{DD}\;\wtilde{\kappa}\;\hat{X}_{DD}^{+}}}}  \right)_{\alpha\beta}^{21}
& \left( \cosh\Big(v\sqrt{\wtilde{\kappa}\;\hat{X}_{DD}^{+}\;\hat{X}_{DD}}\Big) \right)_{\alpha\beta}^{22}
\eea\right)
\eeq
\beq  \lb{A107}
\left(\cosh\Big(v\sqrt{\hat{X}_{DD}\;\wtilde{\kappa}\;\hat{X}_{DD}^{+}}\Big)\right)_{\alpha\beta}^{11}&=&
\left(\cosh\Big(v\sqrt{\wtilde{\kappa}\;\hat{X}_{DD}^{+}\;\hat{X}_{DD}}\Big)\right)_{\alpha\beta}^{22}= \\ \no &=&
\left(\bea{cc}
\cos\Big(v\;|\ovv{c}_{m}|\Big)\;\;\delta_{mn} & 0 \\
0 & \cosh\Big(v\;|\ovv{f}_{r}|\Big)\;\delta_{rr\ppr}\;\delta_{\mu\nu} \eea\right)^{11}
\eeq
\beq \lb{A108}
\lefteqn{\left(\frac{{\ds\sinh\Big(v\sqrt{\hat{X}_{DD}\;
\wtilde{\kappa}\;\hat{X}_{DD}^{+}}\Big)}}{{\ds\sqrt{\hat{X}_{DD}\;
\wtilde{\kappa}\;\hat{X}_{DD}^{+}}}} \;\hat{X}_{DD}\right)_{\alpha\beta}^{12} =} \\ \no &=& \left(\bea{cc}
-\sin\Big(v\;|\ovv{c}_{m}|\Big)\;\;\exp\{\im\;\varphi_{m}\}\;\delta_{mn} & 0 \\
0 & \sinh\Big(v\;|\ovv{f}_{r}|\Big)\;\;\exp\{\im\;\phi_{r}\}\;
\delta_{rr\ppr}\;\big(\tau_{2}\big)_{\mu\nu} \eea\right)^{12}
\eeq
\beq \lb{A109}
\lefteqn{\left(\wtilde{\kappa}\;\hat{X}_{DD}^{+}\frac{{\ds
\sinh\Big(v\sqrt{\hat{X}_{DD}\;\wtilde{\kappa}\;\hat{X}_{DD}^{+}}\Big)}}{{\ds
\sqrt{\hat{X}_{DD}\;\wtilde{\kappa}\;\hat{X}_{DD}^{+}}}}\right)_{\alpha\beta}^{21} =}
\\ \no &=& \left(\bea{cc}
\sin\Big(v\;|\ovv{c}_{m}|\Big)\;\;\exp\{-\im\;\varphi_{m}\}\;\delta_{mn} & 0 \\
0 & \sinh\Big(v\;|\ovv{f}_{r}|\Big)\;\;\exp\{-\im\;\phi_{r}\}\;
\delta_{rr\ppr}\;\big(\tau_{2}\big)_{\mu\nu} \eea\right)^{21}\;\;\;.
\eeq
Finally, we compute in relations (\ref{A110},\ref{A111}) the integrand for (\ref{A112},\ref{A83},\ref{A80})
\beq \lb{A110}
\lefteqn{\Big(\exp\{v\;\hat{Y}_{DD}\}\;\;d\hat{Y}\ppr\;\;\exp\{-v\;\hat{Y}_{DD}\}\Big)^{12}_{\alpha\beta}= }
\\ \no &=& \Bigg(-{\ds
\frac{ \sinh\Big(v\sqrt{\hat{X}_{DD}\;\wtilde{\kappa}\;\hat{X}_{DD}^{+}}\Big)}{
\sqrt{\hat{X}_{DD}\;\wtilde{\kappa}\;\hat{X}_{DD}^{+}}} \;\hat{X}_{DD}\;\wtilde{\kappa}\;d\hat{X}^{\prime +}
\frac{\sinh\Big(v\sqrt{\hat{X}_{DD}\;\wtilde{\kappa}\;\hat{X}_{DD}^{+}}\Big)}{
\sqrt{\hat{X}_{DD}\;\wtilde{\kappa}\;\hat{X}_{DD}^{+}}} \;\hat{X}_{DD}  }+ \\ \no &+& {\ds
\cosh\Big(v\sqrt{\hat{X}_{DD}\;\wtilde{\kappa}\;\hat{X}_{DD}^{+}}\Big)\;d\hat{X}\ppr\;
\cosh\Big(v\sqrt{\wtilde{\kappa}\;\hat{X}_{DD}^{+}\;\hat{X}_{DD}}\Big)}\Bigg)_{\alpha\beta}
\eeq
\beq \lb{A111}
\lefteqn{\Big(\exp\{v\;\hat{Y}_{DD}\}\;\;d\hat{Y}\ppr\;\;\exp\{-v\;\hat{Y}_{DD}\}\Big)^{21}_{\alpha\beta}=
\wtilde{\kappa}\;
\Big(\exp\{v\;\hat{Y}_{DD}\}\;\;d\hat{Y}\ppr\;\;\exp\{-v\;\hat{Y}_{DD}\}\Big)^{12,\mathbf{+}}_{\alpha\beta}= }
\\ \no &=& \Bigg(-{\ds
\wtilde{\kappa}\;\hat{X}_{DD}^{+}\;\frac{
\sinh\Big(v\sqrt{\hat{X}_{DD}\;\wtilde{\kappa}\;\hat{X}_{DD}^{+}}\Big)}{
\sqrt{\hat{X}_{DD}\;\wtilde{\kappa}\;\hat{X}_{DD}^{+}}} \;d\hat{X}\ppr\;\wtilde{\kappa}\;\hat{X}_{DD}^{+}\;
\frac{\sinh\Big(v\sqrt{\hat{X}_{DD}\;\wtilde{\kappa}\;\hat{X}_{DD}^{+}}\Big)}{
\sqrt{\hat{X}_{DD}\;\wtilde{\kappa}\;\hat{X}_{DD}^{+}}}  } +\\ \no &+& {\ds
\cosh\Big(v\sqrt{\wtilde{\kappa}\;\hat{X}_{DD}^{+}\;\hat{X}_{DD}}\Big)\;\wtilde{\kappa}\;d\hat{X}^{\prime
+}\; \cosh\Big(v\sqrt{\hat{X}_{DD}\;\wtilde{\kappa}\;\hat{X}_{DD}^{+}}\Big)}\Bigg)_{\alpha\beta}
\eeq
\beq  \lb{A112}
\Big(d\wtilde{Y}\Big)^{\prime ab}&=&\int_{0}^{1}d v\;
\Big(\exp\{v\;\hat{Y}_{DD}\}\;\;d\hat{Y}\ppr\;\;\exp\{-v\;\hat{Y}_{DD}\}\Big)^{ab}_{\alpha\beta}
\hspace*{1.0cm}(a\neq b)
\\ \no &=& \left(\bea{cc}
\Big(d\wtilde{Y}\Big)^{\prime ab}_{BB;mn} & \Big(d\wtilde{Y}\Big)^{\prime ab}_{BF;m,r\ppr\nu} \\
\Big(d\wtilde{Y}\Big)^{\prime ab}_{FB;r\mu,n} & \Big(d\wtilde{Y}\Big)^{\prime ab}_{FF;r\mu,r\ppr\nu}
\eea\right) \;\;\;.
\eeq
We have to split the super-matrices in (\ref{A112}) into its boson-boson 'BB', fermion-fermion 'FF'
and fermion-boson 'FB', boson-fermion 'BF' parts. This is accomplished in (\ref{A113}-\ref{A118})
where the dependence on the integration variable \(dv\) becomes obvious in the trigonometric- and
hyperbolic-(co)sine functions with the modulus of the eigenvalues \(\ovv{c}_{m}\), \(\ovv{f}_{r}\)
\beq  \lb{A113}
\lefteqn{\Big(\exp\{v\;\hat{Y}_{DD}\}\;d\hat{Y}\ppr\;\exp\{-v\;\hat{Y}_{DD}\}\Big)_{\alpha\beta}^{12}=}
\\ \no \hspace*{-0.5cm}&\hspace*{-0.5cm}=&\hspace*{-0.5cm}
\left(\bea{cc}
\Big(\exp\{v\;\hat{Y}_{DD}\}\;d\hat{Y}\ppr\;\exp\{-v\;\hat{Y}_{DD}\}\Big)_{BB;mn}^{12}  & \hspace*{-0.5cm}
\Big(\exp\{v\;\hat{Y}_{DD}\}\;d\hat{Y}\ppr\;\exp\{-v\;\hat{Y}_{DD}\}\Big)_{BF;m,r\ppr \nu}^{12}  \\
\Big(\exp\{v\;\hat{Y}_{DD}\}\;d\hat{Y}\ppr\;\exp\{-v\;\hat{Y}_{DD}\}\Big)_{FB;r\mu,n}^{12}  & \hspace*{-0.5cm}
\Big(\exp\{v\;\hat{Y}_{DD}\}\;d\hat{Y}\ppr\;\exp\{-v\;\hat{Y}_{DD}\}\Big)_{FF;r\mu,r\ppr\nu}^{12}
\eea\right)_{\alpha\beta}^{12}
\eeq
\beq \lb{A114}
\lefteqn{\Big(\exp\{v\;\hat{Y}_{DD}\}\;d\hat{Y}\ppr\;\exp\{-v\;\hat{Y}_{DD}\}\Big)_{BB;mn}^{12} =}
\\ \no &=&-\bigg(d\hat{c}_{D;mn}^{\prime *}\;e^{\im(\varphi_{m}+\varphi_{n})}\;
\sin\big(v\;|\ovv{c}_{m}|\big)\;\sin\big(v\;|\ovv{c}_{n}|\big)+
d\hat{c}_{D;mn}\ppr\;\cos\big(v\;|\ovv{c}_{m}|\big)\;
\cos\big(v\;|\ovv{c}_{n}|\big)\bigg)_{BB;mn}^{12}
\eeq
\beq \lb{A115}
\lefteqn{\Big(\exp\{v\;\hat{Y}_{DD}\}\;d\hat{Y}\ppr\;\exp\{-v\;\hat{Y}_{DD}\}\Big)_{BF;m,r\ppr\nu}^{12} =}
\\ \no &\hspace*{-0.55cm}=&
\hspace*{-0.5cm}\bigg(d\hat{\eta}_{D;m,r\ppr\lambda}^{\prime *}\;\big(\tau_{2}\big)_{\lambda\nu}\;
e^{\im(\varphi_{m}+\phi_{r\ppr})}\;
\sin\big(v\;|\ovv{c}_{m}|\big)\;\sinh\big(v\;|\ovv{f}_{r\ppr}|\big)+
d\hat{\eta}_{D;m,r\ppr\nu}\ppr\;\cos\big(v\;|\ovv{c}_{m}|\big)\;
\cosh\big(v\;|\ovv{f}_{r\ppr}|\big)\bigg)_{BF;m,r\ppr\nu}^{12}
\eeq
\beq  \lb{A116}
\lefteqn{\Big(\exp\{v\;\hat{Y}_{DD}\}\;d\hat{Y}\ppr\;\exp\{-v\;\hat{Y}_{DD}\}\Big)_{FB;r\mu,n}^{12} =} \\ \no &=&
\bigg(\big(\tau_{2}\big)_{\mu\kappa}\;d\hat{\eta}_{D;r\kappa,n}^{\prime *}\;
e^{\im(\phi_{r}+\varphi_{n})}\;
\sinh\big(v\;|\ovv{f}_{r}|\big)\;\sin\big(v\;|\ovv{c}_{n}|\big)-
d\hat{\eta}_{D;r\mu,n}\ppr\;\cosh\big(v\;|\ovv{f}_{r}|\big)\;
\cos\big(v\;|\ovv{c}_{n}|\big)\bigg)_{FB;r\mu,n}^{12}
\eeq
\beq \lb{A117}
\lefteqn{\Big(\exp\{v\;\hat{Y}_{DD}\}\;d\hat{Y}\ppr\;\exp\{-v\;\hat{Y}_{DD}\}\Big)_{FF;r\mu,r\ppr\nu}^{12} =}
\\ \no &=&\bigg(\Big(\big(\tau_{0}\big)_{\mu\nu}\;d\hat{f}_{D;rr\ppr}^{\prime(0)*}-\sum_{k=1}^{3}
\big(\tau_{k}\big)_{\mu\nu}\;d\hat{f}_{D;rr\ppr}^{\prime(k)*}\Big)\;
e^{\im(\phi_{r}+\phi_{r\ppr})}\;\sinh\big(v\;|\ovv{f}_{r}|\big)\;\sinh\big(v\;|\ovv{f}_{r\ppr}|\big)+
\\ \no &+& \Big(\sum_{k=0}^{3}\big(\tau_{k}\big)_{\mu\nu}\;d\hat{f}_{D;rr\ppr}^{\prime (k)}\Big)\;
\cosh\big(v\;|\ovv{f}_{r}|\big)\;\cosh\big(v\;|\ovv{f}_{r\ppr}|\big) \bigg)_{FF;r\mu,r\ppr\nu}^{12}
\eeq
\beq \lb{A118}
d\hat{f}_{D;q\kappa,q\ppr\lambda}^{\prime +}&=&
\bigg(\sum_{k=0}^{3}\big(\tau_{k}\big)_{\kappa\lambda}\;
d\hat{f}_{D;qq\ppr}^{\prime(k)}\bigg)_{q\kappa,q\ppr\lambda}^{+}=
-\sum_{k=0,1,3}\big(\tau_{k}\big)_{\kappa\lambda}\;
d\hat{f}_{D;qq\ppr}^{\prime(k)*}+
\big(\tau_{2}\big)_{\kappa\lambda}\;d\hat{f}_{D;qq\ppr}^{\prime(2)*}\;\;\;.
\eeq
Using the above results, the coset integration measure
\(\big(d s_{Osp\backslash U}\big(\delta\hat{\lambda}\big)\big)^{2}\) (\ref{A99})
is further reduced to the 'BB', 'FF' and 'FB', 'BF' parts of super-matrices
\((d\wtilde{Y})^{\prime 12}_{\alpha\beta}\) ,
\((d\wtilde{Y})^{\prime 21}_{\alpha\beta}\)
\beq \lb{A119}
\Big(d s_{Osp\backslash U}\big(\delta\hat{\lambda}\big)\Big)^{2}  &=&-2\;
\strab\bigg[\Big(d\wtilde{Y}\Big)^{\prime 12}_{\alpha\beta}\;\;
\Big(d\wtilde{Y}\Big)^{\prime 21}_{\beta\alpha}\;\;
\Big(\delta\wtilde{\lambda}_{\beta}+\delta\wtilde{\lambda}_{\alpha}\Big)^{2}
\bigg]=  \\ \no &=&-2\;\sum_{m,n=1}^{L}
\Big(d\wtilde{Y}\Big)^{\prime 12}_{BB;mn}\;\;
\Big(d\wtilde{Y}\Big)^{\prime 21}_{BB;nm}\;\;
\Big(\delta\wtilde{\lambda}_{B;n}+\delta\wtilde{\lambda}_{B;m}\Big)^{2}+ \\ \no &+&2\;
\sum_{r,r\ppr=1}^{S/2}\sum_{\mu,\nu=1,2}
\Big(d\wtilde{Y}\Big)^{\prime 12}_{FF;r\mu,r\ppr\nu}\;\;
\Big(d\wtilde{Y}\Big)^{\prime 21}_{FF;r\ppr\nu,r\mu}\;\;
\Big(\delta\wtilde{\lambda}_{F;r\ppr\nu}+\delta\wtilde{\lambda}_{F;r\mu}\Big)^{2}+
\\ \no &-&2\;\sum_{m=1}^{L}\sum_{r\ppr=1(\nu=1,2)}^{S/2}
\Big(d\wtilde{Y}\Big)^{\prime 12}_{BF;m,r\ppr\nu}\;\;
\Big(d\wtilde{Y}\Big)^{\prime 21}_{FB;r\ppr\nu,m}\;\;
\Big(\delta\wtilde{\lambda}_{F;r\ppr\nu}+\delta\wtilde{\lambda}_{B;m}\Big)^{2}
\\ \no &+&2\;\sum_{n=1}^{L}\sum_{r=1(\mu=1,2)}^{S/2}
\Big(d\wtilde{Y}\Big)^{\prime 12}_{FB;r\mu,n}\;\;
\Big(d\wtilde{Y}\Big)^{\prime 21}_{BF;n,r\mu}\;\;
\Big(\delta\wtilde{\lambda}_{B;n}+\delta\wtilde{\lambda}_{F;r\mu}\Big)^{2}\;\;\;.
\eeq
In the following, we further analyze the matrices
\((d\wtilde{Y})^{\prime 12}\) , \((d\wtilde{Y})^{\prime 21}\) and perform the
integrations over \(v\in[0,1]\). These integrations over \(dv\) with the modulus of the
eigenvalues are achieved in relations (\ref{A120}-\ref{A126}) for the 'BB' part
of \((d\wtilde{Y})^{\prime 12}_{BB;mn}\) , \((d\wtilde{Y})^{\prime 21}_{BB;mn}\)
with distinction between diagonal \(m=n\) (\ref{A120}-\ref{A122})
and off-diagonal \(m\neq n\) (\ref{A123}-\ref{A126}) elements
\beq \lb{A120}
\Big(d\wtilde{Y}\Big)^{\prime 12}_{BB;mm}&=&\int_{0}^{1}d v\;
\Big(\exp\{v\;\hat{Y}_{DD}\}\;\;d\hat{Y}\ppr\;\; \exp\{-v\;\hat{Y}_{DD}\}\Big)^{12}_{BB;mm} \\ \no &=&
-\Big(d\wtilde{Y}\Big)^{\prime 21,*}_{BB;mm}=-\int_{0}^{1}d v\;
\Big(\exp\{v\;\hat{Y}_{DD}\}\;\;d\hat{Y}\ppr\;\; \exp\{-v\;\hat{Y}_{DD}\}\Big)^{21,*}_{BB;mm}
\\ \no &=&
-d\hat{c}\ppr_{D;mm}\;A_{BB;mm}(|\ovv{c}_{m}|)-d\hat{c}^{\prime *}_{D;mm}\;
\exp\{\im\;2\;\varphi_{m}\}\;B_{BB;mm}(|\ovv{c}_{m}|) \\   \lb{A121}
A_{BB;mm}(|\ovv{c}_{m}|)&=&\int_{0}^{1}d v\;\;\Big(\cos\big(v\;|\ovv{c}_{m}|\big)\Big)^{2} =
\bigg(\frac{1}{2}+\frac{1}{4}\frac{\sin\big(2\;|\ovv{c}_{m}|\big)}{|\ovv{c}_{m}|}\bigg) \\   \lb{A122}
B_{BB;mm}(|\ovv{c}_{m}|)&=&\int_{0}^{1}d v\;\;\Big(\sin\big(v\;|\ovv{c}_{m}|\big)\Big)^{2} =
\bigg(\frac{1}{2}-\frac{1}{4}\frac{\sin\big(2\;|\ovv{c}_{m}|\big)}{|\ovv{c}_{m}|}\bigg)
\eeq

\beq \lb{A123}
m&\neq& n \\  \lb{A124}
\Big(d\wtilde{Y}\Big)^{\prime 12}_{BB;mn}&=&\int_{0}^{1}d v\;
\Big(\exp\{v\;\hat{Y}_{DD}\}\;\;d\hat{Y}\ppr\;\;
\exp\{-v\;\hat{Y}_{DD}\}\Big)^{12}_{BB;mn} \\ \no &=&
\Big(d\wtilde{Y}\Big)^{\prime 12}_{BB;nm}=-\Big(d\wtilde{Y}\Big)_{BB;mn}^{\prime 21,*}=
-\Big(d\wtilde{Y}\Big)_{BB;nm}^{\prime 21,*} \\ \no &=&
-\int_{0}^{1}d v\;\Big(\exp\{v\;\hat{Y}_{DD}\}\;\;d\hat{Y}\ppr\;\;
\exp\{-v\;\hat{Y}_{DD}\}\Big)^{21,*}_{BB;mn}
\\ \no &=&
-d\hat{c}\ppr_{D;mn}\;A_{BB;mn}(|\ovv{c}_{m}|,|\ovv{c}_{n}|)-d\hat{c}^{\prime *}_{D;mn}\;
\exp\{\im\;(\varphi_{m}+\varphi_{n})\}\; B_{BB;mn}(|\ovv{c}_{m}|,|\ovv{c}_{n}|)  \\   \lb{A125}
A_{BB;mn}(|\ovv{c}_{m}|,|\ovv{c}_{n}|)&=&
\int_{0}^{1}d v\;\;\cos\big(v\;|\ovv{c}_{m}|\big)\;\;\cos\big(v\;|\ovv{c}_{n}|\big)
\hspace*{0.5cm}(m\neq n)
\\ \no &=& \bigg(
\frac{|\ovv{c}_{m}|\;\cos(|\ovv{c}_{n}|)\;\sin(|\ovv{c}_{m}|)-
|\ovv{c}_{n}|\;\cos(|\ovv{c}_{m}|)\;\sin(|\ovv{c}_{n}|)}{|\ovv{c}_{m}|^{2}-
|\ovv{c}_{n}|^{2}}\bigg) \\  \lb{A126}
B_{BB;mn}(|\ovv{c}_{m}|,|\ovv{c}_{n}|)&=&
\int_{0}^{1}d v\;\;\sin\big(v\;|\ovv{c}_{m}|\big)\;\;\sin\big(v\;|\ovv{c}_{n}|\big)
\hspace*{0.5cm}(m\neq n)
\\  \no &=& \bigg(
\frac{|\ovv{c}_{n}|\;\cos(|\ovv{c}_{n}|)\;\sin(|\ovv{c}_{m}|)-
|\ovv{c}_{m}|\;\cos(|\ovv{c}_{m}|)\;\sin(|\ovv{c}_{n}|)}{|\ovv{c}_{m}|^{2}-
|\ovv{c}_{n}|^{2}}\bigg)\;\;\;.
\eeq
The corresponding relations for
\(\big(d\wtilde{Y}\big)^{\prime 12}_{FF;r\mu,r\ppr\nu}\) ,
\(\big(d\wtilde{Y}\big)^{\prime 21}_{FF;r\mu,r\ppr\nu}\)
are listed for the 'FF' section in relations (\ref{A127}-\ref{A133})
where one has also to distinguish between quaternionic diagonal entries
with \((\tau_{2})_{\mu\nu}\), (\(r=r\ppr\), (\ref{A127}-\ref{A129})) and non-diagonal
quaternion elements with \(r\neq r\ppr\) (\ref{A130}-\ref{A133}).
The functions \(A_{FF;rr}(|\ovv{f}_{r}|)\) (\ref{A128}),
\(B_{FF;rr}(|\ovv{f}_{r}|)\) (\ref{A129}) and
\(A_{FF;rr\ppr}(|\ovv{f}_{r}|,|\ovv{f}_{r\ppr}|)\) (\ref{A132}),
\(B_{FF;rr\ppr}(|\ovv{f}_{r}|,|\ovv{f}_{r\ppr}|)\) (\ref{A133}) specify the dependence on
the eigenvalues in \(\big(d\wtilde{Y}\big)^{\prime 12}_{FF;r\mu,r\ppr\nu}\) ,
\(\big(d\wtilde{Y}\big)^{\prime 21}_{FF;r\mu,r\ppr\nu}\)
\beq  \lb{A127}
\Big(d\wtilde{Y}\Big)^{\prime 12}_{FF;r\mu,r\nu}&=&\int_{0}^{1}d v\;
\Big(\exp\{v\;\hat{Y}_{DD}\}\;\;d\hat{Y}\ppr\;\;\exp\{-v\;\hat{Y}_{DD}\}\Big)_{FF;r\mu,r\nu}^{12} \\ \no &=&
-\Big(d\wtilde{Y}\Big)^{\prime 12}_{FF;r\nu,r\mu}=
-\Big(d\wtilde{Y}\Big)_{FF;r\mu,r\nu}^{\prime 21,*}=
\Big(d\wtilde{Y}\Big)^{\prime 21,*}_{FF;r\nu,r\mu} \\ \no &=&
-\int_{0}^{1}d v\;
\Big(\exp\{v\;\hat{Y}_{DD}\}\;\;d\hat{Y}\ppr\;\;\exp\{-v\;\hat{Y}_{DD}\}\Big)_{FF;r\mu,r\nu}^{21,*} \\ \no &=&
\big(\tau_{2}\big)_{\mu\nu}\;\bigg(df_{D;rr}^{\prime (2)}\;A_{FF;rr}(|\ovv{f}_{r}|)
+df_{D;rr}^{\prime (2)*}\;
\exp\{\im\;2\;\phi_{r}\}\;B_{FF;rr}(|\ovv{f}_{r}|)\bigg) \\ \lb{A128}
A_{FF;rr}(|\ovv{f}_{r}|)&=&
\int_{0}^{1}d v\;\;\Big(\cosh\big(v\;|\ovv{f}_{r}|\big)\Big)^{2} =
\bigg(\frac{1}{2}+\frac{1}{4}\frac{\sinh(2\;|\ovv{f}_{r}|)}{|\ovv{f}_{r}|}\bigg) \\  \lb{A129}
B_{FF;rr}(|\ovv{f}_{r}|)&=&
-\int_{0}^{1}d v\;\;\Big(\sinh\big(v\;|\ovv{f}_{r}|\big)\Big)^{2} =
\bigg(\frac{1}{2}-\frac{1}{4}\frac{\sinh(2\;|\ovv{f}_{r}|)}{|\ovv{f}_{r}|}\bigg)
\eeq

\beq \lb{A130}
r&\neq& r\ppr \\  \lb{A131}
\Big(d\wtilde{Y}\Big)^{\prime 12}_{FF;r\mu,r\ppr\nu}
&=&\int_{0}^{1}d v\;
\Big(\exp\{v\;\hat{Y}_{DD}\}\;\;d\hat{Y}\ppr\;\; \exp\{-v\;\hat{Y}_{DD}\}\Big)^{12}_{FF;r\mu,r\ppr\nu} \\ \no &=&
-\Big(d\wtilde{Y}\Big)^{\prime 12}_{FF;r\ppr\nu,r\mu}=-
\Big(d\wtilde{Y}\Big)^{\prime 21,*}_{FF;r\mu,r\ppr\nu}=
\Big(d\wtilde{Y}\Big)^{\prime 21,*}_{FF;r\ppr\nu,r\mu} \\ \no &=&
\int_{0}^{1}d v\;
\Big(\exp\{v\;\hat{Y}_{DD}\}\;\;d\hat{Y}\ppr\;\; \exp\{-v\;\hat{Y}_{DD}\}\Big)^{21,*}_{FF;r\ppr\nu,r\mu}
\\ \no &=&
\bigg(\sum_{k=0}^{k=3}\big(\tau_{k}\big)_{\mu\nu}\;df_{D;rr\ppr}^{\prime (k)}\bigg)
\;A_{FF;rr\ppr}(|\ovv{f}_{r}|,|\ovv{f}_{r\ppr}|)+ \\ \no &+&
\bigg(\big(\tau_{0}\big)_{\mu\nu}\;df_{D;rr\ppr}^{\prime (0)*}-
\sum_{k=1}^{k=3}\big(\tau_{k}\big)_{\mu\nu}\;df_{D;rr\ppr}^{\prime (k)*}\bigg)
\;\exp\{\im\;(\phi_{r}+\phi_{r\ppr})\}\; B_{FF;rr\ppr}(|\ovv{f}_{r}|,|\ovv{f}_{r\ppr}|)  \\   \lb{A132}
A_{FF;rr\ppr}(|\ovv{f}_{r}|,|\ovv{f}_{r\ppr}|)\hspace*{-0.25cm}&=&\hspace*{-0.25cm}
\int_{0}^{1}d v\;\;\cosh\big(v\;|\ovv{f}_{r}|\big)\;\;\cosh\big(v\;|\ovv{f}_{r\ppr}|\big)
\hspace*{0.5cm}(r\neq r\ppr)
\\ \no
\hspace*{-0.25cm}&=&\hspace*{-0.25cm}\bigg(
\frac{|\ovv{f}_{r}|\;\cosh(|\ovv{f}_{r\ppr}|)\;\sinh(|\ovv{f}_{r}|)-
|\ovv{f}_{r\ppr}|\;\cosh(|\ovv{f}_{r}|)\;\sinh(|\ovv{f}_{r\ppr}|)}{|\ovv{f}_{r}|^{2}-
|\ovv{f}_{r\ppr}|^{2}}\bigg) \\    \lb{A133}
B_{FF;rr\ppr}(|\ovv{f}_{r}|,|\ovv{f}_{r\ppr}|)\hspace*{-0.25cm}&=&\hspace*{-0.25cm}
\int_{0}^{1}d v\;\;\sinh\big(v\;|\ovv{f}_{r}|\big)\;\;\sinh\big(v\;|\ovv{f}_{r\ppr}|\big)
\hspace*{0.5cm}(r\neq r\ppr)
\\ \no
\hspace*{-0.25cm}&=&\hspace*{-0.25cm}\bigg(
\frac{|\ovv{f}_{r}|\;\cosh(|\ovv{f}_{r}|)\;\sinh(|\ovv{f}_{r\ppr}|)-
|\ovv{f}_{r\ppr}|\;\cosh(|\ovv{f}_{r\ppr}|)\;
\sinh(|\ovv{f}_{r}|)}{|\ovv{f}_{r}|^{2}- |\ovv{f}_{r\ppr}|^{2}}\bigg)\;\;\;.
\eeq
Similar considerations hold for the 'BF', 'FB' sections of the resulting
super-matrices with \(\big(d\wtilde{Y}\big)^{\prime 12}_{BF;m,r\ppr\nu}\) ,
\(\big(d\wtilde{Y}\big)^{\prime 12}_{FB;r\mu,n}\)
\beq \lb{A134}
\Big(d\wtilde{Y}\Big)^{\prime 12}_{BF;m,r\ppr\nu}&=&\int_{0}^{1}d v\;
\Big(\exp\{v\;\hat{Y}_{DD}\}\;\;d\hat{Y}\ppr\;\; \exp\{-v\;\hat{Y}_{DD}\}\Big)^{12}_{BF;m,r\ppr\nu} \\ \no &=&
-\Big(d\wtilde{Y}\Big)^{\prime 12}_{FB;r\ppr\nu,m}=
\Big(d\wtilde{Y}\Big)^{\prime 21,*}_{BF;m,r\ppr\nu}=
\Big(d\wtilde{Y}\Big)^{\prime 21,*}_{FB;r\ppr\nu,m} \\ \no &=&
\int_{0}^{1}d v\;
\Big(\exp\{v\;\hat{Y}_{DD}\}\;\;d\hat{Y}\ppr\;\; \exp\{-v\;\hat{Y}_{DD}\}\Big)^{21,*}_{BF;m,r\ppr\nu}
\\ \no &=&
d\hat{\eta}\ppr_{D;m,r\ppr\nu}\;A_{BF;m,r\ppr}(|\ovv{c}_{m}|,|\ovv{f}_{r\ppr}|)+ \\ \no &+&
d\hat{\eta}_{D;m,r\ppr\lambda}^{\prime *}\;\big(\tau_{2}\big)_{\lambda\nu}\;
\exp\{\im(\varphi_{m}+\phi_{r\ppr})\}\;B_{BF;r\ppr,m}(|\ovv{f}_{r\ppr}|,|\ovv{c}_{m}|)   \\  \lb{A135}
A_{BF;m,r\ppr}(|\ovv{c}_{m}|,|\ovv{f}_{r\ppr}|)\hspace*{-0.25cm}&=&\hspace*{-0.25cm}
\int_{0}^{1}d v\;\;\cos\big(v\;|\ovv{c}_{m}|\big)\;\;\cosh\big(v\;|\ovv{f}_{r\ppr}|\big)
\\ \no \hspace*{-0.25cm}&=&\hspace*{-0.25cm}\bigg(
\frac{|\ovv{c}_{m}|\;\cosh(|\ovv{f}_{r\ppr}|)\;\sin(|\ovv{c}_{m}|)+
|\ovv{f}_{r\ppr}|\;\cos(|\ovv{c}_{m}|)\;\sinh(|\ovv{f}_{r\ppr}|)}{|\ovv{c}_{m}|^{2}+
|\ovv{f}_{r\ppr}|^{2}}\bigg) \\   \lb{A136}
B_{BF;r\ppr,m}(|\ovv{f}_{r\ppr}|,|\ovv{c}_{m}|)\hspace*{-0.25cm}&=&\hspace*{-0.25cm}
\int_{0}^{1}d v\;\;\sin\big(v\;|\ovv{c}_{m}|\big)\;\;\sinh\big(v\;|\ovv{f}_{r\ppr}|\big)
\\ \no \hspace*{-0.25cm}&=&\hspace*{-0.25cm}\bigg(
\frac{|\ovv{f}_{r\ppr}|\;\cosh(|\ovv{f}_{r\ppr}|)\;\sin(|\ovv{c}_{m}|)-
|\ovv{c}_{m}|\;\cos(|\ovv{c}_{m}|)\;\sinh(|\ovv{f}_{r\ppr}|)}{|\ovv{c}_{m}|^{2}+
|\ovv{f}_{r\ppr}|^{2}}\bigg)\;\;\;.
\eeq
Applying relations (\ref{A120}-\ref{A136}), we can insert
\((d\wtilde{Y})^{\prime 12}_{\alpha\beta}\) , \((d\wtilde{Y})^{\prime 21}_{\alpha\beta}\)
without remaining integrations over \(v\in[0,1]\) into the invariant coset length
\(\big(d s_{Osp\backslash U}\big(\delta\hat{\lambda}\big)\big)^{2}\) (\ref{A119})
which can be separated correspondingly into 'BB', 'FF' and 'BF', 'FB' parts (\ref{A137}).
However, one has to refer to different relations concerning diagonal and off-diagonal
elements in the 'BB' and'FF' sections (see Eqs. (\ref{A120}-\ref{A122}) for
\((d\wtilde{Y})^{\prime a\neq b}_{BB;mm}\) and Eqs. (\ref{A127}-\ref{A129}) for
\((d\wtilde{Y})^{\prime a\neq b}_{FF;r\mu,r\nu}\) ; in the case of non-diagonal
elements compare relations (\ref{A123}-\ref{A126}) for
\((d\wtilde{Y})^{\prime a\neq b}_{BB;mn}\), (\(m\neq n\)) and relations (\ref{A130}-\ref{A133})
for \((d\wtilde{Y})^{\prime a\neq b}_{FF;r\mu,r\ppr\nu}\), (\(r\neq r\ppr\)))
\beq \no \lefteqn{
\Big(d s_{Osp\backslash U}\big(\delta\hat{\lambda}\big)\Big)^{2}=
\Big(d s_{Osp\backslash U}\big(\delta\hat{\lambda}\big)\Big)_{BB}^{2}+
\Big(d s_{Osp\backslash U}\big(\delta\hat{\lambda}\big)\Big)_{FF}^{2}+
\Big(d s_{Osp\backslash U}\big(\delta\hat{\lambda}\big)\Big)_{BF,FB}^{2} =}  \\ \lb{A137} &=&
2\sum_{m=1}^{L}\Big(d\wtilde{Y}\Big)^{\prime 12,*}_{BB;mm}\;\;
\Big(d\wtilde{Y}\Big)^{\prime 12}_{BB;mm}\;\;
\big(2\;\delta\wtilde{\lambda}_{B;m}\big)^{2} \\ \no &+&
4\sum_{m=1}^{L}\sum_{n=m+1}^{L}\Big(d\wtilde{Y}\Big)^{\prime 12,*}_{BB;mn}\;\;
\Big(d\wtilde{Y}\Big)^{\prime 12}_{BB;mn}\;\;
\big(\delta\wtilde{\lambda}_{B;n}+\delta\wtilde{\lambda}_{B;m}\big)^{2} \\ \no
&-&2\sum_{r=1}^{S/2}\sum_{\mu,\nu=1,2}
\Big(d\wtilde{Y}\Big)^{\prime 12}_{FF;r\mu,r\nu}\;\;
\Big(d\wtilde{Y}\Big)^{\prime 12,*}_{FF;r\nu,r\mu}\;\;
\big(\delta\wtilde{\lambda}_{F;r\nu}+\delta\wtilde{\lambda}_{F;r\mu}\big)^{2}
\\ \no &-&4\sum_{r=1}^{S/2}\sum_{r\ppr=r+1}^{S/2}\sum_{\mu,\nu=1,2}
\Big(d\wtilde{Y}\Big)^{\prime 12}_{FF;r\mu,r\ppr\nu}\;\;
\Big(d\wtilde{Y}\Big)^{\prime 12,*}_{FF;r\ppr\nu,r\mu}\;\;
\big(\delta\wtilde{\lambda}_{F;r\ppr\nu}+\delta\wtilde{\lambda}_{F;r\mu}\big)^{2}
\\ \no &+&4\sum_{m=1}^{L}\sum_{r\ppr=1}^{S/2}\sum_{\nu=1,2}
\Big(d\wtilde{Y}\Big)^{\prime 12}_{BF;m,r\ppr\nu}\;\;
\Big(d\wtilde{Y}\Big)^{\prime 12,*}_{FB;r\ppr\nu,m}\;\;
\big(\delta\wtilde{\lambda}_{F;r\ppr\nu}+\delta\wtilde{\lambda}_{B;m}\big)^{2} \;\;\;.
\eeq
We exploit the symmetries between
\(\big(\wtilde{T}_{0}^{-1}\;d\wtilde{T}_{0}\big)_{\alpha\beta}^{12}\) and
\(\big(\wtilde{T}_{0}^{-1}\;d\wtilde{T}_{0}\big)_{\alpha\beta}^{21}\), according
to the similarity to the generators \(d\hat{X}\ppr\) , \(\wtilde{\kappa}\;d\hat{X}^{\prime +}\),
and construct the appropriate sub-metrics from the 'BB', 'FF' and 'BF', 'FB' parts
\beq \lb{A138}
\big(\wtilde{T}_{0}^{-1}\;d\wtilde{T}_{0}\big)_{\alpha\beta}^{12}&=&-
\int_{0}^{1}d v\;
\Big(\exp\{v\;\hat{Y}_{DD}\}\;\;d\hat{Y}\ppr\;\; \exp\{-v\;\hat{Y}_{DD}\}\Big)^{12}_{\alpha\beta}=-
\Big(d\wtilde{Y}\Big)^{\prime 12}_{\alpha\beta}  \\ \no &=&-
\left(\bea{cc}
\Big(d\wtilde{Y}\Big)^{\prime 12}_{BB;mn} &
\Big(d\wtilde{Y}\Big)^{\prime 12}_{BF;m,r\ppr\nu} \\
\Big(d\wtilde{Y}\Big)^{\prime 12}_{FB;r\mu,n} &
\Big(d\wtilde{Y}\Big)^{\prime 12}_{FF;r\mu,r\ppr\nu}
\eea\right)_{\alpha\beta}\propto
\underbrace{\left(\bea{cc}
-d\hat{c}_{D;L\times L}\ppr & d\hat{\eta}_{D;L\times S}^{\prime T} \\
-d\hat{\eta}_{D;S\times L}\ppr & d\hat{f}_{D;S\times S}\ppr
\eea\right)_{\alpha\beta}}_{=\;d\hat{X}_{\alpha\beta}\ppr}  \\  \lb{A139}
\big(\wtilde{T}_{0}^{-1}\;d\wtilde{T}_{0}\big)_{\beta\alpha}^{21} &=&-
\int_{0}^{1}d v\;
\Big(\exp\{v\;\hat{Y}_{DD}\}\;\;d\hat{Y}\ppr\;\; \exp\{-v\;\hat{Y}_{DD}\}\Big)^{21}_{\beta\alpha}=-
\Big(d\wtilde{Y}\Big)^{\prime 21}_{\beta\alpha}  \\ \no &=&
\left(\bea{cc}
\Big(d\wtilde{Y}\Big)^{\prime 12,*}_{BB;nm} &
-\Big(d\wtilde{Y}\Big)^{\prime 12,*}_{BF;n,r\mu} \\
\Big(d\wtilde{Y}\Big)^{\prime 12,*}_{FB;r\ppr\nu,m} &
\Big(d\wtilde{Y}\Big)^{\prime 12,*}_{FF;r\ppr\nu,r\mu}
\eea\right)_{\beta\alpha}\propto
\underbrace{\left(\bea{cc}
d\hat{c}_{D;L\times L}^{\prime +} & d\hat{\eta}_{D;L\times S}^{\prime +} \\
d\hat{\eta}_{D;S\times L}^{\prime *} & d\hat{f}_{D;S\times S}^{\prime +}
\eea\right)_{\beta\alpha}}_{=\;\wtilde{\kappa}\;d\hat{X}_{\beta\alpha}^{\prime +}}  \\ \lb{A140} &&
\underbrace{\big(\wtilde{T}_{0}^{-1}\;d\wtilde{T}_{0}\big)^{21}}_{\propto\;\;
\wtilde{\kappa}\;d\hat{X}_{N\times N}^{\prime +}} =
\underbrace{\Big[\big(\wtilde{T}_{0}^{-1}\;d\wtilde{T}_{0}\big)^{12}\;
\wtilde{\kappa}\Big]^{+}}_{\propto\;\;
\big(d\hat{X}_{N\times N}\ppr\;\wtilde{\kappa}\big)^{+}}  \;\;\;.
\eeq
We take into account the detailed relation (\ref{A137}) for
\(\big(d s_{Osp\backslash U}\big(\delta\hat{\lambda}\big)\big)^{2}\) and
consider separately the sub-metric tensors \(\hat{M}_{BB}\) with
Eqs. (\ref{A141}-\ref{A145}) for the 'BB' part and \(\hat{M}_{FF}\) with Eqs. (\ref{A150}-\ref{A159})
for the even 'FF' part. The square root of the determinant of these sub-metric tensors
yields the final coset integration measure or super-Jacobi determinant
for the corresponding subspace with indices \(m,\;n\) and
\(r,\mu\;;\;r\ppr,\nu\) (compare Eqs. (\ref{A146}-\ref{A149}) for the 'BB' part
and Eqs. (\ref{A160}-\ref{A167}) for the 'FF' part).
We catalogue the straightforward relations for determining the invariant
coset integration measure of the even boson-boson and even fermion-fermion
parts with the detailed Eqs. (\ref{A146}-\ref{A167})
where sub-metric tensors are constructed for
the even elements of \(d\hat{Y}\ppr\) with the eigenvalues of \(\hat{Y}_{DD}\)
\beq \lb{A141}
\lefteqn{\Big(d s_{Osp\backslash U}\big(\delta\hat{\lambda}\big)\Big)_{BB}^{2} = }
 \\ \no &=& 2
\sum_{m=1}^{L}
\Big(d\wtilde{Y}\Big)^{\prime 12,*}_{BB;mm}\;\;
\Big(d\wtilde{Y}\Big)^{\prime 12}_{BB;mm}\;\;
\big(2\cdot\delta\wtilde{\lambda}_{B;m}\big)^{2}+   \\ \no &+& 4
\sum_{m=1}^{L}\sum_{n=m+1}^{L}
\Big(d\wtilde{Y}\Big)^{\prime 12,*}_{BB;mn}\;\;
\Big(d\wtilde{Y}\Big)^{\prime 12}_{BB;mn}\;\;
\big(\delta\wtilde{\lambda}_{B;n}+\delta\wtilde{\lambda}_{B;m}\big)^{2} \\ \no &=&
2 \sum_{m=1}^{L}
\left(\bea{c}
d\hat{c}_{D;mm}\ppr \\ d\hat{c}_{D;mm}^{\prime *}
\eea\right)^{+}\left(
\bea{cc}
M_{BB;mm}^{(11)} & M_{BB;mm}^{(12)} \\
M_{BB;mm}^{(21)} & M_{BB;mm}^{(22)}
\eea\right)\left(\bea{c}
d\hat{c}_{D;mm}\ppr \\ d\hat{c}_{D;mm}^{\prime *}
\eea\right)\times\big(2\cdot\delta\wtilde{\lambda}_{B;m}\big)^{2}+ \\ \no &+& 4
\sum_{m=1}^{L}\sum_{n=m+1}^{L}
\left(\bea{c}
d\hat{c}_{D;mn}\ppr \\ d\hat{c}_{D;mn}^{\prime *}
\eea\right)^{+}\left(
\bea{cc}
M_{BB;mn}^{(11)} & M_{BB;mn}^{(12)} \\
M_{BB;mn}^{(21)} & M_{BB;mn}^{(22)}
\eea\right)\left(\bea{c}
d\hat{c}_{D;mn}\ppr \\ d\hat{c}_{D;mn}^{\prime *}
\eea\right)\times
\big(\delta\wtilde{\lambda}_{B;n}+\delta\wtilde{\lambda}_{B;m}\big)^{2}
\eeq
\beq \lb{A142}
M_{BB;mm}^{(11)}&=&M_{BB;mm}^{(22)} = \frac{1}{2}\bigg[
\Big(A_{BB;mm}(|\ovv{c}_{m}|)\Big)^{2}+\Big(B_{BB;mm}(|\ovv{c}_{m}|)\Big)^{2}\bigg] \\ \no &=&
\frac{1}{4}+\frac{1}{16}\bigg(\frac{\sin(2|\ovv{c}_{m}|)}{|\ovv{c}_{m}|}\bigg)^{2} \\ \lb{A143}
M_{BB;mm}^{(12)}&=&\Big(M_{BB;mm}^{(21)}\Big)^{*} =
\exp\{\im\;2\varphi_{m}\}\;\;A_{BB;mm}(|\ovv{c}_{m}|)\;\;B_{BB;mm}(|\ovv{c}_{m}|) \\ \no &=&
\exp\{\im\;2\varphi_{m}\}\;\;\bigg[
\frac{1}{4}-\frac{1}{16}\bigg(\frac{\sin(2|\ovv{c}_{m}|)}{|\ovv{c}_{m}|}\bigg)^{2}\bigg] \\ \lb{A144}
M_{BB;mn}^{(11)}&=&M_{BB;mn}^{(22)} = \frac{1}{2}\bigg[
\Big(A_{BB;mn}(|\ovv{c}_{m}|,|\ovv{c}_{n}|)\Big)^{2}+
\Big(B_{BB;mm}(|\ovv{c}_{m}|,|\ovv{c}_{n}|)\Big)^{2}\bigg] \\ \lb{A145}
M_{BB;mn}^{(12)}&=&\Big(M_{BB;mn}^{(21)}\Big)^{*}  =
\exp\{\im\;(\varphi_{m}+\varphi_{n})\}\;\;
A_{BB;mn}(|\ovv{c}_{m}|,|\ovv{c}_{n}|)\;\;B_{BB;mn}(|\ovv{c}_{m}|,|\ovv{c}_{n}|)
\eeq
\beq  \lb{A146}  2\;
\sqrt{\det\Big(M_{BB;mm}^{(ab)}\Big)}\;\big(2\;\delta\lambda_{B;m}\big)^{2}&=& 2\;
\left|\frac{\sin\big(2\;|\ovv{c}_{m}|\big)}{|\ovv{c}_{m}|}\right|\;\;\;\big(\delta\lambda_{B;m}\big)^{2}
\eeq
\beq \lb{A147}
\lefteqn{\prod_{\{\vec{x},t_{p}\}}\Bigg\{\prod_{m=1}^{L}
\frac{d\hat{c}_{D;mm}^{*}\wedge d\hat{c}_{D;mm}}{2\;\im}\;\;2\;
\sqrt{\det\Big(M_{BB;mm}^{(ab)}\Big)}\;\big(2\;\delta\lambda_{B;m}\big)^{2}\Bigg\}=} \\ \no &=&
\prod_{\{\vec{x},t_{p}\}}\Bigg\{\prod_{m=1}^{L}
\frac{d\hat{c}_{D;mm}^{*}\wedge d\hat{c}_{D;mm}}{2\;\im}\;\;2\;
\left|\frac{\sin\big(2\;|\ovv{c}_{m}|\big)}{|\ovv{c}_{m}|}\right|\;\;\;\big(\delta\lambda_{B;m}\big)^{2}
\Bigg\}
\eeq
\beq \no
\lefteqn{4\;\sqrt{\det\Big(M_{BB;mn}^{(ab)}\Big)}\;\big(\delta\lambda_{B;n}+\delta\lambda_{B;m}\big)^{2}=
\left|\frac{\cos\big(2\;|\ovv{c}_{m}|\big)-\cos\big(2\;|\ovv{c}_{n}|\big)}{
\big(|\ovv{c}_{m}|^{2}-|\ovv{c}_{n}|^{2}\big)}\right|\;
\big(\delta\lambda_{B;n}+\delta\lambda_{B;m}\big)^{2}=  } \\ \lb{A148} &=& 2\;
\left|\frac{\sin\big(|\ovv{c}_{m}|+|\ovv{c}_{n}|\big)}{|\ovv{c}_{m}|+|\ovv{c}_{n}|}\right|\;\;
\left|\frac{\sin\big(|\ovv{c}_{m}|-|\ovv{c}_{n}|\big)}{|\ovv{c}_{m}|-|\ovv{c}_{n}|}\right|\;\;
\big(\delta\lambda_{B;n}+\delta\lambda_{B;m}\big)^{2}
\eeq
\beq \lb{A149}
\lefteqn{\hspace*{-0.7cm}\prod_{\{\vec{x},t_{p}\}}\Bigg\{\prod_{m=1}^{L}\prod_{n=m+1}^{L}
\frac{d\hat{c}_{D;mn}^{*}\wedge d\hat{c}_{D;mn}}{2\;\im}\;\;4\;
\sqrt{\det\Big(M_{BB;mn}^{(ab)}\Big)}\;\big(\delta\lambda_{B;n}+\delta\lambda_{B;m}\big)^{2}=} \\ \no &=&
\prod_{\{\vec{x},t_{p}\}}\Bigg\{\prod_{m=1}^{L}\prod_{n=m+1}^{L}
\frac{d\hat{c}_{D;mn}^{*}\wedge d\hat{c}_{D;mn}}{2\;\im} \\ \no &\times& 2\;
\left|\frac{\sin\big(|\ovv{c}_{m}|+|\ovv{c}_{n}|\big)}{|\ovv{c}_{m}|+|\ovv{c}_{n}|}\right|\;\;
\left|\frac{\sin\big(|\ovv{c}_{m}|-|\ovv{c}_{n}|\big)}{|\ovv{c}_{m}|-|\ovv{c}_{n}|}\right|\;\;
\big(\delta\lambda_{B;n}+\delta\lambda_{B;m}\big)^{2}\Bigg\}
\eeq

\beq \lb{A150}
\lefteqn{\Big(d s_{Osp\backslash U}\big(\delta\hat{\lambda}\big)\Big)_{FF}^{2} = } \\ \no &=& -2
\sum_{r=1}^{S/2}\sum_{(\mu,\nu=1,2)}
\Big(d\wtilde{Y}\Big)^{\prime 12}_{FF;r\mu,r\nu}\;\;
\Big(d\wtilde{Y}\Big)^{\prime 12,*}_{FF;r\nu,r\mu}\;\;
\big(\delta\wtilde{\lambda}_{F;r\nu}+\delta\wtilde{\lambda}_{F;r\mu}\big)^{2}+ \\ \no &-& 4
\sum_{r=1}^{S/2}\sum_{r\ppr=r+1}^{S/2}\sum_{(\mu,\nu=1,2)}
\Big(d\wtilde{Y}\Big)^{\prime 12}_{FF;r\mu,r\ppr\nu}\;\;
\Big(d\wtilde{Y}\Big)^{\prime 12,*}_{FF;r\ppr\nu,r\mu}\;\;
\big(\delta\wtilde{\lambda}_{F;r\ppr\nu}+\delta\wtilde{\lambda}_{F;r\mu}\big)^{2} \\ \no &=& 4
\sum_{r=1}^{S/2}
\left(\bea{c}
d\hat{f}_{D;rr}^{\prime (2)} \\ d\hat{f}_{D;rr}^{\prime(2)*}
\eea\right)^{+}\left(
\bea{cc}
M_{FF;rr}^{(11)} & M_{FF;rr}^{(12)} \\
M_{FF;rr}^{(21)} & M_{FF;rr}^{(22)}
\eea\right)\left(\bea{c}
d\hat{f}_{D;rr}^{\prime (2)} \\ d\hat{f}_{D;rr}^{\prime(2)*}
\eea\right)\big(\delta\wtilde{\lambda}_{F;r1}+
\delta\wtilde{\lambda}_{F;r2}\big)^{2}+ \\   \no &+& 4
\sum_{r=1}^{S/2}\sum_{r\ppr=r+1}^{S/2}\sum_{(k\ppr,k=0,3)}
\left(\bea{c}
d\hat{f}_{D;rr\ppr}^{\prime (k\ppr)} \\ d\hat{f}_{D;rr\ppr}^{\prime(k\ppr)*}
\eea\right)^{+}\left(
\bea{cc}
M_{FF;rr\ppr}^{(11),k\ppr k} & M_{FF;rr\ppr}^{(12),k\ppr k} \\
M_{FF;rr\ppr}^{(21),k\ppr k} & M_{FF;rr\ppr}^{(22),k\ppr k}
\eea\right)\left(\bea{c}
d\hat{f}_{D;rr\ppr}^{\prime (k)} \\ d\hat{f}_{D;rr\ppr}^{\prime(k)*}
\eea\right)  \\  \no &+& 4
\sum_{r=1}^{S/2}\sum_{r\ppr=r+1}^{S/2}\sum_{(k\ppr,k=1,2)}
\left(\bea{c}
d\hat{f}_{D;rr\ppr}^{\prime (k\ppr)} \\ d\hat{f}_{D;rr\ppr}^{\prime(k\ppr)*}
\eea\right)^{+}\left(
\bea{cc}
M_{FF;rr\ppr}^{(11),k\ppr k} & M_{FF;rr\ppr}^{(12),k\ppr k} \\
M_{FF;rr\ppr}^{(21),k\ppr k} & M_{FF;rr\ppr}^{(22),k\ppr k}
\eea\right)\left(\bea{c}
d\hat{f}_{D;rr\ppr}^{\prime (k)} \\ d\hat{f}_{D;rr\ppr}^{\prime(k)*}
\eea\right)
\eeq
\beq \lb{A151}
M_{FF;rr}^{(11)}&=&M_{FF;rr}^{(22)} = \frac{1}{2}\bigg[
\Big(A_{FF;rr}(|\ovv{f}_{r}|)\Big)^{2}+\Big(B_{FF;rr}(|\ovv{f}_{r}|)\Big)^{2}\bigg] \\ \no &=&
\frac{1}{4}+\frac{1}{16}\bigg(\frac{\sinh(2|\ovv{f}_{r}|)}{|\ovv{f}_{r}|}\bigg)^{2} \\  \lb{A152}
M_{FF;rr}^{(12)}&=&\Big(M_{FF;rr}^{(21)}\Big)^{*}  =
\exp\{\im\;2\phi_{r}\}\;\;A_{FF;rr}(|\ovv{f}_{r}|)\;\;B_{FF;rr}(|\ovv{f}_{r}|) \\ \no &=&
\exp\{\im\;2\phi_{r}\}\;\;\bigg[
\frac{1}{4}-\frac{1}{16}\bigg(\frac{\sinh(2|\ovv{f}_{r}|)}{|\ovv{f}_{r}|}\bigg)^{2}\bigg]
\eeq
\beq \lb{A153}
\lefteqn{\sum_{\mu,\nu=1,2}\big(\tau_{k\ppr}\big)_{\mu\nu}\;\;\big(\tau_{k}\big)_{\nu\mu}\;\;
\Big(\delta\wtilde{\lambda}_{F;r\ppr\nu}+\delta\wtilde{\lambda}_{F;r\mu}\Big)^{2} = } \\ \no &=&
\Big(\delta_{k\ppr,0}\;\delta_{k,0}+\delta_{k\ppr,3}\;\delta_{k,3}\Big)\;\;
\bigg[\Big(\delta\wtilde{\lambda}_{F;r1}+\delta\wtilde{\lambda}_{F;r\ppr 1}\Big)^{2} +
\Big(\delta\wtilde{\lambda}_{F;r2}+\delta\wtilde{\lambda}_{F;r\ppr 2}\Big)^{2}\bigg]+
\\ \no &+&
\Big(\delta_{k\ppr,0}\;\delta_{k,3}+\delta_{k\ppr,3}\;\delta_{k,0}\Big)\;\;
\bigg[\Big(\delta\wtilde{\lambda}_{F;r1}+\delta\wtilde{\lambda}_{F;r\ppr 1}\Big)^{2} -
\Big(\delta\wtilde{\lambda}_{F;r2}+\delta\wtilde{\lambda}_{F;r\ppr 2}\Big)^{2}\bigg]+
\\ \no &+&
\Big(\delta_{k\ppr,1}\;\delta_{k,1}+\delta_{k\ppr,2}\;\delta_{k,2}\Big)\;\;
\bigg[\Big(\delta\wtilde{\lambda}_{F;r1}+\delta\wtilde{\lambda}_{F;r\ppr 2}\Big)^{2} +
\Big(\delta\wtilde{\lambda}_{F;r2}+\delta\wtilde{\lambda}_{F;r\ppr 1}\Big)^{2}\bigg]+
\\ \no &+& \im\;\;\ve_{k\ppr k3} \;\;
\bigg[\Big(\delta\wtilde{\lambda}_{F;r1}+\delta\wtilde{\lambda}_{F;r\ppr 2}\Big)^{2} -
\Big(\delta\wtilde{\lambda}_{F;r2}+\delta\wtilde{\lambda}_{F;r\ppr 1}\Big)^{2}\bigg]
\eeq
\beq \lb{A154}
M_{FF;rr\ppr}^{(11),00}&=&M_{FF;rr\ppr}^{(22),00}=M_{FF;rr\ppr}^{(11),33}=M_{FF;rr\ppr}^{(22),33}
\\ \no &=&\frac{1}{2}\;
\Big[\Big(A_{FF;rr\ppr}(|\ovv{f}_{r}|,|\ovv{f}_{r\ppr}|)\Big)^{2}+
\Big(B_{FF;rr\ppr}(|\ovv{f}_{r}|,|\ovv{f}_{r\ppr}|)\Big)^{2}\Big]\times \\ \no &\times&
\Big[\big(\delta\wtilde{\lambda}_{F;r1}+\delta\wtilde{\lambda}_{F;r\ppr 1}\big)^{2}+
\big(\delta\wtilde{\lambda}_{F;r2}+\delta\wtilde{\lambda}_{F;r\ppr 2}\big)^{2}\Big] \\  \lb{A155}
M_{FF;rr\ppr}^{(11),03}&=&M_{FF;rr\ppr}^{(22),30}=M_{FF;rr\ppr}^{(11),30}=M_{FF;rr\ppr}^{(22),03}
\\ \no &=&\frac{1}{2}\;
\Big[\Big(A_{FF;rr\ppr}(|\ovv{f}_{r}|,|\ovv{f}_{r\ppr}|)\Big)^{2}-
\Big(B_{FF;rr\ppr}(|\ovv{f}_{r}|,|\ovv{f}_{r\ppr}|)\Big)^{2}\Big]\times \\ \no &\times&
\Big[\big(\delta\wtilde{\lambda}_{F;r1}+\delta\wtilde{\lambda}_{F;r\ppr 1}\big)^{2}-
\big(\delta\wtilde{\lambda}_{F;r2}+\delta\wtilde{\lambda}_{F;r\ppr 2}\big)^{2}\Big] \\ \lb{A156}
M_{FF;rr\ppr}^{(11),11}&=&M_{FF;rr\ppr}^{(22),11}=M_{FF;rr\ppr}^{(11),22}=M_{FF;rr\ppr}^{(22),22}
\\ \no &=&\frac{1}{2}\;
\Big[\Big(A_{FF;rr\ppr}(|\ovv{f}_{r}|,|\ovv{f}_{r\ppr}|)\Big)^{2}+
\Big(B_{FF;rr\ppr}(|\ovv{f}_{r}|,|\ovv{f}_{r\ppr}|)\Big)^{2}\Big]\times \\ \no &\times&
\Big[\big(\delta\wtilde{\lambda}_{F;r1}+\delta\wtilde{\lambda}_{F;r\ppr 2}\big)^{2}+
\big(\delta\wtilde{\lambda}_{F;r2}+\delta\wtilde{\lambda}_{F;r\ppr 1}\big)^{2}\Big] \\ \lb{A157}
M_{FF;rr\ppr}^{(11),12}&=&M_{FF;rr\ppr}^{(22),21}=\Big(M_{FF;rr\ppr}^{(11),21}\Big)^{*}
=\Big(M_{FF;rr\ppr}^{(22),12}\Big)^{*}
\\ \no &=&\frac{\im}{2}\;\;
\Big[\Big(B_{FF;rr\ppr}(|\ovv{f}_{r}|,|\ovv{f}_{r\ppr}|)\Big)^{2}-
\Big(A_{FF;rr\ppr}(|\ovv{f}_{r}|,|\ovv{f}_{r\ppr}|)\Big)^{2}\Big]\times \\ \no &\times&
\Big[\big(\delta\wtilde{\lambda}_{F;r1}+\delta\wtilde{\lambda}_{F;r\ppr 2}\big)^{2}-
\big(\delta\wtilde{\lambda}_{F;r2}+\delta\wtilde{\lambda}_{F;r\ppr 1}\big)^{2}\Big]
\eeq
\beq \lb{A158}
M_{FF;rr\ppr}^{(12),00}&=&-M_{FF;rr\ppr}^{(12),33}=\Big(M_{FF;rr\ppr}^{(21),00}\Big)^{*}=-
\Big(M_{FF;rr\ppr}^{(21),33}\Big)^{*}
\\ \no &=&\exp\{\im\;(\phi_{r}+\phi_{r\ppr})\}\;\;
A_{FF;rr\ppr}(|\ovv{f}_{r}|,|\ovv{f}_{r\ppr}|)\;\;
B_{FF;rr\ppr}(|\ovv{f}_{r}|,|\ovv{f}_{r\ppr}|)\;\times \\ \no &\times&
\Big[\big(\delta\wtilde{\lambda}_{F;r1}+\delta\wtilde{\lambda}_{F;r\ppr 1}\big)^{2}+
\big(\delta\wtilde{\lambda}_{F;r2}+\delta\wtilde{\lambda}_{F;r\ppr 2}\big)^{2}\Big] \\  \lb{A159}
M_{FF;rr\ppr}^{(12),11}&=&M_{FF;rr\ppr}^{(12),22}=\Big(M_{FF;rr\ppr}^{(21),11}\Big)^{*}=
\Big(M_{FF;rr\ppr}^{(21),22}\Big)^{*}
\\ \no &=& - \;\exp\{\im\;(\phi_{r}+\phi_{r\ppr})\}\;\;
A_{FF;rr\ppr}(|\ovv{f}_{r}|,|\ovv{f}_{r\ppr}|)\;\;
B_{FF;rr\ppr}(|\ovv{f}_{r}|,|\ovv{f}_{r\ppr}|)\;\times \\ \no &\times&
\Big[\big(\delta\wtilde{\lambda}_{F;r1}+\delta\wtilde{\lambda}_{F;r\ppr 2}\big)^{2}+
\big(\delta\wtilde{\lambda}_{F;r2}+\delta\wtilde{\lambda}_{F;r\ppr 1}\big)^{2}\Big]
\eeq
\beq \lb{A160}
M_{FF;rr\ppr}^{(ab);(k\ppr,k=0,3)}&=&\left(
\bea{cccc}
M_{FF;rr\ppr}^{(11),00} & M_{FF;rr\ppr}^{(12),00} & M_{FF;rr\ppr}^{(11),03} & 0 \\
M_{FF;rr\ppr}^{(21),00} & M_{FF;rr\ppr}^{(22),00} & 0 & M_{FF;rr\ppr}^{(22),03} \\
M_{FF;rr\ppr}^{(11),30} & 0 & M_{FF;rr\ppr}^{(11),33} & M_{FF;rr\ppr}^{(12),33} \\
0 & M_{FF;rr\ppr}^{(22),30} & M_{FF;rr\ppr}^{(21),33} & M_{FF;rr\ppr}^{(22),33}
\eea\right) \\   \lb{A161}
M_{FF;rr\ppr}^{(ab);(k\ppr,k=1,2)}&=&\left(
\bea{cccc}
M_{FF;rr\ppr}^{(11),11} & M_{FF;rr\ppr}^{(12),11} & M_{FF;rr\ppr}^{(11),12} & 0 \\
M_{FF;rr\ppr}^{(21),11} & M_{FF;rr\ppr}^{(22),11} & 0 & M_{FF;rr\ppr}^{(22),12} \\
M_{FF;rr\ppr}^{(11),21} & 0 & M_{FF;rr\ppr}^{(11),22} & M_{FF;rr\ppr}^{(12),22} \\
0 & M_{FF;rr\ppr}^{(22),21} & M_{FF;rr\ppr}^{(21),22} & M_{FF;rr\ppr}^{(22),22}
\eea\right)
\eeq
\beq \lb{A162}
4\;\sqrt{\det\Big(M_{FF;rr}^{(ab)}\Big)}\;\big(\delta\lambda_{F;r1}+\delta\lambda_{F;r2}\big)^{2}&=&
\frac{\sinh\big(2\;|\ovv{f}_{r}|\big)}{|\ovv{f}_{r}|}\;\;
\big(\delta\lambda_{F;r1}+\delta\lambda_{F;r2}\big)^{2}
\eeq
\beq \lb{A163}
\lefteqn{\prod_{\{\vec{x},t_{p}\}}\Bigg\{\prod_{r=1}^{S/2}
\frac{d\hat{f}_{D;rr}^{(2)*}\wedge d\hat{f}_{D;rr}^{(2)}}{2\;\im}\;\;4\;
\sqrt{\det\Big(M_{FF;rr}^{(ab)}\Big)}\;
\big(\delta\lambda_{F;r1}+\delta\lambda_{F;r2}\big)^{2} \Bigg\}=} \\ \no &=&
\prod_{\{\vec{x},t_{p}\}}\Bigg\{\prod_{r=1}^{S/2}
\frac{d\hat{f}_{D;rr}^{(2)*}\wedge d\hat{f}_{D;rr}^{(2)}}{2\;\im}\;\;
\frac{\sinh\big(2\;|\ovv{f}_{r}|\big)}{|\ovv{f}_{r}|}\;\;
\big(\delta\lambda_{F;r1}+\delta\lambda_{F;r2}\big)^{2}\Bigg\}
\eeq
\beq \lb{A164}
\lefteqn{\sqrt{4^{4}\;\det\Big(M_{FF;rr\ppr}^{(ab);(k\ppr,k=0,3)}\Big)}=} \\ \no &=& 4
\left(\frac{\cosh\big(2\;|\ovv{f}_{r}|\big)-\cosh\big(2\;|\ovv{f}_{r\ppr}|\big)}{
|\ovv{f}_{r}|^{2}-|\ovv{f}_{r\ppr}|^{2}}\right)^{2}\;
\big(\delta\lambda_{F;r1}+\delta\lambda_{F;r\ppr 1}\big)^{2}\;
\big(\delta\lambda_{F;r2}+\delta\lambda_{F;r\ppr 2}\big)^{2} \\ \no &=& 16
\left(\frac{\sinh\big(|\ovv{f}_{r}|+|\ovv{f}_{r\ppr}|\big)}{
|\ovv{f}_{r}|+|\ovv{f}_{r\ppr}|}\right)^{2}\;
\left(\frac{\sinh\big(|\ovv{f}_{r}|-|\ovv{f}_{r\ppr}|\big)}{
|\ovv{f}_{r}|-|\ovv{f}_{r\ppr}|}\right)^{2}\;
\big(\delta\lambda_{F;r1}+\delta\lambda_{F;r\ppr 1}\big)^{2}\;
\big(\delta\lambda_{F;r2}+\delta\lambda_{F;r\ppr 2}\big)^{2}
\eeq
\beq   \lb{A165}
\lefteqn{\sqrt{4^{4}\;\det\Big(M_{FF;rr\ppr}^{(ab);(k\ppr,k=1,2)}\Big)}=} \\ \no &=& 4
\left(\frac{\cosh\big(2\;|\ovv{f}_{r}|\big)-\cosh\big(2\;|\ovv{f}_{r\ppr}|\big)}{
|\ovv{f}_{r}|^{2}-|\ovv{f}_{r\ppr}|^{2}}\right)^{2}\;
\big(\delta\lambda_{F;r2}+\delta\lambda_{F;r\ppr 1}\big)^{2}\;
\big(\delta\lambda_{F;r1}+\delta\lambda_{F;r\ppr 2}\big)^{2} \\ \no &=& 16
\left(\frac{\sinh\big(|\ovv{f}_{r}|+|\ovv{f}_{r\ppr}|\big)}{
|\ovv{f}_{r}|+|\ovv{f}_{r\ppr}|}\right)^{2}\;
\left(\frac{\sinh\big(|\ovv{f}_{r}|-|\ovv{f}_{r\ppr}|\big)}{
|\ovv{f}_{r}|-|\ovv{f}_{r\ppr}|}\right)^{2}\;
\big(\delta\lambda_{F;r2}+\delta\lambda_{F;r\ppr 1}\big)^{2}\;
\big(\delta\lambda_{F;r1}+\delta\lambda_{F;r\ppr 2}\big)^{2}
\eeq
\beq \lb{A166}
\lefteqn{\hspace*{-0.7cm}\prod_{\{\vec{x},t_{p}\}}\Bigg\{\prod_{r=1}^{S/2}\prod_{r\ppr=r+1}^{S/2}
\frac{d\hat{f}_{D;rr\ppr}^{(0)*}\wedge d\hat{f}_{D;rr\ppr}^{(0)}}{2\;\im}\;
\frac{d\hat{f}_{D;rr\ppr}^{(3)*}\wedge d\hat{f}_{D;rr\ppr}^{(3)}}{2\;\im}\;
\sqrt{4^{4}\det\Big(M_{FF;rr\ppr}^{(ab);(k\ppr,k=0,3)}\Big)}\Bigg\} = } \\ \no &=&
\prod_{\{\vec{x},t_{p}\}}\Bigg\{\prod_{r=1}^{S/2}\prod_{r\ppr=r+1}^{S/2}
\frac{d\hat{f}_{D;rr\ppr}^{(0)*}\wedge d\hat{f}_{D;rr\ppr}^{(0)}}{2\;\im}\;
\frac{d\hat{f}_{D;rr\ppr}^{(3)*}\wedge d\hat{f}_{D;rr\ppr}^{(3)}}{2\;\im}\times \\ \no &\times& 16
\left(\frac{\sinh\big(|\ovv{f}_{r}|+|\ovv{f}_{r\ppr}|\big)}{
|\ovv{f}_{r}|+|\ovv{f}_{r\ppr}|}\right)^{2}\;
\left(\frac{\sinh\big(|\ovv{f}_{r}|-|\ovv{f}_{r\ppr}|\big)}{
|\ovv{f}_{r}|-|\ovv{f}_{r\ppr}|}\right)^{2}\;
\big(\delta\lambda_{F;r1}+\delta\lambda_{F;r\ppr 1}\big)^{2}\;
\big(\delta\lambda_{F;r2}+\delta\lambda_{F;r\ppr 2}\big)^{2}\Bigg\}
\eeq
\beq  \lb{A167}
\lefteqn{\hspace*{-0.7cm}\prod_{\{\vec{x},t_{p}\}}\Bigg\{\prod_{r=1}^{S/2}\prod_{r\ppr=r+1}^{S/2}
\frac{d\hat{f}_{D;rr\ppr}^{(1)*}\wedge d\hat{f}_{D;rr\ppr}^{(1)}}{2\;\im}\;
\frac{d\hat{f}_{D;rr\ppr}^{(2)*}\wedge d\hat{f}_{D;rr\ppr}^{(2)}}{2\;\im}\;
\sqrt{4^{4}\;\det\Big(M_{FF;rr\ppr}^{(ab);(k\ppr,k=1,2)}\Big)}\Bigg\} = } \\ \no &=&
\prod_{\{\vec{x},t_{p}\}}\Bigg\{\prod_{r=1}^{S/2}\prod_{r\ppr=r+1}^{S/2}
\frac{d\hat{f}_{D;rr\ppr}^{(1)*}\wedge d\hat{f}_{D;rr\ppr}^{(1)}}{2\;\im}\;
\frac{d\hat{f}_{D;rr\ppr}^{(2)*}\wedge d\hat{f}_{D;rr\ppr}^{(2)}}{2\;\im}\times \\ \no &\times& 16
\left(\frac{\sinh\big(|\ovv{f}_{r}|+|\ovv{f}_{r\ppr}|\big)}{
|\ovv{f}_{r}|+|\ovv{f}_{r\ppr}|}\right)^{2}\;
\left(\frac{\sinh\big(|\ovv{f}_{r}|-|\ovv{f}_{r\ppr}|\big)}{
|\ovv{f}_{r}|-|\ovv{f}_{r\ppr}|}\right)^{2}\;
\big(\delta\lambda_{F;r2}+\delta\lambda_{F;r\ppr 1}\big)^{2}\;
\big(\delta\lambda_{F;r1}+\delta\lambda_{F;r\ppr 2}\big)^{2}\Bigg\} \;\;\;.
\eeq

The analogous procedure is performed for constructing sub-metric tensors in the 'BF', 'FB'
parts for the odd parameters \(d\hat{\eta}_{D;m,r\ppr\nu}\ppr\), \(d\hat{\eta}_{D;r\mu,n}^{\prime *}\)
(\ref{A84},\ref{A85}) from \(d\hat{Y}\ppr\) in \(\big(d\wtilde{Y}\big)^{\prime 12}_{BF;m,r\ppr\nu}\) ,
\(\big(d\wtilde{Y}\big)^{\prime 12,*}_{FB;r\ppr\nu,m}\) (\ref{A134})
with the trigonometric- and hyperbolic (co)sine-, (co)sinh- functions of the eigenvalues in the coefficients
\(A_{BF;mr\ppr}(|\ovv{c}_{m}|,|\ovv{f}_{r\ppr}|)\) ,
\(B_{BF;r\ppr m}(|\ovv{f}_{r\ppr}|,|\ovv{c}_{m}|)\) (\ref{A135},\ref{A136}).
The {\it inverse} square root of the sub-metric tensors yields the coset integration measure
for the Grassmann variables (\ref{A173},\ref{A174})
because the 'FF' section of a super-Jacobi-matrix is considered with
its sub-determinant in the {\it denominator} of the super-determinant
\beq \lb{A168}
\lefteqn{\Big(d s_{Osp\backslash U}\big(\delta\hat{\lambda}\big)\Big)_{BF,FB}^{2} = 4
\sum_{m=1}^{L}\sum_{r\ppr=1(\nu=1,2)}^{S/2}
\Big(d\wtilde{Y}\Big)^{\prime 12}_{BF;m,r\ppr\nu}\;\;
\Big(d\wtilde{Y}\Big)^{\prime 12,*}_{FB;r\ppr\nu,m}\;\;
\big(\delta\wtilde{\lambda}_{F;r\ppr\nu}+\delta\wtilde{\lambda}_{B;m}\big)^{2}   }
\\ \no &=& -4
\sum_{m=1}^{L}\sum_{r\ppr=1(\nu=1,2)}^{S/2}\sum_{\kappa,\lambda=1,2}
\bigg[d\hat{\eta}_{D;m,r\ppr\lambda}^{\prime *}\;\big(\tau_{2}\big)_{\lambda\nu}\;
\exp\{\im\;(\varphi_{m}+\phi_{r\ppr})\}\;B_{BF;r\ppr m}(|\ovv{f}_{r\ppr}|,|\ovv{c}_{m}|) + \\ \no &+&
d\hat{\eta}_{D;m,r\ppr \nu}\ppr\;A_{BF;mr\ppr}(|\ovv{c}_{m}|,|\ovv{f}_{r\ppr}|)\bigg] \times
\bigg[\big(\tau_{2}\big)_{\nu\kappa}\;d\hat{\eta}_{D;r\ppr\kappa,m}\ppr\;
\exp\{-\im\;(\phi_{r\ppr}+\varphi_{m})\}\;B_{BF;r\ppr m}(|\ovv{f}_{r\ppr}|,|\ovv{c}_{m}|)+ \\ \no &+&
d\hat{\eta}_{D;r\ppr\nu,m}^{\prime *}\;A_{BF;mr\ppr}(|\ovv{c}_{m}|,|\ovv{f}_{r\ppr}|)\bigg]\times
\big(\delta\wtilde{\lambda}_{F;r\ppr\nu}+\delta\wtilde{\lambda}_{B;m}\big)^{2}=
\\ \no &=& -4
\sum_{m=1}^{L}\sum_{r\ppr=1}^{S/2}
\Bigg\{d\hat{\eta}_{D;m,r\ppr 1}^{\prime *}\;d\hat{\eta}_{D;r\ppr 1,m}\ppr\times
\bigg[\Big(B_{BF;r\ppr m}(|\ovv{f}_{r\ppr}|,|\ovv{c}_{m}|)\Big)^{2}\;\;
\big(\delta\wtilde{\lambda}_{F;r\ppr 2}+\delta\wtilde{\lambda}_{B;m}\big)^{2}+ \\ \no   &-&
\Big(A_{BF;mr\ppr}(|\ovv{c}_{m}|,|\ovv{f}_{r\ppr}|)\Big)^{2}\;\;
\big(\delta\wtilde{\lambda}_{F;r\ppr 1}+\delta\wtilde{\lambda}_{B;m}\big)^{2}\bigg] + \\ \no &+&
d\hat{\eta}_{D;m,r\ppr 2}^{\prime *}\;d\hat{\eta}_{D;r\ppr 2,m}\ppr\times
\bigg[\Big(B_{BF;r\ppr m}(|\ovv{f}_{r\ppr}|,|\ovv{c}_{m}|)\Big)^{2}\;\;
\big(\delta\wtilde{\lambda}_{F;r\ppr 1}+\delta\wtilde{\lambda}_{B;m}\big)^{2} + \\ \no &-&
\Big(A_{BF;mr\ppr}(|\ovv{c}_{m}|,|\ovv{f}_{r\ppr}|)\Big)^{2}\;\;
\big(\delta\wtilde{\lambda}_{F;r\ppr 2}+\delta\wtilde{\lambda}_{B;m}\big)^{2}\bigg] +
\\ \no &+&\im\;
d\hat{\eta}_{D;m,r\ppr 2}^{\prime *}\;d\hat{\eta}_{D;r\ppr 1,m}^{\prime *}\;\;
A_{BF;mr\ppr}(|\ovv{c}_{m}|,|\ovv{f}_{r\ppr}|)\;\;
B_{BF;r\ppr m}(|\ovv{f}_{r\ppr}|,|\ovv{c}_{m}|)\;\;\exp\{\im(\varphi_{m}+\phi_{r\ppr})\}\times \\ \no &\times&
\Big[\big(\delta\wtilde{\lambda}_{F;r\ppr 1}+\delta\wtilde{\lambda}_{B;m}\big)^{2}+
\big(\delta\wtilde{\lambda}_{F;r\ppr 2}+\delta\wtilde{\lambda}_{B;m}\big)^{2}\Big] +
\\ \no &-&\im\;
d\hat{\eta}_{D;m,r\ppr 1}^{\prime}\;d\hat{\eta}_{D;r\ppr 2,m}^{\prime}\;\;
A_{BF;mr\ppr}(|\ovv{c}_{m}|,|\ovv{f}_{r\ppr}|)\;\;
B_{BF;r\ppr m}(|\ovv{f}_{r\ppr}|,|\ovv{c}_{m}|)\;\;\exp\{-\im(\phi_{r\ppr}+\varphi_{m})\}\times \\ \no &\times&
\Big[\big(\delta\wtilde{\lambda}_{F;r\ppr 1}+\delta\wtilde{\lambda}_{B;m}\big)^{2}+
\big(\delta\wtilde{\lambda}_{F;r\ppr 2}+\delta\wtilde{\lambda}_{B;m}\big)^{2}\Big]\Bigg\}= \\ \no &=&-4
\sum_{m=1}^{L}\sum_{r\ppr=1}^{S/2}\sum_{(\mu,\nu=1,2)}
\left(\bea{c}
d\hat{\eta}_{D;r\ppr\nu,m}^{\prime} \\ d\hat{\eta}_{D;r\ppr\nu,m}^{\prime *}
\eea\right)^{+}\left(\bea{cc}
M_{BF;m,r\ppr}^{(11),\nu\mu} & M_{BF;m,r\ppr}^{(12),\nu\mu} \\
M_{BF;m,r\ppr}^{(21),\nu\mu} & M_{BF;m,r\ppr}^{(22),\nu\mu}
\eea\right)
\left(\bea{c}
d\hat{\eta}_{D;r\ppr\mu,m}^{\prime} \\ d\hat{\eta}_{D;r\ppr\mu,m}^{\prime *}
\eea\right)
\eeq
\beq \lb{A169}
M_{BF;m,r\ppr}^{(11),11}&=&-M_{BF;m,r\ppr}^{(22),11} =
\frac{1}{2}\bigg[\Big(B_{BF;r\ppr m}(|\ovv{f}_{r\ppr}|,|\ovv{c}_{m}|)\Big)^{2}\;
\big(\delta\wtilde{\lambda}_{F;r\ppr 2}+\delta\wtilde{\lambda}_{B;m}\big)^{2} + \\ \no &-&
\Big(A_{BF;mr\ppr}(|\ovv{c}_{m}|,|\ovv{f}_{r\ppr}|)\Big)^{2}\;
\big(\delta\wtilde{\lambda}_{F;r\ppr 1}+\delta\wtilde{\lambda}_{B;m}\big)^{2}\bigg] \\   \lb{A170}
M_{BF;m,r\ppr}^{(11),22}&=&-M_{BF;m,r\ppr}^{(22),22} =
\frac{1}{2}\bigg[\Big(B_{BF;r\ppr m}(|\ovv{f}_{r\ppr}|,|\ovv{c}_{m}|)\Big)^{2}\;
\big(\delta\wtilde{\lambda}_{F;r\ppr 1}+\delta\wtilde{\lambda}_{B;m}\big)^{2} + \\ \no &-&
\Big(A_{BF;mr\ppr}(|\ovv{c}_{m}|,|\ovv{f}_{r\ppr}|)\Big)^{2}\;
\big(\delta\wtilde{\lambda}_{F;r\ppr 2}+\delta\wtilde{\lambda}_{B;m}\big)^{2}\bigg] \\  \lb{A171}
M_{BF;m,r\ppr}^{(12),21}&=&-M_{BF;m,r\ppr}^{(12),12}=\Big(M_{BF;m,r\ppr}^{(21),12}\Big)^{*}=
-\Big(M_{BF;m,r\ppr}^{(21),21}\Big)^{*}= \\ \no &=&\frac{\im}{2}\;\;
A_{BF;mr\ppr}(|\ovv{c}_{m}|,|\ovv{f}_{r\ppr}|)\;\;
B_{BF;r\ppr m}(|\ovv{f}_{r\ppr}|,|\ovv{c}_{m}|) \times \\ \no &\times&
\exp\{\im(\varphi_{m}+\phi_{r\ppr})\}\;
\Big[\big(\delta\wtilde{\lambda}_{F;r\ppr 1}+\delta\wtilde{\lambda}_{B;m}\big)^{2}+
\big(\delta\wtilde{\lambda}_{F;r\ppr 2}+\delta\wtilde{\lambda}_{B;m}\big)^{2}\Big]
\eeq
\beq \lb{A172}
M_{BF;m,r\ppr}^{(ab);(\nu,\mu=1,2)}&=&\left(
\bea{cccc}
M_{BF;m,r\ppr}^{(11),11} & 0 & 0 & M_{BF;m,r\ppr}^{(12),12} \\
0 & M_{BF;m,r\ppr}^{(22),11} & M_{BF;m,r\ppr}^{(21),12} & 0 \\
0 & M_{BF;m,r\ppr}^{(12),21} & M_{BF;m,r\ppr}^{(11),22} & 0 \\
M_{BF;m,r\ppr}^{(21),21} & 0 & 0 & M_{BF;m,r\ppr}^{(22),22}
\eea\right)
\eeq
\beq \lb{A173}
\lefteqn{\bigg((-4)^{4}\det\Big(M_{BF;m,r\ppr}^{(ab);(\nu,\mu=1,2)}\Big)\bigg)^{\mbox{\boldmath{$-1/2$}}} = }
\\ \no &=&
\left(\frac{\Big(\cosh\big(2\;|\ovv{f}_{r\ppr}|\big)-\cosh\big(2\im\;|\ovv{c}_{m}|\big)\Big)^{2}}{
\Big(|\ovv{f}_{r\ppr}|^{2}-\big(\im\;|\ovv{c}_{m}|\big)^{2}\Big)^{2}}\;\;
\Big(\delta\lambda_{F;r\ppr 1}+\delta\lambda_{B;m}\Big)^{2}\;
\Big(\delta\lambda_{F;r\ppr 2}+\delta\lambda_{B;m}\Big)^{2}\right)^{\mbox{\boldmath{$-1$}}} = \\ \no &=&\left(2\;
\left|\frac{\sinh\big(|\ovv{f}_{r\ppr}|+\im\;|\ovv{c}_{m}|\big)}{|\ovv{f}_{r\ppr}|+\im\;|\ovv{c}_{m}|}\right|\;\;
\left|\frac{\sinh\big(|\ovv{f}_{r\ppr}|-\im\;|\ovv{c}_{m}|\big)}{|\ovv{f}_{r\ppr}|-\im\;|\ovv{c}_{m}|}\right|\;\;
\right)^{\mbox{\boldmath{$-2$}}} \times  \\ \no &\times&
\left(\Big(\delta\lambda_{F;r\ppr 1}+\delta\lambda_{B;m}\Big)^{2}\;
\Big(\delta\lambda_{F;r\ppr 2}+\delta\lambda_{B;m}\Big)^{2}\right)^{\mbox{\boldmath{$-1$}}}
\eeq
\beq  \lb{A174}
\lefteqn{\prod_{\{\vec{x},t_{p}\}}\Bigg\{\prod_{m=1}^{L}\prod_{r\ppr=1}^{S/2}
\frac{d\hat{\eta}_{D;r\ppr 1,m}^{*}\; d\hat{\eta}_{D;r\ppr 1,m}\;\;
d\hat{\eta}_{D;r\ppr 2,m}^{*}\; d\hat{\eta}_{D;r\ppr 2,m}}{
\sqrt{(-4)^{4}\;\det\Big(M_{BF;m,r\ppr}^{(ab);(\nu,\mu=1,2)}\Big)}}\Bigg\}= } \\ \no &=&
\prod_{\{\vec{x},t_{p}\}}\Bigg\{\prod_{m=1}^{L}\prod_{r\ppr=1}^{S/2}
d\hat{\eta}_{D;r\ppr 1,m}^{*}\; d\hat{\eta}_{D;r\ppr 1,m}\;\;
d\hat{\eta}_{D;r\ppr 2,m}^{*}\; d\hat{\eta}_{D;r\ppr 2,m}\;\;\times \\ \no &\times&
\left(2\;\left|\frac{\sinh\big(|\ovv{f}_{r\ppr}|+\im\;|\ovv{c}_{m}|\big)}{|\ovv{f}_{r\ppr}|+\im\;|\ovv{c}_{m}|}\right|\;\;
\left|\frac{\sinh\big(|\ovv{f}_{r\ppr}|-\im\;|\ovv{c}_{m}|\big)}{|\ovv{f}_{r\ppr}|-\im\;|\ovv{c}_{m}|}\right|
\right)^{\mbox{\boldmath{$-2$}}}  \times\\ \no &\times &
\left(\Big(\delta\lambda_{F;r\ppr 1}+\delta\lambda_{B;m}\Big)^{2}\;
\Big(\delta\lambda_{F;r\ppr 2}+\delta\lambda_{B;m}\Big)^{2}\right)^{\mbox{\boldmath{$-1$}}} \Bigg\}
\eeq

\subsection{Invariant integration measure for
$\delta\hat{\Sigma}_{D;2N\times 2N}\;\wtilde{K}=
\hat{Q}_{2N\times 2N}^{-1}\;\delta\hat{\Lambda}_{2N\times 2N}\;\hat{Q}_{2N\times 2N}$
of $U(L|S)$} \lb{sa4}

In this subsection \ref{sa4} we combine the various factors of the integration measure
of the coset decomposition \(Osp(S,S|2L)\backslash U(L|S)\). Apart from the action
\(\mcal{A}_{\hat{J}_{\psi\psi}}[\hat{T}]\) (\ref{s4_34}-\ref{s4_38}), the \(U(L|S)\) subgroup part of densities
with matrices \(\delta\hat{\Sigma}_{D;2N\times 2N}^{aa}\;\wtilde{K}\),
\(\hat{Q}_{\alpha\beta}^{aa}\), \(\delta\hat{\Lambda}_{\alpha}^{a}\) does not appear
in the remaining actions of the effective coherent state path integral (\ref{s5_6}) for the
anomalous fields with coefficients averaged over the background density field
\(\sigma_{D}^{(0)}(\vec{x},t_{p})\) (\ref{s4_90},\ref{s4_95}). Therefore, we neglect the detailed integration
measure of \(U(L|S)\) subgroup or 'hinge' fields for the action
\(\mcal{A}_{\hat{J}_{\psi\psi}}[\hat{T}]\) (\ref{s4_34}-\ref{s4_38}) which determines the creation
of the pair condensate fields \(\hat{Y}(\vec{x},t_{p})\) out of the vacuum.
If one omits this detailed creation process and initial preparation of pair condensate
terms, the derived actions \(\mcal{A}_{\mcal{N}^{-1}}\ppr[\hat{T}]\),
\(\mcal{A}_{\mcal{N}^{0}}\ppr[\hat{T}]\), \(\mcal{A}_{\mcal{N}^{+1}}\ppr[\hat{T}]\) (\ref{s5_7}-\ref{s5_14})
from the gradient expansion remain for investigation of the dynamics of gradually
varying pair condensates, immersed in an environment described by the density
field \(\sigma_{D}^{(0)}(\vec{x},t_{p})\) (\ref{s4_90},\ref{s4_95}).
Therefore, we list in Eqs. (\ref{A175}-\ref{A183})
the \(U(L|S)\) subgroup integration measure without further reduction
\beq \lb{A175}
\Big(d s_{U}\Big)^{2}&=&
2\;\mbox{str}\Big[d\big(\delta\wtilde{\lambda}_{\alpha}\big)\;\;
d\big(\delta\wtilde{\lambda}_{\alpha}\big)\Big] -
2\;\strab\Big[\big(\wtilde{T}_{0}^{-1}\;d\wtilde{T}_{0}\big)_{\alpha\beta}^{11}\;\;
\big(\wtilde{T}_{0}^{-1}\;d\wtilde{T}_{0}\big)_{\beta\alpha}^{11}\;\;
\big(\delta\wtilde{\lambda}_{\beta}-\delta\wtilde{\lambda}_{\alpha}\big)^{2}\Big] \\ \lb{A176}
\Big(\wtilde{T}_{0}^{-1}\;d\wtilde{T}_{0}\Big)_{\alpha\beta}^{aa} &\simeq &
-\Big(\big(\hat{P}\hat{Q}^{-1}\big)\;\;\big(d\hat{Q}\;\hat{Q}^{-1}\big)\;\;
\big(\hat{P}\hat{Q}^{-1}\big)^{-1}\Big)_{\alpha\beta}^{aa} \\   \lb{A177}
d\big(\delta\wtilde{\Lambda}\big)&=&\wtilde{\Lambda}^{(\alpha)}(p_{\kappa},q_{\kappa})\;\;
d\big(\delta\lambda_{\alpha}\big) \\   \lb{A178}
\wtilde{\Lambda}^{(\alpha)}(p_{\kappa},q_{\kappa})&=&
\big(\hat{P}\hat{Q}^{-1}\big)\;\;\hat{\Lambda}^{(\alpha)}\;\;
\big(\hat{P}\hat{Q}^{-1}\big)^{-1}   \\       \lb{A179}
\Big(d s_{U}\Big)^{2}&=&
2\;\mbox{str}\Big[d\big(\delta\hat{\lambda}_{\alpha}\big)\;\;
d\big(\delta\hat{\lambda}_{\alpha}\big)\Big] -
2\;\strab\Big[\big(d\hat{Q}\;\hat{Q}^{-1}\big)_{\alpha\beta}^{11}\;\;
\big(d\hat{Q}\;\hat{Q}^{-1}\big)_{\beta\alpha}^{11}\;\;
\big(\delta\hat{\lambda}_{\beta}-\delta\hat{\lambda}_{\alpha}\big)^{2}\Big]
\eeq
\beq  \lb{A180}
d\Big(\hat{Q}^{-1}\;\delta\hat{\Lambda}\;\hat{Q}\Big)_{2N\times 2N}&=&
d\Big(\delta\hat{\Sigma}_{D;2N\times 2N}\;\wtilde{K}\Big) =
\hat{Q}^{-1}\;\bigg(\Big[\delta\hat{\Lambda}\;,\;d\hat{Q}\;\hat{Q}^{-1}\Big]+d\Big(\delta\hat{\Lambda}\Big)\bigg)
\;\hat{Q} \\ \lb{A181}
\Big(d s_{U}\Big)^{2}&=&\STRAB\bigg[
d\Big(\hat{Q}^{-1}\;\delta\hat{\Lambda}\;\hat{Q}\Big)_{\alpha\beta}^{aa}\;\;
d\Big(\hat{Q}^{-1}\;\delta\hat{\Lambda}\;\hat{Q}\Big)_{\beta\alpha}^{aa}\bigg] \\ \no &=&
\STRAB\bigg[d\Big(\delta\hat{\Sigma}_{D;2N\times 2N}\;\wtilde{K}\Big)_{\alpha\beta}^{aa}\;\;
d\Big(\delta\hat{\Sigma}_{D;2N\times 2N}\;\wtilde{K}\Big)_{\beta\alpha}^{aa}\bigg]
\eeq
\beq \no
\lefteqn{d\big[d\hat{Q}(t_{p})\;\hat{Q}^{-1}(t_{p}),\delta\hat{\lambda}(t_{p})\big]=
d\big[\delta\hat{\Sigma}_{D}(t_{p})\;\wtilde{K}\big]=
\prod_{\{\vec{x},t_{p}\}}\bigg\{2^{(L+S)/2}\;
\bigg(\prod_{m=1}^{L}d\big(\delta\lambda_{B;m}\big)\bigg)
\bigg(\prod_{i=1}^{S}d\big(\delta\lambda_{F;i}\big)\bigg)\bigg\}\times } \\ \no &\times&
\prod_{\{\vec{x},t_{p}\}}\bigg\{
\prod_{m=1}^{L}\prod_{n=m+1}^{L}\bigg(4\;
\frac{\big(d\hat{Q}^{11}\;\hat{Q}^{11,-1}\big)_{BB;mn}\wedge
\big(d\hat{Q}^{11}\;\hat{Q}^{11,-1}\big)_{BB;nm}}{2\;\im}\;
\big(\delta\hat{\lambda}_{B;n}-\delta\hat{\lambda}_{B;m}\big)^{2}\bigg)\bigg\}
\\ \lb{A182} &\times&
\prod_{\{\vec{x},t_{p}\}}\bigg\{
\prod_{i=1}^{S}\prod_{i\ppr=i+1}^{S}\bigg(4\;
\frac{\big(d\hat{Q}^{11}\;\hat{Q}^{11,-1}\big)_{FF;ii\ppr}\wedge
\big(d\hat{Q}^{11}\;\hat{Q}^{11,-1}\big)_{FF;i\ppr i}}{2\;\im}\;
\big(\delta\hat{\lambda}_{F;i\ppr}-\delta\hat{\lambda}_{F;i}\big)^{2}\bigg)\bigg\}
\\ \no &\times&
\prod_{\{\vec{x},t_{p}\}}\bigg\{
\prod_{m=1}^{L}\prod_{i\ppr=1}^{S}\bigg(\frac{1}{4}\;
\big(d\hat{Q}^{11}\;\hat{Q}^{11,-1}\big)_{BF;mi\ppr}\;
\big(d\hat{Q}^{11}\;\hat{Q}^{11,-1}\big)_{FB;i\ppr m}\;
\big(\delta\hat{\lambda}_{F;i\ppr}-
\delta\hat{\lambda}_{B;m}\big)^{\mbox{\boldmath{$^{-2}$}}}\bigg)\bigg\}
\eeq
\beq \no
\lefteqn{d\big[\delta\hat{\Sigma}_{D}(t_{p})\;\wtilde{K}\big]=
d\big[d\hat{Q}(t_{p})\;\hat{Q}^{-1}(t_{p}),\delta\hat{\lambda}(t_{p})\big] =
\prod_{\{\vec{x},t_{p}\}}\bigg\{2^{(L+S)/2}\;
\bigg(\prod_{m=1}^{L}d\big(\delta\hat{B}_{D;mm}\big)\bigg)
\bigg(\prod_{i=1}^{S}d\big(\delta\hat{F}_{D;ii}\big)\bigg)\bigg\}\times} \\ \no &\times&
\prod_{\{\vec{x},t_{p}\}}\left\{\bigg(
\prod_{m=1}^{L}\prod_{n=m+1}^{L}4\;\;
\frac{d\big(\delta\hat{B}_{D;mn}^{*}\big)\;\wedge\;
d\big(\delta\hat{B}_{D;mn}\big)}{2\;\im}\bigg)   \bigg(
\prod_{i=1}^{S}\prod_{i\ppr=i+1}^{S}4\;\;
\frac{d\big(\delta\hat{F}_{D;ii\ppr}^{*}\big)\;\wedge\;
d\big(\delta\hat{F}_{D;ii\ppr}\big)}{2\;\im}\bigg)\right\}
\\ \lb{A183} &\times&
\prod_{\{\vec{x},t_{p}\}}\bigg\{
\prod_{m=1}^{L}\prod_{i\ppr=1}^{S}\bigg(\frac{1}{4}\;\;
d\big(\delta\hat{\chi}_{D;mi\ppr}^{*}\big)\;\;
d\big(\delta\hat{\chi}_{D;i\ppr m}\big)\bigg)\bigg\}\;\;\;.
\eeq
The complete integration measure (\ref{A184}) consists of the background density field
and the coset part \(d\big[\hat{T}^{-1}(t_{p})\;d\hat{T}(t_{p});\delta\hat{\lambda}(t_{p})\big]\)
with a polynomial \(\mcal{P}\big(\delta\hat{\lambda}(t_{p})\big)\) (\ref{A186})
of the eigenvalues of the subgroup part, following from the various sub-metric tensors with
their (inverse) square root of sub-determinants
\beq \no
d\big[\hat{\sigma}_{D}^{(0)}(t_{p})\big]\;\;d\big[\delta\wtilde{\Sigma}(t_{p})\;\wtilde{K}\big]&=&
d\big[\hat{\sigma}_{D}^{(0)}(t_{p})\big]\;\;
d\big[d\hat{Q}(t_{p})\;\hat{Q}^{-1}(t_{p});\delta\hat{\lambda}(t_{p})\big]
\;\;d\big[\hat{T}^{-1}(t_{p})\;d\hat{T}(t_{p});\delta\hat{\lambda}(t_{p})\big] \\  \lb{A184} &=&
d\big[\hat{\sigma}_{D}^{(0)}(t_{p})\big]\;\;d\big[\delta\hat{\Sigma}_{D}(t_{p})\;\wtilde{K}\big]\;\;
d\big[\hat{T}^{-1}(t_{p})\;d\hat{T}(t_{p});\delta\hat{\lambda}(t_{p})\big]
\eeq
\be \lb{A185}
d\big[\hat{T}^{-1}(t_{p})\;d\hat{T}(t_{p});\delta\hat{\lambda}(t_{p})\big]=
\mcal{P}\big(\delta\hat{\lambda}(t_{p})\big)\;\;\;
d\big[\hat{T}^{-1}(t_{p})\;d\hat{T}(t_{p})\big]
\ee
\beq \no
\lefteqn{\mcal{P}\big(\delta\hat{\lambda}(t_{p})\big)=\prod_{\{\vec{x},t_{p}\}}\Bigg\{
\bigg(\prod_{m=1}^{L}\big(\delta\hat{\lambda}_{B;m}\big)^{2}\bigg)
\bigg(\prod_{r=1}^{S/2}\big(\delta\hat{\lambda}_{F;r1}+\delta\hat{\lambda}_{F;r2}\big)^{2}\bigg) \;\;
\bigg(\prod_{m=1}^{L}\prod_{n=m+1}^{L}
\big(\delta\hat{\lambda}_{B;n}+\delta\hat{\lambda}_{B;m}\big)^{2}\bigg) }
\\ \no &\times& \bigg(\prod_{r=1}^{S/2}\prod_{r\ppr=r+1}^{S/2}
\big(\delta\hat{\lambda}_{F;r1}+\delta\hat{\lambda}_{F;r\ppr 1}\big)^{2}\;\;
\big(\delta\hat{\lambda}_{F;r2}+\delta\hat{\lambda}_{F;r\ppr 2}\big)^{2} \;\;
\big(\delta\hat{\lambda}_{F;r2}+\delta\hat{\lambda}_{F;r\ppr 1}\big)^{2}\;\;
\big(\delta\hat{\lambda}_{F;r1}+\delta\hat{\lambda}_{F;r\ppr 2}\big)^{2} \bigg) \\ \lb{A186} &\times&
\bigg(\prod_{m=1}^{L}\prod_{r\ppr=1}^{S/2}
\big(\delta\hat{\lambda}_{F;r\ppr 1}+\delta\hat{\lambda}_{B;m}\big)^{2}\;\;
\big(\delta\hat{\lambda}_{F;r\ppr 2}+\delta\hat{\lambda}_{B;m}\big)^{2}
\bigg)^{\mbox{\boldmath{$-1$}}}\Bigg\} \;\;\; .
\eeq
We separate this polynomial
\(\mcal{P}\big(\delta\hat{\lambda}(t_{p})\big)\) (\ref{A186}) from the coset integration
measure of anomalous fields and move it to the integration measure of the subgroup parts
so that the 'relevant' coset integration measure
\(d\big[\hat{T}^{-1}(t_{p})\;d\hat{T}(t_{p})\big]\) (\ref{A187}) remains for the
pair condensate fields in the coherent state path integral (\ref{s5_6}) with effective actions
\(\mcal{A}_{\mcal{N}^{-1}}\ppr[\hat{T}]\), \(\mcal{A}_{\mcal{N}^{0}}\ppr[\hat{T}]\),
\(\mcal{A}_{\mcal{N}^{+1}}\ppr[\hat{T}]\) (\ref{s5_7}-\ref{s5_14})
\beq  \lb{A187}
\lefteqn{d\big[\hat{T}^{-1}(t_{p})\;d\hat{T}(t_{p})\big]=} \\ \no &=&
\prod_{\{\vec{x},t_{p}\}}\Bigg\{\prod_{m=1}^{L}\bigg(
\frac{d\hat{c}_{D;mm}^{*}\wedge d\hat{c}_{D;mm}}{2\;\im}\;\;2\;
\left|\frac{\sin\big(2\;|\ovv{c}_{m}|\big)}{|\ovv{c}_{m}|}\right|\bigg)\Bigg\}
\prod_{\{\vec{x},t_{p}\}}\Bigg\{\prod_{r=1}^{S/2}\bigg(
\frac{d\hat{f}_{D;rr}^{(2)*}\wedge d\hat{f}_{D;rr}^{(2)}}{2\;\im}\;\;
\frac{\sinh\big(2\;|\ovv{f}_{r}|\big)}{|\ovv{f}_{r}|}\bigg)\Bigg\}
\\ \no &\times&
\prod_{\{\vec{x},t_{p}\}}\Bigg\{\prod_{m=1}^{L}\prod_{n=m+1}^{L}\bigg(
\frac{d\hat{c}_{D;mn}^{*}\wedge d\hat{c}_{D;mn}}{2\;\im}\;\;2\;
\left|\frac{\sin\big(|\ovv{c}_{m}|+|\ovv{c}_{n}|\big)}{|\ovv{c}_{m}|+|\ovv{c}_{n}|}\right|\;\;
\left|\frac{\sin\big(|\ovv{c}_{m}|-|\ovv{c}_{n}|\big)}{|\ovv{c}_{m}|-|\ovv{c}_{n}|}\right| \bigg)\Bigg\}
\\ \no &\times&
\prod_{\{\vec{x},t_{p}\}}\Bigg\{\prod_{r=1}^{S/2}\prod_{r\ppr=r+1}^{S/2}\prod_{k=0}^{3}\bigg(
\frac{d\hat{f}_{D;rr\ppr}^{(k)*}\wedge d\hat{f}_{D;rr\ppr}^{(k)}}{2\;\im}\;4\;
\left|\frac{\sinh\big(|\ovv{f}_{r}|+|\ovv{f}_{r\ppr}|\big)}{
|\ovv{f}_{r}|+|\ovv{f}_{r\ppr}|}\right|\;
\left|\frac{\sinh\big(|\ovv{f}_{r}|-|\ovv{f}_{r\ppr}|\big)}{
|\ovv{f}_{r}|-|\ovv{f}_{r\ppr}|}\right|\bigg)\Bigg\}   \\ \no &\times&
\prod_{\{\vec{x},t_{p}\}}\Bigg\{\prod_{m=1}^{L}\prod_{r\ppr=1}^{S/2}
\frac{d\hat{\eta}_{D;r\ppr 1,m}^{*}\; d\hat{\eta}_{D;r\ppr 1,m}\;\;
d\hat{\eta}_{D;r\ppr 2,m}^{*}\; d\hat{\eta}_{D;r\ppr 2,m}}{ {\ds
\left(2\;\left|\frac{\sinh\big(|\ovv{f}_{r\ppr}|+\im\;|\ovv{c}_{m}|\big)}{|\ovv{f}_{r\ppr}|+\im\;|\ovv{c}_{m}|}\right|\;\;
\left|\frac{\sinh\big(|\ovv{f}_{r\ppr}|-\im\;|\ovv{c}_{m}|\big)}{|\ovv{f}_{r\ppr}|-\im\;|\ovv{c}_{m}|}\right|
\right)^{\mbox{\boldmath{$2$}}}  }}   \Bigg\}_{\mbox{.}}
\eeq

\section{Determination of matrix elements in the gradient expansion} \lb{sb}

\subsection{Simplifying matrix elements of Green functions with
unsaturated spatial derivatives} \lb{sb1}

We transform and reduce the matrix elements \(C_{\beta\alpha;j}^{b\neq a}\),
\(C_{\beta\alpha}^{b\neq a}(\mbox{{\boldmath$\wtilde{\pp}_{i}$}})\),
\(C_{\beta\alpha;j}^{b\neq a}(\mbox{{\boldmath$\wtilde{\pp}_{i}$}},
\mbox{{\boldmath$\wtilde{\pp}_{j}$}})\)
(\ref{s4_134}-\ref{s4_136}) for \(\mcal{A}_{SDET}\ppr\) with
increasing order of unsaturated spatial gradients according to the commutators (\ref{s4_123},\ref{s4_124}) and
with the rules of propagation for Green functions defined in subsection \ref{s44}. In order to apply
the rules (\ref{s4_116},\ref{s4_117}) in subsection \ref{s44}, we consider the separation of
\(\hat{T}^{-1}(\vec{x},T_{p})\;\big(\wtilde{\pp}_{i}\hat{T}(\vec{x},t_{p})\big)\) into the generators
\(\hat{H}^{(\kappa)}\) of the subgroup \(U(L|S)\) and into the generators \(\hat{Y}^{(\kappa)}\)
of the coset space (compare with appendix \ref{sa2}, Eqs. (\ref{A59}-\ref{A74})).
The corresponding functions \(g_{\kappa}(\vec{x},t_{p})\) and \(s_{\kappa}(\vec{x},t_{p})\) depend
on the bosonic and fermionic pair condensate fields. We briefly list in Eqs. (\ref{B1}-\ref{B6})
the results from appendix \ref{sa} for the purpose of
simplifying the matrix elements (\ref{s4_134}-\ref{s4_136})
\beq \lb{B1}
\Big[\hat{T}^{-1}(\vec{x},t_{p})\;\Big(\wtilde{\pp}_{i}\hat{T}(\vec{x},t_{p})\Big)\Big]_{\beta\alpha}^{ba} &=&
\im\;\Big[\Big(\wtilde{\pp}_{i}g_{\kappa}(\vec{x},t_{p})\Big)\;\hat{H}^{(\kappa)}\Big]_{\beta\alpha}^{ba}\;
\delta_{a,b}-\Big[\Big(\wtilde{\pp}_{i}s_{\kappa}(\vec{x},t_{p})\Big)\;
\hat{Y}^{(\kappa)}\Big]_{\beta\alpha}^{ba}\bigg|^{b\neq a}  \\  \lb{B2}
d\hat{T}(\vec{x},t_{p}) &=&
d\wtilde{x}^{i}\;\;\big(\wtilde{\pp}_{i}\hat{T}(\vec{x},t_{p})\big)    \\  \lb{B3}
g_{\kappa}\big(\vec{x},t_{p}\big) &=&g_{\kappa}\Big(\hat{c}_{D}(\vec{x},t_{p}),\hat{f}_{D}(\vec{x},t_{p}),
\hat{c}_{D}^{+}(\vec{x},t_{p}),\hat{f}_{D}^{+}(\vec{x},t_{p}),
\hat{\eta}_{D}(\vec{x},t_{p}),\hat{\eta}_{D}^{+}(\vec{x},t_{p})\Big)  \\ \lb{B4}
s_{\kappa}\big(\vec{x},t_{p}\big) &=& s_{\kappa}\Big(\hat{c}_{D}(\vec{x},t_{p}),\hat{f}_{D}(\vec{x},t_{p}),
\hat{c}_{D}^{+}(\vec{x},t_{p}),\hat{f}_{D}^{+}(\vec{x},t_{p}),
\hat{\eta}_{D}(\vec{x},t_{p}),\hat{\eta}_{D}^{+}(\vec{x},t_{p})\Big)   \\   \lb{B5}
\Big\{\hat{H}^{(\kappa)}\Big\} = \Big\{\hat{\Lambda}^{(\alpha)},\hat{h}^{(\kappa)}\Big\}&:=&
\mbox{generators of subgroup elements }\;U(L|S) \\ \lb{B6}
\Big\{\hat{Y}^{(\kappa)}\Big\} &:=& \mbox{generators of coset elements }\;Osp(S,S|2L)\backslash U(L|S)\;.
\eeq
Using the important integral relation (\ref{B7}) for
\(\hat{T}^{-1}(\vec{x},T_{p})\;\big(\wtilde{\pp}_{i}\hat{T}(\vec{x},t_{p})\big)\)
\beq \lb{B7}
\lefteqn{\hat{T}^{-1}(\vec{x},t_{p})\;\Big(\wtilde{\pp}_{i}\hat{T}(\vec{x},t_{p})\Big)=
\exp\Big\{\hat{Y}(\vec{x},t_{p})\Big\}\;\;\Big(\wtilde{\pp}_{i} \exp\Big\{-\hat{Y}(\vec{x},t_{p})\Big\}\Big)= } \\
\no &=& - \int_{0}^{1}d v\;\;
\exp\Big\{v\;\hat{Y}(\vec{x},t_{p})\Big\}\;\;\Big(\wtilde{\pp}_{i}\hat{Y}(\vec{x},t_{p})\Big)\;\;
\exp\Big\{-v\;\hat{Y}(\vec{x},t_{p})\Big\}\;\;\;,
\eeq
we can derive a formal series of commutators acting onto the derivative of the coset generator
\(\hat{Y}(\vec{x},t_{p})\) (\ref{B8}). This equivalent relation to (\ref{B1}) follows  after integration
over the parameter \(v\in[0,1]\) in (\ref{B7}) and specifies the functions
\(g_{\kappa}(\vec{x},t_{p})\) and \(s_{\kappa}(\vec{x},t_{p})\)
\beq \lb{B8}
\lefteqn{\Big[\Big(\wtilde{\pp}_{i}s_{\kappa}(\vec{x},t_{p})\Big)\;
\hat{Y}^{(\kappa)}\Big]_{\beta\alpha}^{ba}\bigg|^{b\neq a} -
\im\;\Big[\Big(\wtilde{\pp}_{i}g_{\kappa}(\vec{x},t_{p})\Big)\;\hat{H}^{(\kappa)}\Big]_{\beta\alpha}^{ba}\;
\delta_{a,b} =  } \\     \no  &=& -
\Big[\hat{T}^{-1}(\vec{x},t_{p})\;\Big(\wtilde{\pp}_{i}\hat{T}(\vec{x},t_{p})\Big)\Big]_{\beta\alpha}^{ba}
 = \int_{0}^{1}d v\;\;
\bigg[\exp\Big\{v\;\Big[\hat{Y}(\vec{x},t_{p})\;,\;\ldots\Big]\Big\}\;\;
\Big(\wtilde{\pp}_{i}\hat{Y}(\vec{x},t_{p})\Big)\bigg]_{\beta\alpha}^{ba}  =  \\ \no &=&
\left[\frac{\exp\Big\{\Big[\hat{Y}(\vec{x},t_{p})\;,\;\ldots\Big]\Big\}-\hat{1}_{2N\times 2N}}{
\Big[\hat{Y}(\vec{x},t_{p})\;,\;\ldots\Big] }\;\;
\Big(\wtilde{\pp}_{i}\hat{Y}(\vec{x},t_{p})\Big)\right]_{\beta\alpha}^{ba}\;\;\;.
\eeq
The commutator series in (\ref{B8}) can be calculated with the super-generator \(\hat{Y}(\vec{x},t_{p})\)
for the anomalous terms being separated into generators or matrices \(\hat{Y}^{(\kappa)}\) and
corresponding even and odd parameters, abbreviated by \(y_{\kappa}(\vec{x},t_{p})\) for the pair
condensate amplitudes (as e.g. in the fundamental representation of \(Osp(S,S|2L)\)). However, one can
also apply the adjoint representation (\ref{B11}) of \(Osp(S,S|2L)\) with supercommutator
\(\big[\hat{Y}^{(\kappa)},\ldots\}\) in order to determine (\ref{B8}). This is in particular possible
if one has already computed the root vectors of \(Osp(S,S|2L)\) (\cite{luc}, subsection \ref{s53})
\beq \lb{B9}
\hat{Y}(\vec{x},t_{p}) &=&y_{\kappa}(\vec{x},t_{p})\;\;\hat{Y}^{(\kappa)} \\ \lb{B10}
y_{\kappa}(\vec{x},t_{p}) &:=& \mbox{even and odd parameters}   \\   \lb{B11}
\Big[\hat{Y}(\vec{x},t_{p})\;,\;\ldots\Big] &=&
y_{\kappa}(\vec{x},t_{p})\;\;\Big[\hat{Y}^{(\kappa)}\;,\;\ldots\Big\}
 \simeq  y_{\kappa}(\vec{x},t_{p})\;\;\mbox{ad}_{\big[\hat{Y}^{(\kappa)}\;,\;\ldots\big\}}  \;\;\;.
\eeq
In order to apply the rules (\ref{s4_116},\ref{s4_117}) in subsection \ref{s44}, we have to relate
the anomalous-like self-energy \(\delta\hat{\Sigma}_{\beta\alpha}^{b\neq a}(\vec{x},t_{p})\)
and its derivative
\(\big(\wtilde{\pp}_{i}\delta\hat{\Sigma}_{\beta\alpha}^{b\neq a}(\vec{x},t_{p})\big)\)
with \(b\neq a\) to the term
\(\big[\hat{T}^{-1}(\vec{x},T_{p})\;
\big(\wtilde{\pp}_{i}\hat{T}(\vec{x},t_{p})\big)\big]_{\beta\alpha}^{b\neq a}\)
\beq \lb{B12}
\Big[\hat{T}^{-1}(\vec{x},t_{p})\;
\Big(\wtilde{\pp}_{i}\hat{T}(\vec{x},t_{p})\Big)\Big]_{\beta\alpha}^{ba}\bigg|^{b\neq a}&=&
\Big[-\Big(\wtilde{\pp}_{i}s_{\kappa}(\vec{x},t_{p})\Big)\;
\hat{Y}^{(\kappa)}\Big]_{\beta\alpha}^{ba}\bigg|^{b\neq a}
 \simeq  \Big(\wtilde{\pp}_{i}\delta\hat{\Sigma}_{\beta\alpha}^{b\neq a}(\vec{x},t_{p})\Big)
\bigg|^{b\neq a} \\ \lb{B13} \Longrightarrow\;\;\;
\delta\hat{\Sigma}_{\beta\alpha}^{b\neq a}(\vec{x},t_{p})\bigg|^{b\neq a} &\simeq &
\Big[-s_{\kappa}(\vec{x},t_{p})\;\; \hat{Y}^{(\kappa)}\Big]_{\beta\alpha}^{b\neq a}\bigg|^{b\neq a}\;\;\;.
\eeq
Application of Eqs. (\ref{s4_116},\ref{s4_117})
allows to identify the spatial derivative of the anomalous-like
self-energy with the term (\ref{B12}) so that the matrix element \(C_{\beta\alpha;j}^{b\neq a}\)
(\ref{s4_134}) is transformed to Eq. (\ref{B14})
where the part \(\big[-s_{\kappa}(\vec{x},t_{p})\;\; \hat{Y}^{(\kappa)}\big]_{\beta\alpha}^{b\neq a}\)
refers to pair condensate amplitudes without the spatial derivative (\ref{B13})
\beq \lb{B14}
\lefteqn{C_{\beta\alpha;j}^{b\neq a}=
\widehat{\langle\vec{x},t_{p}}^{b}|\;\hat{G}^{(0)}[\hat{\sigma}_{D}^{(0)}]\;\hat{\eta}\;
\Big[\hat{T}^{-1}\;\big(\wtilde{\pp}_{j}\hat{T}\big)\Big]_{\beta\alpha}^{b\neq a}\;
\hat{G}^{(0)}[\hat{\sigma}_{D}^{(0)}]\;|\widehat{\vec{x},t_{p}\rangle}^{a}\bigg|^{b\neq a}
 \approx } \\ \no &=&
\frac{1}{\mcal{N}^{2}}\bigg( \Big[\hat{T}^{-1}(\vec{x},t_{p})\;\big(\wtilde{\pp}_{j}\hat{T}(\vec{x},t_{p})
\big)\Big]_{\beta\alpha}^{b\neq a}+2 \Big(\wtilde{\pp}_{j}\breve{v}(\vec{x},t_{p})\Big)\;
\Big[-s_{\kappa}(\vec{x},t_{p})\;\hat{Y}^{(\kappa)}\Big]_{\beta\alpha}^{b\neq a}\bigg)\bigg|^{b\neq a}
\eeq
\be \lb{B15}
\breve{v}(\vec{x},t_{p})=\breve{u}(\vec{x}) +\breve{\sigma}_{D}^{(0)}(\vec{x},t_{p})\hspace*{1.0cm}
\breve{u}(\vec{x})=\frac{u(\vec{x})}{\mcal{N}}  \hspace*{1.0cm}
\breve{\sigma}_{D}^{(0)}(\vec{x},t_{p}) = \frac{\sigma_{D}^{(0)}(\vec{x},t_{p})}{\mcal{N}}
\;\;\;.
\ee
The trap potential \(u(\vec{x})\) and background field \(\sigma_{D}^{(0)}(\vec{x},t_{p})\) have to be
normalized by the energy parameter \(\mcal{N}=\hbar\Omega\;\mcal{N}_{x}\) (\ref{B15}) because
the operator \(\hat{\mcal{O}}=\hat{\mcal{H}}+\hat{\sigma}_{D}^{(0)}\cdot\hat{1}_{2N\times 2N}+\Delta\hat{\mcal{O}}\),
which determines the gradient expansion with the Green functions \(\hat{G}^{(0)}[\hat{\sigma}_{D}^{(0)}]\),
is defined with this energy scale of the discrete finite intervals for spatial and time-like variables
in the coherent state path integrals (\ref{s4_71},\ref{s4_82}).

In analogy the matrix element
\(C_{\beta\alpha}^{b\neq a}(\mbox{{\boldmath$\wtilde{\pp}_{i}$}})\)
(\ref{s4_135}) is reduced with propagation rule (\ref{s4_117}) where one
has also to take into account that anomalous-like terms
\(f_{\alpha}(\vec{x}_{1},t_{q}\ppr)\otimes f_{\beta}(\vec{x}_{2},t_{q}\ppr)\) only propagate in the
traces '\(\mbox{Tr}\)' of doubled Hilbert space for \(\mcal{A}_{SDET}\ppr\)
\beq \no
\lefteqn{C_{\beta\alpha}^{b\neq a}(\mbox{{\boldmath$\wtilde{\pp}_{i}$}})=
\widehat{\langle\vec{x},t_{p}}^{b}|\;\mbox{{\boldmath$\wtilde{\pp}_{i}$}}\;
\hat{G}^{(0)}[\hat{\sigma}_{D}^{(0)}]\;\hat{\eta}\;
\Big[\hat{T}^{-1}\;\big(\wtilde{\pp}_{j}\hat{T}\big)\Big]_{\beta\alpha}^{b\neq a}\;
\hat{G}^{(0)}[\hat{\sigma}_{D}^{(0)}]\; \Big(\wtilde{\pp}_{j}\hat{u}+
\wtilde{\pp}_{j}\hat{\sigma}_{D}^{(0)}\Big)\;
\hat{G}^{(0)}[\hat{\sigma}_{D}^{(0)}]\; |\widehat{\vec{x},t_{p}\rangle}^{a}\bigg|^{b\neq a}\approx} \\ \lb{B16} &=&
\widehat{\langle\vec{x},t_{p}}^{b}|\;\mbox{{\boldmath$\wtilde{\pp}_{i}$}}\;
\hat{G}^{(0)}[\hat{\sigma}_{D}^{(0)}]\;\hat{\eta}\;
\Big[\hat{T}^{-1}\;\big(\wtilde{\pp}_{j}\hat{T}\big)\Big]_{\beta\alpha}^{b\neq a}\;
\hat{G}^{(0)}[\hat{\sigma}_{D}^{(0)}]\; |\widehat{\vec{x},t_{p}\rangle}^{a}\;\;
\Big(\wtilde{\pp}_{j}\breve{v}(\vec{x},t_{p})\Big)\bigg|^{b\neq a}  \\ \no &=&
\Big(\wtilde{\pp}_{j}\breve{v}(\vec{x},t_{p})\Big)
\int_{C}\frac{d t_{q}\ppr}{\hbar}\eta_{q}\mcal{N}\;\sum_{\vec{x}\ppr}
\frac{1}{2}\bigg(\wtilde{\pp}_{i}
\Big(\widehat{\langle\vec{x},t_{p}}^{b}|\hat{G}^{(0)}[\hat{\sigma}_{D}^{(0)}]
|\widehat{\vec{x}\ppr,t_{q}\ppr\rangle}^{b}\Big)^{2}\bigg)\;\eta_{q}\;
\Big[\hat{T}^{-1}(\vec{x}\ppr,t_{q}\ppr)\;
\big(\wtilde{\pp}_{j}\ppr\hat{T}(\vec{x}\ppr,t_{q}\ppr\big)\Big]_{\beta\alpha}^{b\neq a} \\ \no
&=&\frac{1}{2} \bigg(\wtilde{\pp}_{i}\; \widehat{\langle\vec{x},t_{p}}^{b}|\;
\hat{G}^{(0)}[\hat{\sigma}_{D}^{(0)}]\;\hat{\eta}\;
\Big[\hat{T}^{-1}\;\big(\wtilde{\pp}_{j}\hat{T}\big)\Big]_{\beta\alpha}^{b\neq a}\;
\hat{G}^{(0)}[\hat{\sigma}_{D}^{(0)}]\;|\widehat{\vec{x},t_{p}\rangle}^{a}\bigg)\;\;
\Big(\wtilde{\pp}_{j}\breve{v}(\vec{x},t_{p})\Big)  \\ \no
&=&\!\!\frac{1}{2}\frac{1}{\mcal{N}^{2}}
\Big(\wtilde{\pp}_{j}\breve{v}(\vec{x},t_{p})\Big)
\bigg\{\bigg(\wtilde{\pp}_{i}\Big[\hat{T}^{-1}(\vec{x},t_{p})\;
\big(\wtilde{\pp}_{j}\hat{T}(\vec{x},t_{p})\big)\Big]_{\beta\alpha}^{b\neq a}\bigg)+ 2
\bigg(\wtilde{\pp}_{i}\Big(\wtilde{\pp}_{j}\breve{v}(\vec{x},t_{p})\Big)\;
\Big[-s_{\kappa}(\vec{x},t_{p})\;\hat{Y}^{(\kappa)}\Big]_{\beta\alpha}^{b\neq a}\bigg)\bigg\}.
\eeq
Relation \(C_{\beta\alpha;j}^{b\neq a}(\mbox{{\boldmath$\wtilde{\pp}_{i}$}},
\mbox{{\boldmath$\wtilde{\pp}_{j}$}})\)
(\ref{s4_136},\ref{B17}) with two unsaturated spatial gradients needs the additional commutators
(\ref{B18},\ref{B19}) for further transformations
\beq \lb{B17}
\lefteqn{C_{\beta\alpha;j}^{b\neq a}(\mbox{{\boldmath$\wtilde{\pp}_{i}$}},
\mbox{{\boldmath$\wtilde{\pp}_{j}$}}) =
\widehat{\langle\vec{x},t_{p}}^{b}|\;\mbox{{\boldmath$\wtilde{\pp}_{i}$}}\;
\mbox{{\boldmath$\wtilde{\pp}_{j}$}}\;\hat{G}^{(0)}[\hat{\sigma}_{D}^{(0)}]\;\hat{\eta}\;
\Big[\hat{T}^{-1}\;\big(\wtilde{\pp}_{j}\hat{T}\big)\Big]_{\beta\alpha}^{b\neq a}\;
\hat{G}^{(0)}[\hat{\sigma}_{D}^{(0)}]\; |\widehat{\vec{x},t_{p}\rangle}^{a}\bigg|^{b\neq a} +  } \\ \no &+&
\widehat{\langle\vec{x},t_{p}}^{b}|\;\hat{G}^{(0)}[\hat{\sigma}_{D}^{(0)}]\;\hat{\eta}\;
\Big[\hat{T}^{-1}\;\big(\wtilde{\pp}_{j}\hat{T}\big)\Big]_{\beta\alpha}^{b\neq a}\;
\hat{G}^{(0)}[\hat{\sigma}_{D}^{(0)}]\;\mbox{{\boldmath$\wtilde{\pp}_{i}$}}\;
\mbox{{\boldmath$\wtilde{\pp}_{j}$}}\;|\widehat{\vec{x},t_{p}\rangle}^{a}\bigg|^{b\neq a} = \\ \no &=&
\int_{C}\frac{d t_{q}\ppr}{\hbar}\eta_{q}\sum_{\vec{x}_{1},\vec{x}_{2}}
\bigg(\widehat{\langle\vec{x},t_{p}}^{b}|\;\mbox{{\boldmath$\wtilde{\pp}_{i}$}}\;
\mbox{{\boldmath$\wtilde{\pp}_{j}$}}\;\hat{G}^{(0)}[\hat{\sigma}_{D}^{(0)}]\;
|\widehat{\vec{x}_{1},t_{q}\ppr\rangle}^{b}\;\;
\widehat{\langle\vec{x}_{2},t_{q}\ppr}^{a}|\;\hat{G}^{(0)}[\hat{\sigma}_{D}^{(0)}]\;
|\widehat{\vec{x},t_{p}\rangle}^{a}+ \\ \no &+&
\widehat{\langle\vec{x},t_{p}}^{b}|\;\hat{G}^{(0)}[\hat{\sigma}_{D}^{(0)}]\;
|\widehat{\vec{x}_{1},t_{q}\ppr\rangle}^{b}\;\;
\widehat{\langle\vec{x}_{2},t_{q}\ppr}^{a}|\;\hat{G}^{(0)}[\hat{\sigma}_{D}^{(0)}]\;
\mbox{{\boldmath$\wtilde{\pp}_{i}$}}\;\mbox{{\boldmath$\wtilde{\pp}_{j}$}}\;
|\widehat{\vec{x},t_{p}\rangle}^{a}\bigg)\;\eta_{q}\;
\Big[\hat{T}^{-1}(t_{q}\ppr)\;\big(\wtilde{\pp}_{j}\hat{T}(t_{q}\ppr)
\big)\Big]_{\vec{x}_{1},\beta;\vec{x}_{2},\alpha}^{b\neq a}\;
\bigg|_{\vec{x}_{1}=\vec{x}_{2}}^{b\neq a}
\eeq
\beq \lb{B18}
\lefteqn{\Big[\mbox{{\boldmath$\wtilde{\pp}_{i}$}}\;\mbox{{\boldmath$\wtilde{\pp}_{j}$}}
\;,\;\hat{G}^{(0)}[\hat{\sigma}_{D}^{(0)}]\Big]= -\hat{G}^{(0)}[\hat{\sigma}_{D}^{(0)}]\;\Big(
\wtilde{\pp}_{i}\wtilde{\pp}_{j}\hat{u}+\wtilde{\pp}_{i}\wtilde{\pp}_{j}\hat{\sigma}_{D}^{(0)}\Big)\;
\hat{G}^{(0)}[\hat{\sigma}_{D}^{(0)}]+  } \\ \no &+&\bigg\{ \bigg(\hat{G}^{(0)}[\hat{\sigma}_{D}^{(0)}]\;\Big(
\wtilde{\pp}_{j}\hat{u}+\wtilde{\pp}_{j}\hat{\sigma}_{D}^{(0)}\Big)\;
\hat{G}^{(0)}[\hat{\sigma}_{D}^{(0)}]\;\Big(\wtilde{\pp}_{i}\hat{u}+
\wtilde{\pp}_{i}\hat{\sigma}_{D}^{(0)}\Big)\; \hat{G}^{(0)}[\hat{\sigma}_{D}^{(0)}]+
\\ \no &-&
\hat{G}^{(0)}[\hat{\sigma}_{D}^{(0)}]\;\Big( \wtilde{\pp}_{i}\hat{u}+
\wtilde{\pp}_{i}\hat{\sigma}_{D}^{(0)}\Big)\;
\hat{G}^{(0)}[\hat{\sigma}_{D}^{(0)}]\;\mbox{{\boldmath$\wtilde{\pp}_{j}$}} \bigg)+
\Big[\;i\;\mbox{ exchanged with }\;j\Big]\bigg\}
\eeq
\beq \lb{B19}
\lefteqn{
\Big[\mbox{{\boldmath$\wtilde{\pp}_{i}$}}\;\mbox{{\boldmath$\wtilde{\pp}_{j}$}}
\;,\;\hat{G}^{(0)}[\hat{\sigma}_{D}^{(0)}]\Big]= \hat{G}^{(0)}[\hat{\sigma}_{D}^{(0)}]\;\Big(
\wtilde{\pp}_{i}\wtilde{\pp}_{j}\hat{u}+
\wtilde{\pp}_{i}\wtilde{\pp}_{j}\hat{\sigma}_{D}^{(0)}\Big)\;
\hat{G}^{(0)}[\hat{\sigma}_{D}^{(0)}]+  } \\ \no &-&
\bigg\{ \bigg(\hat{G}^{(0)}[\hat{\sigma}_{D}^{(0)}]\;\Big(
\wtilde{\pp}_{j}\hat{u}+\wtilde{\pp}_{j}\hat{\sigma}_{D}^{(0)}\Big)\;
\hat{G}^{(0)}[\hat{\sigma}_{D}^{(0)}]\;\Big(\wtilde{\pp}_{i}\hat{u}+
\wtilde{\pp}_{i}\hat{\sigma}_{D}^{(0)}\Big)\; \hat{G}^{(0)}[\hat{\sigma}_{D}^{(0)}]+
\\ \no &+&
\mbox{{\boldmath$\wtilde{\pp}_{j}$}}\;\hat{G}^{(0)}[\hat{\sigma}_{D}^{(0)}]\;
\Big(\wtilde{\pp}_{i}\hat{u}+\wtilde{\pp}_{i}\hat{\sigma}_{D}^{(0)}\Big)\;
\hat{G}^{(0)}[\hat{\sigma}_{D}^{(0)}]\bigg)+\Big[\;i\;\mbox{ exchanged with }\;j\Big]\bigg\} \;\;\;.
\eeq
Upon insertion of the commutator relations (\ref{B18},\ref{B19}) into (\ref{B17}), the unsaturated
gradients are shifted in such a manner that they only operate on the pair condensate amplitudes
\(\big[\hat{T}^{-1}\big(\wtilde{\pp}_{j}\hat{T}\big)\big]_{\beta\alpha}^{b\neq a}\), yielding
additional derivatives of the coset matrices
\beq \lb{B20}
\lefteqn{C_{\beta\alpha;j}^{b\neq a}(\mbox{{\boldmath$\wtilde{\pp}_{i}$}},
\mbox{{\boldmath$\wtilde{\pp}_{j}$}}) =
\widehat{\langle\vec{x},t_{p}}^{b}|\;\mbox{{\boldmath$\wtilde{\pp}_{i}$}}\;
\mbox{{\boldmath$\wtilde{\pp}_{j}$}}\;\hat{G}^{(0)}[\hat{\sigma}_{D}^{(0)}]\;\hat{\eta}\;
\Big[\hat{T}^{-1}\;\big(\wtilde{\pp}_{j}\hat{T}\big)\Big]_{\beta\alpha}^{b\neq a}\;
\hat{G}^{(0)}[\hat{\sigma}_{D}^{(0)}]\; |\widehat{\vec{x},t_{p}\rangle}^{a}\bigg|^{b\neq a} +  } \\ \no &+&
\widehat{\langle\vec{x},t_{p}}^{b}|\;\hat{G}^{(0)}[\hat{\sigma}_{D}^{(0)}]\;\hat{\eta}\;
\Big[\hat{T}^{-1}\;\big(\wtilde{\pp}_{j}\hat{T}\big)\Big]_{\beta\alpha}^{b\neq a}\;
\hat{G}^{(0)}[\hat{\sigma}_{D}^{(0)}]\;\mbox{{\boldmath$\wtilde{\pp}_{i}$}}\;
\mbox{{\boldmath$\wtilde{\pp}_{j}$}}\;|\widehat{\vec{x},t_{p}\rangle}^{a}\bigg|^{b\neq a} =
\int_{C}\frac{d t_{q}\ppr}{\hbar}\eta_{q}\mcal{N}\sum_{\vec{x}_{1},\vec{x}_{2}}
\\ \no &\times& \mbox{{\boldmath${\ds\Bigg[}$}}
\widehat{\Big\langle\vec{x},t_{p}}^{b}|\;\hat{G}^{(0)}[\hat{\sigma}_{D}^{(0)}]\;
\mbox{{\boldmath$\wtilde{\pp}_{i}$}}\;\mbox{{\boldmath$\wtilde{\pp}_{j}$}}-
\hat{G}^{(0)}[\hat{\sigma}_{D}^{(0)}]\;\Big(
\wtilde{\pp}_{i}\wtilde{\pp}_{j}\hat{u}+\wtilde{\pp}_{i}\wtilde{\pp}_{j}\hat{\sigma}_{D}^{(0)}\Big)\;
\hat{G}^{(0)}[\hat{\sigma}_{D}^{(0)}]+ \\ \no &+& \bigg\{\bigg(\hat{G}^{(0)}[\hat{\sigma}_{D}^{(0)}]\;\Big(
\wtilde{\pp}_{j}\hat{u}+\wtilde{\pp}_{j}\hat{\sigma}_{D}^{(0)}\Big)\;
\hat{G}^{(0)}[\hat{\sigma}_{D}^{(0)}]\;\Big(
\wtilde{\pp}_{i}\hat{u}+\wtilde{\pp}_{i}\hat{\sigma}_{D}^{(0)}\Big)\;
\hat{G}^{(0)}[\hat{\sigma}_{D}^{(0)}]+ \\ \no &-&
\hat{G}^{(0)}[\hat{\sigma}_{D}^{(0)}]\;
\Big( \wtilde{\pp}_{i}\hat{u}+\wtilde{\pp}_{i}\hat{\sigma}_{D}^{(0)}\Big)\;
\hat{G}^{(0)}[\hat{\sigma}_{D}^{(0)}]\;\mbox{{\boldmath$\wtilde{\pp}_{j}$}}\bigg)+   \\ \no &+&
\Big[\;i\;\mbox{ exchanged with }\;j\Big]\bigg\}\; |\widehat{\vec{x}_{1},t_{q}\ppr\Big\rangle}^{b}\; \times\;
\widehat{\Big\langle\vec{x}_{2},t_{q}\ppr}^{a}|\;\hat{G}^{(0)}[\hat{\sigma}_{D}^{(0)}]\;
|\widehat{\vec{x},t_{p}\Big\rangle}^{a}+  \\  \no &+&
\widehat{\Big\langle\vec{x},t_{p}}^{b}|\;\hat{G}^{(0)}[\hat{\sigma}_{D}^{(0)}]\;
|\widehat{\vec{x}_{1},t_{q}\ppr\Big\rangle}^{b} \times
\widehat{\Big\langle\vec{x}_{2},t_{q}\ppr}^{a}|\;\mbox{{\boldmath$\wtilde{\pp}_{i}$}}\;
\mbox{{\boldmath$\wtilde{\pp}_{j}$}}\;\hat{G}^{(0)}[\hat{\sigma}_{D}^{(0)}] -
\hat{G}^{(0)}[\hat{\sigma}_{D}^{(0)}]\;\Big(
\wtilde{\pp}_{i}\wtilde{\pp}_{j}\hat{u}+\wtilde{\pp}_{i}\wtilde{\pp}_{j}\hat{\sigma}_{D}^{(0)}\Big)\;
\hat{G}^{(0)}[\hat{\sigma}_{D}^{(0)}] + \\ \no &+&
\bigg\{\bigg(\hat{G}^{(0)}[\hat{\sigma}_{D}^{(0)}]\;\Big(
\wtilde{\pp}_{j}\hat{u}+\wtilde{\pp}_{j}\hat{\sigma}_{D}^{(0)}\Big)\; \hat{G}^{(0)}[\hat{\sigma}_{D}^{(0)}]\;\Big(
\wtilde{\pp}_{i}\hat{u}+\wtilde{\pp}_{i}\hat{\sigma}_{D}^{(0)}\Big)\; \hat{G}^{(0)}[\hat{\sigma}_{D}^{(0)}]+
\\ \no &+&
\mbox{{\boldmath$\wtilde{\pp}_{j}$}}\;\hat{G}^{(0)}[\hat{\sigma}_{D}^{(0)}]\;\Big(
\wtilde{\pp}_{i}\hat{u}+\wtilde{\pp}_{i}\hat{\sigma}_{D}^{(0)}\Big)\; \hat{G}^{(0)}[\hat{\sigma}_{D}^{(0)}]\bigg)+
\Big[\;i\;\mbox{ exchanged with }\;j\Big]\bigg\}\; |\widehat{\vec{x},t_{p}\Big\rangle}^{a}
\mbox{{\boldmath${\ds\Bigg]}$}} \\ \no &\times&  \eta_{q}\;\Big[\hat{T}^{-1}(t_{q}\ppr)\;
\big(\wtilde{\pp}_{j}\hat{T}(t_{q}\ppr)\big)\Big]_{\vec{x}_{1},\beta;\vec{x}_{2},\alpha}^{b\neq a}\;\;
\bigg|_{\vec{x}_{1}=\vec{x}_{2}}^{b\neq a} \;\;\;.
\eeq
We further approximate above relation (\ref{B20}) and only consider up to first order spatial derivatives
of the pair condensates or coset matrices
\beq \lb{B21}
\lefteqn{C_{\beta\alpha;j}^{b\neq a}(\mbox{{\boldmath$\wtilde{\pp}_{i}$}},
\mbox{{\boldmath$\wtilde{\pp}_{j}$}}) =
\widehat{\langle\vec{x},t_{p}}^{b}|\;\mbox{{\boldmath$\wtilde{\pp}_{i}$}}\;
\mbox{{\boldmath$\wtilde{\pp}_{j}$}}\;\hat{G}^{(0)}[\hat{\sigma}_{D}^{(0)}]\;\hat{\eta}\;
\Big[\hat{T}^{-1}\;\big(\wtilde{\pp}_{j}\hat{T}\big)\Big]_{\beta\alpha}^{b\neq a}\;
\hat{G}^{(0)}[\hat{\sigma}_{D}^{(0)}]\; |\widehat{\vec{x},t_{p}\rangle}^{a}\bigg|^{b\neq a} +  } \\ \no &+&
\widehat{\langle\vec{x},t_{p}}^{b}|\;\hat{G}^{(0)}[\hat{\sigma}_{D}^{(0)}]\;\hat{\eta}\;
\Big[\hat{T}^{-1}\;\big(\wtilde{\pp}_{j}\hat{T}\big)\Big]_{\beta\alpha}^{b\neq a}\;
\hat{G}^{(0)}[\hat{\sigma}_{D}^{(0)}]\;\mbox{{\boldmath$\wtilde{\pp}_{i}$}}\;
\mbox{{\boldmath$\wtilde{\pp}_{j}$}}\;|\widehat{\vec{x},t_{p}\rangle}^{a}\bigg|^{b\neq a}
\approx \\ \no &=&2\;
\Bigg(2\Big(\wtilde{\pp}_{i}\breve{v}(\vec{x},t_{p})\Big)\Big(\wtilde{\pp}_{j}\breve{v}(\vec{x},t_{p})\Big)
-\Big(\wtilde{\pp}_{i}\wtilde{\pp}_{j}\breve{v}(\vec{x},t_{p})\Big)\Bigg)\;\times \\ \no &\times&
\widehat{\langle\vec{x},t_{p}}^{b}|\;\hat{G}^{(0)}[\hat{\sigma}_{D}^{(0)}]\;\hat{\eta}\;
\Big[\hat{T}^{-1}\;\big(\wtilde{\pp}_{j}\hat{T}\big)\Big]_{\beta\alpha}^{b\neq a}\;
\hat{G}^{(0)}[\hat{\sigma}_{D}^{(0)}]\;|\widehat{\vec{x},t_{p}\rangle}^{a}\bigg|^{b\neq a} =
\\ \no &=&\frac{2}{\mcal{N}^{2}}
\Bigg(2\Big(\wtilde{\pp}_{i}\breve{v}(\vec{x},t_{p})\Big)\Big(\wtilde{\pp}_{j}\breve{v}(\vec{x},t_{p})\Big) -
\Big(\wtilde{\pp}_{i}\wtilde{\pp}_{j}\breve{v}(\vec{x},t_{p})\Big)\Bigg) \times
\\ \no &\times &
\bigg(\Big[\hat{T}^{-1}(\vec{x},t_{p})\;
\big(\wtilde{\pp}_{j}\hat{T}(\vec{x},t_{p})\big)\Big]_{\beta\alpha}^{b\neq a}+
2\Big(\wtilde{\pp}_{j}\breve{v}(\vec{x},t_{p})\Big)\;
\Big[-s_{\kappa}(\vec{x},t_{p})\;\hat{Y}^{(\kappa)}\Big]_{\beta\alpha}^{b\neq a}\bigg)\bigg|^{b\neq a}_{\mbox{.}}
\eeq
Finally, we substitute (\ref{B14},\ref{B16},\ref{B21}) into the trace term '\(\mbox{Tr}\)' with two spatially
unsaturated gradient operators (see Eq. (\ref{s4_133}) with matrix elements (\ref{s4_134}-\ref{s4_136})) and
obtain relation (\ref{B22}) which also contains pair condensate amplitudes
\(\big[-s_{\kappa}(\vec{x},t_{p})\;\hat{Y}^{(\kappa)}\big]_{\beta\alpha}^{b\neq a}\) without any derivative
\beq \lb{B22}
\lefteqn{\bigg\langle\mbox{Tr}\bigg[\mbox{STR}
\bigg(\hat{\eta}\;\hat{T}^{-1}\;\big(\wtilde{\pp}_{i}\hat{T}\big)\;
\mbox{{\boldmath$\wtilde{\pp}_{i}$}}\;\hat{G}^{(0)}[\hat{\sigma}_{D}^{(0)}]\;\hat{\eta}\;
\hat{T}^{-1}\;\big(\wtilde{\pp}_{j}\hat{T}\big)\;
\mbox{{\boldmath$\wtilde{\pp}_{j}$}}\;\hat{G}^{(0)}[\hat{\sigma}_{D}^{(0)}]
\bigg)\bigg]\bigg\rangle_{\hat{\sigma}_{D}^{(0)}} = \int_{C}\frac{d
t_{p}}{\hbar}\;\frac{1}{\mcal{N}}\; \sum_{\vec{x}}\sum_{a,b=1,2}^{(a\neq b)}  }
\\ \no &\times& \Bigg\langle\mbox{{\boldmath${\ds\Bigg[}$}}
-\frac{1}{2} \strab\bigg\{\Big[\hat{T}^{-1}(\vec{x},t_{p})\;
\big(\wtilde{\pp}_{i}\hat{T}(\vec{x},t_{p})\big)\Big]_{\alpha\beta}^{a\neq b} \; \times \;
\bigg[\bigg(\wtilde{\pp}_{i}\wtilde{\pp}_{j}\Big[\hat{T}^{-1}(\vec{x},t_{p})\;
\big(\wtilde{\pp}_{i}\hat{T}(\vec{x},t_{p})\big)\Big]_{\beta\alpha}^{b\neq a}\bigg) +
\\ \no &+&
2\bigg(\wtilde{\pp}_{i}\wtilde{\pp}_{j}
\Big(\wtilde{\pp}_{j}\breve{v}(\vec{x},t_{p})\Big)\;
\Big[-s_{\kappa}(\vec{x},t_{p})\;\hat{Y}^{(\kappa)}\Big]_{\beta\alpha}^{b\neq a}\bigg)\bigg]\bigg\}+
\\  \no &-&\frac{1}{2}
\strab\bigg\{\Big[\hat{T}^{-1}(\vec{x},t_{p})\;
\big(\wtilde{\pp}_{i}\hat{T}(\vec{x},t_{p})\big)\Big]_{\alpha\beta}^{a\neq b} \;\;
\bigg[\bigg(\wtilde{\pp}_{i}\Big[\hat{T}^{-1}(\vec{x},t_{p})\;
\big(\wtilde{\pp}_{j}\hat{T}(\vec{x},t_{p})\big)\Big]_{\beta\alpha}^{b\neq a}\bigg) +  \\ \no &+& 2
\bigg(\wtilde{\pp}_{i}\Big(\wtilde{\pp}_{j}\breve{v}(\vec{x},t_{p})\Big)
\Big[-s_{\kappa}(\vec{x},t_{p})\;\hat{Y}^{(\kappa)}\Big]_{\beta\alpha}^{b\neq a}\bigg) \bigg]\bigg\}
\Big(\wtilde{\pp}_{j}\breve{v}(\vec{x},t_{p})\Big)+
\\ \no &+&
\strab\bigg\{\Big[\hat{T}^{-1}(\vec{x},t_{p})\;
\big(\wtilde{\pp}_{i}\hat{T}(\vec{x},t_{p})\big)\Big]_{\alpha\beta}^{a\neq b}
\bigg[\Big[\hat{T}^{-1}(\vec{x},t_{p})\;
\big(\wtilde{\pp}_{j}\hat{T}(\vec{x},t_{p})\big)\Big]_{\beta\alpha}^{b\neq a} +  \\ \no &+&
2\Big(\wtilde{\pp}_{j}\breve{v}(\vec{x},t_{p})\Big)
\Big[-s_{\kappa}(\vec{x},t_{p})\;\hat{Y}^{(\kappa)}\Big] \bigg]_{\beta\alpha}^{b\neq a}\bigg\}
\Bigg(2\Big(\wtilde{\pp}_{i}\breve{v}(\vec{x},t_{p})\Big)\Big(\wtilde{\pp}_{j}\breve{v}(\vec{x},t_{p})\Big) -
\Big(\wtilde{\pp}_{i}\wtilde{\pp}_{j}\breve{v}(\vec{x},t_{p})\Big)\Bigg)
\mbox{{\boldmath${\ds\Bigg]}$}} \Bigg\rangle_{\hat{\sigma}_{D}^{(0)}}
\eeq
\be \lb{B23}
\hat{T}\;,\;\; \hat{T}^{-1}\;,\;\; (\wtilde{\pp}_{i}\hat{T})\;,\;\;
(\wtilde{\pp}_{i}\wtilde{\pp}_{i}\hat{T})\;,\;\;
(\wtilde{\pp}_{i}\hat{T})\cdot(\wtilde{\pp}_{j}\hat{T})\;,\;\; (E_{p}\hat{T})\;\;\;.
\ee
One usually restricts to coset matrices with at most two spatial derivatives and first order
time derivative (\ref{B23}). We assume that this is sufficient to extract the Goldstone modes
of a spontaneous symmetry breaking of \(Osp(S,S|2L)\backslash U(L|S)\otimes U(L|S)\). Since the
metric tensor \(\hat{G}_{Osp\backslash U}\) appears simultaneously in the action \(\mcal{A}_{SDET}\ppr\)
and in the invariant coset measure with the square root of the super-determinant
\(\big(\mbox{SDET}(\hat{G}_{Osp\backslash U})\big)^{1/2}\), a conformal invariance results {\it with
the inclusion of the coset integration measure} leading to Gaussian integrals concerning the action (\ref{B22}).
However, the resulting relation (\ref{B24}) allows further spatial derivatives to be transferred
by partial integrations from the averages of the background field to the pair condensates
\(\big[-s_{\kappa}(\vec{x},t_{p})\;\hat{Y}^{(\kappa)}\big]_{\beta\alpha}^{b\neq a}\) without any derivatives.
After neglecting the averages of background fields involving three derivatives (\ref{B25}-\ref{B27}),
the second order gradient term (\ref{B24},\ref{s4_133}) reduces to the stated relation (\ref{s4_137})
in subsection \ref{s45} which is further transformed to the coset matrices
\(\hat{Z}(\vec{x},t_{p})=\hat{T}(\vec{x},t_{p})\;\hat{S}\;\hat{T}^{-1}(\vec{x},t_{p})\)
\beq \no
\lefteqn{\bigg\langle\mbox{Tr}\bigg[\mbox{STR}
\bigg(\hat{\eta}\;\hat{T}^{-1}\;\big(\wtilde{\pp}_{i}\hat{T}\big)\;
\mbox{{\boldmath$\wtilde{\pp}_{i}$}}\;\hat{G}^{(0)}[\hat{\sigma}_{D}^{(0)}]\;\hat{\eta}\;
\hat{T}^{-1}\;\big(\wtilde{\pp}_{j}\hat{T}\big)\;
\mbox{{\boldmath$\wtilde{\pp}_{j}$}}\;\hat{G}^{(0)}[\hat{\sigma}_{D}^{(0)}]
\bigg)\bigg]\bigg\rangle_{\hat{\sigma}_{D}^{(0)}} \approx  \int_{C}\frac{d
t_{p}}{\hbar}\;\frac{1}{\mcal{N}}\; \sum_{\vec{x}}\sum_{a,b=1,2}^{(a\neq b)} \times }
\\ \lb{B24} &\times&\hspace*{-0.19cm}
\Bigg\langle\mbox{{\boldmath${\ds\Bigg[}$}} -\strab\bigg\{\Big[\hat{T}^{-1}(\vec{x},t_{p})\;
\big(\wtilde{\pp}_{i}\hat{T}(\vec{x},t_{p})\big)\Big]_{\alpha\beta}^{a\neq b} \times
\bigg(\wtilde{\pp}_{i}\wtilde{\pp}_{j}\Big(\wtilde{\pp}_{j}\breve{v}(\vec{x},t_{p})\Big)\;
\Big[-s_{\kappa}(\vec{x},t_{p})\;\hat{Y}^{(\kappa)}\Big]_{\beta\alpha}^{b\neq a}\bigg)\bigg\} +
\\ \no &-&
\strab\bigg\{\Big[\hat{T}^{-1}(\vec{x},t_{p})\;
\big(\wtilde{\pp}_{i}\hat{T}(\vec{x},t_{p})\big)\Big]_{\alpha\beta}^{a\neq b}
\;\; \Big[\hat{T}^{-1}(\vec{x},t_{p})\;
\big(\wtilde{\pp}_{i}\hat{T}(\vec{x},t_{p})\big)\Big]_{\beta\alpha}^{b\neq a}\bigg\}  \times
\Big(\wtilde{\pp}_{j}\breve{v}(\vec{x},t_{p})\Big)^{2} +
\\ \no &-&  \strab\bigg\{\Big[\hat{T}^{-1}(\vec{x},t_{p})\;
\big(\wtilde{\pp}_{i}\hat{T}(\vec{x},t_{p})\big)\Big]_{\alpha\beta}^{a\neq b} \;\;
\Big[-s_{\kappa}(\vec{x},t_{p})\;\hat{Y}^{(\kappa)}\Big]_{\beta\alpha}^{b\neq a}\bigg\} \times \frac{1}{2}
\bigg(\wtilde{\pp}_{i}\Big(\wtilde{\pp}_{j}\breve{v}(\vec{x},t_{p})\Big)^{2} \bigg)+
\\ \no &+&
\strab\bigg\{\Big[\hat{T}^{-1}(\vec{x},t_{p})\;
\big(\wtilde{\pp}_{i}\hat{T}(\vec{x},t_{p})\big)
\Big]_{\alpha\beta}^{a\neq b} \;\; \Big[\hat{T}^{-1}(\vec{x},t_{p})\;
\big(\wtilde{\pp}_{j}\hat{T}(\vec{x},t_{p})\big)\Big]_{\beta\alpha}^{b\neq a}\bigg\}\;\times
\\ \no &\times &
\Bigg(2\Big(\wtilde{\pp}_{i}\breve{v}(\vec{x},t_{p})\Big)\Big(\wtilde{\pp}_{j}\breve{v}(\vec{x},t_{p})\Big) -
 \Big(\wtilde{\pp}_{i}\wtilde{\pp}_{j}\breve{v}(\vec{x},t_{p})\Big)\Bigg) +
\\ \no &+& 2\;
\strab\bigg\{\Big[\hat{T}^{-1}(\vec{x},t_{p})\;
\big(\wtilde{\pp}_{i}\hat{T}(\vec{x},t_{p})\big)\Big]_{\alpha\beta}^{a\neq b} \;\;
\Big[-s_{\kappa}(\vec{x},t_{p})\;\hat{Y}^{(\kappa)}\Big]_{\beta\alpha}^{b\neq a}\bigg\}\times
\\ \no &\times &
\bigg[2\Big(\wtilde{\pp}_{i}\breve{v}(\vec{x},t_{p})\Big)
\Big(\wtilde{\pp}_{j}\breve{v}(\vec{x},t_{p})\Big)^{2} -
\frac{1}{2} \bigg(\wtilde{\pp}_{i}
\Big(\wtilde{\pp}_{j}\breve{v}(\vec{x},t_{p})\Big)^{2}\bigg)\bigg]
\mbox{{\boldmath${\ds\Bigg]}$}} \Bigg\rangle_{\hat{\sigma}_{D}^{(0)}}
\eeq
\be \lb{B25}
\bigg(\wtilde{\pp}_{i}\bigg\langle
\Big(\wtilde{\pp}_{j}\breve{v}(\vec{x},t_{p})\Big)^{2}
\bigg\rangle_{\hat{\sigma}_{D}^{(0)}}\bigg)  \approx  0 \;\;;\;\;
\bigg\langle\Big(\wtilde{\pp}_{i}\breve{v}(\vec{x},t_{p})\Big)\;\;
\Big(\wtilde{\pp}_{j}\breve{v}(\vec{x},t_{p})\Big)^{2}
\bigg\rangle_{\hat{\sigma}_{D}^{(0)}}  \approx  0    \;\;;\;\;
\bigg(\wtilde{\pp}_{i}\bigg\langle \Big(\wtilde{\pp}_{j}\wtilde{\pp}_{j}\breve{v}(\vec{x},t_{p})\Big)
\bigg\rangle_{\hat{\sigma}_{D}^{(0)}}\bigg)  \approx  0  \; .
\ee
After neglecting consequently averages of background fields with three spatial derivatives in (\ref{B24}),
(\ref{B25}), we achieve final relation (\ref{B28}) for the gradient expansion of \(\mcal{A}_{SDET}\ppr\)
with two unsaturated spatial operators. One might argue that terms like
\be \lb{B26}
\strab\bigg\{\Big[-s_{\kappa}(\vec{x},t_{p})\;\hat{Y}^{(\kappa)}\Big]_{\alpha\beta}^{a\neq b}\;\;
\Big[\hat{T}^{-1}(\vec{x},t_{p})\;
\big(\wtilde{\pp}_{i}\hat{T}(\vec{x},t_{p})\big)\Big]_{\beta\alpha}^{b\neq a}
\bigg\}\times\bigg(\wtilde{\pp}_{i}\Big(\mbox{coefficients }\breve{v}(\vec{x},t_{p})\Big)\bigg)
\ee
should be taken into account after partial integration with \(\wtilde{\pp}_{i}\) of the prevailing coefficients
\(\breve{v}(\vec{x},t_{p})\) so that this derivative moves to the anomalous fields, yielding
second order gradients of them. However, this leads to a total divergence of the pair condensates
\be \lb{B27}
\left(\wtilde{\pp}_{i}\;\;
\strab\bigg\{\Big[-s_{\kappa}(\vec{x},t_{p})\;\hat{Y}^{(\kappa)}\Big]_{\alpha\beta}^{a\neq b}\;\;
\Big[\hat{T}^{-1}(\vec{x},t_{p})\;
\big(\wtilde{\pp}_{i}\hat{T}(\vec{x},t_{p})\big)\Big]_{\beta\alpha}^{b\neq a}
\bigg\}\right)\times\Big(\mbox{coefficients }\breve{v}(\vec{x},t_{p})\Big)\;,
\ee
which can be transformed to a surface integral. These integrals should vanish for appropriate surfaces
on the premise that the coefficients (\ref{B25}) have a negligible, gradually varying dependence
in space. Consequently, we attain the final relation (\ref{B28},\ref{s4_137}) which is the most important contribution
for the extraction of Goldstone modes in the spontaneous symmetry breaking of \(Osp(S,S|2L)\) to
\(Osp(S,S|2L)\backslash U(L|S)\). The pair condensates only remain in the traces '\(\mbox{Tr}\)'
of doubled Hilbert space with an anti-unitary second part whereas the density terms completely disappear
due to the propagation with the product of two Green functions starting and ending at the same contour time point
\beq \lb{B28}
\lefteqn{\bigg\langle\mbox{Tr}\bigg[\mbox{STR}
\bigg(\hat{\eta}\;\hat{T}^{-1}\;\big(\wtilde{\pp}_{i}\hat{T}\big)\;
\mbox{{\boldmath$\wtilde{\pp}_{i}$}}\;\hat{G}^{(0)}[\hat{\sigma}_{D}^{(0)}]\;\hat{\eta}\;
\hat{T}^{-1}\;\big(\wtilde{\pp}_{j}\hat{T}\big)\;
\mbox{{\boldmath$\wtilde{\pp}_{j}$}}\;\hat{G}^{(0)}[\hat{\sigma}_{D}^{(0)}]
\bigg)\bigg]\bigg\rangle_{\hat{\sigma}_{D}^{(0)}} \approx } \\ \no &=&
\int_{C}\frac{d t_{p}}{\hbar}\;\frac{1}{\mcal{N}}\;
\sum_{\vec{x}}\sum_{a,b=1,2}^{(a\neq b)}
\strab\bigg\{\Big[\hat{T}^{-1}(\vec{x},t_{p})\;
\big(\wtilde{\pp}_{i}\hat{T}(\vec{x},t_{p})\big)\Big]_{\alpha\beta}^{a\neq b} \;\;
\Big[\hat{T}^{-1}(\vec{x},t_{p})\;
\big(\wtilde{\pp}_{j}\hat{T}(\vec{x},t_{p})\big)\Big]_{\beta\alpha}^{b\neq a}\bigg\}\;\times
\\ \no &\times &
\bigg\langle2\Big(\wtilde{\pp}_{i}\breve{v}(\vec{x},t_{p})\Big)
\Big(\wtilde{\pp}_{j}\breve{v}(\vec{x},t_{p})\Big) -2
\Big(\wtilde{\pp}_{i}\wtilde{\pp}_{j}\breve{v}(\vec{x},t_{p})\Big)-\delta_{ij}\sum_{k=1}^{d}\bigg(
\Big(\wtilde{\pp}_{k}\breve{v}(\vec{x},t_{p})\Big)^{2}+
\Big(\wtilde{\pp}_{k}\wtilde{\pp}_{k}\breve{v}(\vec{x},t_{p})\Big)\bigg)
\bigg\rangle_{\hat{\sigma}_{D}^{(0)}} \;\;\;.
\eeq
\vspace*{0.46cm}

Apart from the gradient expansion with (\ref{B29}), we have also to take into account the
generating source field \(\wtilde{\mcal{J}}(\hat{T}^{-1},\hat{T})\) (\ref{B30}) whose second
order expansion of the effective actions is listed in relation (\ref{B31}). This generating
source field \(\wtilde{\mcal{J}}(\hat{T}^{-1},\hat{T})\) (\ref{B30}) can be replaced by derivatives with
respect to the pair condensate 'seeds' \(\im\;\hat{J}_{\psi\psi;\alpha\beta}^{a\neq b}(\vec{x},t_{p})\;\wtilde{K}\)
(\ref{s2_63},\ref{s2_64}) of the action \(\mcal{A}_{\hat{J}_{\psi\psi}}[\hat{T}]\) (\ref{s4_34}) for observables
which go beyond the second order expansion of \(\wtilde{\mcal{J}}(\hat{T}^{-1},\hat{T})\) in relation (\ref{B31}).
However, the pair condensate 'seed' fields \(\im\;\hat{J}_{\psi\psi;\alpha\beta}^{a\neq b}(\vec{x},t_{p})\;\wtilde{K}\)
(\ref{s2_63},\ref{s2_64}) of the action \(\mcal{A}_{\hat{J}_{\psi\psi}}[\hat{T}]\) (\ref{s4_34})
do not allow for generating density terms as
the source field \(\wtilde{\mcal{J}}(\hat{T}^{-1},\hat{T})\) (\ref{B30})
\be \lb{B29}
\delta\hat{\mcal{H}}(\hat{T}^{-1},\hat{T}) = -\hat{\eta}\Big(\hat{T}^{-1}\;\hat{S}\;\big(E_{p}\hat{T}\big)
+\hat{T}^{-1}\;\big(\wtilde{\pp}_{i}\wtilde{\pp}_{i}\hat{T}\big)  +
\big(\hat{T}^{-1}\;\hat{S}\;\hat{T}-\hat{S}\big)\;\mbox{{\boldmath$\hat{E}_{p}$}} +
2\;\hat{T}^{-1}\;\big(\wtilde{\pp}_{i}\hat{T}\big)\;\mbox{{\boldmath$\wtilde{\pp}_{i}$}}\Big)
\ee
\beq \lb{B30}
\wtilde{\mcal{J}}_{\vec{x},\alpha;\vec{x}\ppr,\beta}^{ab}(\hat{T}^{-1}(t_{p}),\hat{T}(t_{q}\ppr)) &=&
\hat{T}_{\alpha\alpha\ppr}^{-1;aa\ppr}(\vec{x},t_{p})\;\;
\hat{I}\;\hat{K}\;\eta_{p}\;\frac{\hat{\mcal{J}}_{\vec{x},\alpha\ppr;
\vec{x}\ppr,\beta\ppr}^{a\ppr b\ppr}(t_{p},t_{q}\ppr)}{\mcal{N}_{x}}\;\eta_{q}\;
\hat{K}\;\hat{I}\;\wtilde{K}\;\;\hat{T}_{\beta\ppr\beta}^{b\ppr b}(\vec{x}\ppr,t_{q}\ppr)
\eeq
\beq\lb{B31}
\lefteqn{\mcal{A}\ppr\big[\hat{T};\hat{\mcal{J}}\big]=-\frac{1}{4}\bigg\langle
\mbox{Tr}\;\mbox{STR}\bigg[\wtilde{\mcal{J}}(\hat{T}^{-1},\hat{T})\;\hat{G}^{(0)}[\hat{\sigma}_{D}^{(0)}]\;
\wtilde{\mcal{J}}(\hat{T}^{-1},\hat{T})\;\hat{G}^{(0)}[\hat{\sigma}_{D}^{(0)}]\bigg]
\bigg\rangle_{\hat{\sigma}_{D}^{(0)}} +}  \\ \no &-&
\frac{\im}{2}\frac{1}{\mcal{N}}\widehat{\langle J_{\psi;\beta}^{b}}|\hat{\eta}\Big(
\hat{I}\;\wtilde{K}\;\hat{T}\;\hat{G}^{(0)}[\hat{\sigma}_{D}^{(0)}]\;
\Big(\wtilde{\mcal{J}}(\hat{T}^{-1},\hat{T})\;\hat{G}^{(0)}[\hat{\sigma}_{D}^{(0)}]\Big)^{2}\;
\hat{T}^{-1}\;\hat{I}\Big)_{\beta\alpha}^{ba}\hat{\eta}|\widehat{J_{\psi;\alpha}^{a}\rangle}
\bigg\rangle_{\hat{\sigma}_{D}^{(0)}}  +  \\ \no &-&\frac{1}{2}\bigg\langle
\mbox{Tr}\;\mbox{STR}\bigg[\delta\hat{\mcal{H}}(\hat{T}^{-1},\hat{T})\;
\hat{G}^{(0)}[\hat{\sigma}_{D}^{(0)}]\;\wtilde{\mcal{J}}(\hat{T}^{-1},\hat{T})\;
\hat{G}^{(0)}[\hat{\sigma}_{D}^{(0)}]\bigg]\bigg\rangle_{\hat{\sigma}_{D}^{(0)}} + \\ \no &-&
\frac{\im}{2}\frac{1}{\mcal{N}}\bigg\langle
\widehat{\langle J_{\psi;\beta}^{b}}|\hat{\eta}\bigg(
\hat{I}\;\wtilde{K}\;\hat{T}\;\hat{G}^{(0)}[\hat{\sigma}_{D}^{(0)}]\;\Big(
\delta\hat{\mcal{H}}(\hat{T}^{-1},\hat{T})\;
\hat{G}^{(0)}[\hat{\sigma}_{D}^{(0)}]\;\wtilde{\mcal{J}}(\hat{T}^{-1},\hat{T})+  \\ \no &+&
\wtilde{\mcal{J}}(\hat{T}^{-1},\hat{T})\;\hat{G}^{(0)}[\hat{\sigma}_{D}^{(0)}]\;
\delta\hat{\mcal{H}}(\hat{T}^{-1},\hat{T})\Big)\;\hat{G}^{(0)}[\hat{\sigma}_{D}^{(0)}]\;
\hat{T}^{-1}\;\hat{I}\bigg)_{\beta\alpha}^{ba}\hat{\eta}|\widehat{J_{\psi;\alpha}^{a}\rangle}
\bigg\rangle_{\hat{\sigma}_{D}^{(0)}} +  \\ \no &+& \frac{1}{2}
\mbox{Tr}\;\mbox{STR}\bigg[\wtilde{\mcal{J}}(\hat{T}^{-1},\hat{T})\;
\Big\langle\hat{G}^{(0)}[\hat{\sigma}_{D}^{(0)}]\Big\rangle_{\hat{\sigma}_{D}^{(0)}}\bigg]
+  \\ \no &+&
\frac{\im}{2}\frac{1}{\mcal{N}}\bigg\langle
\widehat{\langle J_{\psi;\beta}^{b}}|\hat{\eta}\Big(
\hat{I}\;\wtilde{K}\;\hat{T}\;\hat{G}^{(0)}[\hat{\sigma}_{D}^{(0)}]\;
\wtilde{\mcal{J}}(\hat{T}^{-1},\hat{T})\;\hat{G}^{(0)}[\hat{\sigma}_{D}^{(0)}]\;
\hat{T}^{-1}\;\hat{I}\Big)_{\beta\alpha}^{ba}\hat{\eta}|\widehat{J_{\psi;\alpha}^{a}\rangle}
\bigg\rangle_{\hat{\sigma}_{D}^{(0)}} \;\;\;.
\eeq

\section{Dependence of gradients $\boldsymbol{\big[\hat{T}^{-1}\;\big(\vec{\pp}
\hat{T}\big)\big]}$ on the coset generator $\boldsymbol{\hat{Y}}$}  \lb{sc}

\subsection{Calculation of $\big[\hat{T}^{-1}(\vec{x},t_{p})\;\big(\wtilde{\pp}_{i}
\hat{T}(\vec{x},t_{p})\big)\big]_{\alpha\beta}^{ab}$ in terms of the independent pair condensates} \lb{sc1}

The \(U(L|S)\) left-invariant current-matrix-fields
\(\big[\hat{T}^{-1}(\vec{x},t_{p})\;\big(\wtilde{\pp}_{i}\hat{T}(\vec{x},t_{p})\big)\big]_{\alpha\beta}^{ab}\)
appear in the calculation of the integration measure of \(Osp(S,S|2L)\backslash U(L|S)\) and also
in the effective actions for the pair condensate fields (compare appendices \ref{sa1}-\ref{sa4} with
subsection \ref{s33} for the change of integration variables in the coherent state path integrals; see
section \ref{s4} with appendix \ref{sb} for the gradient expansion and the effective actions of
pair condensates with the transport coefficients averaged over the scalar density as background field).
These currents depend on the coset super-generator \(\hat{Y}(\vec{x},t_{p})\) which consists of the
independent anomalous fields. Therefore, it is of importance to find a suitable simplifying
representation of the matrix fields
\(\big[\hat{T}^{-1}(\vec{x},t_{p})\;\big(\wtilde{\pp}_{i}\hat{T}(\vec{x},t_{p})\big)\big]_{\alpha\beta}^{ab}\)
in terms of \(\hat{Y}(\vec{x},t_{p})\) in the exponential of
\(\hat{T}(\vec{x},t_{p})=\exp\{-\hat{Y}(\vec{x},t_{p})\}\). We transform the coset generator
\(\hat{Y}(\vec{x},t_{p})\) of the pair condensates with the \(U(L|S)\) matrices
\(\hat{P}_{\alpha\beta}^{aa}(\vec{x},t_{p})\) to its diagonal form \(\hat{Y}_{DD}(\vec{x},t_{p})\) (\ref{C1})
and apply this with the relations (\ref{s1_17}-\ref{s1_24}) for varying the exponentials of the
\(U(L|S)\) left-invariant current-matrix-fields (\ref{C2})\footnote{The detailed form of the \(U(L|S)\)
matrix \(\hat{P}_{\alpha\beta}^{aa}(\vec{x},t_{p})\) is described in subsection \ref{s33} and appendix
\ref{sa1}.}. We introduce the 'rotated' coset generator
\(\big(\wtilde{\pp}_{i}\hat{Y}\big)_{\hat{P}}=\big(\wtilde{\pp}_{i}\hat{Y}\ppr\big)\) (\ref{C3}) which
leads to the same dependence and symmetries in the coset matrix fields
\(\big(\wtilde{\pp}_{i}\hat{X}\ppr\big)_{\alpha\beta}\) and
\(\wtilde{\kappa}\;\big(\wtilde{\pp}_{i}\hat{X}\ppr\big)_{\alpha\beta}^{+}\) due to the properties
of a coset decomposition (\ref{C4}). Accordingly, we consider the 'rotated' boson-boson and fermion-fermion
pair condensate fields \(\big(\wtilde{\pp}_{i}\hat{c}_{D;mn}\ppr\big)\),
\(\big(\wtilde{\pp}_{i}\hat{f}_{D;rr\ppr}^{\prime(k)}\big)\) (\(k=0,1,3\))
which are symmetric and anti-symmetric under transposition, (except for
the symmetric fermion-fermion field \(\big(\wtilde{\pp}_{i}\hat{f}_{D;rr\ppr}^{\prime(2)}\big)\)
which achieves anti-symmetry by the anti-symmetric Pauli-matrix \(\tau_{2}\)) (\ref{C5}-\ref{C7}).
The odd pair condensate fields \(\big(\wtilde{\pp}_{i}\hat{\eta}_{D;r\mu,n}\ppr\big)\) are also effected
by the 'rotation' with \(\hat{P}_{\alpha\beta}^{aa}(\vec{x},t_{p})\) and differ only by sign between
the fermion-boson and boson-fermion blocks (\ref{C4})
\beq \lb{C1}
\hat{Y}&=&\hat{P}^{-1}\;\;\hat{Y}_{DD}\;\;\hat{P} \\ \lb{C2}
\hat{P}\;\hat{T}^{-1}\;\big(\wtilde{\pp}_{i}\hat{T}\big)\;\hat{P}^{-1}&=&
\hat{P}\;\exp\big\{\hat{Y}\big\}\;\Big(\wtilde{\pp}_{i}\exp\big\{-\hat{Y}\big\}\Big)\;\hat{P}^{-1} \\ \no &=&
-\int_{0}^{1}dv\;\;\exp\big\{v\;\hat{Y}_{DD}\big\}\;\;
\hat{P}\;\big(\wtilde{\pp}_{i}\hat{Y}\big)\;\hat{P}^{-1}\;\;\exp\big\{-v\;\hat{Y}_{DD}\big\} \\ \lb{C3}
\hat{P}\;\big(\wtilde{\pp}_{i}\hat{Y}\big)\;\hat{P}^{-1} &=&
\hat{P}\;\Big(\wtilde{\pp}_{i}\;\hat{P}^{-1}\;\hat{Y}_{DD}\;\hat{P}\Big)\;\hat{P}^{-1} =
\Big(\wtilde{\pp}_{i}\hat{Y}\Big)_{\hat{P}}=\big(\wtilde{\pp}_{i}\hat{Y}\ppr\big) = \\ \no &=&
\left(\bea{cc} 0 & \big(\wtilde{\pp}_{i}\hat{X}\ppr\big)_{\alpha\beta}  \\
\wtilde{\kappa}\;\big(\wtilde{\pp}_{i}\hat{X}\ppr\big)_{\alpha\beta}^{+} & 0 \eea\right)_{\alpha\beta}^{ab} \\ \lb{C4}
\big(\wtilde{\pp}_{i}\hat{X}\ppr\big)_{\alpha\beta} &=&\left(\bea{cc}
-\big(\wtilde{\pp}_{i}\hat{c}_{D;mn}\ppr\big) &
\big(\wtilde{\pp}_{i}\hat{\eta}_{D;m,r\ppr\nu}\ppr\big)   \\
-\big(\wtilde{\pp}_{i}\hat{\eta}_{D;r\mu,n}\ppr\big) &
\big(\wtilde{\pp}_{i}\hat{f}_{D;r\mu,r\ppr\nu}\ppr\big)  \eea\right)_{\alpha\beta}  \\  \lb{C5}
\big(\wtilde{\pp}_{i}\hat{c}_{D}\ppr\big)^{T}=\big(\wtilde{\pp}_{i}\hat{c}_{D}\ppr\big) &\Longrightarrow&
\big(\wtilde{\pp}_{i}\hat{c}_{D;mn}\ppr\big)=\big(\wtilde{\pp}_{i}\hat{c}_{D;nm}\ppr\big)\;\;;\hspace*{0.64cm}
m,n=1,\ldots,L   \\  \lb{C6}
\big(\wtilde{\pp}_{i}\hat{f}_{D;r\mu,r\ppr\nu}\ppr\big) &=&\sum_{k=0}^{3}\big(\tau_{k}\big)_{\mu\nu}\;\;
\big(\wtilde{\pp}_{i}\hat{f}_{D;rr\ppr}^{\prime(k)}\big)\;;\hspace*{0.64cm}r,r\ppr=1,\ldots,S/2\;;\hspace*{0.28cm}
\mu,\nu=1,2  \\  \no && \tau_{0}=\hat{1}_{2\times 2}\;;\hspace*{0.37cm}\tau_{1},\;\tau_{2},\;\tau_{3}=
\mbox{ Pauli-matrices }    \\    \lb{C7}
\big(\wtilde{\pp}_{i}\hat{f}_{D}\ppr\big)^{T}=-\big(\wtilde{\pp}_{i}\hat{f}_{D}\ppr\big) &\Longrightarrow&
\big(\wtilde{\pp}_{i}\hat{f}_{D;rr\ppr}^{\prime(2)}\big)=
\big(\wtilde{\pp}_{i}\hat{f}_{D;r\ppr r}^{\prime(2)}\big)   \\ \no&\Longrightarrow&
\big(\wtilde{\pp}_{i}\hat{f}_{D;rr\ppr}^{\prime(k)}\big)=-
\big(\wtilde{\pp}_{i}\hat{f}_{D;r\ppr r}^{\prime(k)}\big)\;\;\;\mbox{ for }\;k=0,1,3 \;\;\;.
\eeq
The eigenvalue matrices \(\hat{Y}_{DD}(\vec{x},t_{p})\) (\ref{C8}) and \(\hat{X}_{DD}(\vec{x},t_{p})\) (\ref{C9})
of the pair condensate fields in \(\hat{Y}(\vec{x},t_{p})\) (\ref{C1})
are determined by the complex fields \(\ovv{c}_{m}(\vec{x},t_{p})\)
with modulus \(|\ovv{c}_{m}(\vec{x},t_{p})|\) and phase \(\varphi_{m}(\vec{x},t_{p})\) for the
boson-boson part (\ref{C10}). Due to the anti-symmetry of BCS-pair condensates, the fermion-fermion part
of \(\hat{X}_{DD}(\vec{x},t_{p})\) (\ref{C9}) contains the anti-symmetric quaternion Pauli-matrix \(\tau_{2}\)
with the complex field \(\ovv{f}_{r}(\vec{x},t_{p})\) consisting of modulus
\(|\ovv{f}_{r}(\vec{x},t_{p})|\) and phase \(\phi_{r}(\vec{x},t_{p})\) (\ref{C11})
\beq \lb{C8}
\hat{Y}_{DD}&=&\left(\bea{cc} 0 & \hat{X}_{DD}   \\
\wtilde{\kappa}\;\hat{X}_{DD}^{+} & 0 \eea\right)   \\  \lb{C9}
\hat{X}_{DD}&=&\left(\bea{cc}-\ovv{c}_{m}\;\;\delta_{mn} & 0 \\
0 & \big(\tau_{2}\big)_{\mu\nu}\;\ovv{f}_{r}\;\;\delta_{rr\ppr} \eea\right)  \\  \lb{C10}
\ovv{c}_{m}&=&|\ovv{c}_{m}|\;\;\exp\{\im\;\varphi_{m}\}\;;\hspace*{0.64cm}m,n=1,\ldots,L \\  \lb{C11}
\ovv{f}_{r}&=&|\ovv{f}_{r}|\;\;\exp\{\im\;\phi_{r}\}\;;\hspace*{0.64cm}r,r\ppr=1,\ldots,S/2\;;
\;\;\;\;\mu,\nu=1,2\;.
\eeq
In relations (\ref{C12}-\ref{C25}), we explicitly list the results of the transformations (\ref{C1}-\ref{C7})
to the eigenvalues for the subgroup parts of
\(\big(\hat{P}\;\hat{T}^{-1}\;\big(\wtilde{\pp}_{i}\hat{T}\big)\;\hat{P}^{-1}\big)_{\alpha\beta}^{aa}\).
We separate the super-matrix fields
\(\big(\hat{P}\;\hat{T}^{-1}\;\big(\wtilde{\pp}_{i}\hat{T}\big)\;\hat{P}^{-1}\big)_{\alpha\beta}^{aa}\)
into its boson-boson, fermion-fermion, fermion-boson and boson-fermion blocks and compare the '11' part
with the '22' part by introducing the hermitian matrices \(\big(\wtilde{\pp}_{i}\hat{D}\big)_{mn}\),
\(\big(\wtilde{\pp}_{i}\hat{G}\big)_{r\mu,r\ppr\nu}\) for the boson-boson and fermion-fermion sections
of the \(U(L|S)\) subgroup matrix parts (\ref{C12},\ref{C13}). The relation among the odd parts of
\(\big(\hat{P}\;\hat{T}^{-1}\;\big(\wtilde{\pp}_{i}\hat{T}\big)\;\hat{P}^{-1}\big)_{\alpha\beta}^{aa}\)
becomes obvious by including the Grassmann fields \(\big(\wtilde{\pp}_{i}\hat{\xi}\big)_{r\mu,n}\),
\(\big(\wtilde{\pp}_{i}\hat{\xi}^{*}\big)_{r\mu,n}\) as abbreviation so that the anti-hermitian
property of the odd sections in the '11' subgroup part is displayed against the hermitian property in the
'22' block
\beq  \lb{C12}
\Big(\hat{P}\;\hat{T}^{-1}\;\big(\wtilde{\pp}_{i}\hat{T}\big)\;\hat{P}^{-1}\Big)_{\alpha\beta}^{11}&=&
\left(\bea{cc}
\Big(\hat{P}\;\hat{T}^{-1}\;\big(\wtilde{\pp}_{i}\hat{T}\big)\;\hat{P}^{-1}\Big)_{BB;mn}^{11} &
\Big(\hat{P}\;\hat{T}^{-1}\;\big(\wtilde{\pp}_{i}\hat{T}\big)\;\hat{P}^{-1}\Big)_{BF;m,r\ppr\nu}^{11} \\
\Big(\hat{P}\;\hat{T}^{-1}\;\big(\wtilde{\pp}_{i}\hat{T}\big)\;\hat{P}^{-1}\Big)_{FB;r\mu,n}^{11} &
\Big(\hat{P}\;\hat{T}^{-1}\;\big(\wtilde{\pp}_{i}\hat{T}\big)\;\hat{P}^{-1}\Big)_{FF;r\mu,r\ppr\nu}^{11}
\eea\right)  \\ \no &\propto& \im\;\left(\bea{cc}
\big(\wtilde{\pp}_{i}\hat{D}\big)_{mn} & \big(\wtilde{\pp}_{i}\hat{\xi}\big)_{m,r\ppr\nu}^{+} \\
\big(\wtilde{\pp}_{i}\hat{\xi}\big)_{r\mu,n} & \big(\wtilde{\pp}_{i}\hat{G}\big)_{r\mu,r\ppr\nu}
\eea\right)^{11}  \\ \lb{C13} \big(\wtilde{\pp}_{i}\hat{D}\big)_{mn}^{+}=\big(\wtilde{\pp}_{i}\hat{D}\big)_{mn}  &&
\big(\wtilde{\pp}_{i}\hat{G}\big)_{r\mu,r\ppr\nu}^{+}=\big(\wtilde{\pp}_{i}\hat{G}\big)_{r\mu,r\ppr\nu} \\ \lb{C14}
\Big(\hat{P}\;\hat{T}^{-1}\;\big(\wtilde{\pp}_{i}\hat{T}\big)\;\hat{P}^{-1}\Big)_{\alpha\beta}^{11,+}&=& -
\Big(\hat{P}\;\hat{T}^{-1}\;\big(\wtilde{\pp}_{i}\hat{T}\big)\;\hat{P}^{-1}\Big)_{\alpha\beta}^{11}
\eeq
\beq  \lb{C15}
\Big(\hat{P}\;\hat{T}^{-1}\;\big(\wtilde{\pp}_{i}\hat{T}\big)\;\hat{P}^{-1}\Big)_{\alpha\beta}^{22}&=&
\left(\bea{cc}
\Big(\hat{P}\;\hat{T}^{-1}\;\big(\wtilde{\pp}_{i}\hat{T}\big)\;\hat{P}^{-1}\Big)_{BB;mn}^{22} &
\Big(\hat{P}\;\hat{T}^{-1}\;\big(\wtilde{\pp}_{i}\hat{T}\big)\;\hat{P}^{-1}\Big)_{BF;m,r\ppr\nu}^{22} \\
\Big(\hat{P}\;\hat{T}^{-1}\;\big(\wtilde{\pp}_{i}\hat{T}\big)\;\hat{P}^{-1}\Big)_{FB;r\mu,n}^{22} &
\Big(\hat{P}\;\hat{T}^{-1}\;\big(\wtilde{\pp}_{i}\hat{T}\big)\;\hat{P}^{-1}\Big)_{FF;r\mu,r\ppr\nu}^{22}
\eea\right)  \\ \no &\propto& -\im\;\left(\bea{cc}
\big(\wtilde{\pp}_{i}\hat{D}\big)_{mn}^{T} & -\big(\wtilde{\pp}_{i}\hat{\xi}\big)_{m,r\ppr\nu}^{T} \\
\big(\wtilde{\pp}_{i}\hat{\xi}^{*}\big)_{r\mu,n} & \big(\wtilde{\pp}_{i}\hat{G}\big)_{r\mu,r\ppr\nu}^{T}
\eea\right)^{22}_{\mbox{.}}
\eeq
In fact, the '11' subgroup part
\(\big(\hat{P}\;\hat{T}^{-1}\;\big(\wtilde{\pp}_{i}\hat{T}\big)\;\hat{P}^{-1}\big)_{\alpha\beta}^{11}\)
follows by super-transposition of the '22' block with an additional minus sign (\ref{C16}). We also have
the anti-hermitian property of the boson-boson and fermion-fermion sections of the '22' part (\ref{C17},\ref{C18})
as in the '11' block (\ref{C14}), but a hermitian relation between boson-fermion and fermion-boson sections
(\ref{C19}) instead of the complete anti-hermitian property in the '11' part (\ref{C14})
\beq \lb{C16}
\Big(\hat{P}\;\hat{T}^{-1}\;\big(\wtilde{\pp}_{i}\hat{T}\big)\;\hat{P}^{-1}\Big)_{\alpha\beta}^{22,st} &=& -
\Big(\hat{P}\;\hat{T}^{-1}\;\big(\wtilde{\pp}_{i}\hat{T}\big)\;\hat{P}^{-1}\Big)_{\alpha\beta}^{11}  \\ \lb{C17}
\Big(\hat{P}\;\hat{T}^{-1}\;\big(\wtilde{\pp}_{i}\hat{T}\big)\;\hat{P}^{-1}\Big)_{BB;mn}^{22,+} &=& -
\Big(\hat{P}\;\hat{T}^{-1}\;\big(\wtilde{\pp}_{i}\hat{T}\big)\;\hat{P}^{-1}\Big)_{BB;mn}^{22}  \\ \lb{C18}
\Big(\hat{P}\;\hat{T}^{-1}\;\big(\wtilde{\pp}_{i}\hat{T}\big)\;\hat{P}^{-1}\Big)_{FF;r\mu,r\ppr\nu}^{22,+} &=& -
\Big(\hat{P}\;\hat{T}^{-1}\;\big(\wtilde{\pp}_{i}\hat{T}\big)\;\hat{P}^{-1}\Big)_{FF;r\mu,r\ppr\nu}^{22}  \\ \lb{C19}
\Big(\hat{P}\;\hat{T}^{-1}\;\big(\wtilde{\pp}_{i}\hat{T}\big)\;\hat{P}^{-1}\Big)_{BF;m,r\ppr\nu}^{22,+} &=&
\Big(\hat{P}\;\hat{T}^{-1}\;\big(\wtilde{\pp}_{i}\hat{T}\big)\;\hat{P}^{-1}\Big)_{FB;r\ppr\nu,m}^{22}\;\;\;.
\eeq
Applying the transformations (\ref{C1}-\ref{C11}) to the eigenvalue matrix \(\hat{Y}_{DD}\) of
the coset super-generator \(\hat{Y}\), we obtain for the boson-boson parts of
\(\big(\hat{P}\;\hat{T}^{-1}\;\big(\wtilde{\pp}_{i}\hat{T}\big)\;\hat{P}^{-1}\big)_{BB;mn}^{aa}\)
the Eqs. (\ref{C20},\ref{C21}) where one has to distinguish between the diagonal \(m=n\) and off-diagonal
\(m\neq n\) matrix elements
\beq \lb{C20}
\Big(\hat{P}\;\hat{T}^{-1}\;\big(\wtilde{\pp}_{i}\hat{T}\big)\;\hat{P}^{-1}\Big)_{BB;mm}^{11} &=& -
\Big(\hat{P}\;\hat{T}^{-1}\;\big(\wtilde{\pp}_{i}\hat{T}\big)\;\hat{P}^{-1}\Big)_{BB;mm}^{22}=  \\  \no &=&
\bigg(\big(\wtilde{\pp}_{i}\hat{c}_{D;mm}^{\prime *}\big)\;e^{\im\;\varphi_{m}}-
\big(\wtilde{\pp}_{i}\hat{c}_{D;mm}^{\prime}\big)\;e^{-\im\;\varphi_{m}}\bigg)
\frac{\Big(\sin\big(|\ovv{c}_{m}|\big)\Big)^{2}}{2\;|\ovv{c}_{m}|}
\eeq
\beq \lb{C21}
\lefteqn{
\Big(\hat{P}\;\hat{T}^{-1}\;\big(\wtilde{\pp}_{i}\hat{T}\big)\;\hat{P}^{-1}\Big)_{BB;mn}^{11}
\stackrel{m\neq n}{=} -
\Big(\hat{P}\;\hat{T}^{-1}\;\big(\wtilde{\pp}_{i}\hat{T}\big)\;\hat{P}^{-1}\Big)_{BB;nm}^{22}
\stackrel{m\neq n}{=} }  \\  \no &=&
\big(\wtilde{\pp}_{i}\hat{c}_{D;mn}^{\prime *}\big)\;e^{\im\;\varphi_{m}}
\bigg(\frac{|\ovv{c}_{m}|-|\ovv{c}_{m}|\;\cos\big(|\ovv{c}_{m}|\big)\;\cos\big(|\ovv{c}_{n}|\big)-
|\ovv{c}_{n}|\;\sin\big(|\ovv{c}_{m}|\big)\;\sin\big(|\ovv{c}_{n}|\big)}
{|\ovv{c}_{m}|^{2}-|\ovv{c}_{n}|^{2}}\bigg) + \\  \no &-&
\big(\wtilde{\pp}_{i}\hat{c}_{D;mn}^{\prime}\big)\;e^{-\im\;\varphi_{n}}
\bigg(\frac{|\ovv{c}_{n}|-|\ovv{c}_{n}|\;\cos\big(|\ovv{c}_{n}|\big)\;\cos\big(|\ovv{c}_{m}|\big)-
|\ovv{c}_{m}|\;\sin\big(|\ovv{c}_{n}|\big)\;\sin\big(|\ovv{c}_{m}|\big)}
{|\ovv{c}_{n}|^{2}-|\ovv{c}_{m}|^{2}}\bigg)_{\mbox{.}}
\eeq
Similar results hold for the fermion-fermion block of
\(\big(\hat{P}\;\hat{T}^{-1}\;\big(\wtilde{\pp}_{i}\hat{T}\big)\;\hat{P}^{-1}\big)_{FF;r\mu,r\ppr\nu}^{aa}\)
where one has to take into account the quaternion structure of the matrix elements with Pauli matrices
\(\tau_{0}\), \(\tau_{1}\), \(\tau_{2}\), \(\tau_{3}\). The case of diagonal entries with
\(r=r\ppr\), \(\mu=\nu\) is given in relation (\ref{C22}) and that of the off-diagonal matrix elements
with \(r\neq r\ppr\), \(\mu,\nu=1,2\) in Eq. (\ref{C23})
\beq \lb{C22}
\Big(\hat{P}\;\hat{T}^{-1}\;\big(\wtilde{\pp}_{i}\hat{T}\big)\;\hat{P}^{-1}\Big)_{FF;r\mu,r\nu}^{11} &=& -
\Big(\hat{P}\;\hat{T}^{-1}\;\big(\wtilde{\pp}_{i}\hat{T}\big)\;\hat{P}^{-1}\Big)_{FF;r\nu,r\mu}^{22}=  \\  \no &=&
\bigg(\big(\wtilde{\pp}_{i}\hat{f}_{D;rr}^{\prime (2)}\big)\;e^{-\im\;\phi_{r}}-
\big(\wtilde{\pp}_{i}\hat{f}_{D;rr}^{\prime (2)*}\big)\;e^{\im\;\phi_{r}}\bigg)\;\delta_{\mu\nu}\;
\frac{\Big(\sinh\big(|\ovv{f}_{r}|\big)\Big)^{2}}{2\;|\ovv{f}_{r}|}
\eeq
\beq \lb{C23}
\lefteqn{
\Big(\hat{P}\;\hat{T}^{-1}\;\big(\wtilde{\pp}_{i}\hat{T}\big)\;\hat{P}^{-1}\Big)_{FF;r\mu,r\ppr\nu}^{11}
\stackrel{r\neq r\ppr}{=} -
\Big(\hat{P}\;\hat{T}^{-1}\;\big(\wtilde{\pp}_{i}\hat{T}\big)\;\hat{P}^{-1}\Big)_{FF;r\ppr\nu,r\mu}^{22}
\stackrel{r\neq r\ppr}{=} }
\\ \no &=& \bigg(\big(\wtilde{\pp}_{i}\hat{f}_{D;rr\ppr}^{\prime(0)}\big)\;\big(\tau_{2}\big)_{\mu\nu}+
\big(\wtilde{\pp}_{i}\hat{f}_{D;rr\ppr}^{\prime(1)}\big)\;\im\;\big(\tau_{3}\big)_{\mu\nu} +
\big(\wtilde{\pp}_{i}\hat{f}_{D;rr\ppr}^{\prime(2)}\big)\;\big(\tau_{0}\big)_{\mu\nu}-
\big(\wtilde{\pp}_{i}\hat{f}_{D;rr\ppr}^{\prime(3)}\big)\;\im\;\big(\tau_{1}\big)_{\mu\nu}\bigg)\;
e^{-\im\;\phi_{r\ppr}}\;\;\times \\ \no &\times&
\bigg(\frac{-|\ovv{f}_{r\ppr}|+|\ovv{f}_{r\ppr}|\;\cosh\big(|\ovv{f}_{r\ppr}|\big)\;\cosh\big(|\ovv{f}_{r}|\big)-
|\ovv{f}_{r}|\;\sinh\big(|\ovv{f}_{r\ppr}|\big)\;\sinh\big(|\ovv{f}_{r}|\big)}
{|\ovv{f}_{r\ppr}|^{2}-|\ovv{f}_{r}|^{2}}\bigg) + \\  \no &+&
\bigg(\big(\wtilde{\pp}_{i}\hat{f}_{D;rr\ppr}^{\prime(0)*}\big)\;\big(\tau_{2}\big)_{\mu\nu}-
\big(\wtilde{\pp}_{i}\hat{f}_{D;rr\ppr}^{\prime(1)*}\big)\;\im\;\big(\tau_{3}\big)_{\mu\nu} -
\big(\wtilde{\pp}_{i}\hat{f}_{D;rr\ppr}^{\prime(2)*}\big)\;\big(\tau_{0}\big)_{\mu\nu}+
\big(\wtilde{\pp}_{i}\hat{f}_{D;rr\ppr}^{\prime(3)*}\big)\;\im\;\big(\tau_{1}\big)_{\mu\nu}\bigg)\;
e^{\im\;\phi_{r}}\;\;\times \\ \no &\times&
\bigg(\frac{-|\ovv{f}_{r}|+|\ovv{f}_{r}|\;\cosh\big(|\ovv{f}_{r}|\big)\;\cosh\big(|\ovv{f}_{r\ppr}|\big)-
|\ovv{f}_{r\ppr}|\;\sinh\big(|\ovv{f}_{r}|\big)\;\sinh\big(|\ovv{f}_{r\ppr}|\big)}
{|\ovv{f}_{r}|^{2}-|\ovv{f}_{r\ppr}|^{2}}\bigg)_{\mbox{.}}
\eeq
The relations for the fermion-boson and boson-fermion blocks are listed in Eqs. (\ref{C24},\ref{C25})
with the anti-commuting pair condensate fields
\(\big(\wtilde{\pp}_{i}\hat{\eta}_{D;r\mu,n}^{\prime}\big)\),
\(\big(\wtilde{\pp}_{i}\hat{\eta}_{D;r\mu,n}^{\prime *}\big)\)
and with the sine-, cosine-functions of the eigenvalues \(|\ovv{c}_{m}|\), \(|\ovv{c}_{n}|\)
from the boson-boson section and the corresponding hyperbolic functions of the eigenvalues
\(|\ovv{f}_{r}|\), \(|\ovv{f}_{r\ppr}|\) from the fermion-fermion part
\beq \lb{C24}
\lefteqn{
\Big(\hat{P}\;\hat{T}^{-1}\;\big(\wtilde{\pp}_{i}\hat{T}\big)\;\hat{P}^{-1}\Big)_{FB;r\mu,n}^{11} =
\Big(\hat{P}\;\hat{T}^{-1}\;\big(\wtilde{\pp}_{i}\hat{T}\big)\;\hat{P}^{-1}\Big)_{BF;n,r\mu}^{22,T}= } \\ \no &=&
\big(\tau_{2}\big)_{\mu\kappa}\;\big(\wtilde{\pp}_{i}\hat{\eta}_{D;r\kappa,n}^{\prime *}\big)\;
e^{\im\;\phi_{r}}\;\bigg(\frac{|\ovv{f}_{r}|-|\ovv{f}_{r}|\;\cos\big(|\ovv{c}_{n}|\big)\;
\cosh\big(|\ovv{f}_{r}|\big)-|\ovv{c}_{n}|\;\sin\big(|\ovv{c}_{n}|\big)\;\sinh\big(|\ovv{f}_{r}|\big)}
{|\ovv{c}_{n}|^{2}+|\ovv{f}_{r}|^{2}}\bigg) +  \\ \no &-&
\big(\wtilde{\pp}_{i}\hat{\eta}_{D;r\mu,n}^{\prime}\big)\;
e^{-\im\;\varphi_{n}}\;\bigg(\frac{|\ovv{c}_{n}|-|\ovv{c}_{n}|\;\cos\big(|\ovv{c}_{n}|\big)\;
\cosh\big(|\ovv{f}_{r}|\big)+|\ovv{f}_{r}|\;\sin\big(|\ovv{c}_{n}|\big)\;\sinh\big(|\ovv{f}_{r}|\big)}
{|\ovv{c}_{n}|^{2}+|\ovv{f}_{r}|^{2}}\bigg)
\eeq
\beq \lb{C25}
\lefteqn{
\Big(\hat{P}\;\hat{T}^{-1}\;\big(\wtilde{\pp}_{i}\hat{T}\big)\;\hat{P}^{-1}\Big)_{BF;m,r\ppr\nu}^{11} = -
\Big(\hat{P}\;\hat{T}^{-1}\;\big(\wtilde{\pp}_{i}\hat{T}\big)\;\hat{P}^{-1}\Big)_{FB;r\ppr\nu,m}^{22,T}= } \\ \no &=&
\big(\wtilde{\pp}_{i}\hat{\eta}_{D;m,r\ppr\nu}^{\prime *}\big)\;
e^{\im\;\varphi_{m}}\;\bigg(\frac{|\ovv{c}_{m}|-|\ovv{c}_{m}|\;\cos\big(|\ovv{c}_{m}|\big)\;
\cosh\big(|\ovv{f}_{r\ppr}|\big)+|\ovv{f}_{r\ppr}|\;\sin\big(|\ovv{c}_{m}|\big)\;\sinh\big(|\ovv{f}_{r\ppr}|\big)}
{|\ovv{c}_{m}|^{2}+|\ovv{f}_{r\ppr}|^{2}}\bigg) +  \\ \no &-&
\big(\wtilde{\pp}_{i}\hat{\eta}_{D;m,r\ppr\lambda}^{\prime}\big)\;\big(\tau_{2}\big)_{\lambda\nu}\;
e^{-\im\;\phi_{r\ppr}}\;\bigg(\frac{|\ovv{f}_{r\ppr}|-|\ovv{f}_{r\ppr}|\;\cos\big(|\ovv{c}_{m}|\big)\;
\cosh\big(|\ovv{f}_{r\ppr}|\big)-|\ovv{c}_{m}|\;\sin\big(|\ovv{c}_{m}|\big)\;\sinh\big(|\ovv{f}_{r\ppr}|\big)}
{|\ovv{c}_{m}|^{2}+|\ovv{f}_{r\ppr}|^{2}}\bigg)_{\mbox{.}}
\eeq
The structure of matrix elements
\(\big(\hat{P}\;\hat{T}^{-1}\;\big(\wtilde{\pp}_{i}\hat{T}\big)\;\hat{P}^{-1}\big)_{\alpha\beta}^{a\neq b}\)
in the coset parts \(a\neq b\) is itemized in relations (\ref{C26}-\ref{C30}) by introducing the symmetric
boson-boson matrix \(\big(\wtilde{\pp}_{i}\hat{B}\big)_{mn}\), the anti-symmetric fermion-fermion
matrix \(\big(\wtilde{\pp}_{i}\hat{A}\big)_{r\mu,r\ppr\nu}\) and the odd parts
\(\big(\wtilde{\pp}_{i}\hat{\zeta}\big)_{r\mu,n}\),
\(\big(\wtilde{\pp}_{i}\hat{\zeta}\big)_{m,r\ppr\nu}^{T}\).
These matrices \(\big(\wtilde{\pp}_{i}\hat{B}\big)_{mn}\), \(\big(\wtilde{\pp}_{i}\hat{A}\big)_{r\mu,r\ppr\nu}\),
\(\big(\wtilde{\pp}_{i}\hat{\zeta}\big)_{r\mu,n}\) and
\(\big(\wtilde{\pp}_{i}\hat{\zeta}\big)_{m,r\ppr\nu}^{T}\)
have symmetries (\ref{C27},\ref{C30}) and respect with their ordering in the '12' and '21' coset parts
the general defining relation (\ref{C28}) of the \(Osp(S,S|2L)\backslash U(L|S)\) coset decomposition
\beq \no
\Big(\hat{P}\;\hat{T}^{-1}\;\big(\wtilde{\pp}_{i}\hat{T}\big)\;\hat{P}^{-1}\Big)_{\alpha\beta}^{12}&=&
\left(\bea{cc}
\Big(\hat{P}\;\hat{T}^{-1}\;\big(\wtilde{\pp}_{i}\hat{T}\big)\;\hat{P}^{-1}\Big)_{BB;mn}^{12} &
\Big(\hat{P}\;\hat{T}^{-1}\;\big(\wtilde{\pp}_{i}\hat{T}\big)\;\hat{P}^{-1}\Big)_{BF;m,r\ppr\nu}^{12} \\
\Big(\hat{P}\;\hat{T}^{-1}\;\big(\wtilde{\pp}_{i}\hat{T}\big)\;\hat{P}^{-1}\Big)_{FB;r\mu,n}^{12} &
\Big(\hat{P}\;\hat{T}^{-1}\;\big(\wtilde{\pp}_{i}\hat{T}\big)\;\hat{P}^{-1}\Big)_{FF;r\mu,r\ppr\nu}^{12}
\eea\right)  \\ \lb{C26} &\propto& \left(\bea{cc}
-\big(\wtilde{\pp}_{i}\hat{B}\big)_{mn}^{T} & \big(\wtilde{\pp}_{i}\hat{\zeta}\big)_{m,r\ppr\nu}^{T} \\
-\big(\wtilde{\pp}_{i}\hat{\zeta}\big)_{r\mu,n} & \big(\wtilde{\pp}_{i}\hat{A}\big)_{r\mu,r\ppr\nu}
\eea\right)^{12}  \\   \lb{C27}
\big(\wtilde{\pp}_{i}\hat{B}\big)_{mn}^{T}=\big(\wtilde{\pp}_{i}\hat{B}\big)_{mn}  &&
\big(\wtilde{\pp}_{i}\hat{A}\big)_{r\mu,r\ppr\nu}^{T}=-\big(\wtilde{\pp}_{i}\hat{A}\big)_{r\mu,r\ppr\nu} \\ \lb{C28}
\bigg(\Big(\hat{P}\;\hat{T}^{-1}\;\big(\wtilde{\pp}_{i}\hat{T}\big)\;\hat{P}^{-1}\Big)^{12}\;\wtilde{\kappa}
\bigg)_{\alpha\beta}^{+}&=&
\Big(\hat{P}\;\hat{T}^{-1}\;\big(\wtilde{\pp}_{i}\hat{T}\big)\;\hat{P}^{-1}\Big)_{\alpha\beta}^{21}  \\ \no
\Big(\hat{P}\;\hat{T}^{-1}\;\big(\wtilde{\pp}_{i}\hat{T}\big)\;\hat{P}^{-1}\Big)_{\alpha\beta}^{21}&=&
\left(\bea{cc}
\Big(\hat{P}\;\hat{T}^{-1}\;\big(\wtilde{\pp}_{i}\hat{T}\big)\;\hat{P}^{-1}\Big)_{BB;mn}^{21} &
\Big(\hat{P}\;\hat{T}^{-1}\;\big(\wtilde{\pp}_{i}\hat{T}\big)\;\hat{P}^{-1}\Big)_{BF;m,r\ppr\nu}^{21} \\
\Big(\hat{P}\;\hat{T}^{-1}\;\big(\wtilde{\pp}_{i}\hat{T}\big)\;\hat{P}^{-1}\Big)_{FB;r\mu,n}^{21} &
\Big(\hat{P}\;\hat{T}^{-1}\;\big(\wtilde{\pp}_{i}\hat{T}\big)\;\hat{P}^{-1}\Big)_{FF;r\mu,r\ppr\nu}^{21}
\eea\right)  \\ \lb{C29} &\propto& \left(\bea{cc}
\big(\wtilde{\pp}_{i}\hat{B}\big)_{mn}^{+} & \big(\wtilde{\pp}_{i}\hat{\zeta}\big)_{m,r\ppr\nu}^{+} \\
\big(\wtilde{\pp}_{i}\hat{\zeta}^{*}\big)_{r\mu,n} & \big(\wtilde{\pp}_{i}\hat{A}\big)_{r\mu,r\ppr\nu}^{+}
\eea\right)^{21}  \\   \lb{C30}
\big(\wtilde{\pp}_{i}\hat{B}\big)_{mn}^{+}=\big(\wtilde{\pp}_{i}\hat{B}^{*}\big)_{mn}  &&
\big(\wtilde{\pp}_{i}\hat{A}\big)_{r\mu,r\ppr\nu}^{+}=-\big(\wtilde{\pp}_{i}\hat{A}^{*}\big)_{r\mu,r\ppr\nu}\;\;\;.
\eeq
We also record the coset currents
\(\big(\hat{P}\;\hat{T}^{-1}\;\big(\wtilde{\pp}_{i}\hat{T}\big)\;\hat{P}^{-1}\big)_{\alpha\beta}^{a\neq b}\)
in terms of the eigenvalues \(\hat{Y}_{DD}\), \(\hat{X}_{DD}\) according to relations (\ref{C1}-\ref{C11}).
The symmetric boson-boson part for \(a\neq b\) is achieved with the 'rotated' fields
\(\big(\wtilde{\pp}_{i}\hat{c}_{D;mn}^{\prime}\big)\),
\(\big(\wtilde{\pp}_{i}\hat{c}_{D;mn}^{\prime *}\big)\) and modulus and phase of the eigenvalues
\(\ovv{c}_{m}(\vec{x},t_{p})\), \(\ovv{c}_{n}(\vec{x},t_{p})\)
which result into sine and cosine functions. One has also to split for the cases of diagonal \(m=n\) and
off-diagonal matrix elements \(m\neq n\) so that we find from Eq. (\ref{C2}) the relations (\ref{C31},\ref{C32})
\beq \lb{C31}
\lefteqn{
\Big(\hat{P}\;\hat{T}^{-1}\;\big(\wtilde{\pp}_{i}\hat{T}\big)\;\hat{P}^{-1}\Big)_{BB;mm}^{12}= -
\Big(\hat{P}\;\hat{T}^{-1}\;\big(\wtilde{\pp}_{i}\hat{T}\big)\;\hat{P}^{-1}\Big)_{BB;mm}^{21,*} =} \\ \no &=&
\big(\wtilde{\pp}_{i}\hat{c}_{D;mm}^{\prime *}\big)\;e^{\im\;2\varphi_{m}}\;
\bigg(\frac{1}{2}-\frac{\sin\big(2\;|\ovv{c}_{m}|\big)}{4\;|\ovv{c}_{m}|}\bigg) +
\big(\wtilde{\pp}_{i}\hat{c}_{D;mm}^{\prime}\big)\;
\bigg(\frac{1}{2}+\frac{\sin\big(2\;|\ovv{c}_{m}|\big)}{4\;|\ovv{c}_{m}|}\bigg)
\eeq
\beq \lb{C32}
\lefteqn{
\Big(\hat{P}\;\hat{T}^{-1}\;\big(\wtilde{\pp}_{i}\hat{T}\big)\;\hat{P}^{-1}\Big)_{BB;mn}^{12}
\stackrel{m\neq n}{=} -
\Big(\hat{P}\;\hat{T}^{-1}\;\big(\wtilde{\pp}_{i}\hat{T}\big)\;\hat{P}^{-1}\Big)_{BB;mn}^{21,*}
\stackrel{m\neq n}{=}  } \\ \no &=&
\big(\wtilde{\pp}_{i}\hat{c}_{D;mn}^{\prime *}\big)\;e^{\im\;(\varphi_{m}+\varphi_{n})}\;
\bigg(\frac{|\ovv{c}_{n}|\;\cos\big(|\ovv{c}_{n}|\big)\;\sin\big(|\ovv{c}_{m}|\big)-
|\ovv{c}_{m}|\;\cos\big(|\ovv{c}_{m}|\big)\;\sin\big(|\ovv{c}_{n}|\big)}
{|\ovv{c}_{m}|^{2}-|\ovv{c}_{n}|^{2}}\bigg)  +   \\ \no &+&
\big(\wtilde{\pp}_{i}\hat{c}_{D;mn}^{\prime}\big)\;
\bigg(\frac{|\ovv{c}_{m}|\;\cos\big(|\ovv{c}_{n}|\big)\;\sin\big(|\ovv{c}_{m}|\big)-
|\ovv{c}_{n}|\;\cos\big(|\ovv{c}_{m}|\big)\;\sin\big(|\ovv{c}_{n}|\big)}
{|\ovv{c}_{m}|^{2}-|\ovv{c}_{n}|^{2}}\bigg)_{\mbox{.}}
\eeq
Corresponding relations of (\ref{C31},\ref{C32}) follow for the fermion-fermion sections in the
coset parts \(a\neq b\) of
\(\big(\hat{P}\;\hat{T}^{-1}\;\big(\wtilde{\pp}_{i}\hat{T}\big)\;\hat{P}^{-1}\big)_{\alpha\beta}^{a\neq b}\);
however, we have to consider the quaternion structure of the matrix elements in the fermion-fermion sections
and have to replace the sine, cosine of the eigenvalues by its hyperbolic functions
of \(|\ovv{f}_{r}|\), \(|\ovv{f}_{r\ppr}|\). The diagonal matrix elements \(r=r\ppr\) with quaternion
Pauli matrix \(\tau_{2}\) are determined by the 'rotated' pair condensate fields
\(\big(\wtilde{\pp}_{i}\hat{f}_{D;rr}^{\prime(2)}\big)\),
\(\big(\wtilde{\pp}_{i}\hat{f}_{D;rr}^{\prime(2)*}\big)\) (\ref{C33}) whereas the off-diagonal
elements \(r\neq r\ppr\) are contained in Eq. (\ref{C34})
\beq \lb{C33}
\lefteqn{
\Big(\hat{P}\;\hat{T}^{-1}\;\big(\wtilde{\pp}_{i}\hat{T}\big)\;\hat{P}^{-1}\Big)_{FF;r\mu,r\nu}^{12}= -
\Big(\hat{P}\;\hat{T}^{-1}\;\big(\wtilde{\pp}_{i}\hat{T}\big)\;\hat{P}^{-1}\Big)_{FF;r\mu,r\nu}^{21,*} =} \\ \no &=&
-\big(\tau_{2}\big)_{\mu\nu}\;
\bigg[\big(\wtilde{\pp}_{i}\hat{f}_{D;rr}^{\prime(2)}\big)\;
\bigg(\frac{1}{2}+\frac{\sinh\big(2\;|\ovv{f}_{r}|\big)}{4\;|\ovv{f}_{r}|}\bigg) +
\big(\wtilde{\pp}_{i}\hat{f}_{D;rr}^{\prime(2)*}\big)\;e^{\im\;2\phi_{r}}\;
\bigg(\frac{1}{2}-\frac{\sinh\big(2\;|\ovv{f}_{r}|\big)}{4\;|\ovv{f}_{r}|}\bigg)\bigg]
\eeq
\beq \lb{C34}
\lefteqn{
\Big(\hat{P}\;\hat{T}^{-1}\;\big(\wtilde{\pp}_{i}\hat{T}\big)\;\hat{P}^{-1}\Big)_{FF;r\mu,r\ppr\nu}^{12}
\stackrel{r\neq r\ppr}{=} -
\Big(\hat{P}\;\hat{T}^{-1}\;\big(\wtilde{\pp}_{i}\hat{T}\big)\;\hat{P}^{-1}\Big)_{FF;r\mu,r\ppr\nu}^{21,*}
\stackrel{r\neq r\ppr}{=}  }    \\ \no &=& -
\bigg(\big(\wtilde{\pp}_{i}\hat{f}_{D;rr\ppr}^{\prime(0)}\big)\;\big(\tau_{0}\big)_{\mu\nu}+
\big(\wtilde{\pp}_{i}\hat{f}_{D;rr\ppr}^{\prime(1)}\big)\;\big(\tau_{1}\big)_{\mu\nu} +
\big(\wtilde{\pp}_{i}\hat{f}_{D;rr\ppr}^{\prime(2)}\big)\;\big(\tau_{2}\big)_{\mu\nu}+
\big(\wtilde{\pp}_{i}\hat{f}_{D;rr\ppr}^{\prime(3)}\big)\;\big(\tau_{3}\big)_{\mu\nu}\bigg)
\;\times \\ \no &\times&
\bigg(\frac{|\ovv{f}_{r}|\;\cosh\big(|\ovv{f}_{r\ppr}|\big)\;\sinh\big(|\ovv{f}_{r}|\big)-
|\ovv{f}_{r\ppr}|\;\cosh\big(|\ovv{f}_{r}|\big)\;\sinh\big(|\ovv{f}_{r\ppr}|\big)}
{|\ovv{f}_{r}|^{2}-|\ovv{f}_{r\ppr}|^{2}}\bigg) + \\  \no &+&
\bigg(-\big(\wtilde{\pp}_{i}\hat{f}_{D;rr\ppr}^{\prime(0)*}\big)\;\big(\tau_{0}\big)_{\mu\nu}+
\big(\wtilde{\pp}_{i}\hat{f}_{D;rr\ppr}^{\prime(1)*}\big)\;\big(\tau_{1}\big)_{\mu\nu} +
\big(\wtilde{\pp}_{i}\hat{f}_{D;rr\ppr}^{\prime(2)*}\big)\;\big(\tau_{2}\big)_{\mu\nu}+
\big(\wtilde{\pp}_{i}\hat{f}_{D;rr\ppr}^{\prime(3)*}\big)\;\big(\tau_{3}\big)_{\mu\nu}\bigg)\;
e^{\im\;(\phi_{r}+\phi_{r\ppr})}\;\times \\ \no &\times&
\bigg(\frac{|\ovv{f}_{r}|\;\cosh\big(|\ovv{f}_{r}|\big)\;\sinh\big(|\ovv{f}_{r\ppr}|\big)-
|\ovv{f}_{r\ppr}|\;\cosh\big(|\ovv{f}_{r\ppr}|\big)\;\sinh\big(|\ovv{f}_{r}|\big)}
{|\ovv{f}_{r}|^{2}-|\ovv{f}_{r\ppr}|^{2}}\bigg)_{\mbox{.}}
\eeq
Finally, we also acquire relations (\ref{C35},\ref{C36}) for the 'rotated' odd pair condensate
fields \(\big(\wtilde{\pp}_{i}\hat{\eta}_{D;r\mu,n}^{\prime}\big)\),
\(\big(\wtilde{\pp}_{i}\hat{\eta}_{D;r\mu,n}^{\prime *}\big)\).
These equations contain the trigonometric sine-, cosine-functions of the boson-boson eigenvalues
\(\ovv{c}_{m}\), \(\ovv{c}_{n}\) and its corresponding hyperbolic functions for the fermion-fermion
eigenvalues \(\ovv{f}_{r}\), \(\ovv{f}_{r\ppr}\) so that the boson-boson sections of the super-pair
condensates have a compact structure whereas the fermion-fermion pair condensates
have non-compact degrees of freedom
\beq \lb{C35}
\lefteqn{
\Big(\hat{P}\;\hat{T}^{-1}\;\big(\wtilde{\pp}_{i}\hat{T}\big)\;\hat{P}^{-1}\Big)_{FB;r\mu,n}^{12}= -
\Big(\hat{P}\;\hat{T}^{-1}\;\big(\wtilde{\pp}_{i}\hat{T}\big)\;\hat{P}^{-1}\Big)_{FB;r\mu,n}^{21,*} =}
\\ \no &=&
\big(\wtilde{\pp}_{i}\hat{\eta}_{D;r\mu,n}^{\prime}\big)\;
\bigg(\frac{|\ovv{c}_{n}|\;\cosh\big(|\ovv{f}_{r}|\big)\;\sin\big(|\ovv{c}_{n}|\big)+
|\ovv{f}_{r}|\;\cos\big(|\ovv{c}_{n}|\big)\;\sinh\big(|\ovv{f}_{r}|\big)}
{|\ovv{c}_{n}|^{2}+|\ovv{f}_{r}|^{2}}\bigg) + \\  \no &-&
\big(\tau_{2}\big)_{\mu\kappa}\;\big(\wtilde{\pp}_{i}\hat{\eta}_{D;r\kappa,n}^{\prime *}\big)\;
e^{\im\;(\phi_{r}+\varphi_{n})}\;
\bigg(\frac{|\ovv{f}_{r}|\;\cosh\big(|\ovv{f}_{r}|\big)\;\sin\big(|\ovv{c}_{n}|\big)-
|\ovv{c}_{n}|\;\cos\big(|\ovv{c}_{n}|\big)\;\sinh\big(|\ovv{f}_{r}|\big)}
{|\ovv{c}_{n}|^{2}+|\ovv{f}_{r}|^{2}}\bigg)
\eeq
\beq \lb{C36}
\lefteqn{
\Big(\hat{P}\;\hat{T}^{-1}\;\big(\wtilde{\pp}_{i}\hat{T}\big)\;\hat{P}^{-1}\Big)_{BF;m,r\ppr\nu}^{12}=
\Big(\hat{P}\;\hat{T}^{-1}\;\big(\wtilde{\pp}_{i}\hat{T}\big)\;\hat{P}^{-1}\Big)_{BF;m,r\ppr\nu}^{21,*} =}
\\ \no &=& -
\big(\wtilde{\pp}_{i}\hat{\eta}_{D;m,r\ppr\nu}^{\prime}\big)\;
\bigg(\frac{|\ovv{c}_{m}|\;\cosh\big(|\ovv{f}_{r\ppr}|\big)\;\sin\big(|\ovv{c}_{m}|\big)+
|\ovv{f}_{r\ppr}|\;\cos\big(|\ovv{c}_{m}|\big)\;\sinh\big(|\ovv{f}_{r\ppr}|\big)}
{|\ovv{c}_{m}|^{2}+|\ovv{f}_{r\ppr}|^{2}}\bigg) + \\  \no &-&
\big(\wtilde{\pp}_{i}\hat{\eta}_{D;m,r\ppr\lambda}^{\prime *}\big)\;\big(\tau_{2}\big)_{\lambda\nu}\;
e^{\im\;(\varphi_{m}+\phi_{r\ppr})}\;
\bigg(\frac{|\ovv{f}_{r\ppr}|\;\cosh\big(|\ovv{f}_{r\ppr}|\big)\;\sin\big(|\ovv{c}_{m}|\big)-
|\ovv{c}_{m}|\;\cos\big(|\ovv{c}_{m}|\big)\;\sinh\big(|\ovv{f}_{r\ppr}|\big)}
{|\ovv{c}_{m}|^{2}+|\ovv{f}_{r\ppr}|^{2}}\bigg)_{\mbox{.}}
\eeq

\end{appendix}


\end{document}